\documentclass[twoside,fleqn,11pt,a4paper]{report}
\usepackage[latin1]{inputenc}
\usepackage[T1]{fontenc}
\usepackage{times}
\usepackage{fancyhdr}
\usepackage{makeidx}
\usepackage{graphicx}
\usepackage{amsmath}
\usepackage{amsfonts}
\usepackage{amssymb}
\usepackage{wasysym}
\usepackage{bm}
\usepackage{enumerate}
\usepackage{ifthen}
\usepackage[noautoscale,vcentermath]{style/youngtab}
\usepackage{style/diss_english}
\usepackage{style/time}
\usepackage[dvips]{color}
\usepackage[font=small,labelfont=bf,justification=raggedright,labelsep=newline]{caption}
\definecolor{darkgreen}{rgb}{0,0.5,0}
\usepackage[OT2,OT1]{fontenc}
\newcommand{\cyr}{%
\renewcommand\rmdefault{wncyr}%
\renewcommand\sfdefault{wncyss}%
\renewcommand\encodingdefault{OT2}%
\normalfont\selectfont}
\DeclareTextFontCommand {\textcyr}{\cyr}

\renewcommand{\eqref}[1]{(\ref{EQ:#1})}

\graphicspath{{./}}
\makeindex

\begin{document}
%
%
%
%
%
%
%
%
%
%
\pagenumbering{roman}
\thispagestyle{empty}
\vspace*{3\baselineskip}

\begin{center}

\noindent
\begin{Huge}
\textbf{
Jet Propagation\\
and Mach-Cone Formation\\ 
in (3+1)-dimensional Ideal Hydrodynamics \\
}
\end{Huge}

\vspace*{4\baselineskip}

\noindent
\begin{Large}
Dissertation\\
zur Erlangung des Doktorgrades\\
der Naturwissenschaften\\
\end{Large}
\vspace*{3\baselineskip}

\noindent
\begin{Large}
vorgelegt beim Fachbereich Physik\\
der Johann Wolfgang Goethe--Universität\\
in Frankfurt am Main\\
\end{Large}
\vspace*{3\baselineskip}

\noindent
\begin{Large}
von\\
Barbara Betz\\
aus Hanau am Main\\
\end{Large}
\vspace*{3\baselineskip}

\noindent
\begin{Large}
Frankfurt am Main 2009\\
(D30)
\end{Large}

\end{center}
\clearpage

%
%
%
%
%
\thispagestyle{empty}
\mbox{}
\vspace*{24\baselineskip}
\vfill 

\begin{large}
\noindent
vom Fachbereich Physik der Johann Wolfgang Goethe--Universität\\
in Frankfurt am Main als Dissertation angenommen\\

\end{large}
\vspace*{3\baselineskip}

\begin{large}
\noindent
\begin{tabular}{rl}
\parbox[t]{0.32\linewidth}{\mbox{} \hfill Dekan}                   &
\parbox[t]{0.80\linewidth}{Prof.\ Dr.\ Dirk H.\ Rischke} \\[1.5em]
\parbox[t]{0.32\linewidth}{\mbox{} \hfill Gutachter}               & 
\parbox[t]{0.80\linewidth}{Prof.\ Dr.\ Dirk H.\ Rischke, Prof.\ Dr.\ Horst Stöcker } \\[1.5em]
\parbox[t]{0.32\linewidth}{\mbox{} \hfill Datum der Disputation }  & 13. Oktober 2009 \\[1.5em]
\end{tabular}\\
\end{large}
\vspace*{3\baselineskip}
\newpage
%

%
%
%

%
%
%
%
%
%
%
%
\thispagestyle{empty}
\vspace*{4\baselineskip}
\vspace*{7cm}
\hspace*{1.5cm}
\begin{minipage}[t]{8cm}
Nicht [die] Kunst und Wissenschaft allein,\\
Geduld will bei dem Werke sein.\\
{---------------------------------------------}\\
{\small {\it Faust I, Hexenküche, 1808}\\
Johann Wolfgang von Goethe (1749-1832)}\\[0.5cm]~
\end{minipage}

\clearpage

\clearpage{\pagestyle{empty}\cleardoublepage}
%
%
%
\pagenumbering{arabic}
\setcounter{page}{1}
\markboth{Dissertation Barbara Betz -- Jet Propagation in einem hydrodynamischen Modell}{Dissertation Barbara Betz -- Jet Propagation in einem hydrodynamischen Modell}
%
%
%
\vspace*{2.6\baselineskip}

\noindent
{\huge\bfseries Übersicht}
\vspace*{2\baselineskip}

\noindent
Diese Arbeit untersucht Jet-Medium-Wechselwirkungen in einem 
Quark-Gluon-Plasma mittels eines hydrodynamischen Modells. Ein solches Quark-Gluon-Plas\-ma 
repräsentiert eine Frühphase unseres Universums und kann in Schwerionenkollisionen 
erzeugt werden. Seine Eigenschaften sind Gegenstand der aktuellen For\-schung. Da 
der Vergleich von Meßdaten und Modellrechnungen nahelegt, dass sich das 
Quark-Gluon-Plasma wie eine nahezu ideale Flüssigkeit verhält, läßt sich das bei einer 
Schwerionenkollision gebildete Medium mittels hydrodynamischer Simulationen beschreiben. Eine 
der in die\-sem Zusammenhang grundlegenden Fragestellungen ist, ob energiereiche Teilchen 
(sogenannte Jets), die zu Beginn einer Kollision erzeugt werden und das Medium durchqueren, zur 
Bildung eines Machkegels führen. Dieser kann theoretisch immer erwartet werden, wenn sich ein Jet 
mit Überschallgeschwindigkeit relativ zum Medium bewegt. Die gemessene Winkelverteilung der aus 
der Kollision hervorgehenden und in den Detektoren gemessenen Teilchen sollte dann eine 
charakteristische Struktur aufweisen, aus der man auf direktem Wege Rückschlüsse auf die 
Zustandsgleichung des Mediums, im Besonderen auf seine Schallgeschwindigkeit, ziehen kann. 
Es werden unterschiedliche Szenarien eines Jetenergieverlustes betrachtet, dessen exakte Form 
und der ihm zugrundeliegenden Wechselwirkungen unbekannt sind. Dazu werden verschiedene Quellterme 
untersucht, die eine solche Wechselwirkung des Jets mit dem Medium repräsentieren und die Abgabe 
von Energie und Impuls an das Medium beschreiben. Dabei werden sowohl Mechanismen einer schwachen 
Wechselwirkung (basierend auf Rechnungen der perturbativen Quantenchromodynamik, pQCD) als auch 
einer starken Wechselwirkung (welche anhand der sogenannten ``Anti-de-Sitter/Conformal Field
Theory''-Korrespondenz, AdS/CFT, ermittelt wird) behandelt. Obwohl diese in unterschiedlichen 
Winkel\-verteilungen resultieren und somit (für Einzeljetereignisse) eine Unterscheidung der 
zugrundeliegenden Pro\-zesse ermöglichen könnten, zeigt sich, dass die für die 
gemessenen Teilchenspektren charakteristische Struktur durch die Überlagerung verschiedener
Jettrajektorien beschrieben werden kann. Eine solche Struktur lässt sich nicht direkt mit der 
Zustandsgleichung in Verbindung bringen. In diesem Zusammenhang werden die Auswirkungen 
eines star\-ken Flusses diskutiert, der sich bei nahezu allen betrachteten 
Jetenergieverlustszenarien entlang der Trajektorie des Jets bildet. Darüber hinaus werden 
die Transportgleichungen der dissipativen Hydrodynamik diskutiert, welche die Grundlage 
einer numerischen Berechung von viskosen Effekten innerhalb eines Quark-Gluon-Plasmas bilden.

\clearpage{\pagestyle{empty}\cleardoublepage}
\pagenumbering{roman}
\setcounter{page}{1}
\thispagestyle{empty}
\markboth{Zusammenfassung}{Zusammenfassung}
%
%
%
\chapter*{Zusammenfassung}
\renewcommand{\figurename}{Abbildung}

\section*{Einleitung}

Von jeher beschäftigte Menschen die Frage nach dem Ursprung des Lebens, 
dem Beginn des Universums und dem Aufbau der Materie. Letzterer wird durch das
sogenannte Standardmodell \cite{Glashow:1961tr,Weinberg:1967tq} 
beschrieben, das die Elementarteilchen und deren Wechselwirkungen 
zusammenfasst. Eine Vielzahl der uns heute umgebenden Materie (wie Protonen 
und Neutronen, aus denen beispielsweise die Atomkerne aufgebaut sind) setzt 
sich aus Quarks zusammen, die durch Gluonen miteinander wechselwirken. \\
Diese (starke) Wechselwirkung wird durch die Quantenchromodynamik (QCD) 
beschrieben, welche eine Besonderheit aufweist, die als ``asymptotische 
Freiheit'' \cite{Gross:1973id,Politzer:1973fx} 
bezeichnet wird: Für hohe Temperaturen und/oder Dichten 
nimmt die Stärke der Wechselwirkung zwischen den Quarks und Gluonen ab, 
so dass sich diese wie nahezu freie Teilchen verhalten und eine eigene 
Phase, das Quark-Gluon-Plasma \cite{Collins:1974ky,Freedman:1976ub,
Shuryak:1977ut}, bilden.\\
Eine solche heiße und dichte Phase existierte wahrscheinlich kurz nach dem 
Ur\-knall, bevor im Prozeß der Ausdehnung und Abkühlung des Universums die Quarks 
und Gluonen zu Teilchen rekombinierten. Man nimmt an, dass das Quark-Gluon-Plasma 
heutzutage im Inneren von dichten Neutronensternen vorhanden 
ist.\\
Bereits in den 1960er Jahren wurde die Möglichkeit diskutiert, Materie 
experimentell unter extremen Bedingungen zu untersuchen \cite{Hofmann:1975by}. Daraus entwickelte 
sich einer der spannendsten Forschungsbereiche der modernen Physik. \\
Hochenergeti\-sche, relativistische Schwerionenkollisionen eröffnen die 
einzigartige Möglichkeit, sehr heiße und dichte Materie im Labor zu erzeugen.
Dabei geht es nicht nur um den eindeutigen Nachweise des Quark-Gluon-Plasmas, der 
die Theorie der Quantenchromodynamik und somit das Standardmodell bestätigen würde, 
sondern auch darum, die Eigenschaften jener Phase mit experimentellen 
Observablen in Verbindung zu bringen.\\
Obwohl die Erzeugung eines Quark-Gluon-Plasmas sowohl bei Messungen am Super Proton 
Synchrotron (SPS, CERN) als auch am Relativistic Heavy Ion Collider (RHIC, Brookhaven 
National Laboratory) offiziell erklärt wurde \cite{CERN,Arsene:2004fa,
Adcox:2004mh,Back:2004je,Adams:2005dq}, bleiben seine Eigenschaften weiterhin 
umstritten, da ein solch experimentell erzeugtes Quark-Gluon-Plasma nur für sehr kurze 
Zeit existiert. In naher Zukunft werden große Beschleunigeranlagen wie der Large Hadron
Collider (LHC, CERN) und die Facility for Antiproton and Ion Research (FAIR, GSI) Materie 
bei noch höheren Energien bzw. Dichten analysieren.\\[2mm]~
Eines der herausragenden Ergebnisse des RHIC-Programms war der Nachweis, dass sich das 
erzeugte Medium annähernd wie eine ideale Flüssigkeit verhält
\cite{Arsene:2004fa,Adcox:2004mh,Back:2004je,Adams:2005dq,Shuryak:2003xe}. Damit scheint 
es gerechtfertigt, hydrodynamische Modelle zur Beschreibung des in einer 
Schwerionenkollision gebildeten Mediums zu verwenden, dessen Eigenschaften sich anhand 
verschiedener Proben (Sonden) ermitteln lassen. \\
Man nimmt an, dass zu Beginn der Kollision energiereiche Teilchen, die sog.\ harten 
Sonden, gebildet werden, welche das sich ausbildende Plasma durchqueren. Dabei 
induzieren sie Teilchenschauer, die Jets genannt werden, und geben Energie sowie Impuls 
an das Medium ab.\\
Durch die Wechselwirkungen des Jets mit dem Medium erhofft man sich Rück\-schlüsse 
auf das Medium ziehen zu können. Unter anderem geht man davon aus, dass ein sich mit 
Überschallgeschwindigkeit relativ zum Plasma bewegender Jet die Bildung eines Machkegels 
hervorruft, welcher wiederum eine charakteristi\-sche Struktur in der Winkelverteilung der durch 
die Kollision gebildeten Teilchen bewirkt. Da diese Struktur direkt mit der 
Schallgeschwindigkeit des Mediums in Verbindung gebracht werden kann, erwartet man, 
anhand entsprechender Messungen nähere Informationen über die Zustandsgleichung des 
Quark-Gluon-Plasmas ableiten zu können.\\
Diese Arbeit untersucht die Propagation von Jets in einem hydrodynamischen Medi\-um für 
unterschiedliche Szenarien der Energieabgabe, wobei sowohl starke als auch schwache 
Wechselwirkungen zwischen Jet und Medium betrachtet werden. Es zeigt sich, dass die 
resultierenden Winkelverteilungen verschiedenen Einflüssen der Kollision unterworfen sind.

\section*{Theoretischer Hintergrund}
\subsection*{Hydrodynamik}

Die Beschreibung der dynamischen Prozesse von Schwerionenkollisionen mit Hilfe von 
hydrodynamischen Modellen hat eine fast 30-jährige Tradition, da die Zu\-standsgleichung 
eine der wenigen essentiell benötigten Informationen darstellt und man somit leicht 
Eigenschaften verschiedener Medien (mit unterschiedlichen Zustandsgleichungen) überprüfen kann. 
Allerdings muss zunächst ein Anfangszu\-stand festgelegt werden, der jedoch mit erheblichen 
theoretischen Unsicherheiten verbunden ist. Es werden sowohl geometrische Ansätze (wie 
das Glauber-Modell, das in der folgenden Betrachtung für die Beschreibung des sich 
expandierenden Mediums angewendet wird) als auch Plasmainstabilitäts-Modelle oder das 
soge\-nannte {\it Colour-Glass Condensate} verwendet.\\
Relativistische Hydrodynamik bedeutet die (lokale) Erhaltung von Energie und Impuls 
(repräsentiert durch den Energie-Impuls-Tensor $T^{\mu\nu}$) sowie der Ladung 
(bzw.\ Baryonenzahl, ausgedrückt durch die Ladungsdichte $N$)
\begin{eqnarray}
\partial_\mu T^{\mu\nu}=0,\hspace*{1cm}
\partial_\mu N^\mu = 0\,.
\end{eqnarray}
Falls sich die Materie im lokalen thermodynamischen Gleichgewicht befindet, lassen sich der 
Energie-Impuls-Tensor und die erhaltene Ladung durch
\begin{eqnarray}
T^{\mu\nu}=(\varepsilon + p)u^\mu u^\nu -pg^{\mu\nu},\hspace*{1cm}
N^\mu = nu^\mu
\end{eqnarray}
ausdrücken und hängen nur von der Energiedichte $\varepsilon$, dem Druck $p$, der Ladungsdichte 
$n$, der Vierergeschwindigkeit des Mediums $u^\mu$ und dem metrischen Tensor 
$g^{\mu \nu}={\rm diag}(+,-,-,-)$ ab, wobei sich alle Größen auf das Ruhesystem des Mediums 
beziehen. Um dieses System der Bewegungsgleichungen zu schließen, benötigt man eine 
Zustandsgleichung der Form
\begin{eqnarray}
p = p(\varepsilon,n)\,.
\end{eqnarray}
Anhand dieser Gleichungen wird die Dynamik der Kollision eindeutig aus dem Anfangszustand 
bestimmt. Hydrodynamik stellt somit eine direkte Verbindung zwischen der Zustandsgleichung 
und dem expandierenden Medium dar, die sich auch im emittierten Teilchenspektrum manifestiert.\\
Bei numerischen Simulationen werden die Erhaltungsgleichungen allerdings meist ins Laborsystem
transformiert und in diskretisierter Version mit unterschiedlichen Algorithmen gelöst. Einer 
davon ist SHASTA (SHarp And Smooth Transport Algorithm), der sich besonders 
gut zur Beschreibung von Stoßwellenphänomenen eignet, zu deren Klasse auch die Machkegel gehören,
und der im Folgenden Verwendung findet.\\
Hydrodynamische Simulationen bestimmen die zeitliche Entwicklung von Feldern wie dem Temperatur- 
oder Geschwindigkeitsfeld. Um das Resultat einer solchen hydrodynamischen Rechnung mit 
experimentellen Daten vergleichen zu können, benötigt es eine Beschreibung zur Umwandlung der 
Flüssigkeit in Teilchen. Somit muss die hydrodynamische Konfiguration in ein Emissionsspektrum 
übersetzt werden, wobei die Erhaltung beispielsweise der Energie, des Impulses und der 
Teil\-chenzahl (sowie aller anderen Quantenzahlen) garantiert sein muss.\\
Ein häufig verwendeter Ansatz ist der sogenannte Cooper-Frye Ausfrierprozeß \cite{Cooper:1974mv}. 
Zunächst wird anhand eines bestimmten Kriteriums (wie etwa einer kritischen Zeit oder einer 
kritischen Temperatur $T_c$) der Ausfrierpunkt auf einer Hyperfläche $\Sigma$ der Raum-Zeit bestimmt. 
Mit einer thermischen Verteilungsfunktion $f$, welche die Teilchenweltlinien zählt, die durch 
$\Sigma$ hindurchtreten, kann das Emissionsspektrum über
\begin{eqnarray}
E\frac{dN}{d^3p} = \int_{\Sigma} d\Sigma_\mu\, p^\mu\, f(u\cdot p/T)
\label{EqZusCF}
\end{eqnarray}
bestimmt werden. Allerdings bedeutet diese Methode, dass das Flüssigkeitsfeld instantan in freie 
Teilchen umgewandelt wird. Da jadoch die Viskosität während der letzten Entwicklungsstufen des 
sich expandierenden Medium zunimmt, führt diese direkte Umwandlung möglicherweise zu 
unphysikalischen Artefakten. Sie ist trotzdem eine anerkannte Methode, da die Entkopplungszeiten 
für die Flüssigkeit und somit für die Umwandlung in Teilchen unbekannt sind.

\subsection*{Viskosität}

Wie oben bereits erwähnt, verhält sich das in einer Schwerionenkollision am RHIC gebildete 
Medium wie eine nahezu ideale Flüssigkeit. Während ideale Hydrodynamik einen kontinuierlichen 
Anstieg der Anisotropien in der Winkelverteil\-ung (dem sogenannten Elliptischen Fluß) mit 
zunehmendem Transversalimpuls ($p_T$) der erzeugten Teilchen vor\-her\-sagen, flachen die 
gemessenen Werte bei ungefähr $p_T\approx 1.5 - 2$~GeV ab, was auf viskose Effekte 
zurückgeführt werden konnte. Da diese Anisotropien in peripheren 
(d.h.\ nicht-zentralen Stößen) und für geringere Kollisionsenergien stärker ausgeprägt sind, 
müssen sie in realistischen Simulationen mit einbezogen werden.\\
In diesem Fall führt die Tensorzerlegung des Energie-Impulstensors $T^{\mu\nu}$ und der 
Ladung $N$ auf
\begin{eqnarray}
T^{\mu\nu} = \varepsilon u^\mu u^\nu - (p+\Pi) \Delta^{\mu\nu} + q^{\mu}u^{\nu} 
+ q^{\nu}u^{\mu} + \pi^{\mu\nu},\hspace*{1cm} N^\mu = nu^\mu + V^\mu\,,
\end{eqnarray}
wobei $\Pi$ den isotropen viskosen Druck ({\it bulk pressure}), $q^\mu$ den Wärmefluß, 
$\pi^{\mu\nu}$ den Schertensor und $V^\mu$ den Diffusionsstrom bezeichnen. 
$\Delta^{\mu\nu}=g^{\mu\nu}-u^\mu u^\nu$ beschreibt eine Projektion auf einen 
$3$-dimen\-siona\-len Unterraum, orthogonal zu $u^\mu$. Mit Hilfe dieser Ausdrücke werden 
erstmals ausgehend von der Boltzmann Gleichung für ein verdünntes Gas
\begin{eqnarray}
k^\mu\partial_\mu f(k,x)&=&C[f]\,,
\end{eqnarray} 
wobei $C$ den Kollisionsterm und $k^\mu$ den Impuls beschreibt, die Transportgleichungen 
bis zur zweiten Ordnung in den auftretenden Gradienten für den isotropen viskosen Druck, den 
Wärmefluß und den Schertensor abgeleitet. Diese sind von grundlegender Bedeutung 
für jede numerische Simulation viskoser Hydrodynamik. 

\subsection*{Jet-Energieverlust und Hydrodynamik}

Aufgrund von Energie- und Impulserhaltung sind die in der Frühphase einer Schwer\-ionenkollision 
gebildeten Jets eigentlich ein System bestehend aus zwei Jets (Di-Jet), die sich in 
entgegengesetzte Richtung bewegen. Man nimmt an, dass sich solche Di-Jets vorwiegend am Rande 
des Medims bilden und einer der beiden Jets (der sogenannte Trigger-Jet) das Medium ohne 
weitere Wechselwirkung verlassen kann, während der andere (der assoziierte oder {\it away-side} Jet) 
die während der Kollision entstandene Materie durchquert und dabei entlang seines Weges Energie 
und Impuls an das Medium abgibt. \\
Eine Messung \cite{Adams:2005ph} beweist, dass sich der mittlere Impuls der emittierten 
Teilchen auf der Seite des Mediums, das der Jet durchquert hat, zumindest für nicht allzu 
periphere Kollisionen dem Wert für ein Medium im thermischen Gleichgewicht annähert. Deshalb 
dürfte die vom Jet in das Medium deponierte Energie schnell thermalisieren, weshalb man die 
vom Jet hervorgerufene Störung hydrodynamisch beschreiben kann.\\
Dazu wird ein Quellterm $J^\nu$ in das System von Energie- und Impulserhaltungsglei\-chungen 
eingebaut,
\begin{eqnarray}
\partial_\mu T^{\mu\nu} =J^\nu\,,
\end{eqnarray}
so dass die zeitliche Entwicklung des vom Jet durchdrungenen Mediums bestimmt werden kann. 
Allerdings ist dieser Quellterm, $J^\mu$, nicht der Quellterm des Jets, sondern das Residuum 
an Energie und Impuls, das der Jet an das Medium abgibt.\\ 
Die Ableitung eines Quellterms, welcher die Wechselwirkung des Jets mit dem Medium
korrekt wiedergibt, ist Gegenstand der aktuellen Forschung, wobei sowohl eine schwache 
Wechselwirkung (die mittels perturbativer Quantenchromodynamik, pQCD, beschrieben werden kann) 
als auch eine starke Wechselwirkung (welche im Rahmen der Anti-de-Sitter/Conformal Field 
Theory-Korrespondenz, AdS/CFT, formuliert ist) zugrundegelegt wird. Neben diesen beiden Arten der 
Jet-Medium-Wechselwirkung wird zudem noch ein schematischer Quellterm betrachtet, der die 
Abgabe von Energie und Impuls entlang einer Trajektorie durch das Medium beschreibt. \\
Aus dem gewonnenen Teilchenspektrum werden jeweils Winkelkorrelationen be\-stimmt, die den
Relativwinkel zwischen Trigger- und einem assoziierten Teilchen beschreiben. Ein solches 
Spektrum sollte im Fall der Ausbildung eines Machkegels zwei Maxima auf der dem 
Triggerteilchen gegenüberliegenden Seite (der Away-side) aufweisen.

\section*{Ergebnisse}

\subsection*{Das statische Medium}

\begin{figure}[t]
\hspace*{-1.0cm}\vspace*{2ex}
\includegraphics[width=15cm]{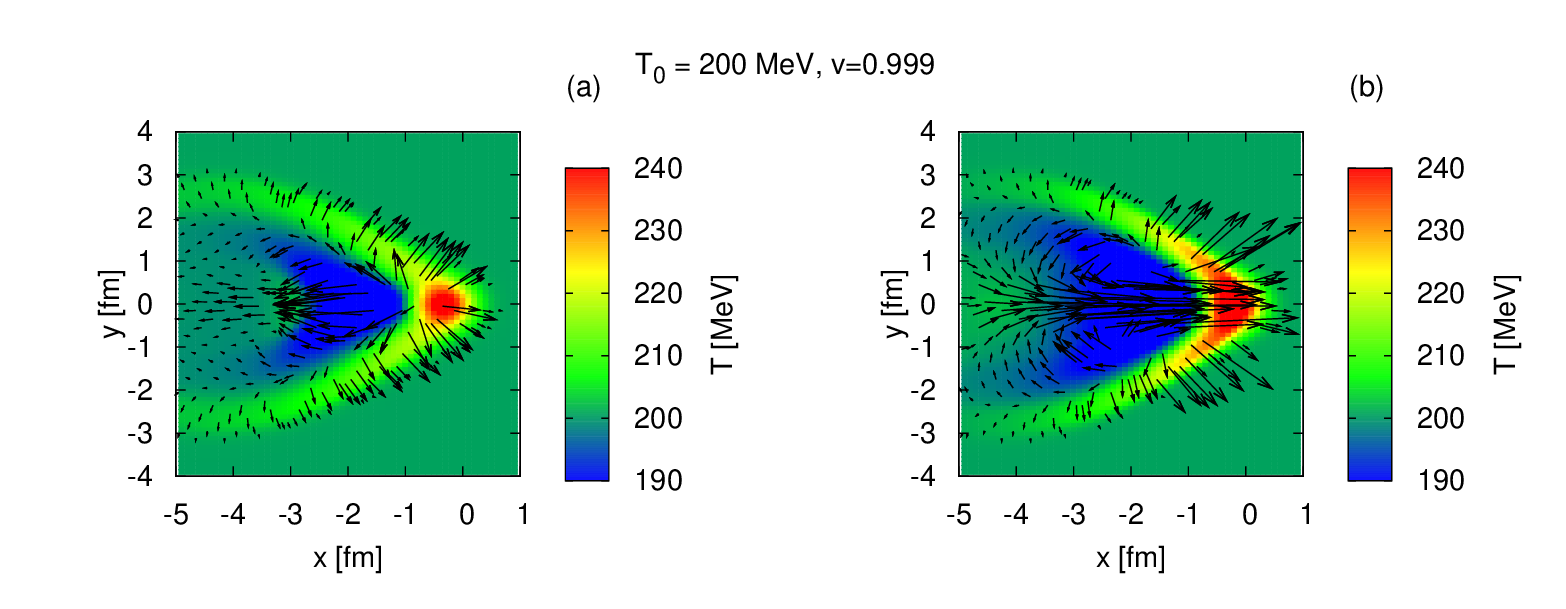}
\caption*
{{\bf Abbildung 1:} Temperatur- und Geschwindigkeitsprofil (Pfeile) im Fall einer 
reinen Energieabgabe (links) und für ein Szenario mit Energie- und Impulsabgabe 
(rechts) für einen Jet, der das Medium mit einer konstanten Geschwindigkeit nahe 
der Lichtgeschwindigkeit durchschlägt.}\vspace*{0.5cm}
\label{FigZus1}
\end{figure}

Zunächst wird ein statisches Medium für verschiedene Szenarien der Energie- und Impulsabgabe 
betrachtet. Da die exakte Form eines entsprechenden Quellterms, der die 
Jet-Medium-Wechselwirkung im Quark-Gluon-Plasma beschreibt, unbekannt ist, 
wird hierbei sowohl auf die Möglichkeit des reinen Energieverlustes 
(welcher unwahrscheinlich ist, jedoch nicht ausgeschlossen werden kann) als auch des 
gleich starken Energie- und Impulsübertrages eingegangen. \\
Es zeigt sich, dass in beiden Fällen eine kegelartige Struktur ausgebildet wird, wie der 
Abbildung 1 zu entnehmen ist. Wird jedoch nicht nur Energie, sondern auch Impuls an das 
Medium abgegeben, so entsteht ein starker Fluß entlang der Trajektorie des Jets, den 
man als {\it Diffusion Wake} (Diffusionswelle) bezeichnet. \\
Erfolgt der Ausfrierprozeß, d.h.\ die Umwandlung der Flüssigkeitsfelder in Teilchen, 
gemäß der Cooper-Frye Formel [vgl.\ Gl.\ (\ref{EqZusCF})], so führt dieser starke Fluß 
zur Aus\-prägung eines Maximums auf der {\it Away-side}. Für den Prozeß der reinen Energieabgabe 
hingegen tritt eine Zweipeak-Struktur auf wie sie im Fall der Ausprägung eines Machkegels 
zu erwarten ist und in den aus experimentellen Daten gewonnenen (und veröffentlichten) 
Winkelverteilungen auftritt.
Diese Ergebnisse stimmen mit früheren Untersuchungen \cite{CasalderreySolana:2004qm} 
unter Verwendung linearisierter Hydrodynamik über\-ein. Inwieweit sie allerdings auf ein 
expandierendes Medium übertragen werden können, läßt sich erst nach einer Betrachtung 
eines solchen Systems sagen (siehe unten). Auf die Bedeutung der {\it Diffusion Wake} wird 
im Folgenden noch mehrfach eingegangen.\\
Darüber hinaus konnte gezeigt werden, dass diese Ergebnisse unabhängig davon sind, 
ob der Impuls longitudinal oder transversal zur Jetrichtung abgegeben wird und ob 
dieser Jet im Medium thermalisiert wird (also stoppt) oder durch das Medi\-um 
durchschlägt. Hierbei wurde die Rückreaktion des Mediums bei Abbremsung eines Jets 
durch ein einfaches Bethe--Bloch-Modell beschrieben, welches zur Ausprägung eines 
sogenannten {\it Bragg-Peaks} führt, also eines Maximums der Energie- und Impulsabgabe 
kurz vor dem Endpunkt der Jettrajektorie.

\subsection*{Teilchenkorrelationen in pQCD und AdS/CFT}

\begin{figure}[t]
\centering
  \includegraphics[scale = 0.71]{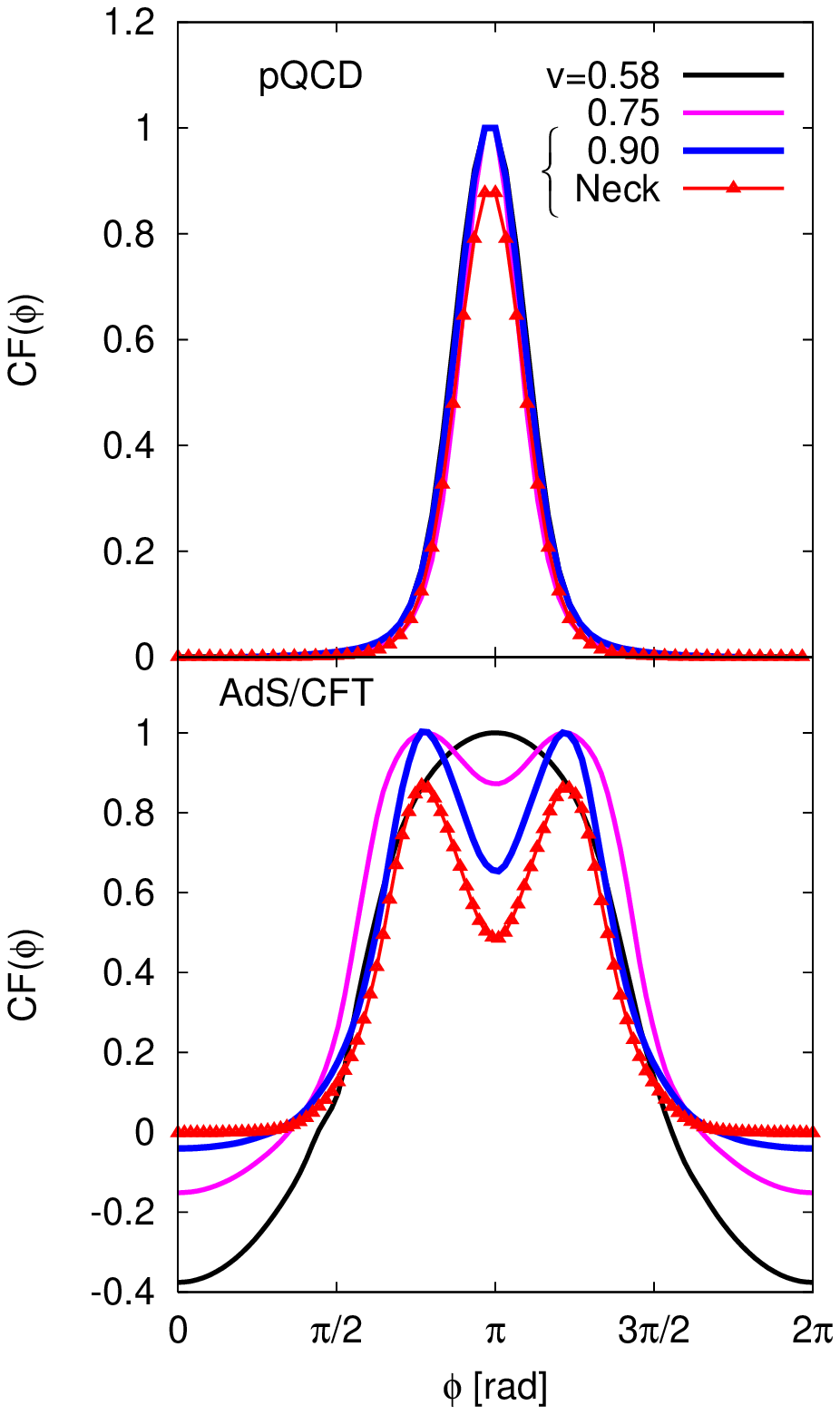}
  \caption*
  {{\bf Abbildung 2:} Normierte Winkelkorrelationen nach Hintergrundsubtraktion, 
  erhalten gemäß dem Cooper-Frye Ausfrierprozeß für einen Quellterm der pQCD 
  (oben) und der AdS/CFT (unten) für $p_T\sim 3.14$~GeV, verschwindende
  Rapiditäten ($y=0$) und verschiedene Geschwindigkeiten des Jets. Die rote 
  Linie mit Dreiecken repräsentiert den Beitrag der sog.\ {\it Neck region} für 
  einen Jet mit $v=0.9$~c.}\vspace*{-2ex}
  \label{FigZus2}
\end{figure}

Bei dem oben betrachteten Quellterm handelt es sich um eine schematische 
Be\-schrei\-bung eines Jetenergie- und Impulsverlustes. Es wurden jedoch auch 
Jet-Me\-dium-Wechselwirkungen sowohl im Rahmen der perturbativen
Quantenchromodynamik (pQCD) \cite{Neufeld:2008fi} als auch unter Anwendung der 
Anti-de-Sitter/Conformal Field Theory-Korrespondenz (AdS/CFT) abgeleitet 
\cite{Gubser:2007ga} und untersucht \cite{Noronha:2008un}.
Diese beiden vollkommen unabhängigen Ansätze beziehen sich auf eine 
schwache bzw.\ starke Wechselwirkung.\\
In der vorliegenden Arbeit konnte durch Vergleich der resultierenden 
Winkelkorrelationen gezeigt werden, dass die experimentelle Bestimmung einer 
Zweiteilchen\-korrelation für einzelne, identifizierte, schwere Quarks am RHIC oder LHC 
die Möglichkeit bieten könnte, eine Aussage über die Stärke der Wechselwirkung zu 
machen, da die berechneten Winkelverteilungen, wie man Abbildung 2 entnehmen kann, 
unterschiedliche Strukturen aufweisen.\\
Hierzu wurde der pQCD-Quellterm in die hydrodynamische Simulation eingebaut 
und das Ergebnis einer AdS/CFT-Rechnung gegenübergestellt (siehe Abbildung 2). 
Man sieht deutlich, dass für den pQCD-Quellterm ein Maximum in der 
dem Trigger-Jet entgegengesetzten Richtung auftritt, während die mittels 
AdS/CFT erhaltenen Korrelationen eine Doppelpeak-Struktur aufweist, 
welche jedoch nicht auf die Ausbildung eines Machkegels zurückgeführt 
werden kann, da für einen solchen Machkegel, im Gegensatz zu den erhaltenen 
Resultaten, die Lage der Maxima von der Ge\-schwin\-dig\-keit des Jets abhängt.\\
Stattdessen führt ein starker Transversalimpuls in der sogenannten 
{\it Neck region}, einem Bereich nahe dem Jet, zur Ausprägung dieser 
Doppelpeak-Struktur. Obwohl in beiden Fällen sowohl eine Machkegel-artige 
Struktur als auch eine {\it Diffusion Wake} auftritt, wird die Winkelkorrelation 
letztendlich eindeutig von dieser {\it Neck region} dominiert.

\subsection*{Das expandierende Medium}

\begin{figure}[t]
\centering
\begin{minipage}[c]{4.2cm}
\hspace*{-3.0cm}
  \includegraphics[scale = 0.58]{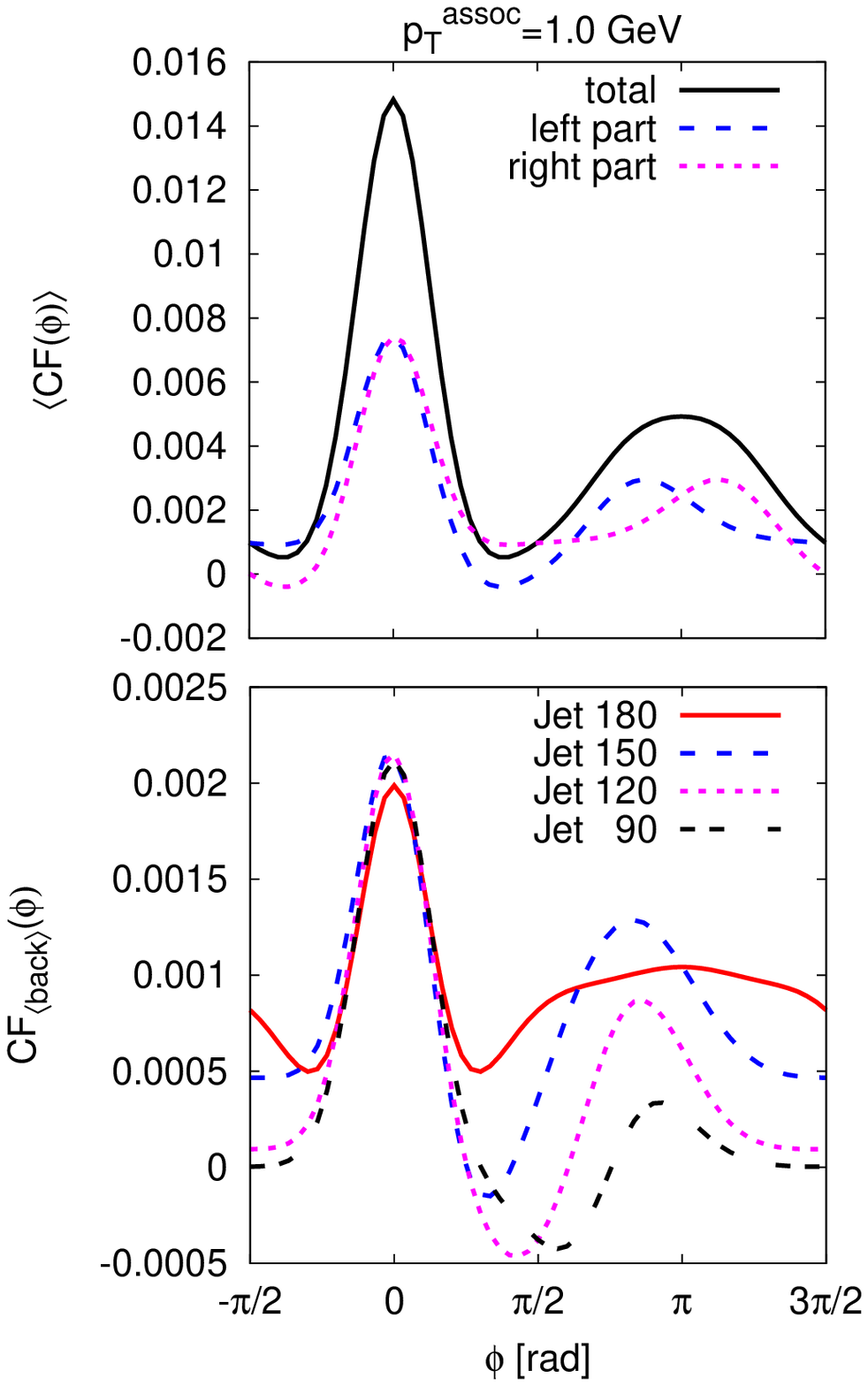}
\end{minipage}
\hspace*{-0.5cm}  
\begin{minipage}[c]{4.2cm} 
  \includegraphics[scale = 0.58]{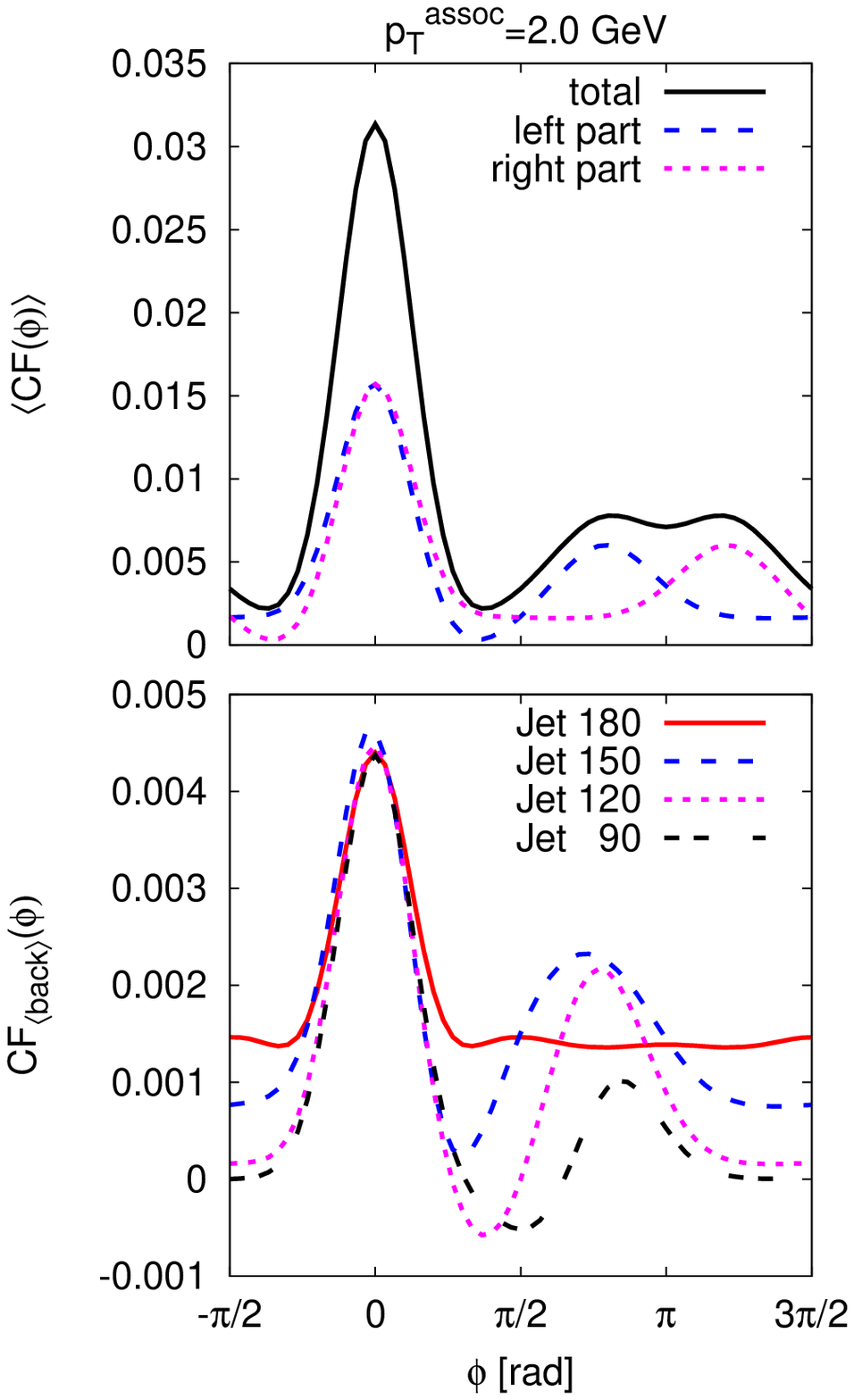}
\end{minipage}
  \caption*
  {{\bf Abbildung 3:} Normierte und über verschiedene Jettrajektorien gemittelte 
  Zweiteilchenkorrelationen nach einem Cooper-Frye Ausfrierprozeß (durchgezogene
  schwarze Linie) für einen Jet, der insgesamt $5$~GeV an Energie und Impuls an das 
  Medium abgibt, wobei für die assoziierten Teilchen ein Transversalimpuls von 
  $p_T^{\rm assoc}=1$~GeV (links) und $p_T^{\rm assoc}=2$~GeV (rechts) angenommen 
  wird. Die gestrichelten Linien (blau und violett) repräsentieren dabei die 
  gemittelten Beiträge von verschiedenen Jets,   unterschieden durch den 
  Azimuthalwinkel ihres Anfangspunktes. Die Beiträge für vier  verschiedene 
  Jettrajektoren, deren Mittelung die blau gestrichelte Linie ergibt,
  sind im unteren Teil der Abbildung dargestellt. Dabei repräsentiert 'Jet 180' 
  denjenigen Jet, der das Medium mittig durchquert.}\vspace*{2ex}
  \label{FigZus3}
\end{figure}

Die realistische Beschreibung einer Schwerionenkollision erfordert natürlich ein expandierendes 
Medium. Zwar ist es unwahrscheinlich (wenn auch nicht un\-mög\-lich), dass während eines sogenannten {\it Events}
(also einer Kollision) mehrere Jets gleichzeitig die bei dem Stoß gebildete Materie durchqueren,
die experimentell ermittelten Zweiteilchenkorrelationen werden jedoch in jedem Fall über mehrere
{\it Events} gemittelt. Somit werden unterschiedliche Jettrajektorien in Betracht gezogen.\\
Wie sich herausstellt (vgl.\ Abbildung 3), führen die Einzelbeiträge verschiedener Jettrajektorien
(siehe Abbildung 3 unten) zu einer Substruktur auf der {\it Away-side} (ge\-strich\-el\-te Linien in den
oberen Abbildungen), welche entweder zu einem breiten Maximum oder einer ursprünglich durch die 
Bildung eines Machkegels erwarteten Doppelpeak-Struktur führen. \\
Dies legt die Schlußfolgerung nahe, dass die experimentell be\-obachte\-te Struktur durch einen 
Mittelungsprozeß verschiedener abgelenkter Jets entsteht und somit nicht direkt mit der 
Zustandsgleichung in Verbindung gebracht werden kann.\\
Es muss allerdings erwähnt werden, dass dies die Bildung eines Machkegels in Schwerionenkollisionen 
nicht ausschließt, jedoch dürfte dessen Emissionswinkel erst durch Untersuchung von 
Einzeljet-Events möglich werden.\\
Es konnte außerdem gezeigt werden, dass die Wechselwirkung zwischen radialem und durch die 
Bewegung des Jets hervorgerufenem Fluß zu einer Reduzierung der {\it Diffusion Wake} und einer Krümmung 
des (Mach-)Kegels führen kann, wie es in einer früheren Arbeit aus theoretischen Überlegungen 
abgeleitet wurde \cite{Satarov:2005mv}. Der Einfluß der {\it Diffusion Wake} auf das resultierende 
Teilchenspektrum bleibt jedoch erhalten, hängt aber von der Weglänge des Jets und somit von der 
Relativbewegung zwischen Jet und Medium ab. Wie man Abbildung 3 entnehmen kann, haben jedoch die 
nicht-zentralen Jets einen großen Einfluß auf die resultierende Struktur, wobei dieser vom 
Betrag und von der zugrundegelegten Energie- und Impulsverlustrate abhängt, deren Bestimmung 
somit, besonders in Abhängigkeit der abgegebenen Energie und der Geschwindigkeit des Jets, 
notwendig ist.\\
Durch eine Vergleichsrechnung, die wie im statischen Fall einen reinen Energie\-ver\-lust 
beschreibt, konnte gezeigt werden, dass dieses Jetenergieverlustszenario im Gegensatz zum 
statischen Fall (siehe oben) nicht zu einer experimentell beobachte\-ten Doppelpeak-Struktur führt und 
somit im Widerspruch zu den experimentellen Ergebnissen steht.

\subsection*{Schlußfolgerungen}

Die Struktur der experimentell gewonnenen Winkelverteilungen läßt sich durch die  
Bewegung von Jets durch ein hydrodynamisches Medium beschreiben. Allerdings muss dabei über 
verschiedene Jettrajektorien gemittelt und die Wechselwirkung mit einem expandierenden Medium 
berücksichtigt werden. Somit ist es von fundamentaler Bedeutung, die Wechselwirkung zwischen Jet 
und Medium mit nicht-linearer Hydrodynamik zu untersuchen. Der Einfluß verschiedener Effekte wie 
der der {\it Diffusion Wake} oder der sog.\ {\it Neck region} kann jedoch nur durch die Analy\-se 
eines statischen Mediums bestimmt werden.\\
Somit ist es nicht möglich, aus den bisher bestimmten Winkelverteilungen 
eindeutig auf die E\-mis\-sions\-winkel und somit auf die (gemittelte) Zustandsgleichung des in 
einer Schwer\-ionenkollision entstandenen Mediums zu schließen.\\
Dazu müßte man, wie es in naher Zukunft am RHIC und LHC möglich sein dürfte, die Winkelverteilung 
für einzelne (energiereiche) Jets extrahieren. Die kürzlich veröffentlichten Daten 
\cite{vanLeeuwen:2008pn,PHENIX_QM09_2} über die Veränderung des Kegelwinkels 
mit der Re\-aktions\-ebene versprechen allerdings bereits weitere Einsichten in die 
Phänome\-nologie der Machkegel in Schwerionenkollisionen, wenngleich neue Untersuchungen 
\cite{STAR_FullJet,PHENIX_FullJet}, die eine Rekonstruktion des Jets umfassen, 
aufzeigen, dass die der Jet-Medium-Wechselwirkung zugrundeliegenden Prozesse weiterer 
Untersuchungen bedürfen. Für den Vergleich der experimentell gemessenen Daten mit hydrodynamischen 
Simulationen ist eine solche Jetrekonstruktion unerläßlich, da im letzteren Fall per Definition 
ein bzw.\ mehrere Jets vorliegen.\\
Zudem bleibt das Verfahren der experimentellen Hintergrundsubtraktion und somit die Frage zu 
klären, ob bzw.\ wie stark die Bewegung des Jets Einfluß auf den Gesamtfluß des Mediums hat. 
Dieses kann und sollte mittels hydrodynamischer Untersuchungen analysiert werden.\\
Um die Beschreibung der Jet-Medium-Wechselwirkungen möglichst genau an die realistischen 
Gegebenheiten anzupassen, ist es weiterhing notwendig, longitudinale Expansion, nicht-zentrale 
Stöße und verschiedene Zustandsgleichungen zu betrachten. Außerdem könnte der Ausfrierprozeß 
bzw.\ die nach der Entstehung der Teilchen zwischen diesen auftretenden Wechselwirkungen noch 
Einfluß auf die Winkelverteilungen haben.\\
Allgemein gesehen bedarf es noch eines detaillierteren Verständnisses des Quellterms, dessen 
Bedeutung auch für die durch die verschiedenen Experimente (RHIC, LHC und FAIR) abgedeckten 
Energiebereiche ermittelt werden muss.

\renewcommand{\figurename}{Figure}

\clearpage{\pagestyle{empty}\cleardoublepage}
%
%
\markboth{Dissertation Barbara Betz -- Jet Propagation in Ideal Hydrodynamics}{Dissertation Barbara Betz -- Jet Propagation in Ideal Hydrodynamics}
%
%
%
\vspace*{2.6\baselineskip}

\noindent
{\huge\bfseries Abstract}
\vspace*{2\baselineskip}

\noindent
This thesis investigates the jet-medium interactions in a Quark-Gluon Plasma using a 
hydrodynamical model. Such a Quark-Gluon Plasma represents a very early stage of our 
universe and is assumed to be created in heavy-ion collisions. Its properties are 
subject of current research. Since the comparison of measured data to model calculations
suggests that the Quark-Gluon Plasma behaves like a nearly perfect liquid, 
the medium created in a heavy-ion collision can be described
applying hydrodynamical simulations. One of the crucial questions in this context is if 
highly energetic particles (so-called jets), which are produced at the beginning of the 
collision and traverse the formed medium, may lead to the creation of a Mach cone. 
Such a Mach cone is always expected to develop if a jet moves with a velocity larger 
than the speed of sound relative to the medium. In that case, the measured angular 
particle distributions are supposed to exhibit a characteristic structure allowing 
for direct conclusions about the Equation of State and in particular about the speed 
of sound of the medium. Several different scenarios of jet energy loss are examined 
(the exact form of which is not known from first principles) and different 
mechanisms of energy and momentum loss are analyzed, ranging from weak interactions 
(based on calculations from perturbative Quantum Chromodynamics, pQCD) to strong 
interactions (formulated using the Anti-de-Sitter/Conformal Field Theory 
Correspondence, AdS/CFT). Though they result in different angular particle correlations 
which could in principle allow to distinguish the underlying processes (if it becomes 
possible to analyze single-jet events), it is shown that the characteristic structure 
observed in experimental data can be obtained due to the different contributions of
several possible jet trajectories through an expanding medium. Such a structure cannot 
directly be connected to the Equation of State. In this context, the impact of a strong 
flow created behind the jet is examined which is common to almost all jet deposition 
scenarios. Besides that, the transport equations for dissipative hydrodynamics are 
discussed which are fundamental for any numerical computation of viscous effects in a 
Quark-Gluon Plasma. 

\clearpage{\pagestyle{empty}\cleardoublepage}
%
%
\thispagestyle{empty}
\tableofcontents
\clearpage{\pagestyle{empty}\cleardoublepage}
%
%
%
\thispagestyle{empty}
\listoffigures
\clearpage{\pagestyle{empty}\cleardoublepage}
\pagenumbering{arabic}
\setcounter{page}{1}
%
%
%
\part[Quark-Gluon Plasma and Heavy-Ion Collisions -- an overview]
{Quark-Gluon Plasma and Heavy-Ion Collisions -- an overview}
\label{part01}
\clearpage{\pagestyle{empty}\cleardoublepage}
%
%
\chapter[Introduction]{Introduction}
\label{Intro}

One of the common features to all civilizations is the development of 
(philosophical) models about the beginning of the universe, the origin
of live and the description of matter. However, testing
these paradigms was always (and is) a demanding problem and just at the beginning of 
the last century, it became possible to prove the existence of various types of 
(short-living) particles.\\
Based on those findings, the {\it standard model} was developed which goes back to the 
work of Glashow, Weinberg, and Salam \cite{Glashow:1961tr,Weinberg:1967tq} and characterizes the 
electroweak as well as strong interactions, but does not include gravity. Thus, the
standard model is not a complete theory of all fundamental interactions.\\
{\it Quantum Chromodynamics} (QCD), the theory of strong interactions, is part
of the standard model and exhibits a characteristic property, named
``asymptotic freedom'' \cite{Gross:1973id,Politzer:1973fx}: For high temperatures and/or densities, 
the strength of the interaction between the fundamental particles of QCD
(the quarks and gluons) decreases, allowing them 
to behave as nearly free particles and to form a particular state of matter, the
{\it Quark-Gluon Plasma} (QGP) \cite{Collins:1974ky,Freedman:1976ub,Shuryak:1977ut}. \\
Such extreme conditions probably existed shortly after the Big Bang at the very hot 
and early stages of the evolution of the universe before in the 
process of expansion and cooling quarks and gluons recombined into compound, 
color-neutral particles, the hadrons. It is assumed that nowadays the QGP is present 
in the cold, but dense inner core of neutron stars. \\
The idea to experimentally study matter under extreme conditions goes back to the 1960's
\cite{Hofmann:1975by} (which was even before QCD was introduced). Today, one of the most 
exciting and challenging research programs is to probe the QGP. Relativistic high-energy heavy-ion
collisions offer the unique possibility to study matter experimentally under extreme
conditions in the laboratory. A formidable problem is to conclusively identify QGP 
production and to relate its properties to experimental observables. An unambiguous evidence 
of the QGP implies a proof of QCD and thus of the standard model.\\
Both, the Super Proton Synchrotron (SPS) program at CERN\footnote{European Organization for 
Nuclear Research close to Geneva, the acrynom is derived from its former french name 
Conseil Européen pour la Recherche Nucléaire.} \cite{CERN} and the 
Relativistic Heavy Ion Collider (RHIC) program at 
BNL\footnote{Brookhaven National Laboratory, Long Island, USA.} 
\cite{Arsene:2004fa,Adcox:2004mh,Back:2004je,Adams:2005dq} 
have announced success in detection 
of the QGP. However, its fundamental properties are still discussed vividly which is
due to the fact that the lifetime of such an experimentally created QGP is extremely short,
making its detection highly challenging.
In the near future, large accelerator facilities, the Large Hadron Collider (LHC) at 
CERN and Facility of Antiproton and Ion Research (FAIR) at GSI\footnote{GSI Helmholtz
Centre for Heavy Ion Research GmbH in Darmstadt, Germany. The acrynom is also derived from 
its former name of Gesellschaft für Schwerionenforschung.}, 
will examine further the matter created in heavy-ion collisions, promising deeper 
insights into fundamental interactions of matter. 
\\
One of the major results of the RHIC program was to show that the medium created during the 
collision behaves like a fluid 
\cite{Arsene:2004fa,Adcox:2004mh,Back:2004je,Adams:2005dq,Shuryak:2003xe}. 
Thus, it seems justified to use hydrodynamic models for the description of the matter formed 
during a heavy-ion collision \cite{Kolb:2003dz,Romatschke:2007mq,Huovinen:2003fa}. 
Moreover, the fluid-like behaviour points to having created a 
new state of matter since normal hadronic matter is very viscous \cite{Muronga:2003tb}.\\
There exist different methods to probe the matter created in a heavy-ion collision. 
One of those are the so-called {\it hard probes} which are partons produced at an early 
stage of the evolution with a large transverse momentum ($p_T$)
that penetrate the system formed, inducing showers of particles ({\it jets}) and
depositing energy in the medium. The interaction of jets with the medium, which significantly 
affects the properties of the jets, are assumed to allow for conclusions about medium properties. 
Amongst other things, a hard-$p_T$ jet travelling through the medium supersonically is supposed 
to create a {\it Mach cone} with an enhanced particle emission into the distinct Mach cone
angle. Since this angle is connected to the speed of sound of the medium, it could
provide a constraint on the average speed of sound of the strongly coupled QGP (sQGP) 
\cite{Shuryak:2004kh}. \\
In this work, we investigate the propagation of a jet through a medium using $(3+1)$-dimensional
ideal hydrodynamics. First, we consider different energy-loss mechanisms in a static medium
and determine the azimuthal particle correlations to see if a conical structure has formed and 
if differences between various energy- and momentum-loss scenarios can be determined. 
Then, for comparison to experimental data, we study an expanding medium and discuss the 
importance of the convolution of different jets produced by experiment for the azimuthal particle 
correlations. \\
The thesis is organized as follows: In the first part, we give an overview
about the properties of matter, present possible signatures of the QGP, and introduce 
important experimental findings. Part \ref{part02} then focuses on the hydrodynamcial prescription 
of heavy-ion collisions. Having obtained a qualitative understanding of the system's evolution, 
we introduce the different energy- and momentum-loss mechanisms. The results
of our simulations are presented in part \ref{part03} for the static as well as the expanding 
medium. After a subsequent discussion, the thesis concludes with a summary and an 
outlook in part \ref{part04}. \\[2cm]~

\section[The Standard Model of Elementary Particle Physics]
{The Standard Model of Elementary Particle Physics}
\label{SecStandardModell}

\begin{table}[b]
\centering
\begin{tabular}{|c|c|c|c|}\hline
\multicolumn {3}{|c|}{particle} & charge \\\cline{1-4}
$\;\;\;e\;\;\;$ & $\;\;\;\mu\;\;\;$ & $\;\;\;\tau\;\;\;$ & -1\\
$\;\;\;\nu_e\;\;\;$ & $\;\;\;\nu_\mu\;\;\;$ & $\;\;\;\nu_\tau\;\;\;$ & 0\\\hline
$\;\;\;u\;\;\;$ & $\;\;\;c\;\;\;$ & $\;\;\;t\;\;\;$ & +1/3\\
$\;\;\;d\;\;\;$ & $\;\;\;s\;\;\;$ & $\;\;\;b\;\;\;$ & -2/3\\\hline
\end{tabular}
\caption{Building blocks of matter in the standard model. Matter consists of
fermions that can be classified into three families of equal electric charge, but
different masses. The mass increases from the left to the right.}
\label{tabparticles}
%
\vspace*{0.4cm}
\begin{tabular}{|c|c|c|c|}\hline
gravitation& weak & electromagnetic & strong \\\cline{1-4}
graviton & $Z^0, W^{\pm}$ & $\gamma$ & g\\\hline
\end{tabular}
\caption{The gauge bosons of the standard model, mediating the four fundamental
interactions.}
\label{tabgaugebosons}
\end{table}

The standard model of elementary particle physics \cite{Glashow:1961tr,Weinberg:1967tq} 
describes the elementary particles, {\it quarks} and {\it leptons}\footnote{Quarks and leptons 
are fermions, i.e., spin $1/2$-particles.}, and their interactions mediated by the corresponding 
gauge bosons\footnote{Gauge bosons are spin $1$-particles.}. There exist six different quarks 
({\it up, down, charm, strange, top, bottom}) and six different lepons ({\it electron, muon, 
tauon, electron-neutrino, muon-neutrino, tauon-neutrino}) as well as four gauge bosons 
({\it graviton\footnote{Though gravitation is not incorporated into the standard model, it is 
commonly also mentioned in this context.}, $Z^0/ W^{\pm}$ bosons, photon, gluons}). \\
Quarks and leptons, just like their antiparticles, can be combined to three doublets due to their 
mass as seen in table \ref{tabparticles}. Each gauge boson mediates (again sorted by increasing 
strength, see table \ref{tabgaugebosons}) either gravitation, weak, electromagnetic, or strong 
interaction. All fermions have an antiparticle with opposite electromagnetic and color charge, 
but same mass and spin.\\
Leptons carry an electromagetic as well as a weak charge, while quarks additionally have a 
so-called color charge that is a quantum number associated with a local SU(3) symmetry. This color 
charge couples to a gluon, the gauge boson of the strong interactions, which carries color charge 
itself. Thus, QCD is a non-Abelian gauge symmetry. In Abelian theories, like Quantum 
Electrodynamics (QED), the gauge bosons (i.e., the photons) do not carry (electric) charge.\\
Below energy densities of $\sim 1$~GeV/fm$^3$ quarks, in contrast to leptons, are always confined 
in color-neutral {\it hadrons} (for a brief history of the formation of particles during the 
evolution if the universe see appendix \ref{EvolutionUniverse}). 
Those are particles consisting either of quark-antiquark pairs 
({\it mesons}) or three quarks or antiquarks ({\it baryons}). If two quarks get separated in 
space, the gluon field between them forms a narrow tube (string) and the potential between the 
quarks increases with distance \cite{Bali:1992ab},
\begin{eqnarray}
V(r) &=& -\frac{4}{3} \frac{\alpha_s(Q^2)}{r} + kr\, ,
\end{eqnarray}
where $\alpha_s$ is the running coupling constant. At a certain distance, the energy becomes 
sufficiently large to create quark-antiquark pairs and thus other hadrons. 
\begin{figure}[t]
\centering
  \includegraphics[scale = 0.52]{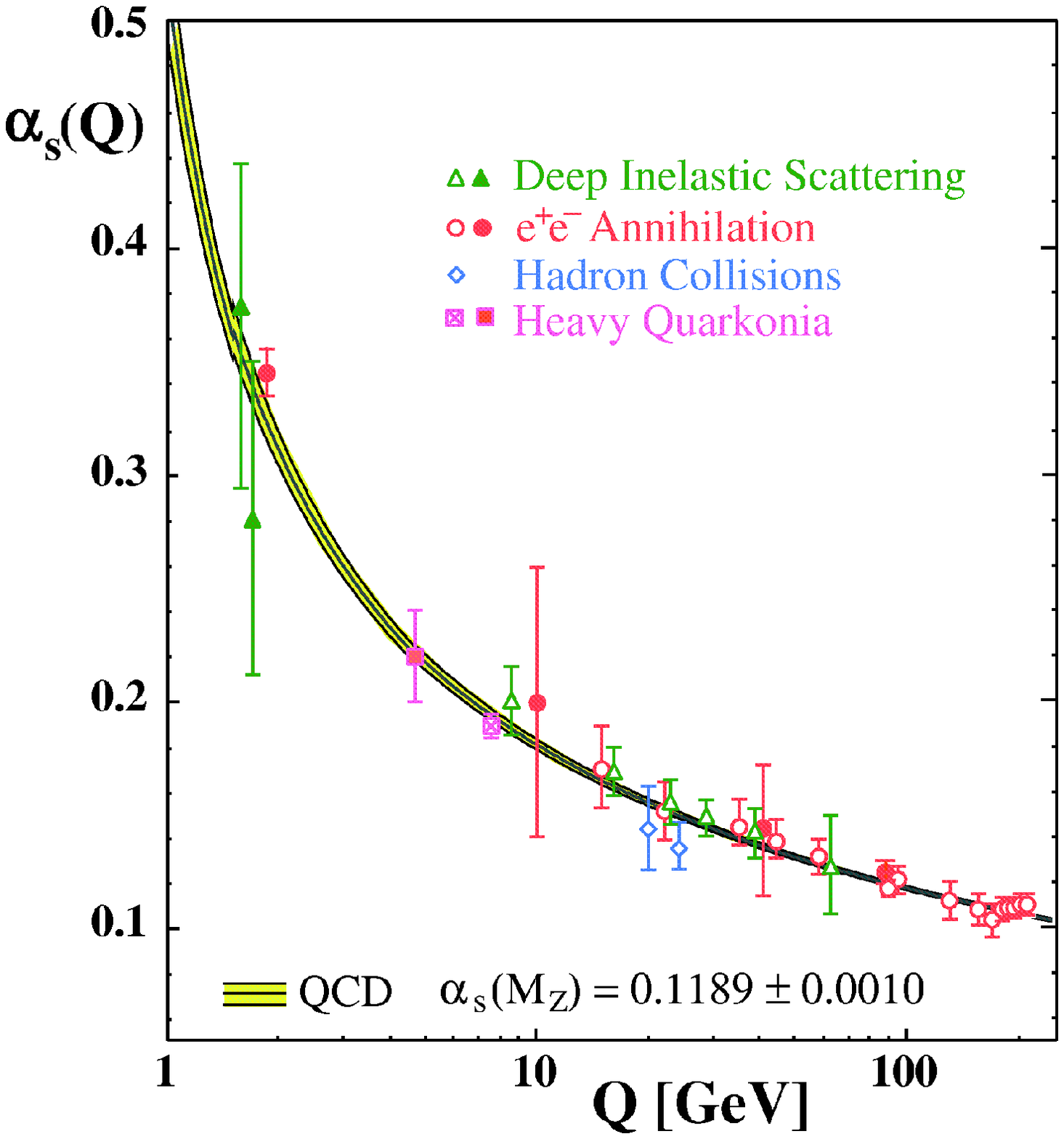}
  \caption[Comparison of different measurements of the coupling constant $\alpha_s$ as a function 
  of the transferred momentum to QCD model calculations.]
  {Comparison of different measurements of the coupling constant $\alpha_s$ as a function of the 
  transferred momentum   to QCD model calculations \cite{Bethke:2006ac}.}
  \label{runningcoupling}\vspace*{2ex}
\end{figure}
\\
This concept of color neutrality is called {\it color confinement} and is one of the two most 
prominent features of QCD. The other one is the above mentioned {\it asymptotic freedom}. \\
Both are due to the fact that the strength of the interaction between quarks and gluons becomes 
weaker as the exchanged momentum $Q^2$ increases or, equivalently, the distance decreases,
as can be seen from Fig.\ \ref{runningcoupling}. The decrease in the coupling constant for 
{\it hard} processes (i.e., for large transferred momentum) allows not only for the creation of a 
QGP, but also to describe the related processes perturbatively, i.e., to low order in the 
coupling constant.\\
 
\section[The Phase Diagram of QCD]
{The Phase Diagram of QCD}
\label{PhaseDiagram}

\begin{figure}[t]
\centering
  \includegraphics[scale = 0.72]{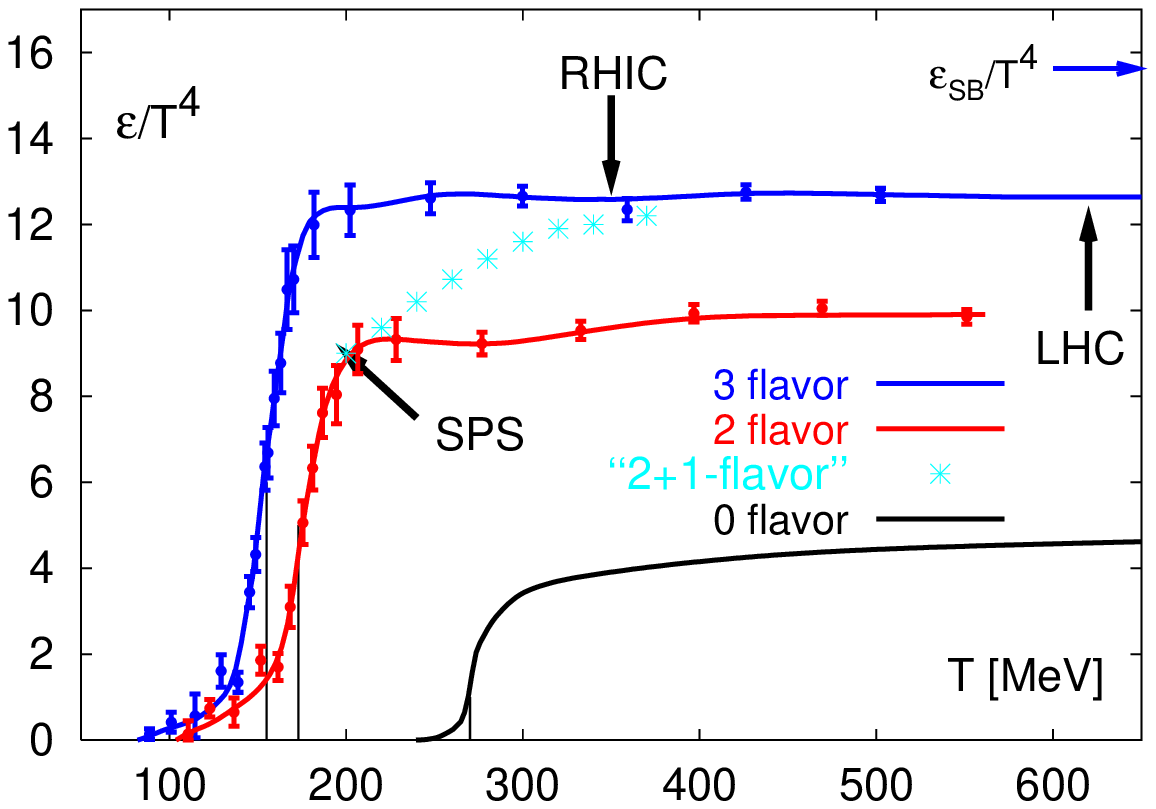}
  \caption[The energy density (scaled by $T^4$) as a function of temperature from lattice QCD 
  calculations.]
  {The energy density (scaled by $T^4$) as a function of temperature from lattice QCD calculations, 
  taken from \cite{Gyulassy:2004zy}, for various numbers of quark species. The figure includes the 
  estimated temperatures reached at SPS, RHIC, and LHC.}
  \label{e_T_Lattice}
\end{figure}

\begin{figure}[t]
\centering
  \includegraphics[scale = 0.67]{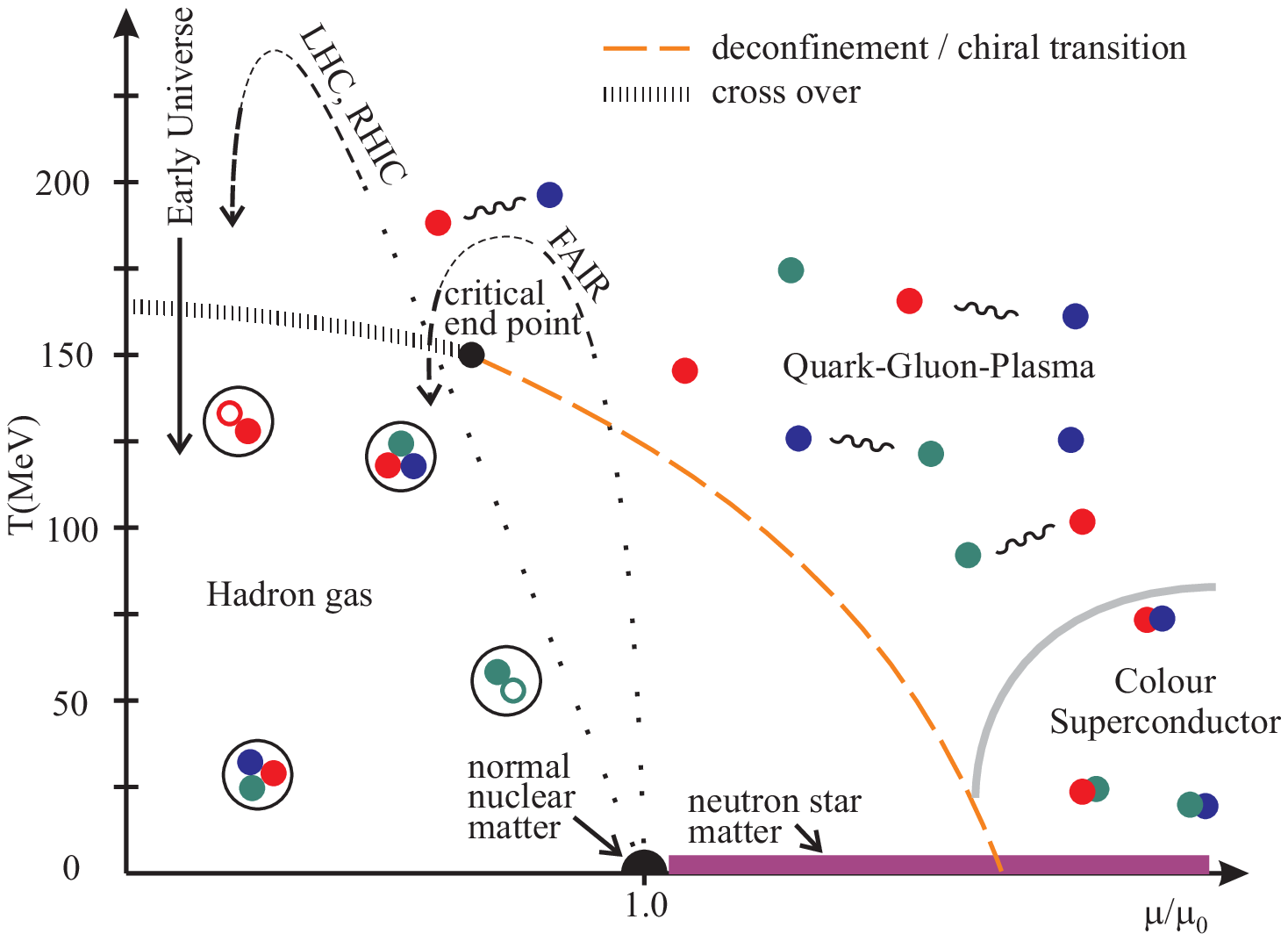}
  \caption[A schematic phase diagram of QCD matter in the $(T,\mu)$-plane.]
  {A schematic phase diagram of QCD matter in the $(T,\mu)$-plane.
   The solid black line represents the chemical freeze-out, while the dashed orange line 
   illustrates the chiral/deconfinement transition. Both end at the critical point
   which is connected to the $\mu=0$ axis by a cross-over around $T\approx
   170$~MeV. The ground state of nuclear matter is at $T=0$~MeV and $\mu=\mu_0$.
   For high chemical potenial and low temperature, there exists a phase of color
   superconductivity. The dashed black lines indicate the estimated properties of 
   the medium created by various experiments.}\vspace*{2ex}
  \label{FigPhaseDiagram}
\end{figure}

As discussed in the previous section, strongly interacting matter can occur in different phases. 
A phase describes a state of a thermally equilibrated system that is characterized by a certain 
set of macroscopic observables. \\
With a change in at least one of those thermodynamic variables, the system undergoes a {\it phase 
transition}. Phase transitions are characterized by the nature of change of the relevant 
thermodynamic variable(s) and are commonly classified into phase transitions of {\it first order}, 
{\it second order}, and {\it cross-over}. \\
In general, there is an $n$th order phase transition if the $(n-1)$th derivative of a thermodynamic
variable (like the energy density or the number density) is discontinous. While a first-oder
transition always implies the existance of a {\it mixed phase}, a second order phase transition 
does not exhibit a mixed phase. \\
If the characteristic observables change rapidly, but without showing a (mathematical) 
discontinuity, the transition is called a {\it cross-over} and the actual transition between 
the phases cannot exactly be specified. A common structure in phase diagrams is that the line of 
a first-order phase transition ends in a critical end point of second order. Beyond that critical 
point, the transition becomes a cross-over. \\
Lattice QCD calculations 
\cite{Stephanov:2004wx,Fodor:2001pe,Fodor:2004mf,Karsch:2003jg,Karsch:2003va,Fodor:2004nz} (see Fig.\ 
\ref{e_T_Lattice}) indicate a rapid increase of the energy density around $T\approx 160 - 170$~MeV.
Since energy density, like pressure or entropy density, is proportional to the number of degrees 
of freedom, this behaviour can be understood as a phase transition between confined and deconfined 
states. The lattice QCD calculation shown in Fig.\ \ref{e_T_Lattice} is performed for a vanishing 
baryo-chemical potential, $\mu_B=0$.\\
The QCD phase diagram (see Fig.\ \ref{FigPhaseDiagram}) is obtained by plotting the relevant 
degrees of freedom for strongly interacting matter. Commonly, these degrees of freedom are studied 
as a function of temperature $T$ and quark chemical potential $\mu$ which is associated with the 
net-quark number. 
Thus, for a vanishing quark chemical potential, an equal number of quarks and antiquarks is 
present in the system. Consequently, a positive quark chemical potential implies an excess of 
quarks over antiquarks. Therefore, a large value of $\mu$ is equivalent to a large net quark density 
$\rho$. Since quarks carry baryon number ($B=1/3$), an equivalent measure to characterize 
the density is the baryochemical potential $\mu_B$ which is just scaled by a factor of $3$
as compared to the quark chemical potential $\mu_B=3\mu$.
\\Nuclear matter at its ground state has a temperature of $T\simeq 0$~MeV and a baryon density of 
$\rho\simeq 0.16$~fm$^{-3}$ which corresponds to a quark chemical potential of $\mu_0=308$~MeV 
\cite{Rischke:2003mt}. For low temperatures and small values of the quark chemical potential,
strongly interacting matter forms a hadron gas. At sufficiently high temperature and/or chemical 
potential, hadrons strongly overlap, creating an extended region of deconfined quark-gluon matter, 
the QGP \cite{Collins:1974ky,Freedman:1976ub,Shuryak:1977ut}. The phase transition is presumed to 
be of first order, ending in a critical point of second order, the location of which is not yet 
determined exactly. Predictions vary widely \cite{Stephanov:2004wx,Fodor:2001pe,Fodor:2004mf,
Karsch:2003jg,Karsch:2003va}, 
but some lattice calculations \cite{Fodor:2004nz}, extended to the region of non-zero 
baryo-chemical potential, predict the critical end point to occur around $T\approx 160$~MeV 
and $\mu\approx 260$~MeV which would be a region accessible to heavy-ion collisions. \\
For cool and dense quark matter another phase transition is proposed. Due to the attractive 
interaction between quarks the formation of another ground state is expected. This phase is 
commonly referred to as color-superconducting \cite{Rischke:2003mt,Bailin:1983bm} and seems to 
contain a variety of additional phases \cite{Ruester:2005jc}. \\
At high temperatures, the nature of the QGP is not yet fully explored. There, the QGP is most 
probably not an ideal gas of non-interacting quarks and gluons \cite{Rischke:1992uv}, but behaves 
like a strongly coupled plasma ({\it strongly coupled QGP}, sQGP) corresponding to an ideal fluid, 
a concept that is supported by RHIC data 
\cite{CERN,Arsene:2004fa,Adcox:2004mh,Back:2004je,Adams:2005dq,Shuryak:2003xe,Kolb:2003dz,
Romatschke:2007mq}.\\[4ex~]
For massless quarks, the QCD Lagrangian\footnote{The QCD Lagrangian is given by 
${\cal L}=\bar{\psi}(i\gamma^\mu D_\mu - m)\psi - 1/4 F_a^{\mu\nu} F^a_{\mu\nu}$,
where $\psi$ ($\bar{\psi}$) denotes the spinor (adjoint spinor) of the quark field, $m$ the quark 
mass, $D_\mu$ the covariant derivative and $F_a^{\mu\nu}$ the field-strength tensor.} 
is chirally symmetric, i.e., invariant under a $U(N_f)_{rl}$ transformation of the right- and 
left-handed components of the quark spinors
\begin{eqnarray}
{\cal L}(\psi_r,\psi_l) &=& {\cal L}(U_r\psi_r,U_l\psi_l)\, .
\end{eqnarray}
Since spontaneous symmetry breaking is supposed to vanish for high temperatures and the mass of the
deconfined quarks\footnote{Bound quarks are described by the so-called {\it constituent quarks} 
with a large effective mass of $m\approx m_{\rm nucleon}/3$ that is mainly caused by the 
interaction of the gluons.} ($m_{u,d}\approx 10$~MeV), breaking the symmetry explicitly, is 
negligible small, chiral symmetry is supposed to be restored in the QGP \cite{Glashow:1967rx}.
However, the chiral phase transition might not coincide with the deconfinement phase transition.\\
Our knowledge about the phase diagram of QCD is currently limited to the region of the ground state 
of nuclear matter. The exploration of the other parts, including the verification of phase 
transitions, the confirmation of the critical point as well as the characterization of the 
QGP and hadronic matter are subject of current research. Two regions of the QCD phase diagram 
were or are realized in nature under extreme conditions: At vanishing quark chemical potential and 
large temperature, the QGP existed to our knowledge within the first microseconds of the universe 
\cite{Schwarz:2003du}, while dense and cool quark matter might exist (today) in the inner 
core of compact stars. 
\begin{figure}[t]
\centering
  \includegraphics[scale = 0.47]{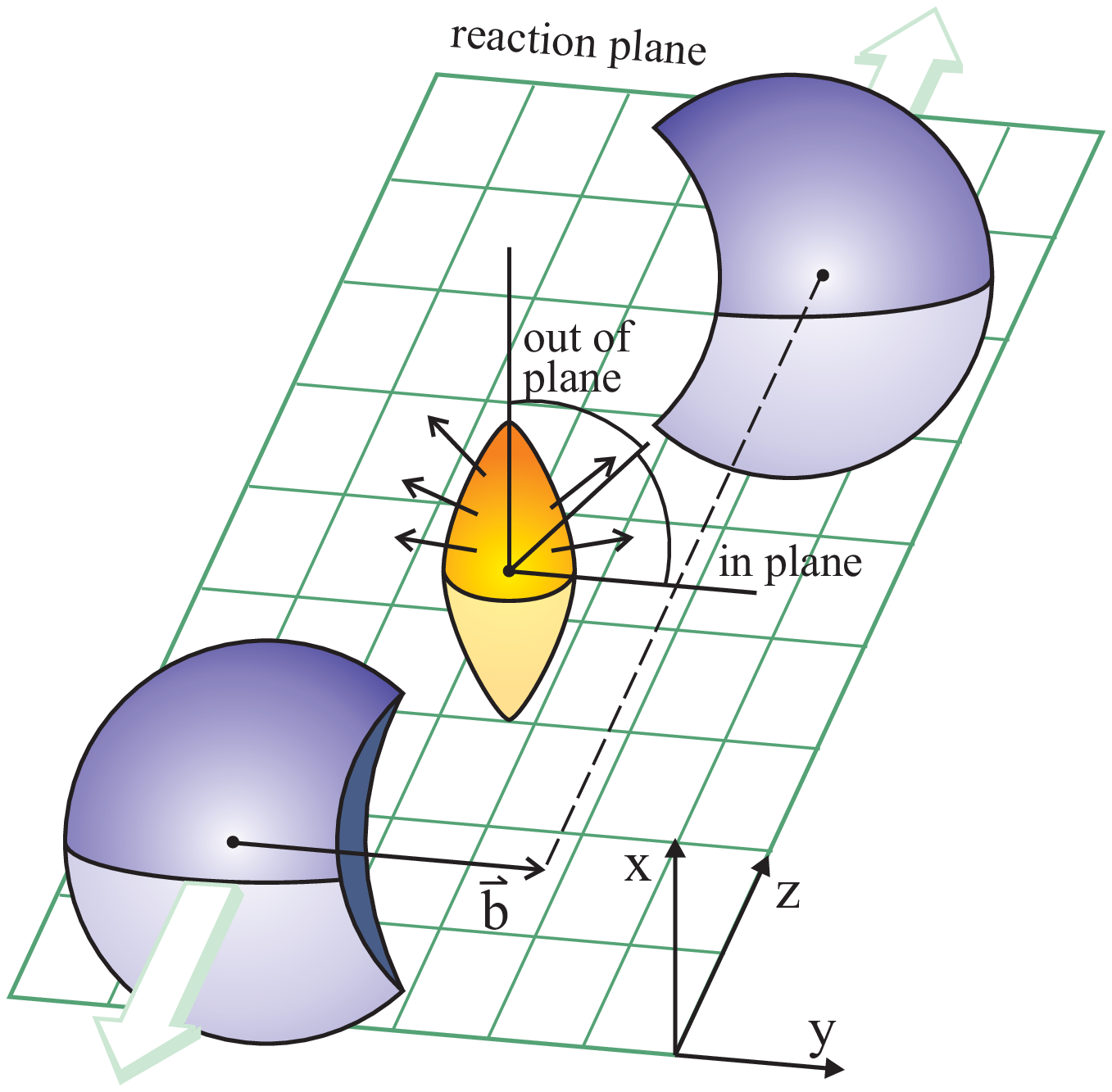}
  \caption[Geometry of a heavy-ion collision.]
  {Geometry of a heavy-ion collision. The two nuclei move along the beam axis ($z$-axis) with 
  an impact parameter $b$, determining the reaction plane. The corresponding definitions of 
  in-plane and out-of-plane are also displayed.}
  \label{reactionplane}
\end{figure}\\
One feasible method to probe the phase diagram of QCD is to study the collisions of heavy nuclei 
at ultra-relativistic energies which offer the possibility to artificially create matter under 
extreme conditions. Different collision energies enable us to test various regions of the phase 
diagram. While the large-energy runs (at RHIC and LHC) explore the region aroung $\mu\sim 0$, the 
lower-energy runs at RHIC and GSI (FAIR) are dedicated to the search for the critical end point.
Moreover, the experimentally collected data allow to draw conclusions about the properties of 
matter.

\section[Probing the QGP: Ultra-relativistic Heavy-Ion Collisions]
{Probing the QGP: Ultra-relativistic Heavy-Ion Collisions}
\label{ProbingQGP}

Though it is assumed that already the collision of two nuclei at a centre-of-mass energy larger 
than $\sqrt{s}_{NN}\approx 8$~GeV leads to such a strong compression of matter that 
colour charges are deconfined, the experimental proof that such a QGP is really created in a 
heavy-ion collision is extremely challenging since a deconfined phase is supposed to exist only 
for a very short time of roughly $\Delta t \leq 10$~fm/c~$\sim 3\cdot 10^{-23}$~s, depending
on the collision energy. \\
The description of such a collision is based on the assumption that ions are composed of nucleons. 
Due to the fact that these ions are accelerated to ultra-relativistic velocities, they are 
Lorentz-contracted, approaching each other along the {\it beam axis} (which is normally taken to 
be the $z$-axis, see Fig.\ \ref{reactionplane}) with an offset that is called the {\it impact 
parameter} $b$. \\
This impact parameter (or more precisely the direction of the strongest momentum gradient) and the 
direction of the beam axis determine the so-called {\it reaction plane}. For $b>0$~fm, some of the 
nucleons do not participate in the collision. They are called {\it spectators} and leave the 
reaction zone immediately, in contrast to the {\it participants} of the reaction.
\begin{figure}[t]
\centering
  \hspace*{-3.0cm}
  \includegraphics[scale = 0.40]{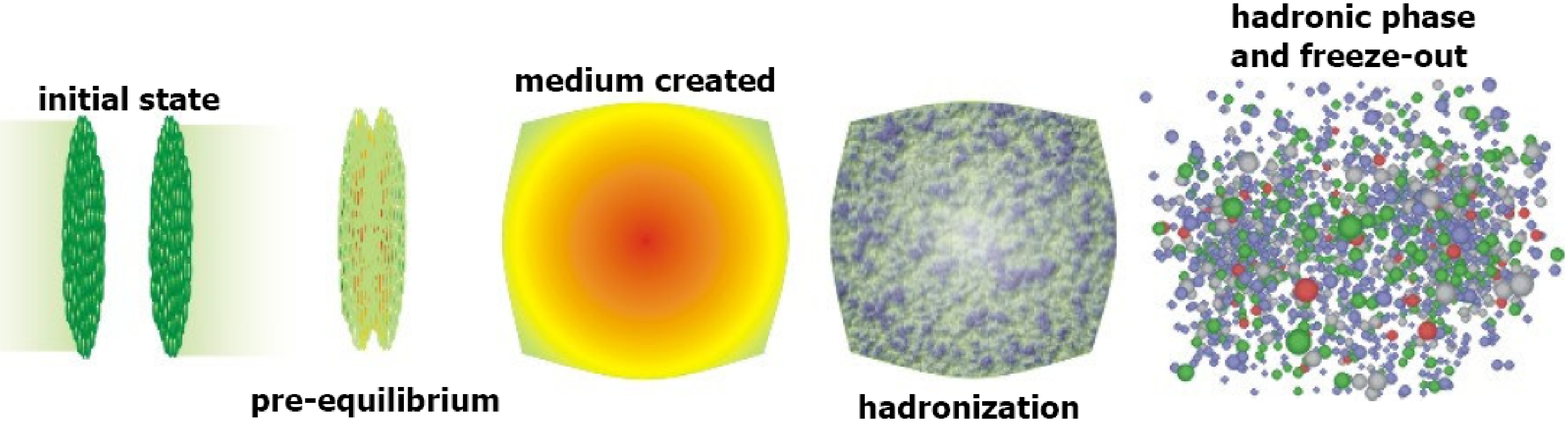}
  \caption[Sequences of a heavy-ion collision.]
  {Sequence of a heavy-ion collision, from Ref.\ \cite{Bass_HIC_Evolution}. The incoming nuclei 
  are Lorentz-contracted since they are accelerated to ultra-relativistic velocities (initial 
  state). At the beginning of the collision, a non-equilibrated phase (pre-equilibrium) is created 
  that develops into a thermalized and expanding fireball. During the cooling, hadrons are 
  formed again (hadronization) which   interact until the system is too dilute (hadronic phase 
  and freeze-out).}
  \vspace*{-2ex}
  \label{HIC_Evolution}
\end{figure}
\\A snapshot of the subsequent collision (which is here assumed to be central, i.e., $b=0$~fm) is 
shown in Fig.\ \ref{HIC_Evolution}. After the impact, in the early phase of the collision, the 
matter is strongly compressed, indicating a {\it pre-equilibrium} state. When compression is 
completed, a phase with extremely high temperatures and densities is created. The system is 
supposed to be able to equilibrate, developing a {\it fireball} which expands and cools rapidly. 
As soon as the temperature drops below the phase transition to the deconfined phase, 
hadrons are formed again ({\it hadronization}). Subsequent interactions of those hadrons (in the 
hadronic phase) will be both elastic and inelastic until {\it chemical freeze-out} is 
reached where {\it inelastic collisions} terminate. 
The expanding system becomes more and more dilute so that finally, at the {\it kinetic freeze-out}, 
all further interactions have ceased. The created hadrons which might be subject to decay
will finally reach the detectors. \\
However, hadrons are always created in a heavy-ion collision, independent of QGP formation. 
Thus, the only way to prove the existence of a QGP experimentally, via a heavy-ion collision, is 
to analyze the particle distributions at the end of the collision and to compare them to 
theoretical predictions assuming the creation of a deconfined phase. Therefore, it is extremely 
important to identify robust critera to distinguish a QGP from a hot and dense hadron gas. \\
Since the system of coordinates (as shown in Fig.\ \ref{reactionplane}) is usually defined such 
that the beam axis is aligned with the $z$-direction, the momentum $\vec{p}$ of a particle can 
easily be split into a longitudinal ($p_L$) and a transversal ($p_T$) component (w.r.t.\ the 
beam axis), indicated by the angle $\theta$
\begin{eqnarray}
p_L&=&p\cdot \cos\theta=p_z\,,\\
p_T&=&p\cdot \sin\theta=\sqrt{p_x^2+p_y^2}\,.
\end{eqnarray}
In contrast to the transverse momentum, the longitudinal component ($p_L$) is not invariant under 
Lorentz transformations. Therefore, the {\it rapidity} is introduced, which describes the velocity 
of a particle $p_L/E$ in longitudinal direction,
\begin{eqnarray}
y&=& {\rm artanh}\left(\frac{p_L}{E}\right)
=\frac{1}{2}\ln\left(\frac{E+p_L}{E-p_L}\right)\,,
\end{eqnarray}
implying the relations\footnote{Additionally, it also implies $\cosh y=\gamma_L$ and 
$\sinh y=\gamma_L v_L$ with $\gamma_L=1/\sqrt{1-v_L^2}$.}
\begin{eqnarray}
E&=&m_T\cdot\cosh y\,,\hspace*{0.75cm}p_L=m_T\cdot\sinh y\,,
\end{eqnarray}
where $m_T=\sqrt{m_0^2+p_T^2}$ denotes the transverse mass of the particle. If it is not possible 
to identify the mass of a particle, the
{\it pseudorapidity},
\begin{eqnarray}
\eta&=& -\ln\left[\tan\left(\frac{\vartheta}{2}\right)\right]
=\frac{1}{2}\ln\left(\frac{p+p_L}{p-p_L}\right)\,,
\end{eqnarray}
is used instead, with $\vartheta$ being the emission angle w.r.t.\ the beam axis. In the 
ultra-relativistic case ($E\approx p\gg m_0$), rapidity and pseudorapidity become equal.

\section[The Signatures of QGP]
{The Signatures of QGP}
\label{SignaturesQGP}

Certainly, one of the main challenges of heavy-ion collision experiments is the identification of 
the QGP phase. However, since the early stages of the evolution are not directly accessible, 
(unique) signatures are needed, some of which are:
\begin{itemize}
\item {\bf Global observables}\\
The rapidity distribution of particles $dN/dy$ and transverse energy $dE_T/dy$ allow for the 
determination of temperature, entropy, and energy density of the system created in a heavy-ion 
collision. These observables need to be compared to model calculations, e.g.\ lattice QCD 
calculations \cite{Fodor:2001pe,Fodor:2004mf,Karsch:2003jg, Fodor:2004nz}, in order to investigate
if the system may have reached the QGP phase.
\item {\bf Early thermalization and flow}\\
Though it is not clear from first principles that a QGP phase should be in thermodynamic 
equilibrium, it could be shown that hydrodynamical models, which are based on the assumption of 
(local) thermodynamic equilibrium, can describe flow observables quite well (as will be discussed 
in detail in the following chapter) \cite{Kolb:2003dz,Romatschke:2007mq}. These flow 
characteristics imply that the interaction of the medium constituents are strong enough to 
translate density gradients into pressure and thus to convert spatial into momentum anisotropies. 
This collective behaviour has to result from the early stages of the collision since the spatial 
anisotropy (in contrast to momentum anisotropy) reduces with the expansion of the system. 
Moreover, the collective behaviour is supposed to change when the system undergoes a 
phase transition \cite{Soff:1999yg}.
\item {\bf Photon and dilepton measurements}\\
Photons and dileptons do not interact strongly. Thus, they leave the medium without being 
influenced by the expanding fireball, carrying information about the initial state. Thermal 
photons may serve as a thermometer, while the reconstruction of the spectral density of vector 
mesons (like the $\rho$-meson) could indicate the restoration of chiral symmetry. In general,
such a restoration should lead to the disappearance of the mass splitting between hadronic states 
of opposite parity \cite{Rapp:1999ej}.
\item {\bf Hard probes, jet quenching and Mach cones}\\
The interaction of particles with high transverse momenta, created at the early stages of the 
collision, with the medium they traverse (leading either to quenching effects or the formation of 
a signal that is interpreted to be due to the formation of a Mach cone) has recently attracted a 
lot of interest and is the main topic of this thesis. It will be discussed in detail in the 
following chapters.
\item {\bf Strangeness Enhancement}\\
Since the production of an $s\bar{s}$-pair is more likely in a deconfined phase (where thermal 
production of $s\bar{s}$-pairs, having a threshold of $\sim 300$~MeV, is supposed to set in), 
more $s$-quarks should be present in the QGP phase, leading to an enhancement in the formation of 
multi-strange baryons \cite{Rafelski:1982pu} which was found experimentally \cite{Andersen:1999ym}.
\item {\bf The $J/\psi$-meson}\\
Originally, it was assumed that the production of $J/\psi$ (which is a bound $c\bar{c}$-state) is 
suppressed in the QGP phase since color screening prevents $c\bar{c}$ binding 
\cite{Matsui:1986dk}. Indeed, it was found that the $J/\psi$ melts in the QGP phase, implying 
rapid production and subsequent destruction. However, lattice calculations indicate that the 
$J/\psi$ may survive at temperatures larger than the transition temperature to the QGP and, due to 
the creation of many $c$-quarks at large (LHC) energies, its yield may even be enhanced at very 
high collision energies \cite{Andronic:2006ky}.
\item {\bf Fluctuations and correlations}\\
Fluctuations are sensitive to the underlying degrees of freedom. Thus, the fluctuations of 
conserved quantities like charge, energy, or transverse momentum, could provide a signal of the 
QGP \cite{Jeon:2003gk}. These effects should be very strong in the vicinity of the critical point. 
Likewise, the correlation of charges and their associated anti-charges seen in late-stage 
hadronization may indicate a deconfined phase \cite{Bass:2000az}.
\end{itemize}

\clearpage{\pagestyle{empty}\cleardoublepage}
%
%
\chapter[The Experimental Search for the QGP]{The Experimental Search for the QGP}
\label{ExperimentsQGP}

\begin{table}[t]
\begin{tabular}{|c|c|c|c|}\hline
Accelerator&  Beam & Max.\ Energies & Period\\\cline{1-4}
Bevalac, LBNL & Au & $< 2$~AGeV & 1984 -- 1993\\
AGS, BNL & Si / Au & 14.5 / 11.5 AGeV & 1986 -- 1994\\
SPS, CERN & S / Pb& 200 / 158 AGeV & 1986 -- 2002\\
SIS, GSI & Au & 2 AGeV & 1992 -- \\
RHIC, BNL & Au & $\sqrt{s}_{NN}$ = 130 / 200 GeV& 2000 --\\\hline
LHC, CERN & Pb & $\sqrt{s}_{NN}$ = 5500 GeV& 2009 --\\
FAIR, GSI & Au & $\sqrt{s}_{NN}$ = 5--10 GeV& 2015 --\\\hline
\end{tabular}
\caption{Review of accelerators, their maximal energies, and commission periods that were
pioneering in the sector of heavy-ion collisions.}
\label{tabaccelerators}\vspace*{-0.1cm}
\end{table}

The possibility to accelerate and collide ions was already discussed in the late
1960's \cite{Hofmann:1975by}. After systematic studies with heavy ions at the 
BEVALAC\footnote{The BEVATRON ({\bf B}illions of {\bf eV} Synchro{\bf tron}) 
was coupled to the SuperHILAC ({\bf Super} {\bf H}eavy-{\bf I}on {\bf L}inear 
{\bf A}ccelerator), and subsequently named BEVALAC. It is located at the Law\-rence 
Berkeley National Laboratory (LBNL), Berkely, USA.} and the 
AGS\footnote{{\bf A}lternating {\bf G}radient {\bf S}ynchrotron, BNL, USA.},
the hunt for the QGP started with the collision of lead ions at the 
SPS\footnote{{\bf S}uper {\bf P}rotron {\bf S}ynchrotron, CERN, Switzerland.} at CERN. Seven
experiments (like NA49 and NA45/CERES), all being fixed target experiments, participated
in the SPS project. The SIS\footnote{{\bf S}chwer{\bf I}onen{\bf S}ynchrotron, GSI, Germany.} 
program at GSI followed and its experiments, like 
HADES\footnote{{\bf H}igh {\bf A}cceptance {\bf D}i-{\bf E}lectron 
{\bf S}pectrometer.}, provide (up until today) information about hadron properties
in a hot and dense environment. \\
The development of experimental tools culminated in the first heavy-ion collider, 
RHIC at BNL, which was commissioned in the year 2000. 
It has a circumference of $3.8$~km and is composed of two independent rings. 
Unlike fixed target experiments, the kinetic energy of the accelerated particles is available 
in the center-of-mass frame. Its four experiments BRAHMS, PHENIX, PHOBOS, and 
STAR\footnote{{\bf B}road {\bf RA}nge {\bf H}adron {\bf M}agnetic {\bf S}pectrometers Experiment 
(BRAHMS), {\bf P}ioneering {\bf H}igh {\bf E}nergy 
{\bf N}uclear {\bf I}nteractions E{\bf X}periment (PHENIX), and {\bf S}olenoidal 
{\bf T}racker {\bf A}t {\bf R}HIC 
(STAR).} are dedicated to heavy-ion physics. While RHIC operates around $\mu\approx 0$~MeV, other 
experiments e.g.\ at the SPS, probe regions with nonzero quark chemical potentials, but lower 
temperatures. \\
In the near future, the LHC (with a circumference of $27$~km) at CERN is expected to commence 
operation. The primary target of its experiments ATLAS, CMS, and 
LHCb\footnote{{\bf A} {\bf T}oroidal {\bf L}HC {\bf A}pparatu{\bf s} (ATLAS), {\bf C}ompact {\bf M}uon 
{\bf S}olenoid (CMS), and {\bf LHC} {\bf b}eauty (LHCb).} is the search for the Higgs 
boson\footnote{The Higgs boson is a massive, scalar elementary exchange particle, predicted 
by the standard model.} in proton+proton (p+p) collisions. The ALICE\footnote{{\bf A} 
{\bf L}arge {\bf I}on {\bf C}ollider {\bf E}xperiment.} experiment however, is 
primarily dedicated to heavy-ion physics. \\
The energy regions (see table \ref{tabaccelerators}) covered by the planned FAIR 
Facility is understood to be adequate to probe highly compressed baryonic matter, 
the deconfinement phase transition, and the critical point. \\
The first phase of the RHIC program has become a great success since the results obtained by the 
four experiments, summarized in a series of so-called white papers 
\cite{Arsene:2004fa,Adcox:2004mh,Back:2004je,Adams:2005dq}, are in remarkable agreement
with each other. RHIC offered the unique possibility to study p+p, d+Au and Au+Au at identical
center-of-mass energies\protect{\footnote{The Mandelstam variable
$s$ is the sum of the $4$-momenta of the scattering particles (before the interaction),
$s=(p_1^\mu+p_2^\mu)^2$. In the centre-of-mass system, the total energy of the reaction is
given in terms of $\sqrt{s}_{NN}$.}} from $19.6$ to $200$~GeV for different collision geometries 
using the same detectors.
The main observations can be summarized as follows:
\begin{itemize}
\item Fast thermalization, indicated by strong momentum anisotropies (elliptic flow),
\item Low viscosity of the produced medium, suggesting that it behaves like a ``nearly ideal 
fluid'',
\item Jet Quenching, implying a large energy loss of partons and the creation of a dense and 
opaque system,
\item Strong suppression of high-$p_T$ heavy-flavour mesons, stressing a large energy loss 
of the heavy $c$ and $b$ quarks,
\item Direct photon emission at high transverse momenta confirms the scaling behaviour of hard 
processes,
\item Charmonium suppression similar to the one observed at lower SPS energies.
\end{itemize}
These results are consistent with models describing the creation of a new state of thermalized 
matter, exhibiting an almost ideal hydrodynamic behaviour. However, they do not provide evidence 
that the QGP has been formed and leave a lot of open questions especially concerning the properties 
of the created medium. 
\\This thesis is based on the idea of coupling two of the major experimental findings (the almost 
ideal hydrodynamic behaviour of the phase created and the energy deposition of hard particles) 
to study jet-medium interactions and to identify matter properties if possible.

\section[Collective Effects]
{Collective Effects}
\label{CollectiveEffects}
\begin{figure}[t]
\centering
  \includegraphics[scale = 0.275]{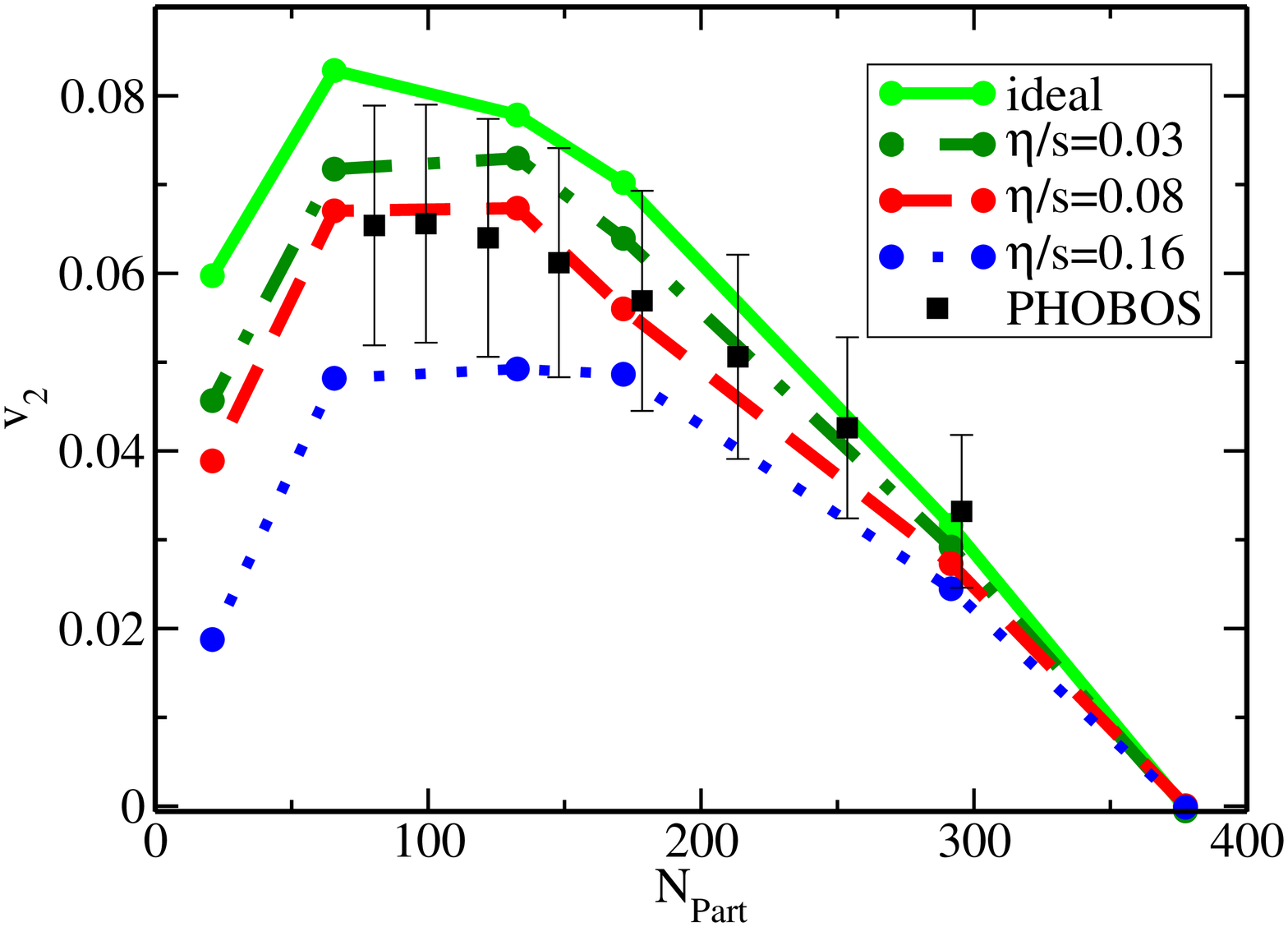}
  \caption[Elliptic flow ($v_2$) for measured PHOBOS data vs.\ a viscous hydrodynamical 
  calculation.]
  {Elliptic flow ($v_2$) from measured PHOBOS data \cite{Alver:2007qw} for charged particles 
  in Au+Au collisions at a center-of-mass energy of $\sqrt{s}_{NN}=200$~MeV compared to a 
  hydrodynamical calculation \cite{Romatschke:2007mq} for various ratios of
  shear viscosity to entropy density, $\eta/s$.}
  \label{v2Romatschke}
\end{figure}
The medium created in a heavy-ion collision at RHIC energies features a strong collective 
behaviour which causes the system to behave like a fluid. For non-central collisions, the initial 
reaction zone reveals an elliptic shape (see Fig.\ \ref{reactionplane}). If the interactions 
between the plasma components are sufficiently strong, the resulting flow will follow the density 
gradients which are stronger along the short side of the reaction zone. Thus, the initial spatial 
density profile becomes an azimuthal momentum anisotropy of the emitted particles that can be 
quantified by the Fourier coefficients of the particle distribution \cite{Poskanzer:1998yz}
\begin{eqnarray}
\frac{dN}{p_Tdp_Tdyd\phi} &=& \frac{1}{2\pi} \frac{dN}{p_Tdp_Tdy}
\left[1+2\sum\limits_{n=1}^\infty v_n (p_T,y,b)\cos(n\phi)\right]\,.
\label{defv2}
\end{eqnarray}
The second coefficient $v_2$, commonly named {\it elliptic flow}, accounts for the largest 
contribution.
Anisotropies arising from the hydrodynamical flow are the prominent indicator of fast 
thermalization in heavy-ion collisions since they require a medium in local thermodynamic 
equilibrium. Appropriate models implicitly assume that the equilibration time is roughly $1$~fm 
\cite{Heinz:2004pj,Xu:2008dv}. It was demonstrated \cite{Romatschke:2007mq} 
(see Fig.\ \ref{v2Romatschke}) that the elliptic flow which characterizes the eccentricity of the 
system can be described by a medium possessing a small amount of viscosity. Ideal fluid calculations 
lead to an overprediction of the data for peripheral collisions, but still show a decent agreement 
for central reactions \cite{Kolb:2003dz,Romatschke:2007mq} . \\
This success of describing the momentum anisotropies which are assumed to be conserved from the 
early stages of the collision using hydrodynamical models lead to the conclusion that the medium 
created at RHIC and thus the medium that was formed at the very early stages of our universe 
behaves like a fluid. This is normally regarded as one of the most important results from RHIC.\\
However, these hydrodynamical calculations strongly depend on the initial conditions and in 
general fail to describe the elliptic flow for charged hadrons away from mid-rapidity. In Ref.\ 
\cite{Arnold:2004ti} it was argued that the observed collective behaviour might not require local 
thermodynamical equilibration. An isotropic momentum distribution of the medium constituents
could be sufficient to obtain the measured data. Indeed, even such a fast isotropization would 
point towards a strongly coupled QGP.

\section[Hard Probes]
{Hard Probes}
\label{HardProbes}

Among all available observables in high-energy nucleus-nucleus collisions, particles with a 
large transverse momentum ($p_T \gtrsim 2$~GeV) are assumed to be a very useful probe to study
the hot and dense phase created in the reaction. Such {\it hard probes} are produced at
the very early stages of the collision ($t\sim 0.1$~fm) in the hard scattering of partons
(quarks, antiquarks or gluons) with large momentum transfer $Q^2$. Subsequent collision and
radiation processes (which can be described using hard QCD processes 
\cite{Gribov:1972ri,Altarelli:1977zs,Dokshitzer:1977sg}) induce 
showers of partices, so-called {\it jets}, that propagate through the system formed. 
\begin{figure}[b]
\centering
  \includegraphics[scale = 0.5,angle=270]{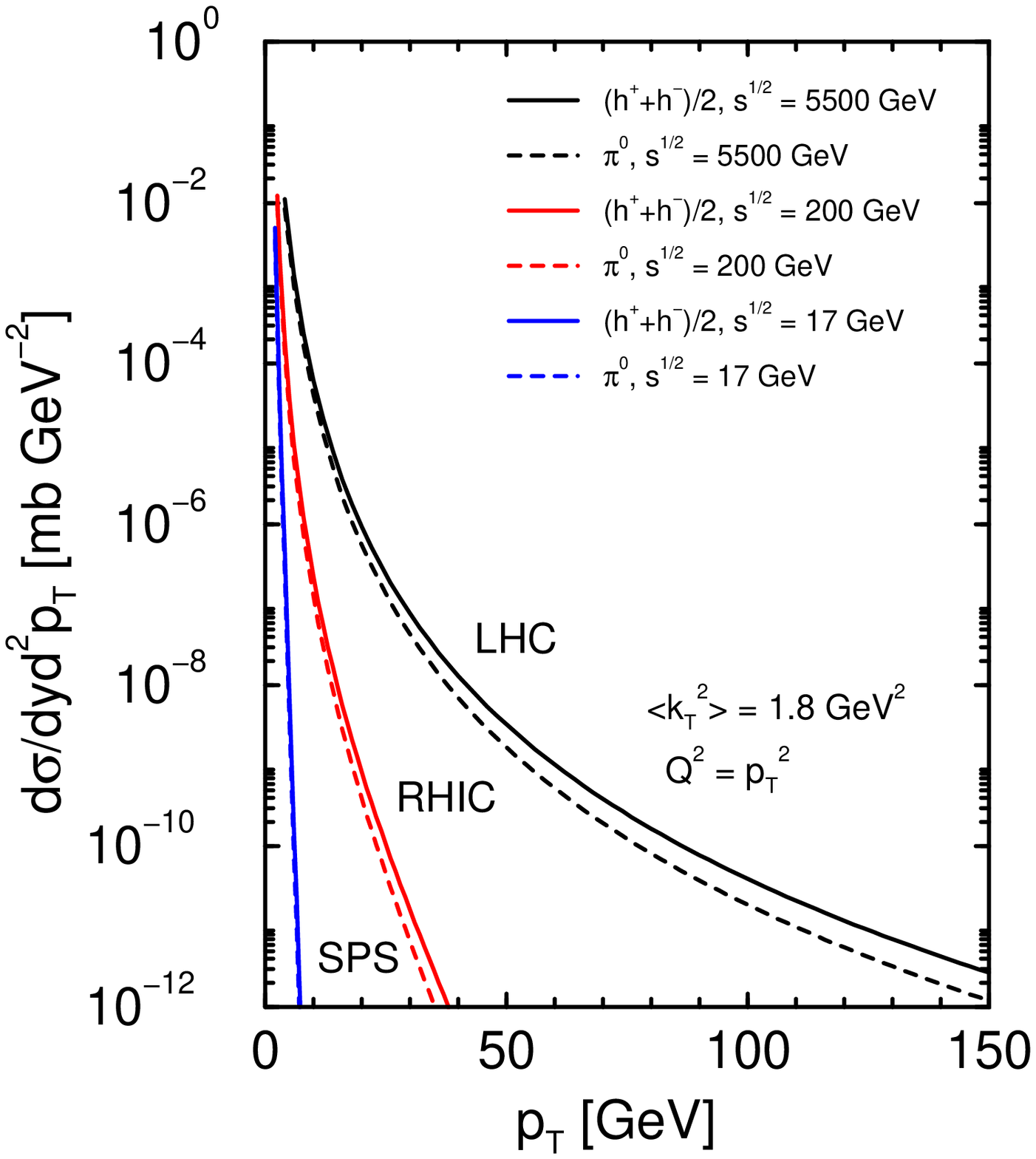}
  \caption[Leading-order differential cross sections for p+p collisions at SPS, RHIC, and LHC 
  energies.]
  {Differential cross section, calculated to leading order (LO) from pQCD, 
  for inclusive neutral pion and charged hadron production at mid-rapidity in p+p reactions at SPS, 
  RHIC, and LHC energies \cite{Accardi:2004gp}.}
  \label{CrossSections}
\end{figure}\\
In contrast to direct photons that should escape the medium unaffectedly, the created jets
will further interact with the medium, resulting in an energy loss of the jet from which 
fundamental information about the jet characteristics (like its energy and mass) and plasma
properties can be deduced. Of course it is preferable to reconstruct single jets. However,
at RHIC individual jet reconstruction is very difficult due to the large background contribution. 
At LHC energies, different cross sections are anticipated (see Fig.\ \ref{CrossSections}) and 
since hard jets with $p_T > 50$~GeV will become accessible, full jet reconstruction will be
possible in the next years.\\
The pattern of particles produced in a heavy-ion collision is examined using particle correlations 
which characterize the jet-medium effects. Several different methods including one, two, and three 
particles have been established within the recent years and will be discussed in the following.

\begin{figure}[t]
\centering
  \includegraphics[scale = 0.45]{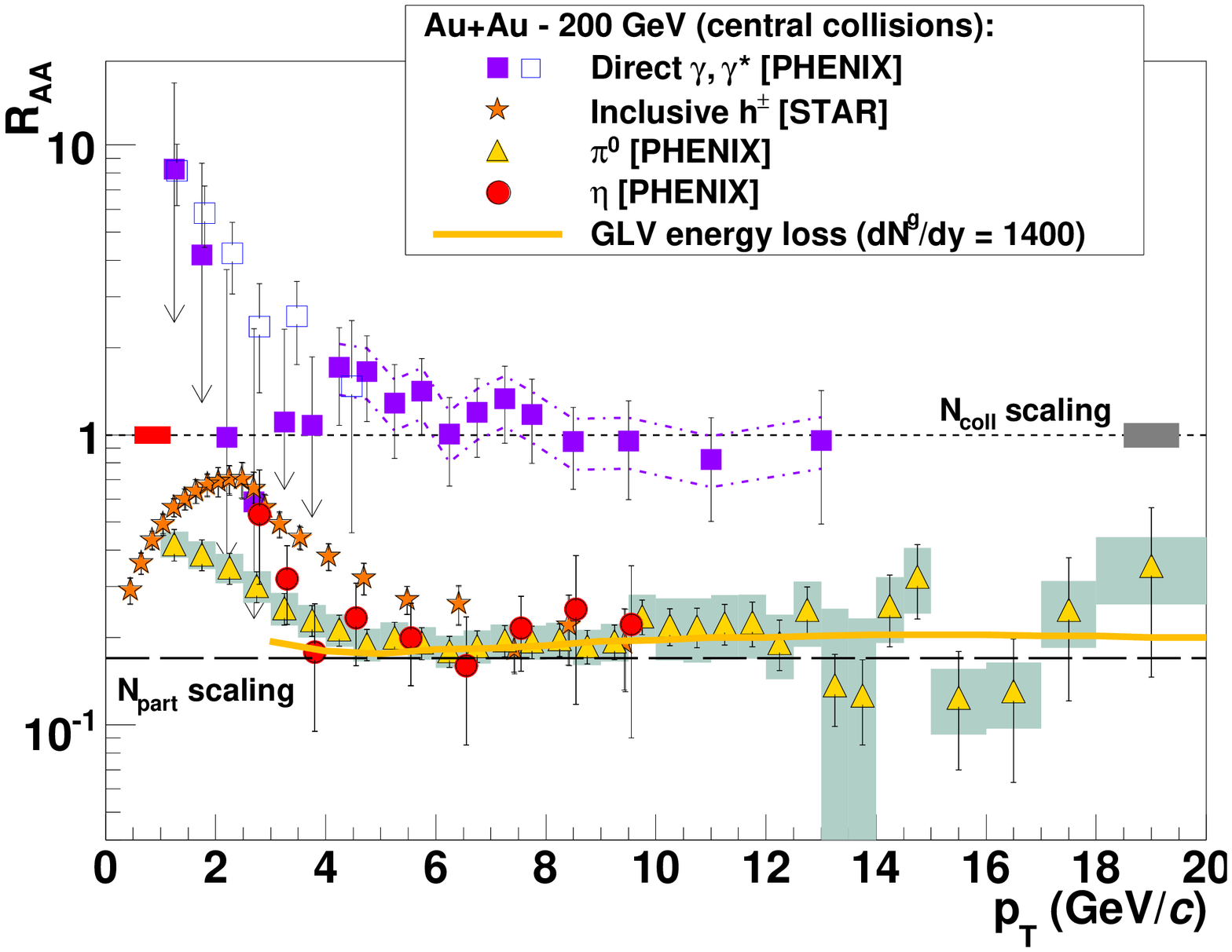}
  \caption[Measured $R_{AA}(p_T)$ in central Au+Au collisions at RHIC.]
  {$R_{AA}(p_T)$ measured in central Au+Au collisions at $\sqrt{s}_{NN}=200$~GeV, taken from 
  \cite{ReviewDavid}, for direct photons \cite{Adler:2005ig}, $\pi^0$ \cite{Adler:2003pb}, $\eta$ 
  mesons \cite{Adler:2006hu}, and charged hadrons \cite{Adams:2003kv,Adler:2003au}, compared to 
  theoretical predictions for parton energy loss in a dense medium with $dN^g/dy = 1400$ (yellow 
  curve) \cite{VitevGyulassy,VitevJet}. }
  \label{RAAPHENIX}
\end{figure}

\subsection[One-particle correlations]{One-particle correlations}
\label{One-particle correlations}

Direct information about thermodynamic and transport properties can be deduced from 
one-particle correlations. They are usually studied applying the nuclear modification factor 
\begin{eqnarray}
R_{AA} &=& \frac{d^2N_{AA}/dp_Tdy}{T_{AA}d\sigma_{NN}/dp_Tdy}\,,
\end{eqnarray}
which is the ratio of the number of particles produced in a nucleus-nucleus (A+A) collision 
to the yield in a p+p collision, scaled by the number of binary collisions, for a certain transverse 
momentum $p_T$ and rapidity $y$. $T_{AA}=\langle N_{\rm coll}\rangle/\sigma_{NN}$ is the nuclear 
overlap function and $\sigma_{NN}$ the nucleon-nucleon cross section. Thus, this measure is 
based on the assumption that the production of high-$p_T$ particles scales with the number 
of binary collisions in p+p reactions. \\
If a jet traverses the medium created without being affected, the nuclear modification factor
will be equal to one. However, initial state effects like the Cronin effect\footnote{The
Cronin effect leads to a $p_T$-broadening and is often attributed to multiple soft parton 
scattering.} \cite{Cronin:1974zm,Antreasyan:1978cw} might result in an enhancement for low $p_T$ 
with $R_{AA}> 1$, while energy loss induced by jet-medium interactions lead to a suppression and 
$R_{AA}< 1$. \\
The nuclear modification factor is usually taken as the experimental measure to determine the 
amount of energy lost by a hard particle. \\
Figure \ref{RAAPHENIX} compiles the measured $R_{AA}(p_T)$ for various hadron species and for
direct photons in central Au+Au collisions at $\sqrt{s}_{NN}=200$~GeV. Above $p_T\sim 5$~GeV, 
direct photons \cite{Adler:2005ig} are in perfect agreement with the binary collision scaling, 
while $\pi^0$ \cite{Adler:2003pb}, $\eta$ \cite{Adler:2006hu}, and charged hadrons 
\cite{Adams:2003kv,Adler:2003au} (dominated by $\pi^\pm$) show a common factor of $\sim 5$ 
suppression. The fact that $R_{AA}\approx 0.2$ irrespective of the nature of the finally produced 
hadrons is consistent with a scenario where the energy loss of the parent parton takes place prior 
to its fragmentation into hadrons. The suppression factor is close to the so-called participant 
scaling which assumes a strong quenching limit where only hadrons formed close to the surface of 
the medium can reach the detector without any further modification \cite{Muller:2002fa}.

\subsection[Two-particle correlations]{Two-particle correlations}
\label{Two-particle correlations}
\begin{figure}[t]
\centering
\begin{minipage}[t]{4.2cm}
\hspace*{-2.0cm}
  \includegraphics[scale = 0.45]{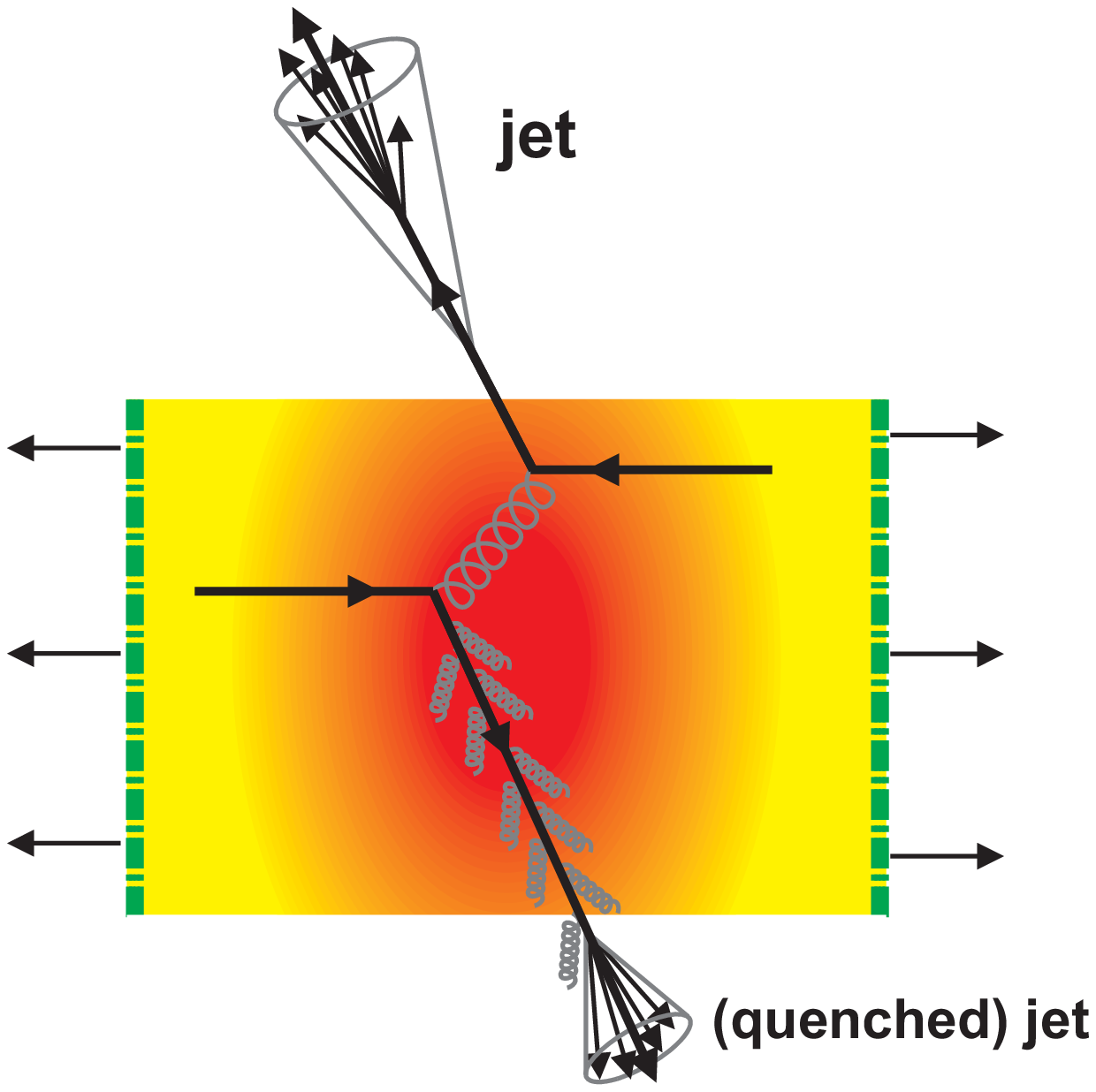}
\end{minipage}
\vspace*{1ex}  
\begin{minipage}[t]{4.2cm}  
  \includegraphics[scale = 0.45]{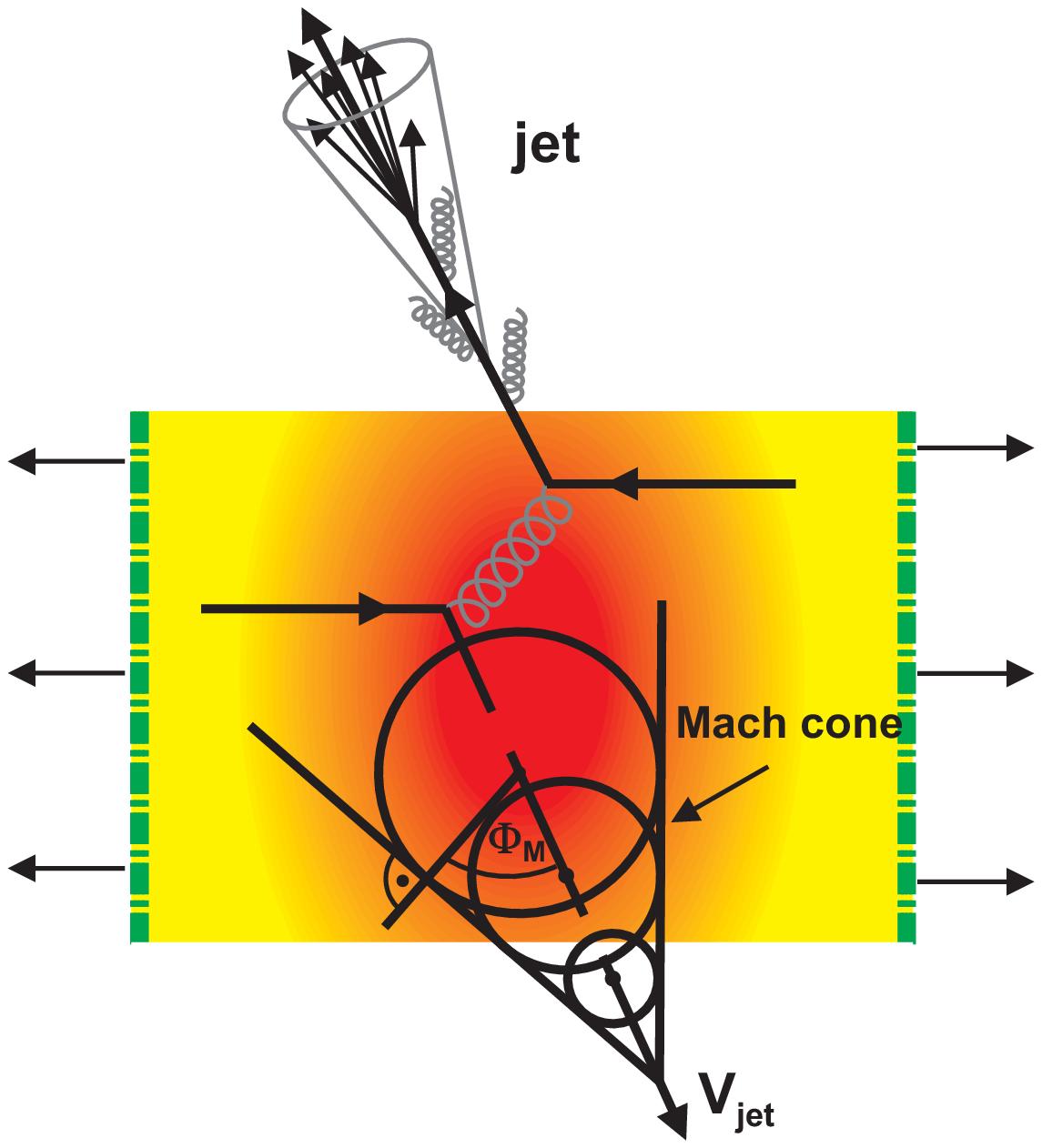}
\end{minipage}
  \caption[Schematic picture for the formation and propagation of a jet.]
  {Formation and propagation of a jet, leading either to a quenched jet (left panel) or the 
  creation of a Mach cone by the interference of sound waves (right panel) according to Ref.\
  \cite{ReviewDavid}.}
  \label{Sketch_JetPropagation}
\end{figure}
\begin{figure}[t]
\centering
\begin{minipage}[c]{4.2cm}
\hspace*{-1.75cm}
\includegraphics[scale = 0.35]{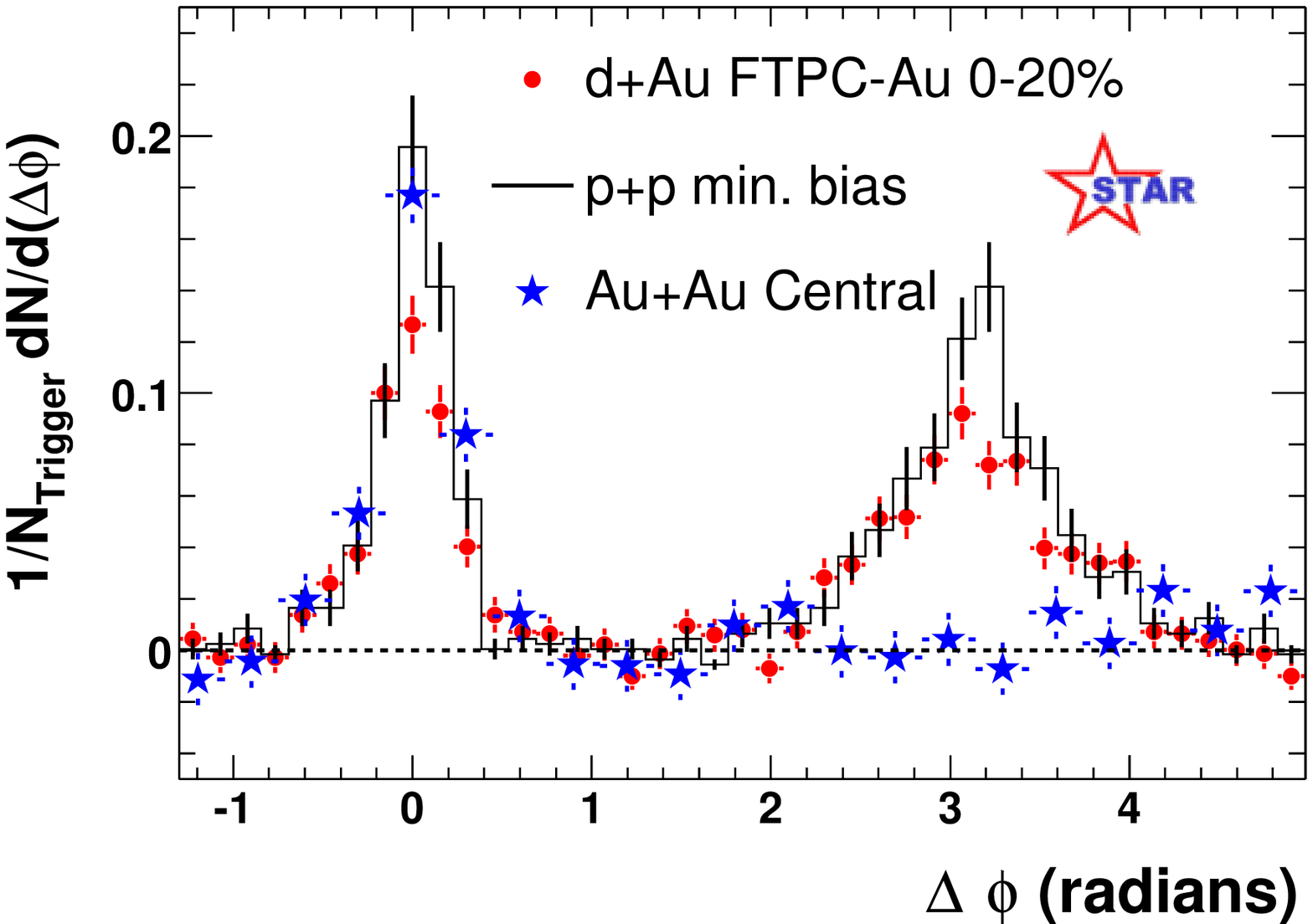}
\end{minipage} 
\hspace*{0.75cm}
\begin{minipage}[c]{4.2cm}
~\\[2mm]
  \includegraphics[scale = 0.34]{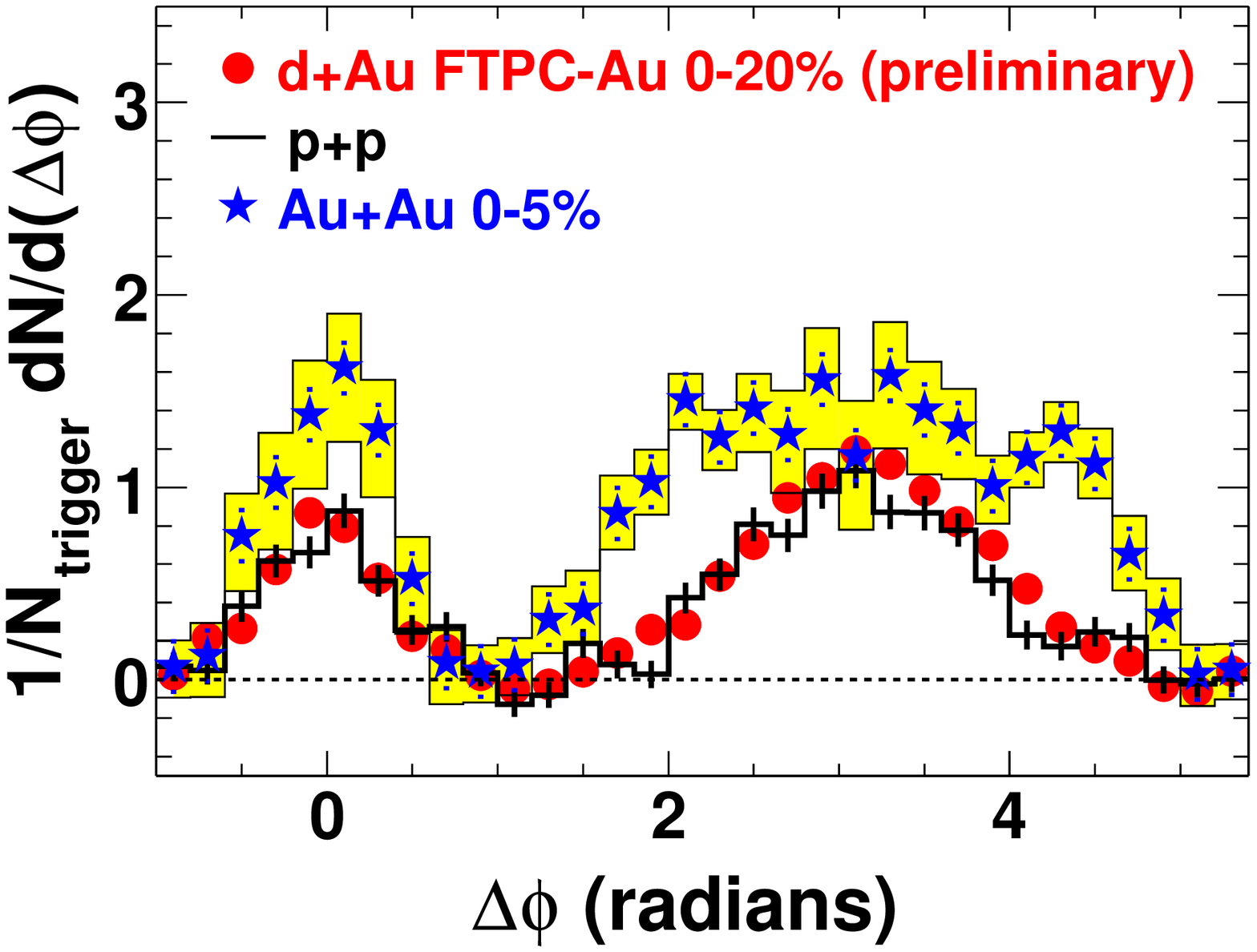} 
\end{minipage}    
  \caption[Two-particle correlation functions for a trigger particle of 
  $4 < p_T^{\rm trig} < 6$~GeV and associated particles with different $p_T$.]
  {Two-particle correlation functions for a trigger particle of $4 < p_T^{\rm trig} < 6$~GeV and 
  associated particles with (left panel) $p_T^{\rm assoc} > 2$~GeV \cite{JetQuenchingSTAR} and 
  (right panel) $0.15 < p_T^{\rm assoc} < 4$~GeV \cite{2pcSTAR} for p+p, d+Au, and central Au+Au 
  collisions.}
  \label{2pc_Experiment}\vspace*{-2ex}
\end{figure}
\begin{figure}[t]
\centering
  \includegraphics[scale = 0.45]{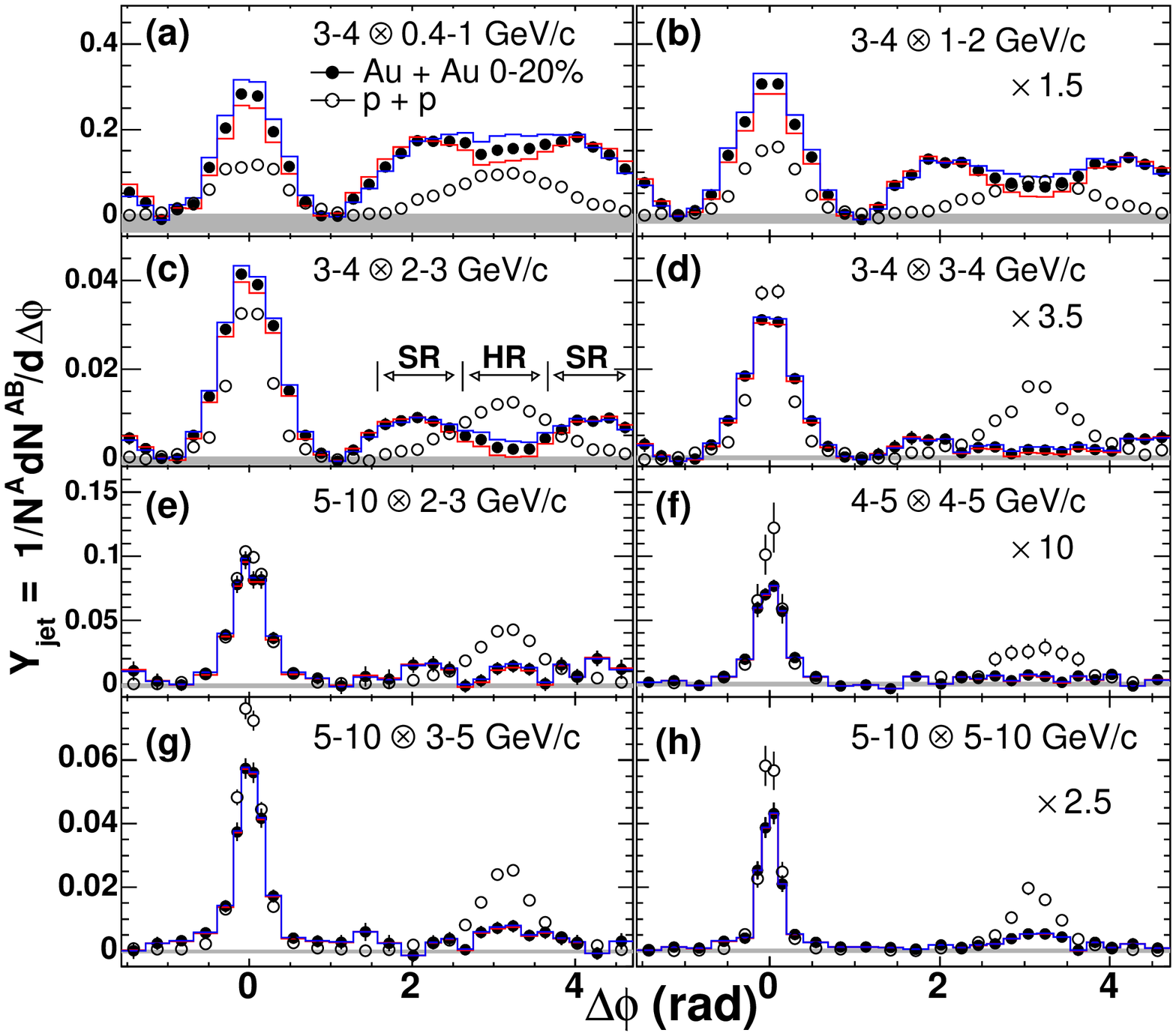}
  \caption[The $\Delta\phi$ distribution for various trigger and (increasing) partner $p_T$ of p+p 
  and $0-20\%$ central Au+Au collisions.]
  {The $\Delta\phi$ distributions for various trigger and (increasing) partner $p_T$ of p+p and 
  ($0-20\%$) central Au+Au collisions. The solid histograms and shaded bands indicate the 
  uncertainties regarding background subtraction \cite{2pcPHENIX}.}
  \label{2pc_PHENIX}
\end{figure}
\begin{figure}[t]
\centering
\begin{minipage}[b]{4.2cm}
\hspace*{-3.0cm}
\includegraphics[height = 9.3 cm, width = 7cm]{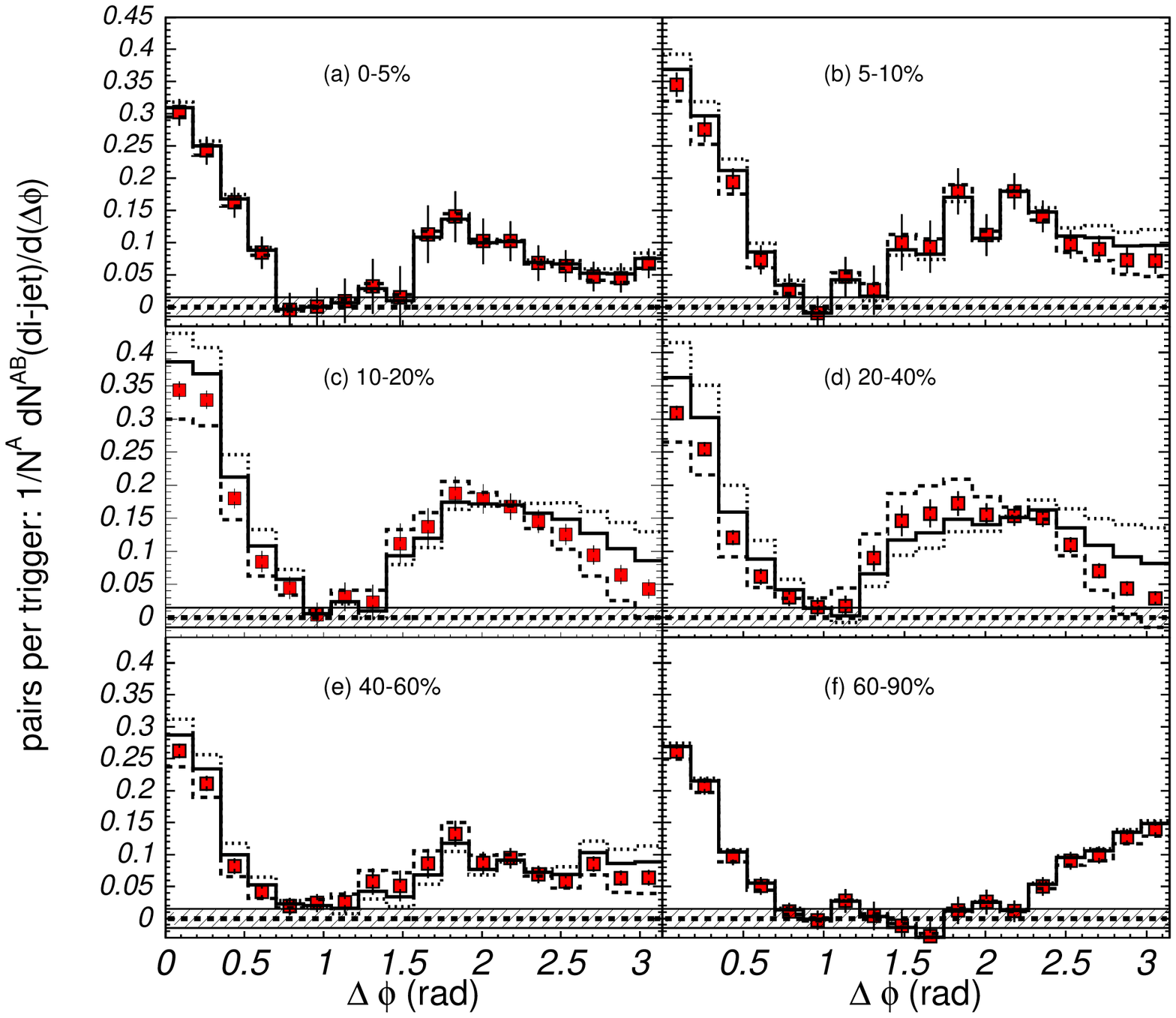}  
\end{minipage} 
\begin{minipage}[b]{4.2cm} 
\includegraphics[scale = 0.40]{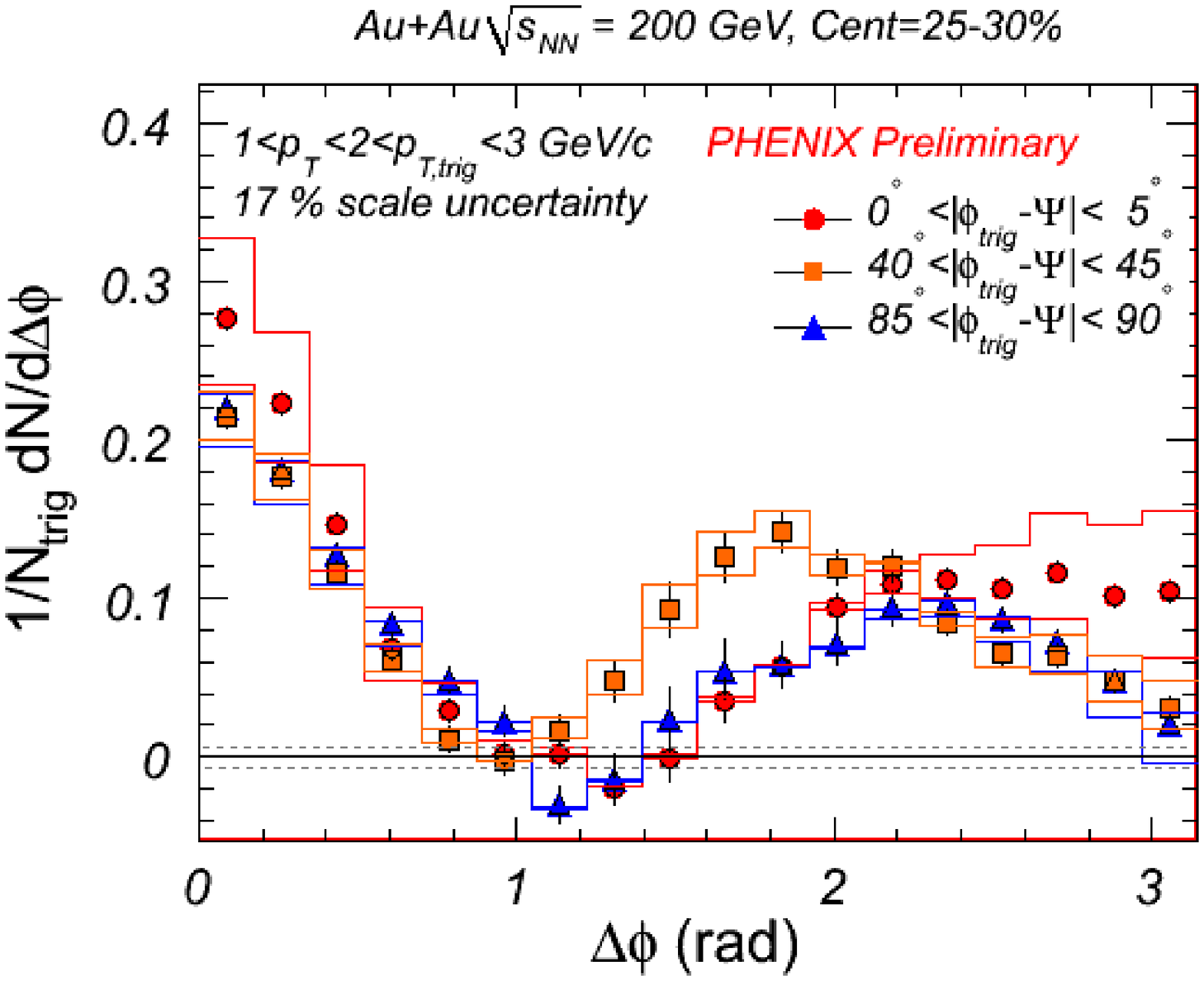}  
\end{minipage}
  \caption[Two-particle correlations for various centrality regions of Au+Au collisions and 
  the reaction plane dependence of the per trigger \newline yield.]
  {(Left panel) Two-particle correlations for a trigger particle with $2.5 < p_T < 4.0$~GeV and 
  associated particles with $1.0 < p_T < 2.5$~GeV for various centrality regions of Au+Au 
  collisions at $\sqrt{s}_{NN}=200$~GeV \cite{Adler:2005ee}.
  (Right panel) Reaction plane dependence of the per trigger yield in $25-30\%$ Au+Au collisions 
  at $\sqrt{s}_{NN}=200$~GeV for trigger particles with $1.0 < p_T < 2.0$~GeV and associated 
  particles of $2.0 < p_T < 3.0$~GeV \cite{PHENIX_QM09_1,PHENIX_QM09_2}.}
  \label{2pc_Paths}
\end{figure}

Due to energy conservation, jets are always produced back-to-back, i.e., separated by an angle of 
$\pi$ in the azimuthal plane. One of these partons, the {\it trigger particle}, is assumed to 
leave the expanding fireball without any further interaction while its partner parton, the 
{\it associated particle}, traverses the medium depositing energy and momentum (see Fig.\ 
\ref{Sketch_JetPropagation}). \\
Two-particle correlations are determined on a statistical basis by selecting high-$p_T$ trigger 
particles and measuring the azimuthal ($\Delta\phi = \phi-\phi_{\rm trig}$) distributions of 
associated hadrons relative to the trigger. A $p_T$-range for the trigger and the associated 
particles has to be defined, taking into account that the hard partons leaving the medium 
unaffected exhibit a larger momentum than those ones propagating through the medium. \\
The fact that a jet, moving through dense matter and depositing its energy, eventually disappears 
is called {\it jet quenching}. This effect was predicted theoretically 
\cite{Bjorken:1982qr,Gyulassy:1990ye,Wang:1991xy} and found experimentally at RHIC 
\cite{Arsene:2004fa,Adcox:2004mh,Back:2004je,Adams:2005dq,JetQuenchingSTAR,JetQuenchingPHENIX}. 
As the left panel of Fig.\ \ref{2pc_Experiment} reveals, the trigger-side
peak (the so-called {\it near-side}) is the same for p+p, d+Au and Au+Au collisions, but the
correlations in the opposite direction of the trigger jet (the {\it away-side}) shows a vanishing 
yield for central Au+Au collisions, demonstrating that the corresponding away-side jet is 
quenched. \\
This observation is considered as a proof that in an Au+Au collision, in contrast to p+p and 
d+Au reactions, a dense and opaque system is formed, indicating the creation of a QGP. 
\\
Since energy and momentum always have to be conserved, the ``missing'' fragments of the away-side 
(quenched) parton are either shifted to lower energies and/or scattered into a broadened angular 
distribution. Both, softening and broadening, are seen in the data when the $p_T$ of the away-side 
associated hadrons is lowered [see right panel of Fig.\ \ref{2pc_Experiment} and also e.g.\ panel 
(c) of Fig.\ \ref{2pc_PHENIX}]. A double or even triple-peaked structure arises. However, the 
main characteristic of the angular distribution is the  ``dip'' on the away-side around 
$\Delta\phi=\pi$, accompanied by the two neighbouring local maxima at 
$\Delta\phi\approx\pi\pm (1.1-1.4)$. Such a structure is interpreted as a signal of the creation
of a Mach cone (see section \ref{SecMC}) due to the preferential 
emission of energy from the quenched parton into a definite angle with respect to the jet axis. \\
Fig.\ \ref{2pc_PHENIX} displays the correlations for two different trigger-$p_T$ ranges 
($3< p_T < 4$~GeV and $5< p_T < 10$~GeV) for increasing $p_T$ of the associated hadrons. The 
decreasing away-side yield for larger $p_T^{\rm assoc}$ and the distortion of the double peak is 
clearly visible. Moreover, for hard trigger and associated particles [see panel (h) of Fig.\ 
\ref{2pc_PHENIX}], an away-side peak around $\Delta\phi=\pi$ is formed again, providing a di-jet 
signal and thus pointing towards highly-energetic partner particles that punch through the medium 
created ({\it punch-through jets}) \cite{2pcPHENIX,PunchThroughSTAR}. \\
The structure of the away-side was examined intensively. A study of the centrality dependence 
(see left panel of Fig.\ \ref{2pc_Paths}) \cite{Ulery:2005cc,Adler:2005ee} reveals that for 
centralities up to $40\%$ a clear double-peaked structure occurs on the away-side while for more 
peripheral collisions an away-side peak around $\Delta\phi=\pi$ appears again, suggesting that 
the developing structure depends on the path length of the jet through the medium. \\
The comparison of different collision energies, from SPS $\sqrt{s}_{NN}=17.2$~GeV to top RHIC 
energies of $\sqrt{s}_{NN}=200$~GeV \cite{Jia:2008kf,CERES_MachCones} showed that the double-peaked
structure gets more pronounced for larger energies. However, still a plateau-like shape with a 
slight peak around $\Delta\phi=2$~rad is visible \cite{CERES_MachCones,Kniege_PhD}.
\\
Recently, the reaction plane dependence of the two-particle correlation was investigated 
\cite{vanLeeuwen:2008pn,PHENIX_QM09_1,PHENIX_QM09_2} (see right panel of Fig.\ 
\ref{2pc_Paths}). For this purpose, different angles of a trigger particle with 
respect to the reaction plane were chosen for a certain centrality of $25-30\%$ 
and the corresponding two-particle correlations were determined to examine the 
influence of different geometries like the path length of the quenched jet. 
As can be seen from the right panel of Fig.\ \ref{2pc_Paths}, the near-side jet contribution is 
nearly the same for all correlations, but the away-side reveals a plateau {\it in-plane} (i.e., 
for a small angle with respect to the reaction plane), while {\it out-of-plane} (thus for large 
angles with respect to the reaction plane) a clear double-peaked structure occurs. The most 
interesting feature is that for an angle of $40 < \vert\phi_{\rm trig}-\psi \vert < 45^0$ 
between trigger jet $\phi_{\rm trig}$ and reaction plane $\psi$, the peak on the away-side is 
shifted to smaller values in $\Delta\phi$.\\
Several explanations for the observation of the double-peaked structure have been discussed:
\begin{itemize}
\item {\bf Large Angle Gluon Radiation}
         \\This medium-induced gluon bremsstrahlung \cite{Qiu:2004id,Vitev:2005yg} is generated 
	 if a (color) charged particle, i.e., a quark, is accelerated or decelerated. The 
	 emitted gluons have a continous spectrum in energy due to the fact that the acceleration 
	 process itself may vary and thus the emitted gluons exhibit different energies/wavelengths.
	 However, since a quark cannot depose more energy into the medium than it got through the 
	 acceleration/deceleration, the energy spectrum is cut at the upper end. Assuming 
	 that the gluon emission only takes places in the direction of the away-side jet 
	 (''large angle'' \cite{Vitev:2005yg}), the gluon radiation causes a deviation of the 
	 jet from being back-to-back.   
\item {\bf Deflected Jets}
        \\The strong interaction of the jet traversing through the medium might cause a deflection 
	of the jet: In non-central collisions the interaction region of the two colliding nuclei 
	has a large eccentricity. Due to this eccentricity, a jet might be carried away with the 
	expanding medium and therefore leave its predetermined direction of propagation.
\item {\bf Mach cone}
        \\If the away-side parton propagates with a velocity $v_{\rm jet}$ larger than the speed 
	of sound $c_s$ of the medium, it is supposed to re-distribute its energy to lower-$p_T$ 
	particles, leading to quenched parton correlations for high-$p_T$. This re-distributed 
	energy might excite sound waves (like a stone sliding through water) which interfere, 
	forming a Mach cone (see right panel of Fig.\ \ref{Sketch_JetPropagation}) with an 
	enhanced particle emission into the distinct Mach cone angle ($\phi_M$)
	\cite{Stoecker:2004qu,CasalderreySolana:2004qm}. This thesis 
	addresses the question if a jet propagating through a hydrodynamical medium, resembling 
	the hot and dense phase created within a heavy-ion collision, can lead to Mach cone 
	formation and if such a Mach cone will result in a measurable signal. Thus, the main 
	discussion of a Mach-cone contribution will follow in the course of the subsequent 
	chapters. For a detailed discussion of Mach cones see chapter \ref{ShockWavePhenomena}.
\item {\bf \u{C}erenkov Gluon Radiation}
         \\
	 \u{C}erenkov gluon radiation\footnote{\u{C}erenkov radiation is electromagnetic 
	 radiation that is emitted when a charged particle passes through an insulator at a speed 
	 greater than the speed of light in that medium. It is named after the Russian scientist 
	 Pavel Alekseyevich \u{C}herekov. Since higher frequencies (and thus shorter wavelengths) 
	 are more intense, the observed radiation is blue. It is often used for particle 
	 identification.} leads to the formation of a cone similar to a Mach cone 
	 \cite{Dremin:2005an}. However, as shown in Refs.\ \cite{Majumder:2005sw,Koch:2005sx}, 
	 there is a strong dependence of the \u{C}erenkov angle [$\cos\Theta_c =1/n(p)$, where $n$ 
	 is the index of reflection] on the momentum $p$ of the emitted particle. This angle 
	 vanishes quickly for larger momenta. Thus, a distinct experimental signature of 
	 \u{C}erenkov gluon radiation is a strong momentum dependence of the cone angle for soft 
	 particles and a disappearance of the cone-like structure for associated particles with 
	 large $p_T$. As can be seen from Fig.\ \ref{2pc_PHENIX}, the position of the peaks within 
	 the away-side do not change strongly with the $p_T$ of the associated particle. Therefore, 
	 it is very unlikely that the observed double-peaked structure results from \u{C}erenkov 
	 Gluon Radiation. A Mach cone however, is independent of the momentum of the emitted 
	 particles. 
\end{itemize}
\begin{figure}[t]
\centering
  \includegraphics[scale = 1.2]{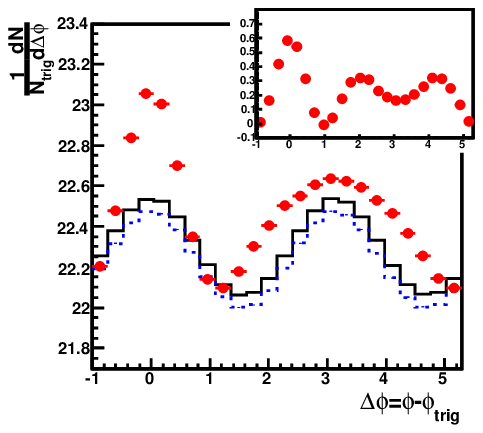}
  \caption[Measured two-particle correlation and background contributions leading to a 
  backgroud-subtracted pattern, applying the ZYAM \newline method.]
  {Measured two-particle correlation (raw data, points) and background from flow modulation 
  (elliptic flow, solid line) as well as scaled according to ZYAM (dashed line) 
  \cite{Ulery:2007ct}. The upper small insert shows the two-particle correlation after background 
  subtraction.}
  \label{ZYAM}
\end{figure}
One of the crucial problems in the context of two-particle correlations is the treatment of the 
background which has to be subtracted. Experimentally, combinatorial background contributions and 
the superimposed effects of collective azimuthal modulations (the elliptic flow) are taken care of 
with different techniques \cite{Adler:2005ee,Adler:2002tq,Adams:2005ph}. The most common method 
applied to experimental data \cite{Ajitanand:2005jj} is to subtract the elliptic flow represented 
by the $v_2$ parameter in Eq.\ (\ref{defv2}). This method (ZYAM) is discussed controversally since it is 
not clear from first principles if the flow is independent of the jet transit. The acronym ZYAM 
(Zero Yield At Minimum) is chosen because the elliptic flow is adjusted to the measured particle 
correlation in such a way that the subtraction leads to a vanishing yield at a certain, freely 
chosen minimum between near-side and away-side. However, since the elliptic flow contribution 
might be much weaker, it is not clear at all that the emerging minimum, which can even be shifted 
by this method, will lead to a vanishing yield. Moreover, as can be seen from Fig.\ \ref{ZYAM}, 
even a measured broad away-side peak might result after background subtraction, applying the 
ZYAM method, in a double-peaked structure. In other words, if the hydrodynamical flow changes due 
to a jet, the particle correlations obtained from experimental measurements should be revised and 
might lead to a different interpretation of the ongoing processes. A theoretical proof for the 
applicability for ZYAM is still missing.

\subsection[Three-particle correlations]{Three-particle correlations}
\label{Three-particle correlations}
\begin{figure}[t]
\centering
  \includegraphics[scale = 0.95]{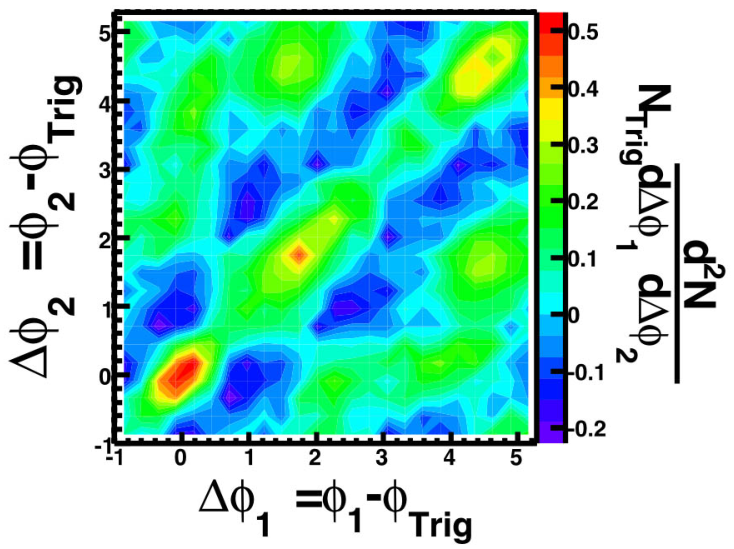}
  \caption[Background subtracted three-particle correlations for a $30-50\%$ central Au+Au 
  collision at top RHIC energies.]
  {Background subtracted three-particle correlations for a $30-50\%$ central $Au+Au$ collision 
  at top RHIC energies \cite{Ulery:2007zb}.}
  \label{3pc_Ulery}
\end{figure}
Two-particle correlations could not clarify the origin of the double-peaked structure on the 
away-side. In particular, since they include a sampling over many different events, it cannot be 
distinguished if the peaks result from randomly deflected jets or if due to the formation of a Mach 
cone two particles are produced on the away-side and emitted into a distinct direction. Thus, for 
a further exploration, three-particle correlations were introduced 
\cite{UleryPRL,Pruneau:2006gj,Ajitanand:2006is}. 
Again, high-$p_T$ trigger particles are chosen on a statistical basis, but for this analysis, 
the azimuthal angle of two associated particles ($\Delta\phi_1 = \phi_1-\phi_{\rm trig}$ and 
$\Delta\phi_2 = \phi_2-\phi_{\rm trig}$) are determined and plotted against each other 
(see Fig.\ \ref{3pc_Ulery}). \\
Randomly deflected jets will lead to a smeared peak around the diagonal axis over a wide range in 
$\Delta\phi_1$ and $\Delta\phi_2$, resulting from the fact that two particles within a deflected 
jet will be redirected into the same direction. On the other hand, if a Mach cone is formed, two 
particles will be produced due to the distinct emission into the Mach-cone angle and thus by 
measuring the relative angle between the trigger particle and its associated hadrons (considering 
the possible interchange of $\Delta\phi_1$ and $\Delta\phi_2$ due to symmetry), four distinct 
peaks are expected \cite{UleryPRL}.\\
The measurement reveals (cf.\ Fig.\ \ref{3pc_Ulery}) that the resulting shape is a superposition 
of deflected jets and Mach-cone contributions. However, since four distinct peaks occur on the 
away-side, the experimentally determined three-particle correlations seem to favour the Mach-cone 
ansatz.

\section[Correlations in Pseudorapditiy: The Near-side Rigde]
{Correlations in Pseudorapditiy: The Near-side Rigde}
\label{Ridge}

\begin{figure}[t]
\centering
\begin{minipage}[t]{4.2cm}
\hspace*{-2.0cm}
  \includegraphics[scale = 0.30]{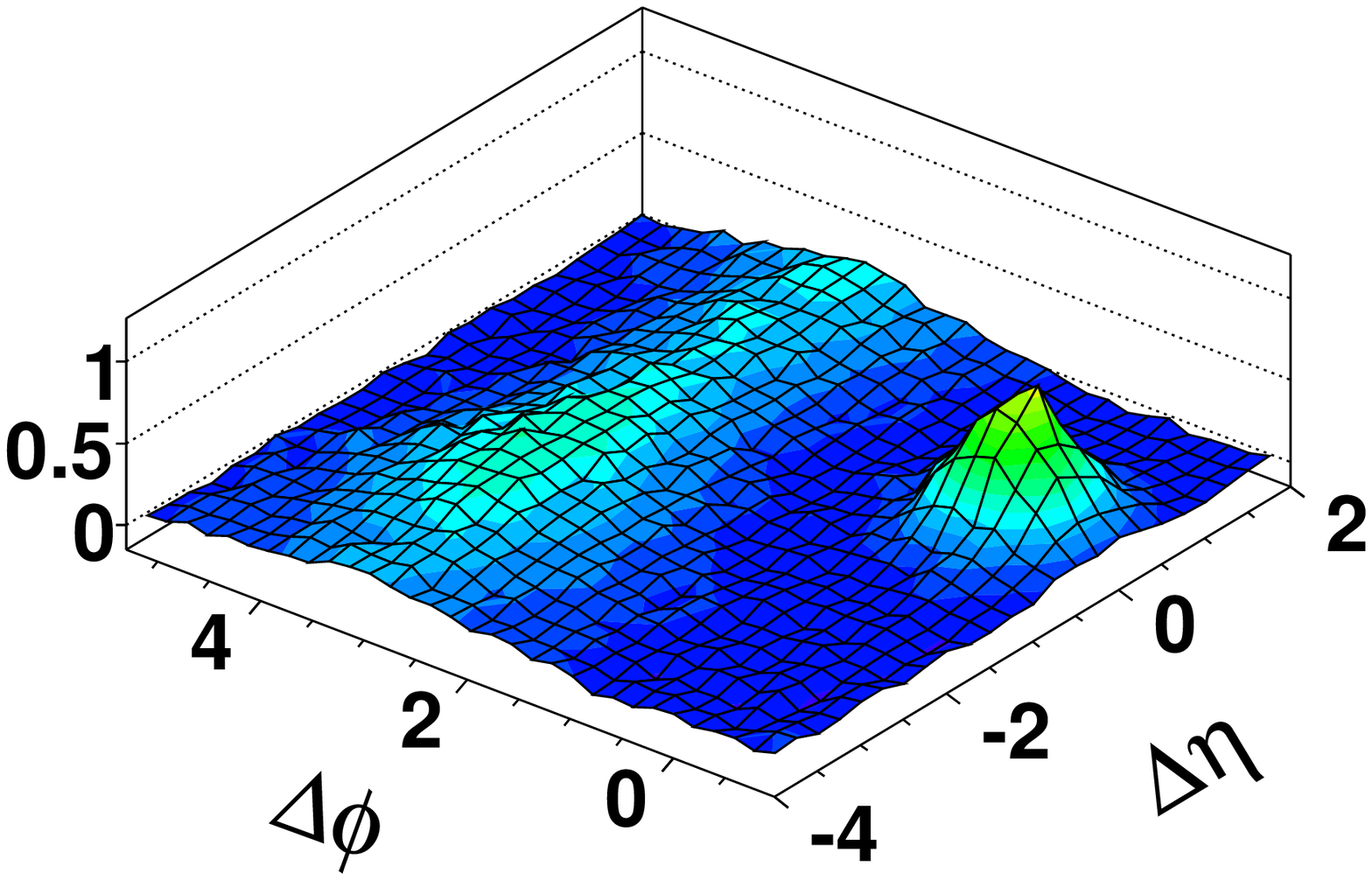}
\end{minipage}
\vspace*{1ex}  
\begin{minipage}[t]{4.2cm}  
  \includegraphics[scale = 0.30]{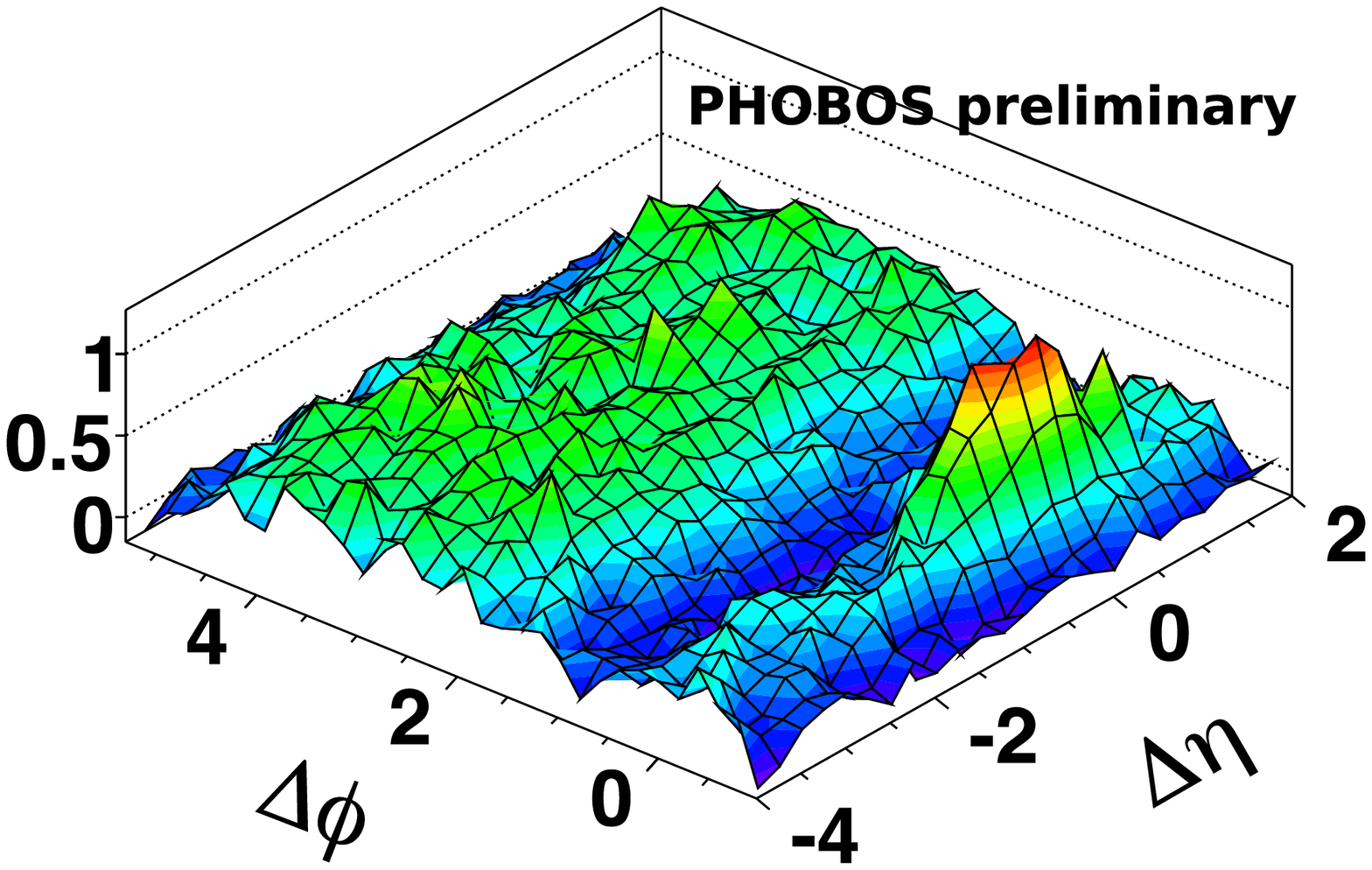}
\end{minipage}
  \caption[Distribution of correlated hadron pairs as a function of $\Delta\phi$ and $\Delta\eta$, 
  leading to the so-called Ridge phenomenon.]
  {Distribution of correlated hadron pairs as a function of $\Delta\phi$ and $\Delta\eta$ for a 
  trigger particle with $p_T>2.5$~GeV in p+p (PYTHIA simulation, left panel) and $0-30\%$ Au+Au 
  collisions \cite{Wenger:2008ts}.}
  \label{PlotRidge}
\end{figure}
The {\it ridge} is a long-ranged structure in $\Delta\eta$ on the near-side. While the near-side 
jet in p+p collisions (see left panel of Fig.\ \ref{PlotRidge}) displays a clear peak at 
$(\Delta\eta,\Delta\phi)\approx(0,0)$, as expected from jet fragmentation, the near-side jet in 
heavy-ion collisions features a peak at the same location, but is elongated over a wide range in 
pseudorapidity (see right panel of Fig.\ \ref{PlotRidge}) 
\cite{2pcPHENIX,Wenger:2008ts,Adams:2005aw,Putschke:2007mi}. \\
The existence of such a long-ranged correlation on the near-side is not understood since by 
construction the trigger parton is the one least affected by the medium. Investigations of 
this structure ensured that its properties like the particle composition, $p_T$-slope, and 
intra-particle correlations are very similar to the soft underlying events of the collision 
\cite{vanLeeuwen:2008pn}. Thus, the ridge seems to be formed by bulk matter and not from jet
fragments. Many different models for explaining the ridge phenomena have been discussed, 
including plasma instabilities \cite{Schenke:2008hw}, glasma flux-tubes \cite{Dumitru:2008wn}, 
and modifications of the two- and three-particle correlations due to radial flow 
\cite{Pruneau:2007ua}. 

%
%
%
\clearpage{\pagestyle{empty}\cleardoublepage}
\part[Hydrodynamics and Jet Energy Loss]{Hydrodynamics and Jet Energy Loss}
\label{part02}
\clearpage{\pagestyle{empty}\cleardoublepage}
%
%
\chapter[Ideal Hydrodynamics]{Ideal Hydrodynamics}
\label{IdealHydrodynamics}

It is a long-standing tradition in heavy-ion physics to model the dynamical evolution of 
heavy-ion collisions using fluid dynamics 
\cite{Landau:1953gs,Stoecker:1986ci,Clare:1986qj,CsernaiBook}, 
mainly because the only essential information needed is the Equation of State (EoS) 
of the matter considered, allowing for straightforward studies of nuclear matter 
properties like the phase transition to the QGP.\\
Once an initial condition is specified, the Equations of Motion (EoM) 
uniquely determine the dynamics of the collision. 
Thus, a detailed knowledge of the microscopic processes on the parton level is
not needed, in contrast to cascade models like UrQMD 
(Ultrarelativistic Quantum Molecular Dynamics) \cite{Bleicher:1999xi} 
or BAMPS (Boltzmann Approach of MultiParton Scatterings) \cite{Xu:2004mz}.\\
The basic requirement for the applicability of hydrodynamics is that the system has
to be in (local) thermodynamic equilibrium. This condition is fulfilled if the mean
free path $\lambda$ of a particle is small compared to the length scale over which the
fluid-dynamical variables vary. In that case, local equilibrium can develop due to the 
scattering of particles. \\
Around 20 years ago, it was already shown that collective effects at the BEVALAC
experiment, like the sideward deflection of matter in the reaction plane [named {\it side-splash} 
or {\it bounce-off}, the coefficient $v_1$ of Eq.\ (\ref{defv2})], and azimuthal deflections out 
of the reaction plane [{\it squeeze-out}, the coefficient $v_2$ of Eq.\ (\ref{defv2})], 
can be described using hydrodynamical models \cite{Gutbrod:1988hh}.
However, the great success of the fluid-dynamical model came with the RHIC experiment. As already 
discussed in section \ref{CollectiveEffects}, the collective flow effects measured at RHIC, 
in particular the elliptic flow ($v_2$), were described by hydrodynamics on a quantitative level 
\cite{Romatschke:2007mq}. \\
While the application of ideal hydrodynamics already leads to qualitative and sometimes even
quantitative agreement with the measured data \cite{Kolb:2003dz}, it is important to 
include viscous effects to check whether their impact is really small. 
Indeed, the matter created in heavy-ion collisions appears to be a ``nearly perfect liquid'', 
characterized by a small amount of viscosity \cite{Romatschke:2007mq}.\\
In a heavy-ion collision, the equilibration time is estimated to be roughly $1$~fm. 
Afterwards, the system can be described using
thermodynamical field quantities like the temperature $T(t,\vec{x})$, the four-velocities 
$u_\mu(t,\vec{x})$, and the chemical potentials $\mu_i(t,\vec{x})$. The temporal evolution of these
variables is determined by the hydrodynamical Equations of Motion until the system is so dilute
that the fluid constituents decouple and can be regarded as free particles, a transition 
called freeze-out (see section \ref{ProbingQGP}).
Therefore, hydrodynamical models can be applied to the expansion stage of heavy-ion collisions.
However, they require a proper choice of the initial conditions and the freeze-out procedure.

\section[The Hydrodynamical Equations of Motion]
{The Hydrodynamical Equations of Motion}
\label{HydroEoM}

Relativistic hydrodynamics implies the (local) conservation of energy-momentum 
(represented by the energy-momentum tensor $T^{\mu\nu}$) and of conserved charge currents like 
(net) baryon number or (net) strangeness (described by $N^\mu_i$) \cite{Landau}.
For $n$ currents, this leads to the hydrodynamical Equations of Motion
\begin{eqnarray}
\partial_\mu T^{\mu\nu}&=&0,\hspace*{0.7cm} \mu,\nu=0,...,3\,;\label{EnMomConservation}\\
\partial_\mu N^\mu_i &=& 0,\hspace*{0.7cm}\,i=1,...,n\label{ChargeConservation}\,.
\end{eqnarray}
They form a system of $4+n$ conservation equations for the energy, the three components of the 
momentum, and the $n$ currents of the system. In general, however, these equations have $10+4n$ 
independent variables, the ten independent components of the symmetric energy-momentum tensor 
$T^{\mu\nu}$ and four components of each current $N_i$. Thus, the system of fluid-dynamical 
equations is not closed and cannot be solved in complete generality. \\
Additional assumptions are needed to close this set of equations. The simplest approximation 
is to consider an ideal gas in local thermodynamical equilibrium. From kinetic theory, 
the energy-momentum tensor and (net) charge currents are \cite{deGroot}
\begin{eqnarray}
T^{\mu\nu}(x) &=& \frac{g}{(2\pi)^3} 
\int \frac{d^3\vec{k}}{E}\,k^\mu k^\nu\,[n(E)+\bar{n}(E)]\,,\label{deGroot1}\\
N^\mu_i(x) &=& 
\frac{g}{(2\pi)^3}\,q_i\int \frac{d^3\vec{k}}{E}\,k^\mu\,[n(E)-\bar{n}(E)]\label{deGroot2}\,,
\end{eqnarray}
where  $g$ counts the particle's degrees of freedom like spin and colour, 
$E=\sqrt{\vec{k}^2+m^2}$ describes the on-shell energy of particles with rest mass $m$, and 
$n(E)$ [$\bar{n}(E)$] denotes the Fermi--Dirac and Bose--Einstein distributions for particles 
(and antiparticles)
\begin{eqnarray}
n(E) &=& \frac{1}{\exp\left\{\left[E-\sum_i \mu_i(x)\right]\Huge/T(x)\right\}\pm1}\,,\\
\bar{n}(E) &=& \frac{1}{\exp\left\{\left[E+\sum_i \mu_i(x)\right]\Huge/T(x)\right\}\pm1}\,.
\end{eqnarray}
Here, $\mu_i(x)$ and $T(x)$ are the local chemical potentials and temperatures. \\
Defining the {\it local rest frame} [LRF, a frame where $u^\mu=(1,\vec{0})$], the
(net) charge density (of type $i$) 
\begin{eqnarray}
n_i&=& gq_i\int\frac{d^3\vec{k}}{(2\pi)^3}[n(E)-\bar{n}(E)]\,,
\end{eqnarray}
the ideal gas energy density
\begin{eqnarray}
\varepsilon &=& g\int\frac{d^3\vec{k}}{(2\pi)^3} E [n(E)+\bar{n}(E)]\,,
\end{eqnarray}
and the ideal gas pressure
\begin{eqnarray}
p&=& g \int\frac{d^3\vec{k}}{(2\pi)^3}\frac{\vec{k}^2}{3E} [n(E)+\bar{n}(E)]\,,
\end{eqnarray}
equations (\ref{deGroot1}) and (\ref{deGroot2}) can be rewritten, resulting in the 
energy-momentum tensor and conserved current for an ideal gas,
\begin{eqnarray}
\label{IdealTmunu}
T^{\mu\nu}&=&(\varepsilon + p)u^\mu u^\nu -pg^{\mu\nu}\label{Tmunu}\,,\\
\label{IdealCharge}
N^\mu_i &=& n_iu^\mu\,.
\end{eqnarray}
Here, $u^\mu$ is the fluid four-velocity, $u^\mu=\gamma(1,\vec{v})$ with $\gamma=(1-v^2)^{-1/2}$,
and $g^{\mu \nu}={\rm diag}(+,-,-,-)$ is the metric tensor. Now, the $4+n$ Equations of Motion 
contain only $5+n$ unknown functions for the energy density $\varepsilon$, the pressure $p$, the 
three components of the $4$-velocity $u^\mu$ and the $n$ conserved currents. To close the system 
an EoS for the fluid has to be specified
\begin{eqnarray}
p=p(\varepsilon,n_i)\,.
\end{eqnarray}
The EoS is the only place where information about the nature of the fluid constituents and their 
macroscopic interactions enters. Though an EoS is normally computed for a system in (local) 
thermodynamic equilibrium, its explicit form is completely unrestricted, allowing e.g.\ the 
inclusion of phase transitions. Thus, the ideal-fluid approximation derived above allows to 
consider a wider class of systems than just an ideal gas in (local) thermodynamic equilibrium.

\section[Numerical Solutions for the Equations of Motion]
{Numerical Solutions for the Equations of Motion}
\label{NumericalSolutions}

For numerical applications, it is convenient to write the conservation equations using 
calculational (i.e.\ laboratory) frame quantities. For the sake of simplicity, we restrict to 
the case of one conserved charge until the end of this chapter. Defining
\begin{eqnarray}
E&\equiv& T^{00}=(\varepsilon +p)\gamma^2-p\label{Lab1}\,,\\
\vec{M}&\equiv&T^{0i}=(\varepsilon+p)\gamma^2\vec{v}\,,\\
R&\equiv&N^0=n\gamma\label{Lab3}\,,
\end{eqnarray}
where $E$ is the energy density, $\vec{M}$ the $3$-momentum and $R$ the charge density,
the conservation equations (\ref{EnMomConservation}) and (\ref{ChargeConservation}) take the form
\begin{eqnarray}
\frac{\partial E}{\partial t} + \vec{\nabla}\cdot(E\vec{v})&=&-\vec{\nabla}\cdot(p\vec{v})\,,
\label{DGL1}\\
\frac{\partial \vec{M}}{\partial t}+\vec{\nabla}\cdot(\vec{M}\vec{v})&=&-\vec{\nabla}p\,,
\label{DGL2}\\
\frac{\partial R}{\partial t}+\vec{\nabla}\cdot(R\vec{v})&=&0\,.\label{DGL3}
\end{eqnarray}
This set of equations can be solved numerically. However, $E$, $\vec{M}$, and $R$ are 
values in the laboratory frame. While in the non-relativistic limit there is no difference 
between $n$ and $R$ or $\varepsilon$ and $E$ and the momentum density of the fluid can be simply 
related to the fluid velocity, the LRF quantities $\varepsilon$, $n$, and $\vec{v}$ 
have to be extracted from the laboratory frame values of $E$, $\vec{M}$, and $R$ in the 
relativistic case that has to be employed for the description of heavy-ion collisions. Such a 
transformation requires the knowledge of the EoS, $p(\varepsilon,n)$, and the velocity $\vec{v}$ 
of the local rest frame.

\section[Transformation between the Laboratory Frame and the Local Rest Frame]
{Transformation between the Laboratory Frame and the Local Rest Frame}
\label{Trafo}

The transformation between the laboratory frame (that is chosen for numerical applications) and 
the LRF is in principle explicitly given by equations (\ref{Lab1}) -- 
(\ref{Lab3}). However, this implies a root-finding algorithm for a system of $5$ equations that 
are nonlinear, since the pressure $p$ is a function of $\varepsilon$ and $n$. This would be very 
time-consuming in a numerical application. Therefore, the complexity of the transformation 
problem is reduced as follows \cite{Rischke:1995ir}.\\
The fundamental observation is that $\vec{M}$ and $\vec{v}$ are parallel. This leads via
\begin{eqnarray}
\vec{M}\cdot\vec{v}=\vert M\vert\cdot \vert v\vert=(\varepsilon + p)\gamma^2v^2=
(\varepsilon+p)(\gamma^2-1)=E-\varepsilon
\end{eqnarray}
[or alternatively using Eqs.\ (\ref{Lab1}) -- (\ref{Lab3})] to the expressions
\begin{eqnarray}
\varepsilon=E-\vert M\vert\cdot \vert v\vert, \hspace*{1cm} n=R/\gamma=R\sqrt{1-v^2}\,.
\label{TrafoExpressions}
\end{eqnarray}
Moreover, using that
\begin{eqnarray}
\vert M\vert = (\varepsilon+p)\gamma^2v=(E+p)v\,,
\end{eqnarray}
a general expression for the velocity $\vec{v}$ in the LRF can be derived
\begin{eqnarray}
v=\frac{\vert M\vert}{E+p(\varepsilon,n)}\,.
\end{eqnarray}
Inserting Eq.\ (\ref{TrafoExpressions}) leads to a fixed point equation for $v$ 
\begin{eqnarray}
v=\frac{\vert M\vert}{E+p(E-\vert M\vert\cdot\vert v\vert, R\sqrt{1-v^2})}\,.
\end{eqnarray}
Thus, for given $E$, $\vec{M}$, and $R$, the modulus of the fluid velocity can be determined. 
From that the fluid velocity vector $\vec{v}$ can be reconstructed using $\vec{v}=v\vec{M}/M$. 
Finally, the expressions of Eq.\ (\ref{TrafoExpressions}) return the energy and charge densities 
$\varepsilon$ and $n$ in the LRF. The pressure $p$ can then be determined using the 
EoS $p(\varepsilon,n)$.

\section[The SHASTA]
{The SHASTA}
\label{SHASTAAlgorithm}

In order to model heavy-ion collisions with ideal fluid dynamics, a discretized version of the five 
conservation equations (\ref{DGL1}) -- (\ref{DGL3}) has to be solved. However, due to the fact 
that a system of coupled, nonlinear equations is studied, complex algorithms have to be applied. 
One of those is the SHASTA (SHarp And Smooth Transport Algorithm) 
\cite{Rischke:1995ir,BorisBook1,BorisBook2,Rischke:1995mt,Rischke:1995pe} that will be used in 
this thesis. Another prominent algorithm is the (relativistic) HLLE 
(Harten--Lax--van Leer--Einfeldt) algorithm, originally proposed in Ref.\ \cite{Schneider:1993gd}. 
Both algorithms are euclidian and explicit, i.e., the local rest frame quantities $E$, $\vec{M}$,
and $R$ are discretized on a fixed, euclidean grid and are calculated at discrete time steps $t_n$, 
involving only quantities from previous time steps $t_{n-1}$.\\
The basic idea is to construct the full $3$-dimensional solution by solving sequentially three 
one-dimensional problems, appling the so-called method of {\it operator splitting}. In general, 
the numerical algorithm solves equations of the type
\begin{eqnarray}
\partial_t U+\partial_x(Uv+f)=0\,,
\label{BasicSHASTA}
\end{eqnarray}
where $U$ is one of the quantities $E,\vec{M}$, or $R$, and $f$ represents one of the source terms 
on the r.h.s.\ of equations (\ref{DGL1}) -- (\ref{DGL3}). For a general review, see e.g.\ Ref.\ 
\cite{Rischke:1998fq}.\\
To smooth over instabilities and oscillations, Boris and Book \cite{BorisBook1,BorisBook2} 
developed the method of {\it flux correction} (FCT, Flux Corrected Transport). An antidiffusion 
term is added to the transport scheme after each timestep.\\
Since the equations of relativistic hydrodynamics are hyperbolic, they respect the causality
requirement. Any algorithm solving the finite difference of a hyperbolic differential equation 
has to fulfill the Courant--Friedrichs--Lewy criterion \cite{CFLTheorem}
\begin{eqnarray}
\frac{\Delta t}{\Delta x}\equiv\lambda<1\,,
\end{eqnarray}  
ensuring that matter is transported causally within each time step, $\Delta t=\lambda\Delta x$. 
Since the algorithm averages the propagated quantities over a cell after each time step, it may 
happen that due to this process matter is distributed over an acausal spatial distance. This 
purely numerical phenomenon is called prediffusion. Some FCT algorithms like the SHASTA require 
that $\lambda < 1/2$. It was shown in Refs.\ \cite{Rischke:1995ir,Rischke:1995pe} that 
$\lambda=0.4$ is a convenient choice for the algorithm used in the following.

\section[Initial Conditions]
{Initial Conditions}
\label{InitialConditions}
\begin{figure}[t]
\centering
\begin{minipage}[t]{4.2cm}
\hspace*{-2.0cm}
  \includegraphics[scale = 0.52]{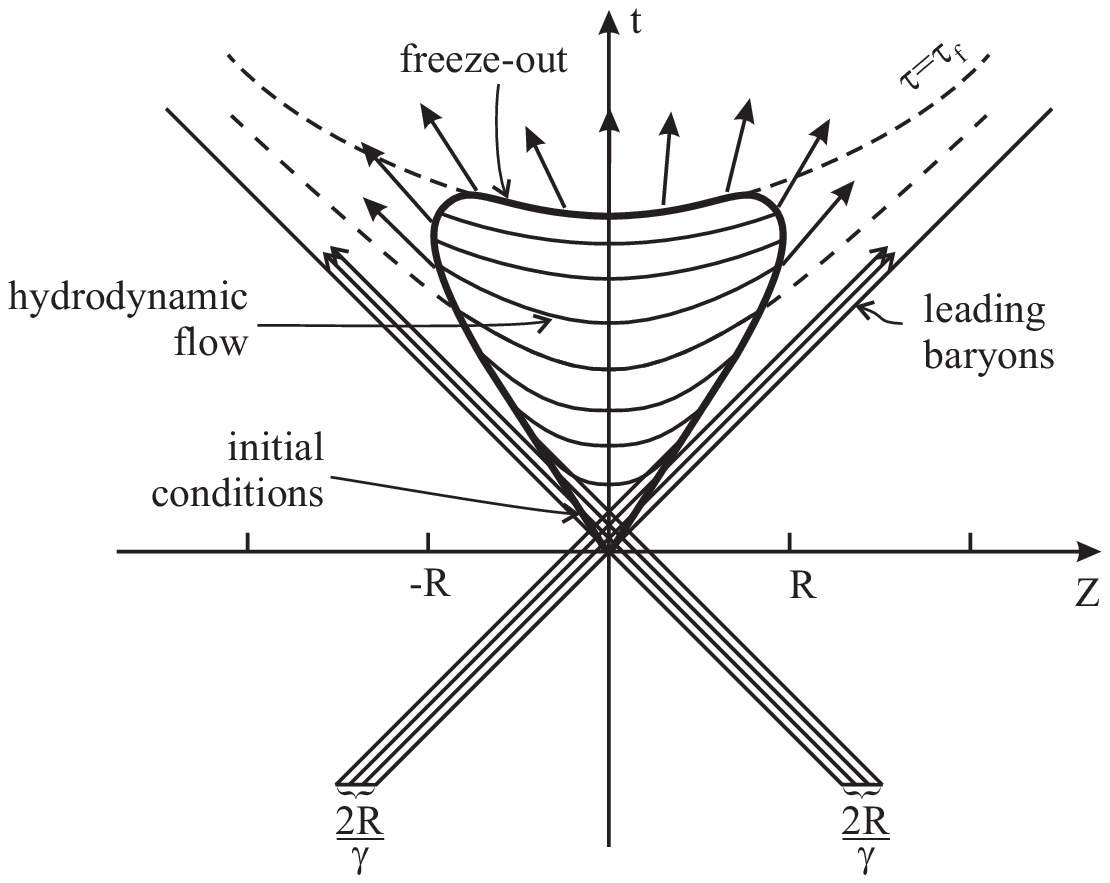}
\end{minipage}
\vspace*{1ex}  
\begin{minipage}[t]{4.2cm}  
  \includegraphics[scale = 0.52]{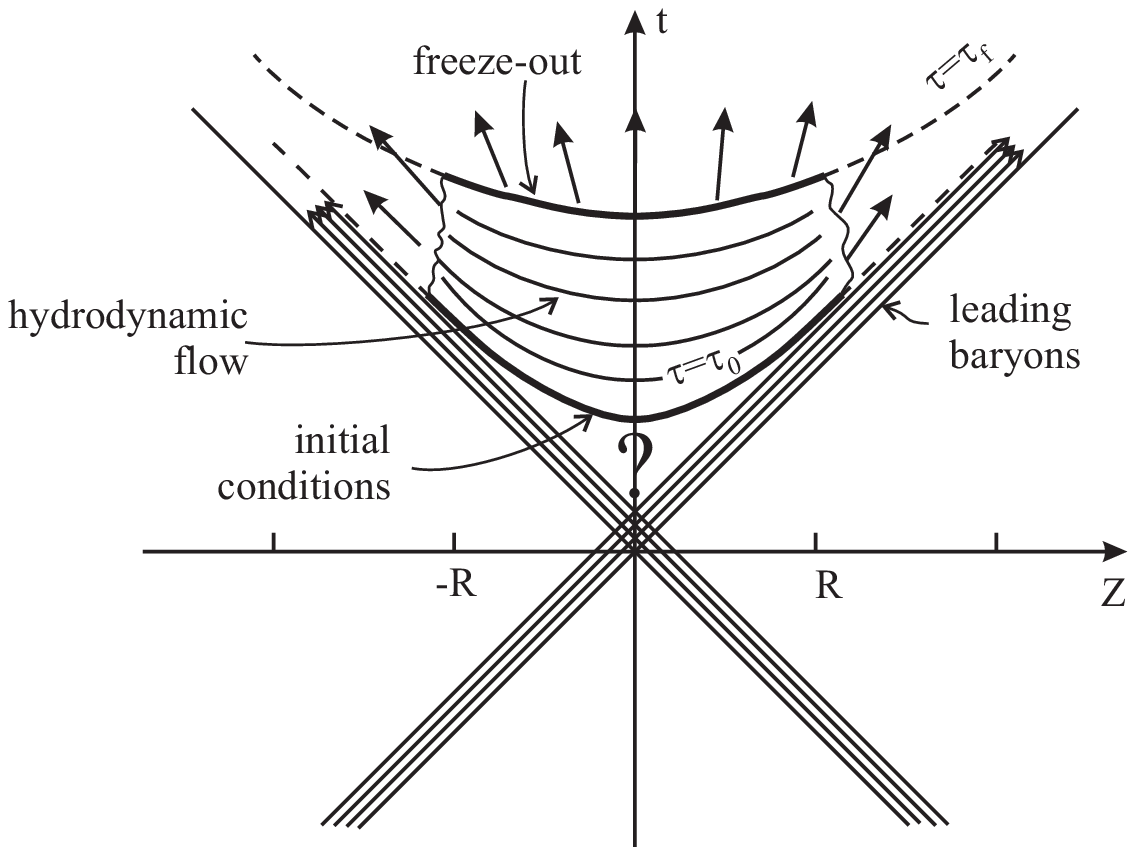}
\end{minipage}
  \caption[Schematic picture of the temporal evolution of a system formed in a heavy-ion
  collision according to the Landau and Bjorken model.]
  {Schematic picture of the temporal evolution of a system formed in a heavy-ion
  collision according to the Landau (left panel) and Bjorken model (right panel) 
  \cite{LectureMishustin}.}
  \label{SketchLandau}
\end{figure}

As mentioned above, the hydrodynamical framework provides a useful tool to study
high-energy nuclear collisions. However, the initial conditions are not specified by hydrodynamics 
and have to be chosen in a proper way, defining the thermodynamic state of matter shortly after 
the impact. The evolution of the collision in longitudinal direction, normally chosen to be along 
the $z$-axis in direction of the beam, is usually described via two idealizations:
\begin{itemize}
\item {\bf Landau Model:} 
Historically, this was the first scenario where fluid dynamics was applied to hadron-hadron
collisions \cite{LandauModel}. It characterizes low-energy collisions and
assumes that the nuclei are completely stopped during the impact, creating a highly 
excited, baryon-free medium at rest which is immediately thermalized. The EoS of this medium is 
supposed to have a simple form of $p=c_s^2\varepsilon$, with the speed of sound $c_s={\rm const.}$
Though Lorentz contraction is weak, the initial extension of the medium in longitudinal direction 
is much smaller than in the transverse plane, resulting in a mainly longitudinal and thus 
one-dimensional expansion. Experimental results are in reasonable agreement with the Landau 
model, however, for unrealistic values of the initial energy.
\item {\bf Bjorken Model:} It describes the penetration of two nuclei through each other and is
therefore applied to collisions with a high center-of-mass energy. Though already investigated 
earlier \cite{Cooper:1974qi}, it was Bjorken who first applied it to hadron-hadron collisions 
\cite{Bjorken:1982qr}. After the impact, again due to the limited amount of nuclear stopping 
power, the baryon charges keep on moving along the light cone, while microscopic processes lead 
to the creation of hot and dense matter in the collision zone that is assumed to thermalize 
within a time of $\tau_0$. In contrast to the Landau model, the collective velocity of this 
medium is supposed to scale like $v_z=z/t$. However, this condition leads to boost 
invariance\footnote{Boost invariance describes the invariance of a system under Lorentz boost which
is a transformation to a system moving with different velocity.}:
Energy density and pressure are independent of the longitudinal coordinate $z$ if it is compared 
at the same proper time $\tau=\sqrt{t^2-z^2}$. Such curves of constant $\tau$ describe hyperbola 
in space-time as can be seen in Fig.\ \ref{SketchLandau}. The disadvantage of this model is 
that it becomes independent of the rapidity since the EoM are boost-invariant along the beam axis. 
In Ref.\ \cite{Huovinen:2003fa} it was shown that at RHIC energies boost-invariant and 
non-boost-invariant calculations lead to very similar results. 
\end{itemize} 
One possibility to determine the initial conditions in the transverse plane is given by the 
{\bf Glauber Model} \cite{Kolb:2001qz} which is described in detail in appendix 
\ref{AppGlauberModel}. It is based on the assumption that in ultra-relativistic heavy-ion
collisions the participating partons can be described as colliding particles. Thus, nucleus-nucleus
collisions are treated as multiple nucleon-nucleon collisions. Starting from a Woods--Saxon 
distribution for a nucleus,
\begin{eqnarray}
\rho_A(\vec{x})&=&\frac{\rho_0}{1+\exp[(\vert\vec{x}\vert-R_A)/d]}\,,
\end{eqnarray}
where $R_A=6.4$~fm and $d=0.54$~fm assuming a gold nucleus with mass number $A=197$, the nuclear 
thickness function
\begin{eqnarray}
T_A(x,y) &=& \int\limits_{-\infty}^{+\infty}\,dz \rho_A(\vec{x})
\end{eqnarray}
is calculated, describing the part of a nucleus A which is seen by a nucleon that passes through 
this nucleus. The parameter $\rho_0$, indicating the mean density in the nucleus, is chosen to
fulfill the condition $\int\,d^3\vec{x}\rho_A(\vec{x})=A$. Subsequently, see appendix 
\ref{AppGlauberModel}, the number density of participating nucleons ($n_{WN}$), also referred to 
as wounded nucleons, and the number density of binary collisions ($n_{BC}$), characterizing 
the number of inelastic nucleon-nucleon collisions, can be calculated via
\begin{eqnarray}
\hspace*{-0.9cm}
n_{WN}(x,y,b)&=&T_A\left(x+\frac{b}{2},y\right)
\left[1-\left(1-\frac{\sigma_{NN}T_A\left(x-\frac{b}{2},y\right)}{A}\right)^A\right]
\nonumber\\
&&\hspace*{-0.3cm}
+T_A\left(x-\frac{b}{2},y\right) 
\left[1-\left(1-\frac{\sigma_{NN}T_A\left(x+\frac{b}{2},y\right)}{A}\right)^A\right]\,,
\\\hspace*{-0.9cm}
n_{BC}(x,y,b)&=&\sigma_{NN}T_A\left(x+\frac{b}{2},y\right)T_A\left(x-\frac{b}{2},y\right)\,,
\end{eqnarray}
where $\sigma_{NN}$ is the nucleon-nucleon cross section which is assumed to be 
$\sigma\simeq 40$~mb for a Au+Au collision at $\sqrt{s}_{NN}=200$~GeV. While the total number of 
participating nucleons $N_{WN}(b)=\int\,dxdy \,n_{WN}(x,y,b)$ can be used to define the centrality 
class of a collision, Glauber-model initial conditions are characterized by the parametrization
\cite{Luzum:2008cw}
\begin{eqnarray}
\varepsilon(\tau=\tau_0,x,y,b) &=& {\rm const}\times n_{BC}(x,y,b)\,,
\label{GlauberIC}
\end{eqnarray}
which leads to a reasonable description of experimentally measured multiplicity distributions. 
The constant is usually chosen to guarantee that the central energy density in the overlap region 
corresponds to a predefined temperature. As proven e.g.\ in Refs.\ \cite{Kolb:2003dz,Luzum:2008cw},
this temperature can eventually be employed as a fitting parameter for the multiplicities.\\
The above expression for the energy density can be used as initial condition for a hydrodynamic 
simulation of a heavy-ion collision since such a numeric application propagates (as shown in the 
previous sections) an initial energy and velocity distribution.\\
When studying jets traversing a hydrodynamical medium in part \ref{part03}, we will first focus 
on a static medium. However, chapter \ref{ExpandingMedium} discusses an expanding system where 
Glauber-model initial conditions [as defined by Eq.\ (\ref{GlauberIC})] are applied, though 
any longitudinal flow will be neglected.\\
However, there are still theoretical uncertainties regarding the Glauber approach and it is 
discussed if plasma-instability models \cite{Arnold:2003rq} or a color-glass condensate 
\cite{Iancu:2002xk} might be more appropriate to specify the initial stage. The latter one is 
based on the idea of gluon saturation at high energies and, as shown in Ref.\ \cite{Luzum:2008cw}, 
does not seem to require early thermalization in order to describe the elliptic flow of charged 
hadrons.

\section[The Equation of State]
{The Equation of State}
\label{EquationofState}

Once an initial condition is specified, the hydrodynamic evolution is uniquely determined by the 
Equations of Motion (see section\ \ref{HydroEoM}). However, the essential information needed is 
the Equation of State (EoS), providing a correlation between the pressure $p(\varepsilon,n_i)$, 
the energy density $\varepsilon$, and the charge densities $n_i$ of the system.\\
Several different EoS are used in hydrodynamical applications, since they also offer the 
possibility to study phase transitions. Thus, some of the applied EoS try to model the phase 
diagram of QCD (see sec.\ \ref{PhaseDiagram}) to check its properties.\\
The simplest EoS is the one for an ideal gas of massless non-interacting massless particles 
given by
\begin{eqnarray}
p(\varepsilon) &=& \frac{\varepsilon}{3}\,,
\end{eqnarray}
which is independent of the charge current. The relation is derived in appendix \ref{EoSIdealGas}. 
This EoS is certainly an idealization, but appropriate to study the intrinsic characteristics of 
certain effects -- as we are going to do in the course of this thesis. 
Therefore, the results presented in part \ref{part03} will focus on the ideal gas EoS.
In particular, a gas of massless gluons will be considered, for which 
\begin{eqnarray}
\varepsilon (T) &=& \frac{g}{30}\pi^2T^4\hspace*{1.0cm}{\rm and}\hspace*{1.0cm}g=2\cdot 8 =16\,.
\end{eqnarray}
Another very common EoS contains a first order phase transition \cite{Kolb:2003dz,Rischke:1995mt}. 
In Ref.\ \cite{Rischke:1995mt}, the MIT-Bag Model\footnote{The MIT Bag 
Model describes a bag in which the quarks are allowed to move around freely.} 
\cite{Chodos:1974je} specifies the QGP phase and is coupled via Gibbs' conditions for phase 
equilibrium\footnote{Gibbs' conditions for phase equilibrium imply that $p_{\rm had}=p_{\rm QGP}$,
$T_{\rm had}=T_{\rm QGP}$, and $\mu_{\rm had}=\mu_{\rm QGP}$.} to a hadronic phase described by a 
modified version of the $\sigma$-$\omega$~model \cite{Serot:1984ey}.\\
Likewise, an EoS characterizing a cross-over transition to a chiral condensate 
\cite{Steinheimer:2007iy} and an EoS based on Lattice QCD calculations 
\cite{Luzum:2008cw,NoronhaHostler:2008ju} were employed to hydrodynamic models. \\
The EoS defines the speed of sound $c_s$ of the medium studied,
\begin{eqnarray}
c_s^2=\left.\frac{\partial p}{\partial\varepsilon}\right\vert_{s/n}\,.
\label{speedofsound}
\end{eqnarray}
For an ideal gas EoS, the speed of sound is $c_s=\sqrt{1/3}\approx 0.577$.

\section[The Freeze-out]
{The Freeze-out}
\label{Freezeout}
\begin{figure}[t]
\centering
  \includegraphics[scale = 0.25]{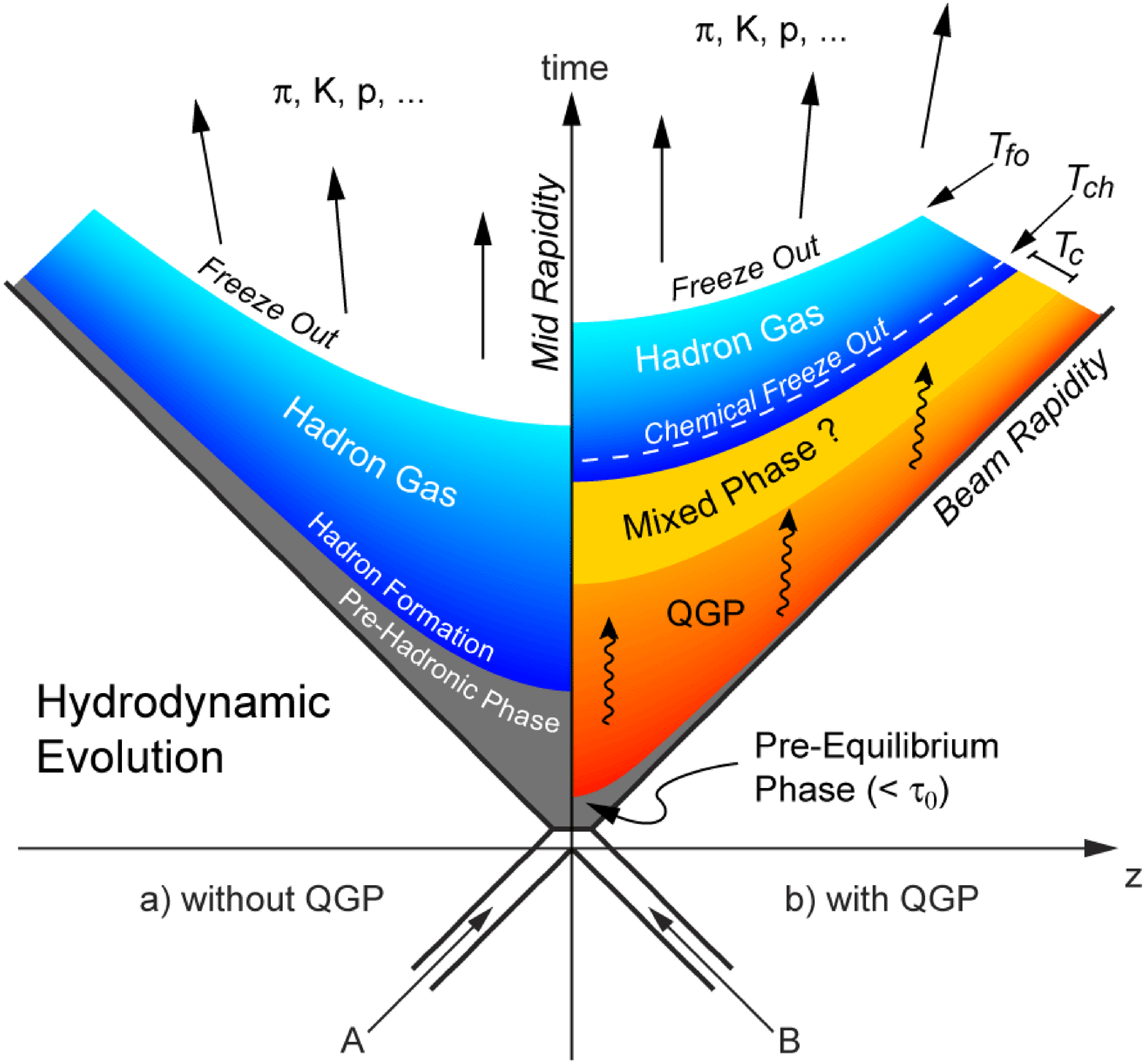}
  \caption[Schematic representation of a heavy-ion collision considering a scenario with and 
  without the creation of a QGP.]
  {Schematic representation of a heavy-ion collision, plotted as a function of $t$ and $z$, 
  considering a scenario without (left panel) and with (right panel) the creation of a QGP 
  \cite{Chemout}.}
  \label{SketchHeavyIon}
\end{figure}

\begin{figure}[t]
\centering
  \includegraphics[scale = 0.7]{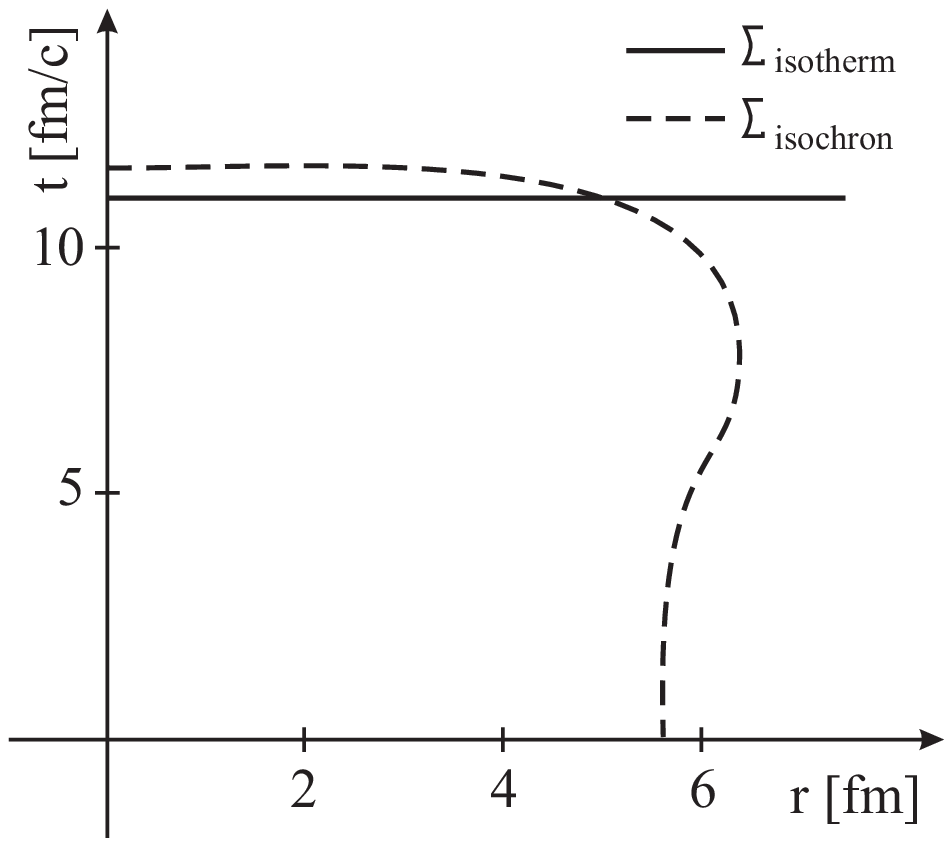}
  \caption[A schematic picture of the space-time distribution of an isochronous and isothermal 
  hypersurface.]
  {A schematic picture of the space-time distribution of an isochronous and isothermal 
  hypersurface.}
  \label{SketchHypersurface}
\end{figure}

Hydrodynamical calculations provide the temporal evolution of fields, like e.g.\ the temperature 
and the velocity (flow) fields. In order to compare the result of such a hydrodynamic evolution 
to experimentally measured observables, a description for the conversion of the fluid into 
particles is needed. \\
Consequently, the hydrodynamical configuration has to be translated into an emission profile, 
providing the average number of particles from the fluid with a momentum $p$. A schematic 
representation of the complete process is given in Fig.\ \ref{SketchHeavyIon}.\\
One of the common approaches used is the {\bf Cooper--Frye} freeze-out \cite{Cooper:1974mv}. 
Here, the above mentioned conversion of the fluid into free particles is achieved instantaneously 
at a critical surface $d\Sigma_\mu$ in space-time (see Fig.\ \ref{SketchHypersurface}). 
Applying a thermal distribution function, the emission pattern can be calculated via
\begin{eqnarray}
E\frac{dN}{d^3\vec{p}} = \int_{\Sigma} d\Sigma_\mu\, p^\mu\, f(u\cdot p/T)\,,
\label{CooperFormula}
\end{eqnarray}
where $T$ is the temperature and $u^\mu$ the flow at the freeze-out position. Here, $f$ denotes 
the Boltzmann, Fermi--Dirac or Bose--Einstein distribution. Since 
\begin{eqnarray}
\frac{dN}{d^3\vec{p}}&=&\frac{1}{E}\frac{dN}{p_Tdp_Td\varphi dy}\,,
\end{eqnarray}
with the rapidity $y$ and the azimuthal angle $\varphi$, the Cooper--Frye formula can be 
rewritten to the form usually applied in heavy-ion physics
\begin{eqnarray}
\frac{dN}{p_Tdp_Td\varphi dy}&=&\int_{\Sigma} d\Sigma_\mu\, p^\mu\, f(u\cdot p/T)\,.
\label{CFFormula}
\end{eqnarray}
In the following, we will focus on the Boltzmann distribution
\begin{eqnarray}
f(u\cdot p/T)&=&\frac{g}{(2\pi)^3}\exp\left[-\frac{u^\mu p_\mu}{T(x)}\right]\,.
\label{BoltzmannDistribution}
\end{eqnarray}

\subsection[The Hypersurface]
{The Hypersurface}
\label{Hypersurface}

The hypersurface indicates the space-time location for particle emission according to some 
criterion which is typically chosen to be a certain time or a certain temperature, specifying an 
{\it isochronous} or {\it isothermal} freeze-out, respectively, see Fig.\ 
\ref{SketchHypersurface}. \\
Since the code \cite{Rischke:1995pe} applied for the hydrodynamical simulations in part 
\ref{part03} is defined in $(t,x,y,z)$, we will use Cartesian coordinates below. For the 
parameterization of a boost-invariant surface, depending on the rapidity, see e.g.\
Ref.\ \cite{KolbThesis}.\\
Thus, the hypersurface is defined as
\begin{eqnarray}
\Sigma^\mu=\Sigma^\mu(t,\vec{x})=\Sigma^\mu[t(\vec{x}),\vec{x}]\,,
\end{eqnarray}
and the normal vector on such a surface is given by
\begin{eqnarray}
d\Sigma^\mu=\varepsilon^{\mu\nu\lambda\rho}
\frac{\partial\Sigma_\nu}{\partial x}\frac{\partial\Sigma_\lambda}{\partial y}
\frac{\partial\Sigma_\rho}{\partial z}d^3\vec{x}\,,
\label{normalhyper}
\end{eqnarray}
where $\varepsilon^{\mu\nu\lambda\rho}$ denotes the Levi--Civita tensor
\begin{eqnarray}
\varepsilon^{\mu\nu\lambda\rho}=-\varepsilon_{\mu\nu\lambda\rho}=
\left( \begin{array}{rc}
1&{\rm even\,\,\, permutation}\\
-1&{\rm odd\,\,\, permutation}\\
0&{\rm otherwise}
\end{array} \right)\,.
\end{eqnarray}
Generally, $d\Sigma^\mu$ can be either space-like ($d\Sigma=g_{\mu\nu}d\Sigma^\mu d\Sigma_\nu < 0$) 
or time-like ($d\Sigma> 0$).

\subsection[Isochronous Freeze-out]
{Isochronous Freeze-out}
\label{IsochronousFO}

An isochronous freeze-out describes the particle emission at a certain time $t={\rm const}$. 
Therefore, the normal vector of the hypersurface, applying Eq.\ (\ref{normalhyper}), becomes
\begin{eqnarray}
d\Sigma_\mu&=&\left(1,\vec{0}\right)d^3\vec{x}\,
\end{eqnarray}
which leads to a straight line in the space-time representation shown in Fig.\ 
\ref{SketchHypersurface}. The general expression\footnote{Here the transformations $E=m_T\cosh y$ 
and $p_z=m_T\sinh y$ are used.} for the four-momentum of the particle is\footnote{Actually 
several particles species have to be considered for a correct prescription of the freeze-out 
leading to various particles with different masses.} 
\begin{eqnarray}
p^\mu &=&\left[m_T\cosh y,\vec{p}_T,m_T\sinh y\right]\,,
\label{fourmomentum}
\end{eqnarray}
where $m_T=\sqrt{\vec{p}_T^{\;2}+m^2}$ denotes the transverse mass of the particle. This relation 
fulfills the condition $p^\mu p_\mu=m^2$. After choosing a transverse momentum distribution, where 
$\vec{p}_T=p_T(\cos\varphi,\sin\varphi)$, and inserting the Boltzmann distribution Eq.\ 
(\ref{BoltzmannDistribution}), the particle emission pattern can be computed via
\begin{eqnarray}
\hspace*{-0.6cm}
\frac{dN}{p_Tdp_Td\varphi dy}&=&\frac{g}{(2\pi)^3}\int\,d^3\vec{x}\,m_T\cosh y\nonumber\\
&&\hspace*{1.0cm}\times\exp\Big\{-\frac{\gamma}{T}\Big[m_T\cosh y-p_T\,v_x\cos\varphi\nonumber\\
&&\hspace*{3.2cm}-p_T\,v_y\sin\varphi - m_T\,v_z\sinh y\Big]\Big\}\,.
\label{CFFOIsochron}
\end{eqnarray}

\subsection[Isothermal Freeze-out]
{Isothermal Freeze-out}
\label{IsothermalFO}
The isothermal freeze-out characterizes the decoupling of the fluid if the temperature has fallen 
below a certain critical temperature $T_{FO}$. In that case, the normal vector of the 
hypersurface [see Eq.\ (\ref{normalhyper})] is 
\begin{eqnarray}
d\Sigma_\mu&=&\left(1,-\partial_x t,-\partial_y t,-\partial_z t\right)d^3\vec{x}\,.
\end{eqnarray}
Thus, for the same assumptions as in the previous section [namely for inserting the four-momentum 
of Eq.\ (\ref{fourmomentum}) and considering a Boltzmann distribution], one obtains
\begin{eqnarray}
\hspace*{-1.0cm}
\frac{dN}{p_Tdp_Td\varphi dy}&=&\frac{g}{(2\pi)^3}\int\,d^3\vec{x}\,
\Big[m_T\cosh y-(\partial_x t)\,p_T\cos\varphi\nonumber\\
&&\hspace*{2.4cm}-(\partial_y t)\,p_T\sin\varphi-(\partial_z t)\,m_T\sinh y\Big]\nonumber\\
&&\hspace*{1.0cm}\times\exp\Big\{-\frac{\gamma}{T}\Big[m_T\cosh y-p_T\,v_x\cos\varphi\nonumber\\
&&\hspace*{3.2cm}-p_T\,v_y\sin\varphi - m_T\,v_z\sinh y\Big]\Big\}\,.
\end{eqnarray}
\\
As mentioned above, the Cooper--Frye method implies an instantaneous conversion of the fluid 
fields into free particles, assuming that the mean-free path immediately changes from zero to 
infinity. Since viscosity (which assumes a non-zero mean-free path within the fluid) increases 
during the last stages of the fireball evolution, it is possible that this direct conversion and 
emission generates considerable unphysical artifacts. \\
Therefore, hydrodynamical evolutions were coupled to hadronic kinetic models like the UrQMD
\cite{Bleicher:1999xi,Steinheimer:2007iy,Nonaka:2005aj,Petersen:2008dd}. 
The transition between these two models, however, is still done using a Cooper--Frye prescription. 
Unfortunately, the decoupling time of the fluid is unknown and because of effects like rescattering
and coalescence \cite{Fries:2003vb,Greco:2003xt} at the late stages of the hadronic evolution
one might have to revise the freeze-out prescription.

\subsection[Bulk Flow Freeze-out]
{Bulk Flow Freeze-out}
\label{BulkFlowFO}
As an alternate freeze-out scheme we will also consider a calorimetric-like observable (called 
{\bf bulk flow freeze-out}) in part \ref{part03}, describing a polar-angle distribution weighted
by the momentum density
\begin{eqnarray}
\frac{d S}{d\cos\theta} &=& \sum_{cells} |\vec{\mathcal{P}}_c|
\delta\left(\cos\theta- \cos\theta_c\right)\nonumber \\
&=& \int d^3 {\bf x}\,\, |\vec{M}(t,\vec{x})|
\delta\left.\left[\cos\theta- \frac{M_x(t,\vec{x})}{|{\vec M}(t,\vec{x})|}\right]\right\vert_{t_f}
\label{bulkeqMiklos}
\end{eqnarray}
that was presented in Ref.\ \cite{Betz:2008wy}. Its basic assumption is that all the particles 
inside a given small subvolume of the fluid will be emitted in the same direction. This quantity 
differs from the Cooper--Frye freeze-out mainly by the neglect of the thermal smearing at the 
freeze-out time. \\
The very strong assumption in this decoupling scheme is that hadrons from each frozen-out cell
emerge parallel to the cell total momentum $\mathcal{P}^i_c=d^3{\vec x}\,\, T^{0i}(t_f,{\vec x})$.
Many other similar purely hydrodynamic measures of bulk flow are possible \cite{Stoecker:2004qu}, 
e.g.\ weighting by energy or entropy density instead of momentum density. 
Indeed, an equivalent measure is given by the energy flow distribution \cite{Betz:2008ka}
\begin{equation}
\frac{d \mathcal{E}}{d\phi dy} = 
\int d^3 {\vec x}\,\, \mathcal{E}(\vec{x})\,
\delta\left[\phi - \Phi(\vec{x})\right]\, 
\delta\left[y-Y(\vec{x})\right]\,,
\label{bulkeq}
\end{equation}
where $\phi$ denotes the azimuthal angle and $y$ the rapidity. 
However, we found no qualitative differences when changing the weight function.

%
%
\chapter[Viscous Hydrodynamics]{Viscous Hydrodynamics}
\label{ViscousHydrodynamics}

One of the major successes at RHIC was to show that the medium created during a heavy-ion collision 
behaves like a ``nearly perfect fluid'' \cite{Arsene:2004fa,Adcox:2004mh,Back:2004je,Adams:2005dq,
Shuryak:2003xe}. This implies that dissipative effects have to be small, but it does not answer 
the question {\it how} perfect the ``nearly perfect liquid'' is, an issue that cannot be answered 
using ideal hydrodynamics. Therefore, dissipative quantities have to be included into the 
description of heavy-ion collisions in order to gain a qualitative understanding. \\
It is necessary to apply viscous hydrodynamics for various realistic initial conditions (which 
are unfortunately not known explicitly for a heavy-ion collision) to confirm the smallness of the 
dissipative quantities. Recently \cite{Luzum:2008cw,Muronga:2001zk,Muronga:2004sf,Heinz:2005bw,
Baier:2006um,Chaudhuri:2006jd,Baier:2006gy,Muronga:2006zw,Muronga:2006zx,Dusling:2007gi,
Molnar:2008xj,El:2008yy}, theoreticians just began to study those viscous effects systematically 
and to apply a corresponding description to the collective flow.  

\section[Tensor Decomposition of $N^\mu$ and $T^{\mu\nu}$]
{Tensor Decomposition of $N^\mu$ and $T^{\mu\nu}$}
\label{TensorDecomposition}
As already discussed in the previous chapter, relativistic hydrodynamics implies the (local) 
conservation of energy-momentum and any (net) charge current, Eqs.\ (\ref{EnMomConservation}) and 
(\ref{ChargeConservation}). The tensor decomposition of the respective quantities, $N^\mu$ and 
$T^{\mu\nu}$, with respect to an arbitrary, time-like, normalized $4$-vector 
$u^\mu=\gamma(1,\vec{v})$ ($u_\mu u^\mu=1$) reads
\begin{eqnarray}
\label{ViscousCharge}
N_i^\mu&=&n_i u^\mu + V_i^\mu\,,\\
\label{ViscousTmunu}
T^{\mu\nu}&=&\varepsilon u^\mu u^\nu -(p+\Pi)\Delta^{\mu\nu}+2q^{(\mu}u^{\nu)}+\pi^{\mu\nu}\,.
\end{eqnarray}
In the LRF, where $u^\mu=(1,\vec{0})$, 
\begin{eqnarray}
n_i&\equiv& N_i^\mu u_\mu 
\end{eqnarray}
is the {\it net charge density},
\begin{eqnarray}
V_i^\mu&\equiv&\Delta^\mu_\nu N^\nu_i
\end{eqnarray}
the {\it diffusion current},
\begin{eqnarray}
\varepsilon&\equiv&u_\mu T^{\mu\nu}u_\nu
\end{eqnarray}
the {\it energy density},
\begin{eqnarray}
p+\Pi&\equiv&-\frac{1}{3}\Delta^{\mu\nu}T_{\mu\nu}
\end{eqnarray}
the sum of the {\it thermodynamical pressure} $p$ and {\it bulk viscous pressure} $\Pi$,
\begin{eqnarray}
\Delta^{\mu\nu}&=&g^{\mu\nu}-u^\mu u^\nu
\end{eqnarray}
denotes the projector onto the $3$-space orthogonal to $u^\mu$,
\begin{eqnarray}
q^\mu&\equiv&\Delta^{\mu\nu}T_{\nu\lambda}u^\lambda
\end{eqnarray}
is the {\it heat flux current} and
\begin{eqnarray}
\pi^{\mu\nu}&\equiv&T^{\langle\mu\nu\rangle}
=\left(\Delta^\mu_{(\alpha}\Delta^\nu_{\beta)}
-\frac{1}{3}\Delta^{\mu\nu}\Delta_{\alpha\beta}\right)\,T^{\alpha\beta}
\end{eqnarray}
the {\it shear stress tensor}. The notation
\begin{eqnarray}
a^{(\mu\nu)}&\equiv&\frac{1}{2}\big(a^{\mu\nu}+a^{\nu\mu}\big)
\end{eqnarray}
stands for the symmetrization in all Lorentz indices and
\begin{eqnarray}
a^{\langle\mu\nu\rangle}&\equiv&\left(\Delta_\alpha^{\,\,(\mu}\Delta^{\nu)}_{\,\,\,\,\beta}
-\frac{1}{3}\Delta^{\mu\nu}\Delta_{\alpha\beta} \right)\, a^{\alpha\beta}
\end{eqnarray}
represents the symmetrized, traceless spatial projection. By construction,
\begin{eqnarray}
\hspace*{-0.98cm}\Delta^{\mu\nu}u_\nu&=&0\,,\hspace*{0.25cm}V^\mu u_\mu=0\,,
\hspace*{0.25cm}q^\mu u_\mu=0\,,\hspace*{0.25cm}
\pi^{\mu\nu}u_\mu=\pi^{\mu\nu}u_\nu=\pi^\mu_\mu=0\,,\\
\hspace*{-0.98cm}\Delta^{\mu\alpha}\Delta_\alpha^{\,\,\,\,\nu}&=&\Delta^{\mu\nu}\,,
\end{eqnarray}
implying that $q^\mu$ has only three and $\pi^{\mu\nu}$ only five independent components. 
Due to the condition for normalization, $u^\mu$ also has only three independent components. The 
space-time derivative decomposes into
\begin{eqnarray}
\partial^\mu a&=&u^\mu \dot{a}+\nabla^\mu a\,,
\end{eqnarray}
where $\dot{a}=u_\mu \partial^\mu a$ is the convective (comoving) time derivative and 
$\nabla^\mu\equiv\Delta^{\mu\nu}\partial_\nu$ the gradient operator. Differentiating
$u^\mu u_\mu=1$ with respect to the space-time coordinates leads to the relation
\begin{eqnarray}
u_\mu\partial_\nu u^\mu=0\,.
\end{eqnarray}
Obviously, setting the dissipative quantities bulk viscous flow $\Pi$, heat flux current 
$q^\mu$, shear stress tensor $\pi^{\mu\nu}$, and net charge density $V_i^\mu$ to zero
\begin{eqnarray}
\Pi=0\,,\hspace*{0.3cm}\,q^\mu=0,\hspace*{0.3cm}\pi^{\mu\nu}=0,\hspace*{0.3cm}V_i^\mu=0\,,
\end{eqnarray}
Eqs.\ (\ref{ViscousCharge}) and (\ref{ViscousTmunu}) reduce to the ideal-fluid limit discussed 
in chapter \ref{IdealHydrodynamics}, Eqs. (\ref{IdealTmunu}) and (\ref{IdealCharge}), describing 
(local) thermodynamical equilibrium.

\section[The Eckart and Landau Frames]{The Eckart and Landau Frames}
\label{EckartLandau}

So far, $u^\mu$ is arbitrary, but one can give it a physical meaning. Then, $u^\mu$ is a 
dynamical quantity whose three independent components have to be determined ($u^\mu u_\mu=0$ is 
still valid).

\begin{itemize}
\item {\bf Eckart (or Particle) frame}:\\
Here, $u^\mu$ is the $4$-velocity of the (net) charge flow, i.e., the velocity of charge 
transport. Thus, $u^\mu$ is parallel to the (net) charge current $N_i^\mu$ 
\begin{eqnarray}
u^\mu&=&\frac{N_i^\mu}{\sqrt{N_i\cdot N_i}}\,,
\end{eqnarray}
and the diffusion current vanishes,
\begin{eqnarray}
 V_i^\mu&=&0\,.
\end{eqnarray}
\item {\bf Landau (or Energy) frame}:\\
This choice describes $u^\mu$ as the $4$-velocity of the energy flow. Therefore, $u^\mu$ is 
parallel to the energy flow 
\begin{eqnarray}
u^\mu&=&\frac{T^\mu_{\,\,\,\nu} \,u^\nu}
{\sqrt{u^\alpha\, T^\beta_\alpha\, T_{\beta\gamma}\,u^\gamma}}\,,
\end{eqnarray}
and obviously the heat flux current $q^\mu$ vanishes,
\begin{eqnarray}
q^\mu&=&0\,.
\end{eqnarray}
\end{itemize}
Both frames have their advantages. The Eckart frame leads to a simple law for charge 
conservation, while the Landau frame reduces the complexity of the energy-momentum tensor. 
However, considering a system without a (net) charge, the $4$-velocity in the Eckart frame is not 
well defined.

\section[The Fluid Dynamical Equations of Motion]{The Fluid Dynamical Equations of Motion}
\label{FluidEoM}
The tensor decomposition of the $n$ (net) charge currents [Eq.\ (\ref{ViscousCharge})] and the 
energy-momentum tensor [Eq.\ (\ref{ViscousTmunu})] contain $11+4n$ unknowns, 
$\varepsilon, p, n_i(n\,{\rm Eqs.}), \Pi$, the three components of $V^\mu_i(3n\,{\rm Eqs.})$, the 
three components of $q^\mu$, and the five components of $\pi^{\mu\nu}$\footnote{An equivalent 
way of counting says that for any given $4$-velocity $u^\mu$ there are the $17+5n$ unknowns 
$\varepsilon(1\,{\rm Eq.}), p(1\,{\rm Eq.}), n_i(n\,{\rm Eqs.}), \Pi(1\,{\rm Eq.}), 
V^\mu_i(4n\,{\rm Eqs.}), q^\mu(4\,{\rm Eqs.})$ and $\pi^{\mu\nu}(10\,{\rm Eqs.})$,
which are reduced to $11+4n$ unknows by the conditions $q^\mu u_\mu=0$ (1 Eq.), $\pi^\mu_\mu=0$ 
(1 Eq.), $u_\mu\pi^{\mu\nu}=0$ (4 Eqs.), and $u_\mu V^\mu_i=0$ ($n$ Eqs.).}.\\
However, the conservation of the $n$ (net) currents, the energy and the $3$-momentum leads to 
$4+n$ Equations of Motion. In particular, these are 
\begin{itemize}
\item {\bf (Net) charge conservation:}
\begin{eqnarray}
\label{ViscousChargeConservation}
\partial_\mu N^\mu_i&=&\dot{n}_i+n_i\theta+ \partial_\mu V_i^\mu=0\,,
\end{eqnarray}
where $\theta=\partial_\mu u^\mu$ is the so-called expansion scalar.
\item {\bf Energy conservation:}
\begin{eqnarray}
\label{ViscousEnergyConservation}
u_\nu\partial_\mu T^{\mu\nu}&=&\dot{\varepsilon}+\big(\varepsilon+p+\Pi \big)\theta
+u_\mu\dot{q}^\mu+\partial_\mu q^\mu-\pi^{\mu\nu}\partial_\nu u_\mu\nonumber\\
&=&0\,.
\end{eqnarray}
\item{\bf Momentum conservation (acceleration equation):}
\begin{eqnarray}
\label{ViscousMomentumConservation}
\Delta^{\mu\nu}\partial^\lambda T_{\nu\lambda}&=&0\,,\\
\big( \varepsilon+p\big)\dot{u}^\mu&=&\nabla^\mu\big(p+\Pi\big) 
-\Pi\dot{u}^\mu-\Delta^{\mu\nu}\dot{q}_\nu\nonumber\\
&&-q^\mu\theta-q^\nu\partial_\nu u^\mu-\Delta^{\mu\nu}\partial^\lambda\pi_{\nu\lambda}\,.
\end{eqnarray}
\end{itemize}
Thus, to close the system, one needs additional $7+3n$ equations, $6+3n$ equations to determine 
the dissipative quantities $\Pi$ (one Eq.), $q^\mu$ (or $V^\mu_i$, $q^\mu$ will be proportional to 
$n_i$, thus $3n$ Eqs.), $\pi^{\mu\nu}$ (five Eqs.) and the EoS. \\
Those equations of dissipative fluid dynamics are either provided by so-called first-order or 
second-order theories (see below). A first-order theory, like the Navier--Stokes approximation 
discussed in the next section, expresses the dissipative quantities $\pi, q^\mu$ (or $V^\mu_i$), and 
$\pi^{\mu\nu}$ in terms of the variables $\varepsilon,p,n_i,u^\mu$, or their gradients. In a 
second-order theory, those variables are treated as independent dynamical quantities whose 
evolution is described by transport equations which are differential equations. In the following, 
we will again restrict to one conserved charge.

\section[The Navier--Stokes Approximation]{The Navier--Stokes Approximation}
\label{NS}

In the NS approximation, the dissipative quantities $\Pi,\, q^\mu,\, \pi^{\mu \nu}$ are given by
\begin{eqnarray}
\label{eq:PiNS}
\Pi_{\rm NS} & = & - \zeta\, \theta\,,\\
\label{eq:qNS}
q^\mu_{\rm NS} & = & \frac{\kappa}{\beta}\,\frac{n}{\beta (\varepsilon +p)}\, 
\nabla^\mu \alpha\,,\\
\label{eq:piNS}
\pi^{\mu \nu}_{\rm NS} & = & 2 \, \eta \, \sigma^{\mu \nu}\,,
\end{eqnarray}
with the definitions $\beta \equiv 1/T$ and $\alpha \equiv \beta \mu$, where $\mu$ is the chemical 
potential associated with the (net) charge density $n$. $\zeta,\, \kappa$, and $\eta$ denote the 
bulk viscosity, thermal conductivity, and shear viscosity coefficients, and 
$\sigma^{\mu \nu} \equiv \nabla^{< \mu} u^{\nu >}$ is the shear tensor.\\
While Eq.\ (\ref{eq:qNS}) is valid in the Eckart frame, it can easily be adjusted to the Landau 
frame by applying the transformation $q^\mu \rightarrow - V^\mu (\varepsilon +p)/n$.\\
Since $\Pi,\, q^\mu$ (or $V^\mu$), and $\pi^{\mu \nu}$ are only given in terms of the primary 
variables, one obtaines a closed set of Equations of Motion by inserting the above expressions 
into Eqs. (\ref{ViscousChargeConservation}) -- (\ref{ViscousMomentumConservation}). \\
However, these equations will lead to unstable solutions and support acausal propagation of 
perturbations \cite{Hiscock:1985zz}.\\
A viable candidate for a relativistic formulation of dissipative fluid dynamics, which does not 
exhibit these problems, is the so-called second-order theory of Israel and Stewart \cite{IS} 
(for a discussion about stable and causal relativistic fluid dynamics 
see Ref.\ \cite{Hiscock:1983zz}). In the course of this 
chapter, we present the full Israel--Stewart equations of relativistic dissipative fluid dynamics 
as they emerge applying Grad's $14$-moment expansion \cite{Grad} to the Boltzmann equation and 
truncating dissipative effects at second order in the gradients.

\section[A Phenomenological Derivation]{A Phenomenological Derivation}

One possibility to obtain the fluid-dynamical Equations of Motion for the bulk pressure, the heat 
flux current, and the shear stress tensor is a phenomenological approach. Here, the derivation 
shall briefly be introduced, before an alternative (using a derivation applying kinetic theory) is
discussed below in detail. It starts from the second law of thermodynamics, the principle of 
non-decreasing entropy,   
\begin{eqnarray}
\partial_\mu S^\mu&\geq&0\,.
\end{eqnarray}
The next step is to find an ansatz for the entropy. In the limit of vanishing $\Pi$, $q^\mu$, and 
$\pi^{\mu\nu}$, the entropy $4$-current should reduce to the one of the ideal fluid 
$S^\mu\rightarrow s u^\mu$. Since the only non-vanishing $4$-vector that can be formed from the 
available tensors $u^\mu$, $q^\mu$, and $\pi^{\mu\nu}$ is $\beta q^\mu$, this leads to
\begin{eqnarray}
S^\mu&=&s u^\mu+\beta q^\mu\,.
\end{eqnarray}
With the help of this ansatz as well as the conservation equations of (net) charge and energy, one 
computes that
\begin{eqnarray}
\hspace*{-0.5cm}
T\partial_\mu S^\mu&=&(T\beta-1)\partial_\mu q^\mu+q^\mu (\dot{u}_\mu+T\partial_\mu\beta)
+\pi^{\mu\nu}\partial_\mu u_\nu+\Pi\theta\geq 0\,.
\end{eqnarray}
This inequality can be ensured by choosing the dissipative quantities to agree with the 
Navier--Stokes equations [Eqs. (\ref{eq:PiNS}) -- (\ref{eq:piNS})]. Moreover, it results in
\begin{eqnarray}
\partial_\mu S^\mu &=&\frac{\Pi^2}{\zeta T}-\frac{q^\mu q_\mu}{\kappa T^2}
+\frac{\pi^{\mu\nu}\pi_{\mu\nu}}{2\eta T}\,.
\end{eqnarray}
However, as already mentioned, these equations contain instabilities. A solution is to implement 
corrections of second order in the dissipative quantities into the entropy current \cite{IS,Mueller},
\begin{eqnarray}
S^\mu=su^\mu +\beta q^\mu +Q^\mu\,,
\end{eqnarray}
where
\begin{eqnarray}
Q^\mu&\equiv&\alpha_0\Pi q^\mu+\alpha_1\pi^{\mu\nu}q_\nu + u^\mu (\beta_0\Pi^2+\beta_1 q^\mu
q_\mu+\beta_2\pi^{\nu\lambda}\pi_{\nu\lambda})\,.
\end{eqnarray}
Inserting this expression into $\partial_\mu S^\mu\geq 0$ leads to differential equations for 
$\Pi$, $q^\mu$, and $\pi^{\mu\nu}$ which depend on the coefficients $\zeta,\eta,\kappa$, 
$\alpha_0,\alpha_1$, $\beta_0,\beta_1$, and $\beta_2$. \\
However, this phenomenological ansatz (which was explained in detail e.g.\ by Muronga 
\cite{Muronga:2003ta}) does not determine the values of these coefficients.

\section[Scales in Fluid Dynamics]{Scales in Fluid Dynamics}
\label{Scales}
In order to derive the equations of dissipative fluid dynamics in terms of a gradient expansion, 
one has to know about the scales in fluid dynamics. In principle, there are three length scales 
in fluid dynamics, two microscopic scales and one macroscopic scale. The two microscopic scales
are the thermal wavelength, $\lambda_{\rm th} \sim \beta$, and the mean free path, 
$\ell_{\rm mfp} \sim (\langle \sigma \rangle n)^{-1}$, where $\langle \sigma \rangle$ is the 
average cross section. The macroscopic scale, $L_{\rm hydro}$, is the scale over which the fluid 
fields (like $\varepsilon,\, n, u^\mu, \ldots$) vary, i.e., gradients of these fields are 
typically of order $\partial_\mu \sim L_{\rm hydro}^{-1}$. Due to the relation 
$n^{-1/3} \sim \beta \sim \lambda_{\rm th}$, the thermal wavelength can be interpreted as the 
interparticle distance. However, the notion of a mean-free path requires the existence of 
well-defined quasi-particles. Since this quasi-particle concept is no longer valid in strongly 
coupled theories, these only exhibit two scales, $\lambda_{\rm th}$ and $L_{\rm hydro}$.\\
The ratios of the transport coefficients $\zeta,\, \kappa/\beta$, and $\eta$ to the entropy 
density $s$ are solely determined by the ratio of the two microscopic length scales, 
$\ell_{\rm mfp}/\lambda_{\rm th}$. For the proof we use that 
$\eta \sim (\langle \sigma \rangle\lambda_{\rm th})^{-1}$ and $n \sim T^3 \sim s$,
\begin{eqnarray}
\label{eq:etas}
\frac{\ell_{\rm mfp}}{\lambda_{\rm th}}
\sim \frac{1}{\langle \sigma \rangle n}\, \frac{1}{\lambda_{\rm th}}
\sim \frac{1}{\langle \sigma \rangle \lambda_{\rm th}}\,
\frac{1}{n} \sim \frac{\eta}{s}\;.
\end{eqnarray}
Similar arguments hold for the other transport coefficients.
There exist two limiting cases, the dilute-gas limit, with 
$\ell_{\rm mfp}/\lambda_{\rm th}\sim \eta/s \rightarrow \infty$, and the ideal-fluid limit, where 
$ \ell_{\rm mfp}/\lambda_{\rm th}\sim \eta/s \rightarrow 0$. Estimating 
$\ell_{\rm mfp} \sim \langle \sigma \rangle^{-1}\lambda_{\rm th}^3$, the first case corresponds to 
$\langle \sigma \rangle /\lambda_{\rm th}^2 \rightarrow 0$, thus, the interaction cross section 
is much smaller than the area given by the thermal wavelength. In other words, the average 
distance between collisions is much larger than the interparticle distance, allowing to interpret 
the dilute-gas limit as a weak-coupling limit. Similarly, the ideal-fluid limit corresponds to 
$\langle \sigma \rangle /\lambda_{\rm th}^2 \rightarrow \infty$, describing the somewhat academic 
case when interactions happen on a scale much smaller than the interparticle distance, defining 
the limit of infinite coupling. The interactions get so strong that the fluid assumes locally and 
instantaneously a state of thermodynamical equilibrium.\\
For any value of $\eta/s$ [and, analogously, $\zeta/s$ and $\kappa/(\beta s)$] between these two 
limits, the equations of dissipative fluid dynamics may be applied for the description of the 
system. The situation is particularly interesting for 
$\ell_{\rm mfp}/\lambda_{\rm th} \sim \eta/s \sim 1$ or, equivalently, 
$\langle \sigma \rangle \sim \lambda_{\rm th}^2 \sim T^{-2}$. Then, the problem reveals only 
one microscopic scale $\lambda_{\rm th}$, as e.g.\ in strongly coupled theories. 

\section[The Knudsen Number]{The Knudsen Number}
\label{KnudsenNumber}

The Knudsen number is defined as
\begin{eqnarray}
K \equiv \ell_{\rm mfp}/L_{\rm hydro}\,.
\label{eqKnudsenNumber}
\end{eqnarray}
Since $L_{\rm hydro}^{-1} \sim \partial_\mu$, an expansion in terms of $K$ is equivalent to a 
gradient expansion, i.e., an expansion in terms of powers of $\ell_{\rm mfp} \, \partial_\mu$ as 
it is done by using Grad's method \cite{Grad}.
One important conclusion is that the ratios of the dissipative quantities $\Pi,\, q^\mu$ 
(or $V^\mu$), and $\pi^{\mu \nu}$, assuming that they do not differ too much from their 
Navier--Stokes values, to the energy density are proportional to the Knudsen number. Applying the 
fundamental relation of thermodynamics, $\varepsilon + p = Ts + \mu n$, to estimate 
$\beta\, \varepsilon \sim \lambda_{\rm th}\, \varepsilon \sim s$ and employing 
Eq.\ (\ref{eq:PiNS}), one can show that 
\begin{eqnarray}
\label{eq:Piepsilon}
\frac{\Pi}{\varepsilon} &\sim &
\frac{\Pi_{\rm NS}}{\varepsilon} \sim
\frac{\zeta \, \theta}{\varepsilon} \sim 
\frac{\zeta}{\lambda_{\rm th}\, \varepsilon}\, \lambda_{\rm th}\, \theta \sim
\frac{\zeta}{s} \, \frac{\lambda_{\rm th}}{\ell_{\rm mfp}}\, \ell_{\rm mfp}\, \partial_\mu
u^\mu \nonumber\\&\sim& \frac{\zeta}{s}\, \left( \frac{\ell_{\rm mfp}}{\lambda_{\rm th}}
\right)^{-1}\, K \, \vert u^\mu\vert \sim K \,.
\end{eqnarray}
In the last step, we have employed Eq.\ (\ref{eq:etas}) and the fact that $u^\mu \sim 1$. The 
result is remarkable in the sense that $\Pi/\varepsilon$ is only proportional to $K$, and 
independent of the ratio of viscosity to entropy density which drops out on account of Eq.\ 
(\ref{eq:etas}). Therefore, we can conclude that if the Knudsen number is small, 
$K \sim \delta \ll 1$, the dissipative quantities are small compared to the primary variables
and the system is close to local thermodynamic equilibrium. \\
Then, the equations of viscous fluid dynamics can be systematically derived in terms of a gradient 
expansion or, equivalently, in terms of a power series in $K$ or, equivalently because of Eq.\ 
(\ref{eq:Piepsilon}), in terms of powers of dissipative quantities. At zeroth order in $K$, 
one obtains the equations of ideal fluid dynamics, at first order in $K$, one finds the 
Navier--Stokes equations, and at second order in $K$, the Israel--Stewart equations emerge. \\
The independence of the ratio of dissipative quantities to primary variables from the viscosity 
to entropy density ratio has an important phenomenological consequence. It guarantees that, 
provided that gradients of the macroscopic fluid fields (and thus $K$) are sufficiently small, 
the Navier--Stokes equations [Eqs. (\ref{eq:PiNS}) to (\ref{eq:piNS})] are valid and applicable 
for the description of systems with large $\eta/s$, like water at room temperate and atmospheric 
pressure.

\section[Transport Equations of Dissipative Quantities To Second Order in the Knudsen Number]
{Transport Equations of Dissipative Quantities To \\Second Order in the Knudsen Number}
\label{TransportEq}

To second order in dissipative quantities [or equivalently, because of Eq.\ (\ref{eq:Piepsilon}), 
to second order in the Knudsen number], the relativistic transport equations for the bulk flow 
$\Pi$, the heat flux current $q^\mu$, and the shear stress tensor $\pi^{\mu\nu}$, derived from the 
Boltzmann equation via Grad's method \cite{Grad}, are given by \cite{Betz:2008me}
\begin{eqnarray}
\hspace*{-1cm}\Pi&=&-\zeta\theta - \tau_\Pi\dot{\Pi}
+\tau_{\Pi q}q_\mu \dot{u}^\mu-l_{\Pi q}\nabla_\mu q^\mu - \zeta\hat{\delta}_0\Pi\theta\nonumber\\
\hspace*{-1.1cm}&&+\lambda_{\Pi q}q^\mu\nabla_\mu \alpha
+\lambda_{\Pi\pi}\pi_{\mu\nu}\sigma^{\mu\nu}\,,
\label{Eq:TransportPi}\\
\hspace*{-1cm}q^\mu&=&\frac{\kappa}{\beta}\frac{n}{\beta(\varepsilon+p)}\nabla^\mu \alpha
-\tau_q\Delta^{\mu\nu}\dot{q}_\nu\nonumber\\
\hspace*{-1.1cm}&&-\tau_{q\Pi}\Pi\dot{u}^\mu-\tau_{q\pi}\pi^{\mu\nu}\dot{u}_\nu
+l_{q\Pi}\nabla^\mu\Pi-l_{q\pi}\Delta^{\mu\nu}\partial^{\lambda}\pi_{\nu\lambda}
+\tau_q\omega^{\mu\nu}q_\nu\nonumber\\
\hspace*{-1.1cm}&&-\frac{\kappa}{\beta}\hat{\delta}_1
q^\mu\theta-\lambda_{qq}\sigma^{\mu\nu}q_\nu+\lambda_{q\Pi}\Pi\nabla^\mu\alpha
+\lambda_{q\pi}\pi^{\mu\nu}\nabla_\nu\alpha\,,
\label{Eq:Transportq}\\
\hspace*{-1cm}\pi^{\mu\nu}&=&
2\eta\sigma^{\mu\nu}-\tau_\pi \dot{\pi}^{\langle\mu\nu\rangle}\nonumber\\
\hspace*{-1.1cm}&&+2\tau_{\pi q}q^{\langle\mu}\dot{u}^{\nu\rangle}
+2l_{\pi q}\nabla^{\langle\mu}q^{\nu\rangle}+4\tau_\pi
\pi_\lambda^{\langle\mu}\omega^{\nu\rangle\lambda}
-2\eta\hat{\delta}_2\theta\pi^{\mu\nu}\nonumber\\
\hspace*{-1.1cm}&&
-2\tau_\pi \pi_\lambda^{\langle\mu}\sigma^{\nu\rangle\lambda}
-2\lambda_{\pi q}q^{\langle\mu}\nabla^{\nu\rangle}\alpha+2\lambda_{\pi\Pi}\Pi\sigma^{\mu\nu}\,.
\label{Eq:Transportpimunu}
\end{eqnarray}
For the details of the derivation, see appendix \ref{AppViscousHydro} and Ref.\ \cite{ViscousBetz}.
The expressions for the relaxation times $\tau_\pi, \tau_q$, and $\tau_\pi$ as well as the 
coefficients $\tau_{\pi q}$, $\ell_{\Pi q}$, $\ell_{q \Pi}$, $\ell_{q \pi}$, $\ell_{\pi q}$, 
$\lambda_{\Pi q}$, $\lambda_{\Pi \pi}$, $\lambda_{qq}$, $\lambda_{q \Pi}$, $\lambda_{q \pi}$, 
$\lambda_{\pi q}$, $\lambda_{\pi \Pi}$, and $\hat{\delta}_0,\,\hat{\delta}_1,\, \hat{\delta}_2$ 
will be given in Ref.\ \cite{ViscousBetz}. They are complicated functions of $\alpha$ and $\beta$, 
divided by tensor coefficients of the second moment of the collision integral. \\
The form of the transport equations is invariant of the calculational frame (Eckart, Landau, 
\ldots), however, the values of the coefficients are frame-dependent, since the physical 
interpretation of the dissipative quantities varies with the frame. For instance, $q^\mu$ 
denotes the heat flux current in the Eckart frame, while in the Landau frame, 
$q^\mu \equiv - V^\mu (\varepsilon +p)/n$ is the (negative of the) diffusion current, multiplied 
by the specific enthalpy. For details see again appendix \ref{AppViscousHydro}.\\
While the Navier--Stokes equations [Eq.\ (\ref{eq:PiNS}) -- (\ref{eq:piNS})] are obtained by 
neglecting all terms to second order of the Knudsen number (i.e., by considering solely terms 
to first order in $K$ which are the first terms on the r.h.s.), the so-called simplified 
Israel--Stewart equations (a term taken from Ref.\ \cite{Song:2008si}) emerge by keeping only the 
first two terms on the r.h.s.\ of Eqs.\ (\ref{Eq:TransportPi}), (\ref{Eq:Transportq}), and 
(\ref{Eq:Transportpimunu}). The resulting equations have the simple interpretation that the 
dissipative quantities $\Pi,\, q^\mu$, and $\pi^{\mu \nu}$ relax to their corresponding 
Navier--Stokes values on time scales of $\tau_\Pi,\, \tau_q$, and $\tau_\pi$.\\
For times $t<\tau_i$ ($i=\Pi,q,\pi$) the dissipative quantities $\Pi,\, q^\mu,\, \pi^{\mu \nu}$ 
are driven towards their Navier--Stokes values. Once they are reasonably close to these, the first 
terms on the r.h.s.\ largely cancel against the l.h.s. The further evolution, for times 
$t > \tau_i$, is then determined by the remaining, second-order terms. Thus, these terms 
constitute important corrections for times $t> \tau_i$ and should not be neglected.\\
The third and fourth term in the first line of Eq.\ (\ref{Eq:TransportPi}), the second line of 
Eq.\ (\ref{Eq:Transportq}), and the three first terms in the second line of Eq.\ 
(\ref{Eq:Transportpimunu}) were also obtained by Israel and Stewart \cite{IS}, while the remaining 
second-order terms were missed or neglected. Presumably, Israel and Stewart made the assumption 
that second-order terms containing $\theta$, $\sigma^{\mu \nu}$, or $\nabla^\mu \alpha$ are even 
smaller than suggested by power counting in terms of $K$. Also, the last two terms in the first 
line of Eq.\ (\ref{Eq:TransportPi}), the last three terms of the second line and the first term of 
the third line of Eq. (\ref{Eq:Transportq}) as well as the last three terms of the second line of 
Eq. (\ref{Eq:Transportpimunu}) were obtained by Muronga \cite{Muronga:2006zw}, while the 
other second-order terms do not appear in that publication. A possible reason is that the 
corresponding treatment is based on the phenomenological approach to derive the Israel--Stewart 
equations and terms not generating entropy are absent. \\
The third line of Eq.\ (\ref{Eq:TransportPi}), as well as the last three terms in the third line 
of Eqs.\ (\ref{Eq:Transportq}) and (\ref{Eq:Transportpimunu}) were (with one exception discussed 
below) neither given by Israel and Stewart \cite{IS} nor by Muronga \cite{Muronga:2006zw}, and are 
thus completely new in the discussion of the transport equations for the dissipative quantities 
\cite{Betz:2008me}.\\
If we set $\Pi = q^\mu=0$ in Eq.\ (\ref{Eq:Transportpimunu}), the resulting equation for 
$\pi^{\mu \nu}$ is identical to that found in Ref.\ \cite{Baier:2007ix}. In particular, the first 
term in the third line was already obtained in that paper, where it appeared in the form 
$(\lambda_1/\eta^2)\, \pi_\lambda^{\hspace*{0.1cm}<\mu} \pi^{\nu> \lambda}$. Using the NS value 
(\ref{eq:piNS}) for $\pi^{\nu \lambda}$, which is admissible because we are computing to second 
order in $K$, to this order this is identical to 
$2 (\lambda_1/\eta)\,\pi_\lambda^{\hspace*{0.1cm}<\mu} \sigma^{\nu> \lambda}$. By comparison with 
Eq.\ (\ref{Eq:Transportpimunu}), we thus get a prediction for the coefficient $\lambda_1$ from 
kinetic theory, $\lambda_1 \equiv \tau_\pi\, \eta$, in agreement with Ref.\ \cite{Baier:2007ix}. \\
Note, however, that this discussion so far neglects additional terms which arise at second order
in $K$ when expanding the second moment of the collision integral. (This was already noted in 
Ref.\ \cite{Baier:2007ix}.) This will change the coefficient of the respective term such that it 
is no longer equal to $\tau_\pi$. It will therefore also lead to a different result for 
$\lambda_1$. In a recent study \cite{York:2008rr} a complete calculation was performed. \\
Thus, the treatment discussed above leads to transport equations of the bulk flow $\Pi$, heat flux 
current $q^\mu$, and shear stress tensor $\pi^{\mu\nu}$ to second order in the Knudsen number. It 
is also applicable to non-conformal systems with non-vanishing (net) charge density. In the 
future, it will be necessary to work on a generalization of those equations to a system of various 
particle species \cite{Prakash:1993bt} as well as on the numerical implementation and application 
to modelling the dynamics of heavy-ion collisions.

\section[The Shear Viscosity over Entropy Ratio]
{The Shear Viscosity over Entropy Ratio}
\label{SecEtaOverS}

One measure for the viscosity of a system is the shear viscosity to entropy ratio, $\eta/s$, 
reflecting the ``degree of thermalization'' since for low $\eta$ a large $s$ means that a hot 
system thermalizes quickly. More than 20 years ago, this ratio was already estimated to be 
\cite{Danielewicz:1984ww}
\begin{eqnarray}
\frac{\eta}{s}&\geq&\frac{1}{15}=0.067\,.
\end{eqnarray} 
Since the contribution from shear viscosity to viscous hydrodynamics is larger than from bulk
viscosity or thermal conductivity, progress has mainly been made in 
performing viscous hydrodynamical calculations including the shear terms. From comparing those 
elliptic flow calculations \cite{Luzum:2008cw,Heinz:2001xi,Song:2007fn} to experimentally measured 
data, it was shown that
\begin{eqnarray}
0<\eta/s\, \raisebox{-1.1mm}{$\stackrel{<}{\sim}$}\, 0.2\,.
\end{eqnarray} 
Recently \cite{Xu:2008dv}, a calculation using the parton cascade BAMPS led to 
$0.08 \leq \eta/s \leq 0.15$.

\clearpage{\pagestyle{empty}\cleardoublepage}
%
%
\chapter[Shock Wave Phenomena]{Shock Wave Phenomena}
\label{ShockWavePhenomena}

\begin{figure}[t]
\centering
\begin{minipage}[c]{4.2cm}
\hspace*{-4.5cm}
  \includegraphics[scale = 0.65]{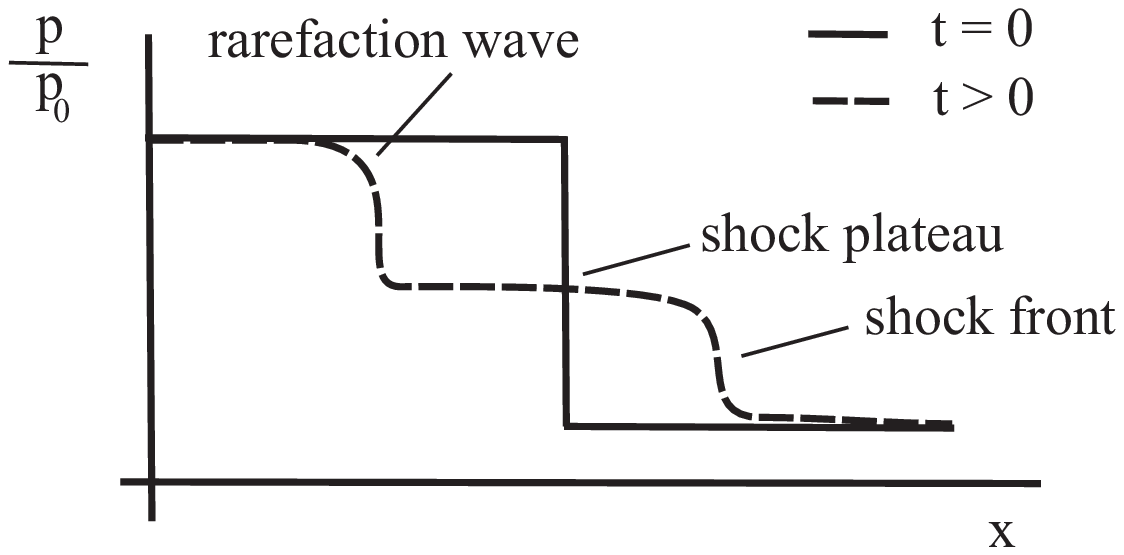}  
\end{minipage}
\hspace*{-1.5cm}  
\begin{minipage}[c]{4.2cm} 
  \includegraphics[scale = 0.57]{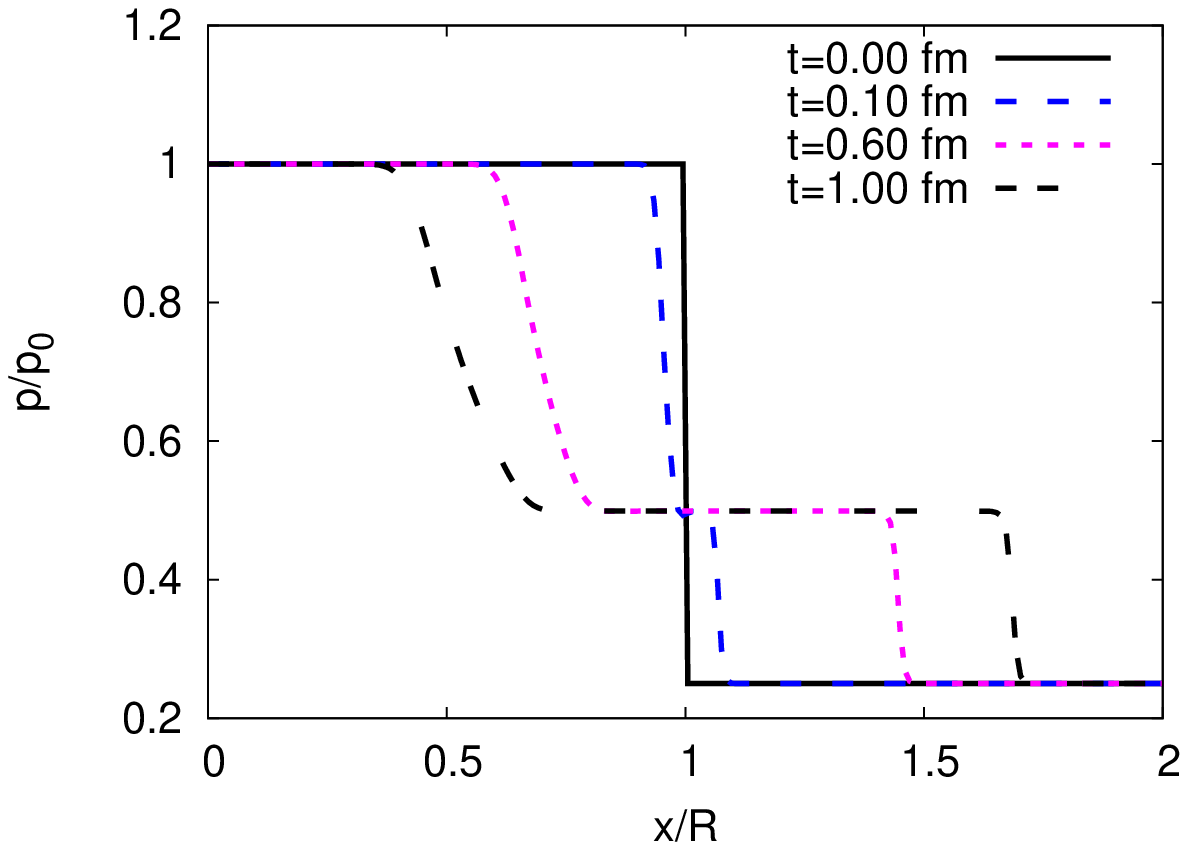}
\end{minipage}
  \caption[The evolution of a shock wave.]
  {The evolution of a discontinuity leading to a shock front as a sketch (left
  panel) and calculated with a $1$-dimensional hydrodynamical algorithm 
  \cite{Rischke:1995ir}.}
  \label{PlotRiemann}
\end{figure}

Fluid dynamics exhibits a special feature: Shock waves. They are 
discontinuities and characterize that part of the medium which will
subsequently move with a velocity $v_{\rm shock}\neq v_{\rm medium}$. \\
The easiest way to introduce shock waves is by discussing the {\it Riemann
problem} which describes the decay of a discontinuity between two regions of
constant flow. It is a well-known problem in fluid dynamics and can be solved
analytically in one dimension for an ideal fluid (for a review see e.g.\ \cite{Rischke:1998fq}).\\
The corresponding initial conditions are two
different regions, both in thermodynamical equilibrium, that differ by their
pressures (see Fig.\ \ref{PlotRiemann}). During the following hydrodynamic evolution,
a shock wave will develop that travels supersonically ($v_{\rm shock} > c_s$)
into that part of the system with the lower pressure (i.e., to the right in Fig.\ 
\ref{PlotRiemann}), while simultaneously a rarefaction wave moves with a velocity 
equal to the speed of sound into matter with larger pressure 
(to the left in Fig.\ \ref{PlotRiemann}). During that process, a region
of constant pressure evolves behind the shock wave which is called the shock plateau.\\
Applying the conservation equations of (net) charge, energy, and momentum, it is possible to 
show that the velocity of the shock front \cite{Landau,Rischke:1995ir,Rischke:1998fq}
\begin{eqnarray}
v_{\rm shock}&=&\sqrt{\frac{(p_2-p_1)(\varepsilon_2+p_1)}
{(\varepsilon_1+p_2)(\varepsilon_2-\varepsilon_1)}}
\end{eqnarray}
can be determined from the relativistic Rankine--Hugonoiot--Taub equation (RHTA) \cite{Taub:1948zz}
\begin{eqnarray}
\frac{w_1^2}{\rho_1^2}-\frac{w_2^2}{\rho_2^2}+(p_2-p_1)\left(
\frac{w_1}{\rho_1^2}+\frac{w_2}{\rho_2^2}\right)&=&0\,,
\end{eqnarray}
where $w=\varepsilon+p$ denotes the enthalpy, $\rho$ the particle density, and $p_1$ ($p_2$) the
pressure in front of (behind) the relativistic shock front so that $p_2 > p_1$.\\
Theoretically, it was already predicted in the 1970's that shock waves should occur
in collisions of heavy nuclei \cite{Scheid:1974zz} due to the strong compression of the medium.
Later on, this phenomenon was observed experimentally 
\cite{Baumgardt:1975qv,Hofmann:1976dy,Gutbrod:1989wd}. Of course, one of the basic questions is 
the origin of the discontinuity. The early works \cite{Baumgardt:1975qv} investigated the 
penetration of a smaller nucleus through a larger one. In that case, a region of extremely dense 
matter is formed (mainly from the remnant of the smaller nuclei) which moves in beam 
direction with a velocity larger than the speed of sound leading to a discontinuity due 
to the superposition of a large number of infinitesimal sonic perturbations. Such perturbations can
superimpose to the characteristic pattern of a Mach cone, which is discussed in the next section. 
Currently, the penetration of a smaller nucleus through a larger one is 
investigated for FAIR conditions \cite{Philip}.\\
It was suggested \cite{Stoecker:2004qu,CasalderreySolana:2004qm,Satarov:2005mv,
CasalderreySolana:2006sq,Renk:2005si,Betz:2006ds} that such Mach cones could also
be created in heavy-ion collisions. As already mentioned in section \ref{ExperimentsQGP},
hard probes formed in the very early stages of the collision can propagate through the medium,
acting as perturbations of the medium.

\section[Mach cones]{Mach cones}
\label{SecMC}

\begin{figure}[t]
\centering
  \includegraphics[scale = 0.55]{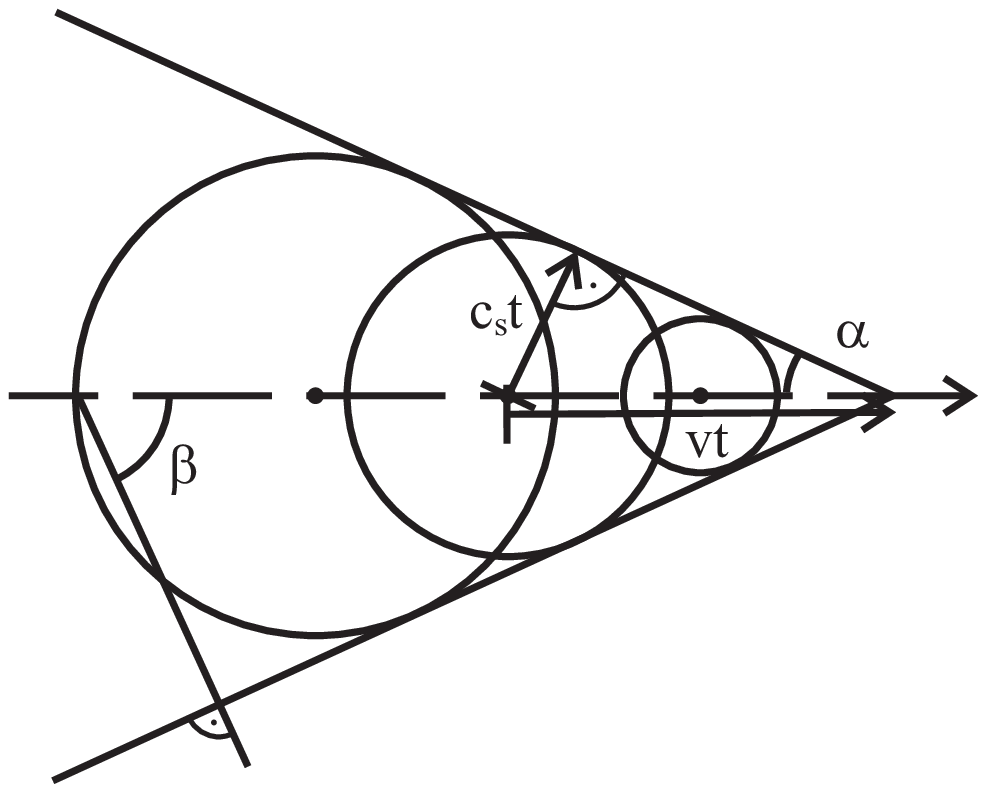}
  \caption[The interference pattern of spherical waves leading to the formation of a Mach cone.]
  {The interference pattern of spherical waves leading to the formation of a Mach cone. The source 
  emitting sound waves propagates with a velocity larger than the speed of sound through the 
  medium so that a conical discontinuity emerges with an opening angle $\alpha$. Particles created 
  in this discontinuity are always emitted into an angle $\beta=\pi-\alpha$.}
  \label{SketchMachCone}\vspace*{-2ex}
\end{figure}

Sound waves can be created due to a perturbation (like a jet) and propagate through 
a fluid. If the origin of this perturbation (i.e., the source) does not move, it emits
spherical waves which travel (isotropically) with the speed of sound through the medium. For a 
moving source, however, the spherical waves interfere, leading to a compression zone in the
direction of motion as described by the Doppler effect. As soon as the velocity of the source is 
larger than the speed of sound, the source moves even faster than the emitted sound waves. 
Thus, as shown in Fig.\ \ref{SketchMachCone}, the interference pattern results in a conical 
compression zone, a discontinuity called {\bf Mach cone}\footnote{Ernst Mach (1838--1916) was an 
Austrian physicist who was a professor in Graz and Prague.}. The opening angle of such a Mach cone 
is given by 
\begin{eqnarray}
\sin\alpha&=&\frac{c_s}{v}\,,
\label{openingangle}
\end{eqnarray}
(as one can easily deduce from Fig.\ \ref{SketchMachCone} applying simple trigonometric functions)
where $v$ denotes the velocity of the source through the medium. If particles are created in 
the discontinuity, they are mostly emitted perpendicularly to the Mach cone. \\
Thus, the emission angle (often also referred to as Mach angle $\phi_M$) is 
$\beta=\phi_M=\pi-\alpha$, and therefore
\begin{eqnarray}
\cos\beta&=&\frac{c_s}{v}\,.
\label{Machangle}
\end{eqnarray}
Since Mach cones are discontinuities, they belong to the category of shock waves. 
If one wants to study Mach cones with a certain numerical alorithm, it has to be first checked that 
it is possible to describe the propagation of shock waves using this particular  
application. For the SHASTA, this was done in Ref.\ \cite{Rischke:1995ir}. Recently, as 
demonstrated in Ref.\ \cite{Bouras:2009nn}, it was shown that it is possible to describe shock 
waves using the parton cascade BAMPS \cite{Xu:2004mz}. \\
The basic idea when applying the concept of Mach cones to high-energy heavy-ion collisions is 
that a hard-$p_T$ particle, travelling with a velocity $v_{\rm jet}$ (close to the speed of light) 
through the medium, re-distributes its energy to lower-$p_T$ particles. Thus, it acts as a 
source in the fluid probably exciting sound waves which interfere and form a Mach cone. 
\\Subsequently, there should be an enhanced particle emission under a distinct Mach-cone angle. 
If this idea is correct and it is possible to extract the emission angle from the measured 
particle distributions, it would give direct access to the (averaged) speed of sound of 
the medium via Eq.\ (\ref{Machangle}). Then, as suggested in Refs.\ 
\cite{Stoecker:2004qu,CasalderreySolana:2004qm}, it might be possible to draw conclusions about 
the properties of the medium, especially about the speed of sound and to extract information about 
the EoS.
\\Assuming that the medium can be described by an ideal gas EoS with $c_s=1/\sqrt{3}\sim 0.57$, 
the Mach-cone emission angle is $\beta=\phi_M=0.96$~rad, which agrees with the measured particle 
distributions introduced in chapter \ref{ExperimentsQGP}. Therefore, the medium formed in a 
heavy-ion collision might possibly behave like an ideal gas which thermalizes quickly, allowing 
for the creation of Mach cones.\\
Of course, it is tacitly assumed here that the velocity of the jet, which is usually estimated to 
be close to the speed of light, is constant and known. However, even in that case it is only 
possible to extract a mean value for the speed of sound. If the medium properties change, e.g.\
due to a phase transition (as it is expected when the medium cools down), the speed of sound 
changes (and might actually go to zero in case of a first-order phase transition), affecting the 
Mach-cone angle as well.\\
Strictly speaking, Eq.\ (\ref{Machangle}) only applies for weak, sound-like perturbations. 
Certainly there is no reason to assume that in physical systems the amplitude of the perturbation 
is really small. Thus, to calculate the opening angle of a cone that is created due to the pile-up 
of matter at the head of the jet which creates a shock perturbation, a more general problem 
has to be adressed. It was shown in Ref.\ \cite{Rischke:1990jy} that this problem splits up into
\begin{itemize}
\item the oblique-shock-wave problem \cite{Landau,Courant} close to the head of the jet 
leading to the cone angle that results from the shock wave, taking into account the influence of flow,
\item and the Taylor--Maccoll problem \cite{TaylorMaccoll1,TaylorMaccoll2} behind the shock 
which calculates the flow of matter along the cone.
\end{itemize}
The first one determines the opening angle of the cone in the vicinity of the shock to be
\begin{eqnarray}
\sin\phi_M&=&\frac{\gamma_{cs}c_s}{\gamma_{v}v}\,,
\end{eqnarray}
with $\gamma_{cs}=1/\sqrt{1-c_s^2}$. Only in the non-relativistic limit, this results in the 
above Eq.\ (\ref{openingangle}). Thus, just (far) behind the head of the 
jet, in the so-called {\it far zone}, where the perturbations due to the jet are much weaker
than close to the head of the jet, Mach-cone angles computed according to Eq.\ (\ref{Machangle}) 
are quite well reproduced. As can be seen from the figures of part \ref{part03}, especially 
slow moving partons create a pile-up of matter leading to a bow shock which changes the opending 
angle of the Mach cone close to the head of the jet. \\
It should be mentioned in this context that the creation of a Mach cone is a general phenomenon 
possible in any plasma. Thus, Ref.\ \cite{Betz:2006ds} discusses the possibility that shock waves 
and Mach cones might also appear in a plasma formed due to mono-jets that are emitted by 
radiating mini-black holes which can be produced in large collider facilities like the LHC. 
\\It was already discussed in chapter \ref{ExperimentsQGP} that conical emission patterns were 
found in the two- and three-particle correlations measured at RHIC. Nevertheless, it could not 
be proven that this shape results from the creation of Mach cones (see section 
\ref{Two-particle correlations}). The main issue of this thesis is to test whether hydrodynamic 
models support the Mach-cone hypothesis. A crucial ingredient is the mechanism of energy and 
momentum loss by the jet which will be addressed in chapter \ref{JetEnergyLoss}.

\section[The Influence of Radial Flow]{The Influence of Radial Flow}
\label{SecRadialFlow}
\begin{figure}[t]
\centering
  \includegraphics[scale = 0.45]{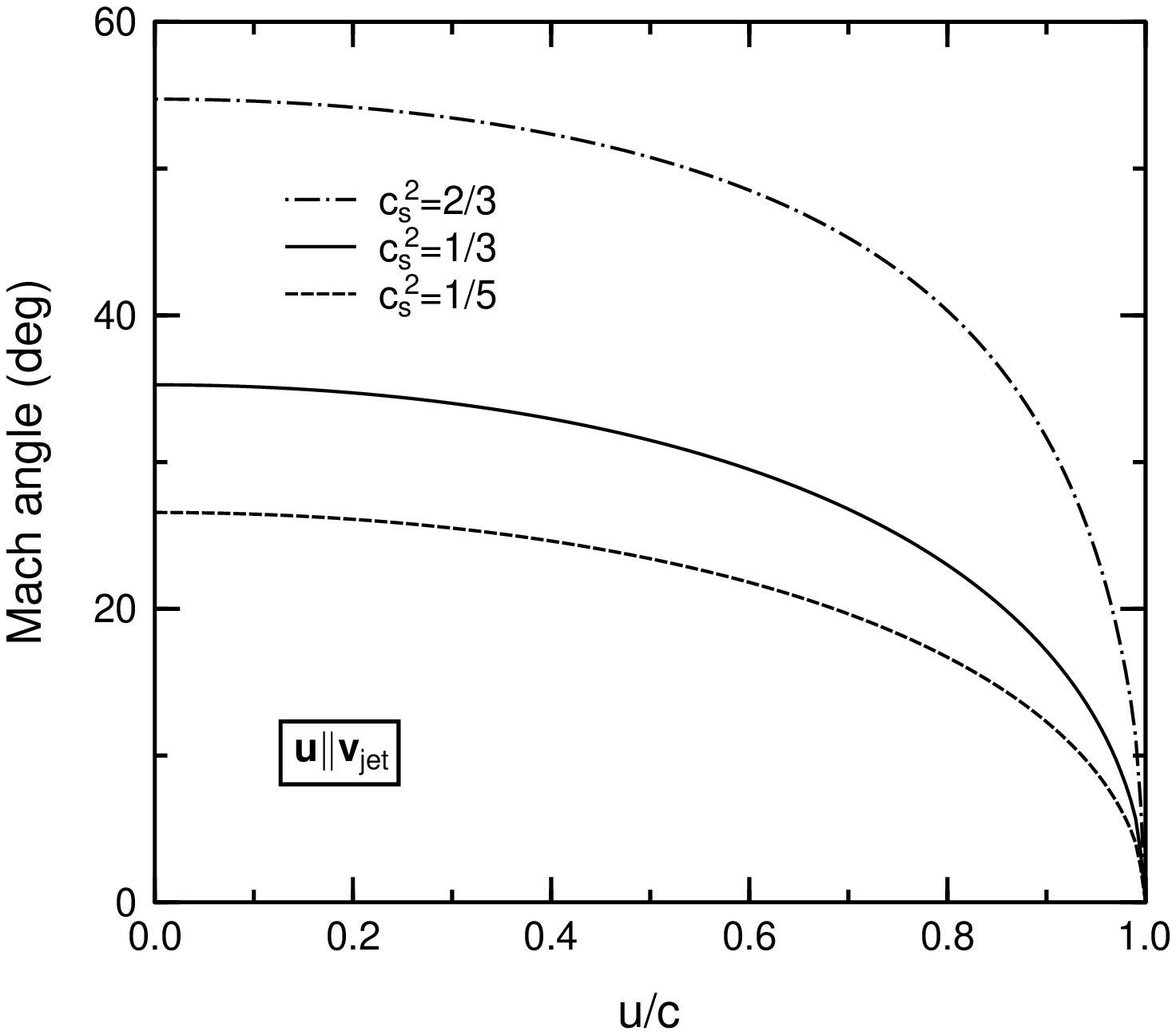}
  \caption[The change in the Mach angle as a function of the background flow.]
  {The change in the Mach angle as a function of the background flow $v_{\rm flow}=u/c$ for 
  hadronic   matter $c_s^2=1/5$, an ideal gas $c_s^2=1/3$ and a strongly coupled QGP (sQGP) with 
  $c_s^2=2/3$ \cite{Satarov:2005mv}. }
  \label{SatarovChangeMachAngle}
\end{figure} 
The medium created in a heavy-ion collision, through which the jet propagates probably exciting 
sound waves, expands rapidly. Consequently, there will be a strong flow ($v_{\rm flow}$) relative 
to the velocity of the jet that certainly influences its propagation, the interference pattern of 
the sound waves, and hence the Mach cone angle. 
\begin{figure}[b]
\centering
  \includegraphics[scale = 0.57]{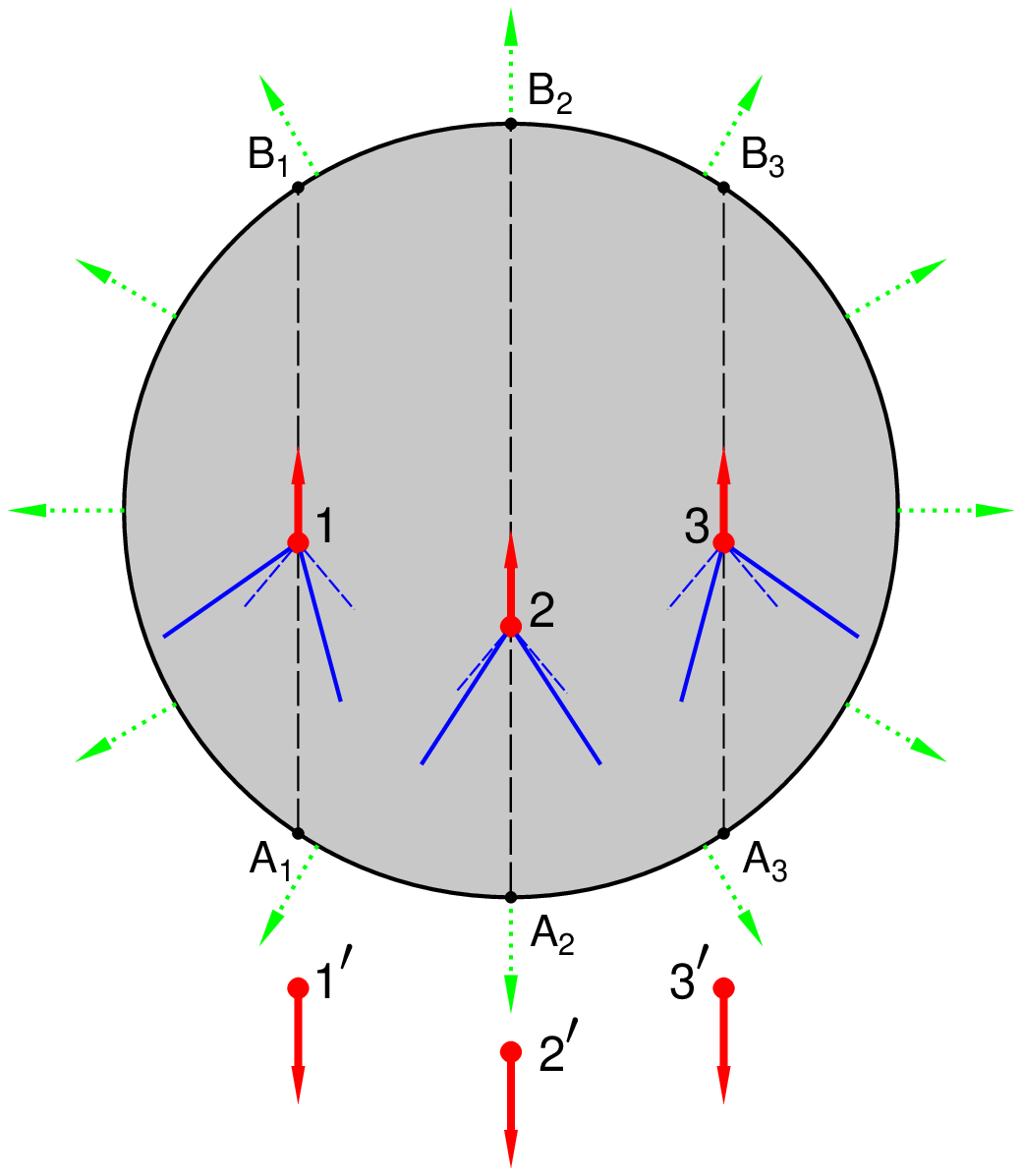}
  \caption[Sketch of the deformation of Mach cones in an expanding medium.]
  {Sketch of the deformation of Mach cones in an expanding medium (solid lines). The thick 
  arrows denote the trigger jets and the expected Mach-cone angles in a static medium are shown 
  by dashed lines \cite{Satarov:2005mv}.}
  \label{MachConeDeformation}
\end{figure} 
It was shown in Ref.\ \cite{Satarov:2005mv} that a background flow parallel to the velocity of 
the jet ($v_{\rm flow}\vert\vert v_{\rm jet}$) leads to a narrowing of the Mach angle in the 
center-of-mass frame which has a purely relativistic origin. Fig.\ \ref{SatarovChangeMachAngle} 
depicts this change in the Mach angle for different flow velocities and various EoS. It illustrates 
that a larger background flow results in a smaller Mach angle. However, assuming e.g.\ an ideal 
gas EoS and estimating the background flow to be of the order of the speed of sound, 
$c_s\sim 0.57$, the change of the Mach cone angle is rather small.
\\Obviously, it is very unlikely that a jet propagates collinearly with the flow. If the 
background flow has a transverse component w.r.t.\ the velocity of the jet, this should cause a 
deformation of the Mach cone as displayed in Fig.\ \ref{MachConeDeformation}. Unfortunately, the 
derivation of a general solution for the deflected Mach cone angle is far from trivial. A 
qualitative discussion of this effect is given in appendix \ref{AppendixDistortion}. \\
Moreover, one always has to consider that the experimentally measured particle distributions 
sample over many events. As can be seen from Fig.\ \ref{MachConeDeformation}, the radial expansion 
of the fireball should lead to a broadening of the measured cone angles when averaging over the 
different possible jet trajectories. However, this effect might get diminished due to the momentum
spread of the initial parton distribution (as shown in Fig.\ \ref{Sketch_JetPropagation}) which 
leads to a systematic error in determining the relative azimuthal angle $\Delta\phi$.\\
Nevertheless, it is necessary to investigate the deformation of Mach cones applying full 
($3+1$)-dimensional hydrodynamical simulations in order to prove the interaction of jet and 
background flow and to study its impact on the particle distributions.

%
%
\chapter[Jet-Energy Loss]
{Jet-Energy Loss}
\label{JetEnergyLoss}

A fast moving particle interacts with the medium it traverses, losing energy. In general, the 
mechanisms by which this energy is lost as well as the amount of energy deposited depends on the 
particle characteristics and on the matter properties. Considering certain kinds of particles, 
this provides fundamental information about the medium itself.\\
Therefore, such jet-medium interactions are significant for the study of a jet moving through a 
plasma and its impact on the measured particle distributions. In the following, we discuss 
different mechanisms that were developed to describe the energy, but not the momentum loss of 
partons within the QGP. Here, we focus on models based on QCD. The next chapter provides a 
detailed introduction to the approach obtained by the gauge/string duality, the AdS/CFT 
correspondence, which can also be used to investigate jet energy loss. \\
The crucial ingredient to study the propagation of jets in heavy-ion collision is the source 
term describing the energy and momentum that is lost by the hard probes and thermalizes in 
the medium. Those source terms will be discussed in the subsequent sections.

\section[In-medium Energy Loss]
{In-medium Energy Loss}
\label{InMediumELoss}
\begin{figure}[t]
\centering
  \includegraphics[scale = 0.7]{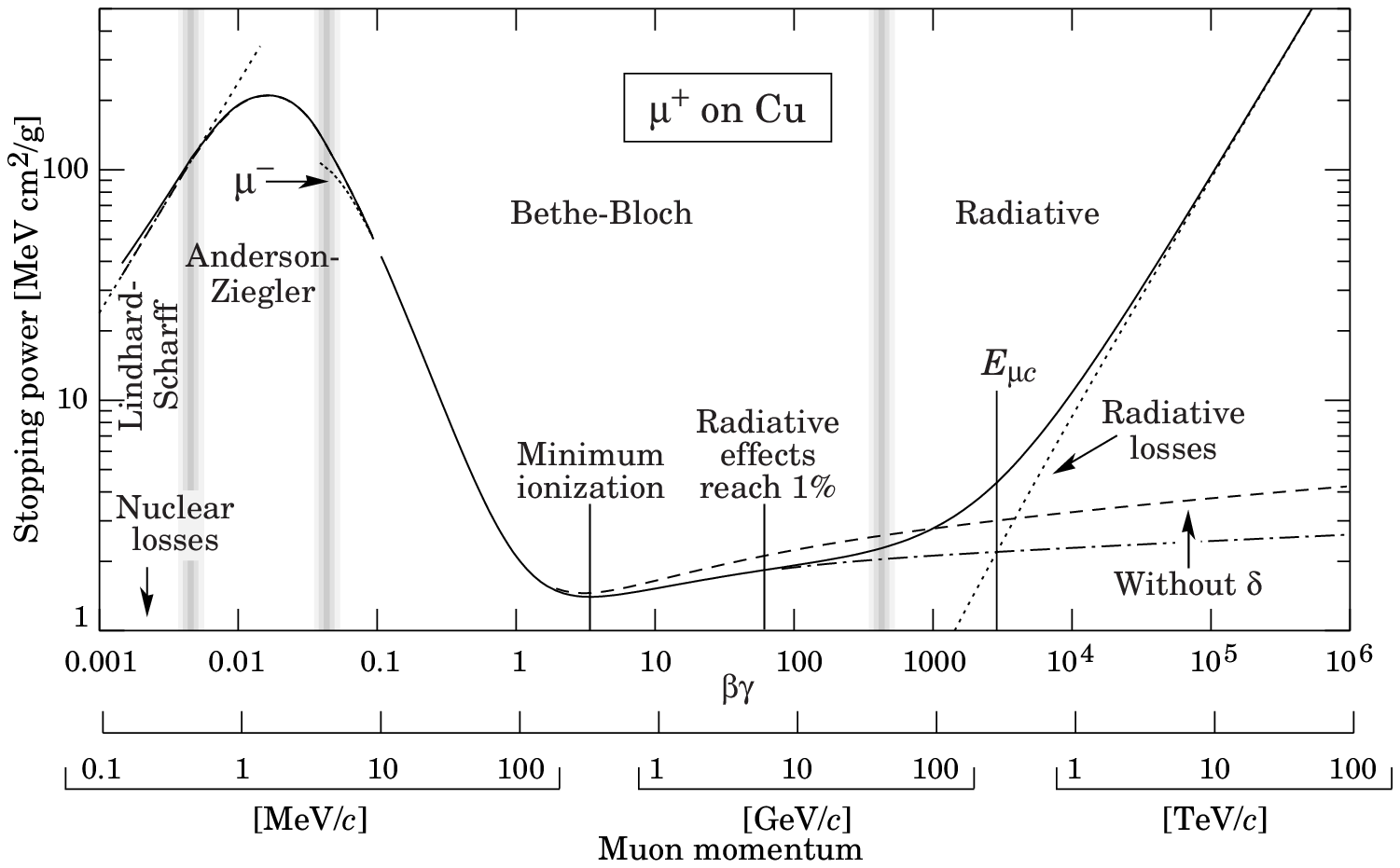}
  \caption[The stopping power, i.e., the energy loss per unit length $dE/dx$, for muons in copper.]
  {The stopping power, i.e., the energy loss per unit length $dE/dx$, for muons in copper as a 
  function of $\beta\gamma=p/Mc$ \cite{Yao:2006px}. }
  \label{BetheBloch}\vspace*{1.5ex}
\end{figure}
\begin{figure}[t]
\centering
\begin{minipage}[c]{4.2cm}
\hspace*{-2.0cm}
  \includegraphics[scale = 0.27]{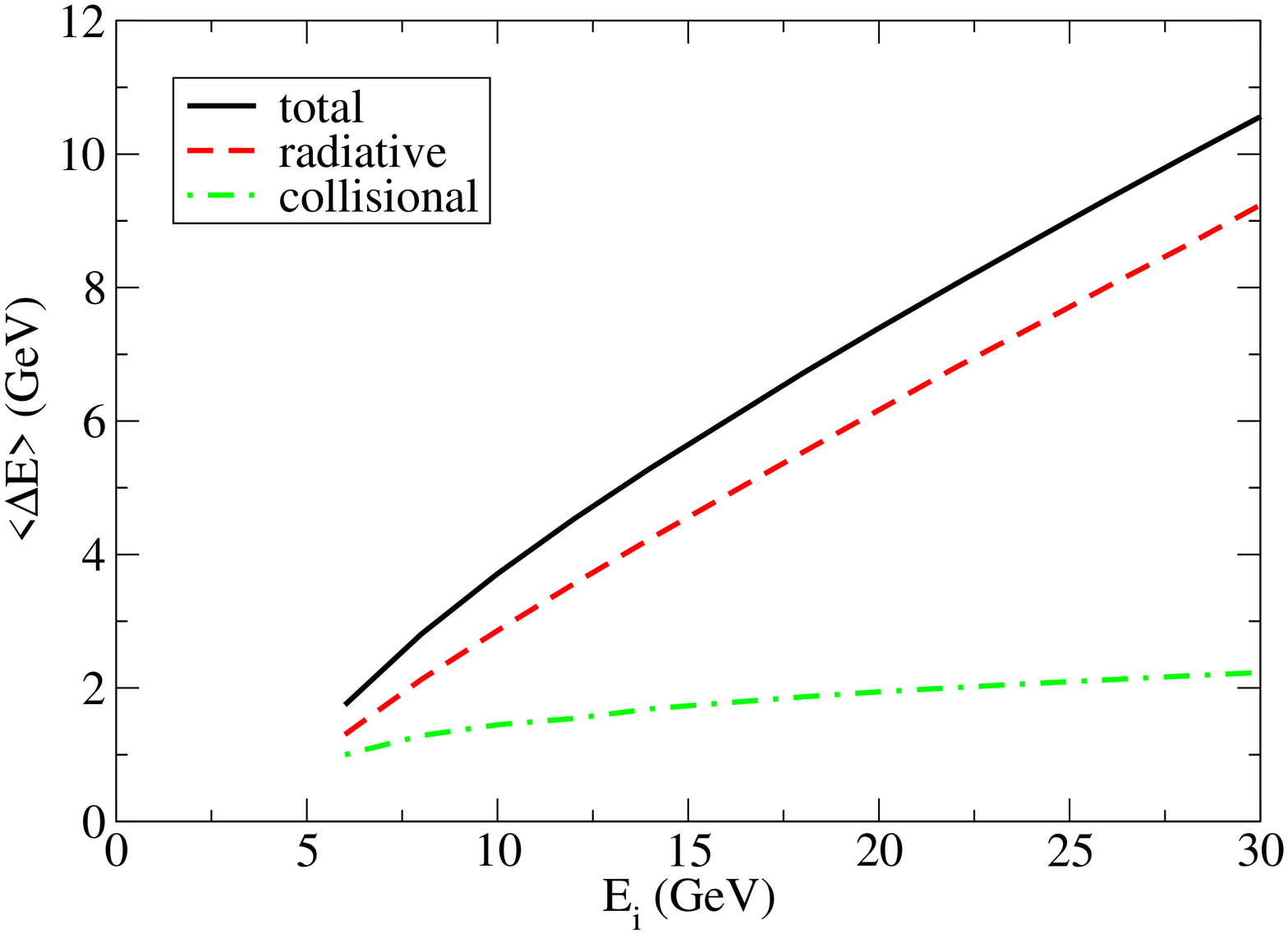}
\end{minipage}
\hspace*{0.6cm}  
\begin{minipage}[c]{4.2cm} 
  \includegraphics[scale = 0.3]{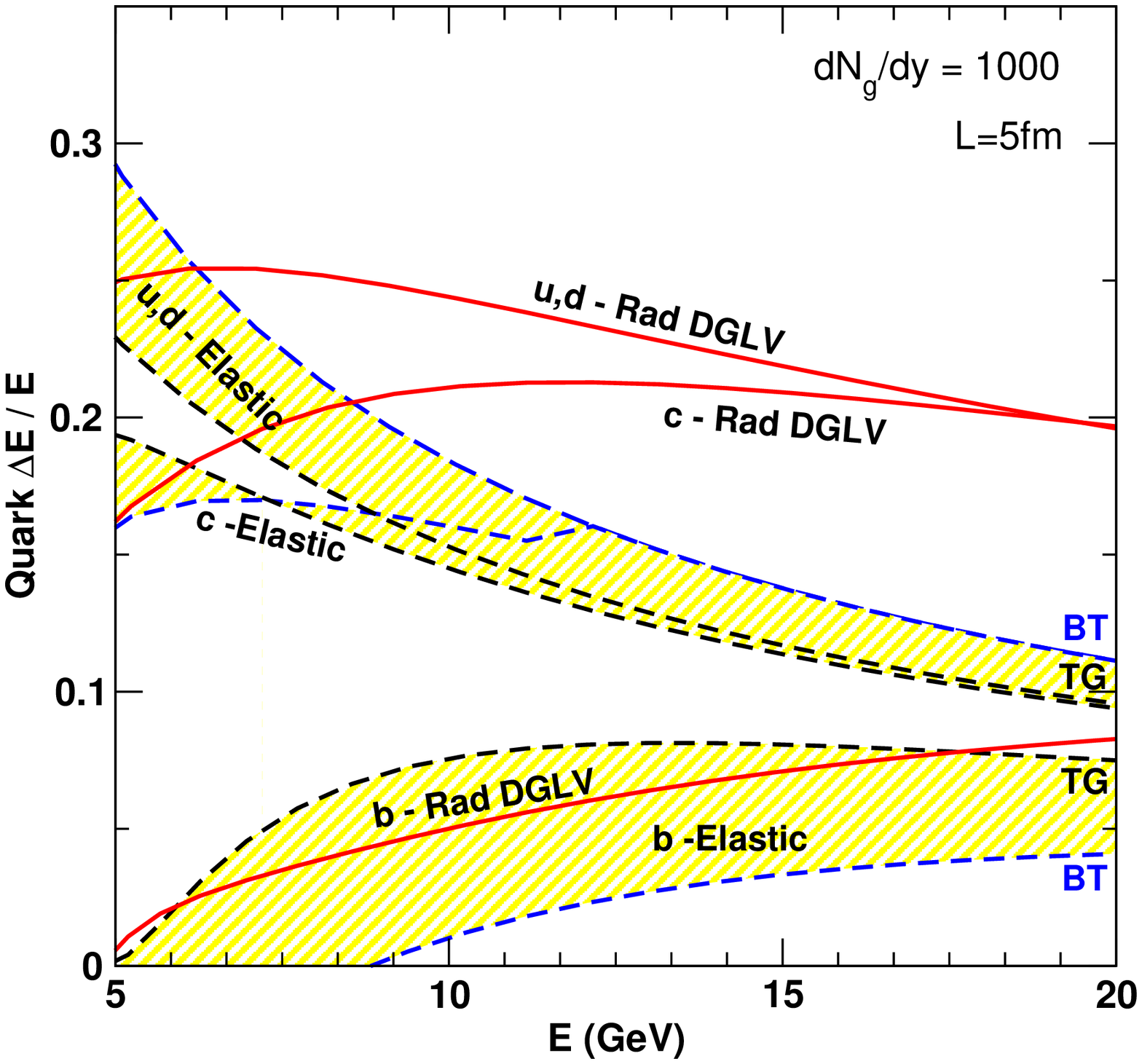}
\end{minipage}
  \caption[Collisional (elastic) and radiative energy loss of light quarks as well as light and 
  heavy quarks passing through a medium produced in central Au+Au collisions at RHIC.]
  {Collisional (elastic) and radiative energy loss of light quarks (left panel) \cite{Qin:2007rn} 
  as well as light and heavy quarks (right panel) \cite{Wicks:2007am} passing through a medium 
  produced in central Au+Au collisions at RHIC.}
  \label{PlotSimon}
\end{figure}

\begin{figure}[t]
\centering
  \includegraphics[scale = 0.55]{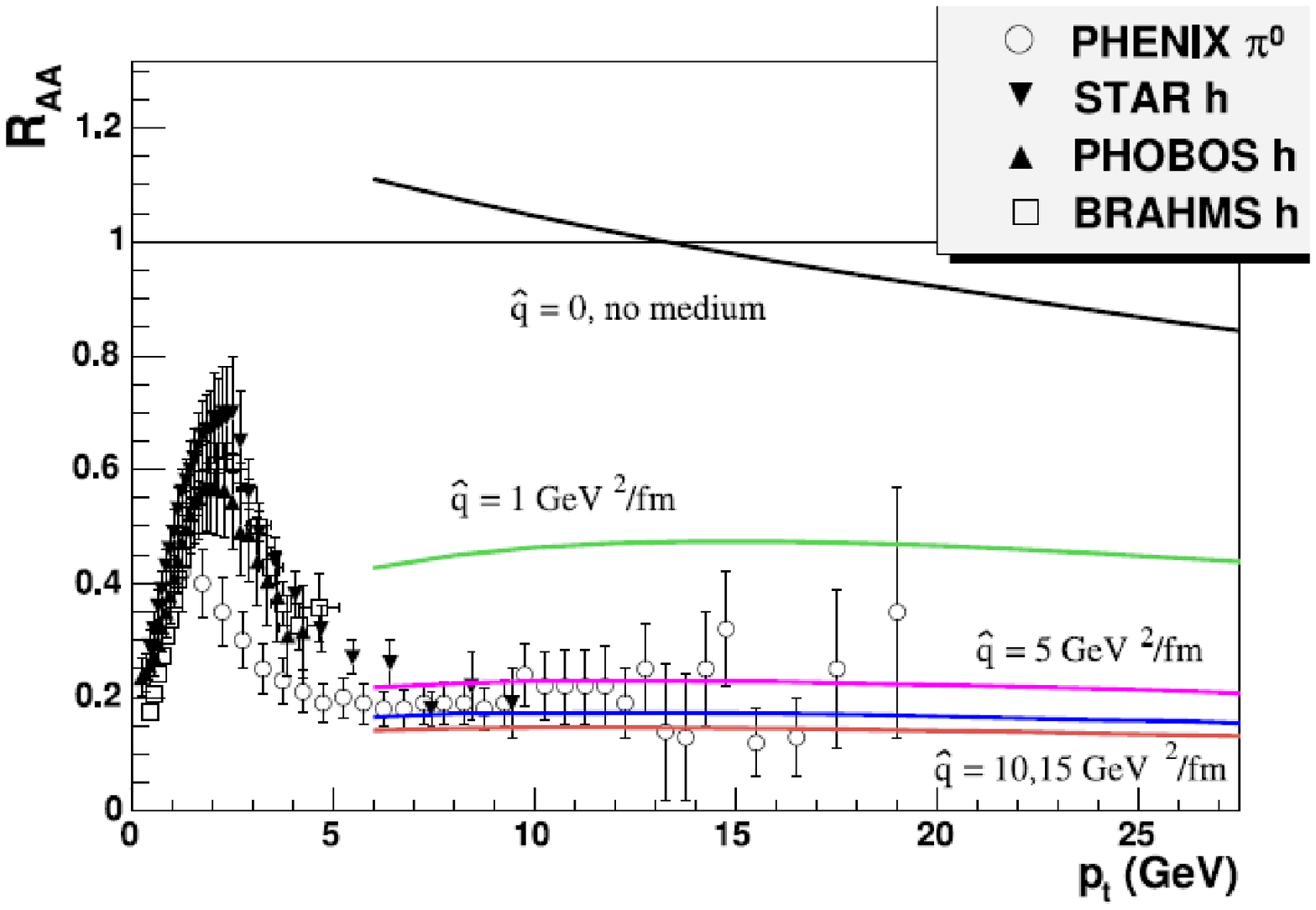}
  \caption[$R_{AA}$ measured for central Au+Au collisions at $\sqrt{s}_{NN}=200$~GeV compared to 
  model calculations for varying values of the $\hat{q}$ \newline parameter.]
  {$R_{AA}$ measured for central Au+Au collisions at $\sqrt{s}_{NN}=200$~GeV and compared to 
  model calculations for varying values of the $\hat{q}$-parameter, based on Ref.\ 
  \cite{Eskola:2004cr}.}
  \label{eskola}
\end{figure}

 
In the most general case, the total energy loss of a particle moving through a medium is the sum of
\begin{itemize}
\item {\bf Collisional energy loss} through elastic scatterings with the medium constituents and
\item {\bf Radiative energy loss} via inelastic scatterings in the medium, determined by the 
corresponding Bremsstrahlung spectrum.\\
For incoherent scatterings, the total energy loss is given by 
$\Delta E^{\rm tot}=N\cdot\Delta E^{1 \rm scat}$, where $N=L/\ell_{\rm mfp}$ is the
opacity\footnote{The opacity describes the number of scattering centers in a medium of thickness 
$L$.} with $L$ being the characteristic length of the medium and $\ell_{\rm mfp}$ the mean-free 
path. Thus, the energy loss per unit length, also called stopping power, is given by
\begin{eqnarray}
\frac{dE}{dx}&\equiv&\frac{dE}{dl}=\frac{\langle \Delta E^{1 \rm scat}\rangle}{\ell_{\rm mfp}}
=\frac{\langle \Delta E\rangle}{L}\,.
\end{eqnarray}
Usually, this quantity is provided with a minus sign demonstrating that the particle has lost 
energy. Since we will not consider the energy that is lost by the jet in our hydrodynamic 
simulations, but rather the energy that is given by the jet to the medium, we will assign a 
positive sign to $dE/dx$. \\
The radiative energy loss itself depends on the thickness of the plasma. For a thin medium 
$(L\ll \ell_{\rm mfp})$, the traversing particle suffers at most one single scattering. This
limit is called the Bethe--Heitler (BH) regime. For thick media $(L\gg \ell_{\rm mfp})$ however, 
the multiple scattering reduces the amount of radiation, an effect called 
Landau--Pomeranchuk--Migdal (LPM) effect\footnote{The LPM effect, originally deduced from photon 
emission \cite{LPM}, describes the fact that multiple scattering/interaction causes destructive 
interference which suppresses the radiative spectrum.}.\\
Moreover, the total amount of radiation emitted from a heavy quark is suppressed at angles smaller 
than the ratio of the quark mass $M$ to its energy $E$ ($\theta=M/E$). This {\it dead cone} 
effect \cite{Vitev:2005yg} leads to a damping of emission by a factor of $m_D^2/M^2$.
\end{itemize}
As shown in Fig.\ \ref{BetheBloch}, which displays the energy loss of a muon in copper, the 
collisional (Bethe--Bloch) energy loss dominates at low energies, while at high energies the 
radiative loss increases linearly and dominates over the logarithmic growth of the collisional 
energy loss. \\
However, this behaviour strongly depends on the properties of the medium as well as on the 
kinematic region considered. Therefore, it is not possible to universally determine the dominant 
energy-loss mechanism. \\
For a long time, it was assumed that the main contribution at high energies results from radiative 
energy loss. While this appears to be correct for light quarks, a proper prescription of heavy 
quark energy loss seems to require both, collisional and radiative energy loss 
\cite{Wicks:2007am}, see Fig.\ \ref{PlotSimon}.
\\All medium modifications are often encoded in the {\it transport coefficient}, the so-called 
$\hat{q}$-parameter, characterizing a fundamental quantity of QCD that is defined as the average 
transverse momentum squared transferred per unit path length,
\begin{eqnarray}
\hat{q}\equiv\frac{\langle Q^2\rangle}{\ell_{\rm mfp}}=\frac{m_D^2}{\ell_{\rm mfp}}\,,
\end{eqnarray}
where the Debye mass $m_D=gT$ describes the lowest momentum exchange with the medium. The values 
of this transport coefficient vary for the various approaches reviewed below, 
$\hat{q}\sim 5-25$~GeV$^2$/fm, exhibiting a large uncertainty when comparing to
experimental data \cite{Eskola:2004cr,Loizides:2006cs,Adare:2008cg}. For a comparison of different
$\hat{q}$-parameters to experimental data see Fig.\ \ref{eskola}. Consequently, the TECHQM program 
\cite{TECHQM} has agreed upon a systematic study of the different mechanisms to clarify the 
ambiguities concerning the various approaches as well as the physical processes during a jet-medium 
interaction.\\
The explicit expressions for the collisional and radiative energy loss depend on the nature of the 
projectile via the colour factor $C_R$, which is the quadratic Casimir of the respective 
representation,
\begin{eqnarray}
C_R=
\left\{ \begin{array}{ll}
C_F=\frac{N_c^2-1}{2N_c}=\frac{4}{3}&{\rm fundamental\;representation\;for\;quarks\,,}\\
C_A=N_c=3&{\rm adjoint\;representation\;for\;gluons\,,}
\end{array} \right.
\label{CasimirOperator}
\end{eqnarray}
implying that $C_A/C_F=9/4$. Therefore, a gluon jet loses roughly twice the energy of a quark 
jet and exhibits a larger hadron multiplicity.

\section[Mechanisms of Jet Energy Loss from QCD]
{Mechanisms of Jet Energy Loss from QCD}
\label{MechanismsELoss}
\begin{figure}[t]
\centering
  \includegraphics[scale = 0.36]{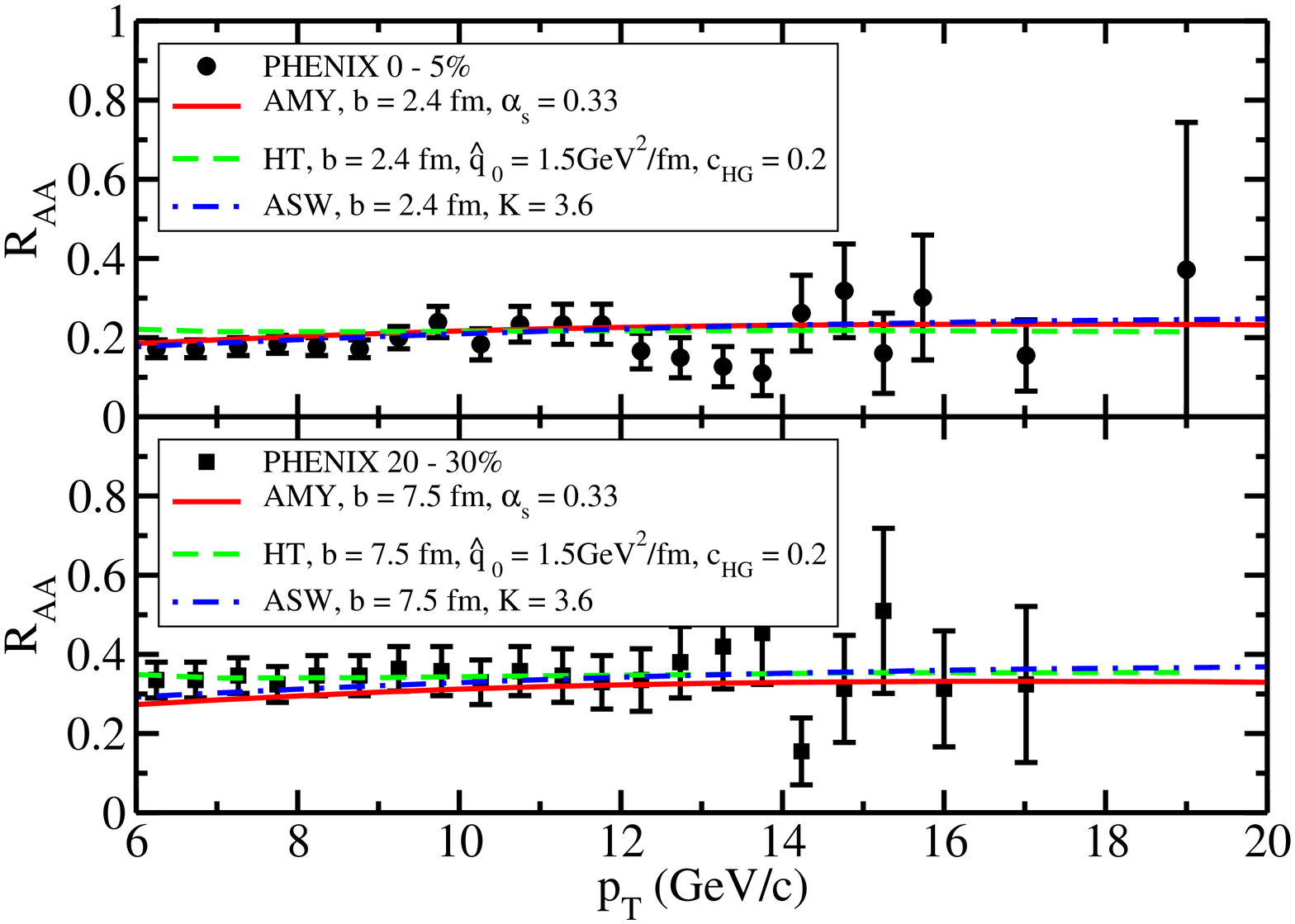}
  \caption[Nuclear modification factor for high-$p_T$ pions in central and semi-central Au+Au 
  collisions at RHIC, compared to AMY, HT and ASW energy loss calculations.]
  {Nuclear modification factor $R_{AA}$ for high-$p_T$ pions in central (upper panel) and 
  semi-central (lower panel) Au+Au collisions at RHIC, compared to AMY, HT and ASW energy loss 
  calculations \cite{Bass:2008ch}.}
  \label{RAA_3Dhydro_Bass}
\end{figure}

Four major phenomenological approaches (often identified with the initials of their authors) 
have been developed to connect the QCD energy loss calculations with the experimental observables. 
Those are
\begin{itemize}
\item The {\bf GLV}\footnote{Gyulassy--Levai--Vitev} approach \cite{Gyulassy:1999zd}, nowadays 
also known as {\bf DGLV}\footnote{Djordjevic--Gyulassy--Levai--Vitev} \cite{Djordjevic:2003zk}, 
calculates the parton energy loss in a dense deconfined medium consisting of almost static (i.e.\ 
heavy) scattering centers which produce a screened (Yukawa) potential. A single hard radiation 
spectrum is expanded to account for gluon emission from multiple scatterings via a recursive 
diagrammatic procedure. This allows to determine the gluon distribution to finite order in opacity.
\item The {\bf BDMPS}\footnote{Baier--Dokshitzer--Müller--Peign\'{e}--Schiffer} scheme 
\cite{Baier:1996kr,Baier:1996sk}, similarily established by Zakharov \cite{Zakharov:1996fv} and 
used in the {\bf ASW}\footnote{Armesto--Salgado--Wiedemann} \cite{Wiedemann:2000za} procedure,
calculates the energy loss in a co\-loured medium for a hard parton interacting with various 
scattering centers that splits into an outgoing parton as well as a radiated gluon. The propagation
of these partons and gluons are expressed using Green's functions which are obtained by path 
integrals over the fields. Finally, a complex analytical expression for the radiated gluon 
distribution function is obtained as a function of the transport coefficient $\hat{q}$. 
\item The {\bf Higher Twist} (HT) approximation \cite{Qiu:1990xxa,Qiu:1990xy} describes the 
multiple scattering of a parton as power corrections to the leading-twist cross section. These 
corrections are enhanced by the medium length $L$ and suppressed by the power of the hard scale 
$Q^2$. Originally, this scheme was applied to calculate the medium corrections to the total cross 
section in nuclear deep-inelastic electron+nucleon scatterings.
\item The {\bf AMY}\footnote{Arnold--Moore--Yaffe} \cite{Arnold:2001ba} approach describes the 
parton energy loss in a hot equilibrated QGP. Multiple scatterings of the partons and their 
radiated gluons are combined to determine the leading-order gluon radiation rate.
\end{itemize}
For a detailed review see e.g.\ Ref.\ \cite{Majumder:2007iu}. All four schemes have made 
successful comparisons to the available data when tuning one distinct model parameter which is 
the initial gluon density in the GLV approach, the $\hat{q}$-parameter for the BDMPS/ASW scheme, 
the initial energy loss in the HT approximation and the temperature in the AMY procedure. 
However, all approaches are based on certain model assumptions, limiting their scope of 
application (see Ref.\ \cite{ReviewDavid,Majumder:2007iu}).\\
The quantitative consistency of the different schemes has been investigated within a 
$3$-dimensional hydrodynamical approch (see Fig.\ \ref{RAA_3Dhydro_Bass}) \cite{Bass:2008ch}
using the same space-time evolution. Recently, the nuclear modification factor was also studied 
using a pQCD-based parton cascade including radiative processes \cite{Fochler:2008ts}.
\\However, while those energy-loss mechanisms predict the amount of energy lost to the medium, 
expressed e.g.\ by the transport coefficient $\hat{q}$, they do not adress the question  
{\bf how} the energy and momentum deposited by the jet affect the medium. First calculations of a 
medium response were presented in Ref.\ \cite{CasalderreySolana:2004qm}, applying a schematic 
source and linearized hydrodynamics as will briefly be reviewed below. A source term expected 
from a parton moving through the QCD plasma was recently derived by Neufeld et al.\ 
\cite{Neufeld:2008fi}, see section \ref{pQCDSourceTerm}.

\section[First Studies of Jet-Energy Transfer to the Medium]
{First Studies of Jet-Energy Transfer to the Medium}
\label{FirstStudies}
\begin{figure}[t]
\centering
\begin{minipage}[c]{4.2cm}
\hspace*{-5.0cm}
  \includegraphics[scale = 0.33,angle=270]{./part02/CasalderreySolanaDiff.epsi}
\end{minipage}
\hspace*{-1.7cm}  
\begin{minipage}[c]{4.2cm} 
  \includegraphics[scale = 0.33,angle=270]{./part02/CasalderreySolanaMach.epsi}
\end{minipage}
  \caption[Azimuthal particle distributions for non-isentropic and isentropic excitations.]
  {Azimuthal particle distributions for non-isentropic (left panel) and isentropic (right panel) 
  excitations \cite{CasalderreySolana:2007km}. The solid black line in the left panel represents 
  an energy loss of $dE/dx=12.6$~GeV/fm, while the red dashed line is obtained for 
  $dE/dx=2$~GeV/fm. In the right panel this energy loss is fixed to $dE/dx=12.6$~GeV/fm and the 
  differenc curves represent the various $p_T$-cuts of $0.2\leq p_T\leq 1$ (solid black line), 
  $1\leq p_T\leq 2$ (dashed red line), $2\leq p_T\leq 3$ (dotted green line) and $3\leq p_T\leq 4$ 
  (dashed-dotted blue line).}
  \label{CasalderreySolana}
\end{figure}
Casalderrey--Solana et al.\ 
\cite{CasalderreySolana:2004qm,CasalderreySolana:2006sq,CasalderreySolana:2007km} 
first examined the problem of {\bf where} the energy of the quenched jets is transferred to. 
They did not focus on the calculation of the amount of energy deposited to the medium, but rather 
on the evolution of (some) re-distributed energy and momentum. \\
Their work is based on two major assumptions. First, they consider a di-jet pair that is created 
back-to-back close to the surface of the medium. While one jet (the trigger jet) escapes, the 
other one penetrates the plasma. Second, they assume that the energy and momentum density 
deposited by the jet (into a homogeneous medium at rest) is a small perturbation compared to the 
total amount of energy stored in the medium which allows to use {\it linearized hydrodynamics}. \\
In such a linearized form, the hydrodynamic equations decouple, and with the definition of 
$\vec{g}=g_L(\vec{k}/k)+\vec{g}_T$ for the momentum density, including the longitudinal $(L)$ and 
transversal part $(T)$, those equations can be written as
\begin{eqnarray}
\label{linhydro1}\partial_t\varepsilon + ikg_L&=&0\,,\\
\label{linhydro2}\partial_t g_L+i c_s^2 k\varepsilon 
+\frac{4}{3}\frac{\eta}{\varepsilon_0+p_0}k^2 g_L&=&0\,,\\
\label{linhydro3}\partial_t g_T+\frac{\eta}{\varepsilon_0+p_0}k^2 g_T &=&0\,.
\end{eqnarray} 
Here, $c_s$ denotes the speed of sound and $\eta$ the viscosity of the medium. Certainly, 
this ansatz breaks down close to the jet (as will be discussed later in detail), a region that is 
not describable with this approach.\\
To adress the issue of matter excitation which is unknown in detail since the interaction and 
thermalization processes of the lost energy are unclear, they studied two different scenarios 
\cite{CasalderreySolana:2004qm,CasalderreySolana:2006sq,CasalderreySolana:2007km}:
\begin{itemize}
\item Local energy and momentum distribution (modelled by Gauss functions) along the path of the 
jet propagating through the medium (non-isentropic excitation),
\item excitation of sound waves due to gradients in the momentum distribution, but vanishing 
energy deposition (isentropic excitation).
\end{itemize}
Performing an isochronous Cooper--Frye freeze-out, they obtained the particle distributions of 
Fig.\ \ref{CasalderreySolana}. Those figures clearly show that in case of energy and momentum 
deposition (see left panel of Fig.\ \ref{CasalderreySolana}), independent of the amount of energy 
loss (which are displayed for $dE/dx=2$~GeV/fm and $dE/dx=12.6$~GeV/fm), a peak occurs in the 
direction of the jet moving through the plasma (which is located here at $\Delta\phi=\pi$). 
However, the excitation of sound waves in the second deposition scenario (see right panel of 
Fig.\ \ref{CasalderreySolana}) leads to a conical structure as anticipated from an interference 
pattern of sound waves (see chapter \ref{ShockWavePhenomena}, especially Fig. 
\ref{SketchMachCone}), though a large $dE/dx$ has to be chosen to observe the effect. Moreover, 
the peaks obtained on the away-side get more pronounced for larger values of the applied 
$p_T$-cuts. \\
Energy and momentum loss (left panel of Fig.\ \ref{CasalderreySolana}) 
do not result in a conical structure. The momentum deposited 
causes additional ``kicks'' in direction of the moving jet, forming a strong flow behind 
the jet \cite{CasalderreySolana:2004qm}, named the {\bf diffusion wake}. Given that a Cooper--Frye 
freeze-out is mainly flow driven, cf.\ Eq.\ (\ref{CFFOIsochron}), a peak occurs in the direction of jet 
motion. It has to be stressed that the energy loss described in this scenario leads by itself to 
the formation of sound waves the interference pattern of which results in a concial structure.
However, this structure is not seen in the freeze-out patterns since 
the flow due to the momentum distribution superimposes and dominates the particle 
distributions. Thus, the deposition of energy and momentum results in a peak in jet-direction.
This is explicitly shown in Figs.\ \ref{FigPunchThrough1} and 
\ref{FigPunchThrough2}.\\
The location of the conical peaks in the second scenario (see right panel of Fig.\ 
\ref{CasalderreySolana}) is in good agreement with the experimental data (presented in chapter 
\ref{ExperimentsQGP}) showing away-side peaks at $\pi\pm 1.1$~rad. This suggested to conclude that 
the second scenario describes the data and reveals some insights into the excitation mechanism. 
Certainly, one has to keep in mind that the whole approach, though very instructive, is an 
appoximation and it will turn out (as we demonstrate in chapter \ref{ExpandingMedium}) that the 
second scenario eventually has to be ruled out.\\
Beyond the question of energy deposition, the linearized hydrodynamic appoximation offers the 
advantage that the propagation of sound waves can be traced back to Eq.\ (\ref{linhydro2}), 
while Eq.\ (\ref{linhydro3}) describes the diffusive component. Remarkably, the viscous terms in 
both equations are of the same order. Thus, dissipative corrections will have about the same 
effects on both, sound waves and diffusion wake. Therefore, the interplay between both effects 
does not change in viscous media.\\
Similar findings about the impact of a jet moving through an expanding medium were shown in Ref.\ 
\cite{Renk:2005si,Chaudhuri:2005vc,Renk:2006mv}. Chaudhuri and Heinz demonstrated that, using 
$(2+1)$-dimensional ideal hydrodynamics, the particle correlations obtained from a jet depositing 
the same amount of energy and momentum, do not lead to a double-peaked structure on the away-side 
\cite{Chaudhuri:2005vc}. Renk and Ruppert, however \cite{Renk:2005si}, reproduced the measured data 
when assuming that a certain fraction of energy is transferred into the sound modes while the 
remaining (smaller amount of) energy is released into the diffusion wake. 
\begin{figure}[t]
\centering
  \includegraphics[scale = 0.33]{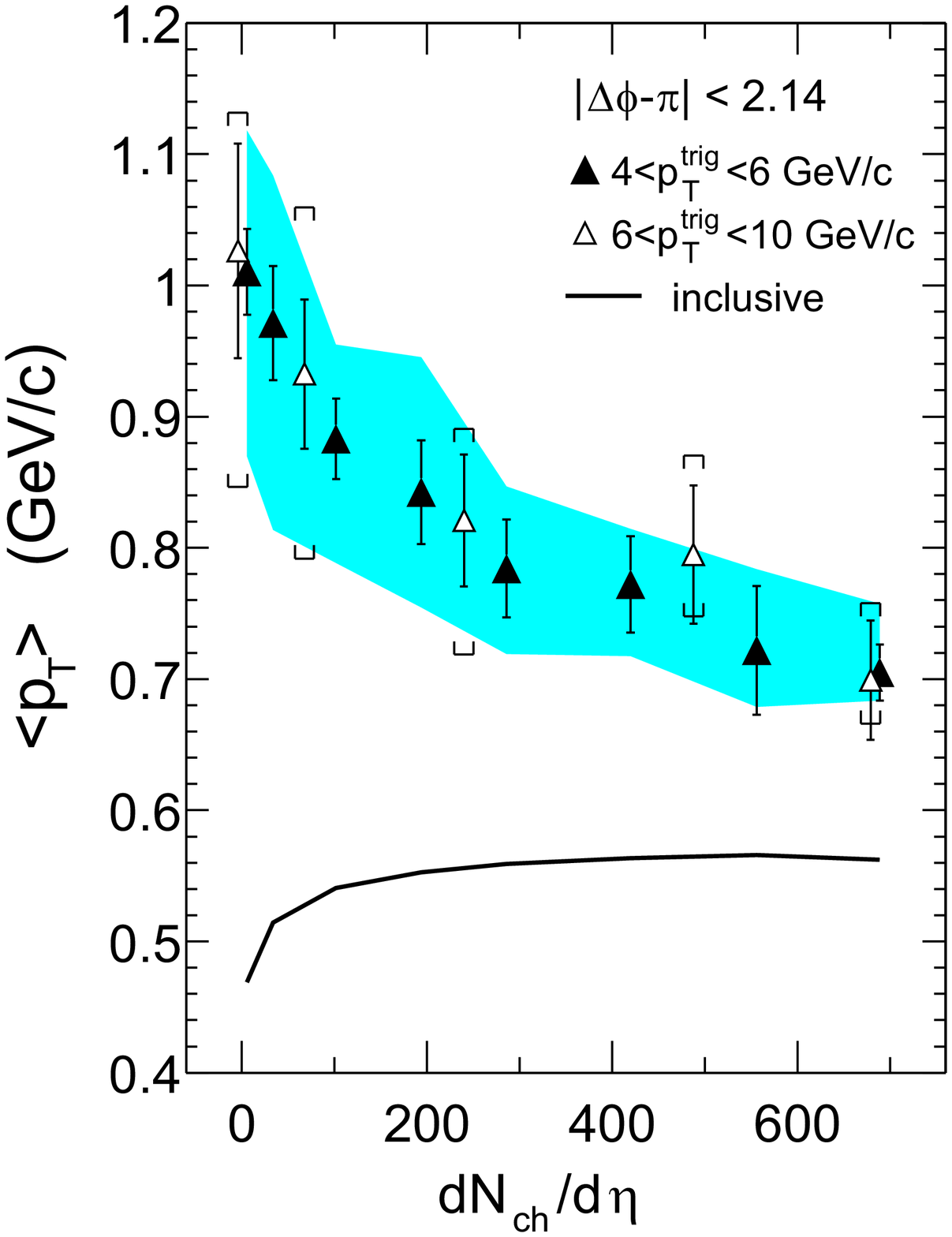}
  \caption[Mean transverse momentum of associated hadrons for different trigger-$p_T$'s as a 
  function of centrality.]
  {Mean transverse momentum $\langle p_T\rangle$ of associated hadrons for different trigger-$p_T$'s
  of $p_T=4-6$~GeV and $p_T=6-10$~GeV as a function of centrality \cite{Adams:2005ph}.}
  \label{mean_pT_STAR}
\end{figure} 
\section[Jet Energy Loss in Hydrodynamics]
{Jet Energy Loss in Hydrodynamics}
\label{ELossHydro}
Each parton propagating through a medium, depositing energy (and momentum), acts as a source to the 
medium. It is not clear from first principles that the energy lost quickly thermalizes and 
therefore can be incorporated into a hydrodynamic description, as it was done above. However, a 
measurement \cite{Adams:2005ph} proved that the average momentum of particles emitted on the same 
side as the jet that passes through the medium approaches the value of a thermalized medium with
decreasing impact parameter, see Fig.\ \ref{mean_pT_STAR}. Thus, a hydrodynamical prescription 
seems to be justified and a source term can be added to the equations for energy and momentum 
conservation [cf.\ Eq.\ (\ref{EnMomConservation})]
\begin{eqnarray}
\partial_\mu T^{\mu\nu}&=&J^\nu\,.
\label{sourceterm}
\end{eqnarray}  
This source term $J^\nu$ is not the source term of the jet, but the 
residue of energy and momentum given by the jet to the medium. The source term that correctly 
depicts the interaction of the jet with the QGP is under current investigation using the
two completely independent and different approaches of pQCD and AdS/CFT. Both source terms 
will be introduced below and their impact will be studied in part \ref{part03}.\\
However, before implementing a rather complicated source term into a hydrodynamic algorithm, it 
might be instructive to first consider a schematic source term that describes the energy and 
momentum deposition of the jet along a trajectory 
\begin{eqnarray}
x^\mu(\tau)&=&x_0^\mu+u^\mu_{\rm jet}\tau\,,
\end{eqnarray}  
where $x_0^\mu$ denotes the formation point and 
$u^\mu_{\rm jet}=(\gamma_{\rm jet},\gamma_{\rm jet}\vec{v}_{\rm jet})$ the $4$-velocity 
of the jet. Such a source term is given by
\begin{eqnarray}
J^\nu&=&\int\limits_{\tau_i}^{\tau_f}d\tau \frac{dM^\nu}{d\tau}\delta^{(4)}
\left[x^\mu-x^\mu_{\rm jet}(\tau) \right]\,,
\label{PebbleSource}
\end{eqnarray}  
with the energy and momentum loss rate $dM^\nu/d\tau$, the proper time interval $\tau_f - \tau_i$ 
associated with the jet evolution and the location of the jet $x_{\rm jet}(\tau)$. Certainly 
this source term comprises a significant simplification of the ongoing processes. Nevertheless, as 
will be demonstrated in part \ref{part03}, it correctly depicts the behaviour of the particle 
correlations compared to those obtained e.g.\ from the pQCD source term introduced in the 
following section.

\section[A pQCD Source Term]
{A pQCD Source Term}
\label{pQCDSourceTerm}
\begin{figure}[t]
\centering
\begin{minipage}[c]{4.2cm}
\hspace*{-1.5cm}
  \includegraphics[scale = 0.48]{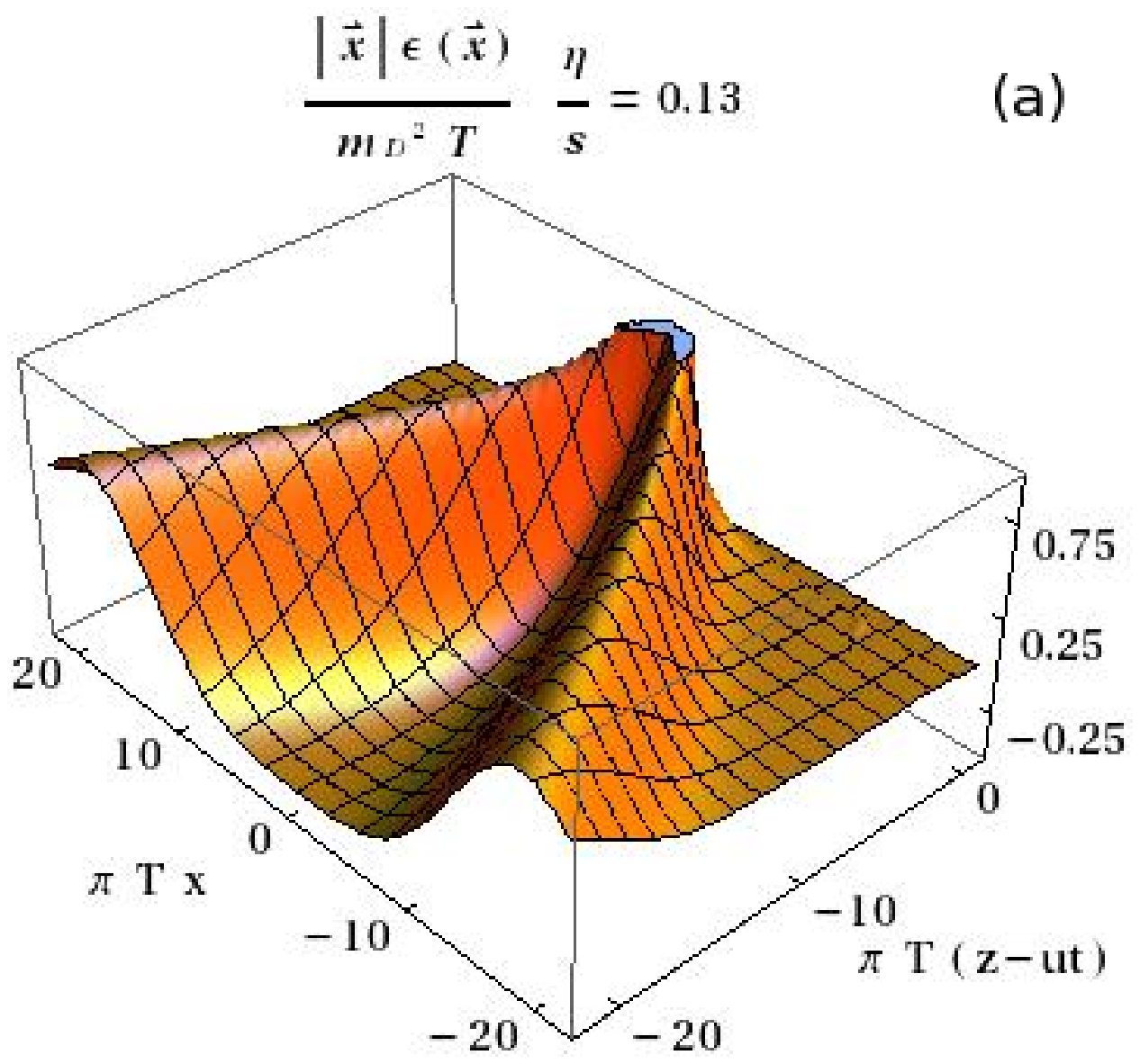}
\end{minipage}
\hspace*{0.6cm}  
\begin{minipage}[c]{4.2cm} 
  \includegraphics[scale = 0.48]{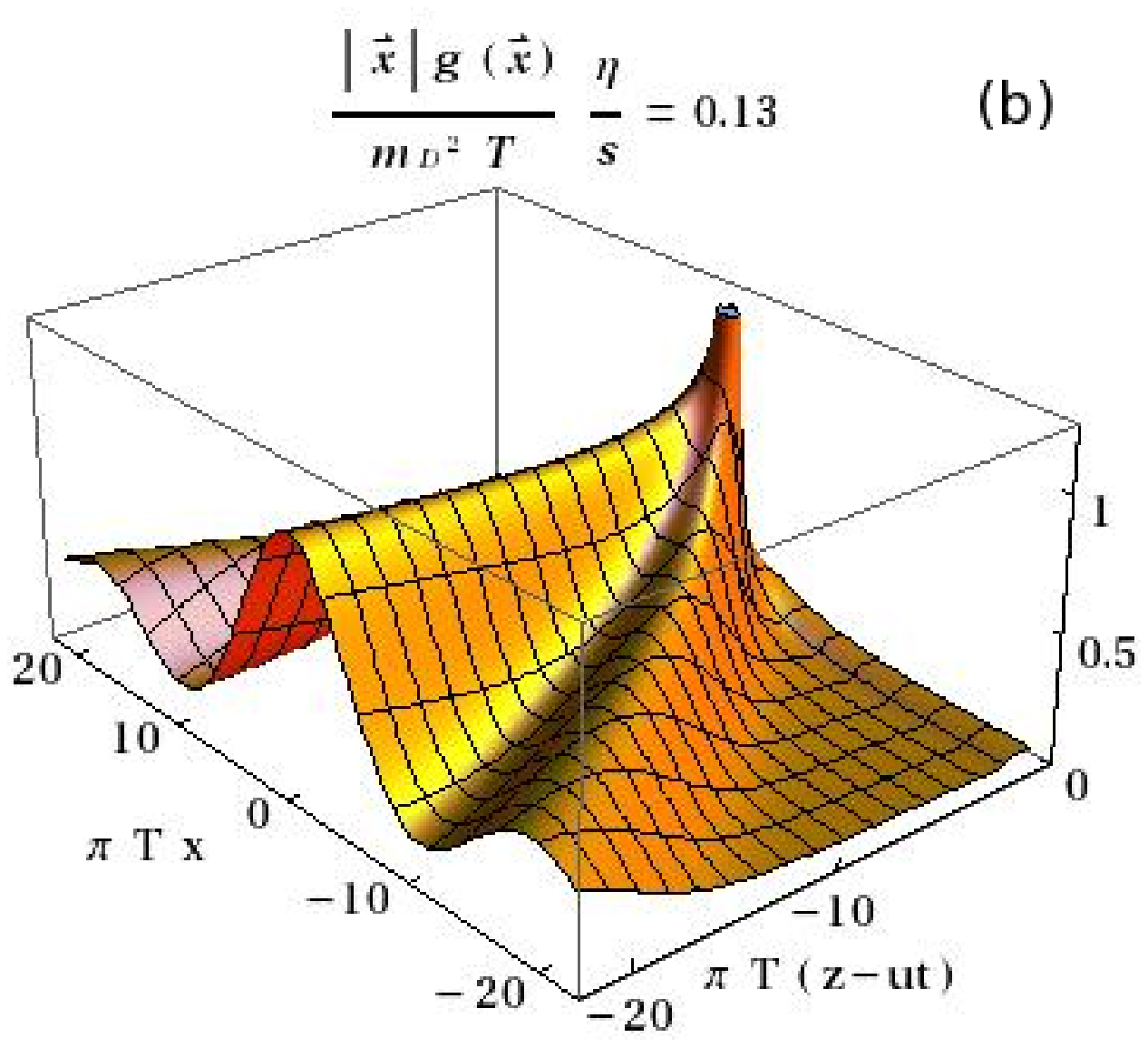}
\end{minipage}
  \caption[Perturbed energy and momentum densities for a gluon moving with a velocity of 
  $u=0.99955$ along the positive $z$-axis of a QCD medium with $\eta/s=0.13$.]
  {Perturbed energy (left) and momentum (right) densities for a gluon moving with a velocity of 
  $u=0.99955$ along the positive $z$-axis of a QCD medium with $\eta/s=0.13$
  \cite{Neufeld:2008fi}.}
  \label{NeufeldMach}
\end{figure}
The first investigations about the medium response to the passage of a fast parton were based on 
schematic source terms (see above), mainly representing the moving jet as a $\delta$-function. 
Recently, Neufeld et al.\ \cite{Neufeld:2008fi,Neufeld:2008hs} derived a source term for a 
supersonic parton popagating through a QCD plasma. Their approach considers the fast parton as a 
source of an external color field that can be described via perturbative QCD applying a 
collisionless Boltzmann equation. For a system of partons in an external color field, described by 
the distribution $f(\vec{x},\vec{p},t)$, this Boltzmann equation is defined as 
\cite{Asakawa:2006jn}
\begin{eqnarray}
\left(\frac{\partial}{\partial t}+\frac{\vec{p}}{E}
\cdot\vec{\nabla}_x - \nabla_{pi}D_{ij}(\vec{p},t)\nabla_{pj}\right) f(\vec{x},\vec{p},t)&=&0\,,
\label{BoltzmannNeufeld}
\end{eqnarray}  
where $\vec{p}/E$ is the velocity of a parton with momentum $\vec{p}$ and energy $E$. Here, 
$D_{ij}(\vec{p},t)=\int_{-\infty}^tdt^\prime\,F_i(\vec{x},t)F_j(\vec{x}^\prime,t^\prime)$ 
describes the integral over Lorentz forces 
$F_i(\vec{x},t)=gQ^a(t)\left[E_i^a(\vec{x},t)+(\vec{v}\times\vec{B})_i^a(\vec{x},t)\right]$ 
[$gQ^a(t)$ denotes the charge] on a medium particle caused by the moving parton, acting 
until a certain time $t$. These Lorentz forces are considered to lowest order in the coupling 
constant $g$.\\
Since the equations for energy and momentum conservation can be derived from the second moment 
of the Boltzmann equation $p^\mu\partial_\mu f(\vec{p},t)=0$ (see appendix \ref{AppViscousHydro}), 
the source term can be defined as
\begin{eqnarray}
J^\nu&\equiv&\int\frac{d\vec{p}}{(2\pi)^3}p^\nu
\big[\nabla_{pi}D_{ij}(\vec{p},t)\nabla_{pj}f(\vec{x},\vec{p},t)\big]\,.
\label{EnMomSource}
\end{eqnarray}  
Omitting dielectric screening and assuming an energy as well as momentum deposition of a 
parton with constant velocity $\vec{u}=u\vec{e}_z$ at a position $\vec{x}=ut\vec{e_z}$, 
this source term is evaluated to be \cite{Neufeld:2008hs}
\begin{figure}[t]
\centering
\hspace*{-3.2cm}
  \includegraphics[scale = 0.5]{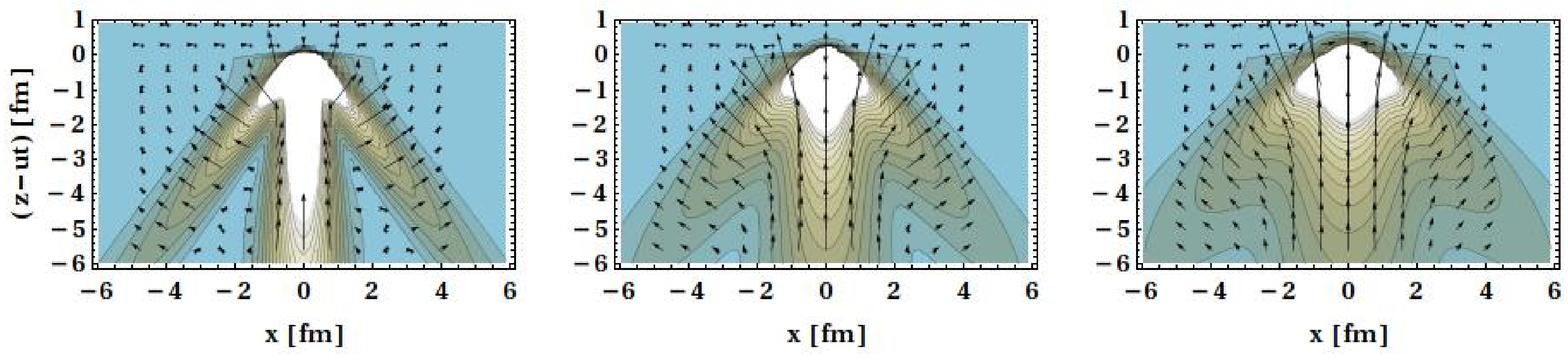}
  \caption[Perturbed momentum density for different values of the shear viscosity to entropy 
  ratio $\eta/s$.]
  {Perturbed momentum density for different values of the shear viscosity to entropy ratio, 
  $\eta/s=1/(4\pi), 3/(4\pi),6/(4\pi)$ \cite{Neufeld:2008dx}.}
  \label{NeufeldViscous}
\end{figure} 
\begin{eqnarray}
J^\nu(\rho,z,t)&=&\left[J^0(\rho,z,t),\vec{u}J^0(\rho,z,t)-\vec{J}^v(\rho,z,t)\right]\,,
\end{eqnarray}  
where $\rho=\sqrt{x^2+y^2}$ denotes the transverse component and
\begin{eqnarray}
\hspace*{-0.9cm}J^0(\rho,z,t)&=&d(\rho,z,t)\gamma u^2\nonumber\\
\hspace*{-0.9cm}&&
\times\left[1-\frac{z_-}{z_-^2+\rho^2}\left(z_-
+\frac{\gamma u\rho^2}{\sqrt{\rho^2+z_-^2\gamma^2}} \right) \right]\,,\\
\hspace*{-0.9cm}J^v(\rho,z,t)&=&(\vec{x}-\vec{u}t)
\frac{\alpha_s Q_p^2 m_D^2}{8\pi(z_-^2+\rho^2)^2}\nonumber\\
&&\hspace*{-1.7cm}\times
\left[\frac{u^4\rho^4+(z_-^2\gamma^2+\rho^2)\left(2z_-^2+(u^2+2)\frac{\rho^2}{\gamma^2}\right)}
{(z_-^2\gamma^2+\rho^2)^2}
-\frac{2uz_-}{\gamma\sqrt{z_-^2\gamma^2+\rho^2}}
\right]\,,
\end{eqnarray}  
with the abbrevation $z_-=(z-ut)$, $\alpha_s=g^2/(4\pi)$, the Casimir operator $Q_p$ 
[see Eq.\ (\ref{CasimirOperator})] and the function $d(\rho,z,t)$
\begin{eqnarray}
d(\rho,z,t)&=&\frac{\alpha_s Q_p^2 m_D^2}{8\pi[\rho^2+\gamma^2(z-ut)^2]^{3/2}}\,.
\end{eqnarray}  
It exhibits a singularity at the point of the jet. Thus, an ultraviolet cut-off is needed which 
was chosen to be $\rho_{\rm min}=1/(2\sqrt{E_p T})$ in Ref.\ \cite{Neufeld:2008hs}, where $E_p$ 
denotes the energy of the fast parton. An infraread cut-off, however, is given by 
$\rho_{\rm max}=1/m_D$.
Neufeld et al.\ \cite{Neufeld:2008fi} applied, as Casalderrey--Solana et al.\ (cf.\ section 
\ref{FirstStudies}), linearized hydrodynamics (including viscous terms) to solve Eqs.\ 
(\ref{sourceterm}). Since the equations decouple in the linearized approach, it allows to 
separately display the perturbed energy and momentum distributions, excited by a gluon which 
moves with a velocity of $u=0.99955$ along the positive $z$-axis through a QCD medium 
(see Fig.\ \ref{NeufeldMach}). While the energy density features the shape of a Mach cone, whose 
intensity is peaked close to the source, the momentum density additionally shows a distinct 
maximum along the positive $z$-axis. Plotting the momentum density (see Fig.\ 
\ref{NeufeldViscous}), it becomes obvious that this maximum is due to a strong flow created
along the trajectory of the jet. Again, like in the scenario studied by  Casalderrey--Solana et al.\ 
\cite{CasalderreySolana:2004qm} applying a schematic source term, this flow is due to the momentum 
deposition $J^i$ of the jet. Therefore, the momentum density resulting from the pQCD source term 
also displays a diffusion wake.\\
Moreover, Fig.\ \ref{NeufeldViscous} illustrates that this diffusion wake gets broader for larger 
values of the shear viscosity to entropy ratio $\eta/s$. Similarly, the intensity of the Mach 
cone decreases.\\
Recently, a calculation carried out in the Higher Twist (HT) formalism, including elastic energy 
loss, for a virtual jet propagating through a medium \cite{Qin:2009uh} showed that a Mach-cone 
structure is formed as well, when plotting the energy-density perturbation.
\begin{figure}[t]
\centering
  \includegraphics[scale = 0.5]{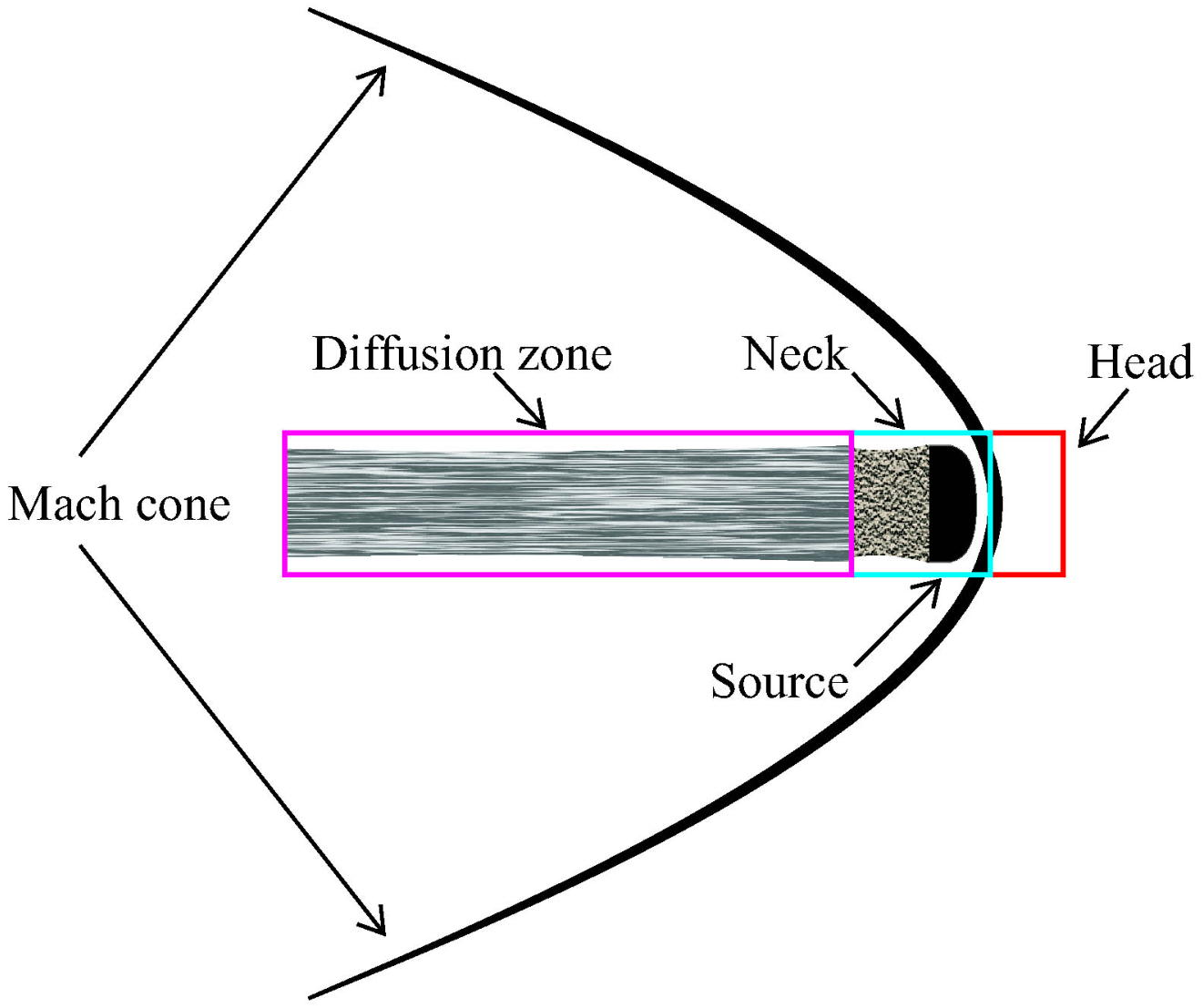}
  \caption[Schematic representation of the different regions associated with a jet event.]
  {Schematic representation of the different regions associated with a jet event.}
  \label{SketchRealJetDeposition}
\end{figure} 
\\However, in physical systems the amount of energy and momentum deposited by a propagating jet 
may be too large to consider them as small perturbations and thus it might not be appropriate to 
use linearized hydrodynamics. In particular, as Fig.\ \ref{SketchRealJetDeposition} reveals, the 
source itself (which needs a special treatment since it is non-thermalized) as well as the so-called head 
and neck regions (close to the position of the jet, discussed in detail in the next chapters) are not 
incorporated in a description applying linearized hydrodynamics. Therefore, something more 
sophisticated is needed for a correct prescription of the ongoing processes in heavy-ion 
collisions. In part \ref{part03}, we use a $(3+1)$-dimensional hydrodynamical approach to 
describe the propagation of a jet through a medium in local thermodynamical equilibrium.

\clearpage{\pagestyle{empty}\cleardoublepage}
%
%
\chapter[From String to Field Theory: The AdS/CFT Correspondence]
{From String to Field Theory: \\The AdS/CFT Correspondence}
\label{AdS/CFT}
\vspace*{-0.5cm}
Nature, as presently understood, can be described using quantum field theories like QCD. These
theories are able to explain experiments at observable distances. However, it is very likely that
at extremely short distances or equivalently at high energies (i.e., at the order of the Planck
scale\footnote{The Planck scale is an energy scale at which gravity becomes 
strong. It is characterized by the Planck length $l_{\rm Pl}=10^{-35}$~m and by the Planck 
mass $m_{\rm Pl}=\sqrt{\hbar c/G}\sim 1\cdot 10^{19}$~GeV/$c^2$ which may be defined 
as the mass for which the Compton wavelength of a particle 
$\lambda=\hbar/(Mc)$ equals its Schwarzschild radius $r_s=(GM)/c^2$.})
the effect of gravity, which is in general not included in quantum field theories, will 
become important.\\
Nevertheless, it is possible to include quantum gravity in a consistent quantum theory by 
assuming that the fundamental particles are not point-like, but extended objects, {\it strings} 
\cite{Strings}.\\
Depending on their state of oscillation, strings are able to give rise to different kinds of
particles and they are able to interact. String theory can only be defined in a certain number of
dimensions and geometries. For a flat space, it can only exist in $10$ dimensions 
\cite{Aharony:1999ti}. Such a $10$-dimensional theory describes strings with fermionic excitations
and gives rise to a supersymmetric theory. It is a candidate for a quantum theory of gravity 
\cite{Aharony:1999ti} since it predicted massless spin 2 particles, gravitons.\\
Originally, string theory was developed in the 1960's to describe the large number of mesons 
that were found experimentally at that time, a characteristic feature of the hadron
spectrum. It was derived to describe the dynamics of mesons propagating in space-time, but 
later on it was discovered that the strong interaction is very successfully described by the 
gauge theory QCD. \\
QCD exhibits the feature of asymptotic freedom, i.e., its effective coupling constant decreases 
with increasing energy. At low energies it becomes strongly coupled, so that perturbation theory 
is no longer valid, complicating any analytic calculations. Lattice gauge theory seems to be the best 
available tool in this energy range.\\
It was 'tHooft \cite{'tHooft:1973jz} who suggested that QCD is simpler in the limit of an 
infinite number of colors, $N_c\rightarrow\infty$. Subsequently, this would allow for an expansion 
in $1/N_c=1/3$. Since a diagrammatic expansion of QCD indicates that in the large 
$N_c$-limit QCD is a weakly interacting string theory, this $N_c\rightarrow\infty$ limit might connect 
gauge and string theories. \\Thus, there may be a correspondence between the large $N_c$-limit 
of field theories and string theories which is rather general.\\
In particular, it was shown by Maldacena \cite{Maldacena:1997re,Witten:1998qj,Witten:1998zw} that 
there is a correspondence between a supersymmetric $SU(N_c)$ (non-Abelian, i.e., Yang--Mills) 
conformal field theory\footnote{Conformal theories are invariant under conformal transformations, 
i.e., dilatation (scale invariance), inversion and so-called special conformal transformations. QCD
is not a conformal invariant theory since there is a scale in the theory, $\Lambda_{QCD}\approx 0.2$~GeV, 
and its coupling constant is runs.} in $4$ dimensions (${\cal N}=4$) which has a constant coupling of 
$g_{YM}^2/(4\pi)=g_s\sim 1/N_c$ ($\lambda\equiv g_{YM}^2N_c$, $g_s$ is the string coupling)
and a $10$-dimensional string theory on the product space $AdS_5\times S^5$. 
The ($5$-dimensional) Anti-de-Sitter space ($AdS_5$) is the space which characterizes the 
maximally symmetric solution of Einstein's equations with negative curvature. 
In general, the strings are propagating in a curved $10$-dimensional background of the form 
$AdS_5\times X^5$ where $X^5$ has to be a compact curved Einstin space. The simplest example 
is when $X^5$ is a $5$-sphere $S^5$ 
\cite{Klebanov:2000me}.
The conjecture is remarkable in a sense that it relates a $10$-dimensional theory of gravity to a 
$4$-dimensional theory without gravity, which is defined at the boundary of the $AdS_5$ space. This is a 
realization of the so-called {\it holographic principle}\footnote{In a theory of quantum gravity
the holographic principle states that all physics in some volume can be thought of as encoded on the boundary 
of that region, which should contain at most one degree of freedom per Planck
area \cite{Aharony:1999ti}.}. For a review, see e.g.\ Ref.\ \cite{Aharony:1999ti}.\\
The advantage of this {\it AdS/CFT correspondence} is that it offers direct access to 
the strong-coupling region. Since the gauge theory has a coupling constant of $\lambda$, which
is connected to the coupling constant of the gravitational theory $\alpha^\prime/R^2$ (with
$R$ being the radius of the AdS space and $\alpha^\prime\sim l_{Pl}^2$) by the relation
$\sqrt{\lambda}=R^2/\alpha^\prime$, the conjecture relates either a weakly coupled
gauge theory with a strongly coupled string theory or vice versa.\\
Moreover, nonzero temperatures can be studied by introducing a black-hole horizon. This allows for 
an extraction of transport coefficients, such as viscosity 
(discussed below) and heat diffusion. However, there is yet {\bf no} dual for QCD itself and most 
calculations performed using the AdS/CFT correspondence are only done to leading order in the 
limit of strong (i.e., infinite) coupling. \\
Hence, there is a caveat in applying the AdS/CFT duality to study properties of QCD, 
especially to describe experimental observables of heavy-ion collisions. One has to be aware of 
the fact that string theory is compared to a gauge theory with a constant coupling and 
that the duality for an infinite numbers of colors is  
already applicable in the case of $N_c=3$, as in QCD.

\section[Hard Probes in AdS/CFT]{Hard Probes in AdS/CFT}
\label{HardProbesAdSCFT}
\begin{figure}[t]
\centering
  \includegraphics[scale = 0.48]{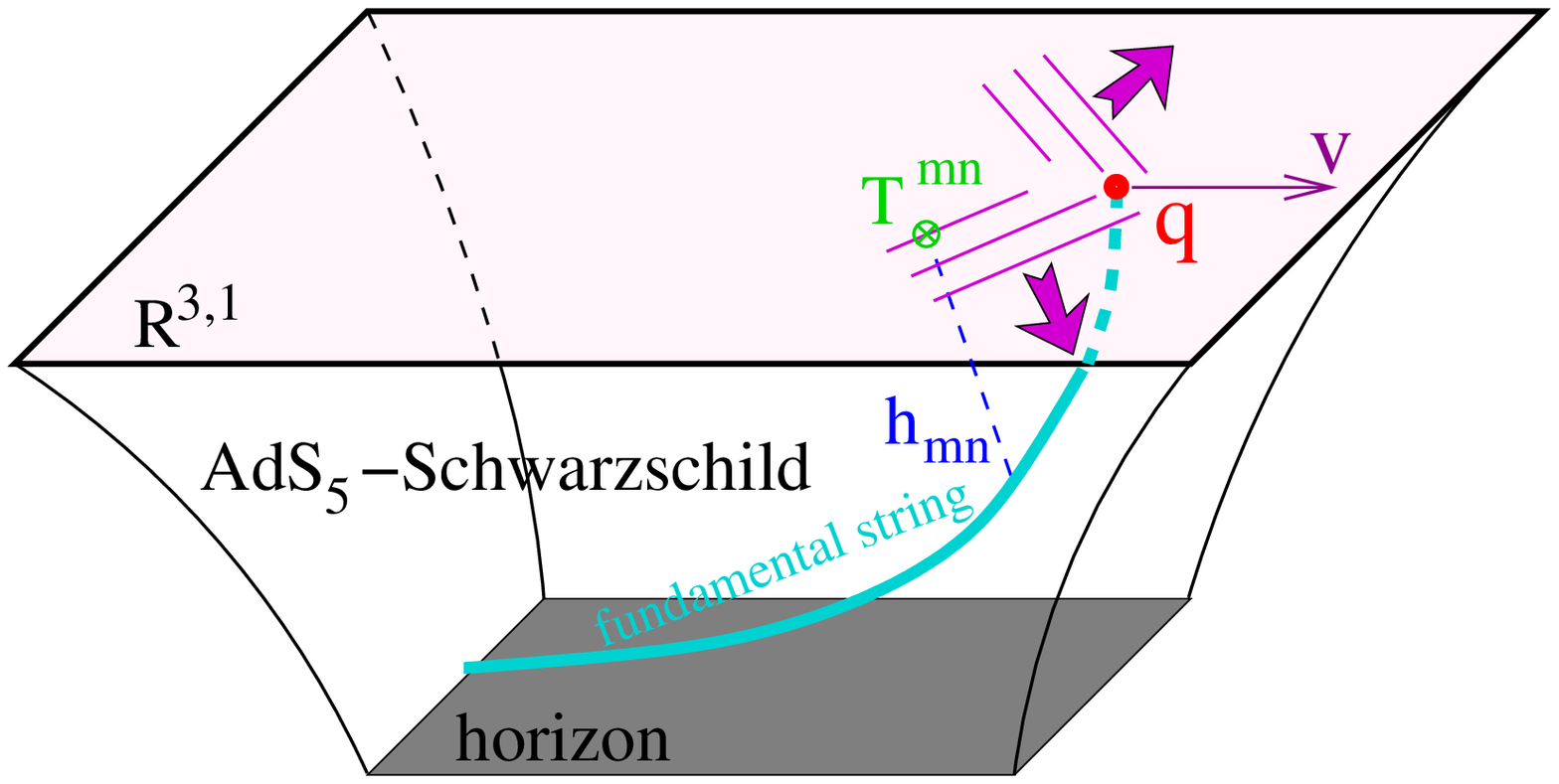}
  \caption[A schematic picture of a moving quark using the AdS/CFT correspondence.]
  {A schematic picture of a moving quark using the AdS/CFT correspondence. A string trails  
  the jet that is moving at the boundary of the $5$-dimensional $AdS_5$-Schwarzschild space 
  \cite{Friess:2006fk}.}
  \label{WakeAdSCFT}
\end{figure}
Hard probes are an excellent tool to study the matter created in a heavy-ion collision since 
their rate of energy and momentum deposition and their interactions with the medium allow for 
conclusions about the properties of the plasma formed. \\
At weak coupling, the energy loss of both light and heavy quarks can be well described using 
perturbative QCD (pQCD). However, in the regime of strong coupling, so far no reliable theoretical 
descriptions are available, but the AdS/CFT correspondence introduced above may offer some 
guidance. \\
Usually the investigation of hard probes using AdS/CFT is more complicated for light \cite{Chesler:2008uy}
than for heavy quarks. The problem of a heavy quark moving at a constant speed through a strongly coupled, 
nonzero temperature ${\cal N}=4$ Super--Yang--Mills (SYM) medium can be analyzed by considering 
metric fluctuations due to a string hanging down from the boundary of an $AdS_5$-Schwarzschild 
background geometry \cite{Herzog:2006gh,Gubser:2006bz,Friess:2006fk}, see Fig.\ \ref{WakeAdSCFT}. 
Here, it is assumed that the quark has been moving since $t\rightarrow -\infty$, thus it 
is a steady state solution.\\
This model is also called the {\it trailing string}. The heavy quark can be seen as
a hard probe if its mass is sufficiently large compared to the temperature of the background. 
To be precise the mass has to be larger than the typical scale of the medium, thus
$M\gg m_D=gT$ for pQCD and $M\gg\sqrt{\lambda}T$ for an ${\cal N}=4$ SYM theory.\\
Energy and momentum flow down the string at a constant rate, corresponding to the re-distribution 
of energy along the string. As a consequence of this energy loss, due to the drag 
(i.e., fluid resistance) force of the pasma, it is necessary to include background electric 
fields on the brane in order to ensure the constant velocity of the quark. Alternatively, it 
can be assumed that the quark is infinitely heavy so that it maintains its velocity despite 
the energy loss. The trailing string provides a source term for the Einstein equations that can be solved
to obtain the full energy-momentum tensor in the wake of the jet.\\
An example of the formidable analytical power of AdS/CFT calculations was given by Yarom in Ref.\ 
\cite{Yarom:2007ni} (see also Ref.\ \cite{Gubser:2007nd}). The total energy-momentum tensor describing 
the near-quark region in the laboratory frame [for cylindrical coordinates $u^\mu=(t,x_1,r,\theta)$
quoted in dimensionless units, thus devided by $\pi T$] is determined to be
\begin{equation}
\ T_{\mu\nu}^{Y}=P_0\,{\rm diag}\{3,1,1,r^2\}+\xi\,P_0 \,\Delta T_{\mu\nu}(x_{1},r)\,,
\label{energymomentumtensor1}
\end{equation}
where $P_0=N_{c}^2 \pi^2 T^4/8$ is the pressure of the ideal 
Super--Yang--Mills (SYM) plasma \cite{Gubser:1996de} and $\xi=8\sqrt{\lambda}\,\gamma/N_{c}^2$. 
The explicit form of $\Delta T_{\mu \nu}$ (which scales as $1/x^2$ with the distance $x$
from the quark) is
\begin{eqnarray}
 \Delta T_{tt} & =& \alpha \frac{v  \left[r^2(-5+13v^2-8v^4)+(-5+11v^2)x_1^2\right]x_1}
 {72\left[r^2(1-v^2)+x_1^2\right]^{5/2}}, \nonumber
 \end{eqnarray}
\begin{eqnarray} 
 \Delta T_{t x_1} & =& -\alpha \frac{v^2 \left[2 x_1^2+(1-v^2)r^2\right]x_1}
 {24 \left[r^2(1-v^2)+x_1^2\right]^{5/2}},\nonumber\\
 \Delta T_{tr} & = &-\alpha \frac{(1-v^2) v^2 \left[11 x_1^2+8r^2(1-v^2)\right] r}
 {72 \left[r^2(1-v^2)+x_1^2\right]^{5/2}}, \nonumber\\
 \Delta T_{x_1 x_1} & =& \alpha \frac{v  \left[r^2(8-13 v^2 +5v^4)+(11-5v^2)x_1^2\right]x_1}
 {72 \left[r^2(1-v^2)+x_1^2\right]^{5/2}},
 \nonumber\\
 \Delta T_{x_1 r} & = &\alpha \frac{ v (1-v^2) \left[8 r^2(1-v^2)+11 x_1^2\right]r}
 {72 \left[r^2(1-v^2)+x_1^2\right]^{5/2}}, \nonumber\\
 \Delta T_{r r} & = &-\alpha \frac{v (1-v^2)  \left[5 r^2(1-v^2) + 8 x_1^2\right]x_1}
 {72 \left[r^2(1-v^2)+x_1^2\right]^{5/2}}, \nonumber\\
 \Delta T_{\theta \theta} & =& -\alpha \frac{v (1-v^2) x_1 }
 {9 \left[r^2(1-v^2)+x_1^2\right]^{3/2}}\,,
\end{eqnarray}
with $\alpha=\gamma\sqrt{\lambda} T^2$. However, it was assumed throughout the derivation of 
Eq.\ (\ref{energymomentumtensor1}) that the metric disturbances caused by the moving string are 
small in comparison to the $AdS_{5}$ background metric, resulting in the condition that 
$\xi\,\Delta T_{\mu \nu} < 1$. Applying this approach to heavy-ion collisions with a proper choice 
of $N_c,\lambda$, and $\gamma$ \cite{Torrieri:2009mv} leads to a condition about the minimum 
distance from the quark where the above result is applicable. This energy-momentum tensor is the 
full non-equilibrium result in the strong-coupling limit and not just a solution of the hydrodynamic 
equations.\\
However, it is important to mention that up until now the experimentally measured particle 
distributions, see chapter \ref{ExperimentsQGP}, are obtained by triggering on {\bf light} 
quark and gluon jets. Thus, the calculations for heavy quark jets based on AdS/CFT have to be 
considered as a prediction.\\
Moreover, an idealized scenario is assumed in that context in which the probe travels through 
an infinitely extended, spatially uniform plasma. Though this is an approximation, it is always 
advisable to first investigate an idealized condition. The resulting properties can be included 
into hydrodynamic models or they can (as will be shown in chapter \ref{pQCDvsAdSCFT}) be compared 
to pQCD calculations for jet energy loss.

\section[Shear Viscosity from the AdS/CFT Correspondence]
{Shear Viscosity from the AdS/CFT Correspondence}
\label{ViscosityInAdSCFT}

One of the most appealing properties of the AdS/CFT correspondence is that it reflects the 
hydrodyamic behaviour of field theories. Thus, it was possible to extract the shear viscosity to 
entropy ratio which was shown to be \cite{Policastro:2001yc}
\begin{eqnarray}
\frac{\eta}{s}&\geq&\frac{1}{4\pi}\,.
\end{eqnarray}
This limit is supposed to be universal for all theories including gravitational duals 
in the supergravity limit ($N_c\rightarrow\infty$ and $\lambda\equiv g_{YM}^2 N_c\rightarrow\infty$)
\cite{Buchel:2003tz,Kovtun:2004de} and implies that a fluid of a given volume and entropy density 
cannot be a perfect liquid (with vanishing viscosity). It means that in all theories with
gravity duals, even in the limit of an infinite coupling, $\eta/s$ is larger 
than $\sim 0.08$.\\
This result is especially interesting because it was shown by calculations based on pQCD 
\cite{Xu:2008dv} and by comparison of elliptic flow measurements to theoretical predictions 
\cite{Luzum:2008cw} that the medium created in a heavy-ion collision can be well described by a 
fluid with a small $\eta/s$ ratio (depending on initial conditions, 
$0<\eta/s $\raisebox{-1.1mm}{$\stackrel{<}{\sim}$} $0.2$). \\
This range is clearly consistent with the value of $\eta/s$ calculated from the AdS/CFT 
correspondence. It was shown that within Gauss--Bonnet theory 
higher order derivatives may lead to a shear viscosity to entropy ratio of
\cite{Kats:2007mq,Brigante:2007nu,Brigante:2008gz} 
\begin{eqnarray}
\frac{\eta}{s}&\geq&\frac{16}{25}\left(\frac{1}{4\pi}\right)\sim 0.05\,,
\end{eqnarray}
smaller than conjectured by the Kovtun--Starinets--Son viscosity bound.

\section[Comparison of AdS/CFT to Hydrodynamics]{Comparison of AdS/CFT to Hydrodynamics}
\label{ComparisonHydro}
\begin{figure}[t]
\centering
  \includegraphics[scale = 0.3]{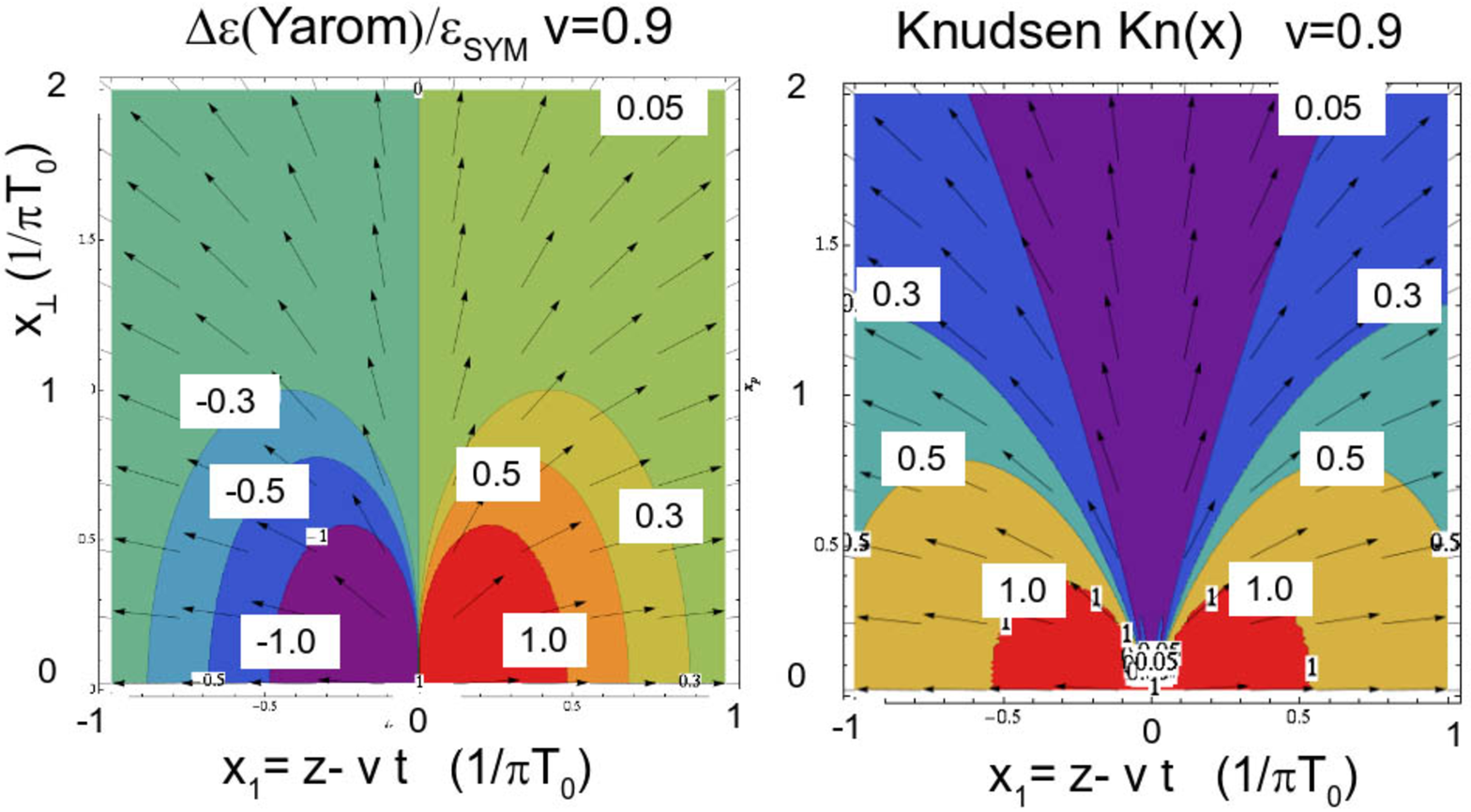}
  \caption[The fractional energy density perturbation and the field of the local Knudsen number 
  for the disturbance caused by a heavy quark jet using the AdS/CFT correspondence.]
  {The fractional energy density perturbation (left panel) and the field of the local Knudsen 
  number (right panel) for the disturbance caused by a heavy quark jet as calculated from the total 
  energy-momentum tensor describing the near-quark region using the AdS/CFT correspondence 
  \cite{Yarom:2007ni}, taken from \cite{Gyulassy:2008fa}.}
  \label{PlotComparisonKnudsen}
\end{figure}
The disturbances in the fluid caused by a moving jet are expected to behave hydrodynamically in 
the region sufficiently far from the present position of the jet ({\it far zone}). Close to the 
jet (in the so-called {\it near zone}) these disturbances should be large and thus 
hydrodynamics is supposed to break down. \\
In fluid dynamics, the Knudsen number $K$ is defined as the ratio between the mean-free path 
and the characteristic macroscopic length of the system [see Eq.\ (\ref{eqKnudsenNumber})]. 
Hydrodynamics is applicable when $K \ll 1$. \\
In conformal field theories at nonzero temperatures, the only dimensionful parameter is 
given by the temperature $T$ and, thus, both the mean-free path and the characteristic length 
should be proportional to $1/T$. However, the mean-free path is not a well-defined quantity in 
${\cal N}=4$ Super--Yang--Mills (SYM)  theories for very strong coupling. Nevertheless, one 
can still establish an effective Knudsen field (that is well defined in the supergravity limit) 
in terms of the sound attenuation (or absorption) length $\Gamma=1/3\pi T$ and the Poynting 
vector\footnote{The Poynting vector represents the energy flux vector of electromagnetic energy.} 
$S^i=T^{0i}$ \cite{Noronha:2007xe}
\begin{equation}
Kn = \Gamma \left| \frac{\nabla\cdot\vec{S}}{S} \right|.
\label{defknudsen}
\end{equation}
As Fig.\ \ref{PlotComparisonKnudsen} shows, 
for a jet moving with $v=0.9$, the region characterized by a large energy perturbation of 
$\Delta\varepsilon/\varepsilon_{\rm SYM}> 0.3$ [computed from the energy-momentum tensor given 
above, see Eq.\ (\ref{energymomentumtensor1})] corresponds approximately with the 
locus defined by the local Knudsen number of $Kn\geq 1/3$. Therefore, it was argued in Refs.\
\cite{Gyulassy:2008fa,Noronha:2008un} that for $Kn\geq 1/3$ the system can no longer be described by
linearized first-oder Navier--Stokes hydrodynamics. A direct comparison of the total energy-momentum tensor describing 
the near-zone energy-momentum tensor of Ref.\ \cite{Yarom:2007ni} [cf.\ see again Eq.\ 
(\ref{energymomentumtensor1})] and a first-order Navier--Stokes ansatz showed \cite{Noronha:2007xe} that a 
hydrodynamic description of the disturbances caused by the heavy quark is valid down to distances 
of about $1/T$ from the heavy quark.
Later on, we will use the condition for applicability of hydrodynamics ($Kn\leq 1/3$) to analyze 
AdS/CFT results.

\section[Mach-Cone-like Correlations in AdS/CFT]
{Mach-Cone-like Correlations in AdS/CFT}
\label{JetsInAdSCFT}
\begin{figure}[t]
\centering
  \includegraphics[scale = 0.35]{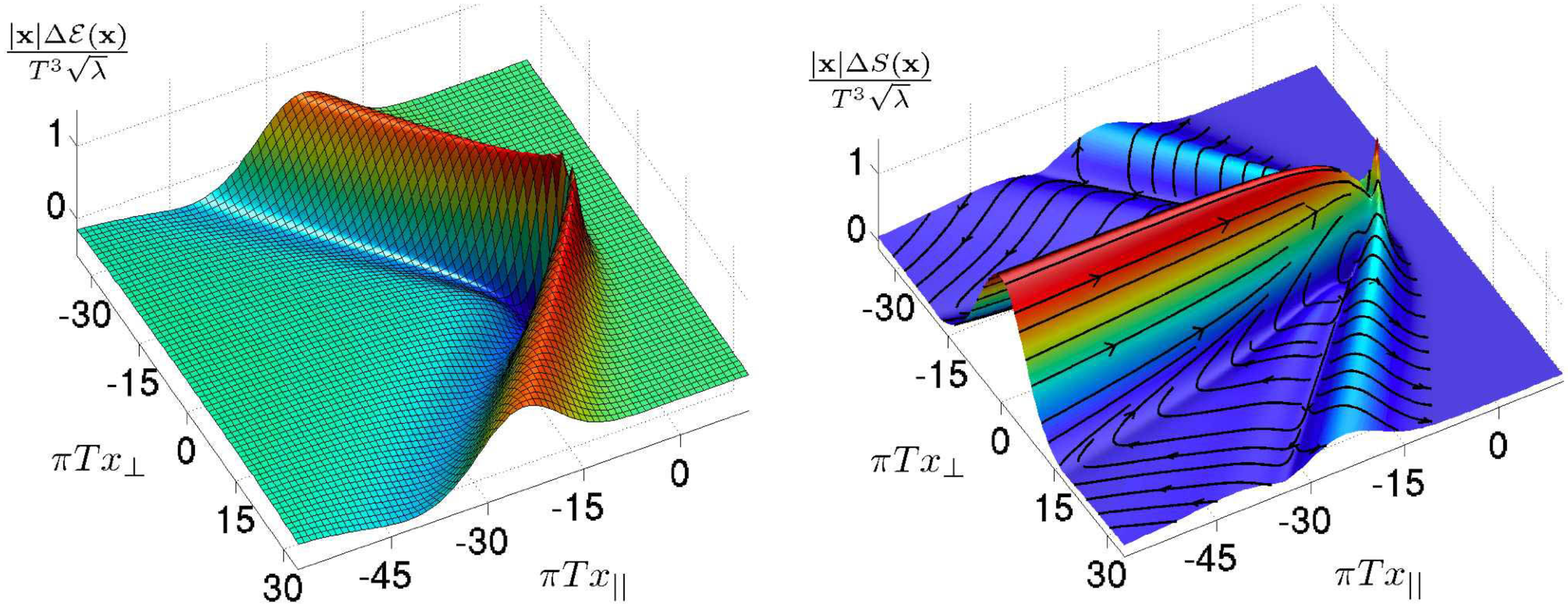}
  \caption[Energy and momentum (Poynting vector) distribution calculated for a jet moving with 
  $v=0.75$ using the AdS/CFT correspondence.]
  {Energy and momentum (Poynting vector) distribution calculated for a jet moving with $v=0.75$ 
  using the AdS/CFT correspondence. A Mach cone angle with $\phi_M\sim 50°$ is visible in both 
  patterns, however, a strong flow along the trajectory of the jet (diffusion wake) is also 
  visible in the right plot \cite{Chesler:2007sv}.}
  \label{AdSCFTMachDiffusion}\vspace*{-2mm}
\end{figure}
\begin{figure}[t]
\centering\hspace*{-0.2cm}
  \includegraphics[scale = 1.6]{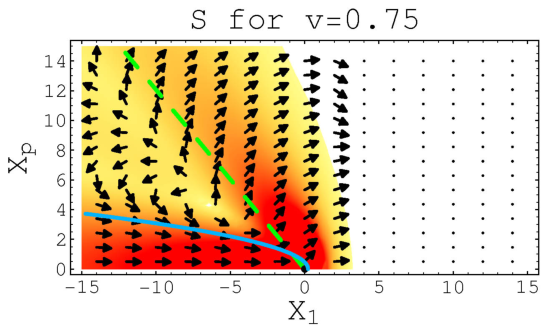}
  \caption[The Poynting vector distribution calculated from the gauge/string duality.]
  {The Poynting vector distribution (the magnitude of $\vert\vec{S}\vert$ is color-coded with red
  indicating a large and white zero magnitude) calculated from the gauge/string duality 
  \cite{Gubser:2007ga}. The green dashed line indicates the Mach cone while the solid blue line 
  estimates where the diffusion-wake profile reaches half of its maximal value.}
  \label{GubserMachDiffusion}
\end{figure}
\begin{figure}[b]
\centering
  \includegraphics[scale = 0.28]{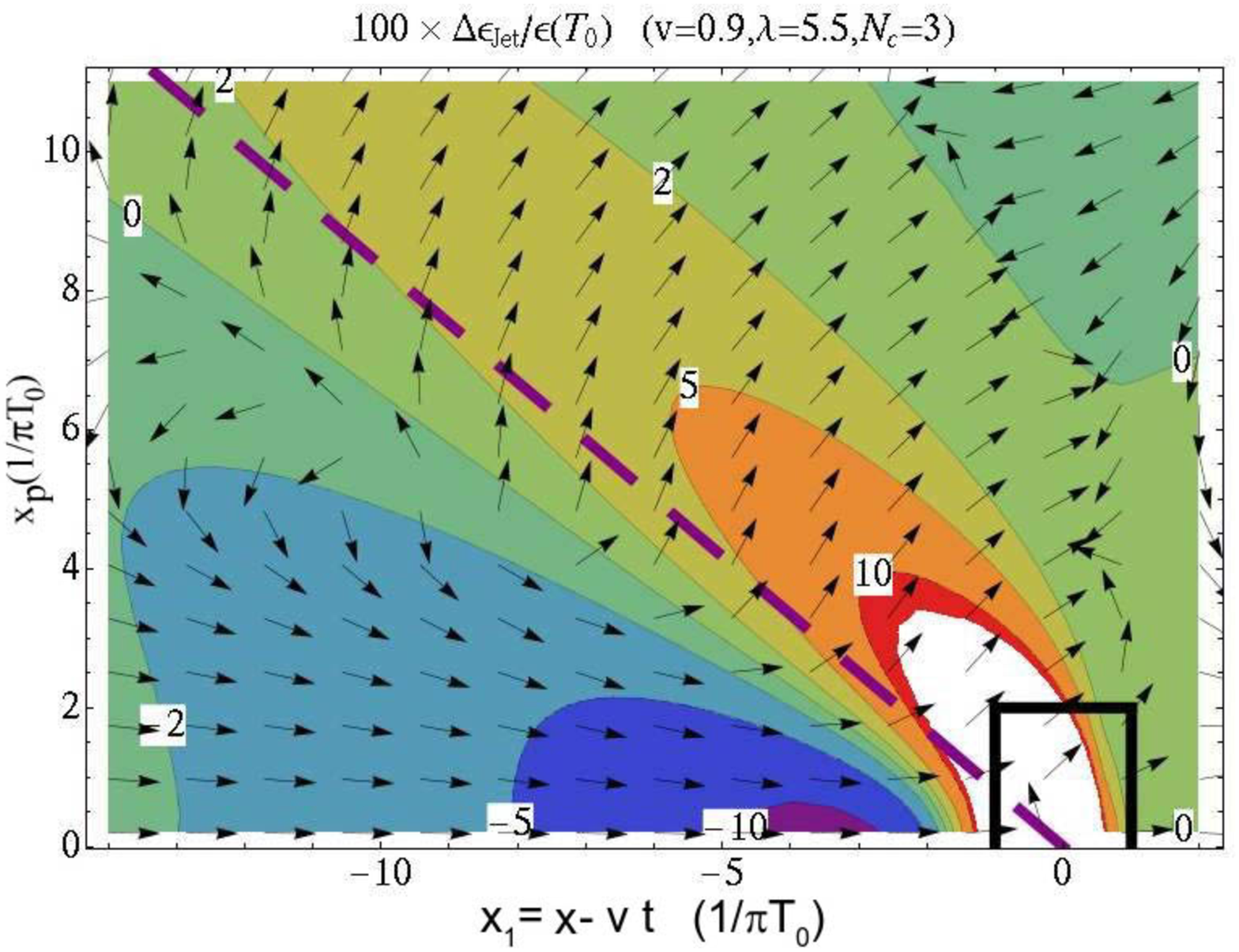}
  \caption[The relative energy-density perturbation due to a heavy quark propagating with $v=0.9$ 
  through an ${\cal N}=4$ Super--Yang--Mills (SYM) plasma.]
  {The relative energy-density perturbation due to a heavy quark propagating with $v=0.9$ through 
  an ${\cal N}=4$ Super--Yang--Mills (SYM) plasma with $N_c=3$ \cite{Gubser:2007ga}. The arrows 
  indicate the Poynting vector flow (momentum flux) directions and the dashed line denotes the 
  Mach cone with $\cos\phi_M=1/(\sqrt{3}v)$. The box indicates the neck zone \cite{Noronha:2008un}.}
  \label{JorgeNeck}
\end{figure}
\begin{figure}[t]
\centering
  \includegraphics[scale = 0.36]{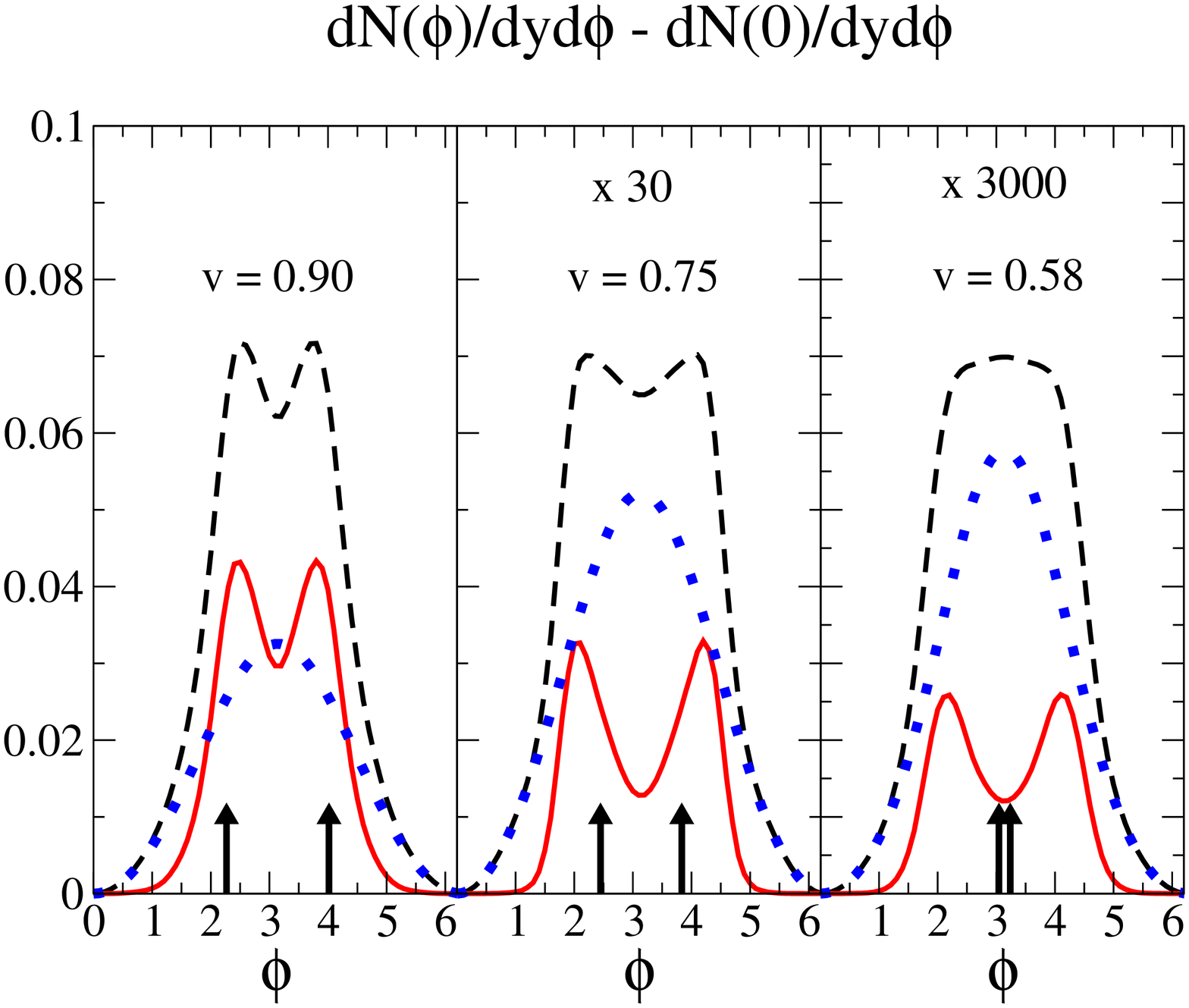}
  \caption[Background-subtracted, azimuthal angular distribution for massless particles at 
  mid-rapidity obtained from an isochronous Cooper--Frye freeze-out for the temperature and 
  velocity fields obtained from the AdS/CFT string-drag model.]
  {Background-subtracted, azimuthal angular distribution for massless particles at mid-rapidity 
  obtained from an isochronous Cooper--Frye freeze-out for the temperature and velocity fields 
  obtained from the AdS/CFT string-drag model \cite{Gubser:2007ga} for the three different 
  scenarios of $v=0.9,p_T/(\pi T_0)=4-5$, $v=0.75,p_T/(\pi T_0)=5-6$ and $v=0.58,p_T/(\pi T_0)=6-7$.
  The red curves show the contribution from the neck region, defined by 
  $\Delta\varepsilon/\varepsilon_{\rm SYM}$, the blue curves result from integrating the far zone 
  (excluding the neck region) and the black line display the total yields. The expected Mach 
  angles are indicated by the arrows \cite{Noronha:2008un}.}
  \label{JorgeParticleDistribution}
\end{figure}

The propagation of a supersonic jet through a fluid, associated with energy deposition, is supposed to lead 
to the creation of Mach cones (see chapter \ref{ShockWavePhenomena}). Since, as discussed above, 
the far zone of the jet can be described using hydrodynamics, the question arises if Mach-cone 
formation also emerges when calculating the disturbance of the medium due to a jet moving faster 
than the speed of sound\footnote{In ${\cal N}=4$ Super--Yang--Mills (SYM) theory the speed 
of sound is $c_s=1/\sqrt{3}$.} applying the AdS/CFT correspondence.\\
In Ref.\ \cite{Chesler:2007sv} (see Fig.\ \ref{AdSCFTMachDiffusion}), it was demonstrated that 
using the gauge/string duality, the perturbation of the energy (left panel of Fig.\ 
\ref{AdSCFTMachDiffusion}) and momentum distributions (often also referred to as Poynting-vector
distribution, right panel of Fig.\ \ref{AdSCFTMachDiffusion}) result in a sonic disturbance 
leading to a shock front with a characteristic opening angle [see Eq.\ (\ref{openingangle})] which 
therefore can be identified with as Mach cone. The same was also shown in Ref.\ 
\cite{Gubser:2007ga}, however it was stressed in this publication that a strong flow is created 
behind the jet which is called the {\bf diffusion wake} (see Fig.\ \ref{GubserMachDiffusion}). 
The impact of this diffusion wake will be discussed in detail in chapter \ref{DiffusionWake}.
\\Those energy and momentum distribution patterns cannot directly be compared to the 
particle correlations obtained from experiment. Thus, it is necessary to perform a convolution 
into particles. 
However, a conical signal based on a hydrodynamic perturbation may be washed out by 
thermal smearing once the system breaks up into particles. Hence, it might also be possible that 
a detectable Mach-cone-like signal could come from the region where the linearized hydrodynamical 
approximation is not valid. It was shown in Ref.\ \cite{Gyulassy:2008fa,Noronha:2008un} that a 
region close to the location of the jet, the so-called {\it neck zone} (see the framed area in 
Fig.\ \ref{JorgeNeck}) reveals a strong transverse component that leads to a double-peaked 
structure in the calculated two-particle distribution (see Fig.\ \ref{JorgeParticleDistribution}).
\\The energy-momentum tensor that was used in this investigation and describes both 
the near and the far-zone was computed by Gubser, Yarum, and Pufu in Ref.\ \cite{Gubser:2007ga} 
for a jet moving with $v=0.9$, applying the string-drag model. Plotting the 
energy-density perturbation $\Delta\varepsilon/\varepsilon_{\rm SYM}$ (cf.\ Fig.\ 
\ref{JorgeNeck}), different regions may be introduced:
\begin{itemize}
\item The Mach cone, indicated by the dashed magenta line,
\item the diffusion zone below that line, characterized by the strong flow along the jet axis 
($x_1$-axis in Fig.\ \ref{JorgeNeck}),
\item the neck zone (denoted by the box in Fig.\ \ref{JorgeNeck}), specified by the condition that 
$\Delta\varepsilon/\varepsilon_{\rm SYM}>0.3$, 
\item the head zone which is an inner region of the neck area\footnote{The head zone
can also be defined as distinct area following the neck
region (see Fig.\ \ref{SketchRealJetDeposition}).}, 
very close to the location of the jet.
\end{itemize}
In the supergravity limit the energy-momentum tensor can be decomposed accordingly (into 
the string-drag zones)
\begin{eqnarray}
\hspace*{-0.9cm}
T^{\mu\nu}(t,\vec{x})&=&T^{\mu\nu}_0(t,\vec{x})+\delta T^{\mu\nu}_{Mach}(t,\vec{x}) 
+ \delta T^{\mu\nu}_{neck}(t,\vec{x})+ \delta T^{\mu\nu}_{Coul}(t,\vec{x})\,.
\label{EqnStressZones}
\end{eqnarray}
Here, $x_1=x-vt$ denotes the direction along the (away-side) jet and $x_T$ the transverse 
coordinate. Then, the Mach part for which the condition $Kn < 1/3$ has to be fulfilled in order to 
guarantee the applicability of linearized first-order Navier--Stokes hydrodynamics, can be 
expressed in terms of the local temperature and flow velocity fields through \cite{Gyulassy:2008fa}
\begin{eqnarray}
\hspace*{-0.8cm}\delta T_{Mach}^{\mu\nu}(x_1,x_\perp) &=& 
\frac{3}{4} K\left\{T^4\left[\frac{4}{3}u^\mu u^\nu-\frac{1}{3}g^{\mu\nu}
+ \frac{\eta}{sT} \partial^{( \mu}u^{\nu)} \right]- T^{\mu\nu}_0\right\} \nonumber \\ 
\hspace*{-0.8cm}&& \times \theta(1-3Kn)\,,
\label{EqMachZone}
\end{eqnarray}
where $\partial^{( \mu}U^{\nu)}$ is the symmetrized, traceless flow velocity gradient, $Kn$ 
is the local Knudsen number, $K=(N_c^2-1)\pi^2/2$ the Stefan Boltzmann constant for the 
${\cal N}=4$ Super--Yang--Mills (SYM) plasma, and $T^{\mu\nu}_0$ the energy-momentum tensor 
of the (static) background. The theta function defines the far zone where the equilibration 
happens quickly enough enough to ensure a hydrodynamical prescription. \\
This far zone excludes the neck region close to the jet which is a non-equilibrium zone that 
can be described in ${\cal N}=4$ SYM via \cite{Yarom:2007ni,Gubser:2007nd}
\begin{equation}
\delta T_{neck}^{\mu\nu}(x_1,x_\perp)\approx \theta(3Kn-1)
\frac{\sqrt{\lambda} T_0^2}{x_\perp^2 +\gamma^2 x_1^2}Y^{\mu\nu}(x_1,x_\perp)
\end{equation}
Here, $Y^{\mu\nu}$ is a dimensionless ``angular'' tensor field, which reduces to the 
energy-momentum tensor given in Eq.\ (\ref{energymomentumtensor1}) at very small distances 
from the jet.\\
The head region is that part of the neck zone where the energy-momentum tensor becomes dominated 
by the contracted Coulomb field of the quark. Thus, as it was shown in Ref.\ 
\cite{Dominguez:2008vd}, the head can be defined by equating the Coulomb energy density to the 
near-zone energy density given by Eq.\ (\ref{energymomentumtensor1}). Its boundary is 
approximately given by \cite{Noronha:2008un}
\begin{equation}
x_{\perp}^2 + \gamma^2 x_{1}^2 =
\frac{1}{(\pi T_0)^4}\frac{(2x_{\perp}^2+x_{1}^2)^2}
{\gamma^4 x_{1}^2(x_{\perp}^2/2 + \gamma^2 x_{1}^2)^2}.
\end{equation}
The observable consequences can be investigated by applying a Cooper--Frye ha\-dronization procedure 
\cite{Gyulassy:2008fa,Noronha:2008un} with an isochronous freeze-out as discussed in chapter 
\ref{IdealHydrodynamics}. \\
This is a very strong model assumption since the freeze-out
is applied to the total volume, including the non-equilibrated neck zone (where the Cooper--Frye
freeze-out may not be applicable, see also chapter \ref{pQCDvsAdSCFT}) which is roughly the 
region between $-1<x_1\,(\pi T_0)<1$ and $0-x_T\,(\pi T_0)<2$ (see Fig.\ \ref{JorgeNeck}).\\
The background-subtracted, azimuthal angular distribution for massless particles at mid-rapidity 
is shown in Fig.\ \ref{JorgeParticleDistribution} for three different velocities of the jet and 
$p_T$-ranges, $v=0.9,p_T/(\pi T_0)=4-5$, $v=0.75,p_T/(\pi T_0)=5-6$ and $v=0.58,p_T/(\pi T_0)=6-7$. 
The black lines display the total yield that consists of the contribution from the neck (red line) 
and from the far region excluding the neck (blue line).
The figure reveals that the double-peaked structure occurring for $v=0.9$ and $v=0.75$ is completely 
due to the (non-equilibrated) neck region that always develops a dip around $\phi=\pi$. 
\\However, this double-peaked structure of the neck region is completely uncorrelated to a Mach cone since 
the location of the peaks do not follow the expected Mach-cone angles, which are indicated by the 
arrows in Fig.\ \ref{JorgeParticleDistribution} and change with the velocity of the jet. 
For $v=0.9$, the two peaks of the neck region incidentally agree with the Mach angle.\\
Thus, the predictive power of the AdS/CFT string drag model could be tested 
\cite{Gyulassy:2008fa,Noronha:2008un} by looking for conical distributions that deviate from 
the expected Mach angle for supersonic, but not ultra-relativistic {\it identified heavy quark 
jets}.

\section[The No-Go Theorem]{The No-Go Theorem}
\label{NoGoTheorem}

Figure \ref{JorgeParticleDistribution} reveals that the far zone, which includes the Mach contribution 
but excludes the neck region, always shows a broad away-side peak instead of an expected
double-peaked structure. It can be shown \cite{Gyulassy:2008fa,Noronha:2008un}, as reviewed 
below, that this result is universal in the large-$N_c$ limit and independent of the strength 
of the diffusion wake formed behind the jet.\\
For associated away-side particles with $p^{\mu}=(p_{T},p_{T}\cos (\pi-\phi),p_{T}\sin(\pi-\phi),0)$
the azimuthal distribution at mid-rapidity is given by [see Eq.\ (\ref{CFFormula})]\vspace*{1.5ex}
\begin{equation}
\frac{dN}{p_Tdp_T d\phi dy}\Big
|_{y=0}=\int_{\Sigma}d\Sigma_{\mu}p^{\mu}\left[f(u\cdot p/T)
-f_{eq} \left( E_0/T_0  \right)\right]\,,
\end{equation}
where a constant, thermal background $f_{eq}$ is subtracted. In the supergravity approximation of 
$N_c\rightarrow \infty$, or equivalently for $\Delta T\ll T_0$ and $\vec{p}\cdot\vec{u}\ll T_0$
(both, $\Delta T\sim\sqrt{\lambda}$ and $u\sim\sqrt{\lambda}/N_c^2$ are small, 
since $\lambda \gg 1$,but $N_c\gg \lambda$), the Boltzmann exponent can be expanded up to 
corrections of $\mathcal{O} (\lambda/N_c^4)$. 
\\Choosing an isochronous ansatz with $d\Sigma^\mu= x_T d x_T dx_1 d\varphi\,\left(1,0,0,0\right)$ 
and coordinates of $u^\mu(x_1,x_T)=(u^0,u_1,u_T\sin\varphi,u_T\cos\varphi,0)$, 
after integration over $\varphi$ in we obtain the distribution
\begin{eqnarray}
\frac{dN}{p_{T}dp_{T}d\phi dy}\Big |_{y=0}&=&
2\pi\,p_T\,\int_{\Sigma} dx_1 dx_T x_T \nonumber\\
&&\hspace*{-0.6cm}\times
\left(\exp\left\{-\frac{p_T}{T}\left[u_0-u_1\cos(\pi-\phi)\right]\right\}\, I_0(a_T)
-e^{-p_T/T_0} \right)\,.\nonumber\\
\label{fulldistrib}
\end{eqnarray}
Here, $a_T=p_T u_\perp\sin(\pi-\phi)/T$ and $I_0$ denotes the modified Bessel function. 
However, in the supergravity approximation $a_{\perp} \ll 1$, and therefore one can expand the 
Bessel function
\begin{equation}
\lim_{x\to 0}\,I_0 (x) =1+\frac{x^2}{4}+\mathcal{O}(x^4)
\end{equation}
to get an approximate expression for the distribution \cite{Noronha:2008un}
\begin{equation}
\frac{dN}{p_Tdp_T d\phi dy}\Big
|_{y=0}\simeq e^{-p_{T}/T_0}\frac{2\pi\,p_{T}^2}{T_0}
\left[\frac{\langle \Delta T\rangle}{T_0} + \langle u_{1}\rangle \cos
(\pi-\phi) \right]\,,
\end{equation}
where $\langle \Delta T\rangle= \int_{\Sigma} dx_1 dx_\perp x_\perp \,\Delta T$ and 
$\langle u_1\rangle= \int_{\Sigma} dx_1 dx_\perp x_\perp \,u_1$. Generally speaking, the above 
equation holds as long as the approximation 
$p_T/T=p_T/(T_0+\Delta T)\approx p_T/T_0(1-\Delta T/T_0)$ is valid. Therefore, in the 
$N_c\rightarrow\infty$ limit, the associated away-side distribution reveals a 
broad peak around $\phi=\pi$. A double-peaked structure in the away-side of the jet correlation 
function can only arise when the approximations made become invalid.

\clearpage{\pagestyle{empty}\cleardoublepage}
%
%
%
\part[Jet Propagation and Mach-Cone Formation]{Jet Propagation and Mach-Cone Formation}
\label{part03}
%
%
\thispagestyle{plain}
\vspace*{19\baselineskip}
\hspace*{-0.75cm}
In the following chapters, we present our investigations about the propagation of a jet through
a strongly-coupled medium, using $(3+1)$-dimensional ideal hydrodynamics. We apply and compare
different source terms describing the energy and momentum deposition of a jet as well as its 
interaction with the QGP, starting with a schematic source term, but also for the two independent 
approaches of pQCD and AdS/CFT. \\
While the pQCD scenario is certainly the correct description in the hard-momentum region where jets 
are produced ($Q \gg T_0$), in the soft part of the process ($Q \sim T_0$) non-perturbative effects, 
covered by the AdS/CFT approach, may become relevant. Here, the most striking differences will
become apparent due to the influence of the neck region as already discussed in chapter 
\ref{AdS/CFT}.\\
We start with a study on the away-side angular correlations for punch-through and
fully stopped jets applying a simple ansatz for the source term. Surprisingly enough, we 
find that the medi\-um's response as well as the corresponding correlations are largely 
insensitive to whether the jet punches through or stops inside the medium. 
The existence of the diffusion wake is shown to be universal to all scenarios 
where momentum as well as energy is deposited into the medium. In ideal hydrodynamics,
this can be readily understood through vorticity conservation which is examined separately in 
chapter \ref{PolarizationProbesofVorticity}.\\
Though the above mentioned analyses are done for a static medium, chapter \ref{ExpandingMedium}, 
which discusses an expanding system, shows that while radial flow may for some jet trajectories
diminish the impact of the diffusion wake \cite{Neufeld:2008eg}, it still strongly influences 
the final angular particle distributions. However, the main characteristics of those distributions are 
due to the different contributions of several possible jet-trajectories through the expanding medium
\cite{Chaudhuri:2006qk,Chaudhuri:2007vc} leading to a double-peaked structure which is therefore 
not directly connected to the emission angles of a Mach cone.

\chapter[The Diffusion Wake]
{The Diffusion Wake}
\label{DiffusionWake}

In general, a fast moving parton (which could be a light quark/gluon or a heavy quark) will
lose a certain amount of its energy and momentum along its path through the medium and consequently 
decelerate. Thus, the fate of the parton jet strongly depends on its initial energy: 
if the parton has enough energy it can punch through the medium and fragment in the vacuum
(a {\it punch-through jet}) or it can be severely quenched until it becomes part of the thermal 
bath (a {\it stopped jet}). \\
Of course, the amount of initial energy required for the parton to punch through depends on the 
properties of the medium (a very large energy loss per unit length $dE/dx$ means that most of the
jets will be quenched while only a few jets would have enough energy to leave the plasma). 
In the following, we solve the $(3+1)$-dimensional ideal hydrodynamical equations including source 
terms (see chapter \ref{ELossHydro}) that describe those two scenarios in order to 
compare the final away-side angular correlations produced by a punch-through and a fully stopped 
jet in a static medium with a background temperature $T_0$. \\
For simplicity, our medium is considered to be a gas of massless $SU(3)$ gluons for which $p=\varepsilon/3$, 
where $p$ and $\varepsilon$ are the pressure and the energy density, respectively.
Two different freeze-out procedures, briefly reviewed below (see also section \ref{Freezeout}), 
are employed in order to obtain the angular distribution of particles 
associated with the away-side jet. \\
We use a simplified Bethe--Bloch model \cite{Bethe} 
to show that the explosive burst of energy and momentum (known as the Bragg peak 
\cite{Bragg,Chen,Sihver,Kraft}) deposited by a fully quenched jet immediately
before it thermalizes does not stop the strong flow behind the jet (the diffusion wake) and, 
thus, the structures on the away-side of angular correlation functions are very similar to those 
of punch-through jets. This explosive release of
energy before complete stopping is a general phenomenon which has been employed, 
for instance, in applications of particle beams for cancer therapy 
\cite{WilsonPRL,pshenichnov1,pshenichnov2}.\\
In our system of coordinates, the beam axis is aligned with the $z$-direction and
the associated jet moves along the $x$-direction with velocity ${\vec v}=v\,\hat{x}$. We
take the net baryon density to be identically to zero. Moreover, we omit the near-side correlations
associated with the trigger jet and assume that the away-side jet travels through the medium 
according to a source term that depends on the jet velocity profile which shall be discussed below 
for the case of punch-through and stopped jets.\\
The away-side jet is implemented in the beginning of the hydrodynamical evolution at 
$x=-4.5$~fm, and its motion is followed until it reaches $x=0$. For a jet moving with a constant 
velocity $v_{\rm jet}$ this happens at $t_f = 4.5/v_{\rm jet}$~fm. \\
In order to obtain the away-side angular correlations, we use the two different methods introduced
in section \ref{Freezeout}. Applying the isochronous Cooper--Frye (CF) \cite{Cooper:1974mv}, 
the fluid velocity $u^\mu(t_f,\vec{x})$ and temperature $T(t_f,\vec{x})$ fields are converted 
into free particles at a freeze-out surface $\Sigma$ at constant time $t_f$. In principle, 
one has to ensure that energy and momentum are conserved during the freeze-out procedure \cite{CsernaiBook}. 
However, the associated corrections are zero if the equation of state is the same before and 
after the freeze-out, as it is assumed in the present considerations. In this case, the 
momentum distribution for associated (massless) particles 
$p^\mu = \left(p_T, p_T\cos(\pi - \phi),p_T\sin(\pi - \phi)\right)$ 
at mid-rapidity $y=0$ is computed via
\begin{equation}
\frac{dN_{\rm ass}}{p_T dp_T dy d\phi}\Big
|_{y=0}=\int_{\Sigma}d\Sigma_{\mu}p^{\mu}
\left[f_0(u\cdot p/T)-f_{eq}\right]\,.
\label{cooperfrye}
\end{equation}
Here, $\phi$ is the azimuthal angle between the emitted particle and the trigger, $p_T$
is the transverse momentum, $f_0 = \exp[-u^\mu(t,\vec{x}) p_\mu/T(t,\vec{x})]$ the local 
Boltzmann equilibrium distribution, and $f_{eq}\equiv f|_{u^{\mu}=0,T=T_0}$ 
denotes the isotropic background yield. We checked that our results do not change significantly 
if we use a Bose--Einstein distribution instead of the Boltzmann distribution. 
The background temperature is set to $T_0=0.2$ GeV. \\
Following Refs.\ 
\cite{CasalderreySolana:2004qm,CasalderreySolana:2006sq,Noronha:2008un,Betz:2008js},
we define the angular function
\begin{equation}
CF(\phi)=\frac{1}{N_{max}}\left.\left(\frac{dN_{\rm ass}(\phi)}{
p_T dp_T dy d\phi}
-\frac{dN_{\rm ass}(0)}{
p_T dp_T dy d\phi}
\right)
\right\vert_{y=0}\, ,
\label{CFfunction}
\end{equation}
where the constant $N_{max}$ is used to normalize the plots. It is important to mention that 
in the associated $p_T$-range of interest a coalescence/recombination hadronization scenario
\cite{Fries:2003vb,Greco:2003xt,Fries:2003kq,Fries:2004hd,Greco:2003mm}
may be more appropriate than CF freeze-out. Since it might well be possible
that procedures following the recombination of partons into hadrons influence the final
particle distributions, this issue deserves further scrutiny.\\
The other freeze-out prescription (called bulk-flow freeze-out) was introduced in Ref.\ 
\cite{Betz:2008wy}. The main assumption behind the bulk-flow freeze-out is that all the particles 
inside a given small sub-volume of the fluid will be emitted into the same direction as the 
average local energy flow
\begin{equation}
\frac{d \mathcal{E}}{d\phi dy} = 
\int d^3 {\vec x}\,\, \mathcal{E}(\vec{x})\,
\delta\left[\phi - \Phi(\vec{x})\right]\, 
\delta\left[y-Y(\vec{x})\right]\,.
\label{Pebbleeq}
\end{equation}
Here, $\phi$ is again the azimuthal angle between the detected particle and the trigger jet and 
$y$ is the particle rapidity. Only the $y=0$ yield is considered. The cells are selected 
according to their local azimuthal angle
$\Phi(\vec{x})=\arctan \left[\mathcal{P}_y(\vec{x})/\mathcal{P}_x(\vec{x})\right]$
and rapidity
$Y(\vec{x})={\rm Artanh}\left[\mathcal{P}_z(\vec{x})/\mathcal{E}(\vec{x})\right]$. 
The local momentum density of the cell is $T^{0i}(\vec{x})=\mathcal{P}_{i}(\vec{x})$, 
while its local energy density in the lab frame is $\mathcal{E}(\vec{x})=T^{00}(\vec{x})$. 
The $\delta$--functions are implemented using a Gaussian representation as in Ref.\ 
\cite{Betz:2008wy}. \\
Due to energy and momentum conservation, this quantity should be conserved after freeze-out. 
Eq.\ (\ref{Pebbleeq}) is not restricted to a certain $p_T$ and does not include the thermal 
smearing that is always present in the CF freeze-out.

\section{Punch-Through Jets in a Static Medium}
\label{PunchThroughJets}

In this section, we consider a jet moving with a uniform velocity of $v_{\rm{jet}}=0.999$
through the medium. The source term is given by [cf.\ Eq.\ (\ref{PebbleSource})]
\begin{eqnarray}
\label{source}
J^\nu = \int\limits_{\tau_i}^{\tau_f}d\tau 
\frac{dM^\nu}{d\tau}\delta^{(4)}
\left[ x^\mu - x^\mu_{\rm jet}(\tau) \right],
\end{eqnarray}
where $\tau_{f}-\tau_i$ denotes the proper time interval associated with the jet evolution.
We further assume a constant energy and momentum loss rate 
$\hspace*{0.5cm}dM^\nu/d\tau = (dE/d\tau,d\vec{M}/d\tau)$ along the trajectory of the jet 
$x^\mu_{\rm jet} (\tau) = x_0^\mu + u^\mu_{\rm jet}\tau$.\\
In non-covariant notation, this source term has the form
\begin{figure}[t]
\hspace*{-0.9cm}
\includegraphics[width=15.5cm]{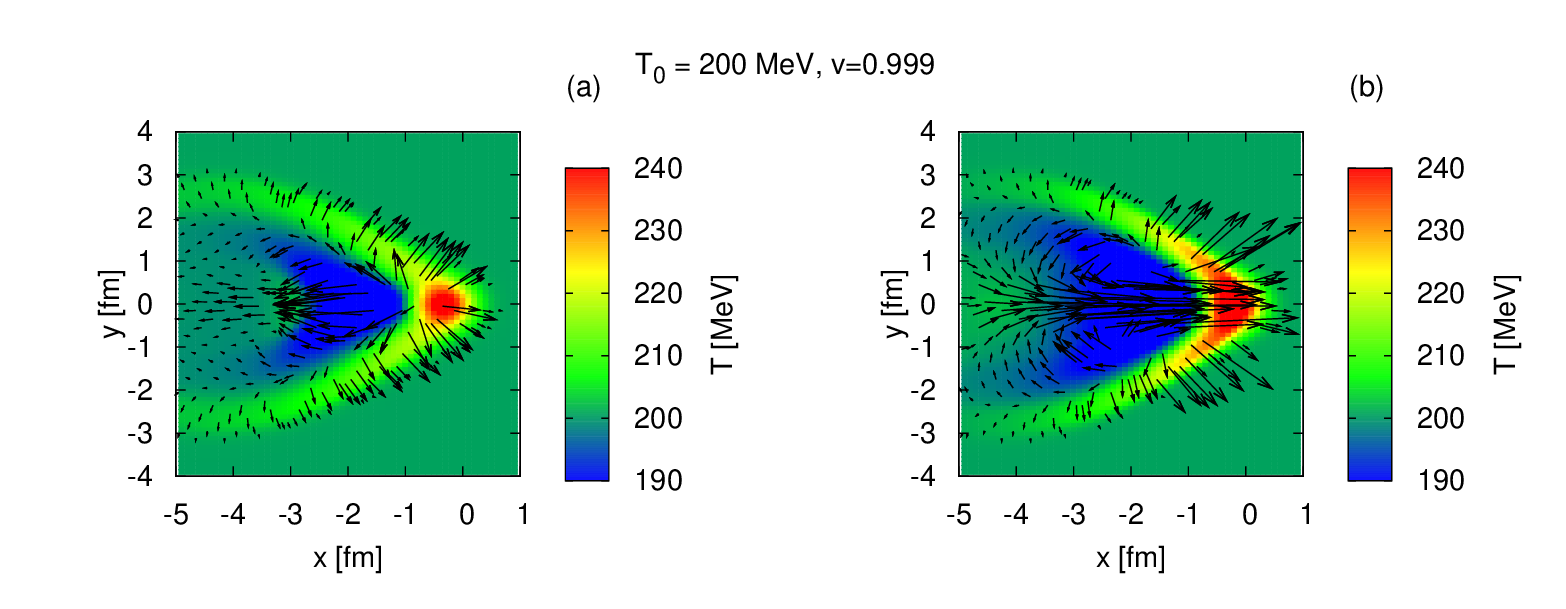}
\caption[Temperature pattern and flow velocity profile for a pure energy deposition as well
as an energy and momentum deposition scenario in a static medium.]
{Temperature pattern and flow velocity profile (arrows) after a hydrodynamical 
evolution of $t=4.5/v_{\rm jet}$~fm, assuming (a) an energy loss rate of $dE/dt = 1.5$~GeV/fm 
for a vanishing momentum loss rate and (b) an energy and momentum loss rate of 
$dE/dt = dM/dt = 1.5$~GeV/fm for a punch-through jet moving with a constant velocity of 
$v_{\rm{jet}}=0.999$ along the $x$-axis through a static background plasma with temperature 
$T_0=200$~MeV. The jet is sitting at the origin of the coordinates at the time of freeze-out 
\cite{Betz:2008ka}.}
\label{FigPunchThrough1}
\end{figure}
\begin{eqnarray}
\label{sourcenoncovariant}
\hspace*{-0.6cm}
J^\nu(t,\vec{x}) &=& \frac{1}{(\sqrt{2\pi}\,\sigma)^3}
\exp\left\{ -\frac{[\vec{x}-\vec{x}_{\rm jet}(t)]^2}{
2\sigma^2}\right\} \left(\frac{dE}{dt},\frac{dM}{dt},0,0\right)\,,
\end{eqnarray}
where $\vec{x}_{\rm jet}$ describes the location of the jet, $\vec{x}$ is the position on the 
computational grid, and $\sigma=0.3$~fm. The system plasma+jet evolves according to Eq.\ 
(\ref{sourceterm}) until the freeze-out time $t_f=4.5/v_{\rm jet}$ fm is reached.\\
The temperature and flow velocity profiles created by a punch-through jet with a constant energy 
loss rate of $dE/dt = 1.5$~GeV/fm and vanishing momentum deposition are shown in Fig.\ 
\ref{FigPunchThrough1} (a). In Fig.\ \ref{FigPunchThrough1} (b) the jet has lost the same amount 
of energy and momentum. One can clearly see that the space-time region close to 
the jet, where the temperature disturbance is the largest, is bigger than in the pure energy 
deposition scenario. \\
The creation of a diffusion wake behind the jet in case of equal energy and momentum deposition 
is clearly visible, which is indicated by the strong flow observed in the forward direction 
(at $\phi=\pi$).
\begin{figure}[t]
\hspace*{0.5cm}
\includegraphics[scale=0.65]{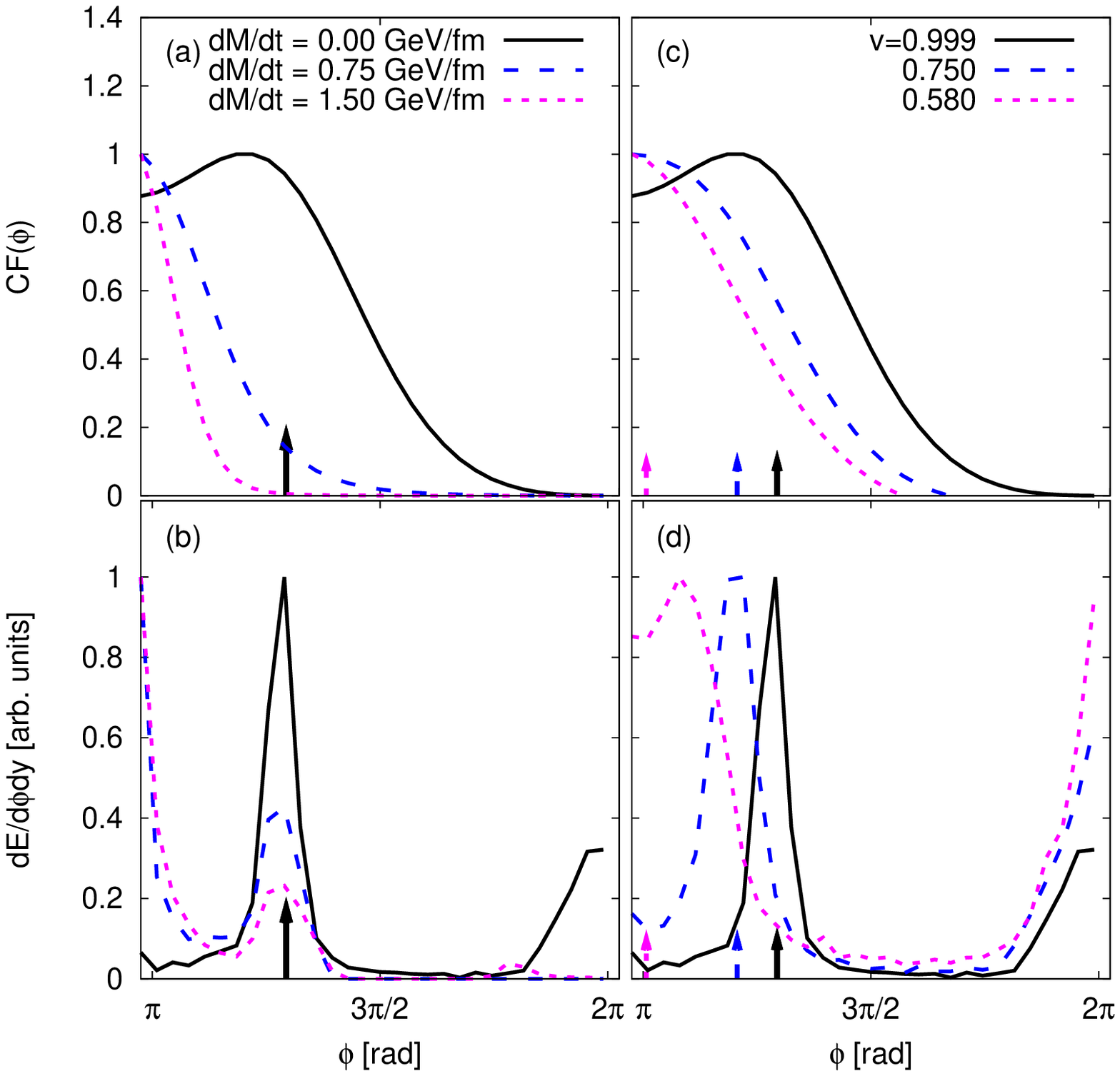}
\caption[Normalized angular distribution created by a punch-through jet 
at mid-rapidity with a fixed energy loss and different momentum loss rates as well as
for a jet moving at different velocities associated with a pure energy deposition scenario.]
{The left panels show the normalized angular distribution created by a punch-through jet 
at mid-rapidity with a fixed energy loss of $dE/dt = 1.5$~GeV/fm and different momentum loss rates.
The jet moves at a constant velocity $v_{\rm jet}=0.999$ through the medium. The right panels 
show the angular distributions associated with jets with $dE/dt = 1.5$~GeV/fm and vanishing 
momentum loss ($dM/dt=0$). Here, the jets move with different velocities through the medium: 
$v_{\rm jet}=0.999$ (black), $v_{\rm jet}=0.75$ (blue), and $v_{\rm jet}=0.58$ (magenta). In the 
upper panels, an isochronous Cooper--Frye freeze-out at $p_T=5$~GeV is used while in the lower 
panels we employed the bulk-flow freeze-out procedure \cite{Betz:2008wy}. The arrows indicate the 
angle of the Mach cone as computed via Mach's law \cite{Betz:2008ka}.}
\label{FigPunchThrough2}
\end{figure}
\\It is important to mention that, see Fig.\ \ref{FigPunchThrough2} (a), for the punch-through 
jet deposition scenario with equal energy and momentum loss one always obtains a peak in the 
associated jet direction after performing the freeze-out using the two prescriptions described 
above.
\begin{figure}[t]
\begin{minipage}[b]{4.2cm}
\hspace*{-0.8cm}
\includegraphics[scale=0.67]{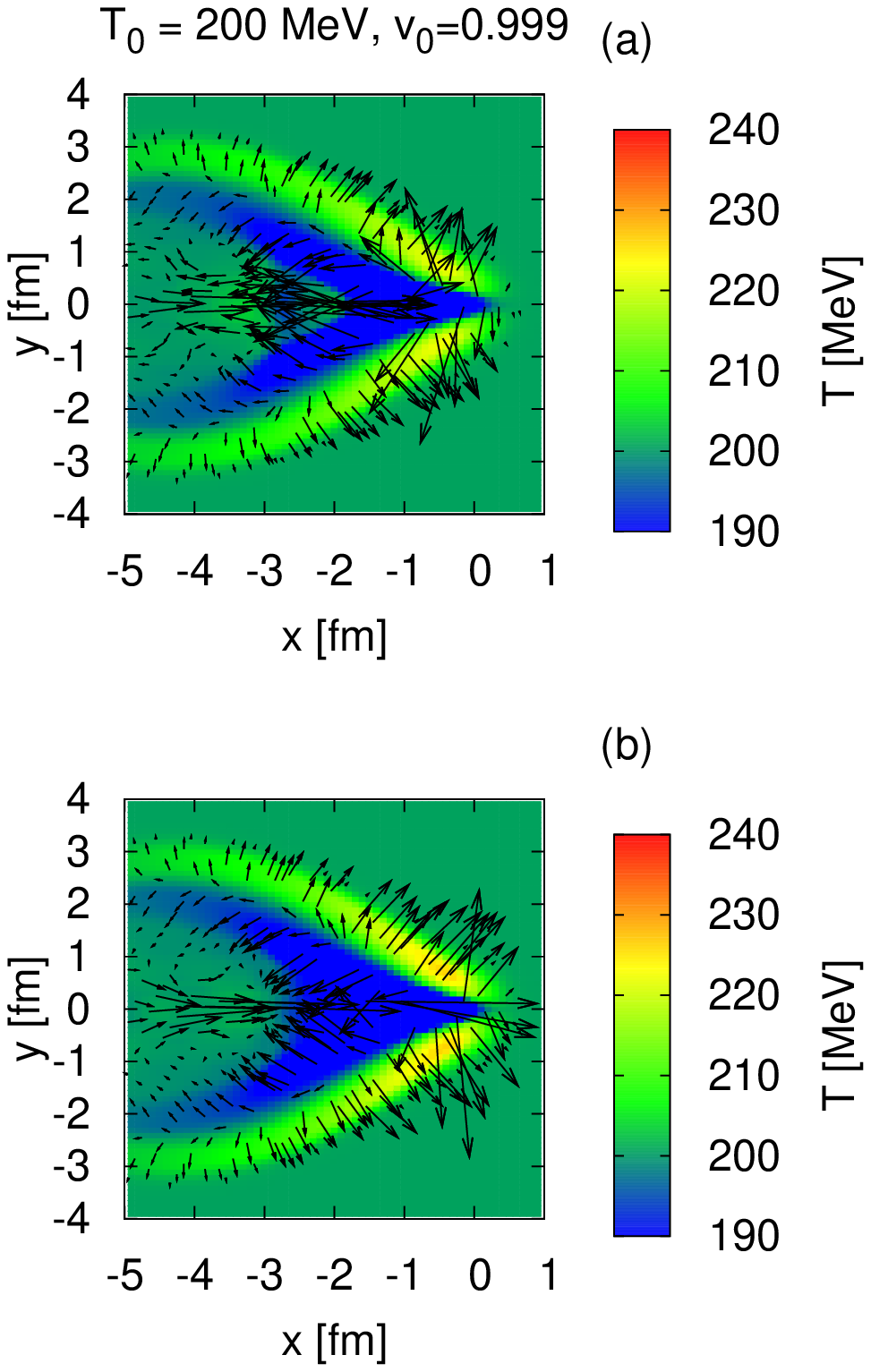}
\\[-1ex]~
\end{minipage}
\begin{minipage}[b]{4.2cm}
\hspace*{2.6cm}
\includegraphics[scale=0.65]{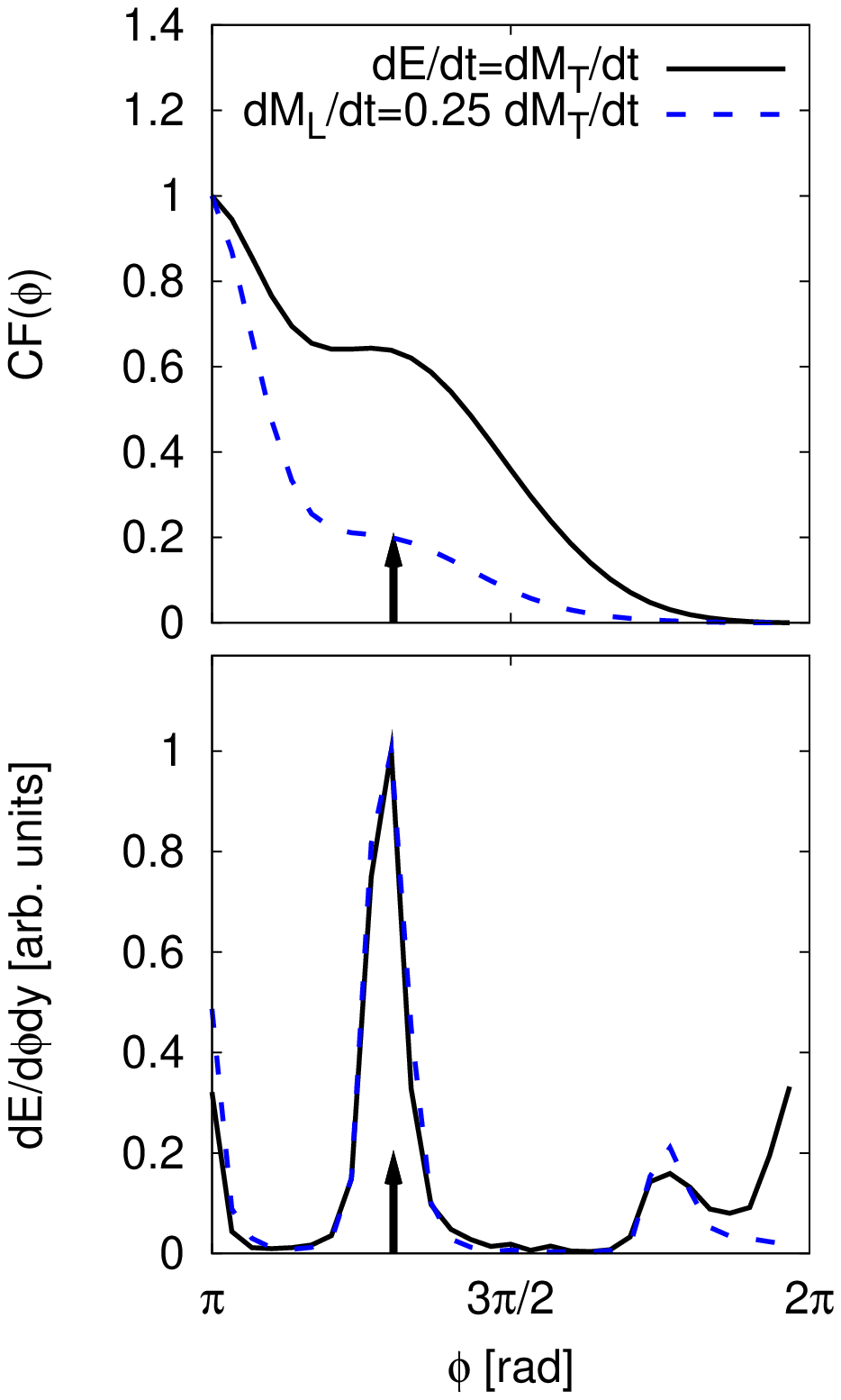}
\end{minipage}
\caption[Temperature pattern and flow velocity profile assuming a constant energy loss 
rate for full transverse momentum deposition and longitudinal as well as transverse momentum 
deposition. Additionally, the corresponding normalized angular distributions are shown.]
{Left panel: Temperature pattern and flow velocity profile (arrows) after a hydrodynamical 
evolution of $t=4.5/v_{\rm jet}$~fm, assuming an energy loss rate of $dE/dt = dM/dt = 1.5$~GeV/fm
for (a) full transverse momentum deposition and (b) longitudinal as well as transverse momentum 
deposition with a ratio of $dM_L/dt = 0.25 \;dM_T/dt$. Right panel: The normalized angular 
distribution created by a punch-through jet at mid-rapidity for the two above mentioned transverse 
momentum deposition scenarios. In the upper panel, an isochronous Cooper--Frye freeze-out at 
$p_{\bot}=5$~GeV is used while in the lower panel the bulk-flow freeze-out procedure 
\cite{Betz:2008wy} is employed. The arrows indicate the ideal Mach-cone angle \cite{Betz:2008ka}.}
\label{figtrans}
\end{figure}\\
However, the energy-flow distribution in Fig.\ \ref{FigPunchThrough2} (b) displays an additional 
small peak at the Mach cone angle indicated by the arrow. This Mach signal cannot be seen in the
Cooper--Frye freeze-out because of thermal smearing \cite{CasalderreySolana:2004qm,Betz:2008wy,
Noronha:2008un,Betz:2008js} and the strong influence of the diffusion wake, which leads to the 
strong peak around $\phi\sim\pi$ in the bulk energy-flow distribution.
However, given that the exact form of the source term in the sQGP is unknown, one may want to 
explore other energy-momentum deposition scenarios where the jet deposits more energy than 
momentum along its path. While this may seem unlikely, such a situation cannot be ruled out. 
Thus, for the sake of completeness, we additionally consider in Fig.\ \ref{FigPunchThrough2} (a) 
the case where the jet source term is described by a fixed energy loss of $dE/dt = 1.5$~GeV/fm 
and different momentum loss rates. \\
In the bulk-flow distribution in Fig.\ \ref{FigPunchThrough2} (b), one can see that the peak at 
the Mach-cone angle is more pronounced for smaller momentum loss while the contribution of the 
diffusion wake (indicated by the peak in forward direction) is reduced. The associated particle 
distribution from the CF freeze-out in Fig.\ \ref{FigPunchThrough2} (a) reveals a peak at 
$\phi \neq \pi$ for pure energy deposition (solid black line), however, the opening angle is 
shifted to a value smaller than the Mach cone angle due to thermal smearing \cite{Betz:2008js}.\\
In Figs.\ \ref{FigPunchThrough2} (c,d) we consider $dM/dt=0$ jets that move through the medium 
with different velocities $v_{\rm jet}=0.999, 0.75$, and $0.58$. In Fig.\ 
\ref{FigPunchThrough2} (d) the peak position changes in the bulk-flow distribution according 
to the expected Mach-cone angles (indicated by the arrows). However, due to the strong bow shock 
created by a jet moving at a slightly supersonic velocity of $v_{\rm jet}=0.58$, there is a strong 
contribution in the forward direction and the peak position is shifted from the 
expected value. In the CF freeze-out shown in Fig.\ \ref{FigPunchThrough2} (c), the peak from the 
Mach cone can again be seen for the jet moving nearly at the speed of light ($v_{\rm jet}=0.999$), 
but for slower jets thermal smearing again leads to a broad distribution peaked in the direction 
of the associated jet. It can indeed be shown that those broad distributions can
be obtained by modelling the away-side distribution with two Gaussians, the peak locations of which 
follow Mach's law if the width of the distributions, and thus the thermal smearing, is suffiently 
large.\\
It is apparently surprising that the above mentioned results are independent of whether the 
momentum deposited by the particle is in the longitudinal (along the motion of the jet) or 
transversal (perpendicular) direction. Repeating the calculation shown in Fig.\ 
\ref{FigPunchThrough1} including transverse momentum deposition
\begin{equation}
J^\nu(t,\vec{x}) \propto \left[ \begin{array}{c} dE/dt \\
dM_L/dt\\
 \left( dM_T/dt \right) \cos\varphi \\
 \left( dM_T/dt \right) \sin\varphi
 \end{array}  \right]\; ,
\label{sourcetrans}
\end{equation}
where $\varphi$ is the latitude angle in the $(y-z)$-plane with respect to the jet motion and 
the magnitude of $J^\nu(t,\vec{x})$ is the same as Eq.\ (\ref{source}), shows that transverse 
momentum deposition will not alter the results presented in this section (see Fig.\ 
\ref{figtrans}). A longitudinal diffusion wake still forms during the fluid evolution stage, 
and its contribution will still dominate the resulting angular inter-particle correlations though 
a peak occurs around the expected Mach cone angle in the CF freeze-out. \\
The reason is that transverse momentum deposition will force the fluid around the jet to expand, 
and the empty space left will be filled by matter flowing in a way that behaves 
much like a diffusion wake. In terms of ideal hydrodynamics, this universality of the diffusion 
wake can be understood in the context of vorticity conservation since momentum deposition, whether 
transverse or longitudinal, will add vorticity to the system. This vorticity will always end up 
behaving as a diffusion wake \cite{Betz:2007kg}. In the next section, we demonstrate that these 
results are largely independent of whether the jet is fully quenched or survives as a hard trigger.

\section{Stopped Jets in a Static Medium}
\label{Section4}

In the previous section we considered a uniformly moving jet that deposited energy and/or momentum 
in the medium at a constant rate. However, due to its interaction with the plasma, the jet will 
decelerate and its energy and/or momentum loss will change. Thus, the deceleration roughly 
represents the response of the medium. In general, a decelerating jet should have a peak in the 
energy loss rate because the interaction cross section increases as the parton's energy decreases. 
In other words, when the particle's velocity goes to zero there appears a peak in $dE/dx$ known as 
the Bragg peak \cite{Bragg}.\\
The question to be considered in this section is whether this energy-deposition scenario might be 
able to somehow stop the diffusion wake and, thus, change the angular distributions shown in 
Fig.\ \ref{FigPunchThrough2}. The source term in this case is still given by Eq.\ 
(\ref{sourcenoncovariant}) and, according to the Bethe--Bloch formalism 
\cite{Bragg,Chen,Sihver,Kraft}, one assumes that
\begin{figure}[t]
\hspace*{2.0cm}
\includegraphics[scale=0.70]{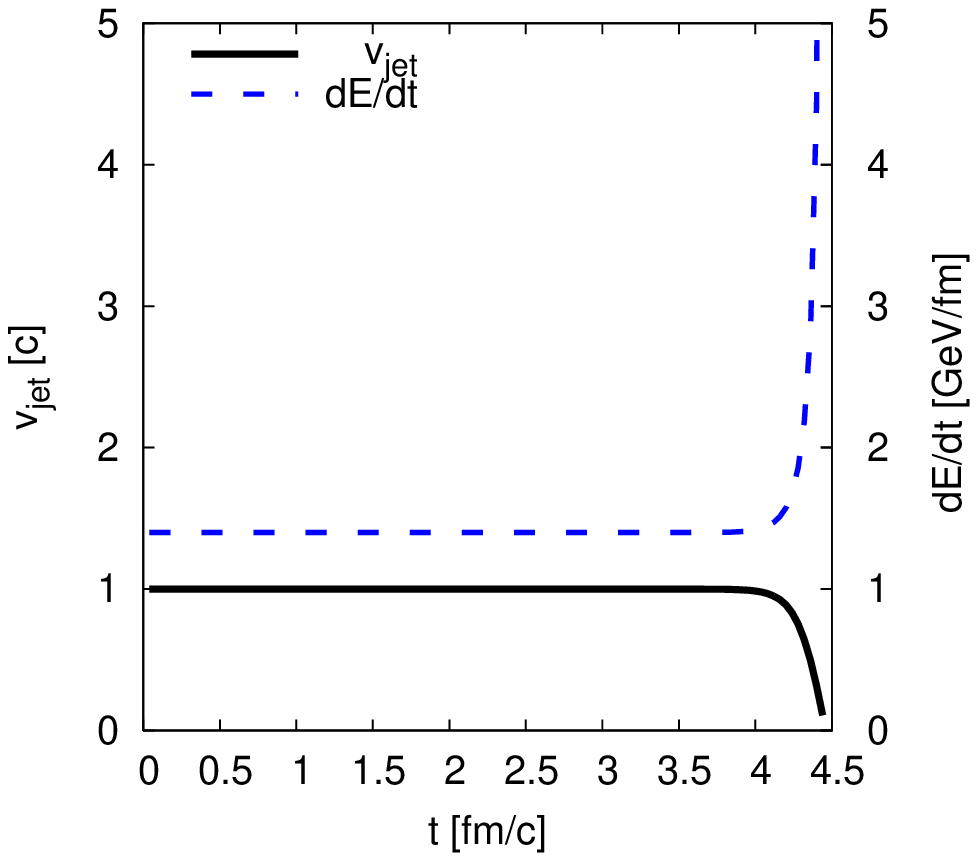}
\caption[The jet velocity $v_{\rm jet}(t)$ and energy-deposition rate $dE(t)/dt$ as a function of
time for a jet decelerating according to the Bethe--Bloch formula.]
{The jet velocity $v_{\rm jet}(t)$ (solid black line) and energy deposition rate $dE(t)/dt$ 
(dashed blue line) according to Eq.\ (\ref{jetbragg}). The initial jet velocity and energy-loss 
rate are $v_{\rm jet}=0.999$ and $a\simeq-1.3607$~GeV/fm, respectively \cite{Betz:2008ka}.}
\label{FigPunchThrough3}
\end{figure}
\begin{eqnarray}
\frac{dE(t)}{dt}=a\frac{1}{v_{\rm jet}(t)}\,,
\label{braggeqn}
\end{eqnarray}
which shows that when the jet decelerates the energy loss rate increases and has a peak as
$v_{\rm jet}\to 0$. Here, $dE/dt$ is the energy lost by the jet, which is the negative of 
the energy given to the plasma. Using this ansatz for the velocity dependence of the energy-loss 
rate and the identities $dE/dt =v_{\rm jet}\, dM/dt$ and $dM/dy_{\rm jet}= m\cosh y_{\rm jet}$ 
(as well as $v_{\rm jet}=\tanh y_{\rm jet}$), one can rewrite Eq.\ (\ref{braggeqn}) as
\begin{eqnarray}
\label{BetheBloch1}
t(y_{\rm jet}) & = & \frac{m}{a}\, \left[ \sinh y_{\rm jet} 
- \sinh y_0 \right. \nonumber \\
&  & \left. - \arccos \frac{1}{\cosh y_{\rm jet}}
+ \arccos \frac{1}{\cosh y_0} \right]\hspace*{0.2ex},
\end{eqnarray}
where $y_0$ is the jet's initial rapidity. A detailed derivation of this equation is given in
appendix \ref{FormulasBetheBloch}. The equation above can be used to determine the 
time-dependent velocity $v_{\rm jet}(t)$. The initial velocity is taken to be 
$v_0= {\rm Artanh} y_0 =0.999$. The mass of the moving parton is assumed to be of the order
of the constituent quark mass, $m=0.3$~GeV. Moreover, the initial energy loss rate 
$a\simeq -1.3607$~GeV/fm is determined by imposing that the jet stops after $\Delta x=4.5$~fm 
(as in the previous section for a jet with $v_{\rm jet}=0.999$). Thus, the jet location as well as 
the energy and momentum deposition can be calculated (see again appendix \ref{FormulasBetheBloch}) 
as a function of time via the following equations
\begin{eqnarray}
\label{BetheBloch2}
x_{\rm jet}(t) &=& x_{\rm jet}(0) +
\frac{m}{a}\, \left[ (2-v_{\rm jet}^2)\gamma_{\rm jet} -
(2-v_{0}^2)\gamma_{0}
 \right] \;, \nonumber \\
\frac{dE}{dt} & =& a\frac{1}{v_{\rm jet}}, \hspace*{1ex}
\frac{dM}{dt} = a\frac{1}{v_{\rm jet}^2}\,,
\label{jetbragg}
\end{eqnarray}
which can be used to determine the corresponding source term for the energy-momentum conservation 
equations. The change of the jet velocity $v_{\rm jet}(t)$ and energy deposition $dE(t)/dt$ are 
displayed in Fig.\ \ref{FigPunchThrough3}. The strong increase of energy deposition shortly before 
the jet is completely stopped corresponds to the well-known Bragg peak \cite{Bragg}.
\begin{figure*}[t]
\hspace*{-3.7cm}
\includegraphics[scale=0.92]{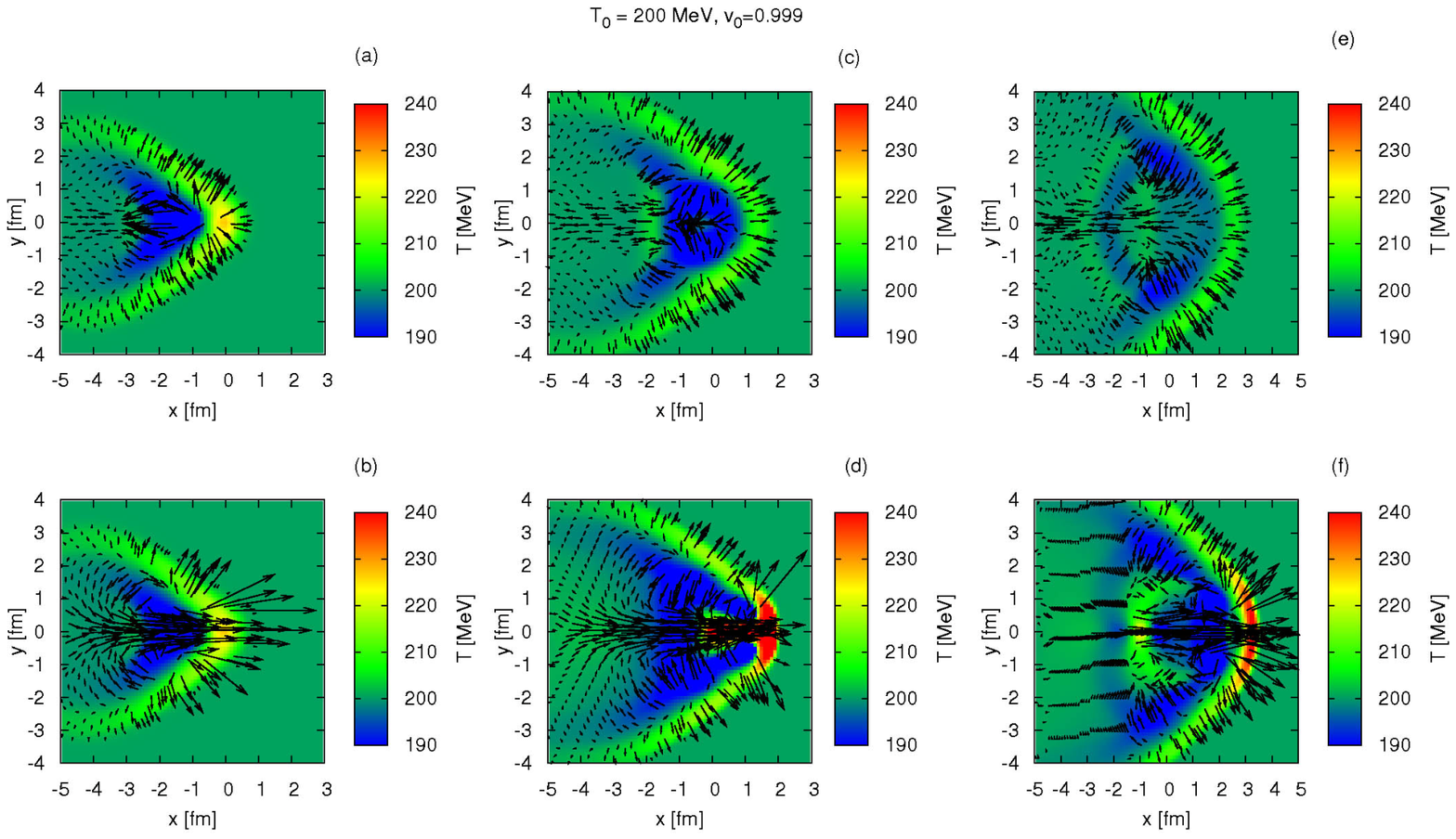}
\caption[Temperature pattern and flow-velocity profile after a hydrodynamical evolution of 
$t=4.5$~fm, $t=6.5$~fm, and $t=8.5$~fm for a jet that decelerates according to the Bethe--Bloch 
formula and stops after $\Delta x=4.5$~fm for a vanishing-momentum loss rate as well
as energy and momentum loss.]
{Temperature pattern and flow-velocity profile (arrows) after a hydrodynamical evolution of 
$t=4.5$~fm (left panel), $t=6.5$~fm (middle panel), and $t=8.5$~fm (right panel) for a jet that 
decelerates according to the Bethe--Bloch formula and stops after $\Delta x=4.5$~fm. The jet's 
initial velocity is $v_{\rm{jet}}=0.999$. In the upper panel a vanishing momentum-loss rate is 
assumed while in the lower panel the momentum loss is related to the energy loss by 
Eq.\ (\ref{jetbragg}) \cite{Betz:2008ka}.}
\label{FigPunchThrough4}
\end{figure*}
\\The main difference between the ansatz described here and the Bethe--Bloch equation is that the 
momentum deposition is longitudinal (parallel to the motion of the jet) rather than transverse 
(perpendicular to the motion of the jet). According to most pQCD calculations, this is true in 
the limit of an infinite energy jet \cite{Gyulassy:1990ye,Gyulassy:1999zd,Baier:1996kr,
Wiedemann:2000za,Gyulassy:1993hr,Wang:1994fx,Wang:2001ifa,Arnold:2001ms,Arnold:2002ja,Liu:2006ug,
Majumder:2007zh}, but it is expected to break down in the vicinity of the Bragg peak where the 
jet energy is comparable to the energy of a thermal particle. However, as we demonstrated in the 
previous section, the freeze-out phenomenology is rather insensitive to whether the momentum 
deposition is transverse or longitudinal. \\
Fig.\ \ref{FigPunchThrough4} displays the temperature and flow velocity profiles of a jet that 
stops after $\Delta x=4.5$~fm, with an energy loss according to Eq.\ (\ref{braggeqn}) and 
vanishing momentum deposition (upper panel) as well as an energy and momentum deposition following 
Eq.\ (\ref{jetbragg}) (lower panel). The left panel shows the hydrodynamical evolution after
$t=4.5$~fm, immediately after the jet is stopped, while in the middle and right panel the evolution
is continued until $t=6.5$~fm and $t=8.5$~fm, respectively.
\begin{figure*}[t]
\hspace*{-0.1cm}
\includegraphics[scale=0.55]{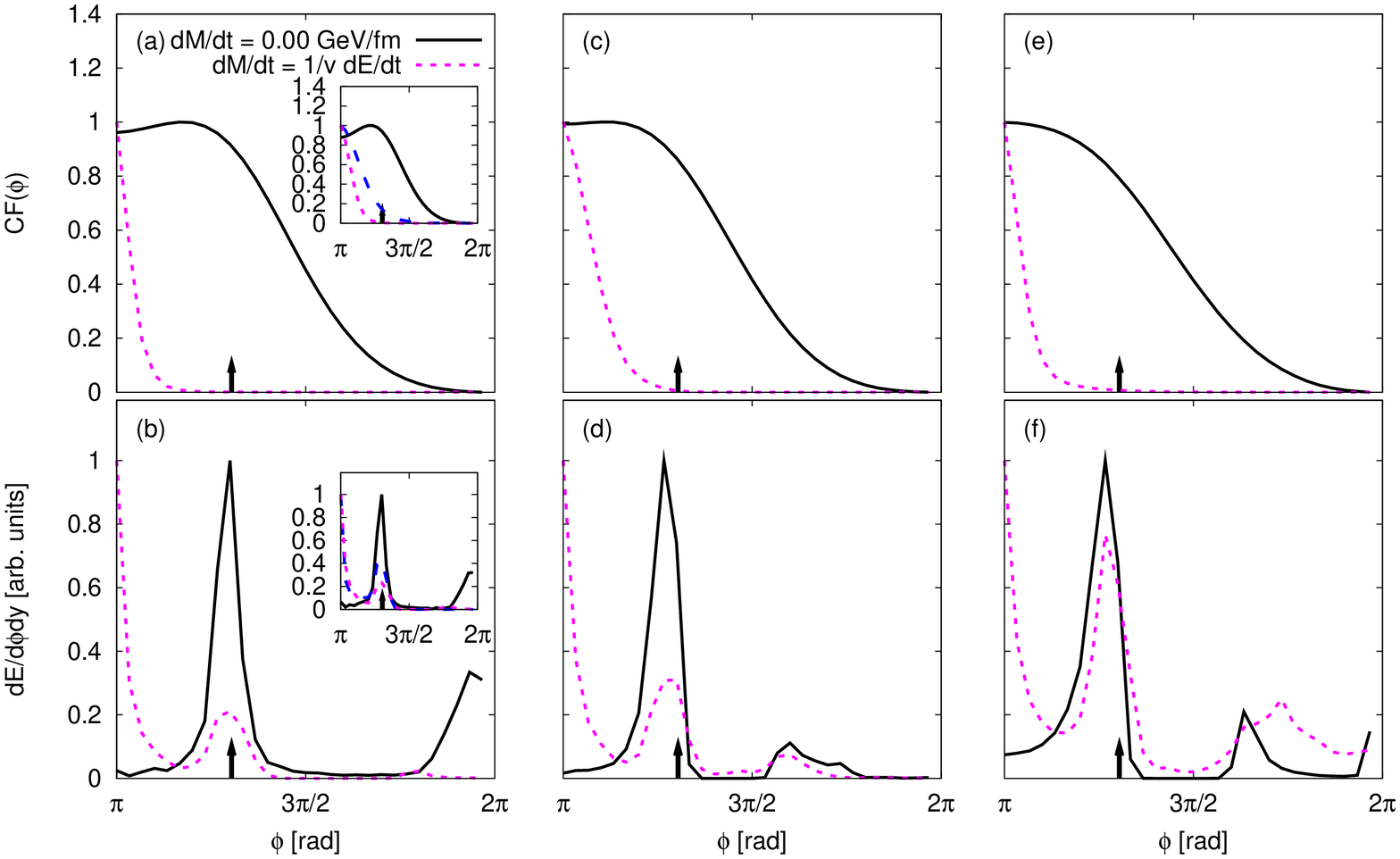}
\caption[The normalized angular distribution generated by a decelerating jet at mid-rapidity
that stops after $\Delta x=4.5$~fm for a hydrodynamical evolution of $t=4.5$~fm, $t=6.5$~fm, and 
$t=8.5$~fm.]
{The normalized angular distribution generated by a decelerating jet (cf.\ also Fig.\ 
\ref{FigPunchThrough4}) at mid-rapidity is shown (upper panel) according to an isochronous 
Cooper--Frye freeze-out at $p_T=5$~GeV for a jet that stops after $\Delta x=4.5$~fm and a 
hydrodynamical evolution of $t=4.5$~fm (left panel), $t=6.5$~fm (middle panel), and $t=8.5$~fm 
(right panel). The corresponding bulk-flow pattern \cite{Betz:2008wy} is shown in the lower panel. 
The solid black line in all plots depicts the pure energy-deposition case while the dashed magenta 
line corresponds to the energy and momentum deposition scenario given by Eq.\ (\ref{jetbragg}). 
The arrows indicate the angle of the Mach cone as computed via Mach's law. The inserts repeat 
Fig.\ \ref{FigPunchThrough2} (a) and (b) for comparison \cite{Betz:2008ka}.}
\label{FigPunchThrough5}
\end{figure*}
\\Comparing this result to Fig.\ \ref{FigPunchThrough1} leads to the conclusion that the diffusion 
wake is present independently of whether the jet is quenched or survives until freeze-out. In the 
former case, the diffusion wake is only weakly sensitive to the duration of the subsequent 
evolution of the system.\\
Within ideal hydrodynamics this can be understood via vorticity conservation. 
The vorticity-dominated diffusion wake will always be there in the ideal fluid, whether the source 
of vorticity has been quenched or not. The only way this vorticity can disappear is via viscous 
dissipation. While a $(3+1)$-dimensional viscous hydrodynamic calculation is needed to quantify 
the effects of this dissipation, linearized hydrodynamics predicts that both Mach cones and 
diffusion wakes are similarly affected 
\cite{CasalderreySolana:2004qm,Landau,CasalderreySolana:2006sq} as already mentioned in section 
\ref{FirstStudies}.
\begin{figure}[t]
\hspace*{1.0cm}
\includegraphics[scale=0.68]{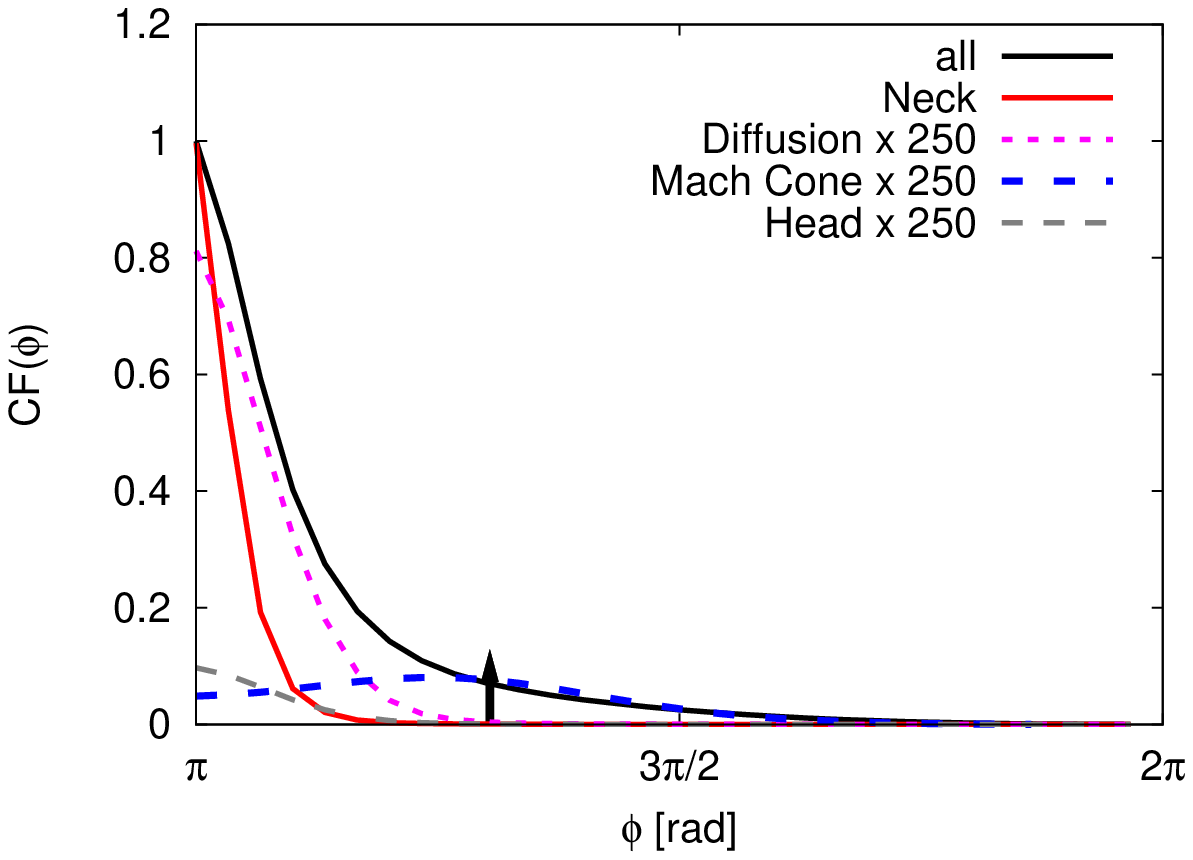}
\caption[Decomposition of a jet event into the regions of head, neck, diffusion wake, and Mach cone, 
applying a Cooper--Freye freeze-out at $p_T=5$~GeV for a jet depositing energy and momentum, and 
stopping according to the Bethe--Bloch formalism.]
{Decomposition of a jet event into the regions of head, neck, diffusion wake, and Mach cone, 
applying a Cooper--Freye freeze-out at $p_T=5$~GeV for a jet depositing energy and momentum, and 
stopping according to the Bethe--Bloch formalism. The arrow indicates the expected Mach cone angle 
for a jet moving with $v_{\rm jet}=0.999$.}
\label{HeadNeckDiff}
\end{figure}
\\The angular distribution associated with the decelerating jet (which stops after 
$\Delta x=4.5$~fm), shown in Fig.\ \ref{FigPunchThrough5}, is determined according to the two 
freeze-out prescriptions described above. When the energy and momentum loss rates are determined 
by Eq.\ (\ref{jetbragg}) (magenta line), both freeze-out procedures display a feature discussed 
in the previous section for the case of punch-through jets: the formation of a strong diffusion 
wake which leads to a strong peak in the associated jet direction. The results after the 
isochronous CF freeze-out are shown in the upper panel of Fig.\ \ref{FigPunchThrough5}. 
Here (cf.\ Fig.\ \ref{FigPunchThrough4}), the medium decouples after $t=4.5$~fm (left panel), $t=6.5$~fm 
(middle panel), and $t=8.5$~fm (right panel). Only the pure energy-deposition scenario produces a 
peak at an angle close to the Mach angle [see Fig.\ \ref{FigPunchThrough5} (a)] which is smeared 
out thermally for larger decoupling times [cf.\ Fig.\ \ref{FigPunchThrough5} (b) and (c)]. On 
the other hand, the bulk-energy-flow freeze-out displayed (lower panel) shows in all cases a peak 
at the Mach-cone angle. In this case the peak becomes more pronounced when $dM/dt=0$. While the 
Mach cone signal increases with the decay time, the signal is still smaller than the forward 
yield of the diffusion wake. \\
Clearly it would be interesting to study other models that describe decelerating jets in 
strongly coupled plasmas. Nevertheless, the simple Bethe--Bloch model used here displays the main 
qualitative features relevant for the hydrodynamic wake associated with decelerating jets. 
\begin{table}[t]
\begin{tabular}{|c|c|c|c|}\hline
$E_{\rm Mach}$  &  53.9\%  &  $M^x_{\rm Mach}$  &  6.5\%\\
$E_{\rm diffusion}  $  &-12.3\%  &  $M^x_{\rm diffusion}$  &  18.7\%\\
$E_{\rm neck}$  &  57.4\%  &  $M^x_{\rm neck}$  &  73.7\%\\
$E_{\rm head}$  &  1.0\%  & $M^x_{\rm head}$  &  1.0\%\\\hline
\end{tabular}
\caption[Relative energy and momentum stored in the different regions of a jet event.]
{Relative energy and momentum stored in the different regions of a jet event. For details see text.}
\label{TableRegions}
\end{table}
\\So far, we have not yet distinguished different regions of the flow in forward direction behind the jet
which we addressed as diffusion wake. However, as already discussed in chapter \ref{AdS/CFT}, 
the jet event can be decomposed into several regions. Since the physical regimes
change close to the jet (expressed by the Knudsen number in chapter \ref{AdS/CFT}), is it
possible and probably inevitable to consider a neck zone (close to the location of the jet)
and a head region (following the neck area), besides a diffusion zone that is characterized
as a region where linearized first-order Navier--Stokes hydrodynamics can be applied
(see Fig.\ \ref{SketchRealJetDeposition}).\\
In the following, we will assume that the transverse expansion of the head, neck, and diffusion 
zone is $\vec{x}_T<1$~fm and define that for the present calculations (see e.g.\ Fig.\ 
\ref{FigPunchThrough1}) $0.5<x_{\rm head}<1$~fm, $-1.5 < x_{\rm neck}< 0.5$~fm and 
$x_{\rm diffusion}<-1.5$~fm. This decomposition reveals (see Fig.\ \ref{HeadNeckDiff}) that for 
a jet depositing energy and momentum, and decelerating according to the Bethe--Bloch formalism, 
the away-side distribution can clearly be attributed to the strong flow behind the jet. Moreover, 
it shows that the main contribution is actually due to the neck region, while the large $p_T$-cut 
results in a strong suppression of the far zone (Mach-cone regions). This means that, as can be seen
from the flow-velocity profiles (cf.\ e.g.\ Fig.\ \ref{FigPunchThrough1}), a Mach-cone contribution
is indeed formed, but it is strongly overwhelmed by the flow along the jet trajectory. \\
However, even the exclusion of the head and neck zones would not change this result since
(see Fig.\ \ref{HeadNeckDiff}) the contribution of the diffusion zone largely overwhelmes 
that one of the Mach region.\\
It is also possible to calculate the amount of energy and momentum stored in these regions from
the hydrodynamic evolution. The neck region (cf.\ Table \ref{TableRegions}) by far
contains the most energy and momentum along the jet trajectory (which is the $x$-axis). The
contribution from the Mach region is large in energy, but small for momentum and the
diffusion zone even reveals a negative energy content which is due to the fact that behind 
the jet there is a region with lower temperature/energy than the background (see again e.g.\ 
Fig.\ \ref{FigPunchThrough1}) 
\\ 
Our results confirm previous studies \cite{CasalderreySolana:2004qm,Chaudhuri:2005vc,Renk:2006mv} 
in the sense that a single peak in the away-side of the associated dihadron correlations occurs 
unless the total amount of momentum loss experienced by the jet is much smaller than the 
corresponding energy loss. Thus, it is clear that in Ref.\ \cite{Chaudhuri:2005vc} the existence 
of conical flow effects could not be proven since it was assumed that $dM/dx=dE/dx$, while as 
seen in Refs.\ \cite{Renk:2005si} the double-peaked structure found by experiment can
be reproduced assuming that most of the energy lost excites sound modes which is equivalent to
$dM/dx \ll dE/dx$.\\[0.2cm]~
We would like to stress that the formation of a diffusion wake (i.e., the strong flow behind
the jet) is a generic phenomenon \cite{Landau} and, thus, its phenomenological consequences must be 
investigated and cannot simply neglected. \\Considering already the above mentioned 
vorticity conservation, the diffusion wake cannot be an artifact of the calculation. \\
In the present study, the path lengths of both types of jets were taken to be the same. A different
scenario in which the light jets are almost immediately stopped in the medium while the heavy quark
jets are still able to punch through may lead to different angular correlations. However, as Fig.\
\ref{HeadNeckDiff} reveals, the main contribution to the strong flow of the diffusion wake comes
from the regions close to the jet and therefore, the diffusion wake is produed on the same time
scale as the Mach cone which needs $t\sim 2$~fm to develope. \\
Nevertheless, one can expect that the strong forward-moving fluid represented by the
diffusion wake can be considerably distorted in an expanding medium by the presence of a large
radial flow. The interplay between radial flow and away-side conical correlations in an expanding
three-dimensional ideal fluid will be presented in chapter \ref{ExpandingMedium}.

\clearpage{\pagestyle{empty}\cleardoublepage}
%
%
\chapter[Polarization Probes of Vorticity]
{Polarization Probes of Vorticity}
\label{PolarizationProbesofVorticity}

We demonstrated in the previous chapter that the creation of a diffusion wake is universal to all 
jet-deposition scenarios implying energy and momentum loss \cite{Betz:2008ka}. However, as can be 
seen from the flow-velocity profiles of Figs.\ \ref{FigPunchThrough1} and \ref{FigPunchThrough4}, 
the formation of such a strong flow behind the jet is connected to the formation of vortex-like 
structures which also persist due to vorticity conservation after the jet is fully thermalized 
in the medium, i.e., after energy and momentum deposition have ceased.\\
Certainly this raises the question if such a conserved structure may have experimentally 
observable consequences, probably connected to polarization effects\footnote{Polarization 
implies the presence of a nonzero expectation value for the particle spin.} 
(due to spin-orbit coupling) of particles formed in such a jet event.\\
More than 30 years ago \cite{Bunce:1976yb}, hyperon\footnote{Hyperons are baryons containing 
at least one $s$-quark.} polarization was discovered. It was suggested in Refs.\ \cite{Hoyer,Panagiotou:1986zq} 
that the disappearance of polarization could signal the onset of an isotropized system where 
locally no reference frame is preferred, a state close to the sQGP (strongly-coupled Quark-Gluon 
Plasma).\\
In the following sections, we will describe the sensitivity of polarization to initial conditions, 
hydrodynamic evolution (including jet events), and mean-free path.

\section[Hyperon Polarization]{Hyperon Polarization}
\label{HyperonPolarization}

\begin{figure}[t]
\centering
  \includegraphics[scale = 0.9]{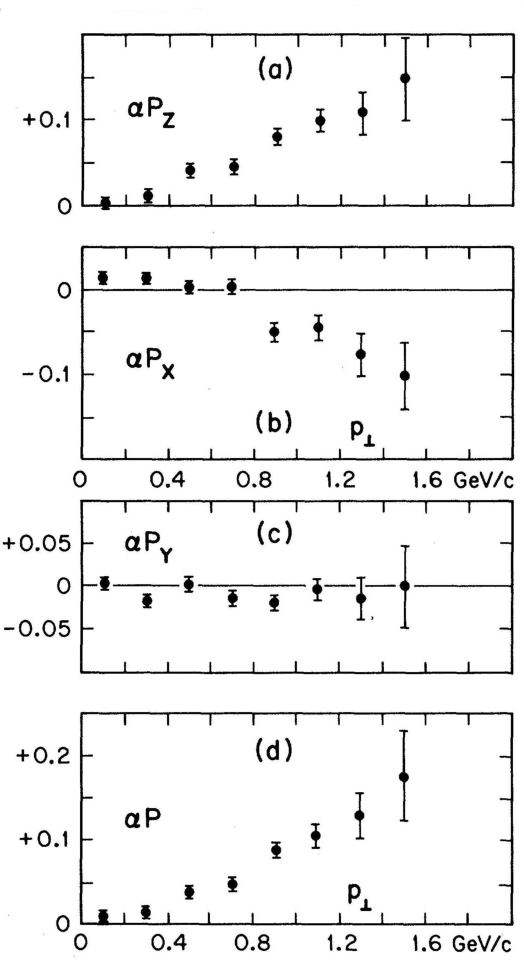}
  \caption[The magnitude of polarization in $x,y$ and $z$-direction ($P_x,P_y$, and $P_z$) from the 
  $\Lambda^0\rightarrow p+\pi^-$ decay as a function of the $\Lambda^0$ transverse momentum.]
  {The magnitude of polarization in $x,y$ and $z$-direction ($P_x,P_y$, and $P_z$) 
  from the $\Lambda^0\rightarrow p+\pi^-$ decay as a function of the $\Lambda^0$ transverse 
  momentum. Here, the beam direction is chosen to be aligned with the $y$-axis 
  \cite{Bunce:1976yb}.}
  \label{HyperonPolarization1}
\end{figure}

\begin{figure}[t]
\centering
  \includegraphics[scale = 0.5]{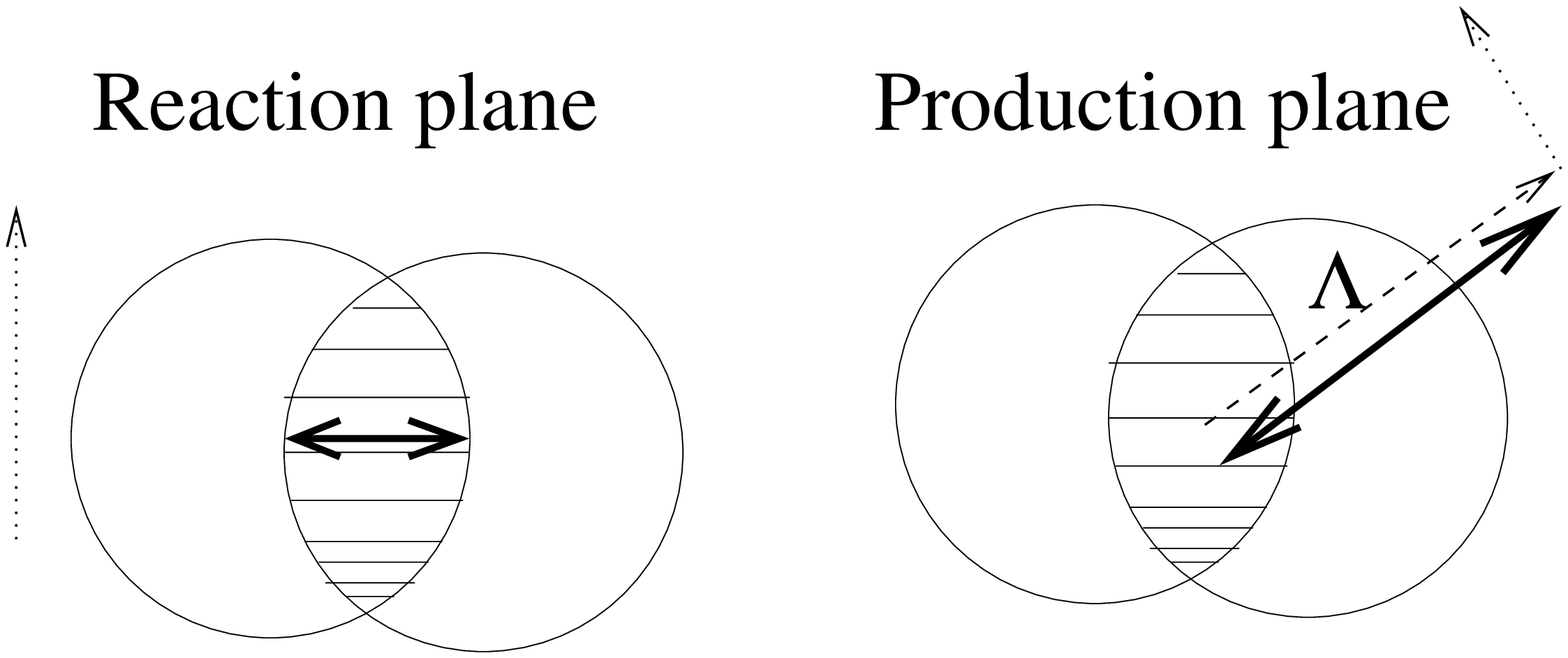}
  \caption[Definition of production and reaction plane.]
  {Definition of production and reaction plane. The beam line (traditionally the $z$-axis) 
  is perpendicular to the sheet. The dotted line, with arrow, indicates the direction of 
  polarization of the produced $\Lambda$ \cite{Betz:2007kg}.}
  \label{planes}
\end{figure}

The effect of hyperon polarization was first discovered in the channel 
$p+Be\rightarrow\Lambda^0+X$, where $X$ is the sum over the 
unobserved states and the polarization itself was measured through the $\Lambda^0$-decay 
$\Lambda^0\rightarrow p+\pi^-$ (see Fig.\ \ref{HyperonPolarization1}). Various models 
\cite{Bunce:1976yb,Hoyer,Andersson:1979wj,Bensinger:1983vc} explain the polarization of 
$\Lambda$-particles, which is a ($uds$)-quark state, based on the assumption that the spin of this 
particle is determined by the spin of the $s$-quark.
Generally, in the rest frame of the hyperon $Y$, the angular decay distribution w.r.t.\ the 
polarization plane is
\cite{Bunce:1976yb}
\begin{equation}
\label{poldef}
\frac{dN}{d \theta} = 1 + \alpha_Y P_Y \cos \theta\,,
\end{equation}
where $P_Y$ is the hyperon polarization, $\theta$ the angle between the proton momentum and the 
$\Lambda$-polar axis, and $\alpha_Y$ a hyperon-specific constant that is measured in elementary 
processes \cite{Bunce:1976yb}. It has a value of $\alpha_Y=0.647\pm0.013$ for $\Lambda^0$'s. 
\\Hyperons are polarized perpendicular to the hyperon {\it production plane} (see right panel of 
Fig.\ \ref{planes}) \cite{Liang:2004ph}
\begin{equation}
\label{equationpolarization}
\vec{n}=\frac{\vec{p}\times\vec{p}_H}{\vert\vec{p}\times\vec{p}_H \vert}\,,
\end{equation}
with $\vec{p}$ and $\vec{p}_H$ being the beam and hyperon momenta. This polarization can also be 
seen from Fig.\ \ref{HyperonPolarization1} which shows the hyperon polarizations $P_x,P_y$ and $P_z$
in $x,y$ and $z$-direction. Here, the beam direction was chosen along the $y$-axis. Thus, the
polarization in $y$-direction vanishes. \\
The initially generated amount of the {\it reaction-plane} polarization (see left panel of Fig.\ 
\ref{planes}) \cite{Liang:2004ph} is given by 
\begin{equation}
\langle P_q^R \rangle \sim \int d^2 \vec{x}_T \,\rho(\vec{x}_T)\, 
\vec{p}\cdot(\vec{x}_T \times \vec{n}) \sim - 
\langle p_z x_T \rangle \,,  
\label{pqint}
\end{equation}
where  $\rho(\vec{x}_T)  =\int d^3 \vec{p} f(\vec{x}_T,\vec{p})$ is the participant transverse 
density, $\vec{x}_T$ the directions perpendicular to the beam axis, 
and $\vec{n}$ a unit vector perpendicular to both $\vec{x}_T$ and $\vec{p}$. This expression
strongly depends on the initial density-momentum correlation within the system. Thus, the 
reaction-plane polarization can be a useful signature for probing the initial conditions of the 
system created in a heavy-ion collision.
\\If a specific net $\langle P_Y\rangle \ne 0$ exists for any axis definable event-by-event, it is 
in principle possible to measure this polarization, using Eq.\ (\ref{poldef}), in the observed 
spectra of $\Lambda$ and other hyperon (like $\Xi$ or $\Omega$) decay products. This opens a 
new avenue to investigate heavy-ion collisions which has been proposed both to determine 
confined/deconfined regimes \cite{Hoyer,Panagiotou:1986zq,rafpol}, and to mark global 
properties of the event \cite{Liang:2004ph,liangvector,Becattini:2007zn,Liangqm2006}.\\
While a strong transverse polarization of hyperons in the {\it production plane} (see right panel 
of Fig.\ \ref{planes}) is observed in (unpolarized) p+p and p+A collisions 
\cite{Bravar:1999rq,Liang:1997rt}, a disappearance of this polarization is assumed to be a 
signal of deconfinement \cite{Hoyer,Panagiotou:1986zq}. So far, no such measurement exists at 
RHIC energies, but a measurement at AGS shows that the transverse polarization is comparable to 
those of p+A collisions, thus at AGS the signal does not disappear.\\
It has also been suggested \cite{Liang:2004ph} to use hyperon polarization in the
{\em reaction plane} (left panel, Fig. \ref{planes}) to test for the vorticity of
matter produced in heavy-ion collisions. The idea is that the initial momentum gradient in 
non-central collisions should result in a net angular momentum (shear) in this direction 
which will be transferred to the hyperon spin via spin-orbit coupling. This polarization direction 
will be called $P_{Y}^R$. A similar, though quantitatively different result, can be obtained 
from a microcanonical ensemble with a net angular momentum \cite{Becattini:2007zn}.\\
The STAR collaboration measured the reaction-plane polarization \cite{Abelev:2007zk}, 
reporting results consistent with zero. The above mentioned production-plane polarization 
measurement is also planned.\\
In the following, we will present a few general considerations regarding the insights that can be 
gained from polarization measurements. We examine how the polarization, in both production and 
reaction plane, is sensitive to initial conditions, hydrodynamic evolution (in particular the 
vortex structure of a jet event), and mean-free path. We suggest that measuring the polarization in 
different directions (production, reaction, and jet axis) could provide a way to go beyond 
model dependence.\\
While we use the $\Lambda$ polarization as our signature of choice throughout this chapter, the 
points made here can easily be generalized to the detection of vector-meson polarization 
\cite{jacob} which is also used as a polarization probe in a way very similar to hyperons 
\cite{liangvector}.

\section[Initial Conditions and Reaction-Plane Polarization]
{Initial Conditions and Reaction-Plane Polarization}
\label{IniCondPol}

\begin{figure}[t]
\centering
  \includegraphics[scale = 0.27]{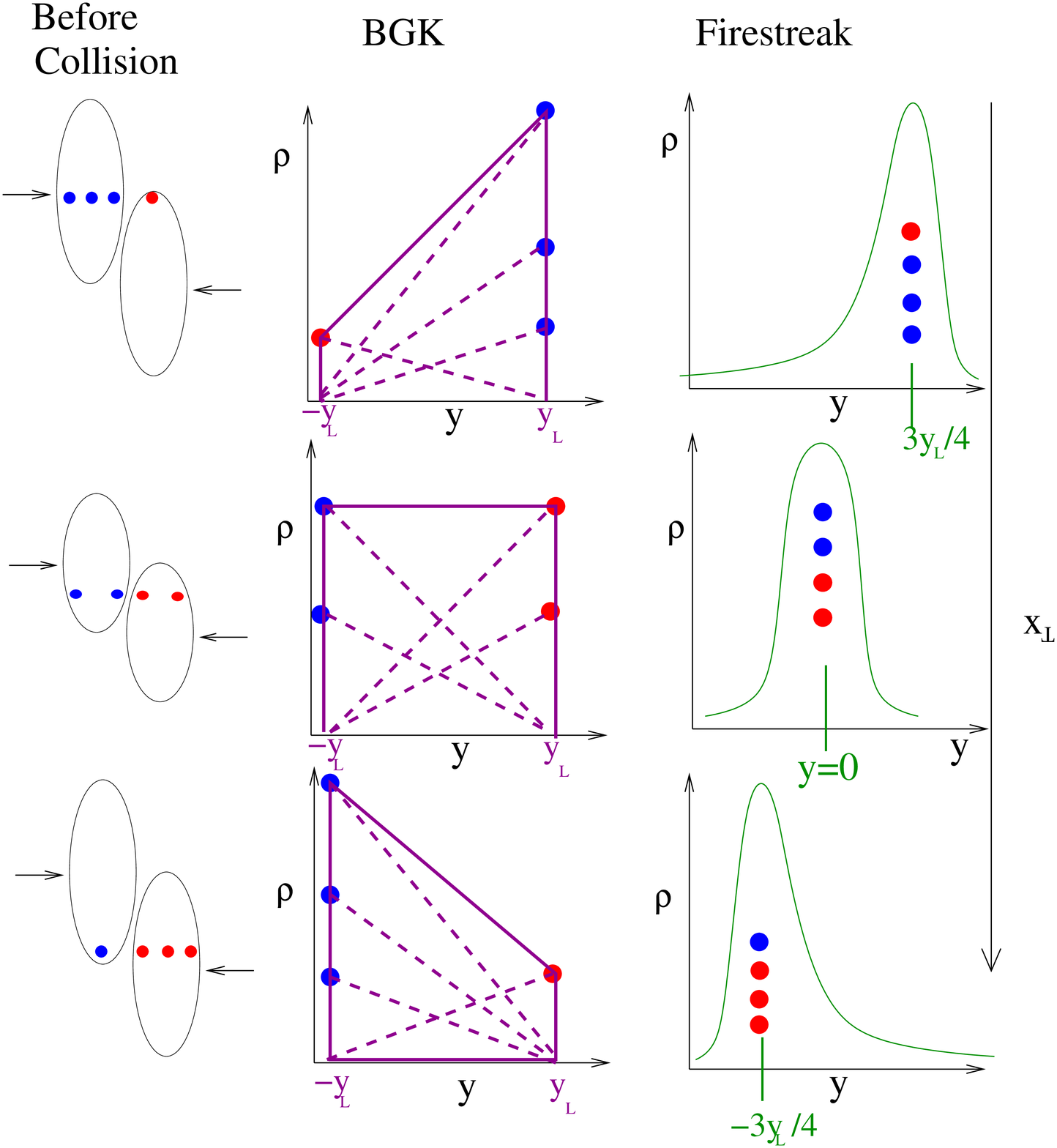}
  \caption[Initial densities in the Brodsky--Gunion--Kuhn model as well as the firestreak model.]
  {Initial densities in the BGK model (left panel) as well as the firestreak model used in Ref.\ 
  \cite{Liang:2004ph} (right panel). In the BGK case, dashed lines represent the rapidity extent 
  of individual nucleons, while solid lines correspond to 
  the cumulative density \cite{Betz:2007kg}. For detailed model definitions and further 
  explanations see text.}
  \label{initial_density}
\end{figure}

\begin{figure}[t]
\centering
  \includegraphics[scale = 0.25]{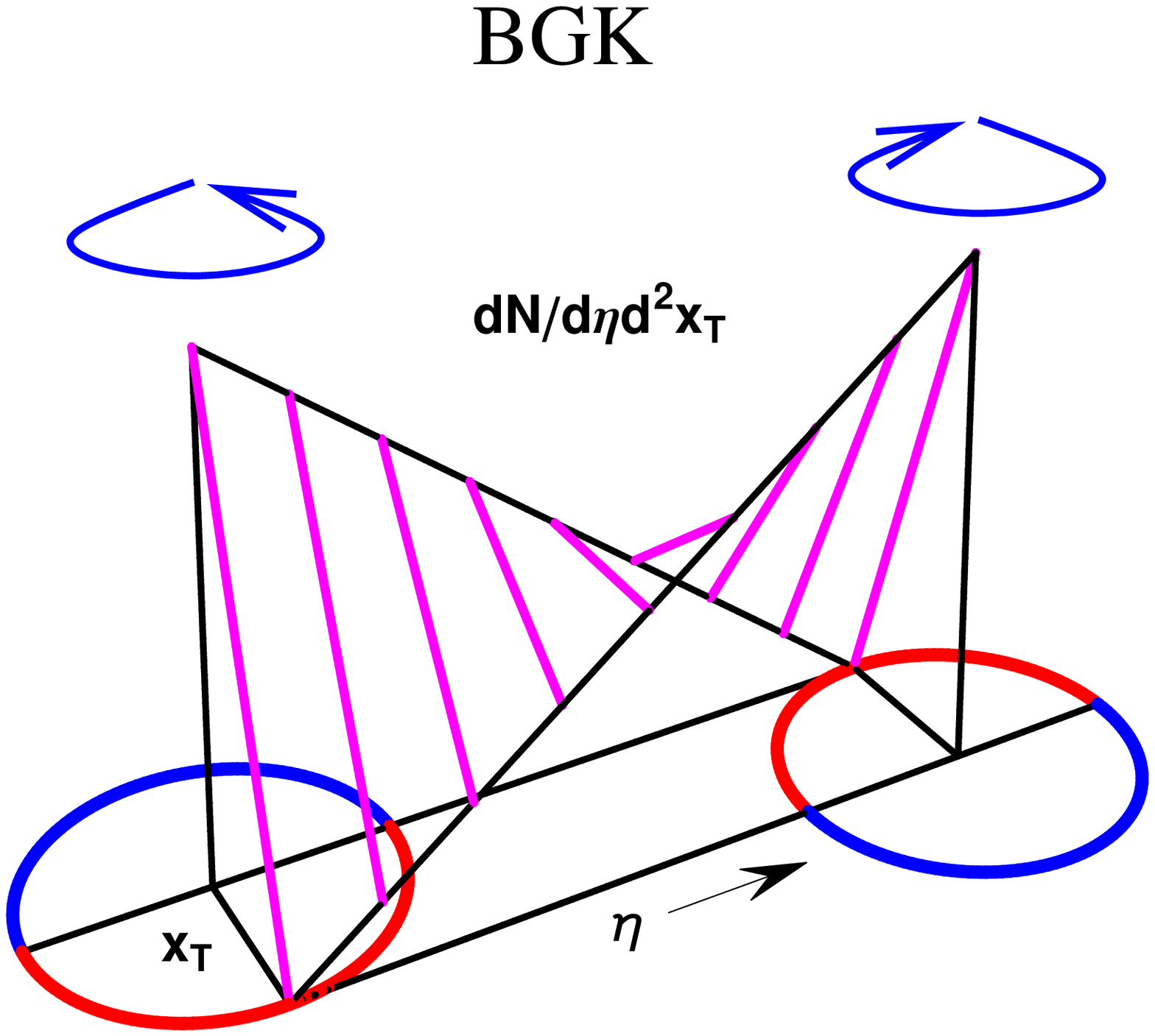}
  \includegraphics[scale = 0.45]{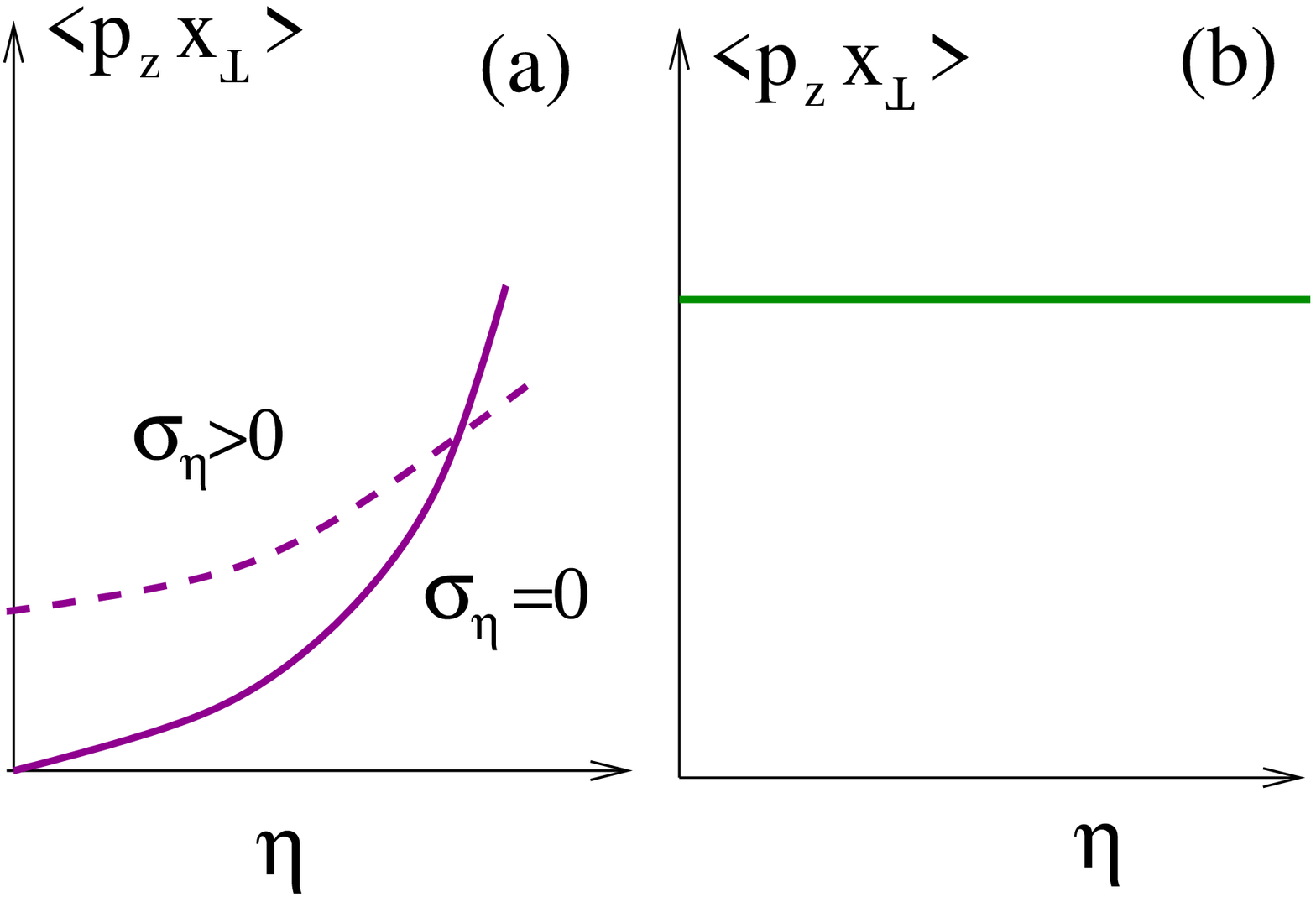}
  \caption[Initial shear in the Brodsky--Gunion--Kuhn model as well as the firestreak model.]
  {Initial shear in the BGK model (a), as well as the firestreak model used in Ref.\ 
  \cite{Liang:2004ph} (b) \cite{Betz:2007kg}.}
  \label{initial_shear}
\end{figure}

The QCD spin-orbit coupling is capable of transforming the total orbital angular momentum 
$\langle \vec{x} \times \vec{p}\,\rangle$ into spin.\\
For a large system, such as a heavy nucleus, we have to convolute the net polarizing interaction 
cross section per unit of transverse nuclear surface ($ d \Delta \sigma/d^2 \vec{x}_T$), 
calculated in Ref.\ \cite{Liang:2004ph}, with the (initial) parton phase-space distributions 
$f(\vec{x}_T,\vec{p})$ to obtain the net local polarized parton phase-space density 
$\rho_{P^R_q}$ produced in the first interactions
\begin{equation}
\rho_{P^R_q} (\vec{x}_T,\vec{p}) =  
\int d^2 \vec{x}\,'_T d^3 \vec{p}\,' f(\vec{x}_T-\vec{x}\,'_T,\vec{p}-\vec{p}\,') 
\frac{ d \Delta \sigma}{d^2 \vec{x}\,'_T} 
\left(\vec{p}\,' \right) \,.
\end{equation}
Here, $f(\vec{x}_T,\vec{p})$ is the local parton distribution of the medium.\\
Provided that the initial Debye mass and constituent quark mass is small, the quark polarization 
in the reaction plane 
\begin{equation}
\langle P_q^R\rangle = \int d^2 \vec{x}\, d^3 \vec{p}\, \rho_{P^R_q} (\vec{x},\vec{p}) 
\end{equation}
becomes, as demonstrated in Ref.\ \cite{Liang:2004ph} [see also Eq.\ (\ref{pqint})],
\begin{equation}
\langle P_q^R\rangle \sim \int d^2 \vec{x}_T\, \rho(\vec{x}_T)\, \vec{p}\cdot(\vec{x}_T \times \vec{n}) \sim 
- \langle p_z x_T\rangle \,. 
\end{equation}
In ultra-relativistic collisions all significant initial momentum goes into the beam 
($z$-) direction.\\
For non-central collisions with a nonzero impact parameter $\vec{b}$ however, $\langle \vec{p} 
\times \vec{x}_T\rangle \propto \vec{b} \ne 0$ results in a net polarization. \\ 
Thus, the initially generated amount of reaction-plane polarization strongly depends on the 
initial density-momentum correlation within the system. In other words, the reaction-plane 
polarization could be a useful signature for probing the initial conditions of the system 
created in heavy-ion collisions.\\
According to the Glauber model, the initial density distribution transverse to the beam axis 
is given by the sum of the participant and target densities $\rho_P$ and $\rho_T$,
\begin{equation}
\rho(\vec{x}_T) = \left[ \rho_{P} (\vec{x}_T) + \rho_T (\vec{x}_T) \right] \phi (y,\eta)
\end{equation}
where
\begin{equation}
\rho_{p,T} = T_{P,T} \left( x_T \mp \frac{b}{2} \right) 
\left\{ 1 - \exp \left[ -\sigma_N T_{T,p} \left( x_T \pm \frac{b}{2} \right) \right]  \right\}
\end{equation}
and $\sigma_N$, $T_{P,T}$ as well as $b$ refer, respectively, to the nucleon-nucleon cross section, 
the nuclear (projectile and target) density, and the impact parameter.
\begin{figure}[t]
\centering
  \includegraphics[scale = 0.62]{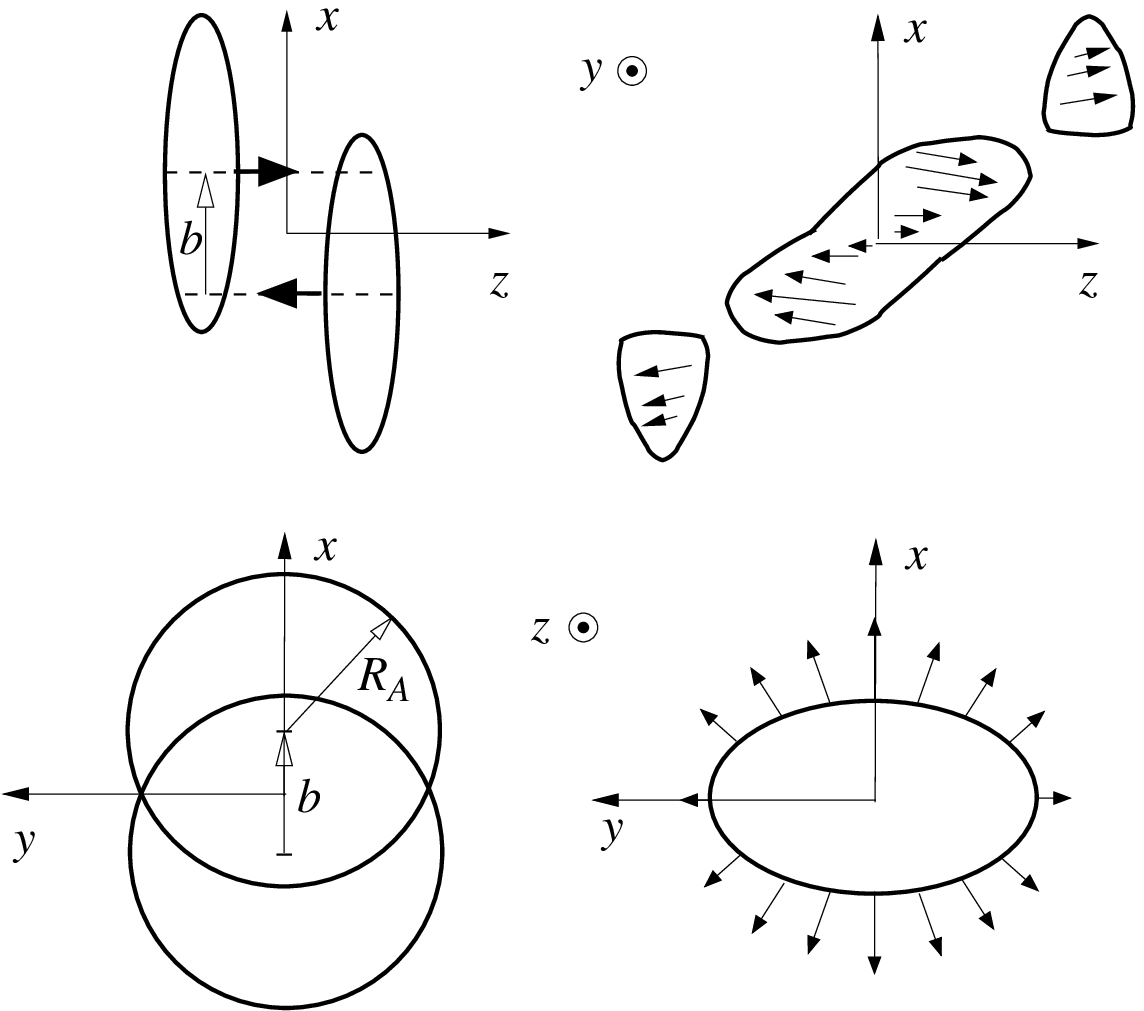}
  \caption[Schematic illustration of non-central heavy-ion collisions. Partons produced in the 
  overlap region carry global angular momentum along the $(-y)$-axis, opposite to the reaction 
  plane.]
  {Schematic illustration of non-central heavy-ion collisions. Partons produced in the overlap 
  region carry global angular momentum along the $(-y)$-axis, opposite to the reaction plane 
  \cite{Liang:2004ph}.}
  \label{geometryWang}
\end{figure}
\\A crucial model parameter is given by the rapidity which is longitudinally distributed in 
space-time 
\begin{equation}
\label{spacerap}
\eta = \frac{1}{2} \ln \left( \frac{t+z}{t-z}  \right)\,,
\end{equation}
and flow rapidity
\begin{equation}
\label{flowrap}
y =  \frac{1}{2} \ln \left( \frac{E+p_z}{E-p_z}  \right) = 
\frac{1}{2} \ln \left( \frac{1+v_z}{1-v_z}  \right)\,,
\end{equation}
which influence the form of $\phi (y,\eta)$.
The calculation of the hyperon polarization in the reaction plane \cite{Liang:2004ph} depends on 
an assumption of an initial condition described in Fig.\ 1 of Ref.\ \cite{Liang:2004ph} (which 
is reviewed in Fig.\ \ref{geometryWang} for convenience). Such an initial condition 
(generally referred to as the {\it firestreak model}), is roughly equivalent to two ``pancakes'', 
inhomogeneous in the transverse coordinate $\vec{x}_T$, sticking together inelastically. Each 
element of this system then streams in the direction of the local net momentum (Fig.\ 
\ref{initial_density}, right column).\\   
Since projectile and target have opposite momenta in the center-of-mass frame, assuming 
projectile and target nuclei to be identical, $\phi (y,\eta)$ can be approximated via
\begin{equation}
\phi (y,\eta) \simeq \delta(\eta) \delta \left[ y - y_{cm} (\vec{x}_T) \right]\,,
\end{equation}
where $y_{cm}$ is the local (in transverse space) longitudinal rapidity, corresponding to the 
flow velocity $v_{cm}$. Then, it is possible to rewrite the polarization in the reaction plane for
the firestreak model as
\begin{equation}
\langle p_z x_T\rangle \sim \frac{\sqrt{s}}{c(s)\,m_N} \langle D_\rho\rangle\,,
\label{csdef}
\end{equation}
where
\begin{equation}
\label{drho}
\langle D_\rho\rangle= \int d^2 \vec{x}_T \vec{x}_T \left[ \rho_P (\vec{x}_T) 
- \rho_T (\vec{x}_T) \right] \,.
\end{equation}
Here, $c(s)$ is an energy-dependent parameter \cite{Liang:2004ph} which can be estimated 
from final multiplicity using phenomenological formulas \cite{Milov:2004sv},
\begin{equation}
\label{csmilov}
c(s) \sim \frac{2 y}{N} \frac{dN}{dy} 
  \simeq \frac{1}{1.5} 
  \ln \left( \frac{\sqrt{s}}{1.5 GeV} \right)\ln \left( \frac{2 \sqrt{s}}{ GeV} \right)\,,
\end{equation}
assuming that all partons receive an equal share of momentum.
Since all nuclei have the same $\sqrt{s}$, $\langle P_q^R\rangle$ should be finite and constant 
over rapidity.
\\
The physical validity of this firestreak model is compelling at low energies, when the baryon 
stopping of nuclear matter is large. \\
At high energies and initial transparencies, however, a more generally 
accepted ansatz for the initial condition is that the initial partons are produced all throughout 
the longitudinal flow rapidity spanned between the forward-travelling projectile and the 
backward-travelling target (middle panel of Fig.\ \ref{initial_density}). This approximation is
named Brodsky--Gunion--Kuhn (BGK) \cite{bgk,Adil:2005qn} model.\\
If participant and target densities are equal, $\rho_P (\vec{x}_T)=\rho_T (\vec{x}_T)$, this 
ansatz reduces to a boost-invariant initial condition. 
For a non-central collision, however, such equality will only hold at the midpoint in $\vec{x}_T$ 
of the collision region. Interpolating linearly in rapidity between participant $\rho_P$ 
(at $y=y_L$) and target densities $\rho_T$ (at $y=-y_L$), one gets
\begin{eqnarray}
\label{rhoybgk}
\phi (y,\eta) &=&  \left( A  +  y B \right) \delta(y-\eta)\,,
\end{eqnarray}
with
\begin{eqnarray}
A&=& \frac{1 }{2},\;\;B=
\frac{\rho_P (\vec{x}_T) - \rho_T (\vec{x}_T)}
{\rho_P (\vec{x}_T) + \rho_T (\vec{x}_T)}\frac{1}{2 y_L}\,,
\end{eqnarray}
which leads to a polarization in the recation plane for the BGK model
\begin{eqnarray}
\label{pzbgk}
\langle p_z x_T\rangle  &\propto& \int d y \sinh(y) \vec{x}_T \rho(\vec{x}_T,y) d \vec{x}_T  
\nonumber \\ 
&\propto& \langle D_\rho\rangle \left[ y_L \cosh(y_L) - \sinh(y_L) \right]\,.
\end{eqnarray}
For this initial condition, the axial symmetry of the initial pancakes forces the net 
polarization to be zero at mid-rapidity.\\
However, it is also reasonable to assume that the density of matter (in $\eta$) flowing with 
rapidity $y$ is
\begin{equation}
\phi(y,\eta) \sim \exp\left[ \frac{-(\eta - y)^2}{2 \sigma_\eta^2} \right]\,,
\end{equation}
where $\sigma_\eta$ is a parameter which has to be determined. Applying this distribution 
instead of the $\delta$--function used above yields, at $\eta=0$, to another expression for the
polarization in the reaction plane for the BGK model
\begin{eqnarray}
\label{bgkfluct}
\hspace*{-1.0cm}\langle p_z x_T\rangle &\sim& \frac{1}{2 \sqrt{ 2 \pi}} 
\left[ B e^{\frac{1}{2} y_L ( -2 - \frac{y_L}{\sigma_\eta^2})} 
\sigma_\eta \;\times \right. \nonumber\\
&&\hspace*{1.0cm}\left. \left\{ 2 - 2 e^{2 y_L} +  
e^{\frac{(\sigma_\eta^2 - y_L)^2}{\sigma_\eta^2}} \sqrt{2 \pi} \sigma_\eta 
\left[ -\mathrm{erf} 
\left( \frac{\sigma_\eta^2 - y_L}{\sqrt{2} \sigma_\eta} \right) \right. \right. \right.
\nonumber\\
&&\hspace*{3.0cm}+ \mathrm{erf} 
\left. \left.\left.\left( \frac{\sigma_\eta^2 - y_L}
{\sqrt{2} \sigma_\eta} \right) \right] \right\} \right]\,,
\end{eqnarray}
which simplifies at mid-rapidity to 
\begin{equation}
\label{sigmasimple}
\langle p_z x_T\rangle \propto B e^{\sigma_\eta^2/2} \sigma_\eta^2\,.
\end{equation}
The rapidity distributions are summarized in Fig.\ \ref{initial_density}, and the corresponding 
shear created is displayed in Fig.\ \ref{initial_shear}.
\\Thus, at high (RHIC and LHC) energies we expect that the net polarization around the reaction 
plane of A+A collisions should vanish at mid-rapidity and re-appear in the target and projectile 
regions [Fig.\ \ref{initial_shear} (a)]. At lower energies, on the other hand, reaction plane 
$\Lambda$ or $\overline{\Lambda}$ polarization should be more uniform in rapidity space, and be 
significantly above zero at mid-rapidity [Fig. \ref{initial_shear} (b)].\\
More realistic nuclear geometries should not alter these very basic considerations, although for 
the BGK case they might considerably slow down the shear rise in rapidity. This is also true for 
corrections to linear interpolation in rapidity space. Detailed hydrodynamic simulations 
\cite{Romatschke:2007mq} also reinforce the conclusion that within the boost-invariant limit 
vorticity is negligible.\\
Given that the two expressions for the polarization in the reaction plane are known for the
firestreak model [see Eq.\ (\ref{csdef})] and the BGK model [see Eq.\ (\ref{sigmasimple})],
the measurement of the $\Lambda$ polarization in the reaction plane could be a valuable 
tool of investigating the initial longitudinal geometry of the system. At the moment, the 
longitudinal geometry, and in particular the longitudinal scale variation with energy (i.e., if, 
how, and at at what energy initial conditions change from ``firestreak'' to ``BGK'') is not well 
understood \cite{Busza:2004mc}. This understanding is crucial for both the determination of the 
EoS and the viscosity, since longitudinal geometry is correlated with the initial energy density, 
and hence to the total lifetime of the system and the time in which flow observables can form 
\cite{Torrieri:2007qy}.\\
The measurement of the energy and system-size dependence of $\Lambda$ polarization in the reaction 
plane at mid-rapidity could be a significant step in qualitatively assessing the perfection of 
the fluid, and determining at what energy the system enters a fluid-like behaviour.\\
To investigate the difference between the two above discussed models, it is useful to calculate the
ratio
\begin{equation}
\label{ratio}
\frac{\left. \langle P_q^R\rangle \right|_{BGK}}
{\left. \langle P_q^R\rangle \right|_{firestreak}} 
= c(s)\frac{m_N e^{\sigma_\eta^2/2} \sigma_\eta^2}{\sqrt{s}}\,.
\end{equation}
In the limit of $\sigma_\eta \rightarrow 0$ the system has no vorticity and thus, due to spin-orbit
coupling, no polarization. While only at very low 
energies [where Eq.\ (\ref{sigmasimple}) and the BGK picture are untenable as approximations] 
the BGK and firestreak pictures are comparable, vorticity at BGK could still be non-negligible, 
provided that $\sigma_\eta \sim 1$.
\begin{figure}[t]
\centering
  \includegraphics[scale = 0.42]{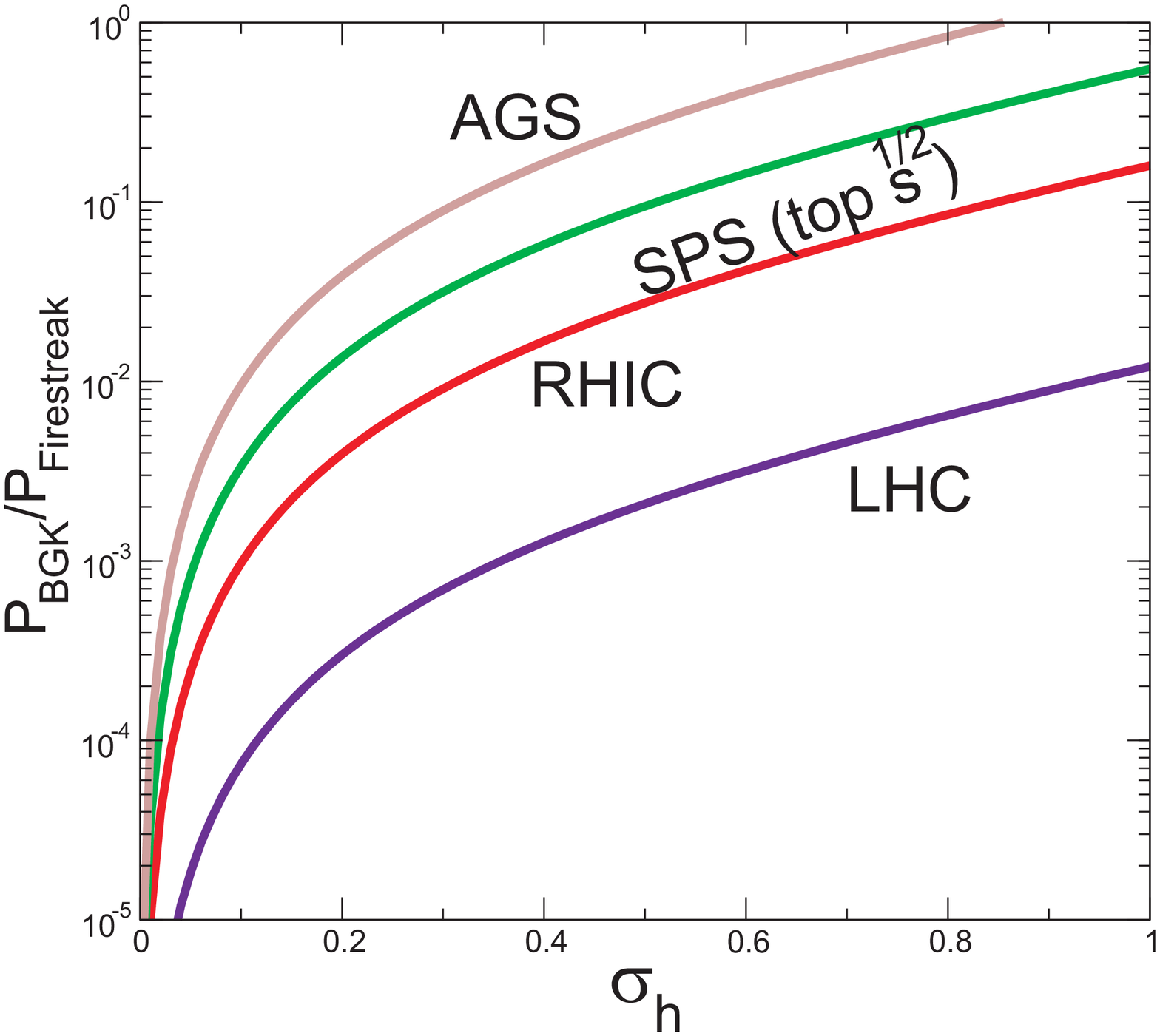}
  \caption[Ratio of Brodsky--Gunion--Kuhn to firestreak predictions as a function of $\sqrt{s}$ 
  and $\sigma_\eta$, the correlation length between space-time and flow rapidity.]
  {Ratio of BGK to firestreak predictions as a function of $\sqrt{s}$ and $\sigma_\eta$, the 
  correlation length between space-time and flow rapidity, calculated using Eqs.\ (\ref{csmilov}) 
  and (\ref{ratio}) \cite{Betz:2007kg}.}
  \label{bgkfire}
\end{figure}
\\It should be stressed that $c(s)$ and $\sigma_\eta$ contain very different physics: In Ref.\ 
\cite{Liang:2004ph}, $c(s)$ is interpreted as the number of partons into which the energy of the 
initial collision is distributed. On the other hand, $\sigma_\eta$ depends on the 
imperfection of ``Bjorken'' expansion (the correlation between space-time and flow rapidity). 
These two effects, however, go in the same direction, although $c(s) \sim \left( \ln  \sqrt{s} 
\right)^2 $ is much less efficient at diminishing polarization than a small $\sigma_\eta$.\\
Combining $c(s)$ of Eq.\ (\ref{csmilov}) with Eq.\ (\ref{ratio}), one obtains the ratio between 
BGK and firestreak expectations, and its dependence on energy and the parameter $\sigma_\eta$.  
The result is shown in Fig.\ \ref{bgkfire}, assuming $\sigma_\eta \ll y_{L}$. This 
figure should be taken as an illustration for the sensitivity of the polarization measure to the 
longitudinal structure of the initial condition, rather than as a prediction of the polarization 
in the two models (as shown in Ref.\ \cite{Liangqm2006}, the small-angle approximation used in 
Ref.\ \cite{Liang:2004ph} is in any case likely to be inappropriate). As can be seen, the effects 
of $c(s)$ in the firestreak picture are only at low energies comparable to the effects of a 
non-negligible $\sigma_\eta$ in the BGK picture (where the firestreak picture is thought to work 
better). At top RHIC energy, even at $\sigma_\eta$ of one unit, the BGK polarization should be 
suppressed with respect to the firestreak expectation with about two orders of magnitude. This 
grows to several orders of magnitude for LHC energies.\\
Connecting the experimental measurement of the $\Lambda$ polarization to the initial condition is, 
however, non-trivial, as this observable is sensitive not just to the initial stage but also to 
the subsequent evolution of the system, up to the final freeze-out.\\
In the following two sections we will qualitatively discuss the effect the later stages will have 
on the final observable. We will argue that, while the observable is likely to be modified by the 
subsequent evolution, a comparison of several kinds of polarization could be useful in obtaining 
information about initial conditions, the mean-free path and the freeze-out scenario.

\section[Hydrodynamic Evolution, Polarization, and Jets]
{Hydrodynamic Evolution, Polarization, and Jets}
\begin{figure}[t]
\centering
\begin{minipage}[c]{4.2cm}
\hspace*{-4.0cm}
  \includegraphics[scale = 0.33]{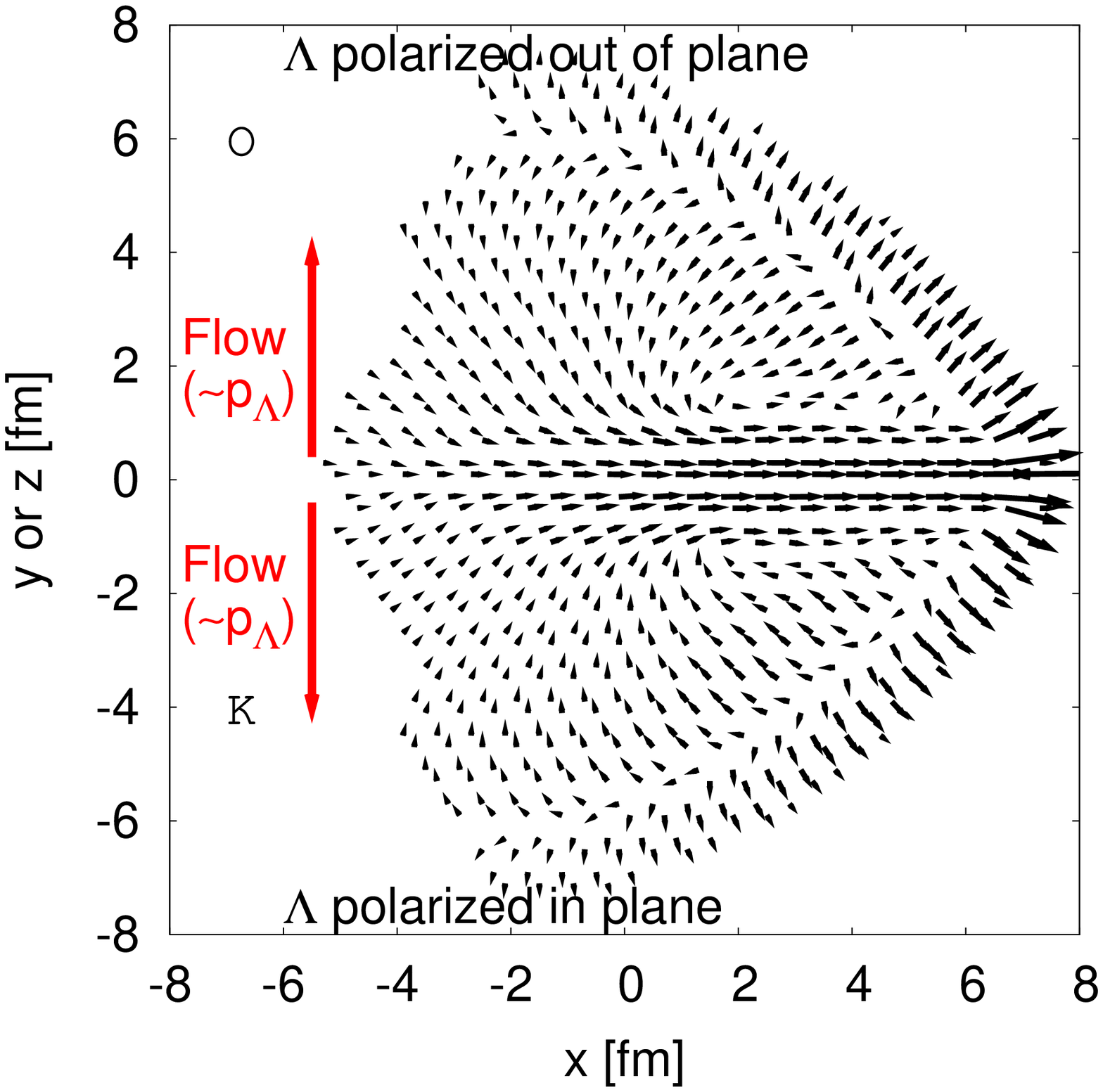}
\end{minipage}
\hspace*{-2.0cm}  
\begin{minipage}[c]{4.2cm} 
  \includegraphics[scale = 0.44,angle=270]{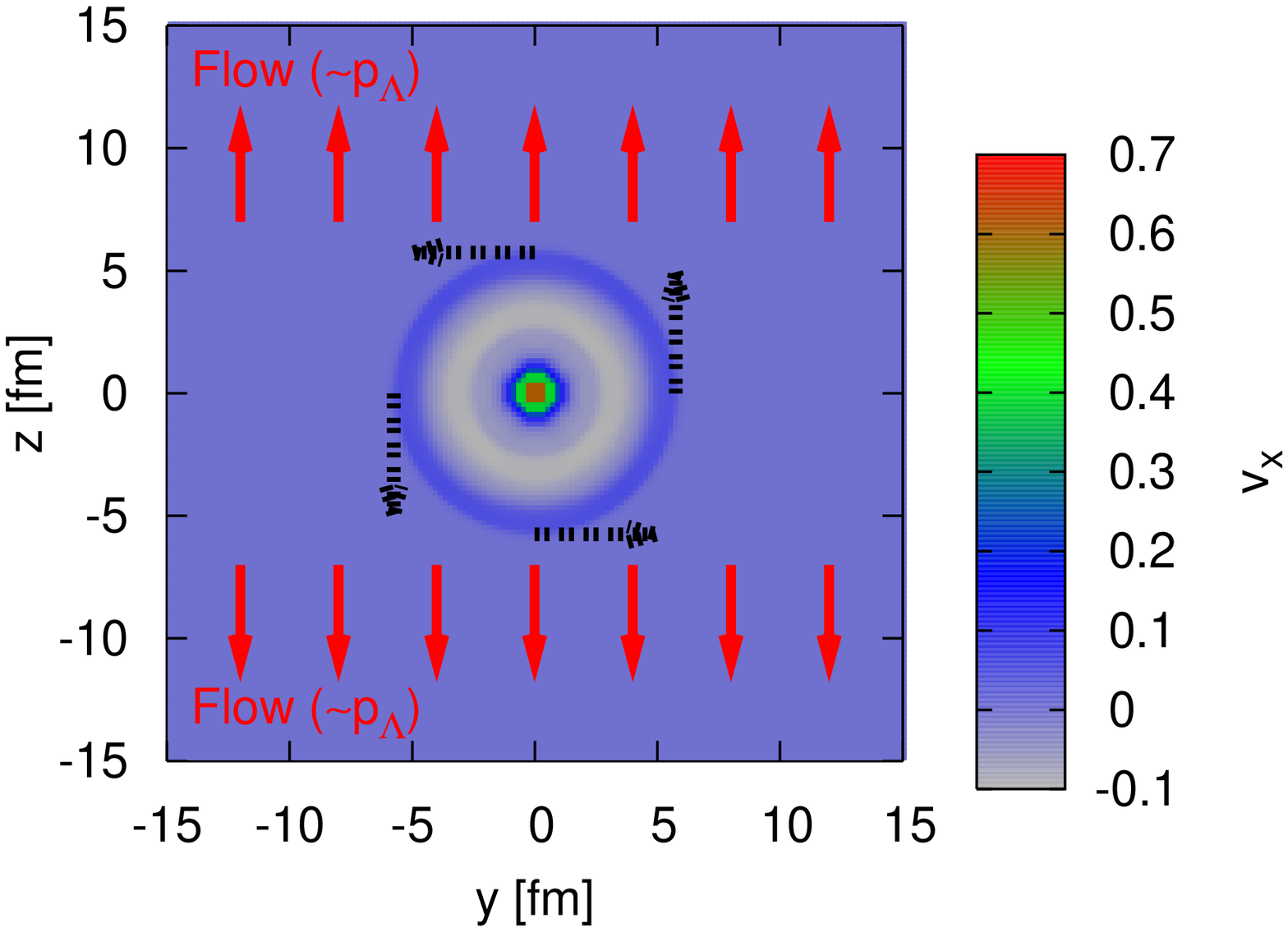}
\end{minipage}
  \caption[Vorticity generated by a fast jet traversing the system in the positive $x$-direction.]
  {Vorticity generated by a fast jet traversing the system in the positive $x$-direction. The 
  arrows in the left panel show the momentum density of fluid elements in the $(x,y)$-plane, while 
  the contour in the right panel displays the $x$-component of the velocity in the $(y,z)$-plane. 
  The jet has been travelling for $t = 11.52$~fm through a static medium \cite{Abreu:2007kv}. The 
  dashed arrows in the right panel indicate the expected direction of polarization of a 
  $\Lambda$-particle (out of plane for left panel, tangentially in right panel). If the medium 
  undergoes transverse and longitudinal expansion, the $\Lambda$ position within the 
  smoke ring is correlated with its mean momentum.  Thus, measuring $\Lambda$ polarization 
  in the plane defined by its momentum and the jet momentum should yield a positive net result 
  \cite{Betz:2007kg}.}
  \label{smokering}
\end{figure}
In relativistic hydrodynamics, vorticity works somewhat differently than in the non-relativistic 
limit \cite{Taub,Florkowski:1992yd}. While in non-relativistic ideal hydrodynamics, the conserved 
circulation is simply defined as $\vec{\nabla} \times \vec{v}$, relativistically the conserved 
vorticity is
\begin{equation}
\vec{\Omega} = \vec{\nabla} \times w \gamma \vec{v}\, ,
\end{equation}
where $w$ is the enthalpy per particle.
In the non-relativistic limit, where $ w \simeq m$ and $\gamma=1$, the usual limit is recovered. 
In a relativistic fluid with strong pressure and energy-density gradients, on the other hand, 
vortices can be created and destroyed even in perfectly smooth initial conditions, such as the 
BGK case described in the previous section.\\
Since vorticity development is a highly non-linear phenomenon, quantitative details require 
numerical simulations. Here, we consider the vorticity which develops when a momentum source moving 
at the speed of light traverses a uniform relativistic fluid.  This could be an appropriate 
description of the thermalized jet-energy loss (see also section \ref{FirstStudies} and chapter 
\ref{DiffusionWake}), if the jet loses energy fast and locally. The calculation is done using a 
$(3+1)$-dimensional hydrodynamical prescription \cite{Rischke:1995pe}. The flow vector in the 
$(x,y,z)$-coordinate system, where the fluid is at rest (if co-moving with the collective flow), 
is shown in Fig.\ \ref{smokering}, parallel (left panel) and perpendicular (right panel) to the 
jet direction.\\
The simulation shown in Fig.\ \ref{smokering} is based on a jet energy-loss model that assumes a 
high momentum gradient. It is not surprising that a large initial momentum gradient, such as that 
produced by a jet quickly losing energy, can introduce vorticity into the system. \\
As shown in the simulation, these vortices are stable enough to last throughout the 
lifetime of the fluid. Therefore, an interesting polarization measurement is to trigger on jet 
events and measure $\Lambda$ polarization $(P_\Lambda^{J})$ in the plane perpendicular to the 
{\it jet} production plane. Since vorticity in such events exists independently of the global 
initial conditions, this measurement is sensitive only to the mean-free path and perhaps 
final-state effects.\\
Fig.\ \ref{smokering} also illustrates how such a measurement could be performed: the polarization 
axis is defined based on the (high-$p_T$ trigger) jet direction. Since vortices above and below 
the jet move in opposite directions, it would be impossible to detect vorticity via polarization 
measurements in a {\it static} medium.\\
If, however, a smoke ring occurs in a medium undergoing transverse or longitudinal expansion, the 
flow introduces a correlation between the $\Lambda$ position within the smoke ring (and hence its 
polarization) and its average momentum $\langle p_\Lambda\rangle$. Measuring the polarization of 
moderately high momentum but thermal $\Lambda$-particles ($\sim 700 MeV$) in the plane defined by 
the $\Lambda$ momentum and the jet direction should thus yield a non-zero result.\\
We therefore propose to measure the polarization $P_i^J$ of jet-associated moderate momentum 
particles, in the plane defined by the jet direction and the direction of the particle. \\
The observation of this polarization would be a strong indication of collective behavior, since 
it signifies jet-induced vorticity.\\
Unlike production-plane vorticity, jet vorticity does not depend on initial conditions, but should 
hold for a wide variety of jet energy-loss scenarios, provided the coupling between the system and 
the jet is strong. It is not at all clear, however, whether in the strongly-coupled regime 
(rather than the perturbative one, on which the calculations of Ref.\ \cite{Liang:2004ph} are 
based) vorticity will readily transform into quark polarization. The next section is devoted to 
this topic.

\section[Mean-Free Path and Polarization]
{Mean-Free Path and Polarization}

In a perfect fluid, angular momentum should not go into a {\it locally} preferred direction, but 
into vortices where each volume element is locally isotropic in the frame co-moving with the flow.\\
Such vortices should imprint final observables via longitudinal collective flow 
(e.g.\ odd $v_n$ coefficients away from mid-rapidity), but not via polarization since any
polarization created would immediately be destroyed by subsequent re-interactions if the 
equilibration between gain and loss terms happens instantaneously (``a perfect fluid''). 
In this regime, the equations derived in section \ref{IniCondPol} are no longer tenable because 
they assume unpolarized incoming particles and a coupling constant small enough for perturbative 
expansion.\\
Keeping the first of these assumptions would violate detailed balance\footnote{Detailed balance 
means that $P_{ij}q_i=P_{ji}q_j$, i.e., that there is a balance between the states $i$ and $j$. 
Here, $P_{ij}$ is the transition probability and $q_i$ the equilibrium probability of being in a 
state $i$.}, while the second assumption is probably incompatible with a strong collective 
behavior. \\
The local isotropy of a perfectly thermalized system was used \cite{Hoyer} to suggest that 
the disappearance of the production-plane polarization observed in elementary collisions could 
be a signature of deconfinement. A first-order correction becomes necessary if the size of the 
radius of curvature within the vortex becomes comparable to the mean-free path $\ell_{mfp}$. 
The anisotropy would then be given by the deformation of a volume element of this size. 
In general, the polarization in any direction $i$ (production, reaction, or jet) can be expressed as
\footnote{Here it is assumed that 
$\langle P\rangle = [e^{E/T}-e^{-E/T}]/[e^{E/T}+e^{-E/T}]=\tanh E/T\sim E/T$, leading via
spin-orbit coupling with a first-order approximation for the spin (that is of the order $\ell_{mfp}$)
and the local angular momentum density to Eq.\ (\ref{lmfppol}).}
\begin{eqnarray}
\langle P_q^{i}\rangle &\sim& \tanh\left[ \vec{\zeta}_i \right] \sim \vec{\zeta}_i 
\label{lmfppol1} \\
\vec{\zeta}_i &=& \frac{\ell_{mfp}}{T} \left( \epsilon_{i j k} 
\frac{d\langle \vec{p}_k\rangle}{d \vec{x}_j} \right)\,,
\label{lmfppol}
\end{eqnarray}
where $\langle \vec{p}_j \rangle$ is the local direction of momentum in the laboratory frame and 
$T$ the temperature. Thus, potentially, the amount of residual polarization which survives a 
hydrodynamic evolution (whether from initial geometry or from deformation of the system due to 
jets) is directly connected to the system's mean-free path.\\
Therefore, determining the polarization rapidity dependence (in any plane, production, reaction, 
or jet, where it could be expected to be produced) could perhaps ascertain the rapidity domain 
of the QGP. If a (s)QGP is formed at central rapidity, while the peripheral regions consist of 
a hadron gas, one should observe a sharp rise in production, reaction- and jet-plane polarization 
in the peripheral regions.\\
The problematic aspect of using polarization for such a measurement is that it is sensitive to the
late-stage evolution, including hadronization and the interacting hadron gas phase.\\
As shown in Ref.\ \cite{Barros:2005cy}, an unpolarized QGP medium at freeze-out will produce a 
net production-plane polarization due to hadronic interactions. Similarily, unpolarized p+p and 
p+A collisions result in net hyperon polarization. While local detailed balance inevitably 
cancels out such local polarization, the rather large mean-free path of an interacting hadron 
gas, and the considerable pre-existing flow ensure that any interacting hadron gas phase should 
be well away from detailed balance, and hence likely to exhibit residual polarization.\\
It then follows that the absence of a production-plane polarization is not only a strong 
indication of sQGP formation, but of a ``sudden'' freeze-out where particles are emitted 
directly from the QGP phase.\\
The evidence of quark coalescence even at low momentum \cite{Abelev:2007qg}, together with 
sudden-freeze-out fits \cite{Baran:2003nm,Torrieri:2000xi}, supports further investigation using 
the polarization observable in any plane (reaction, production, and jet) where the vorticity in the 
hot phase is expected to be non-zero.\\
If polarization in {\it all} directions is consistently measured to be zero, including events 
with jets and within high rapidity bins, it would provide strong evidence that the mean-free 
path of the system is negligible, and final-state hadronic interactions are not important 
enough to impact flow observables. However, a measurement of production plane, but not 
reaction-plane polarization proves that the initial state of the system is BGK-like and that the 
interacting hadron gas phase leaves a significant imprint on soft observables. (The BGK nature of 
the initial condition can then be further tested by scanning reaction-plane polarization 
in rapidity.)\\
An observation of polarization in the jet plane could result in a further estimate of the 
mean-free path, showing that the jet-degrees of freedom are thermalized and part of the collective 
medium. \\
A sudden jump in any of these polarizations at a critical rapidity might signal a sharp increase 
in the mean-free path, consistent with the picture of a mid-rapidity QGP and a longitudinal 
hadronic fragmentation region. Analogously, a drop of polarization while scanning in energy and 
system size could signal the critical parameters required for a transition from a very viscous 
hadron gas to a strongly interacting quark-gluon liquid. \\
Thus, motivated by the question if the vortex-like structures produced in a jet event (if 
momentum loss is sufficiently large) may lead to observable consequences, we discussed the 
connection of hyperon polarization to initial geometry and microscopic transport properties. 
However, since the polarization observable can be significantly altered by all stages of the 
evolution, a quantitative description remains problematic. Nevertheless, the proposed 
measurements (reaction plane, production plane, and jet plane polarization) might shed
some light on several aspects of heavy-ion collisions that are not well understood yet.\\
In the following, we will again focus on the mechanisms (i.e., the source term) of jet energy 
and momentum loss by comparing two completely independent and different approaches describing 
the interaction of the jet with a QGP, pQCD, and AdS/CFT.

\clearpage{\pagestyle{empty}\cleardoublepage}
%
%
\chapter[Di-Jet Correlations in pQCD vs.\ AdS/CFT]
{Di-Jet Correlations in pQCD \\ vs.\ AdS/CFT}
\label{pQCDvsAdSCFT}

As already discussed in part \ref{part02}, the Anti-de-Sitter/Conformal Field Theory (AdS/ CFT) 
correspondence \cite{Aharony:1999ti,Maldacena:1997re,Witten:1998qj,Witten:1998zw} is considered 
to be a feasible approach for describing a strongly-coupled medium (like possibly the QGP) although 
QCD, the theory appropriate for characterizing the QGP, is not a conformal field theory.\\
In chapter \ref{AdS/CFT} we reviewed that the propagation of a supersonic heavy quark moving at a 
constant velocity through a static strongly-coupled ${\cal N}=4$ Supersymmetric Yang--Mills (SYM) 
background plasma at nonzero temperature $T_0$ can be formulated using AdS/CFT which provides a detailed 
energy-momentum (stress) tensor \cite{Friess:2006fk,Yarom:2007ni,Gubser:2007ga}. 
It features the expected Mach-cone region (commonly named {\it far zone}) as well as the strong 
forward-moving diffusion wake (see Fig.\ \ref{GubserMachDiffusion}). The far-zone response is well 
described in the strong-coupling limit of the ${\cal N}=4$ SYM 
plasma with a ``minimal'' shear viscosity to entropy density ratio $\eta/s=1/4\pi$ 
[which is close to the uncertainty principle limit \cite{Danielewicz:1984ww}]
\cite{Policastro:2001yc,Kovtun:2004de}. \\
However, we showed in the context of a hydrodynamic prescription (cf.\ chapter 
\ref{DiffusionWake} and Ref.\ \cite{Betz:2008ka}) that the strong forward diffusion wake spoils 
the signature of the Mach cone which would result in a double-peaked structure.\\ 
For AdS/CFT, such diffusion wakes were already presented in Refs.\ 
\cite{Gubser:2007ga,Gubser:2007ni,Gubser:2008vz} and it was demonstrated in Refs.\ 
\cite{Gyulassy:2008fa,Noronha:2008un} (as reviewed in section \ref{NoGoTheorem}) that in the 
strict supergravity limit, $N_c\gg 1,\, g_{SYM}^2\ll 1$ but $\lambda=g_{SYM}^2 N_c\gg 1$, the 
far-zone wakes have such small amplitudes that they only lead to a single broad peak in the 
away-side hadronic correlation after a Cooper--Frye (CF) freeze-out of the fluid 
\cite{Cooper:1974mv}. \\
The azimuthal correlations of associated hadrons after applying such a CF freeze-out to an AdS/CFT 
source term calculated in Ref.\ \cite{Gubser:2007ga} (shown in Ref.\ \cite{Noronha:2008un}, 
see Fig.\ \ref{JorgeParticleDistribution}) exhibit, however, an apparent conical signal which does 
not obey Mach's law and is due to the non-equilibrium neck zone, introduced in section 
\ref{JetsInAdSCFT}. This neck zone is distinguished by a strong transverse flow relative to the 
jet axis (see Fig.\ \ref{JorgeNeck}), inducing the conical correlation that even prevails after a 
CF thermal broadening at freeze-out. \\
In section \ref{JetsInAdSCFT}, based on the considerations of Refs.\ 
\cite{Noronha:2007xe,Gyulassy:2008fa,Noronha:2008un}, the neck zone was defined as the region 
near the heavy quark where the local Knudsen number is 
$Kn= \Gamma \,{|\nabla\cdot {\vec{M}}|}/{|{\vec{M}}|} >1/3$ [see Eq.\ (\ref{defknudsen}) and 
Fig.\ \ref{PlotComparisonKnudsen}], where $M^i (t,x)=T^{0i}(t,x)$ is the momentum flow field 
of matter and $\Gamma\equiv 4\eta/\left(3sT_0\right)\ge 1/\left(3\pi T_0\right)$ is the sound 
attenuation length which is bounded from below for ultra-relativistic systems
\cite{Danielewicz:1984ww,Policastro:2001yc}. \\
In AdS/CFT, this neck region is the field-plasma coupling zone where the stress tensor has a 
characteristic interference form depending on the coordinates, following 
$\mathcal{O}(\sqrt{\lambda} T_0^2/R^2)$ \cite{Yarom:2007ni,Gubser:2007nd}, with $R$ denoting the distance to 
the heavy quark in its rest frame. In contrast, the stress in the far zone has the characteristic 
$\mathcal{O}(T_0^4)$ form. In addition, very near the quark the self-Coulomb field of the heavy quark 
contributes with a singular stress $\mathcal{O}(\sqrt{\lambda}/R^4)$ \cite{Friess:2006fk}. \\
The above strong coupling AdS/CFT results served as a motivation to study whether similar novel 
near-zone field-plasma dynamical coupling effects also arise in weak\-ly-coupled perturbative Quantum 
Chromodynamics (pQCD) \cite{Betz:2008wy}. \\
In Refs.\ \cite{Neufeld:2008fi,Neufeld:2008hs,Neufeld:2008dx} the heavy quark jet induced stress 
in a weakly-coupled QGP (wQGP in contrast to sQGP) generated by the passage of a fast parton 
moving with a constant velocity was computed analytically in the linear-response approximation 
based on the Asakawa--Bass--M\"uller (ABM) \cite{Asakawa:2006jn} generalization of chromo-viscous 
hydrodynamics \cite{Heinz:1985qe}. The ABM generalization concentrates on the ``anomalous 
diffusion'' limit, where the conductivity is dominated by field rather than stochastic dissipative 
scattering dynamics.\\
As in the AdS/CFT string-drag model, the generic far-zone Mach and diffusion wakes are also 
clearly predicted in the pQCD-based ABM formulation \cite{Neufeld:2008fi,Neufeld:2008dx}, 
exhibiting extremely similar patterns of the energy and momentum-density perturbations as 
displayed in Fig.\ \ref{NeufeldMach}.\\
In the following, we compare the azimuthal correlations obtained from AdS/ CFT 
\cite{Gubser:2007ga} with the ones calculated from pQCD by solving numerically the full 
non-linear $(3+1)$-dimensional relativistic hydrodynamic equations using the SHASTA  
\cite{Rischke:1995ir}, supplemented with the chromo-viscous stress source derived in Refs.\ 
\cite{Neufeld:2008fi,Neufeld:2008hs} and reviewed in section \ref{pQCDSourceTerm}.\\
We specialize to the ideal-fluid case of vanishing viscosity to minimize the dissipative 
broadening of any conical correlations and therefore maximizing the signal-to-noise ratio.\\
We emphasize that our aim here is not to address the current light quark/gluon jet RHIC 
correlation data, but to point out the significant differences between weakly and 
strongly-coupled models concerning the mechanisms of heavy quark energy loss that can be tested 
experimentally when identified heavy quark (especially bottom quark) jet correlations will become 
feasible to measure. We limit this study to the most favorable idealized conditions (uniform 
static plasma coupled to the external Lorentz-contracted color fields). Distortion effects due 
to evolution in finite expanding plasma geometries will be discussed in the next chapter.\\
Since the models for the weakly-coupled pQCD and strongly-coupled AdS/CFT prescriptions are 
shown to predict qualitatively different associated hadron correlations with respect to tagged 
heavy quark jets \cite{Betz:2008wy}, we propose that an identified heavy quark jet observable 
may discriminate between those approaches for the QGP dynamics in ultra-relativistic 
nuclear collisions at RHIC and LHC energies.\\
It will be shown that while both models feature similar Mach and diffusion zones (see Fig.\ 
\ref{SketchRealJetDeposition}), they differ significantly in the neck region where strong 
chromo-fields originating from the heavy quark jet couple to the polarizable plasma. 
The associated conical correlations from AdS/CFT are dominated 
(as demonstrated in Ref.\ \cite{Noronha:2008un}) by the jet-induced transverse flow in the neck zone. 
However, in pQCD, the induced transverse flow in the neck region is too weak to produce conical 
correlations after a CF freeze-out. \\
Thus, the observation of conical correlations violating Mach's law would favor the strongly-coupled
AdS/CFT string-drag dynamics, while their absence would support weakly-coupled pQCD-based 
chromo-hydrodynamics.

\section[The Stress Zones]{The Stress Zones}
\begin{figure}[t]
\centering
  \includegraphics[scale = 0.56]{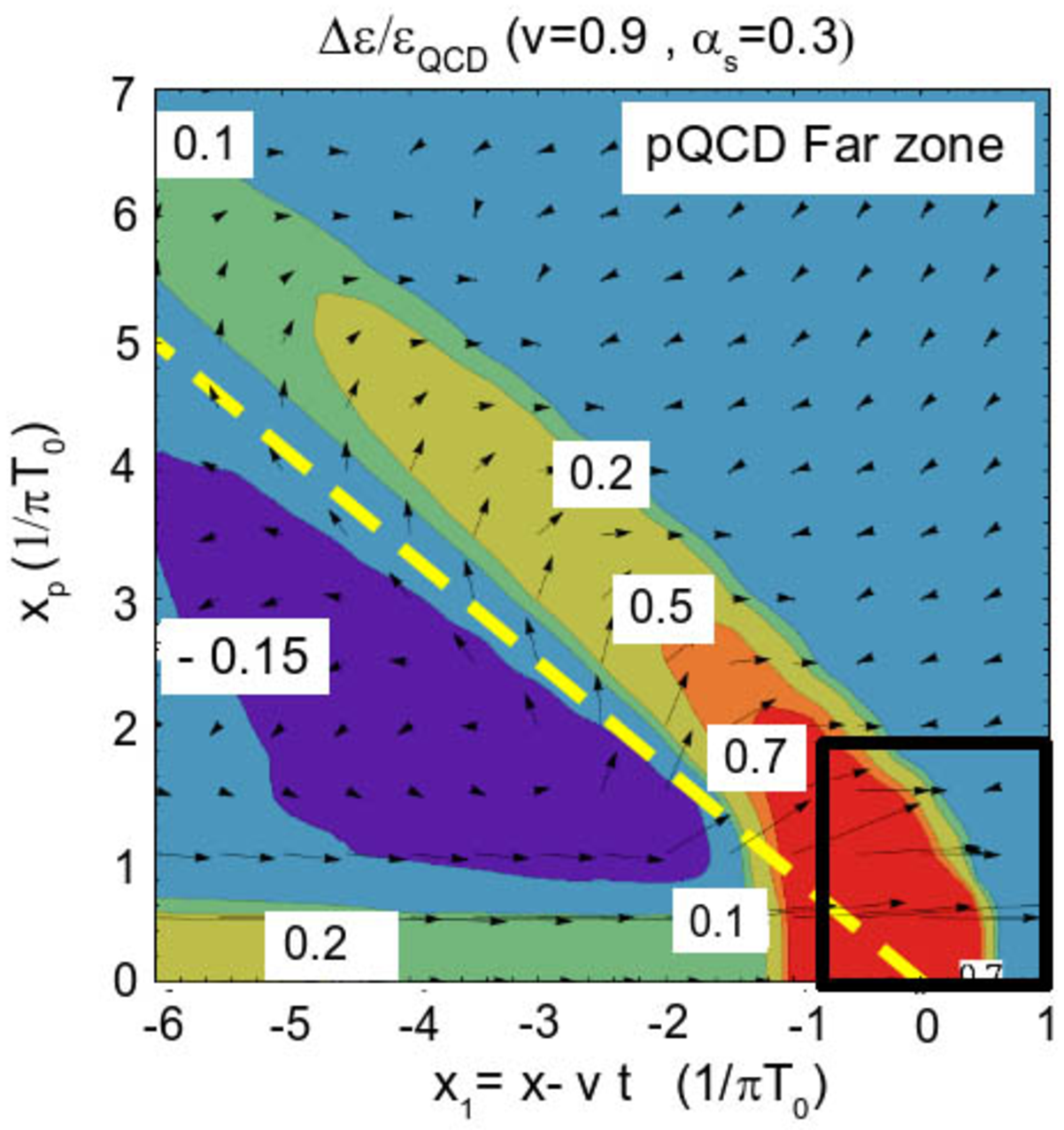}
  \caption[The fractional-energy density perturbation due to a heavy quark with $v=0.9$ in a QCD 
  plasma of temperature $T_0=200$ MeV. ]
  {The fractional-energy density perturbation 
  $\Delta \varepsilon/\varepsilon_0\equiv \varepsilon(x_1,x_p)/\varepsilon_0-1$ (in the laboratory 
  frame) due to a heavy quark with $v=0.9$ in a QCD plasma of temperature $T_0=200$ MeV. The 
  induced fluid stress was calculated using ($3+1$)-dimensional hydrodynamics \cite{Rischke:1995ir}
  with the anomalous pQCD source of Neufeld \cite{Neufeld:2008hs}. A trigger jet (not shown) 
  moves in the $(-\hat{x})$-direction. The away-side jet moves in $\hat{x}$-direction and contours 
  of $\Delta \varepsilon/\varepsilon_{0}=-0.15,0.1,0.2,0.5,0.7$ are labeled in a comoving 
  coordinate system with $x_1=x-vt$ and the transverse radial coordinate $x_p$ in units of 
  $1/\pi T_0\approx 0.3$ fm after a total transit time $t=5$fm/c$=14.4/(\pi T_0)$. The ideal Mach 
  cone for a point source is indicated by the yellow dashed line in the $x_1-x_p$ plane. See Fig.\ 
  \ref{pQCDzones} for a zoom of the neck region inside of the black box \cite{Betz:2008wy}.}
  \label{Neufeld_farFig.1}
\end{figure}
Jet physics, below some transverse-momentum saturation scale \cite{Mueller:2008zt}, depends on the 
properties of the medium and allows for testing different models of jet-medium coupling dynamics 
by studying the detailed angular and rapidity correlations.\\
In the following, we demonstrate a striking difference between strongly-coupled AdS/CFT and 
moderate-coupling, multiple-collision pQCD transport models, and study their experimentally 
measurable consequences.\\
Ignoring again the near-side associated correlations, the energy-density perturbation for a jet 
described by the pQCD source term \cite{Neufeld:2008fi,Neufeld:2008hs} (see section 
\ref{pQCDSourceTerm}) clearly exhibits the Mach-cone region and diffusion wake (including the neck 
zone) as already discussed in the context of the schematic source term of chapter 
\ref{DiffusionWake}. For the corresponding plot with the AdS/CFT source of Ref.\ 
\cite{Gubser:2007ga}, see Fig.\ \ref{JorgeNeck}.\\
The energy-momentum (stress) tensor induced by the away-side heavy quark jet in both pQCD and 
AdS/CFT can be conveniently decomposed, as introduced in section \ref{JetsInAdSCFT} 
[see Eq.\ (\ref{EqnStressZones})] into four separate contributions 
\cite{Noronha:2007xe,Gyulassy:2008fa,Noronha:2008un} and is repeated here for convenience
\begin{equation}
\hspace*{-0.9cm}
T^{\mu\nu}(t,\vec{x})=T^{\mu\nu}_0(t,\vec{x})+\delta T^{\mu\nu}_{Mach}(t,\vec{x}) + 
\delta T^{\mu\nu}_{neck}(t,\vec{x})+ \delta T^{\mu\nu}_{Coul}(t,\vec{x})\,.
\label{energymomentum}
\end{equation}
The static isotropic background stress tensor is assumed to be 
$T^{\mu\nu}_{0}={\rm diag}(\varepsilon,p,p,p)$, where $\varepsilon=K T_0^4$ is the background 
energy density of a gas of massless $SU(3)$ gluons with $K_{QCD}=8\pi^2/15$ (which we use for the 
hydrodynamical calculations applying the pQCD source term) whereas for $SU(N_c)$  
SYM $K_{SYM}=3\pi^2 (N_c^2-1)/8$. In both cases, $\varepsilon=3\,p\,$ and 
the background temperature is $T_0$.\\
The Coulomb contribution to the energy-momentum tensor $\delta T^{\mu\nu}_{Coul}(t,x)$ arises from 
the near-zone Lorentz-contracted Coulomb field that remains attached to the heavy quark since we 
consider only moderate but supersonic velocities $c_s\le v\le 0.9$. We can therefore neglect 
radiative energy loss that dominates in the ultrarelativistic case. The bare comoving Coulomb 
self-field stress has the singular form $\delta T^{\mu\nu}_{Coul}\propto 1/R^4$ in the quark
rest frame. \\
In both pQCD and AdS/CFT cases we subtract this vacuum self-field stress as in Refs.\ 
\cite{Gubser:2007ga,Gubser:2007ni}. In other words, the zero-temperature contribution to the 
in-medium stress tensor is always subtracted. While in AdS/CFT the form of the Coulomb tensor 
is known exactly \cite{Friess:2006fk}, in pQCD this contribution can only be calculated 
perturbatively. The leading-order expression for the chromo-fields produced by the source in 
pQCD, in the limit where the dielectric functions are set to unity, displays the same 
Lienard--Wiechert behavior as in AdS/CFT.\\
The far-zone ``Mach'' part of the stress can be expressed in terms of the  local temperature 
$T(t,\vec{x})$ and fluid flow velocity fields $u^\mu(t,\vec{x})$ through the first-order 
Navier--Stokes stress form, given by Eq.\ (\ref{EqMachZone}) 
\begin{eqnarray}
\hspace*{-0.8cm}\delta T_{Mach}(x_1,x_\perp) &=& 
\frac{3}{4} K\left\{T^4\left[\frac{4}{3}u^\mu u^\nu-\frac{1}{3}g^{\mu\nu}
+ \frac{\eta}{sT} \partial^{( \mu}u^{\nu)} \right]- T^{\mu\nu}_0\right\} \nonumber \\ 
\hspace*{-0.8cm}&& \times \theta(1-3Kn)\,,
\label{hydrotensor}
\end{eqnarray}
The theta function in this equation defines the ``far zone'' that includes the Mach and diffusion 
linearized hydrodynamic sound waves. In the far zone, the equilibration rate is over three times 
the local stress gradient scale, and first-order Navier--Stokes dissipative hydrodynamics provides 
an adequate formulation of the evolution in that zone.\\
In AdS/CFT, the hydrodynamic description was shown to be valid down to distances of roughly 
$3/(\pi T)$ \cite{Noronha:2007xe} (see also \cite{Chesler:2007sv} for an analysis of the far 
zone).\\
Within a thermal Compton wavelength of the heavy quark the near neck zone is in general a 
non-equilibrium region strongly influenced by the coupling of the heavy quark's bare classical 
Coulomb field to the polarizable plasma.

\section[The pQCD Source Term]{The pQCD Source Term}

We use the nonlocal pQCD source term, $\mathcal{J}^\nu(x)$, derived in Ref.\ \cite{Neufeld:2008hs}, 
to drive the perfect fluid response of a pQCD fluid assuming $\eta=0$ in order to 
maximize any pQCD transport-induced azimuthal conical signature. Nonzero viscosity, of course, 
washes out some of the induced correlations as shown in Ref.\ \cite{Neufeld:2008dx}. \\
However, our main finding below is that even in this perfect $\eta=0$ hydrodynamic limit the 
pQCD induced correlations are too weak to generate conical correlations after 
freeze-out.\\
Thus, we consider only the $\eta/s=0$ limit of the full anomalous chromo-viscous equations 
derived in Refs.\ \cite{Asakawa:2006jn} and retain the anomalous diffusion-stress Neufeld source 
\cite{Neufeld:2008hs}. We can rewrite Eqs.\ (6.2) -- (6.11) of Ref.\ \cite{Asakawa:2006jn} in the 
more familiar covariant Joule heating\footnote{Joule heating is the process by which the passage 
of an electric current through a conductor releases heat.} form (for details of this analogy see 
appendix \ref{JouleHeating})
\begin{eqnarray}
\partial_\mu T^{\mu\nu}=\mathcal{J}^\nu= F^{\nu\alpha\,a}J_{\alpha}^a&=&
(F^{\nu\alpha\,a}\sigma_{\alpha\beta\gamma}*F^{\beta\gamma\,a})\,,
\label{eqjouleheating}
\end{eqnarray}
where $F^{\mu\nu\,a}(t,x)$ is the external Yang--Mills field tensor and 
\begin{eqnarray}
J^a(t,x)=\int \frac{d^4 k}{(2\pi)^4}  e^{ik^\mu x_\mu} J^a(k)
\end{eqnarray}
is the color current that is related via Ohm's law to $F^{\mu\nu\,a}(k)$ through the (diagonal in 
color) conductivity rank-three tensor $\sigma_{\mu\alpha\beta}(k)$. The $*$ denotes a 
convolution\footnote{$\sigma_{\alpha\beta\gamma}*F^{\beta\gamma\,a}=J^a_\alpha(t,x)=
\int d^4k/(2\pi)^4\,e^{ik^\mu u_\mu}\sigma_{\alpha\mu\nu}(k)F^{\mu\nu\,a}(k).$} 
over the nonlocal non-static conductive dynamical response of the polarizable plasma.\\
The covariant generalization of Neufeld's source is most easily understood through its Fourier 
decomposition, 
\begin{eqnarray}
J_{\nu}^a(k)=\sigma_{\nu\mu\alpha}(k) F^{\mu\alpha\,a}(k)\,,
\end{eqnarray}
with the color conductivity expressed as in Refs.\ 
\cite{Selikhov:1993ns,Selikhov:1994xn,Eskola:1992bd}
\begin{equation}
\sigma_{\mu\alpha\beta}(k)=i g^2\int d^4 p\frac{p_\mu p_\alpha\,\partial^p_\beta}{
p^\sigma k_\sigma+i \,p^\sigma u_\sigma/\tau^*}f_0(p)\,,
\label{conduct}
\end{equation}
where $f_0(p)=2\left(N_c^2-1\right) \,G(p)$ is the effective plasma equilibrium distribution with 
$G(p)=(2\pi^3)^{-1}\theta(p_0)\delta(p^2)/(e^{p_0/T}-1)$. Here, $u^\mu$ is the 4-velocity of the 
plasma as in Eq.\ (\ref{hydrotensor}).\\
For an isotropic plasma $u^{\beta}\sigma_{\mu\alpha\beta}(k)=-\sigma_{\mu\alpha}(k)$. In the 
long-wavelength limit, 
$u_{\beta}\sigma^{\mu\alpha\beta}(k\rightarrow 0) =-\tau^* m_D^2 \,g^{\mu\alpha} /3$, 
where $m_D^2=g^2 T^2$ is the Debye screening mass for a non-interacting plasma of massless $SU(3)$ 
gluons in thermal equilibrium.\\
The relaxation  or decoherence time $\tau^*$ in Eq.\ (\ref{conduct}) has the general form noted in 
Ref.\ \cite{Asakawa:2006jn}
\begin{equation}
\frac{1}{\tau^*}= \frac{1}{\tau_p} + \frac{1}{\tau_c} + \frac{1}{\tau_{an}}\,,
\end{equation}
with the collisional momentum-relaxation time \cite{Danielewicz:1984ww,Heinz:1985qe}
\begin{equation}
\tau_p\propto \left[\alpha_s^2 T\ln \left(\frac{1}{\alpha_s}\right)\right]^{-1}\,,
\end{equation}
the color-diffusion time defined in Ref.\ \cite{Selikhov:1993ns}
\begin{equation}
\tau_c=\left[\alpha_s N_c T\ln \left(\frac{1}{g}\right)\right]^{-1}\,,
\end{equation}
and the anomalous strong electric and magnetic field relaxation time derived in Eq.\ (6.42) of 
Ref.\ \cite{Asakawa:2006jn}
\begin{equation}
\tau_{an}\propto \left(m_D\sqrt{\frac{\eta|\vec{\nabla}\cdot \vec{U}|}{Ts}}\right)^{-1}\,.
\end{equation}
We note that this expression can be written in terms of the local Knudsen number $Kn=\Gamma_s/L$ 
used in Eq.\ (\ref{hydrotensor})
\begin{equation}
\tau_{an}\propto\frac{1}{gT} \frac{1}{\sqrt{Kn(t,x)}}\,,
\label{anom}
\end{equation}
with $L$ being a characteristic stress gradient scale. However, because of $\eta\propto\tau^* sT$, 
Eq.\ (\ref{anom}) is really an implicit equation for $\tau_{an}$. \\
Combining these relations and taking into account the uncertainty-principle constraint 
\cite{Danielewicz:1984ww} that bounds $\tau^*\stackrel{>}{\sim} 1/\left(3T\right)$ for an 
ultra-relativistic (conformal) plasma, we have
\begin{equation}
 \frac{1}{\tau^*} \propto T \left(a_1\; g^4 \ln g^{-1}
+a_2\; g^2\ln g^{-1} + a_{3}\; g \sqrt{Kn} \right) \stackrel{<}{\sim} 3T\,,
\label{rate}
\end{equation}
where $a_1\,,a_2\,,a_3$ are numerical factors. In the near zone close to the quark, $Kn$ becomes 
large, and $\tau_{an}$ can be the dominant contribution to the relaxation time $\tau^*$  in the 
presence of strong classical field gradients.\\
There is a subtle point in the application of Eq.\ (\ref{rate}) to our heavy quark jet problem. 
In order to neglect viscous dissipation in the pQCD response, the relaxation rate must be very 
large compared to the characteristic gradient scale. Hence, in the far zone at least the imaginary 
part of the conductivity denominator in Eq.\ (\ref{conduct}) must be large and dominant. However, 
in the neck region the field gradients become very large and the relevant wave numbers of the 
hydrodynamic response, $K\gg 3T$, exceed the uncertainty-limited equilibration rate. Since we 
only need to consider the conductivity in the asymptotic large $K$ limit in the near zone, 
it becomes possible to neglect the $\sim i 3T$ maximal relaxation rate in the energy denominator 
and to formally set  $1/\tau^* \rightarrow 0^+$ - {\it as if}  the coupling were parametrically 
small [which is assumed in Eqs.\ (53) -- (56) of Ref.\ \cite{Neufeld:2008hs}].\\
Only in this high-frequency, high-wave number limit, relevant for the neck zone physics, the 
color conductivity is computable as in Ref.\ \cite{Neufeld:2008hs}.\\
The neglect of dissipation in the neck zone maximizes the acceleration of the plasma partons, 
which can subsequently generate transverse collective plasma flow relative to the jet axis. We 
have to check numerically whether this maximum transfer of field energy-momentum from the field 
to the plasma is sufficiently anisotropic to generate a conical correlation of the associated 
hadron fragments.

\section[Freeze-out Procedures]{Freeze-out Procedures}

As mentioned previously, we consider here only the idealized static medium to maximize the plasma 
response signal. Distortion effects due to e.g.\ transverse expansion, while important for 
phenomenological comparisons to heavy-ion data, however obscure the fundamental differences 
between weakly and strongly-coupled dynamics that is our focus here. Given the large theoretical 
systematic uncertainty inherent in any phenomenological model of non-perturbative 
hadronization, we consider two simple limits for modeling the fluid decoupling and freeze-out. \\
In one often used limit, the fluid cells are
frozen-out via the CF prescription on an isochronous hypersurface. This scheme takes into account 
maximal thermal broadening effects. In the opposite limit, we assume an isochronous sudden breakup 
or shattering of fluid cells conserving only energy and momentum and avoiding hadronization 
altogether as described in detail in section \ref{Freezeout} and is shortly reviewed
below for convenience. The difference between the two schemes provides a 
measure of the systematic theoretical uncertainty associated with the unsolved problem of 
hadronization.\\
In the CF method, the conversion of the fluid into free particles is achieved instantaneously at a 
critical surface $d \Sigma_\mu$ \cite{Cooper:1974mv}. If we assume such a freeze-out scheme, the 
particle distributions and correlations can be obtained from the flow velocity field 
$u^{\mu}(t,\vec{x})$ and temperature profile $T(t,\vec{x})$. \\
For associated (massless) particles with 
$p^{\mu}=(p_{T},p_{T}\cos (\pi-\phi),p_{T}\sin (\pi-\phi),0)$ the momentum distribution 
at mid rapidity $y=0$ is [cf.\ Eq.\ (\ref{CFFormula})]
\begin{equation}
\frac{dN}{p_Tdp_Tdy d\phi}\Big
|_{y=0}=\int_{\Sigma}d\Sigma_{\mu}p^{\mu}\left[f_0(u^{\mu},p^{\mu},T)-f_{eq}\right]\,,
\label{cooperfryenew}
\end{equation}
where $f_0=\exp[-u^{\mu}p_{\mu}/T(t,x)]$ is a local Boltzmann equilibrium distribution. No viscous 
corrections to Eq.\ (\ref{cooperfryenew}) are included since we are working here in the 
perfect-fluid limit with $\eta=0$. We subtract the isotropic background yield via 
$f_{eq}\equiv f|_{u^{\mu}=0,T=T_0}$. Moreover, we follow Refs.\ 
\cite{CasalderreySolana:2004qm,Noronha:2008un,Betz:2008js} 
and perform an isochronous freeze-out where $d\Sigma^\mu= d^3 \,{\vec x} \left(1,0,0,0\right)$.\\
We remark that the absence of well-defined quasi-particle states in AdS/CFT plasmas at large 
t'Hooft coupling indicates that CF can, at best give a qualitative idea of the observable 
hadron-level angular correlations \cite{Noronha:2007xe,Gyulassy:2008fa,Noronha:2008un}. Moreover, 
even in the pQCD quasiparticle limit, CF freeze-out remains a strong model assumption. In the 
pQCD case, in the associated $p_T$-range of interest a coalescence/recombination hadronization 
scenario \cite{Fries:2003vb,Fries:2004hd} may be more appropriate. However, we expect similar CF 
thermal broadening effects if coalescence hadronization is assumed and full three-momentum 
conservation is taken into account.\\
As an alternate freeze-out scheme we consider a calorimetric-like observable given by the 
momentum density weighted polar angle distribution relative to the jet axis:
\begin{eqnarray}
\hspace{-0.5cm}\frac{d S}{d\cos\theta} &=& \sum_{cells} |\vec{\mathcal{P}}_c|
\delta\left(\cos\theta- \cos\theta_c\right)\nonumber \\
&=& \int d^3 \vec{x}\,\, |\vec{M}(t,\vec{x})|
\left.\delta\left[\cos\theta- \frac{M_x(t,\vec{x})}
{|\vec{M}(t,\vec{x})|}\right]\right\vert_{t_f}\,.
\label{bulkeqAdS}
\end{eqnarray}
\begin{figure}[t]
\centering
  \includegraphics[scale = 0.50]{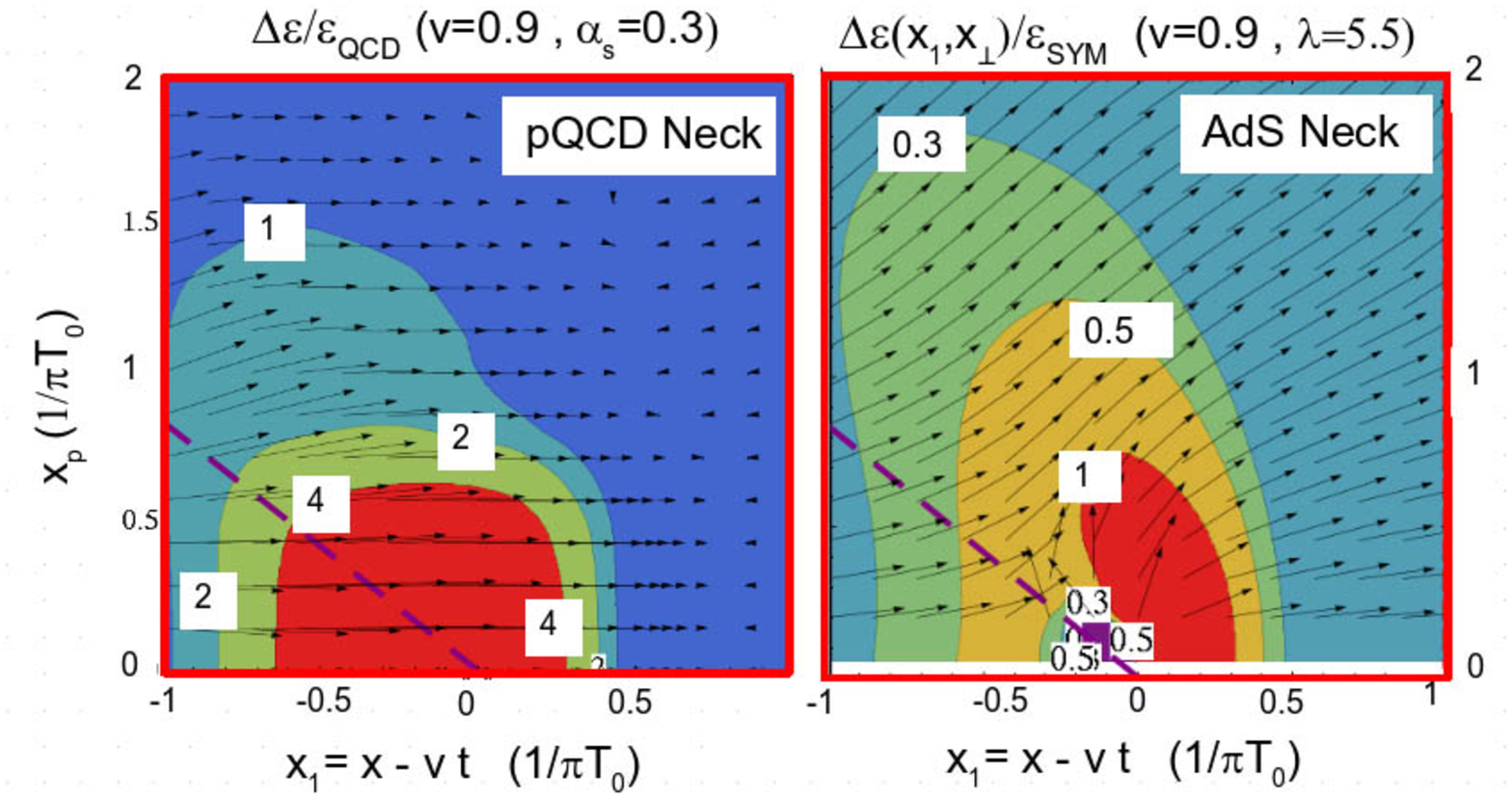}
  \caption[A magnified view of the near ``neck'' zone shows the relative local-energy density 
  perturbation $\Delta \varepsilon/\varepsilon_0$ and fluid-flow directions induced by a 
  heavy supersonic quark jet moving with $v=0.9$.]
  {A magnified view of the near ``neck'' zone shows the relative local-energy density perturbation 
  $\Delta \varepsilon/\varepsilon_0$ and fluid-flow directions induced by a heavy supersonic quark 
  jet moving with $v=0.9$. As in Fig.\ \ref{Neufeld_farFig.1}, the pQCD contours were computed 
  using ($3+1$)-dimensional hydrodynamics \cite{Rischke:1995ir}, including the source term from 
  Ref.\ \cite{Neufeld:2008hs} (left panel). The AdS/CFT neck zone 
  \cite{Noronha:2007xe,Gyulassy:2008fa,Noronha:2008un} (right panel) uses numerical tables from 
  \cite{Gubser:2007ga}. The purple dashed line indicates the ideal far-zone point-source shock 
  angle. The heavy quark is at the origin of these comoving coordinates. The arrows indicate both 
  direction and relative magnitude of the fluid flow velocity. The numbers in the plot label the 
  contours of constant $\Delta \varepsilon/\varepsilon_0$. Note that 
  $\Delta \varepsilon/\varepsilon_0$ is larger in pQCD but that the transverse flow generated 
  near the quark is much stronger in the AdS/CFT model \cite{Betz:2008wy}.}
  \label{pQCDzones}
\end{figure}
\hspace*{-3mm}
This quantity differs from CF mainly by the neglect of the thermal smearing at the freeze-out time, 
and thus it maximally amplifies the angular anisotropy of the associated hadrons. The very strong 
assumption in this decoupling scheme is that hadrons from each frozen-out cell emerge parallel 
to the total momentum of the cell $\mathcal{P}^i_c=d^3\vec{x}\,\, T^{0i}(t_f,\vec{x})$. Here 
$\theta=\pi-\theta_{trigger}$ is the polar angle with respect to the away-side heavy quark jet. \\
Many other similar purely hydrodynamic measures of bulk flow are possible \cite{Stoecker:2004qu}, 
e.g.\ weighted by entropy instead of momentum density. However, we found no qualitative 
differences when the weight function is changed. We used a narrow Gaussian approximation to the 
Dirac delta in Eq.\ (\ref{bulkeqAdS})  with a $\Delta\cos\theta=0.05$ width and checked 
that the results did  not change significantly if the width was varied by $50\%$.\\
Our intention is not to decide which hadronization scheme is preferred but to apply these two 
commonly used measures to help quantify the observable differences between two 
very different approaches to jet-plasma interactions (pQCD x AdS/CFT). \\
The CF freeze-out employed here and in Refs.\ 
\cite{CasalderreySolana:2004qm,Noronha:2008un,Betz:2008js} is
especially questionable in the non-equilibrium neck region but provides a rough estimate of 
intrinsic thermal smearing about the local hydrodynamic flow.\\
In Fig.\ \ref{pQCDzones} we show the relative local energy disturbance and flow profile in the 
neck region created by a $v=0.9$ jet in both pQCD and AdS/CFT. The relative transverse flow in 
the neck zone in AdS/CFT is significantly larger than in pQCD and as we show below this is 
reflected in the final angular correlations from that region in both hadronization schemes.

\section[Freeze-out Results in pQCD]{Freeze-out Results in pQCD}

\begin{figure}[t]
\centering
  \includegraphics[scale = 0.75]{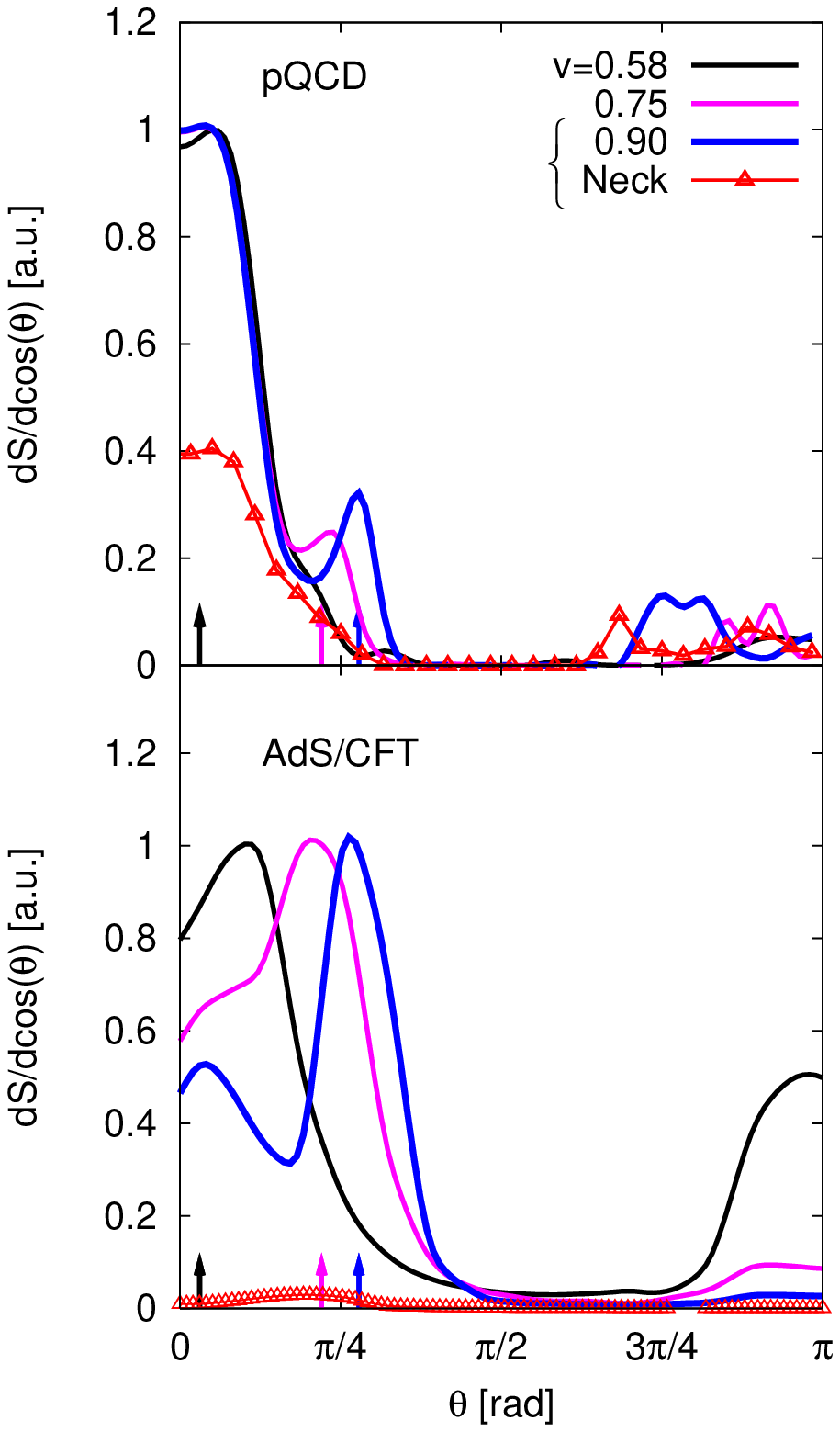}
  \caption[The (normalized) momentum-weighted bulk-flow angular distribution as a function of 
  polar angle with respect to the away-side jet comparing pQCD and AdS/CFT calculations.]
  {The (normalized) momentum-weighted bulk-flow angular distribution as a function of polar angle 
  with respect to the away-side jet is shown for $v=0.58$ (black), $v=0.75$ (magenta), and 
  $v=0.90$ (blue) comparing pQCD anomalous chromo-hydrodynamics to the AdS/CFT string-drag model 
  \cite{Herzog:2006gh,Gubser:2007ga}. The red line with triangles represents the neck contribution 
  for a jet with $v=0.9$ and the arrows indicate the location of the ideal Mach-cone angle given by 
  $\cos\theta_M=c_s/v$, where $c_s=1/\sqrt{3}$ \cite{Betz:2008wy}. }
  \label{Pointingall}
\end{figure}

The initial away-side heavy quark jet is assumed to start at $t=0$ and $x_1 =-4.5$ fm. The 
freeze-out is done when the heavy quark reaches the origin of the coordinates at time 
$t_f=4.5/v$~fm. This provides a rough description of the case in which a uniformly moving heavy 
quark punches through the medium after passing through $4.5$ fm of plasma.\\
The numerical output of the SHASTA code is the temperature and fluid flow velocity fields 
$T(t,\vec{x})$ and $\vec{u}(t,\vec{x})$. The hydrodynamic equations were solved in the presence 
of the source term $\mathcal{J}^{\mu}(t,\vec{x})$ computed analytically in Ref.\ 
\cite{Neufeld:2008hs} in the limit where the dielectric functions that describe the medium's 
response to the color fields created by the heavy quark were set to unity. The effects from 
medium screening on $\mathcal{J}^{\mu}$ were studied in detail there. In our numerical calculations 
we used $x_{p\,max}=1/m_D$ as an infrared cutoff while the minimum lattice spacing naturally 
provided an ultraviolet cutoff. The background temperature was set to $T_0=0.2$ GeV and we assumed 
$\alpha_s=g^2/\left(4\pi\right)=1/\pi$ in our calculations involving the pQCD source.\\
The results for the bulk flow according to Eq.\ (\ref{bulkeqAdS}) in pQCD are shown in the upper 
panel in Fig.\ \ref{Pointingall}. The curves are normalized in such a way that the largest 
contributions are set to unity. For all velocities studied here, the pQCD bulk energy-flow 
distribution has a large forward-moving component in the direction of the jet. In the far 
zone, this forward-moving energy flow corresponds to the diffusion wake studied in Refs.\ 
\cite{CasalderreySolana:2004qm,Betz:2008ka}. The red curve with triangles in the upper panel of 
Fig.\ \ref{Pointingall} corresponds to the yield solely from the neck region for $v=0.9$. The 
very small dip at small $\theta=0$ is mostly due to the weak pQCD neck zone but most of the 
momentum flow from both neck and diffusion zones is directed around the jet axis. \\
The relatively small transverse-energy flow in the neck region is evident on the left panel of 
Fig.\ \ref{pQCDzones} in contrast to the much larger transverse flow predicted via AdS in 
that near zone. The Mach cone emphasized in Refs.\ \cite{Neufeld:2008fi,Neufeld:2008dx} is also 
clearly seen but its amplitude relative to the mostly forward diffusion zone plus neck contribution 
is much smaller than in the AdS/CFT case. The weak Mach peak roughly follows Mach's law as $v$
approaches $c_s$.\\
In Fig.\ \ref{CFplot}, our CF freeze-out results for the associated away-side azimuthal 
distribution for $v=0.58,0.75,0.9$ at mid-rapidity and $p_T=5\pi T_0\sim 3.14$ GeV light 
hadrons are shown. The pQCD case, computed from the hydrodynamical evolution, is shown in 
the upper panel. We again define the angular function
\begin{equation}
 CF(\phi)=\frac{1}{N_{max}}\left.\left[\frac{dN(\phi)}{p_Tdp_Tdy d\phi}\right]\right\vert_{y=0}\,,
\end{equation}
where $N_{max}$ is a constant used to normalize the plots (this function is not positive-definite). 
The pQCD angular distribution shows only a sharp peak at $\phi=\pi$ for all velocities. \\
The red curve with triangles denotes the contribution from the pQCD neck region for $v=0.9$. 
The different peaks found in the bulk-flow analysis of the pQCD data shown in the upper panel in 
Fig. \ref{Pointingall} do not survive CF freeze-out. It was checked that no other structures 
appear if we either double $p_T$ to 5 GeV or increase $\alpha_s$ to 0.5 (see below). We conclude 
that the strong forward-moving diffusion zone as well as the mostly forward bow shock neck zone 
dominate the away-side peak and that the thermal broadened Mach correlations are too weak in pQCD 
to contribute to the final angular correlations.

\section[Freeze-out Results in AdS/CFT]{Freeze-out Results in AdS/CFT}
\begin{figure}[t]
\centering
  \includegraphics[scale = 0.75]{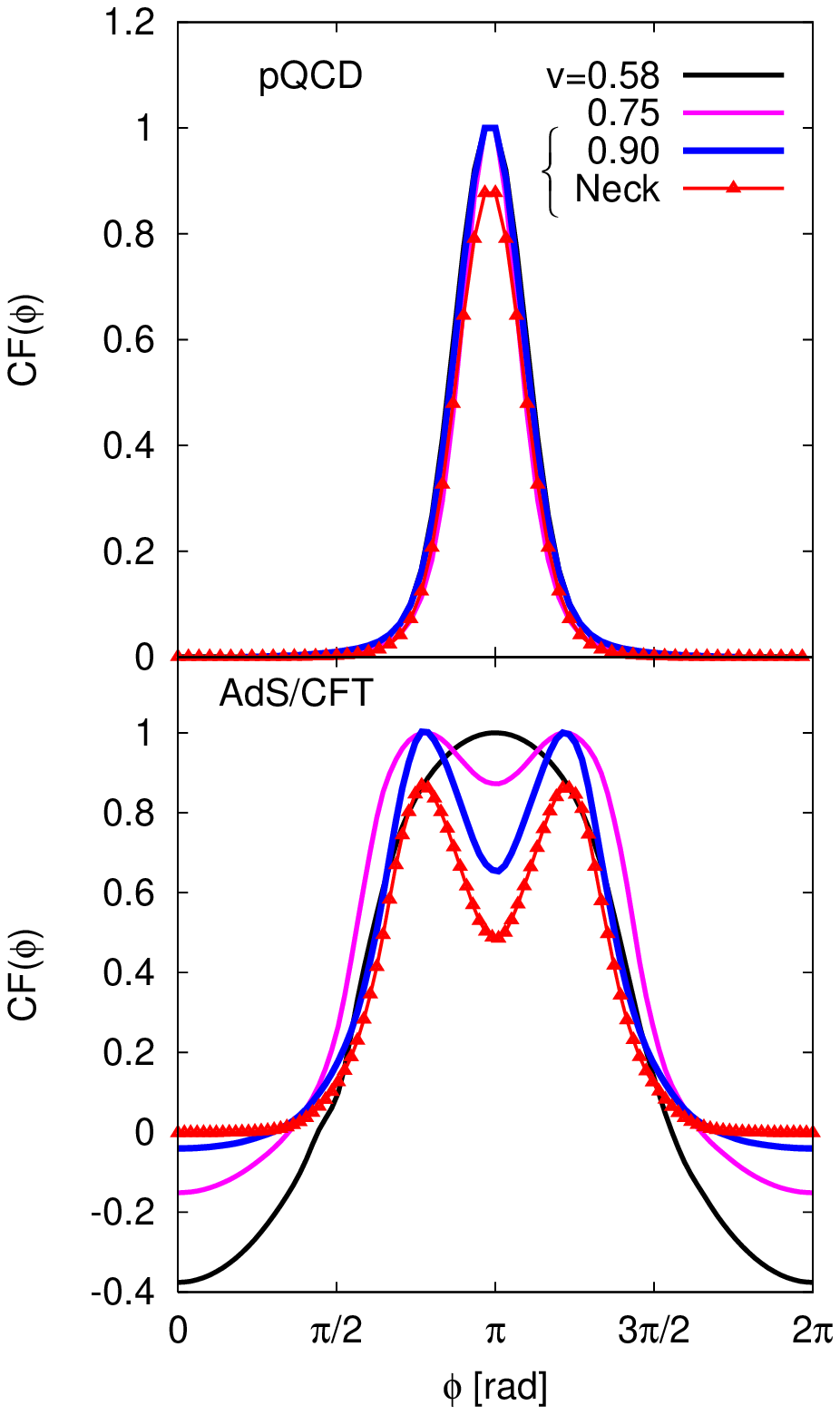}
  \caption[Normalized (and background subtracted) azimuthal away-side jet associated 
  correlation after a Cooper--Frye freeze-out for pQCD and AdS/CFT calculations.]
  {Normalized (and background subtracted) azimuthal away-side jet associated correlation after a 
  Cooper--Frye freeze-out $CF(\phi)$ for pQCD (top) and AdS/CFT from 
  \cite{Gyulassy:2008fa,Noronha:2008un} (bottom). Here $CF(\phi)$ is evaluated at 
  $p_T=5\pi T_0\sim 3.14$ GeV and  $y=0$. The black line is for $v=0.58$, the magenta line for 
  $v=0.75$, and for the blue line $v=0.9$. The red line with triangles represents the neck 
  contribution for a jet with $v=0.9$ \cite{Betz:2008wy}.}
  \label{CFplot}
\end{figure}

We used the same setup employed in Ref.\ \cite{Noronha:2007xe,Gyulassy:2008fa,Noronha:2008un} to 
perform the CF freeze-out of the $\mathcal{N}=4$ SYM AdS/CFT data 
computed by Gubser, Pufu, and Yarom in Ref.\ \cite{Gubser:2007ga}. They calculated the 
energy-momentum disturbances caused by the heavy quark, which in this steady-state solution was 
created at $t\to -\infty$ and has been moving through the infinitely extended $\mathcal{N}=4$ 
SYM static background plasma since then. The freeze-out is computed 
when the heavy quark reaches the origin of the coordinates. The mass of the heavy quark $M$ in the 
AdS/CFT calculations is such that $M/T_0\gg \sqrt{\lambda}$, which allows us to neglect 
the fluctuations of the string \cite{Herzog:2006gh,CasalderreySolana:2007qw}. \\
At $N_c=3$ the simplifications due to the supergravity approximation are not strictly valid but it 
is of interest to extrapolate the numerical solutions to study its phenomenological applications. 
We choose the plasma volume to be the forward light-cone that begins at $x_1=-4.5$ fm and has a 
transverse size of $x_p< 4.5$ fm at $T_0=0.2$ GeV (our background-subtracted results do not 
change when larger volumes were used). This choice implies that we assumed the same background 
temperature for both pQCD and AdS/CFT. \\
The mapping between the physical quantities in $\mathcal{N}=4$ SYM 
and QCD is a highly non-trivial open problem (see, for instance, the discussion in Ref.\ 
\cite{Gubser:2006qh}). We therefore again use CF and bulk-momentum flow as two extreme limits to 
gauge possible systematic uncertainties.\\
The (normalized) bulk-momentum flow associated with the AdS/CFT data, computed using Eq.\ 
(\ref{bulkeqAdS}), is shown in the lower panel of Fig.\ \ref{Pointingall}. It demonstrates that in 
AdS/CFT there are more cells pointing into a direction close to the Mach-cone angle than in the forward 
direction (diffusion zone) when $v=0.9$ and $v=0.75$, unlike in the pQCD case displayed in the 
upper panel. However, when $v=0.58$, the finite angle from the Mach cone is overwhelmed by the 
strong bow shock formed in front of the quark, which itself leads to a small conical dip not at the 
ideal Mach angle (black arrow). \\
The red line with triangles in the bottom panel of Fig.\ \ref{Pointingall} shows that the relative 
magnitude of the contribution from the neck region to the final bulk-flow result in AdS/CFT is 
much smaller than in pQCD. However, the small amplitude peak in the AdS/CFT neck curve is 
located at a much larger angle than the corresponding peak in the pQCD neck, as one would expect 
from the transverse flow shown Fig.\ \ref{pQCDzones}. Moreover, one can see that a peak in the 
direction of the trigger particle can be found for all the velocities studied here. This peak 
represents the backward flow that is always present due to vortex-like structures created by 
the jet as discussed in detail in Ref.\ \cite{Betz:2007kg} and chapter 
\ref{PolarizationProbesofVorticity}.\\
Our results for the CF freeze-out of the AdS/CFT solution for $v=0.58,0.75,0.9$ at mid-rapidity 
and $p_T=5\pi T_0\sim 3.14$ GeV are shown in the lower panel in Fig.\ \ref{CFplot}. A double-peak 
structure can be seen for $v=0.9$ and $v=0.75$. However, the peaks in the AdS/CFT correlation 
functions do not obey Mach's law. The reason is that these correlations come from the neck region 
with a strong transversal non-Mach flow \cite{Noronha:2007xe,Gyulassy:2008fa,Noronha:2008un}. 
This is explicitly shown by the red curve with triangles that represents the neck contribution for 
a jet with $v=0.9$ as in Fig.\ \ref{Pointingall}. For $v=0.58$, the resulting flow is not strong 
enough to lead to non-trivial angular correlations. The negative yield present in the CF curves 
for $v=0.58$ and $v=0.75$ results from the presence of the vortices discussed above and in chapter 
\ref{PolarizationProbesofVorticity}.\\
In general \cite{Noronha:2007xe,Gyulassy:2008fa,Noronha:2008un}, the weak sound waves produced by 
a jet do not lead to a cone-like signal independently of the detailed flow and interference 
patterns because thermal smearing washes out the signal. Formally, if linearized hydrodynamics 
applies and in the low-momentum limit ($\vec{u}\cdot \vec{p}<<T$), the associated hadron away-side 
distribution is only a very broad peak about $\phi=\pi$ regardless of the detailed combination 
of Mach wakes, diffusion wakes, or vortex circulation
\cite{CasalderreySolana:2004qm,Noronha:2007xe,Gyulassy:2008fa,Noronha:2008un,Betz:2007kg}. This 
result involving CF freeze-out can only be circumvented either in regions with high flow 
velocities and large gradients as in the neck zone 
\cite{Noronha:2007xe,Gyulassy:2008fa,Noronha:2008un}, or by increasing $p_T$ to unrealistically 
high values \cite{CasalderreySolana:2004qm,Betz:2008js}.\\
One of the main differences between the two freeze-out procedures employed above (in both AdS/CFT 
and pQCD) concerns the relative magnitude of the contribution from the neck region to the final 
angular correlations: the neck region is much more important in CF than in the bulk-flow measure 
computed via Eq.\ (\ref{bulkeqAdS}). This is due to the exponential factor in CF, which largely 
amplifies the contribution from the small region close to the jet where the disturbances caused by 
the heavy quark become relevant.
\begin{figure}[t]
\centering
  \includegraphics[scale = 0.7]{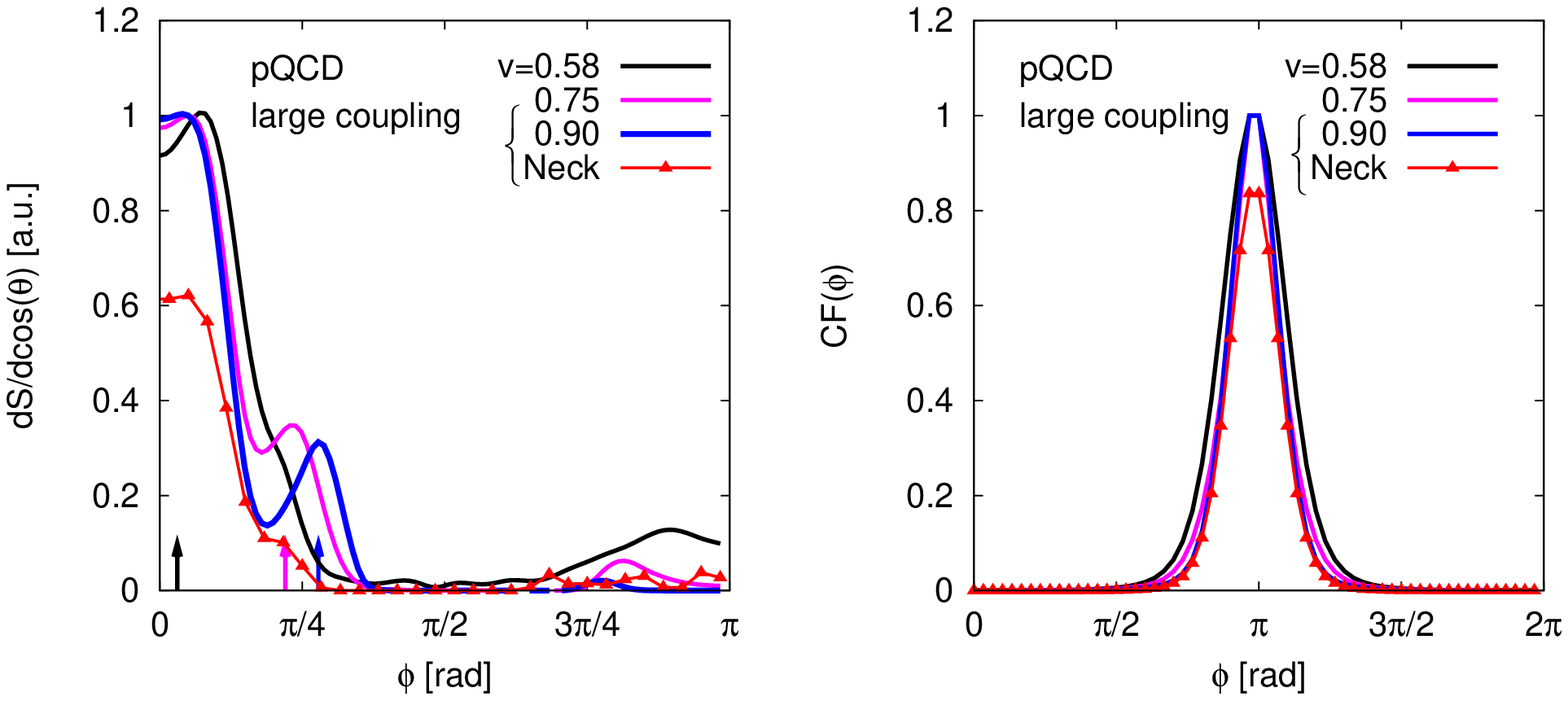}
  \caption[The (normalized) momentum weighted bulk-flow angular distribution and the normalized 
  (and background-subtracted) azimuthal away-side jet-associated correlation after a Cooper--Frye 
  freeze-out using the pQCD source term for $\alpha_s\approx 0.5$.]
  {The (normalized) momentum weighted bulk-flow angular distribution (left panel) and the 
  normalized (and background-subtracted) azimuthal away-side jet-associated correlation after a 
  CF freeze-out (right panel) using the pQCD source term for $\alpha_s\approx 0.5$ (in contrast to 
  the previous plots with $\alpha_s=1/\pi$) for $v=0.58$ (black), $v=0.75$ (magenta), and $v=0.9$ 
  (blue). The red line with triangles represents the neck contribution for a jet with $v=0.9$. 
  The $CF(\phi)$ is evaluated at $p_T=5\pi T_0\sim 3.14$~GeV.}
  \label{StrongerCoupling}
\end{figure}
\\We checked, as already mentioned previously, that the result is independent of $\alpha_s$ (see 
Fig.\ \ref{StrongerCoupling}) by setting the coupling constant to $g=2.5$, or equivalently $\alpha_s=0.5$. 
Since it is known \cite{Fleischer:1984dj} that $\alpha_s=1/2$ is the critical coupling for charged 
bosons (in a $\alpha_s/r$ potential), it can be considered as critical pQCD case. \\
Summarizing, we showed that the angular correlations obtained after an isochronous Cooper--Frye 
(CF) freeze-out of the wake induced by punch-through heavy quark jets (in a static medium) in the 
Neufeld {\it et al.} pQCD model of anomalous chromo-viscous hydrodynamics do not display a conical 
structure. This should be compared to the conical-like structures seen after CF freeze-out of the 
strongly-coupled AdS/CFT string-drag model which, however, are unrelated to Mach's law and result 
from a strong flow in the transverse direction that is absent for the pQCD source term (see 
Fig.\ \ref{pQCDzones}).\\
Unlike AdS/CFT, the conical flow from the associated non-equilibrium neck zone in pQCD (see the 
red region in the left panel of Fig.\ \ref{pQCDzones}) and the red curve in Fig.\ \ref{Pointingall})
is too weak to survive CF freeze-out. In both cases, the actual Mach wakes do not appear after 
standard CF freeze-out. Mach-like peaks are only observable in the sudden shattering  freeze-out 
scenario described in Eq.\ (\ref{bulkeqAdS}) in both pQCD and AdS/CFT, in which thermal 
broadening is entirely neglected.\\
The neck region (in both pQCD and AdS/CFT) gives the largest contribution to the total yield in CF 
freeze-out while its contribution in the other extreme case involving the bulk-flow hadronization 
is not as relevant. This indicates that the magnitude of the neck's contribution to the final 
angular correlations is still strongly model-dependent. Nevertheless, our results suggest that 
conical but non-Mach law correlations are much more likely to appear in AdS/CFT than in pQCD.\\
We therefore propose that the measurement of the jet velocity dependence of the associated 
away-side correlations with identified heavy quark triggers at RHIC and LHC might provide 
important constraints on possible pQCD versus AdS/CFT dynamical non-Abelian field-plasma 
(chromo-viscous) coupling models.\\
The isochronous hypersurface we used is needed in order to compare AdS/CFT to pQCD since AdS/CFT 
heavy quark solutions have only been computed so far in a static medium. For realistic simulations 
that can be compared to data, effects from the medium's longitudinal, transverse, and elliptic 
flow must be taken into account which will be discussed in the next chapter.

%
%
\chapter[Conical Correlations in an Expanding Medium]
{Conical Correlations \\in an Expanding Medium}
\label{ExpandingMedium}

\begin{figure}[t]
\centering
\begin{minipage}[c]{4.2cm}
\hspace*{-2.5cm}
  \includegraphics[scale = 0.65]{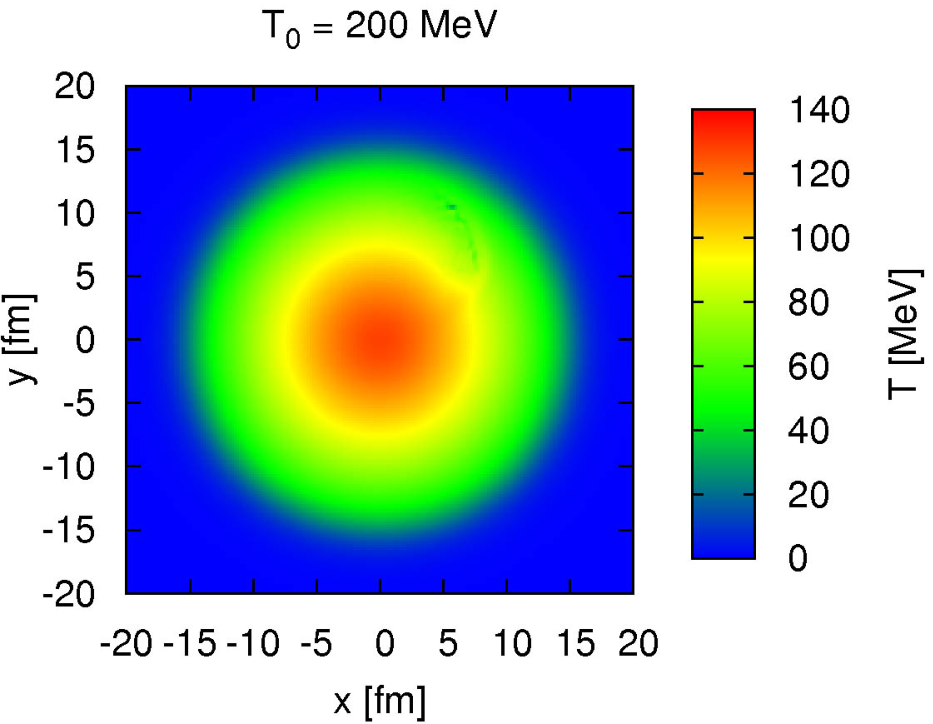}
\end{minipage}
\hspace*{-0.0cm}  
\begin{minipage}[c]{4.2cm} 
  \includegraphics[scale = 0.65]{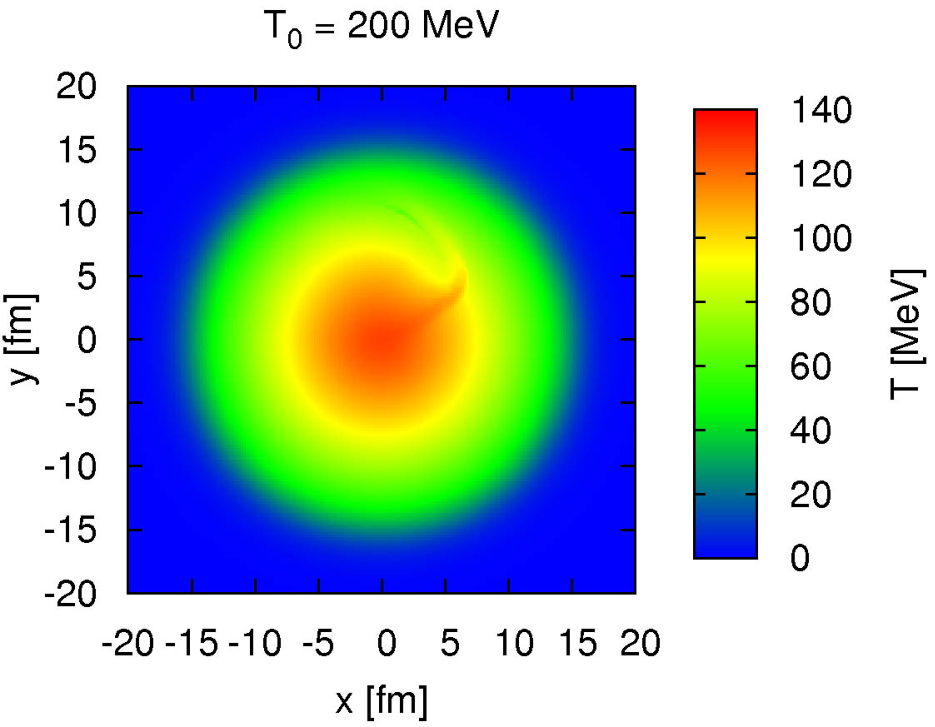}
\end{minipage}\\
\begin{minipage}[c]{4.2cm}
\hspace*{-2.5cm}
  \includegraphics[scale = 0.65]{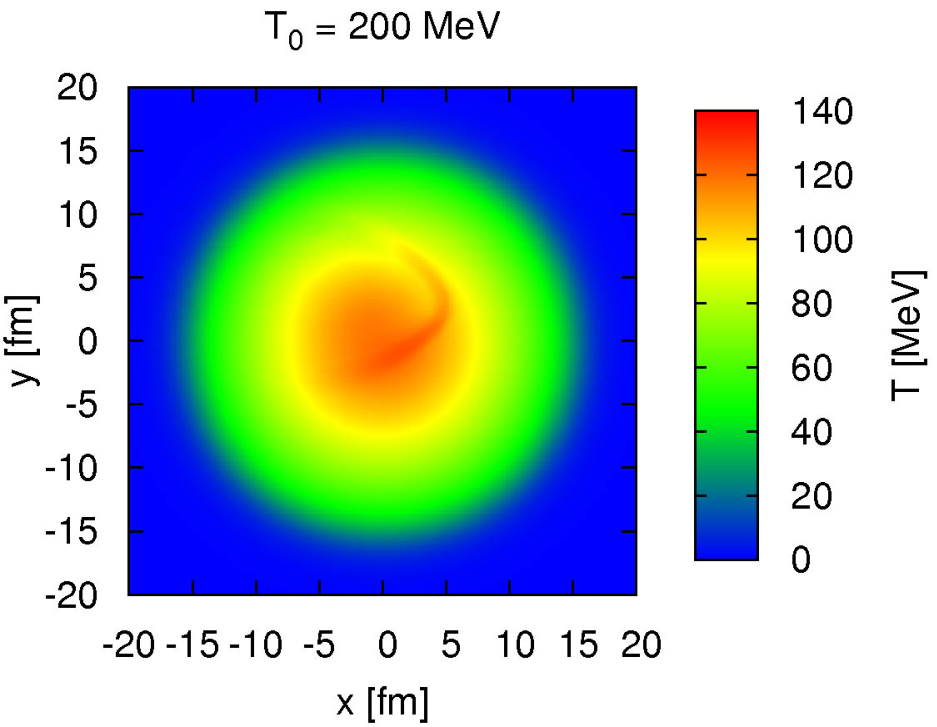}
\end{minipage}
\hspace*{-0.0cm}  
\begin{minipage}[c]{4.2cm} 
  \includegraphics[scale = 0.65]{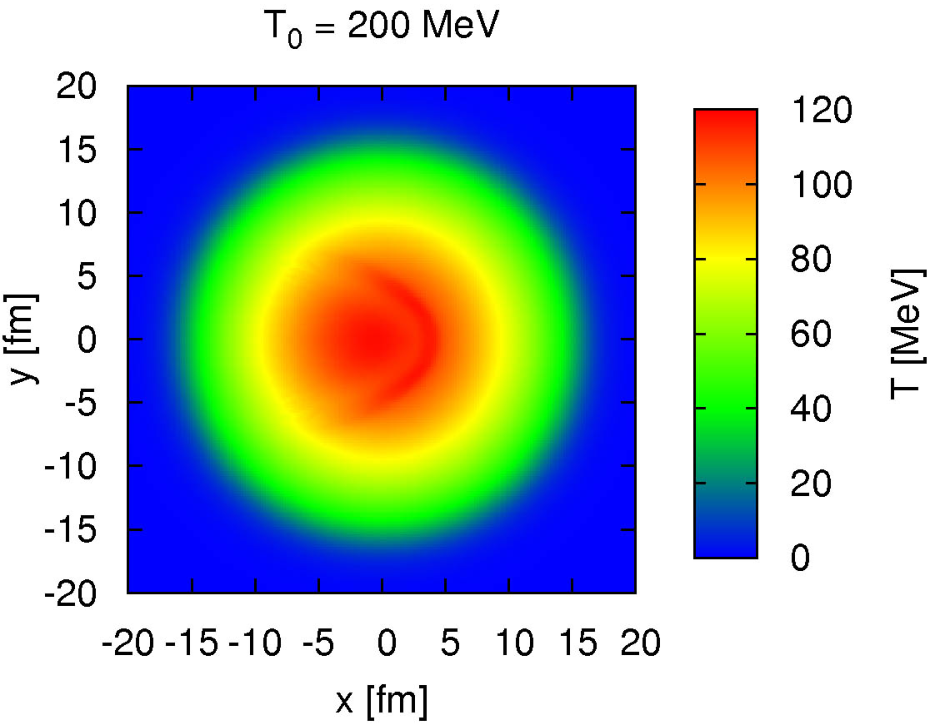}
\end{minipage}
  \caption[The temperature pattern of four different jets with varying origins at the moment of 
  freeze-out.]
  {The temperature pattern of four different jets (Jet 90 upper left panel, Jet 120 upper right 
  panel, Jet 150 lower left panel, and Jet 180 lower right panel) with varying origins is shown 
  at the moment of freeze-out, i.e., when the temperature of all cells has fallen below 
  $T_c=130$~MeV. The distortion of the conical structure is clearly visible which produces 
  different contributions to the away-side correlations displayed below.}
  \label{TempEnd}
\end{figure}

The investigations discussed in the previous chapters focused 
on the prescription of jet energy and momentum loss by reducing the
problem to the most simple case of a static background. This simplification
allowed to elaborate detailed reactions of the medium to the jet deposition, like the 
formation of the diffusion wake (see chapter \ref{DiffusionWake}). Moreover, 
a pQCD source term could be compared to a jet scenario
using AdS/CFT (see chapter \ref{pQCDvsAdSCFT}) since the latter prescription 
always considers an infinite, static, and homogeneous background.\\
However, the real experimental situation is different in two ways. First,
the medium created in a heavy-ion collision expands rapidly. Thus, even
assuming that the system equilibrates very quickly \cite{CERN,Arsene:2004fa,
Adcox:2004mh,Back:2004je,Adams:2005dq,Xu:2008dv}, i.e., behaves like a ``perfect fluid'' 
\cite{Kolb:2003dz,Romatschke:2007mq}, and a Mach cone is formed, the elliptic and radial 
expansion will interact with the flow profile created by the jet and cause 
a distortion of this Mach cone as predicted in Ref.\ \cite{Satarov:2005mv}.
Hence, Mach cones are sensitive to the background flow. This is a qualitative 
model-independent effect.\\
Second, it turned out \cite{Arsene:2004fa,Adcox:2004mh,Back:2004je,Adams:2005dq} that
the number of observed jets at RHIC is by an order of magnitude lower
than expected (also pointing into the direction that the matter created
is an opaque, high-density medium where jets quickly thermalize). This 
means that it is unlikely, but not impossible to have more than one jet in a single event
which by itself might lead to new effects \cite{Tomasik:2008dy}. However, the crucial point is that
the experimentally determined two- and three-particle correlations 
(see chapter \ref{ExperimentsQGP}) consider many different events and thus different
jet trajectories through the medium. Due to the interaction with radial flow, these 
jet paths may result in different contributions to the azimuthal correlations. Those patterns show that 
(the issue of background subtraction is discussed controversally) a two-peaked structure appears on 
the away-side which is interpreted in a way that a Mach cone is formed during 
the process of the collision, completely neglecting the multi-event situation. \\
Therefore, as already discussed by Chaudhuri \cite{Chaudhuri:2006qk,Chaudhuri:2007vc}, 
many different possible paths of a jet through the medium have to be taken into account. \\
As demonstrated in Refs.\ 
\cite{CasalderreySolana:2004qm,Betz:2008ka,Gyulassy:2008fa,Noronha:2008un,Betz:2008js}, 
two factors strongly go against a conical correlation even in a perfect fluid. 
The thermal broadening intrinsic to a Cooper--Frye freeze-out results to first order in $p_T/T$
(see the No-Go Theorem of section \ref{NoGoTheorem}) in a broad away-side peak. Also, the
diffusion wake formed (if the momentum deposition is larger than a certain threshold 
\cite{Betz:2008ka,Betz:2008js}) contributes to one away-side peak opposite to the trigger-jet 
direction, generally overwhelming any conical signal (see chapter \ref{DiffusionWake}).\\
Nevertheless, while the conclusive detection of Mach cones would provide evidence for the 
perfect-fluid behavior, the opposite is not the case though it implies that the double-peaked
structure cannot directly be compared to the EoS.\\
As already mentioned above, the expansion of a system created in a heavy-ion collision
certainly influences any kind of jet-deposition scenario. It was suggested that the 
diffusion-wake contribution can be reduced by radial flow \cite{Neufeld:2008eg}, and 
it was shown that longitudinal flow may cause a broadening of the away-side peaks 
\cite{Renk:2006mv}. Thus, for a realistic simulation, all those effects have to be considered.\\
In the following, we study an expanding medium with Glauber initial conditions
\cite{Kolb:2001qz} (see appendix \ref{AppGlauberModel}), corresponding
to a $Au$-nucleus with $r=6.4$~fm. We focus on radial flow only, i.e., the medium 
we investigate is elongated over the whole grid, forming a cylinder expanding in radial direction. 
Moreover, we only consider most central collisions with an impact parameter of $b=0$~fm (thus 
neglecting any elliptic flow contribution), and assume that the maximum temperature of the medium 
is $200$~MeV. \\
Like in previous chapters, we solve the ideal hydrodynamic equations 
applying an ideal gas EoS for massless gluons. We choose the following ansatz for the energy and 
momentum deposition of the jet [cf.\ Eq.\ (\ref{source})]
\begin{eqnarray}
\label{SourceExpandingMedium0}
J^\nu = \int\limits_{\tau_i}^{\tau_f}d\tau 
\frac{dM^\nu}{d\tau}\Big\vert_0\left[\frac{T(t,\vec{x})}{T_{\rm max}}\right]^3
\delta^{(4)} \left[ x^\mu - x^\mu_{\rm jet}(\tau) \right],
\end{eqnarray}
with the proper-time interval of the jet evolution $\tau_f - \tau_i$, 
the energy and momentum loss rate $dM^\nu/d\tau=(dE/d\tau, d\vec{M}/d\tau)$, 
the location of the jet $x_{\rm jet}$, and $\sigma=0.3$~fm. 
In non-covariant notation, this source term reads
\begin{eqnarray}
\label{SourceExpandingMedium}
J^\nu(t,\vec{x}) &=& \frac{1}{(\sqrt{2\pi}\,\sigma)^3}
\exp\left\{ -\frac{[\vec{x}-\vec{x}_{\rm jet}(t)]^2}{
2\sigma^2}\right\} \nonumber\\
& \times &\left(\frac{dE}{dt}\Big\vert_0,\frac{dM}{dt}\Big\vert_0,0,0\right)
\left[\frac{T(t,\vec{x})}{T_{\rm max}}\right]^3\,,
\end{eqnarray} 
where we use $dE/dt_0 = 1$~GeV/fm and $dM/dt_0 = 1/v dE/dt_0$.
Like in chapter \ref{DiffusionWake} and Refs.\ \cite{Betz:2008ka,Betz:2008js},
we study both jet energy and momentum loss as well as pure energy loss 
in the context of an expanding medium to investigate the impact
of the (radial) expansion and thus the significance of linearized hydrodynamical approaches
as used in Refs.\ \cite{CasalderreySolana:2004qm,Neufeld:2008fi}. \\
Since we are using a dynamical background, the source term, unlike in Eqs.\ 
(\ref{source}) and (\ref{sourcenoncovariant}), scales with the temperature, based on the 
assumption that the source is proportional to the density of colored particles. Additionally, a 
temperature cut is applied to ensure that no deposition takes place outside the medium, i.e.,
below a temperature of $T_c=130$~MeV. In Ref.\ \cite{Betz:2008ka} it was shown that the 
azimuthal correlations obtained after freeze-out do not change significantly when jets decelerate 
during their propagation according to the Bethe--Bloch formalism, creating a Bragg peak. Given that
we are interested in the modifications of such correlations due to an expanding background, we may
therefore simplify to the study of jets propagating with a constant velocity.\\
However, each parton moving through a medium will eventually be  
thermalized after the deposition of all its initial energy. Since jets are always 
created back-to-back, this energy is equal to the one of the trigger jet. 
\begin{figure}[t]
\centering
\vspace*{-2ex}
  \includegraphics[scale = 0.6]{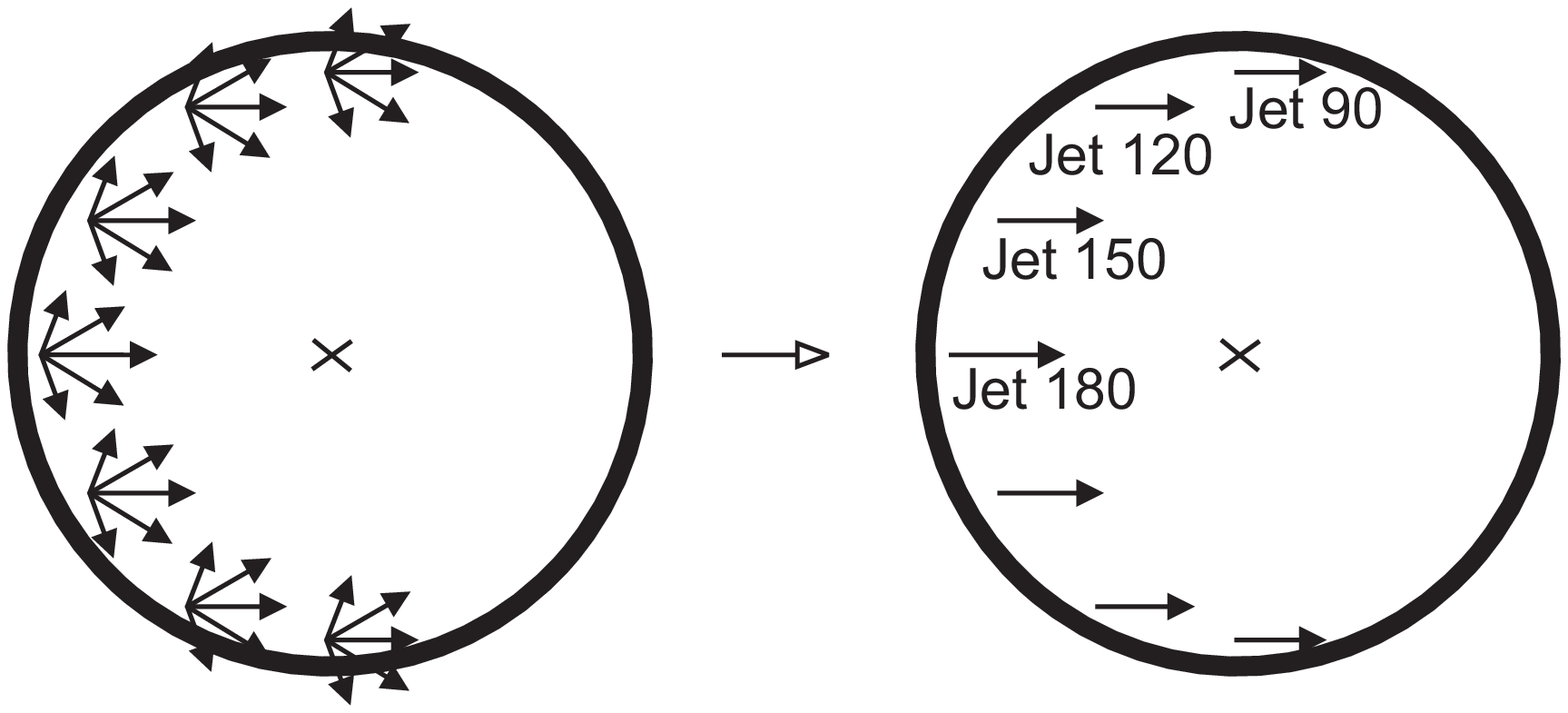}
  \caption[Schematic representation of different jet paths assuming that surface emission is the 
  dominant effect in heavy-ion collisions.]
  {Schematic representation of different jet paths assuming that surface emission is the dominant 
  effect in heavy-ion collisions. The right panel shows the reduction of paths due to reasons of 
  symmetry for a medium with vanishing impact parameter, $b=0$~fm. Those jet trajectories will be 
  studied below.}
  \label{JetPaths}
\end{figure}
\\
Below, we consider a $5$~GeV and a $10$~GeV trigger parton which corresponds 
to trigger-$p_T$'s of $p_T^{\rm trig}=3.5$ and $7.5$~GeV assuming that a 
fragmenting jet creates particles with $\sim 70$\% of its energy, allowing for 
an easier comparison to experimental data (see e.g.\ Fig.\ \ref{2pc_PHENIX}).\\
Experiment can only trigger on the jet direction, but not on the origin of 
the jets formed. Thus, one has to consider different starting points for the jet 
(see Fig.\ \ref{JetPaths}). Due to reasons of symmetry (cf.\ Fig.\ 
\ref{JetPaths}), the number of paths that need be studied in most central 
collisions reduces drastically. The jet itself is always taken to propagate along 
a certain direction, here the positive $x$-axis, but its origin is varied 
according to
\begin{eqnarray}
\label{paths}
x &=& r\cos\phi\,, \hspace*{1.5cm} y = r\sin\phi\,,
\end{eqnarray}
where $r=5$~fm is chosen which is close to the surface of the medium. 
We assume that surface emission is the dominant effect of jet events in heavy-ion 
collisions, i.e., we suppose that jets are mainly created close to the surface of the 
the expanding medium (see Fig.\ \ref{JetPaths}). Here, we 
consider $\phi=90$, $120$, $150,$ $180$, $210$, $240$, $270$~degrees. According to azimuthal 
symmetry, the results for $\phi = 210, 240, 270$~degrees are obtained by 
reflecting the corresponding data for $\phi = 90, 120, 150$~degrees. \\
As can be seen from Fig.\ \ref{FigEloss}, the energy and thus the momentum deposition 
is not constant, but varies because of the temperature dependence in Eq.\ 
(\ref{SourceExpandingMedium0}) with the jet path. For the $5$~GeV jet, the cut for 
depositing the total amount of energy is clearly visible for the most central 
jets (see e.g.\ solid red line in Fig.\ \ref{FigEloss}). 
Of course, it is a strong model assumption that $dE/dt\vert_0$ and 
$dM/dt\vert_0$ are the same for a $5$ and a $10$~GeV jet.\\
There is a qualitative difference between a $5$~GeV and a $10$~GeV jet.
The $5$~GeV jet will stop around the middle of the medium. Thus, it will mainly
move against the background flow. A $10$~GeV jet, on the contrary, will cross 
the middle of the medium and be thermalized before it reaches the opposite 
surface. Therefore, it also propagates a larger distance parallel to the flow. 
\begin{figure}[t]
\centering
\begin{minipage}[c]{4.2cm}
\hspace*{-5.0cm}
  \includegraphics[scale = 0.6]{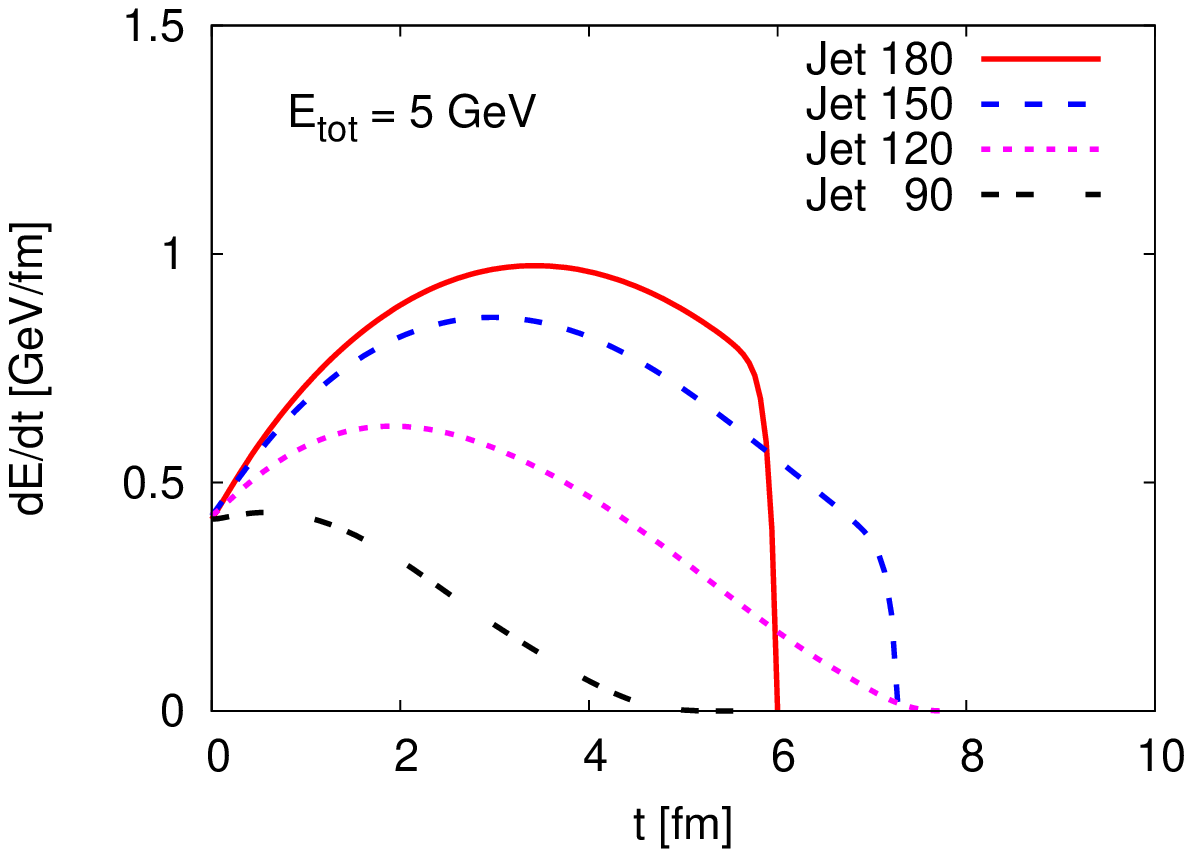}
\end{minipage}
\hspace*{-1.5cm}  
\begin{minipage}[c]{4.2cm} 
  \includegraphics[scale = 0.6]{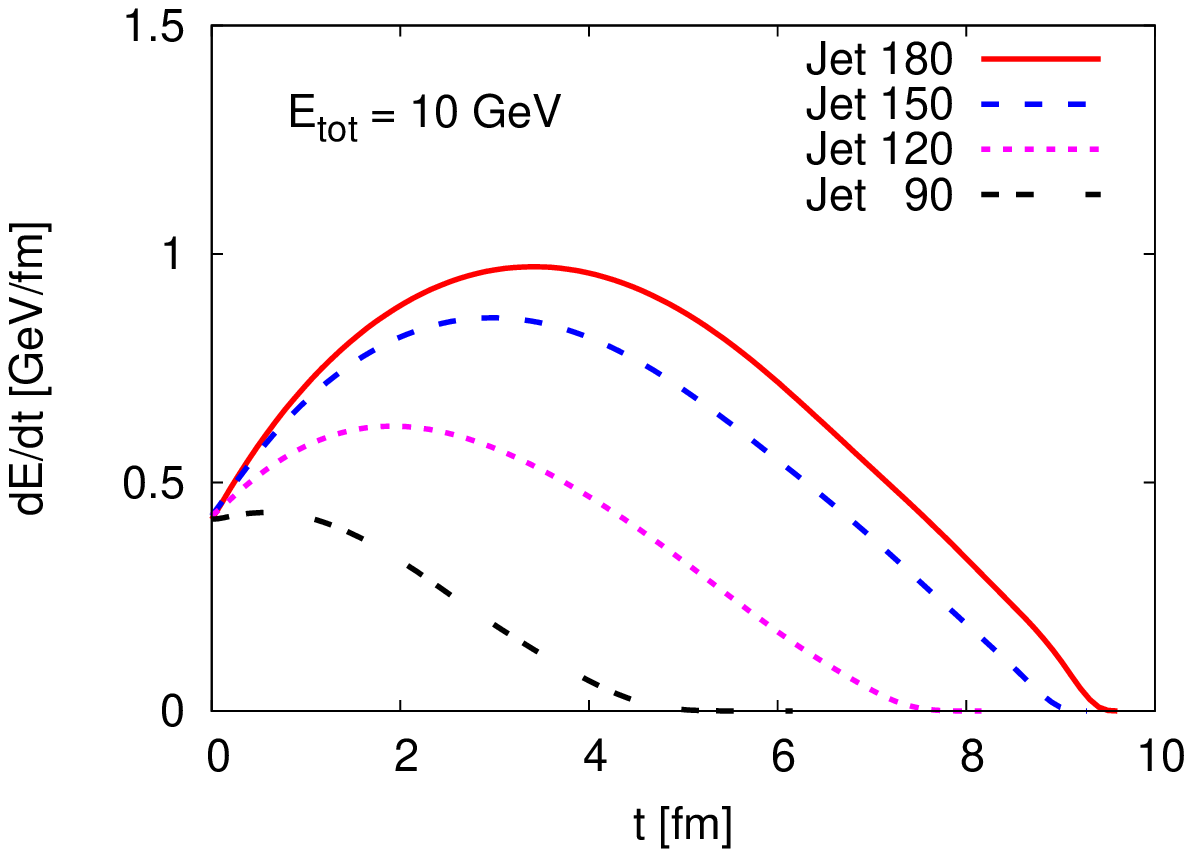}
\end{minipage}
  \caption[Jet energy deposition as a function of time for jets depositing $5$~GeV and
  $10$~GeV along different trajectories.]
  {Jet energy deposition as a function of time for jets depositing $5$~GeV (left panel) and
  $10$~GeV (right panel) along the different trajectories introduced in Fig.\ \ref{JetPaths}. For
  the $5$~GeV jet, the cut in the total amount of deposited energy is clearly visible for the most
  central jets.}
  \label{FigEloss}
\end{figure}
\\The temperature pattern of four different jets (Jet 90, Jet 120, Jet 150, and Jet 180 according to
Fig.\ \ref{JetPaths}) at the moment of freeze-out, i.e., when the temperature of all cells has fallen
below $T_c=130$~MeV, are shown in Fig.\ \ref{TempEnd}. 
The background flow leads to a distortion of the conical structure, resulting in different
contributions to the away-side correlations as will be discussed below in detail.
\\A crucial aspect of hydrodynamical applications to 
heavy-ion collisions is the conversion of the fluid into particles, the 
freeze-out (see e.g.\ section \ref{Freezeout}). While an isochronous freeze-out, which 
means that the conversion appears at a certain time $t_f$, can be an 
adequate prescription for a static medium, this assumption might not be 
reasonable for an expanding medium. Nevertheless, it should be mentioned that 
the ``blast wave'' model \cite{Schnedermann:1993ws} which is based on the 
isochronous freeze-out reproduces experimental results quite well.\\
In appendix \ref{IsochronIsothermFreezeout} we show that the space-time distribution of temperature 
is severely deformed by the jet. This can be seen by plotting the isothermal hypersurface,
i.e., the space-time profile of the fluid cells falling below a certain critical temperature (that 
was chosen to be $T_c=130$~MeV in the present study, see Fig.\ \ref{IsoHypersurface}). Moreover, 
(cf.\ Figs.\ \ref{CFIsochronIsotherm1} and \ref{CFIsochronIsotherm2}), we demonstrate that the 
particle distributions obtained for a jet passing through the middle of the medium (Jet 180 in 
Fig.\ \ref{JetPaths}) obtained via an isochronous and an isothermal CF freeze-out are very 
similar.\\
This suggests that an isochronous prescription, though a strong model assumption, does 
not severely alter the results as compared to an isothermal freeze-out. Such an isochronous CF 
conversion is used below, performed when all cells are below the critical temperature of 
$T_c=130$~MeV, for both the jet+medium and, independently, the background simulations.

\section[Two-Particle Correlations in an Expanding Medium]
{Two-Particle Correlations in an Expanding Medium}
\label{2pcExpandingMedium}
\begin{figure}[t]
\centering
\begin{minipage}[c]{4.2cm}
\hspace*{-3.0cm}
  \includegraphics[scale = 0.6]{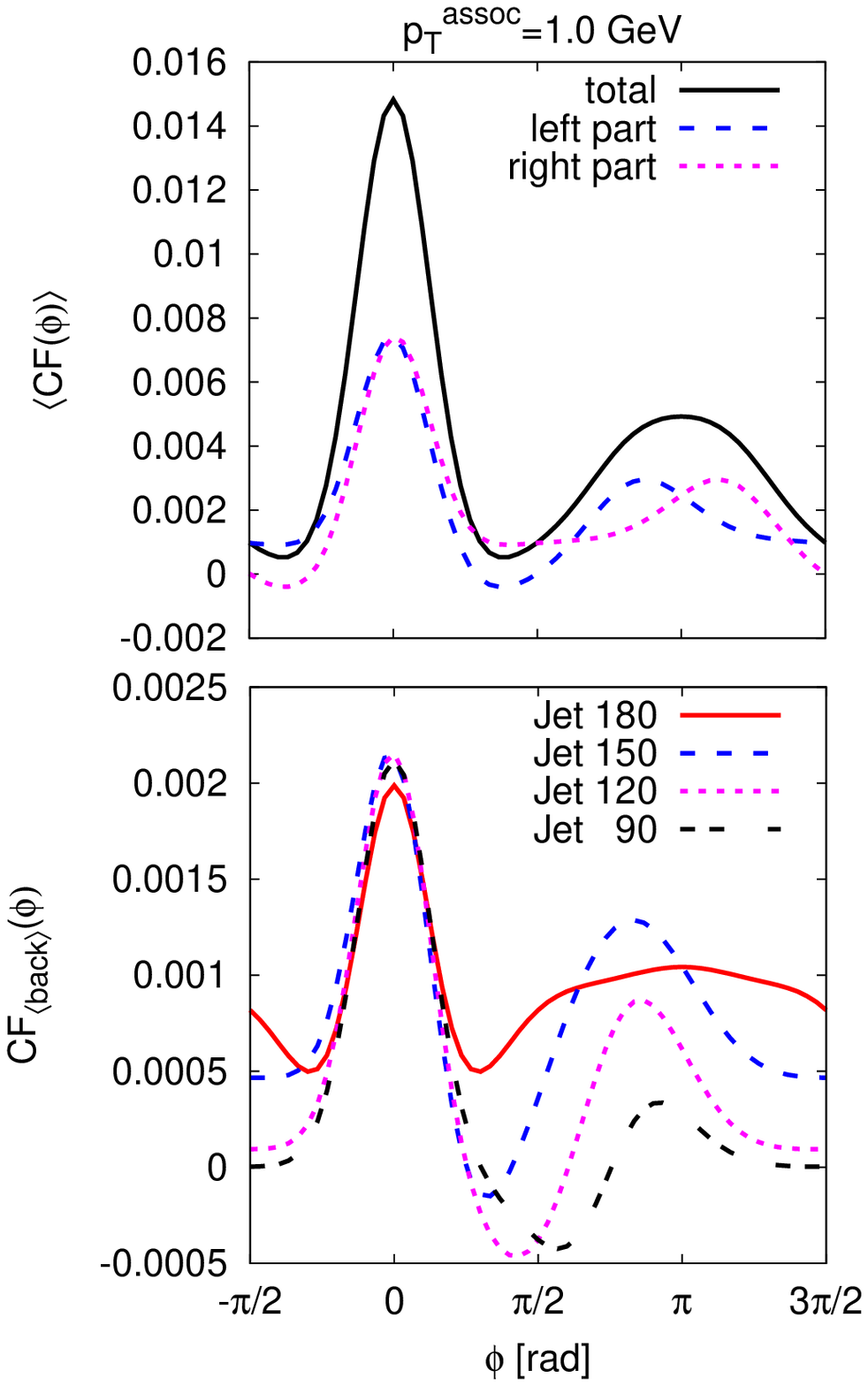}
\end{minipage}
\hspace*{-0.5cm}  
\begin{minipage}[c]{4.2cm} 
  \includegraphics[scale = 0.6]{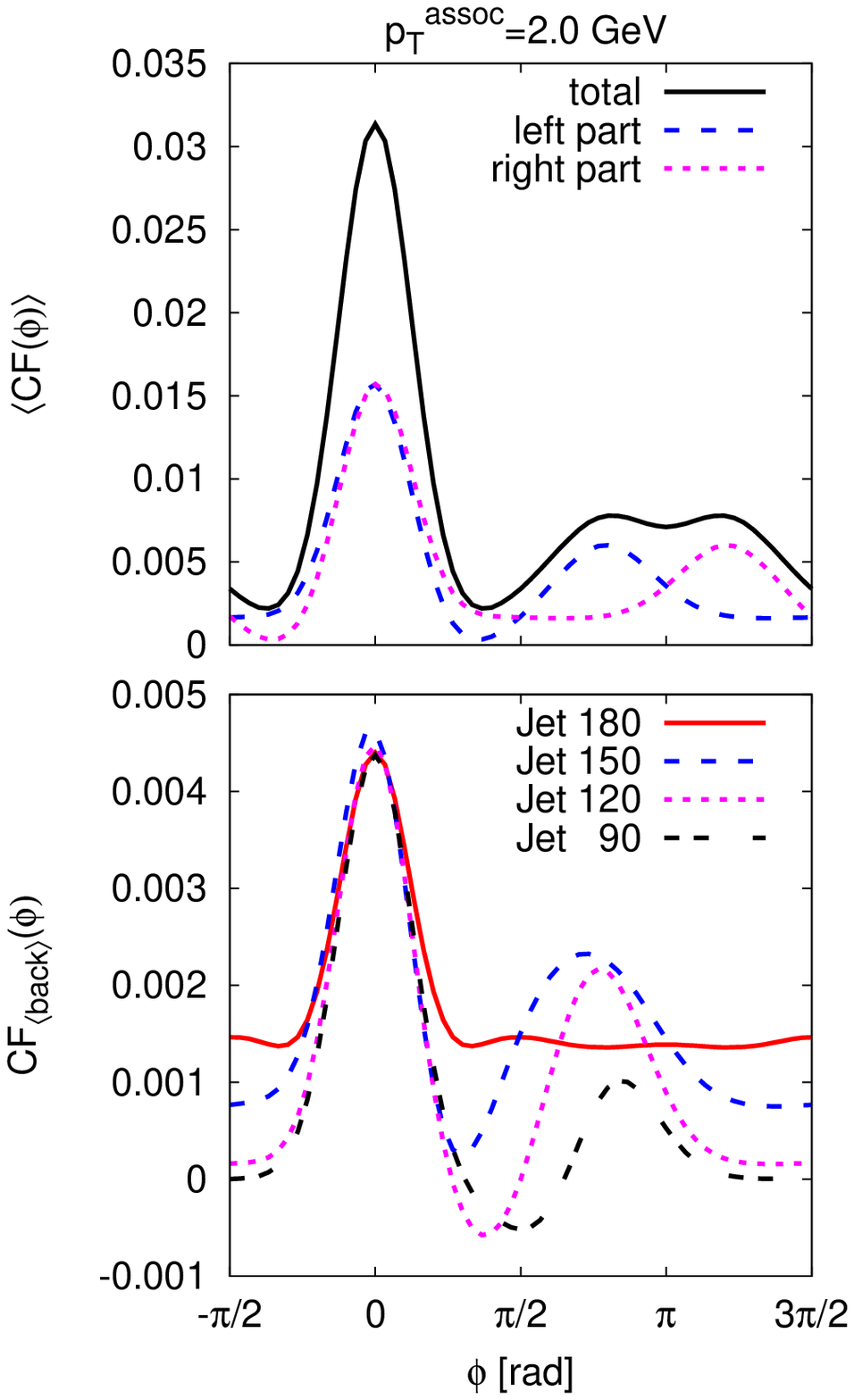}
\end{minipage}
  \caption[The normalized, background-subtracted and path-averaged azimuthal two-particle 
  correlation after performing an isochronous Cooper--Frye freeze-out for jets depositing $5$~GeV 
  of energy and momentum and $p_T^{\rm assoc}=1$~GeV as well as $p_T^{\rm assoc}=2$~GeV.]
  {The normalized, background-subtracted and path-averaged azimuthal two-particle correlation
  after performing an isochronous CF freeze-out (solid black line) for $5$~GeV jets depositing 
  energy and momentum for $p_T^{\rm assoc}=1$~GeV (left panel) and $p_T^{\rm assoc}=2$~GeV 
  (right panel). The dashed blue and magenta lines in the upper panels represent the averaged 
  contributions from the different jet paths in the upper and lower half of the medium 
  (cf.\ Fig.\ \ref{JetPaths}), respectively. The lower panel displays the contribution from 
  the different jet trajectories in the upper half of the medium.}
  \label{CFIsochronous}
\end{figure}

One major difference between the experimental situation and the hydrodynamical calculation
proposed above is that the trajectory of the jet is known by definition in the latter case. 
However, it is possible to simulate the experimental procedure by convoluting the CF 
freeze-out results, which only consider the away-side jets, with a function representing the 
near-side jet
\begin{eqnarray}
f(\phi^\star) &=& \frac{1}{\sqrt{2\pi\sigma^2}}\exp\left(\frac{-\phi^\star}{2\sigma^2}\right)\,,
\end{eqnarray}
(here $\sigma=0.4$~fm), resulting in a two-particle correlation
\begin{eqnarray}
\frac{dN_{\rm con}}{p_T dp_T dy d\phi}&=& A f(\phi)+\int\limits_0^{2\pi}
d\phi^\star \frac{dN(\phi-\phi^\star)}{p_T dp_T dy d\phi} f(\phi^\star)\,,
\end{eqnarray}
where $A$ is an arbitrary amplitude chosen to adjust the heights of the near-side jet. This signal
is then background-subtracted and normalized via
\begin{eqnarray}
CF(\phi)&=&\frac{1}{\int_0^{2\pi}N_{\rm back}(\phi) d\phi}
\left[\frac{dN_{\rm con}(\phi)}{p_T dp_T dy d\phi}-\frac{dN_{\rm back}(\phi)}{p_T dp_T dy d\phi}
\right]\,.
\end{eqnarray}
The different jet paths are implemented by averaging 
\begin{eqnarray}
\frac{d\langle N\rangle}{p_T dp_T dy d\phi}&=&\frac{1}{N_{\rm paths}}
\sum\limits_{i_{\rm paths=1}}^{N_{paths}}\frac{dN^i}{p_T dp_T dy d\phi}\,,
\end{eqnarray}
leading to the definition of the {\it averaged} two-particle correlation
\begin{eqnarray}
\langle CF(\phi)\rangle&=&\frac{1}{\int_0^{2\pi}\langle N_{\rm back}(\phi)\rangle d\phi}
\left[\frac{d\langle N_{\rm con}\rangle(\phi)}{p_T dp_T dy d\phi}-
\frac{d\langle N_{\rm back}\rangle(\phi)}{p_T dp_T dy d\phi}
\right]\,.
\label{def2pc}
\end{eqnarray}
In order to present the contribution of the different paths, we will also depict
\begin{eqnarray}
\hspace*{-0.5cm}
CF_{\langle back \rangle}(\phi)&=&\frac{1}{\int_0^{2\pi}\langle N_{\rm back}(\phi)\rangle d\phi}
\left[\frac{dN_{\rm con}(\phi)}{p_T dp_T dy d\phi}-
\frac{d\langle N_{\rm back}\rangle(\phi)}{p_T dp_T dy d\phi}
\right]\,.
\end{eqnarray}
The results show that for a $5$~GeV jet, corresponding to a trigger-$p_T$ of $3.5$~GeV, 
the normalized, background-subtracted, and path-averaged CF signal (see solid black line in the
upper panel of Fig.\ \ref{CFIsochronous}) displays a broad away-side peak for 
$p_T^{\rm assoc}=1$~GeV (left panel of Fig.\ \ref{CFIsochronous}), while a double-peaked 
structure occurs for $p_T^{\rm assoc}=2$~GeV. The reason is that the contribution of 
the different paths (given in the lower panel of Fig.\ \ref{CFIsochronous} for the paths 
in the upper half of the medium, named according to Fig.\ \ref{JetPaths}), add up to a 
peak in the left part of the away-side (see blue dashed line in the
upper panel of Fig.\ \ref{CFIsochronous}), while the contributions of the paths in the lower half
of the medium (which are not shown here in detail) produce a peak in the right part of the
away-side (see magenta dashed line in the upper panel of Fig.\ \ref{CFIsochronous}).
\\
Depending on the gap between those peaks on the away-side, the different contributions result
either in a broad away-side peak (like for $p_T^{\rm assoc}=1$~GeV) or a double-peaked structure
(as for $p_T^{\rm assoc}=2$~GeV).\\
It is important to notice that the main contributions to the peaks in the left and right part of
the away-side come from jets not propagating through the middle of the medium (see lower panel of 
Fig.\ \ref{CFIsochronous}).\\
Though these simulations do not completely agree with experimental data [comparing with Fig.\ 
\ref{2pc_PHENIX} the cases studied above would roughly correspond to part (b) and (c)], they are
qualitatively similar. For a $p_T^{\rm trig}=3.5$~GeV, the double-peaked structure on the
away-side gets more pronounced with larger $p_T^{\rm assoc}$. However, the yield is different
since experimentally the number of jets decreases with increasing transverse momentum.
\begin{figure}[t]
\centering
\begin{minipage}[c]{4.2cm}
\hspace*{-3.0cm}
  \includegraphics[scale = 0.6]{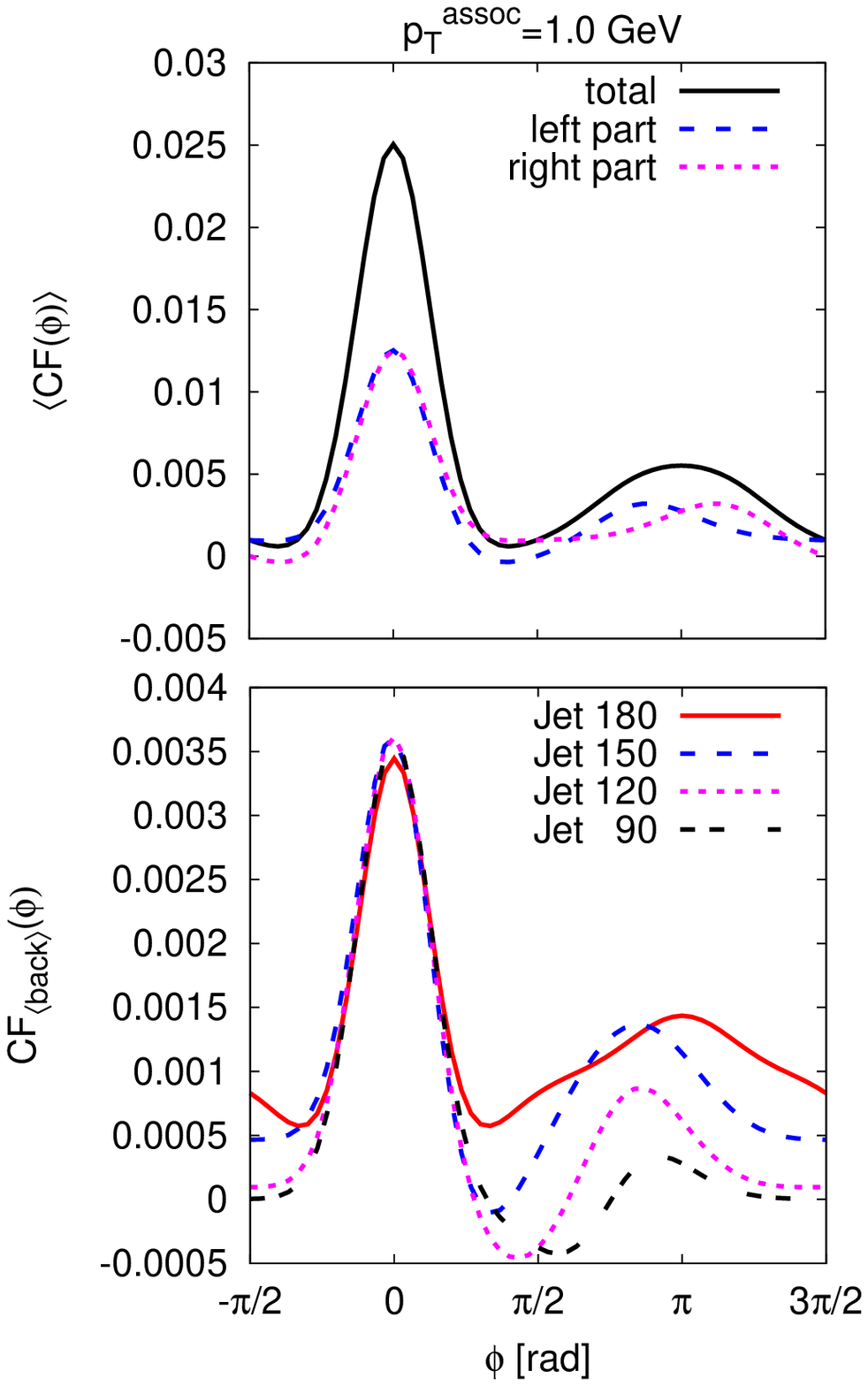}
\end{minipage}
\hspace*{-1.0cm}  
\begin{minipage}[c]{4.2cm} 
  \includegraphics[scale = 0.6]{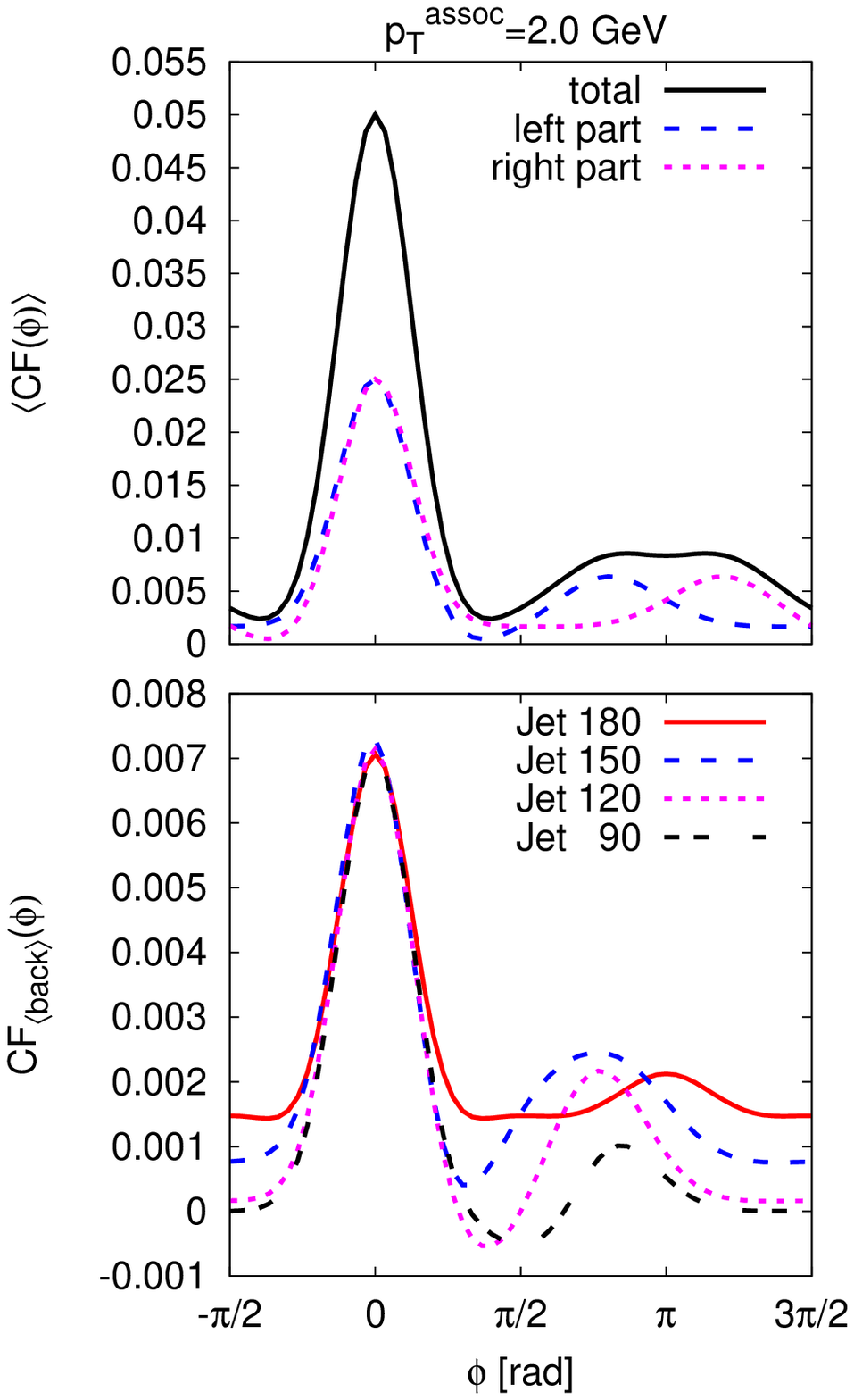}
\end{minipage}
  \caption[The normalized, background-subtracted and path-averaged azimuthal two-particle 
  correlation after performing an isochronous CF freeze-out for jets depositing $10$~GeV of
  energy and momentum and $p_T^{\rm assoc}=1$~GeV as well as $p_T^{\rm assoc}=2$~GeV.]
  {The normalized, background-subtracted and path-averaged azimuthal two-particle correlation
  after performing an isochronous CF freeze-out (solid black line) for $10$~GeV jets depositing 
  energy and momentum for $p_T^{\rm assoc}=1$~GeV (left panel) and $p_T^{\rm assoc}=2$~GeV 
  (right panel). The dashed blue and magenta lines in the upper panels represent the averaged 
  contributions from the different jet paths in the upper and lower half of the medium 
  (cf.\ Fig.\ \ref{JetPaths}), respectively. The lower panel displays the contribution from 
  the different jet trajectories in the upper half of the medium. The impact of the diffusion wake 
  is clearly visible for the jet 180 in the lower panel.}
  \label{CFDiffWake}
\end{figure}
\\Considering now a $10$~GeV jet (see Fig.\ \ref{CFDiffWake}), 
describing a $p_T^{\rm trig}=7.5$~GeV [which can approximately
be compared to part (e) of Fig.\ \ref{2pc_PHENIX}], the resulting shape is still the same.
There is a broad away-side peak for $p_T^{\rm assoc}=1$~GeV and a double-peaked structure for 
$p_T^{\rm assoc}=2$~GeV. However, since the energy and momentum deposition continued for later
times (see Fig.\ \ref{FigEloss}), the jet reaches that part of the medium where the background
flow is parallel to its trajectory. Thus, there is again a strong impact of the diffusion wake as
can be seen from the solid red line in the lower panel of Fig.\ \ref{CFDiffWake} which represents
the jet propagating through the middle of the medium. Compared to the $5$~GeV jet, which stops
about the center of the medium (see solid red line in the lower panel of Fig.\ 
\ref{CFIsochronous}), the diffusion wake contribution is enhanced, resulting in a broader
away-side peak for $p_T^{\rm assoc}=1$~GeV and a smaller dip for $p_T^{\rm assoc}=2$~GeV. This
strengthens the conclusion that a conical signal can be obtained by averaging over different jet
paths.
\begin{figure}[t]
\centering
\begin{minipage}[c]{4.2cm}
\hspace*{-3.0cm}
  \includegraphics[scale = 0.6]{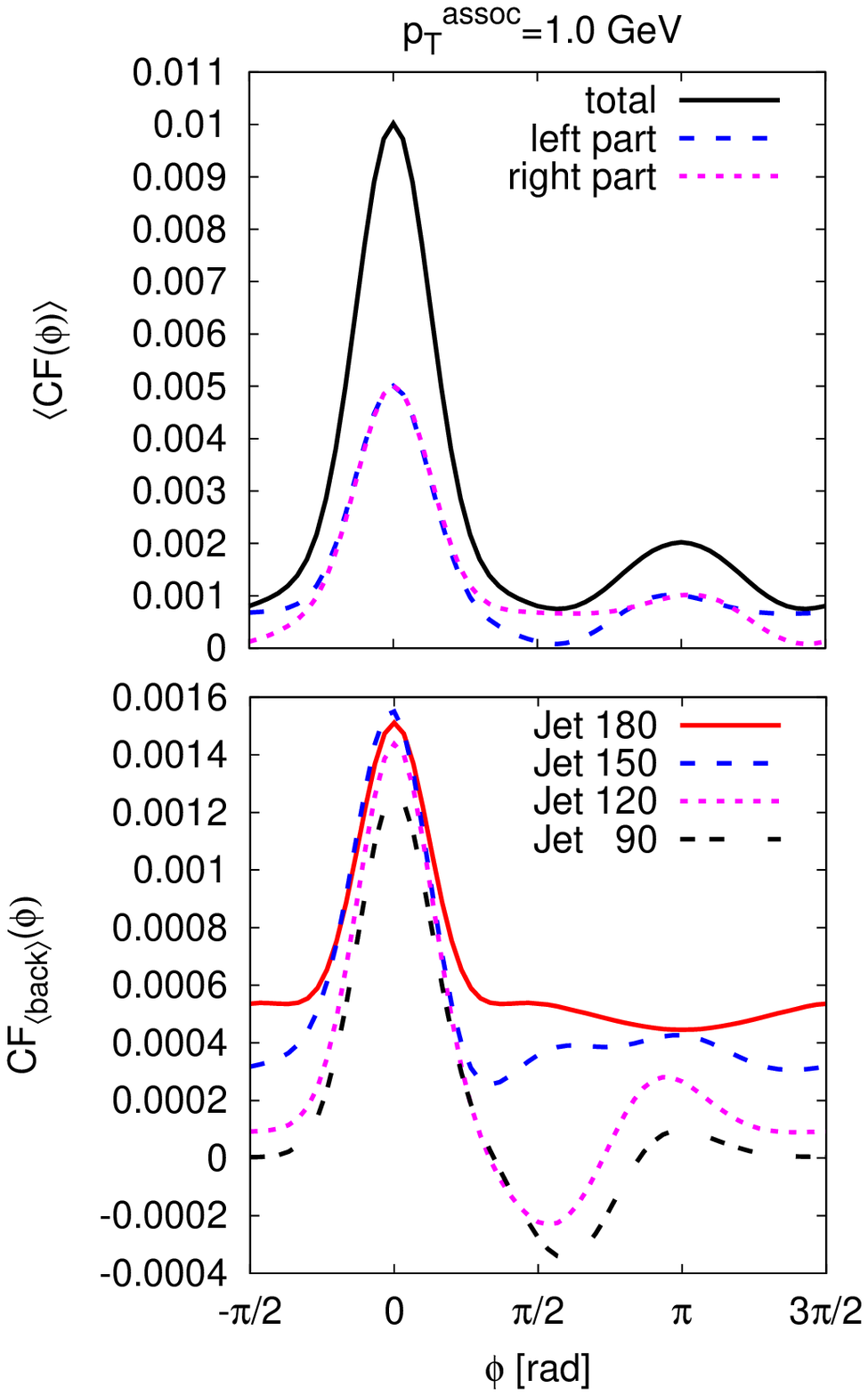}
\end{minipage}
\hspace*{-0.5cm}  
\begin{minipage}[c]{4.2cm} 
  \includegraphics[scale = 0.6]{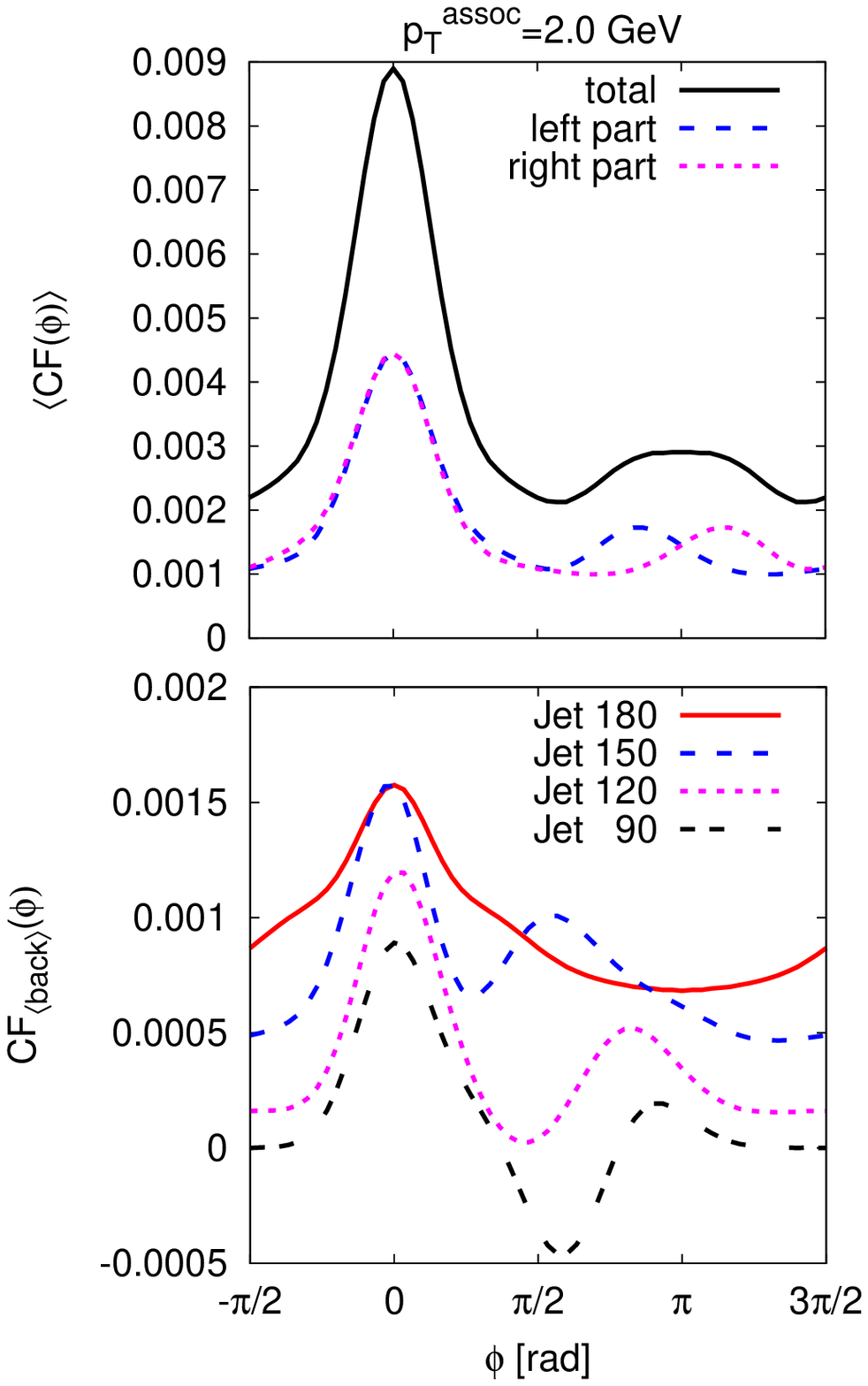}
\end{minipage}
  \caption[The normalized, background-subtracted and path-averaged azimuthal two-particle 
  correlation after performing an isochronous CF freeze-out for jets depositing and energy of 
  $5$~GeV and assuming a vanishing momentum loss rate for $p_T^{\rm assoc}=1$~GeV and 
  $p_T^{\rm assoc}=2$~GeV.]
  {The normalized, background-subtracted and path-averaged azimuthal two-particle correlation
  after performing an isochronous CF freeze-out (solid black line) for $5$~GeV jets assuming a
  vanishing momentum loss rate for $p_T^{\rm assoc}=1$~GeV (left panel) and 
  $p_T^{\rm assoc}=2$~GeV (right panel). The dashed blue and magenta lines in the upper panels 
  represent the averaged contributions from the different jet paths in the upper and lower half 
  of the medium (cf.\ Fig.\ \ref{JetPaths}), respectively. The lower panel displays the 
  contribution from the different jet trajectories in the upper half of the medium. The Jet 180
  in the lower panel exhibits large emission angles.}
  \label{CFEloss}
\end{figure}
\\References \cite{CasalderreySolana:2004qm,Betz:2008ka} showed that, considering a source term 
with a vanishing momentum-deposition rate in a static medium, a conical signal arises from a 
CF freeze-out. However, since the deposition of energy and momentum does 
{\it not} yield a  double-peaked structure on the away-side for a static medium, 
but rather a peak in the opposite trigger-jet direction 
\cite{CasalderreySolana:2004qm,Betz:2008ka}, the question arises if this result may allow for 
conclusions about the properties of the source term.\\
Fig.\ \ref{CFEloss} demonstrates that pure jet energy loss in an expanding medium does {\it not} 
lead to a double-peaked structure on the away-side for the normalized, background-subtracted and 
path-aver\-aged azimuthal correlation, neither for $p_T^{\rm assoc}=1$~GeV nor for 
$p_T^{\rm assoc}=2$~GeV (though there is a plateau region for $p_T^{\rm assoc}=2$~GeV ). 
No peaks occur on the away-side of a jet traversing the middle of the medium (solid red line in the
lower panel of Fig.\ \ref{CFEloss}). However, there appear small 
peaks around $\phi\sim\pi/2$ and $\phi\sim3\pi/2$ for a $p_T^{\rm assoc}=1$~GeV, consistent with 
the prediction of Ref.\ \cite{Satarov:2005mv} that the emission angle increases with increasing
background flow (see also Fig.\ \ref{SatarovChangeMachAngle}\footnote{It is important to mention 
here that Satarov et al.\ predicted a decreasing opening angle of the Mach cone with increasing 
background flow (parallel or anti-parallel to the trajectory of the jet). Clearly, such a 
decreasing opening angle results in an increasing emission angle.} ).
Thus, there is a clear difference between a static and an expanding medium. Nevertheless, it is also
necessary to study the propagation of a jet through a uniform background since otherwise various 
effects (like e.g.\ the impact of the diffusion wake) cannot be understood.\\
In conclusion, we have shown using a full $(3+1)$-dimensional ideal hydrodynamic prescription 
that a double-peaked away-side structure can be formed by averaging over different
contributions of several possible jet trajectories through an expanding medium, as already 
discussed in Refs.\ \cite{Renk:2006mv,Neufeld:2008eg,Chaudhuri:2006qk}. Therefore, it seems natural 
to conclude that this shape, interpreted as a conical signal, does not result from a ``true'' 
Mach cone, but is actually generated by a superposition of distorted wakes. Clearly, the 
emission angle of such a structure is not connected to the EoS. \\
We verified that the interplay between radial flow and diffusion wake may lead to an 
annihilation of the latter as long as the jet trajectory is opposite to the background flow 
(see Fig.\ \ref{CFIsochronous}), as suggested in Ref.\ \cite{Neufeld:2008eg}. Nevertheless, 
the contribution of this diffusion wake, which depends on the path length of the jet, may
strongly influence the final azimuthal distributions obtained after freeze-out from averaging over
different possible jet trajectories (see Fig.\ \ref{CFDiffWake}). \\
However, the main contribution to the away-side correlation is due to jets that do not propagate
through the middle of the medium, depending on the jet energy and momentum loss rate as well as 
the amount of deposited energy (see e.g.\ the lower left panel of Fig.\ \ref{CFIsochronous}). 
Obviously, it is necessary to determine the jet energy and momentum 
loss rates as well as their variance in time depending on the  initial energy and 
velocity of the jet.\\
These results do not exclude Mach-cone formation in heavy-ion collision. As
already proven in Ref.\ \cite{Betz:2008ka} (see chapter \ref{DiffusionWake}, in particular Fig.\ 
\ref{FigPunchThrough1}), a conical shape occurs in all jet-deposition scenarios including energy 
loss, leading to a constructive interference of the outward-moving sound waves, but the signal 
is usually too weak to survive a CF freeze-out scenario 
\cite{Betz:2008wy,Betz:2008ka,Noronha:2008un}. Due to the interaction with radial flow, it might
still be possible to observe a Mach-cone signal in single-jet events. \\
Moreover, these findings are not in contradiction to the measured three-particle correlations. 
As Fig.\ \ref{3pc_Ulery} shows, such pattern exhibits a peak along the diagonal axis, 
supposed to arise from deflected jets, as well as four off-diagonal peaks on the away-side, 
presumably resulting from a Mach-cone contribution. Such a structure may indeed be due to the 
different contributions of the various jet trajectories considered. \\
We also demonstrated that for the present study (at least for very central jets) 
an isochronous and an isothermal CF freeze-out 
prescription (see appendix \ref{IsochronIsothermFreezeout}) leads to very similar azimuthal 
particle correlations. Furthermore, we elaborated the difference between a static and an 
expanding medium by showing that a jet deposition scenario assuming a vanishing momentum-loss 
rate, which results in a conical signal in a uniform background, does not lead to a 
conical structure in an expanding medium for low values of $p_T^{assoc}$ (see Fig.\ \ref{CFEloss}) 
and is therefore not in accordance with experiment \cite{2pcPHENIX} (see Fig.\ \ref{2pc_PHENIX}). Thus, it 
is not possible to directly extrapolate from a static to an expanding medium and
non-linear hydrodynamics is fundamental for quantitative studies of jets in a medium. \\
In addition, we could verify the predictions by Satarov et al.\ \cite{Satarov:2005mv} that 
a background flow antiparallel to the jet direction leads to larger emission angles of a Mach 
cone, see Fig.\ \ref{CFEloss}. For a general, though qualitative, discussion of the distortion
effects in an expanding medium see appendix \ref{AppendixDistortion}. \\
The effects of longitudinal expansion as well as finite impact parameter remain to be considered. 
Moreover, a different freeze-out prescription (like coalescence \cite{Greco:2003xt}) might 
alter the azimuthal correlations. The recent observation of cone-angle variation with respect 
to the reaction plane (see right panel of Fig.\ \ref{2pc_Paths} and 
\cite{vanLeeuwen:2008pn,PHENIX_QM09_1,PHENIX_QM09_2}), which has to be checked 
applying hydrodynamical prescriptions, raises the prospect that the phenomenology of generating 
conical signals in heavy-ion collisions could be tested soon. However, the novel results concerning
the two-particle correlations obtained from full jet reconstructions
\cite{STAR_FullJet,PHENIX_FullJet} reveal that the propagation of a jet through a medium that can 
be described by hydrodynamics deserves further scrutiny and is far from being resolved yet.

%
%
%
\clearpage{\pagestyle{empty}\cleardoublepage}
\thispagestyle{empty}
\part[Jets in Heavy-Ion Collisions: Conclusions and Outlook]{Jets in Heavy-Ion Collisions: Conclusions and Outlook}
\label{part04}
\clearpage{\pagestyle{empty}\cleardoublepage}
\chapter*{}
\markboth{\it Conclusions and Outlook}{}
\hspace*{6.8cm}
\begin{minipage}[t]{6cm}
Non-trivial things are the\\
sum of trivial things.\\
{---------------------------------------------}\\
{\small (Unknown)}\\[0.5cm]~
\end{minipage}

\noindent
The main topic of this thesis was to investigate the jet-medium interactions in a Quark-Gluon 
Plasma (QGP) that is assumed to be created in heavy-ion collisions and to behave like a ``nearly 
perfect fluid''. For this purpose, we studied the propagation of a fast particle, the so-called jet,
through a medium (which can be described by hydrodynamics) using a
$(3+1)$-dimensional ideal hydrodynamic algorithm.\\
The basic question was if the propagation of such a jet leads to the creation of a Mach cone and 
if the disturbance caused by this Mach cone is large enough to be seen in the final particle
distributions that are measured by experiment. Such a result would not only confirm fast 
thermalization, it would also allow to study the Equation of State (EoS) of the medium formed 
in a heavy-ion collision (which is suppossed to resemble the one created shortly after the Big Bang),
since the (particle) emission angles caused by the Mach cone are directly related to the speed of 
sound of the medium.\\
After giving a general overview about the basic properties and the phase diagram of
Quantum Chromodynamics (QCD), which can properly be tested by heavy-ion collisions, we described 
the evolution of such a heavy-ion collision and discussed different probes for the QGP. Then, we 
reviewed the particle correlations measured by experiment. Motivated by the fact
that jets are produced back-to-back, two-particle correlations were introduced considering the
relative azimuthal angle between two particles. One of these particles, the trigger jet, is 
assumed to leave the expanding fireball without any further interaction while the 
associated particle on the away-side, traverses the medium depositing energy and 
momentum. Such (published) two-particle correlations obtained from experimental data exhibit a 
clear double-peaked structure on the away-side which is suggested to be caused by the creation of a 
Mach cone. However, other explanations are also possible and were discussed briefly.\\
In part \ref{part02}, we introduced the concept of ideal hydrodynamics and its applicability
to heavy-ion collisions, starting from a (proper) choice of the initial conditions,
via the numerical solution for the Equations of Motion, to the conversion of the fluid fields of 
temperature and velocities into particles, the so-called freeze-out. \\
Moreover, we discussed the importance and the effects of viscosity in heavy-ion collisions
and deduced the viscous transport equations (for the bulk viscous pressure, the heat flux current, and the 
shear stress tensor) from kinetic theory. Those equations are the basis for any numerical 
treatment of viscous hydrodynamics, a prescription that is important to gain a quantitative 
understanding of the underlying processes in heavy-ion collisions.\\
Subsequently, we discussed shock-wave phenomena in general, reviewed the predictions concerning 
the influence of an expanding background on the formation of Mach cones, and outlined different 
jet energy-loss mechanisms ranging from theories based on weak interactions (as described by  
perturbative QCD, pQCD) to strong interactions (which can be formulated using the 
Anti-de-Sitter/Conformal Field Theory correspondence, AdS/CFT). A basic introduction to the 
latter prescription is given in a separate chapter. Addionally, we also summarized previous 
studies of jet-energy transfer to the medium using the linearized hydrodynamical approach.\\
The results of our numerical simulations were summarized in part \ref{part03}. Here, we distinguished 
a static and an expanding medium and considered different energy and momentum-loss scenarios for a 
source term schematically characterizing the deposition along the trajectory of the jet 
as well as another one derived from pQCD. The latter source term is also directly compared to 
results obtained from a jet propagation described within the framework of AdS/CFT.\\
One of the basic results is the creation of a strong flow (named the {\it diffusion wake}) behind 
the jet common to all different scenarios studied in a static medium if the momentum deposition 
is larger than a certain threshold. The particle yield coming from this strong forward-moving 
diffusion wake always overwhelms the weak Mach-cone signal after freeze-out and leads to one single 
peak on the away-side which is not in agreement with the shape of the published two-particle
correlations obtained by experiment. \\
This peak is found for the schematic as well as for the pQCD source term, but not when evaluating
a jet event applying the AdS/CFT correspondence. In that case, the impact of the transverse 
flow from the so-called neck region (which is an area close to the head of the jet) lead to a 
double-peaked structure in the final particle correlations that is unelated to a Mach-cone 
signal, providing different freeze-out patterns for pQCD and AdS/CFT. 
Thus, the measurement of identified heavy quark jets at RHIC and LHC might provide an important 
constraint on possible jet-medium coupling dynamics. \\
Moreover, we showed that the medium's response and the corresponding away-side angular 
correlations are largely insensitive to whether the jet punches through or stops inside the medium. 
We described the backreaction of the medium for the fully stopped jet by a simple Bethe--Bloch-like
model which causes an explosive burst of energy and momentum (Bragg peak) close to the end of the 
jet's evolution through the medium. 
The resulting correlations are also independent of whether the momentum deposition is longitudinal 
(as generic to pQCD energy loss models) or transverse.\\
The universal existence of the diffusion wake in all scenarios where energy as well as momentum is
deposited into the medium can readily be understood in ideal hydrodynamics through vorticity
conservation. We discussed experimentally observable consequences of such a conserved structure,
probably connected to polarization effects of hadrons (like hyperons or vector mesons). Since
polarization is sensitive to initial conditions, hydrodynamic evolution, and mean-free path, 
corresponding measurements might shed some light on several aspects of heavy-ion collisions that 
are not well understood yet.\\ 
Certainly a realistic description of a heavy-ion collision requires an expanding background. 
We demonstrated that the interaction of the radial and jet flow may lead to the reduction of the 
diffusion wake and to a deflection of the Mach cone as predicted in an earlier work. The strength 
of the diffusion-wake contribution clearly depends on the path length of the jet. \\
However, the correlation patterns obtained from experiment consider several jet 
events. We showed that a conical signal can be created by 
averaging over different possible jet paths. Therefore, it seems natural to conclude that the 
experimentally observed shape does not result from a ``true'' Mach cone, but is actually generated 
by a superposition of distorted wakes. Clearly, the emission angle of such a structure is not 
related to the EoS.\\
We illustrated that the diffusion wake nevertheless strongly influences the structure of the final 
particle correlations obtained from averaging over possible jet trajectories. Depending on the 
energy and momentum-loss rate as well as on the amount of deposited 
energy, the main contribution to the away-side correlation may be due to non-central jets. 
Consequently, it is necessary to determine the jet energy and momentum-loss rates as well as their 
variance in time depending on the initial energy and velocity of the jet.\\
Additionally, we elaborated a clear difference between a static and an expanding background. 
While an energy-loss scenario assuming a vanishing momentum-loss rate results in a conical signal 
for a uniform background, such a structure is not obtained in an expanding medium which 
is in disagreement with the experimental data. Thus, it is fundamental to apply non-linear hydrodynamics for 
quantitative studies of jets in a medium.\\
The results discussed above are not in contradiction to the measured three-particle correlations that display a 
distinct emission pattern if Mach cones are produced since these correlations also show a 
contribution from deflected jets which might be due to the non-central jets mentioned before.\\
Therefore, an unambiguous proof of a Mach cone which can be related to the EoS requires the 
investigation of single-jet events (as it should become possible at LHC) for different trajectories 
through the medium. \\
The recent observation of cone-angle variation with respect to the 
reaction plane, which has to be analyzed using a full hydrodynamical prescription, promises 
further insight into the phenomenology of generating conical interference patterns in heavy-ion 
collisions. Novel results, however, obtained from full jet reconstruction reveal that the 
jet-medium interactions deserve further scrutiny both from the experimental and from the theoretical side. 
For a comparison to hydrodynamic simulations, full jet reconstruction will be essential.\\
Apparently, the effects of longitudinal expansion, nonzero impact parameter, and 
phase transitions (connected to a change in the speed of sound of the medium) remain to be 
considered. \\
It will also be important to test the experimental procedure of background 
subtraction. The common method applied is to subtract the elliptic flow arising from angular
anisotropies of the expanding system. Since it is not clear from first principles that the flow is 
independent of the jet transit, this method (ZYAM) is discussed controversally.\\
Moreover, different freeze-out prescriptions as well as the interaction of the created hadrons 
(like coalescence or resonance decays) might alter the azimuthal correlations and need to be 
examined. \\
To advance this research area, a more detailed understanding of the source for the hard probes 
is needed, ranging from weak interactions (as described by pQCD) to strong 
interactions (formulated by the AdS/CFT correspondence). It is necessary to develop a detailed 
space-time energy-loss model, considering all different contributions (i.e., radiative and 
collisional energy loss) to high order in opacity. By coupling quantum transport to hydrodynamics, 
it will be possible to further investigate the jet-medium interactions for the different energy 
regions supported by the various experiments like RHIC, LHC, and FAIR.

%
%
%
\clearpage{\pagestyle{empty}\cleardoublepage}
\addcontentsline{toc}{part}{Appendices}
\part*{Appendices}
\clearpage{\pagestyle{empty}\cleardoublepage}
\begin{appendix}
%
%
\chapter[The Evolution of the Universe]
{The Evolution of the Universe}
\label{EvolutionUniverse}

\begin{figure}[b]
\centering
  \includegraphics[scale = 0.36]{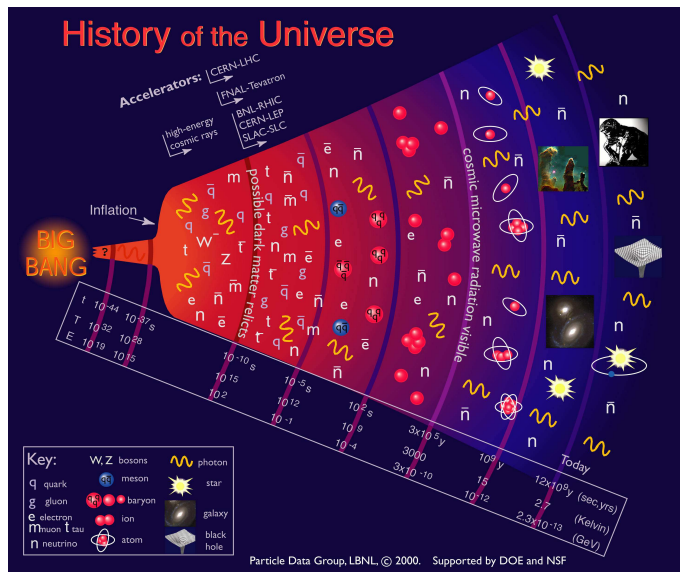}
  \caption[The evolution of the universe.]
  {The evolution of the universe, starting with the Big Bang 13.7
  billion years ago and illustrating the period from the formation of the
  building blocks of matter to the creation of galaxies.}
  \label{PictureEvolutionUniverse}
\end{figure}

About 13.7 billion years ago a hot and dense phase was formed out of a singularity, the so-called 
{\it Big Bang}, where quarks, antiquarks, and gluons could move around as free particles. This 
state of matter is usually called the Quark-Gluon Plasma (QGP). Due to the high pressure gradients,
expansion and cooling set in and below a certain critical value of the energy density 
($\varepsilon_{\rm crit}\approx 1$~GeV/fm$^3$), quarks and gluons combined into color-neutral 
objects, the {\it hadrons}, a process being called confinement. \\
Subsequently, the unstable hadrons decayed so that mainly protons as well as neutrons accumulated. 
Around $1$~s after the Big Bang, the main fraction of the building blocks of matter were already 
formed. Later, heavy elements emerged from fusion processes and supernova explosions. \\[1mm]~
After the temperature had fallen below $T\approx 3000$~K, neutral atoms were generated. Since 
the interaction of the photons with these neutral atoms was very weak, the universum got 
transparent. Then, due to the expansion of the universe, the wavelength of the disconnected 
background radiation increased, resulting in a redshift that corresponds to a temperature of 
$2.73$~K which is measurable today. This decoupling of radiation caused the gravity to gain more 
influence and because of spatial density fluctuations, spacious structures like galaxies were 
created roughly 1 million years after the Big Bang. For a synopsis see Fig.\ 
\ref{PictureEvolutionUniverse}.

\clearpage{\pagestyle{empty}\cleardoublepage}
%
%
\chapter[Glauber Model]
{Glauber Model}
\label{AppGlauberModel}

\begin{figure}[b]
\centering
\includegraphics[scale = 0.4]{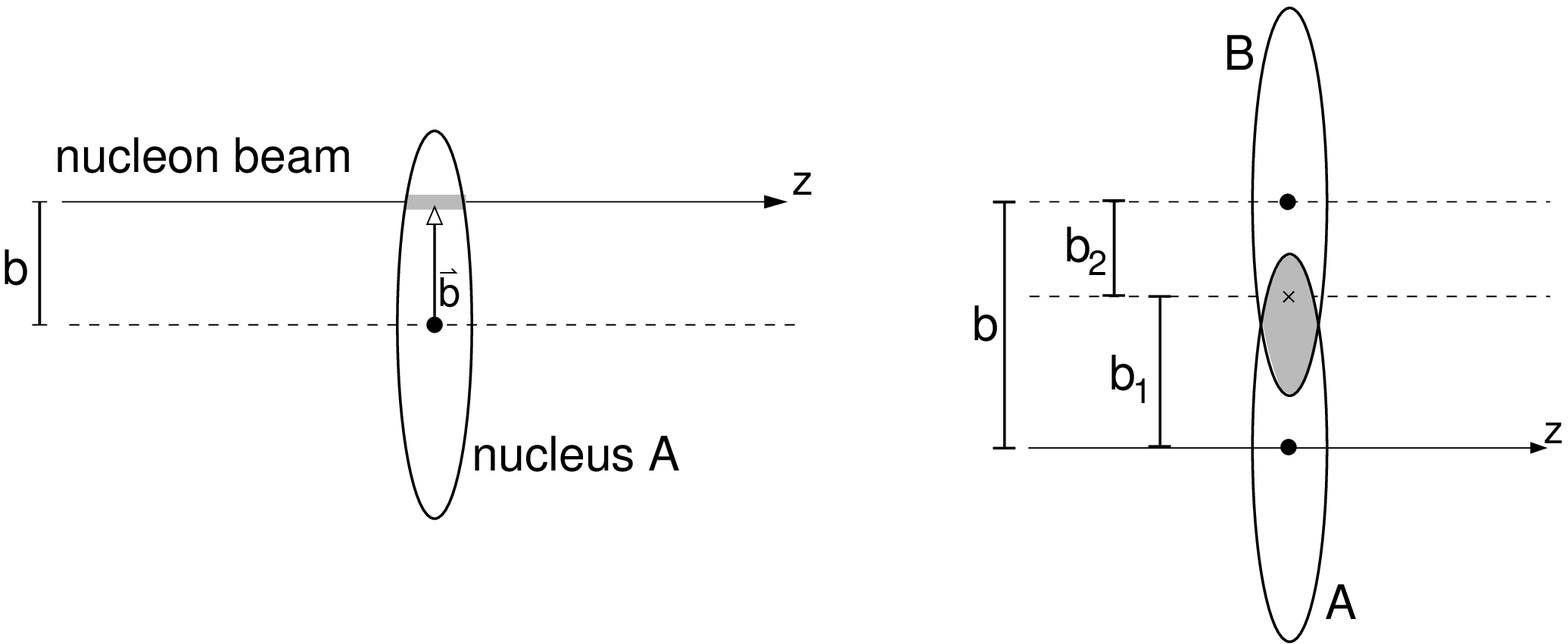}
  \caption[Schematic representation of the Glauber Model geometry.]
  {Schematic representation of the Glauber Model geometry for a nucleon-nucleus collision 
  (left panel) and for a nucleus-nucleus collision (right panel) \cite{DiplomFochler}.}
  \label{SketchGlauber}
\end{figure}

\begin{figure}[t]
\centering
  \includegraphics[scale = 0.5]{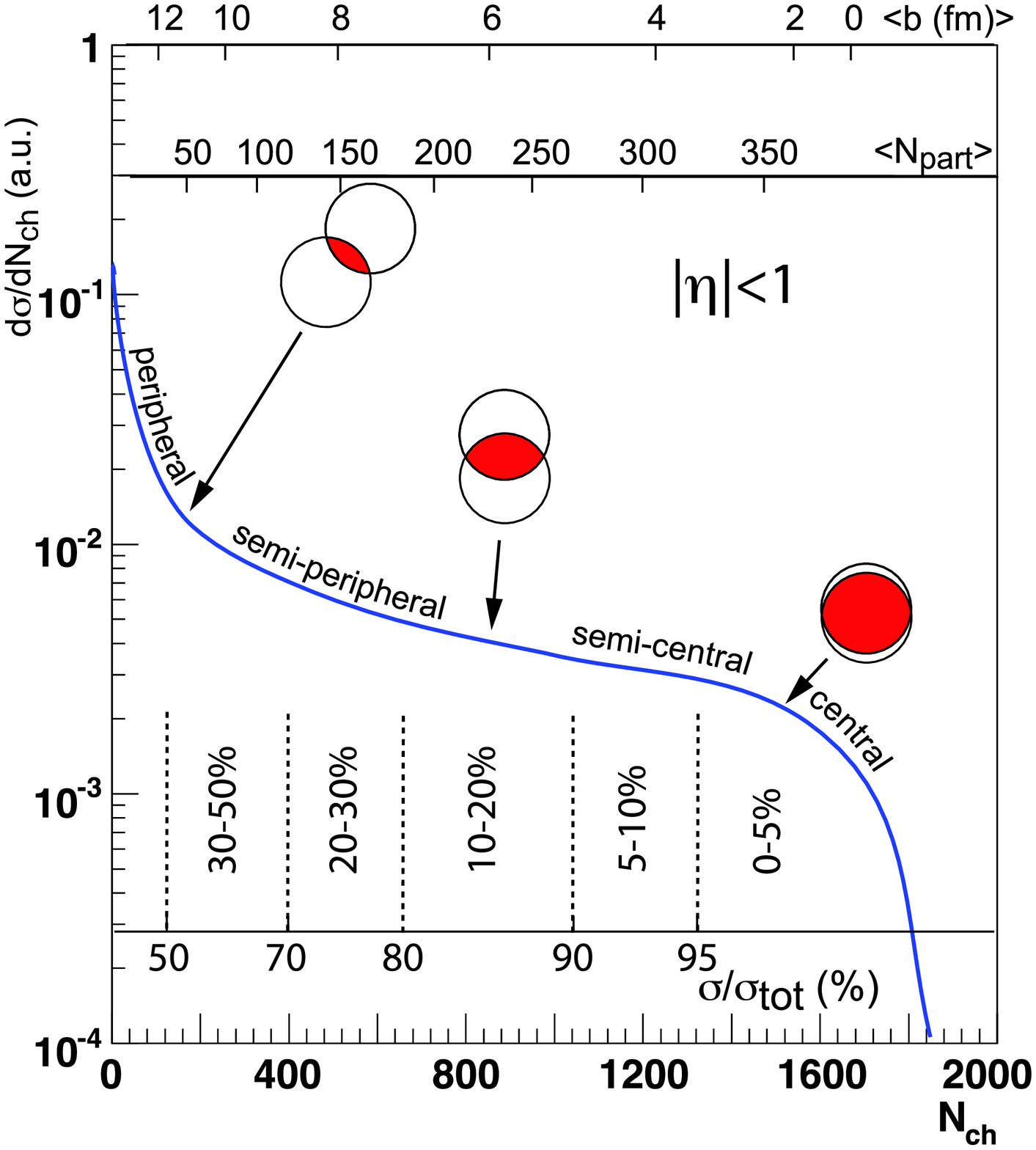}
  \caption[The correlation between the number of participating nucleons in a heavy-ion collision, 
  their cross section and the impact parameter $b$, defining the centrality classes.]
  {The correlation between the number of participating nucleons in a heavy-ion collision, 
  their cross section and the impact parameter $b$, defining the centrality classes 
  \cite{Miller:2007ri}.}
  \label{SketchCentralityClass}
\end{figure}
The Glauber Model \cite{Kolb:2001qz} provides initial conditions for the transverse plane in a 
heavy-ion collision that can be used by models describing the expansion of a hot and dense 
medium (like hydrodynamic applications). It describes a nucleus-nucleus collision in terms of 
multiple nucleon-nucleon interactions. The underlying geometry is shown in Fig.\ 
\ref{SketchGlauber}). It leads to the number of wounded (or participating) nucleons in a 
heavy-ion collision which can be used to specify the centrality class of a collision (see Fig.\ 
\ref{SketchCentralityClass}) and to determine the initial distribution of energy density. The 
latter one is mandatory in the hydrodynamical framework.\\
Since the model is based on nucleon-nucleon collisions, it requires the knowlegde of the inelastic 
cross section $\sigma_{NN}$ and a density profile of the nucleus which is specified to be the 
Woods--Saxon distribution,
\begin{eqnarray}
\rho_A(\vec{x})&=&\frac{\rho_0}{1+\exp[(\vert\vec{x}\vert-R_A)/d]}\,.
\end{eqnarray}
The mean radius of the nucleus $R_A$ can be calculated via 
\begin{eqnarray}
R_A&=&\left(1.12A^{1/3}-0.86A^{-1/3} \right){\rm fm}\,,
\end{eqnarray}
where $d$ is the skin depth and $\rho_0$ has to be chosen such that an integration over the 
Woods--Saxon distribution leads to the number of nucleons $A$. \\
Though it is possible to calculate this profile analytically, Monte Carlo simulation are often 
used to sample $\rho_A$ as probability distribution for nucleons within a nucleus. The model 
distinguishes:
\begin{itemize}
\item {\bf Nucleon-nucleus collision:} 
The nuclear thickness function, describing that part of the nucleus A with which the nucleon 
passing through may interact (see Fig.\ \ref{SketchGlauber}), is given by
\begin{eqnarray}
T_A(x,y) &=& \int\limits_{-\infty}^{+\infty}\,dz \rho_A(\vec{x})\,.
\end{eqnarray}
The probability that two nucleons collide is $T_A(x,y,b)\sigma_{NN}$. From this, the number 
density of binary collisions in a nucleon-nucleus collision can be calculated using the 
probability of $n$ binary collisions
\begin{eqnarray}
P(n;A,b) &=& \left( \begin{array}{c}
A\\
n
\end{array} \right) \left[T_A(x+b,y)\sigma_{NN}\right]^n \nonumber\\
&&\hspace*{0.9cm}\times\left[1-T_A(x+b,y)\sigma_{NN}\right]^{A-n} 
\end{eqnarray}
which leads to
\begin{eqnarray}
n_{BC}(x,y,b) &=& \sum\limits_{n=1}^A\, nP(n;A,b) = AT_A(x+b,y)\sigma_{NN}\,.
\end{eqnarray}
\item {\bf Nucleus-nucleus collision:} 
In analogy to the nucleon-nucleus collision, a density distribution $T_{AB}(x,y,b)$ is defined, 
called nuclear overlap function,
\begin{eqnarray}
T_{AB}(x,y,b) &=& \int dx\, dy\, T_A\left(x+\frac{b}{2},y\right)T_B\left(x-\frac{b}{2},y\right)\,.
\end{eqnarray}
Again, a binomial distribution
\begin{eqnarray}
P(n;AB,b) &=& \left( \begin{array}{c}
AB\\
n
\end{array} \right) \left[T_{AB}(x,y,b)\sigma_{NN}\right]^n \nonumber\\
&&\hspace*{1.25cm}\times
\left[1-T_{AB}(x,y,b)\sigma_{NN}\right]^{AB-n} 
\end{eqnarray}
characterizes the probability for $n$ binary collisions from which the mean number density 
of binary collisions is determined to be 
\begin{eqnarray}
n_{BC}(x,y,b) &=& AB\,T_{AB}(x,y,b)\sigma_{NN}\,.
\end{eqnarray}
In a similar way, the number density of wounded (i.e.\ participating) nucleons can be derived:
\begin{eqnarray}
&&\hspace*{-1.7cm}
n_{WN}(x,y,b)=T_A\left(x+\frac{b}{2},y\right)
\left[1-\left(1-\frac{\sigma_{NN}T_B\left(x-\frac{b}{2},y\right)}{B}\right)^B\right]
\nonumber\\
&&\hspace*{0.55cm}+T_B\left(x-\frac{b}{2},y\right) 
\left[1-\left(1-\frac{\sigma_{NN}T_A\left(x+\frac{b}{2},y\right)}{A}\right)^A\right]
\hspace*{-0.15cm}.
\end{eqnarray}
The number of binary collisons $N_{BC}$ and wounded nucleons $N_{WN}$ is eventually obtained 
by integration
\begin{eqnarray}
N_{BC}(b) &=& \int dx\, dy\, n_{BC}(x,y)\,, \nonumber\\
N_{WN}(b) &=& \int dx\, dy\, n_{WN}(x,y) \,.
\end{eqnarray} 
\end{itemize}

\clearpage{\pagestyle{empty}\cleardoublepage}
%
%
\chapter[The EoS for an Ideal Gas]
{The EoS for an Ideal Gas}
\label{EoSIdealGas}

A gas of massles particles with a degeneracy factor $g$ can be described using the grand-canonical
partition function
\begin{eqnarray}
Z^{\rm gr}(T,V,\mu) &=&
\sum\limits_{N=0}^{\infty}\left[\left(e^{\beta\mu}\right)^N 
Z^{\rm can}(T,V,N)\right]
\end{eqnarray}
with $\beta=1/T$ and the canonical partition function $Z^{\rm can}(T,V,N)$.
For non-interacting systems (for simplicity the Maxwell--Boltzmann limit is
assumed, i.e., there will be no distinction between bosons and fermions), 
this expression simplifies since
\begin{eqnarray}
Z^{\rm can}(T,V,N) &=& \frac{1}{N!}[Z^{\rm can}(T,V,1)]^N\,.
\end{eqnarray}
Thus, the grand-canonical partition function can be rewritten as
\begin{eqnarray}
Z^{\rm gr}(T,V,\mu) &=& \sum\limits_{N=0}^{\infty}
\frac{1}{N!}\left[e^{\beta\mu} Z^{\rm can}(T,V,1)\right]^N\nonumber\\
&=&\exp\left[\exp(\beta\mu) Z^{\rm can}(T,V,1) \right]\,.
\end{eqnarray}
This expression can also be formulated in terms of the fugacity $z=e^{\beta\mu}$
\begin{eqnarray}
Z^{\rm gr}(T,V,z) &=& \exp\left[z\,Z^{\rm can}(T,V,1) \right]\,.
\end{eqnarray}
$Z^{\rm can}(T,V,1)$ describes the canonical partition function for one particle given by
\begin{eqnarray}
Z^{\rm can}(T,V,1) &=& gV \int\frac{d^3k}{(2\pi)^3}e^{-\beta\,k}\nonumber\\
&=& g\frac{V}{\pi^2\beta^3}=g\frac{VT^3}{\pi^2}\,,
\end{eqnarray}
which eventually leads for a gas with vanishing chemical potential $\mu=0$ to a grand-canonical 
partition function of
\begin{eqnarray}
Z^{\rm gr}(T,V,z)&=&\exp\left(g\frac{VT^3}{\pi^2}\right) \,.
\label{GCP}
\end{eqnarray}
The particle density as well as the energy density can now be computed via
\begin{eqnarray}
n&=&\frac{\langle N\rangle}{V}
=\frac{1}{V}\left.\left[z\frac{\partial}{\partial z}\ln Z^{\rm
gr}(T,V,z)\right]\right\vert_{z=1}=g\frac{T^3}{\pi^2}\,,\nonumber\\
\varepsilon &=& \frac{1}{V}\left[-\frac{\partial}{\partial \beta}\ln Z^{\rm
gr}(T,V,z=1)\right]=g\frac{3T^4}{\pi^2}\,.
\label{CGPE}
\end{eqnarray}
Using the grand canonical potential $\Omega=-T\ln Z^{\rm gr}=-pV$ and Eq.\
(\ref{GCP}), the pressure is determined to be
\begin{eqnarray}
p&=& g\frac{T^4}{\pi^2}\,,
\end{eqnarray}
and a comparison of this expression with Eq.\ (\ref{CGPE}) yields the EoS for
an ideal gas of massles, non-interacing particles 
\begin{eqnarray}
\varepsilon&=& \frac{1}{3}p\,.
\end{eqnarray}
For a gas of massless bosons (like gluons), however, the energy density
and pressure can be calculated in the Stefan--Boltzmann limit for $m/T\ll 1$ 
(see e.g.\ Ref.\ \cite{LeBellac}) and are given by
\begin{eqnarray}
\varepsilon&=& \frac{g}{30}\pi^2T^4\,,\hspace*{2cm}p= \frac{g}{90}\pi^2T^4\,.
\end{eqnarray}

%
%
\chapter[A Microscopic Derivation of Relativistic Fluid Dynamics]
{A Microscopic Derivation of Relativistic Fluid Dynamics}
\label{AppViscousHydro}

It has been proven at RHIC that the medium created in a heavy-ion collision can be described as a
``nearly perfect liquid'' \cite{Kolb:2003dz,Romatschke:2007mq}, i.e., dissipative effects 
seem to be small, but nevertheless important for a quantitative description of the measured data.\\
The equations of relativistic fluid dynamics, in particular the transport equations for the
bulk viscous pressure $\Pi$, the heat flux current $q^\mu$, and the shear stress tensor 
$\pi^{\mu\nu}$, entering the energy-momentum tensor
\begin{eqnarray}
T^{\mu\nu}&=&\varepsilon u^\mu u^\nu - \left(p+\Pi \right)\Delta^{\mu\nu}
+q^\mu u^\nu+q^\nu u^\mu + \pi^{\mu\nu}\,,
\end{eqnarray} 
which is propagated (in time) using numerical applications, have been discussed vividly 
\cite{Luzum:2008cw,Muronga:2001zk,Muronga:2004sf,Heinz:2005bw,
Baier:2006um,Chaudhuri:2006jd,Baier:2006gy,Muronga:2006zw,Muronga:2006zx,Dusling:2007gi,
Molnar:2008xj,El:2008yy}.\\
This section discusses the derivation of the fluid-dynamical equations including
dissipative effects up second order in gradients from kinetic theory, using the Boltzmann equation
and Grad's 14-momentum \cite{Grad}.

\section[The Boltzmann Equation]{The Boltzmann Equation}

The Boltzmann equation \cite{deGroot} applies for a sufficiently dilute gas,
\begin{eqnarray}
\label{eqboltzmann}
k^\mu\partial_\mu f(k,x)&=&C[f]\,.
\end{eqnarray} 
Choosing the units to be $k_B=c=1$ and $\hbar^3=g/(2\pi)^3$ (with the degeneracy factor $g$), 
the two-particle collision term reads
\begin{eqnarray}
\label{eqcollterm}
C[f]&=&\frac{1}{2}\int\frac{d^3\vec{k}_1}{E_1}
\frac{d^3\vec{k}^\prime}{E^\prime}\frac{d^3\vec{k}^\prime_1}{E^\prime_1}\,
\sigma(s,\theta)s\,\delta^{(4)}(k+k_1-k^\prime-k^\prime_1)\nonumber\\
&&\hspace*{0.2cm}\times\{ f(k^\prime,x)f(k^\prime_1,x)[1-af(k,x)][1-af(k_1,x)]\nonumber\\
&&\hspace*{0.4cm}-f(k,x)f(k_1,x)[1-af(k^\prime,x)][1-af(k^\prime_1,x)]\}\,,
\end{eqnarray} 
where $a=\pm 1$ for fermions and bosons, respectively, and $a=0$ for classical particles.
The differential cross section of the two-particle collision, $\sigma(s,\theta)$,
only depends on the Mandelstam varable $s=(k+k_1)^2$ and the scattering angle $\theta$ in the
center-of-mass system.
The Boltzmann equation is a complicated integro-differential expression in phase space,
which however can be solved by an expansion in terms of 
its {\it moments}. These moments are obtained by multiplying the Boltzmann equation with the
factors ($1,k^\alpha,k^\alpha k^\beta$,...) and subsequent integration over momentum, leading
to different conservation equations (for details see Ref.\ \cite{DissDirk}):
\begin{itemize}
\item {\bf First moment:} The conservation of a $4$-current $N^\mu$
\begin{eqnarray}
\partial_\mu N^\mu &=& 0\,.
\end{eqnarray} 
Integration of Eq.\ (\ref{eqboltzmann}) over momentum space leads to
\begin{eqnarray}
\int\frac{d^3\vec{k}}{E}k^\mu\partial_\mu f(k,x)=
\partial_\mu\int\frac{d^3\vec{k}}{E}k^\mu f(k,x) = \int\frac{d^3\vec{k}}{E}C[f]\,.
\end{eqnarray} 
The r.h.s.\ vanishes due to the $\delta$-function in Eq.\ (\ref{eqcollterm})
and the (net) charge current is given by
\begin{eqnarray}
N^\mu &=&\int\frac{d^3\vec{k}}{E}k^\mu f(k,x)\,.
\label{currentkinetic}
\end{eqnarray} 
\item {\bf Second moment:} The conservation of the energy-momentum tensor
\begin{eqnarray}
\partial_\mu T^{\mu\nu}&=&0\,,
\end{eqnarray} 
which can be deduced from
\begin{eqnarray}
\hspace*{-0.8cm}
\int\frac{d^3\vec{k}}{E}k^\nu\,k^\mu\partial_\mu f(k,x)=
\partial_\mu\int\frac{d^3\vec{k}}{E}k^\nu\,k^\mu f(k,x) = \int\frac{d^3\vec{k}}{E}k^\nu\,C[f]\,,
\end{eqnarray} 
since the r.h.s.\ again vanishes due to the $\delta$-function in Eq.\ (\ref{eqcollterm}) 
and the definition for the energy-momentum tensor is given by
\begin{eqnarray}
T^{\mu\nu}&=&\int\frac{d^3\vec{k}}{E}k^\mu\,k^\nu\,f(k,x)\,.
\label{enmomkinetic}
\end{eqnarray} 
\item {\bf Third moment:} An Equation of Motion of
\begin{eqnarray}
\partial_\mu S^{\mu\nu\lambda}&=&C^{\nu\lambda}\,,
\end{eqnarray} 
with 
\begin{eqnarray}
\hspace*{-0.8cm}
S^{\mu\nu\lambda}=\int\frac{d^3\vec{k}}{E}k^\mu\,k^\nu\,k^\lambda\,f(k,x)
\hspace*{0.2cm}{\rm and}\hspace*{0.2cm}
C^{\nu\lambda}=\int\frac{d^3\vec{k}}{E}k^\mu\,k^\lambda\,C[f]\,.
\end{eqnarray}
\end{itemize}

\section[Ideal Hydrodynamics from Kinetic Theory]
{Ideal Hydrodynamics from Kinetic Theory}
\label{AppIdHydro}
It was already shown in section \ref{HydroEoM} that the tensor decomposition of 
the charge current $N^\mu$ and the energy-momentum tensor $T^{\mu\nu}$ can be derived by applying 
the expressions from kinetic theory. However, due to the fact that the local equilibrium 
distribution only depends on Lorentz scalars and the $4$-velocity $u^\mu$, the decomposition can 
also be done via
\begin{eqnarray}
\hspace*{-1cm}
N^\mu_0(x)&=&\int\frac{d^3\vec{k}}{E}k^\mu\,f_0(k,x)\equiv I_{10}(x)u^\mu(x)\,,\nonumber\\
\hspace*{-1cm}
T^{\mu\nu}_0(x) &=& \int\frac{d^3\vec{k}}{E}k^\mu\,k^\nu\,f_0(k,x)\equiv 
I_{20}(x)u^\mu(x)\,u^\nu(x)+I_{21}(x)\Delta_{\mu\nu}(x)\,.
\end{eqnarray}
The first index of the scalar $I_{nq}(x)$ characterizes the rank of the moment 
considered and the second one counts how often the projector onto the $3$-dimensional 
subspace 
\begin{eqnarray}
\Delta^{\mu\nu}(x)&=&g^{\mu\nu}-u^\mu(x)\,u^\nu(x)
\end{eqnarray}
appears in the expression. In particular, applying $u^\mu u_\mu=1$, $k^\mu u_\mu=E$, 
$k^\mu k_\mu=m^2$, and $m^2=E^2-\vec{k}^2$, one can show that \cite{DissDirk}
\begin{eqnarray}
I_{10}&=&N^\mu u_\mu = \int d^3\vec{k}\, f_0(k,x)\equiv n_0\,,\nonumber\\
I_{20}&=&T^{\mu\nu}u_\mu u_\nu = \int d^3\vec{k}\,E\, f_0(k,x)\equiv \varepsilon_0\,,\nonumber\\
I_{21}&=&\frac{1}{3} T^{\mu\nu}\Delta_{\mu\nu}
= -\frac{1}{3}\int d^3\vec{k}\;\frac{\vec{k}^2}{E}\,f_0(k,x) \equiv-p_0\,.
\label{decompositionidealhydro}
\end{eqnarray}
This decomposition will be important for the following calculation of the dissipative effects.

\section[Viscous Hydrodynamics from Kinetic Theory]
{Viscous Hydrodynamics from Kinetic Theory}

We apply Grad's 14-moment method \cite{Grad} for the derivation of the relativistic dissipative 
fluid-dynamical Equations of Motion which parametrizes the deviations of the one-particle
distribution function from (local) thermodynamic equilibrium. While for (local) equilibrium this
distribution function reads
\begin{eqnarray}
f_0&=&\left(e^{-y_0}+a \right)^{-1}, 
\end{eqnarray}
where $a$ again characterizes the fermion/boson or Boltzmann statistics and
\begin{eqnarray}
y_0&=&\alpha_0-\beta_0\,k^\lambda\,u_\lambda\,,
\end{eqnarray}
with $\alpha_0\equiv \mu/T$ as well as $\beta_0\equiv 1/T$, 
we assume that the following distribution function 
\begin{eqnarray}
f &=& \left(e^{-y}+a\right)^{-1}\,,
\label{DistributionFunctionGrad}\\
y&=&\alpha-\beta\,k^\lambda u_\lambda - k^\lambda v_\lambda + k^\lambda k^\rho \omega_{\lambda\rho}
\label{DistributionGrad}
\,,
\end{eqnarray}
describing the most general ellipsoidal deformation of the equilibrium distribution, 
is applicable to describe the fluid-dynamical system. Here, $v^\lambda$ is orthogonal to 
$u^\lambda$, thus $v^\lambda u_\lambda=0$, and
\begin{eqnarray}
\omega_{\lambda\rho}&=&\omega\, u_\lambda u_\rho + \omega_\lambda u_\rho
+\omega_\rho u_\lambda + \hat{\omega}_{\lambda\rho}\,,\nonumber\\
\omega &\equiv& \omega_{\lambda\rho}u^\lambda u^\rho\,,\nonumber\\
\omega_\lambda&\equiv&\Delta_{\lambda\alpha}\omega^{\alpha\rho}u_\rho\,,\nonumber\\
\hat{\omega}_{\lambda\rho}&\equiv&\Delta_{\lambda\alpha}\omega^{\alpha\beta}\Delta_{\beta\rho}\,,
\end{eqnarray}
leading to
\begin{eqnarray}
\omega_\lambda u^\lambda&=&0\,,\nonumber\\
\hat{\omega}_{\lambda\rho}u^\lambda&=&0\,,\nonumber\\
\omega^\lambda_\lambda &=&0\,,\nonumber\\
\omega&=&-\hat{\omega}^\lambda_\lambda\,.
\label{defomega}
\end{eqnarray}
Inserting these definitions into Eq.\ (\ref{DistributionGrad}), one obtains
\begin{eqnarray}
y&=&\alpha - \beta E+\vec{v}\cdot\vec{k}+\omega E^2 - 2\,E\,\vec{\omega}\cdot\vec{k}
+\sum\limits_{i,j=x,y,z} k^i \hat{\omega}_{ij}\,k^j
\end{eqnarray}
in the local rest frame, $u^\mu=(1,\vec{0})$, where $\vec{v}$ denotes the vector component of 
the $4$-vector $v^\lambda$. \\
Hence, Eq.\ (\ref{DistributionFunctionGrad}) is a function of 17 independent
parameters (including the constraints for $u_\lambda$ and $\omega_{\lambda\rho}$). 
To connect these quantities to the 14 independent parameters of the hydrodynamical
equations\footnote{The 14 independent parameters are $\varepsilon(1\,{\rm Eq.})$, 
$n(1\,{\rm Eq.})$, $\Pi(1\,{\rm Eq.})$, $V^\mu(3\,{\rm Eqs.})$, $q^\mu(3\,{\rm Eqs.})$, and 
$\pi^{\mu\nu}(5\,{\rm Eqs.})$, see also chapter 
 \ref{ViscousHydrodynamics}.}, we assume a small deformation of the equilibrium distribution
\begin{eqnarray}
\hspace*{-0.5cm}
y-y_0&=&(\alpha-\alpha_0)-(\beta-\beta_0)k^\lambda u_\lambda-k^\lambda v_\lambda 
+ k^\lambda k^\rho\omega_{\lambda\rho} \equiv O(\delta)\,,
\end{eqnarray}
allowing for a linearization around equilibrium 
\begin{eqnarray}
f(y)&=& f(y_0)+f^\prime (y_0)(y-y_0)+O(\varepsilon^2)\nonumber\\
&=&f_0+f_0(1-af_0)(y-y_0)+O(\varepsilon^2)\,.
\label{flinearization}
\end{eqnarray}
Here, the fluid velocity $u^\lambda$ of the system in (local) thermodynamical equilibrium 
is supposed to be the same as the velocity of the dissipative system.\\
Inserting these expressions into Eqs.\ (\ref{currentkinetic}) and (\ref{enmomkinetic}), the
(net) charge current and the energy-momentum tensor become
\begin{eqnarray}
\hspace*{-1cm}
N^\mu&=& N_0^\mu + (\alpha-\alpha_0) J_0^\mu-(\beta-\beta_0)J_0^{\mu\lambda}u_\lambda
-J_0^{\mu\lambda}v_\lambda+J_0^{\mu\lambda\rho}\omega_{\lambda\rho}\,,
\label{tensordecN}\\
\hspace*{-1cm}
T^{\mu\nu}&=&T_0^{\mu\nu}+(\alpha-\alpha_0)J_0^{\mu\nu}-(\beta-\beta_0)J_0^{\mu\nu\lambda}u_\lambda
-J_0^{\mu\nu\lambda}v_\lambda+J_0^{\mu\nu\lambda\rho}\omega_{\lambda\rho}\,,
\label{tensordecTmunu}
\end{eqnarray}
with the equilibrium distribution function for the (net) charge current $N_0^\mu$ and the 
energy-momentum tensor $T_0^{\mu\nu}$. The tensors
\begin{eqnarray}
J_0^{\alpha_1...\alpha_n}&\equiv&\int\frac{d^3\vec{k}}{E}k^{\alpha_1}...\; k^{\alpha_n}\,f_0(1-af_0)
\end{eqnarray}
can be decomposed into invariant subspaces according to
\begin{eqnarray}
J_0^\mu&=&J_{10}u^\mu\,,\\
J_0^{\mu\lambda}&=&J_{20}u^\mu u^\lambda + J_{21}\Delta^{\mu\lambda}\,,\\
J_0^{\mu\lambda\rho}&=&J_{30}u^\mu u^\lambda u^\rho + 3 J_{31} u^{(\mu}\Delta^{\lambda\rho)}\,,\\
J_0^{\mu\nu\lambda\rho}&=&J_{40}u^\mu u^\nu u^\lambda u^\rho 
+ 6 J_{41}u^{(\mu}u^\nu \Delta^{\lambda\rho)}
+3J_{42}\Delta^{\mu(\nu}\Delta^{\lambda\rho)}\,,
\end{eqnarray}
with the following abbreviations characterizing symmetry properties
\begin{eqnarray}
3 u^{(\mu}\Delta^{\lambda\rho)}&=& u^\mu\Delta^{\lambda\rho}+u^\lambda\Delta^{\rho\mu}
+u^\rho\Delta^{\mu\lambda}\,,\\
6 u^{(\mu}u^\nu \Delta^{\lambda\rho)}&=&u^\mu u^\nu\Delta^{\lambda\rho}
+u^\nu u^\lambda\Delta^{\rho\mu}+u^\lambda u^\rho\Delta^{\mu\nu}\nonumber\\
&&\hspace*{-0.3cm}+u^\mu u^\lambda\Delta^{\nu\rho}+u^\mu u^\rho\Delta^{\nu\lambda}
+u^\nu u^\rho\Delta^{\mu\lambda}\,.
\end{eqnarray}
These expressions are a generalization of 
\begin{eqnarray}
2u^{(\mu}u^{\nu)}&\equiv& u^\mu u^\nu + u^\nu u^\mu\,.
\end{eqnarray}
The functions $J_{nq}$ again denote the projections onto the corresponding subspaces (see section
\ref{AppIdHydro}). They only depend on the fluid velocity $u^\lambda$ and equilibrium quantities $T$
and $\mu$. \\
Applying the above tensor decompositions together with the expressions for $N_0^\mu$, 
$T_0^{\mu\nu}$, and $\omega_{\lambda\rho}$, it follows that the (net) charge current $N^\mu$ and 
the energy-momentum tensor $T^{\mu\nu}$ [Eqs.\ (\ref{tensordecN}) and (\ref{tensordecTmunu})] are 
given by
\begin{eqnarray}
N^\mu&=&\big[n_0+(\alpha-\alpha_0)J_{10}-(\beta-\beta_0)J_{20}+\omega(J_{30}-J_{31})\big]u^\mu 
\nonumber\\
&&\hspace*{-0.3cm}
- J_{21}v^\mu + 2J_{31}\omega^\mu\,,
\label{NDec}\\
T^{\mu\nu}&=&\big[\varepsilon_0+(\alpha-\alpha_0)J_{20}-
(\beta-\beta_0)J_{30}+\omega(J_{40}-J_{41})\big]u^\mu u^\nu\nonumber\\
&&\hspace*{-0.3cm}
+\big[-p_0+(\alpha-\alpha_0)J_{21}-(\beta-\beta_0)J_{31}+\omega(J_{41}-J_{42})\big]
\Delta^{\mu\nu}
\nonumber\\
&&\hspace*{-0.3cm}
-2\left(J_{31}v^{(\mu}-2J_{41}\omega^{(\mu} \right)u^{\nu)}+2J_{42}\hat{\omega}^{\mu\nu}
\label{TmunuDec}\,.
\end{eqnarray}
These equations contain 19 unknown quantities, the 17 originating from the above
parametrization of the distribution function as well as $\alpha_0$ and $\beta_0$. The latter ones
are normally determined by assuming that the charge density $n$ and the energy density 
$\varepsilon$ characterize a system in thermodynamic equilibrium,
\begin{eqnarray}
\label{energydensitylocequ}
\varepsilon &=& T^{\mu\nu}u_\mu u_\nu\equiv\varepsilon_0\\
\label{chargedensitylocequ}
n&=&N^\mu u_\mu\equiv n_0\,.
\end{eqnarray}
Then, the extra terms in the first lines of Eqs.\ 
(\ref{NDec}) and (\ref{TmunuDec}) have to vanish, leading via
\begin{eqnarray}
(\alpha-\alpha_0)J_{10}-(\beta-\beta_0)J_{20}+\omega(J_{30}-J_{31})&=&0\,,\\
(\alpha-\alpha_0)J_{20}-(\beta-\beta_0)J_{30}+\omega(J_{40}-J_{41})&=&0\,,
\end{eqnarray}
to the expressions for $(\alpha-\alpha_0)$ and $(\beta-\beta_0)$,
\begin{eqnarray}
\alpha-\alpha_0
&=&-\omega\left[m^2-4\frac{J_{30}J_{31}-J_{20}J_{41}}{J_{30}J_{10}-(J_{20})^2}\right]\,,
\label{alphaalpha0}\\
\beta-\beta_0&=&-4\omega\frac{J_{10}J_{41}-J_{20}J_{31}}{J_{30}J_{10}-(J_{20})^2}\,.
\label{betabeta0}
\end{eqnarray}
Here, the relation \cite{DissDirk}
\begin{eqnarray}
J_{n+2,q}&=&m^2J_{nq}-(2q+3)J_{n+2,q+1}
\label{DefinitionJ}
\end{eqnarray}
needs to be applied. In order to proceed, one has to determine the fluid velocity which we will
first consider in the Eckart frame.

\subsection[Transport Equations in the Eckart Frame]
{Transport Equations in the Eckart Frame}
\label{TransportEqEckart}

In the Eckart system, $u^\mu$ defines the velocity of charge transport. Thus, the (net) charge
density $n$ is equivalent to the one in thermodynamic equilibrium, $n=n_0$, resulting in
\begin{eqnarray}
N^\mu&=&n u^\mu
\end{eqnarray}
and [cf.\ Eq.\ (\ref{NDec})] a condition for $v^\mu$
\begin{eqnarray}
v^\mu&=&2\frac{J_{31}}{J_{21}}\omega^\mu\,,
\label{vmu}
\end{eqnarray}
defining three of the 17 free parameters. The remaining 14 quantities, that are 
determined below, can be connected to the unknowns of the hydrodynamic equations. \\
Inserting $(\alpha-\alpha_0)$ [Eq.\ (\ref{alphaalpha0})], $(\beta-\beta_0)$ 
[Eq.\ (\ref{betabeta0})], and $v^\mu$ [Eq.\ (\ref{vmu})] into the energy-momentum tensor 
$T^{\mu\nu}$ [Eq.\ (\ref{TmunuDec})], the dissipative quantities can be defined as
\begin{eqnarray}
\hspace*{-0.6cm}
\Pi &=& 4\omega\Big[J_{21}\frac{J_{20}J_{41}-J_{30}J_{31}}{J_{30}J_{10}-(J_{20})^2}
+J_{31}\frac{J_{20}J_{31}-J_{10}J_{41}}{J_{30}J_{10}-(J_{20})^2}
+\frac{5}{3}J_{42}\Big]\,,\label{dissPi}\\
\hspace*{-0.6cm}
q^\mu&=&2\frac{J_{41}J_{21}-(J_{31})^2}{J_{21}}\omega^\mu\,,\label{dissq}\\
\hspace*{-0.6cm}
\pi^{\mu\nu}&=& 2J_{42}\tilde{\omega}^{\mu\nu}\label{disspimunu}\,,
\end{eqnarray}
having in mind that 
\begin{eqnarray}
T^{\mu\nu}&=&\varepsilon u^\mu u^\nu - p\Delta^{\mu\nu} +2 q^{(\mu}u^{\nu)}+\pi^{\mu\nu}\,,\\
p&\equiv&-\frac{1}{3}T^{\mu\nu}\Delta_{\mu\nu}\equiv p_0+\Pi\,,\\
q^\mu&\equiv&\Delta^{\mu\alpha}T_{\alpha \beta}u^\beta\,,\\
\pi^{\mu\nu}&\equiv&T^{\langle\mu\nu\rangle}\equiv
\left(\Delta_\alpha^{\,\,\,(\mu}
\Delta^{\nu)}_{\,\,\,\beta}-\frac{1}{3}\Delta_{\alpha\beta}\Delta^{\mu\nu}
\right)T^{\alpha\beta}\,,
\end{eqnarray}
and using the relations
\begin{eqnarray}
\tilde{\omega}^{\mu\nu}&=&\hat{\omega}^{\mu\nu}+\frac{\omega}{3}\Delta^{\mu\nu}\,,\\
\tilde{\omega}^{\lambda\rho}&=&\hat{\omega}^{\langle\lambda\rho\rangle}\,.
\end{eqnarray}
Consequently, there is a direct connection between the nine free parameters of $\omega$,
$\omega^\lambda$, and $\tilde{\omega}^{\lambda\rho}$ as well as the hydrodynamic variables 
$\Pi$, $q^\mu$, and $\pi^{\mu\nu}$ which are defined by nine equations. $\alpha$, $\beta$, and 
$u^\lambda$ have already been determined via the charge density, the energy density, and the 
fluid velocity.\\
Therefore, one has to derive nine additional equations for those undefined para\-meters. This is
done by calculating the second moment of the Boltzmann equation that can be written in terms of
\begin{eqnarray}
\label{Smunulambda}
S^{\mu\nu\lambda}&=&S_0^{\mu\nu\lambda}+(\alpha-\alpha_0)J_0^{\mu\nu\lambda}
-(\beta-\beta_0)J_0^{\mu\nu\lambda\rho}u_\rho\nonumber\\
&&-J_0^{\mu\nu\lambda\rho}v_\rho
+J_0^{\mu\nu\lambda\alpha\beta}\omega_{\alpha\beta}\,,
\end{eqnarray}
applying the linearization of Eq.\ (\ref{flinearization}),
where $J_0^{\mu\nu\lambda\alpha\beta}$ denotes the decomposition
\begin{eqnarray}
J_0^{\mu\nu\lambda\alpha\beta}&=&J_{50}u^\mu u^\nu u^\lambda u^\alpha u^\beta
+10 J_{51}u^{(\mu} u^\nu u^\lambda \Delta^{\alpha\beta)}\nonumber\\
&&15J_{52}u^{(\mu}\Delta^{\nu\lambda}\Delta^{\alpha\beta)}\,.
\end{eqnarray}
Using again the expressions for $(\alpha-\alpha_0)$ [Eq.\ (\ref{alphaalpha0})],
$(\beta-\beta_0)$ [Eq.\ (\ref{betabeta0})], Eqs.\ (\ref{dissPi}) -- (\ref{disspimunu})
as well as the relation
\begin{eqnarray}
S_0^{\mu\nu\lambda}&=&I_{30}u^\mu u^\nu u^\lambda + 3I_{31}u^{(\mu}\Delta^{\nu\lambda)}\,,
\end{eqnarray}
one obtains
\begin{eqnarray}
S^{\mu\nu\lambda}&=& S_1 u^\mu u^\nu u^\lambda + 3S_2 u^{(\mu}\Delta^{\nu\lambda)}
+3\psi_1 q^{(\mu}\left[u^\nu u^{\lambda)}-\frac{1}{5}\Delta^{\nu\lambda)}\right]\nonumber\\
&&+3\psi_2\pi^{(\mu\nu}u^{\lambda)}\,,\label{Smunulambda1}\\
S_1&=&I_{30}+\phi_1\Pi\,,\\
S_2&=&I_{31}+\phi_2\Pi\,,\\[2ex]~
\label{eqphi1}
\phi_1&=&
\frac{J_{30}(J_{30}J_{31}-J_{20}J_{41})+J_{40}(J_{10}J_{41}-J_{20}J_{31})
-J_{51}(J_{30}J_{10}-(J_{20})^2)}
{J_{21}(J_{20}J_{41}-J_{30}J_{31})+J_{31}(J_{20}J_{31}-J_{10}J_{41})
+\frac{5}{3}J_{42}(J_{30}J_{10}-(J_{20})^2)}\,,\nonumber\\
\nonumber\\\\
\phi_2&=&-\frac{\phi_1}{3}\,,\\
\psi_1&=&\frac{J_{21}J_{51}-J_{31}J_{41}}{J_{41}J_{21}-(J_{31})^2}\,,\\
\psi_2&=&\frac{J_{52}}{J_{42}}\,.
\end{eqnarray}
Now, the nine requested equations can be derived by evaluating $\partial_\mu S^{\mu\nu\lambda}$
\begin{eqnarray}
\partial_\mu S^{\mu\nu\lambda}&=&
u^\nu u^\lambda\left[\left(\dot{S}_1 -2\dot{S}_2\right)+\left(S_1-2S_2\right)\theta
+\partial_\mu\left(\psi_1 q^\mu\right)\right]\nonumber\\
&&+\Delta^{\nu\lambda}\left[\dot{S}_2
+S_2\theta-\frac{1}{5}\partial_\mu\left(\psi_1 q^\mu\right) \right]\nonumber\\
&&+2(S_1-3S_2)\dot{u}^{(\nu}u^{\lambda)}+2u^{(\nu}\partial^{\lambda)}S_2
+2S_2\partial^{(\nu}u^{\lambda)}\nonumber\\
&&+\frac{6}{5}\psi_1 q^\mu\partial_\mu(u^\nu u^\lambda)\nonumber\\
&&+\frac{12}{5}\left\{q^{(\nu}u^{\lambda)}\left[\dot{\psi}_1+\psi_1\theta\right]
+\psi_1\left[\dot{u}^{(\lambda}q^{\nu)}+u^{(\lambda}\dot{q}^{\nu)}\right]\right\}\nonumber\\
&&-\frac{2}{5}\left[q^{(\nu}\partial^{\lambda)}\psi_1 + \psi_1\partial^{(\lambda}q^{\nu)} \right]
+\dot{\psi}_2\pi^{\nu\lambda}+\psi_2 \left(\dot{\pi}^{\nu\lambda}+\pi^{\nu\lambda}\theta \right)
\nonumber\\
&&+2\pi^{\mu(\nu}u^{\lambda)}\partial_\mu\psi_2+2\psi_2\pi^{\mu(\nu}\partial_\mu u^{\lambda)}
+2\psi_2 u^{(\nu}\partial_\mu \pi^{\lambda )\mu}\nonumber\\
&\equiv& C^{\nu\lambda}\,,
\end{eqnarray}
with $\dot{S}=u^\mu\partial_\mu S$ and $\theta=\partial_\mu u^\mu$. In principle, these are $10$ 
equations, but the trace
\begin{eqnarray}
\partial_\mu S^{\mu\nu}_{\nu}&=&m^2\partial_\mu N^\mu =C^\nu_\nu=m^2\int\frac{d^3\vec{k}}{E}C[f]=0
\end{eqnarray}
leads to the equation of charge conservation. \\
Projecting $\partial_\mu S^{\mu\nu\lambda}$ onto the invariant subspaces results in the 
following Equations of Motions:
\begin{eqnarray}
\label{PiPrePreliminary}
\hspace*{-1.0cm}
C^{\nu\lambda}u_\nu u_\lambda&=&
\dot{I}_{30}+\phi_1\dot{\Pi}+\dot{\phi}_1\Pi+\left( I_{30}-2I_{31}
+\frac{5}{3}\phi_1\Pi\right)\theta\nonumber\\\hspace*{-1.0cm}
&&+\partial_\mu\left(\psi_1 q^\mu\right)
-2\psi_1 q_\mu \dot{u}^\mu-2\psi_2 \pi^{\lambda\mu}\partial_\mu u_\lambda\,,\\
\hspace*{-1.0cm}
u_\nu\Delta^\mu_{\;\;\lambda} C^{\nu\lambda}\hspace*{-0.25cm}&=&
\left(I_{30}-2I_{31}+\frac{5}{3}\phi_1\Pi\right)\dot{u}^\mu+q^\mu\left(\dot{\psi}_1
+\frac{6}{5}\psi_1\theta
\right)\nonumber\\\hspace*{-1.0cm}
&&+\Delta^\mu_\lambda\left[\partial^\lambda\left(I_{31}+\phi_2\Pi\right)
+\frac{1}{5}\psi_1 q^\nu\partial^\lambda u_\nu
+\psi\dot{q}^\lambda+\psi_2\partial_\nu \pi^{\lambda\nu}\right]\nonumber\\\hspace*{-1.0cm}
&&+\frac{6}{5}\psi_1q^\nu\partial_\nu u^\mu
+\left(-\psi_2\dot{u}_\nu+\partial_\nu\psi_2\right)\pi^{\mu\nu}\,,\\
\label{pimunuPrePreliminary}
\hspace*{-1.0cm}
C^{\langle\nu\lambda\rangle}&=&
2\left(I_{31}+\phi_2\Pi\right)\partial^{\langle\nu}u^{\lambda\rangle}
+\frac{12}{5}\psi_1\dot{u}^{\langle\nu}q^{\lambda\rangle}\nonumber\\\hspace*{-1.0cm}
&&-\frac{2}{5}\left(q^{\langle\nu}\partial^{\lambda\rangle}\psi_1
+\psi_1\partial^{\langle\lambda}q^{\nu\rangle}\right)
+\dot{\psi}_2\pi^{\nu\lambda}\nonumber\\\hspace*{-1.0cm}
&&+\psi_2\left(\dot{\pi}^{\langle\nu\lambda\rangle}+\pi^{\nu\lambda}\theta\right)
+2\psi_2\pi^{\mu\langle\nu}\partial_\mu u^{\lambda\rangle}\,.
\end{eqnarray}
For the evaluation of the l.h.s., the collision term $C^{\nu\lambda}$ has to be linearized
\begin{eqnarray}
C^{\nu\lambda}&=&\frac{1}{2}
\int\frac{d^3\vec{k}}{E}\,\int\frac{d^3\vec{k}_1}{E_1}\,
\int\frac{d^3\vec{k}^\prime}{E^\prime}\,\int\frac{d^3\vec{k}^\prime_1}{E^\prime_1}
\,\sigma(s,\theta)\,s\, \nonumber\\
&&\times \delta^{(4)}(k+k_1-k^\prime-k^\prime_1)\,k^\nu\,k^\lambda\nonumber\\&&
\times\big\{f^\prime(k,x)f^\prime_1(k,x)[1-af(k,x)][1-af_1(k,x)]\nonumber\\
&&\hspace*{0.3cm}-f(k,x)f_1(k,x)[1-af^\prime(k,x)][1-af^\prime_1(k,x)]\big\}\nonumber\\
&\equiv&\omega_{\alpha\beta}C^{\nu\lambda\alpha\beta}+O(\varepsilon^2)\,.
\end{eqnarray}
The most general decomposition of $C^{\nu\lambda\alpha\beta}$ is given by
\begin{eqnarray}
C^{\nu\lambda\alpha\beta}&=&\frac{A}{3}\left(3u^\nu u^\lambda u^\alpha u^\beta
-u^\nu u^\lambda\Delta^{\alpha\beta}-u^\alpha u^\beta\Delta^{\nu\lambda}
+\frac{1}{3}\Delta^{\nu\lambda}\Delta^{\alpha\beta}\right)\nonumber\\
&&\hspace*{-0.3cm}+\frac{B}{5}\Delta^{\alpha\langle\lambda}\Delta^{\nu\rangle\beta}
+4Cu^{(\nu}\Delta^{\lambda)(\alpha}u^{\beta)}\,,
\end{eqnarray}
allowing for the calculation of $C^{\nu\lambda}$
\begin{eqnarray}
C^{\nu\lambda}&=&\omega_{\alpha\beta}C^{\nu\lambda\alpha\beta}\nonumber\\
&=&\frac{4}{3}A\omega\left(u^\nu u^\lambda-\frac{1}{3}\Delta^{\nu\lambda}\right)
+\frac{B}{5}\tilde{\omega}^{\nu\lambda}+4Cu^{(\nu}\omega^{\lambda)}\,.
\end{eqnarray}
Thus, the left-hand sides give
\begin{eqnarray}
u_\nu u_\lambda C^{\nu\lambda}&=&\frac{4}{3}A\omega\equiv A^\prime\Pi\,,\\
u_\nu\Delta^\mu_\lambda C^{\nu\lambda}&=&2C\omega^{\mu}\equiv C^\prime q^\mu\,,\\
C^{\langle\nu\lambda\rangle}&=&\frac{B}{5}
\tilde{\omega}^{\nu\lambda}\equiv B^\prime\pi^{\nu\lambda}\,,
\end{eqnarray}
resulting in 
\begin{eqnarray}
A^\prime&=&\frac{A}{3}\left[J_{31}J_{10}-(J_{20})^2\right]\times\label{eqAprime}\\
&&\left\{J_{21}(J_{20}J_{41}-J_{30}J_{31})+J_{31}(J_{20}J_{31}-J_{10}J_{41})
+\frac{5}{3}J_{42}[J_{30}J_{10}-(J_{20})^2]\right\}^{-1}\hspace*{-0.2cm},\nonumber\\
C^\prime&=&C\frac{J_{21}}{J_{41}J_{21}-(J_{31})^2}\,,\label{DefCPrime}\\
B^\prime&=&\frac{B}{10 J_{42}}\,.
\end{eqnarray}
Therefore, the {\it Equations of Motions for the dissipative variables} $\Pi$, $q^\mu$, and 
$\pi^{\mu\nu}$ are
\begin{eqnarray}
\hspace*{-0.9cm}
A^\prime\Pi&=&
\dot{I}_{30}+\phi_1\dot{\Pi}+\dot{\phi}_1\Pi+\left( I_{30}
-2I_{31}+\frac{5}{3}\phi_1\Pi\right)\theta\nonumber\\\hspace*{-0.9cm}
&&+\partial_\mu\left(\psi_1 q^\mu\right)
-2\psi_1 q_\mu \dot{u}^\mu-2\psi_2 \pi^{\lambda\mu}\partial_\mu u_\lambda\,,
\label{PiPreliminary}\\
\hspace*{-0.9cm}
C^\prime q^\mu&=&
\left(I_{30}-2I_{31}
+\frac{5}{3}\phi_1\Pi\right)\dot{u}^\mu+q^\mu\left(\dot{\psi}_1+\frac{6}{5}\psi_1\theta
\right)\nonumber\\\hspace*{-0.9cm}
&&+\Delta^\mu_\lambda\left[\partial^\lambda\left(I_{31}
+\phi_2\Pi\right)+\frac{1}{5}\psi_1 q^\nu\partial^\lambda u_\nu
+\psi_1\dot{q}^\lambda+\psi_2\partial_\nu \pi^{\lambda\nu}\right]\nonumber\\\hspace*{-0.9cm}
&&+\frac{6}{5}\psi_1q^\nu\partial_\nu u^\mu
+\left(-\psi_2\dot{u}_\nu+\partial_\nu\psi_2\right)\pi^{\mu\nu}\,,
\label{qPreliminary}\\
\hspace*{-0.9cm}
B^\prime \pi^{\mu\nu}&=&
2\left(I_{31}+\phi_2\Pi\right)\partial^{\langle\mu}u^{\nu\rangle}
+\frac{12}{5}\psi_1\dot{u}^{\langle\mu}q^{\nu\rangle}\nonumber\\\hspace*{-0.9cm}
&&-\frac{2}{5}\left(q^{\langle\mu}\partial^{\nu\rangle}\psi_1
+\psi_1\partial^{\langle\mu}q^{\nu\rangle}\right)
+\dot{\psi}_2\pi^{\mu\nu}\nonumber\\\hspace*{-0.9cm}
&&+\psi_2\left(\dot{\pi}^{\langle\mu\nu\rangle}+\pi^{\mu\nu}\theta\right)
+2\psi_2\pi_\lambda^{\langle\mu}\partial^\lambda u^{\nu\rangle}\,.
\label{pimunuPreliminary}
\end{eqnarray}
To rewrite these equations into a more common form containing the relaxation times 
($\tau_\Pi, \tau_q, \tau_\pi$), the coupling lengths coefficients 
($l_{\Pi q}, l_{\pi q}, l_{\Pi\pi}, ...$) as well as the thermodynamic quantities bulk viscosity 
$\zeta$, thermal conductivity $\kappa$, and shear viscosity $\eta$, several mathematical 
operations are necessary.\\
A crucial step is to rewrite the scalar functions $I_{nq}(x)$. Following Israel and Stewart 
\cite{IS}, their derivatives are connected to the $J_{nq}$ via
\begin{eqnarray}
dI_{nk}&=&J_{nk}d\alpha - J_{n+1,k}d\beta\,.
\end{eqnarray}
The expressions for $d\alpha$ and $d\beta$ can be determined using 
the equations for the (net) charge as well as energy and momentum conservation,
Eqs.\ (\ref{ViscousChargeConservation}) to (\ref{ViscousMomentumConservation}), since \cite{IS}
\begin{eqnarray}
dn&=&J_{10}d\alpha - J_{20}d\beta\,,\\
d\varepsilon&=&J_{20}d\alpha - J_{30}d\beta\,.
\end{eqnarray}
Applying the definition \cite{IS}
\begin{eqnarray}
D_{nq}&=&J_{n+1,q}J_{n-1,q}-(J_{nq})^2
\end{eqnarray}
it follows that
\begin{eqnarray}
d\beta&=&\frac{1}{D_{20}}\Big(J_{20}dn-J_{10}d\varepsilon \Big)\,,\\
d\alpha&=&\frac{1}{D_{20}}\Big(J_{30}dn-J_{20}d\varepsilon \Big)\,.
\end{eqnarray}
Making use of the relations
\begin{eqnarray}
\pi_\lambda^{\langle\mu}\partial^\lambda u^{\nu\rangle}&=&
-2\pi_\lambda^{\langle\mu}\omega^{\nu\rangle\lambda}
+\pi_\lambda^{\langle\mu}\sigma^{\nu\rangle\lambda}\,,\\
\sigma^{\mu\nu}&=&\nabla^{\langle\mu}u^{\nu\rangle}\,,
\end{eqnarray}
the equation for the vorticity,
\begin{eqnarray}
\omega^{\mu\nu}&=&\frac{1}{2}\Delta^{\mu\alpha}
\Delta^{\nu\beta}(\partial_\alpha u_\beta - \partial_\beta u_\alpha)\,,
\end{eqnarray}
and the fact that the relaxation times as well as coupling lengths are connected to the bulk 
viscosity $\zeta$, the thermal conductivity $\kappa$, and the shear viscosity $\eta$ 
\cite{Muronga:2003ta}, the {\bf transport equations for the bulk viscous pressure $\Pi$, for 
the heat flux current $q^\mu$, and the shear stress tensor $\pi^{\mu\nu}$} are given by
\begin{eqnarray}
\hspace*{-1cm}\Pi&=&-\zeta\theta - \tau_\Pi\dot{\Pi}
+\tau_{\Pi q}q_\mu \dot{u}^\mu-l_{\Pi q}\nabla_\mu q^\mu - \zeta\hat{\delta}_0\Pi\theta\nonumber\\
\hspace*{-1.1cm}&&+\lambda_{\Pi q}q^\mu\nabla_\mu \alpha
+\lambda_{\Pi\pi}\pi_{\mu\nu}\sigma^{\mu\nu}\,,
\label{TransportPi}\\
\hspace*{-1cm}q^\mu&=&\frac{\kappa}{\beta}\frac{n}{\beta(\varepsilon+p)}\nabla^\mu \alpha
-\tau_q\Delta^{\mu\nu}\dot{q}_\nu\nonumber\\
\hspace*{-1.1cm}&&
-\tau_{q\Pi}\Pi\dot{u}^\mu-\tau_{q\pi}\pi^{\mu\nu}\dot{u}_\nu
+l_{q\Pi}\nabla^\mu\Pi-l_{q\pi}\Delta^{\mu\nu}\partial^{\lambda}\pi_{\nu\lambda}
+\tau_q\omega^{\mu\nu}q_\nu\nonumber\\
\hspace*{-1.1cm}&&-\frac{\kappa}{\beta}\hat{\delta}_1
q^\mu\theta-\lambda_{qq}\sigma^{\mu\nu}q_\nu+\lambda_{q\Pi}\Pi\nabla^\mu\alpha
+\lambda_{q\pi}\pi^{\mu\nu}\nabla_\nu\alpha\,,
\label{Transportq}\\
\hspace*{-1cm}\pi^{\mu\nu}&=&2\eta\sigma^{\mu\nu}
-\tau_\pi \dot{\pi}^{\langle\mu\nu\rangle}\nonumber\\
\hspace*{-1.1cm}&&+2\tau_{\pi q}q^{\langle\mu}\dot{u}^{\nu\rangle}
+2l_{\pi q}\nabla^{\langle\mu}q^{\nu\rangle}+4\tau_\pi
\pi_\lambda^{\langle\mu}\omega^{\nu\rangle\lambda}
-2\eta\hat{\delta}_2\theta\pi^{\mu\nu}\nonumber\\
\hspace*{-1.1cm}&&-2\tau_\pi \pi_\lambda^{\langle\mu}\sigma^{\nu\rangle\lambda}
-2\lambda_{\pi q}q^{\langle\mu}\nabla^{\nu\rangle}\alpha+2\lambda_{\pi\Pi}\Pi\sigma^{\mu\nu}\,.
\label{Transportpimunu}
\end{eqnarray}
The transport coefficients $\zeta, \kappa, \eta$, the relaxation times 
$\tau_\Pi, \tau_q, \tau_\pi$, 
and the coefficients $\tau_{\Pi q},\tau_{q \Pi}, \tau_{q \pi},\tau_{\pi q}$, 
$\ell_{\Pi q}, \ell_{q \Pi}, \ell_{q \pi},\ell_{\pi q}, \lambda_{\Pi q}$, 
$\lambda_{\Pi \pi}, \lambda_{q q}, \lambda_{q \Pi}, \lambda_{q \pi}, \lambda_{\pi q},
\lambda_{\pi \Pi}$, $\hat{\delta}_0$, $\hat{\delta}_1$, $\hat{\delta}_2$
are (complicated) functions of $\alpha,\beta$ and will be presented in detail in Ref.\ 
\cite{ViscousBetz}.\\
It can indeed be shown that using [see Eq.\ (\ref{eq:etas})]
\begin{eqnarray}
\frac{\ell_{\rm mfp}}{\lambda_{\rm th}}\sim\ell_{\rm mfp}\partial_\mu \sim\delta\ll 1\,,
\label{PowerCounting1}
\end{eqnarray}
and [cf.\ Eq.\ (\ref{eq:Piepsilon})]
\begin{eqnarray}
\frac{\Pi}{\varepsilon}\sim\frac{q^\mu}{\varepsilon}\sim\frac{\pi^{\mu\nu}}{\varepsilon}\sim 
K\sim\delta\ll 1\,,
\label{PowerCounting2}
\end{eqnarray}
where the Knudsen number $K$ is defined via Eq.\ (\ref{eqKnudsenNumber}), all terms in the above 
equations are at most of order $\delta^2$. Therefore, the set of equations (\ref{TransportPi}) 
to (\ref{Transportpimunu}) is of {\bf second order in the Knudsen number}.

\subsection[The Tsumura-Kunihiro-Ohnishi Matching Condition]
{The Tsumura-Kunihiro-Ohnishi Matching Condition}

The above mentioned transport equations for the bulk viscous pressure $\Pi$, 
the heat flux current $q^\mu$, and the shear stress tensor $\pi^{\mu\nu}$ 
were derived by assuming [see Eqs.\ (\ref{energydensitylocequ}) and (\ref{chargedensitylocequ})] 
that the actual charge density $n$ and the energy density $\varepsilon$ are equivalent to the 
respective quantities in (local) thermodynamic equilibrium. However, Tsumura, Kunihiro, and 
Ohnishi \cite{Tsumura:2007ji} proposed that
\begin{eqnarray}
\varepsilon=T^{\mu\nu}u_\mu u_\nu+3 \Pi\,,
\end{eqnarray}
leading to a matching condition of
\begin{eqnarray}
\varepsilon&=&\varepsilon_0-3\Pi\,.
\end{eqnarray}
In that case, ($\alpha-\alpha_0$) and ($\beta-\beta_0$) are not given by Eqs.\ (\ref{alphaalpha0}) 
and (\ref{betabeta0}), but can be calculated to be
\begin{eqnarray}
\hspace*{-0.5cm}\alpha-\alpha_0 &=&-\omega\left[m^2-4\frac{J_{30}J_{31}-J_{20}J_{41}}
{J_{30}J_{10}-(J_{20})^2}\right]+\frac{J_{20}}{J_{30}J_{10}-(J_{20})^2}3\Pi\,,\\
\hspace*{-0.5cm}\beta-\beta_0&=&-4\omega\frac{J_{10}J_{41}-J_{20}J_{31}}
{J_{30}J_{10}-(J_{20})^2}+\frac{J_{10}}{J_{30}J_{10}-(J_{20})^2}3\Pi\,,
\end{eqnarray}
resulting in a bulk viscous pressure of
\begin{eqnarray}
\Pi&=&4\omega\left\{\frac{J_{21}(J_{20}J_{41}-J_{30}J_{31})+J_{31}(J_{20}J_{31}-J_{10}J_{41})
+\frac{5}{3}J_{42}[J_{30}J_{10}-(J_{20})^2]}
{J_{10}J_{30}-(J_{20})^2+3J_{20}J_{21}-3J_{10}J_{31}}\right\}\,\nonumber\\
\end{eqnarray}
However, Eqs.\ (\ref{dissq}) and (\ref{disspimunu}) for the heat flux current $q^\mu$ and 
shear stress tensor $\pi^{\mu\nu}$ still apply. Nevertheless, as one can easily prove, the second 
moment of the Boltzmann equation, Eq.\ (\ref{Smunulambda1}), as well as all those ones derived 
from this expression, in particular the transport equations Eqs.\ (\ref{TransportPi}) -- 
(\ref{Transportpimunu}), are still valid using the above definition for the bulk viscous pressure 
$\Pi$. Of course, $\phi_1$ [see Eq.\ (\ref{eqphi1})] and $A^\prime$ [see Eq.\ (\ref{eqAprime})] 
have to be adjust accordingly.

\subsection[Transport Equations in the Landau Frame]
{Transport Equations in the Landau Frame}
So far, we considered $u^\lambda$ as the charge flow velocity, i.e., we worked in the Eckart 
frame. Choosing $u^\lambda$ to be the energy flow velocity, the heat flux current vanishes 
($q^\mu=0$ since the spectator moves with the energy flow), while the (net) charge flow will be 
$V^\mu\neq 0$. Therefore the conservation equations change to
\begin{eqnarray}
N^\mu&=&n u^\mu+V^\mu\\
T^{\mu\nu}&=&\varepsilon u^\mu u^\nu-(p+\Pi)\Delta^{\mu\nu}+\pi^{\mu\nu}\,.
\end{eqnarray}
Thus, it follows from Eq.\ (\ref{TmunuDec}) that
\begin{eqnarray}
0&=&q^\mu=J_{31}v^\mu - 2J_{41}\omega^\mu\,,
\end{eqnarray}
leading to a different condition for $v^\mu$ 
\begin{eqnarray}
v^\mu&=&\frac{2J_{41}}{J_{31}}\omega^\mu\,.
\end{eqnarray}
Using the same procedure as explained above, one obtains the transport equations for the bulk 
viscous pressure $\Pi$, the heat flux current $q^\mu$, and the shear stress tensor 
$\pi^{\mu\nu}$ the form of which is the same as in Eqs.\ (\ref{TransportPi}) -- 
(\ref{Transportpimunu}) with $q^\mu\rightarrow -V^\mu (\varepsilon+p)/n$.\\
However, the values of the dissipative quantities are frame-dependent and will be given in 
Ref.\ \cite{ViscousBetz}.

\subsection[The Israel--Stewart Equations]
{The Israel--Stewart Equations}

It is possible to show that the expressions
\begin{eqnarray}
A^\prime\Pi&=&\dot{I}_{30}+\phi_1\dot{\Pi}+\left( I_{30}-2I_{31}\right)\theta
+\psi_1 \partial_\mu q^\mu\,,\\
C^\prime q^\mu&=&
\left(I_{30}-2I_{31}\right)\dot{u}^\mu\nonumber\\&&
+\Delta^\mu_\lambda\left[\partial^\lambda I_{31}+\psi_1\dot{q}^\lambda+\phi_2\partial^\lambda\Pi
+\psi_2\partial_\nu\pi^{\lambda\nu}\right]\,,\\
B^\prime \pi^{\mu\nu}&=&
2 I_{31}\partial^{\langle\mu}u^{\nu\rangle}
-\frac{2}{5} \psi_1\partial^{\langle\mu}q^{\nu\rangle}+\psi_2\dot{\pi}^{\mu\nu}\,,
\end{eqnarray}
[cf.\ Eqs.\ (\ref{PiPreliminary}) -- (\ref{pimunuPreliminary})], which contain terms partially of
first and second order in the Knudsen number, are equivalent to 
the equations derived by Israel and Stewart [Eq. (2.41) of \cite{IS}]
\begin{eqnarray}
\Pi &=&-\frac{1}{3}\zeta_V \left(\partial_\mu u^\mu + \beta_0 \dot{\Pi}
-\bar{\alpha}_0 \partial_\mu q^\mu\right)\\
q^\lambda &=& -\kappa T\Delta^{\lambda \mu}\left(T^{-1}\partial_\mu T 
-\dot{u}_\mu +\bar{\beta}_1\dot{q}_\mu -
\bar{\alpha}_0\partial_\mu\Pi -\bar{\alpha}_1\partial_\nu\pi^\nu_{\mu}\right)\\
\pi^{\mu\nu}&=&-2\zeta_S\left(\partial^{\langle\mu}u^{\nu\rangle}
+\beta_2\dot{\pi}_{\lambda\mu}-\bar{\alpha}_1 \partial^{\langle\mu}q^{\nu\rangle}\right)\,,
\end{eqnarray}
which however, differ in metric. To prove this, one has to use the conservation 
equations for (net) charge density, energy and momentum, as well the relation (\ref{DefinitionJ}) 
and the definitions
\begin{eqnarray}
\bar{\alpha}_0 -\alpha_0 &=& \bar{\alpha}_1 -\alpha_1 = -(\bar{\beta}_1 -\beta_1) 
= - [\beta J_{31}]^{-1}\,,\\
\alpha_0 &=& \frac{D_{41}D_{20}-D_{31}D_{30}}{\Lambda\zeta\Omega J_{21}J_{31}D_{20}}\,,\\
\alpha_1 &=& \frac{J_{41}J_{42}- J_{31}J_{52}}{\Lambda\zeta J_{21}J_{31}}\,,\\
\beta_0 &=& \frac{3\beta\{5J_{52}-(3/D_{20})
[J_{31}(J_{31}J_{30}-J_{41}J_{20})+ J_{41}(J_{41}J_{10}-J_{31}J_{20})\}}{\zeta^2\Omega^2}\,,
\nonumber\\\\
\beta_1 &=& \frac{D_{41}}{\Lambda^2 \beta J^2_{21}J_{31}}\,,\\\hspace*{-3.0cm}
\beta_2 &=&\frac{1}{2}\beta \frac{J_{52}}{\zeta^2}\,,
\end{eqnarray}
\begin{eqnarray}
\hspace*{-3.0cm}
\zeta_V &=& 3\frac{(\zeta\Omega)^2}{\beta A}\,,\hspace*{0.5cm} \zeta_S = 
10\frac{\zeta^2}{\beta B}\,,\hspace*{0.5cm} \kappa = \frac{(\Lambda\beta J_{21})^2}{C}\,,\\
\hspace*{-3.0cm}
\zeta &=& \beta J_{42}=I_{31}\,, \hspace*{0.5cm} I_{10}=\beta J_{21}\,, \hspace*{0.5cm} \Lambda = 
\frac{D_{31}}{(J_{21})^2}\,,\\[2ex]~\hspace*{-3.0cm}
\Omega&=&\frac{3J_{21}(J_{30}J_{31}-J_{20}J_{41})+3J_{31}(J_{10}J_{41}-J_{20}J_{31})-5J_{42}D_{20}}
{J_{42}D_{20}}\,.\nonumber\\
\end{eqnarray}

\clearpage{\pagestyle{empty}\cleardoublepage}
%
%
\chapter[Energy and Momentum Deposition in the Bethe--Bloch Formalism]
{Energy and Momentum Deposition in the Bethe--Bloch Formalism}
\label{FormulasBetheBloch}

The Bethe--Bloch formalism \cite{Bragg,Chen,Sihver,Kraft} assumes that [see Eq.\ (\ref{braggeqn})]
\begin{eqnarray}
\frac{dE(t)}{dt}=a\frac{1}{v_{\rm jet}(t)}\,.
\end{eqnarray}
In order to determine the location $x_{\rm jet}(t)$ as well as the time dependence of the velocity
$v_{\rm jet}(t)$, from which one can calculate the energy loss using the above formula, we  
start by proving the general expression
\begin{eqnarray}
\frac{dE}{dt}&=&v_{\rm jet}\frac{dM}{dt}\,.
\end{eqnarray}
Applying the relativistic equations ($\beta=v_{\rm jet}/c$) $E=\gamma m$ and 
$M=\beta\gamma m=\beta E$, where $m$ denotes the mass of the jet, the momentum deposition can be 
rewritten as
\begin{eqnarray}
\frac{dM}{dt}&=&\frac{dM}{dx}\frac{dx}{dt}=\beta\frac{dM}{dx}=\beta\left(
\frac{dE}{dx}\beta+E\frac{d\beta}{dx}\right)\,.
\end{eqnarray}
The last term can be further evaluated since $\beta^2=1-1/\gamma^2$ which leads (taking the
derivative w.r.t.\ $x$ and taking into account that $m={\rm const.}$) to the relation
\begin{eqnarray}
\frac{d\beta}{dx}&=&\frac{1}{\beta\gamma^3}\frac{dE}{dx}\frac{1}{m}\,.
\end{eqnarray}
Therefore, the momentum deposition is given by
\begin{eqnarray}
\frac{dM}{dt}&=&\beta^2\frac{dE}{dx}\left(1+\frac{1}{\gamma^2\beta^2}\right)
=\beta^2\frac{dE}{dx}\left(1+\frac{1}{\gamma^2-1}\right)\nonumber\\
&=&\frac{dE}{dx}
=\frac{dE}{dt}\frac{1}{v_{\rm jet}}\,.\hspace*{6cm}\Box
\end{eqnarray}
This expression can now be used to calculate the time as a function of the jet's rapidity $y_{\rm
jet}$, using $dM/dy_{\rm jet}=m\cosh y_{\rm jet}$ and $v_{\rm jet}=\tanh y_{\rm jet}$:
\begin{eqnarray}
\frac{dE}{dt}=v_{\rm jet}\frac{dM}{dt}&=&\frac{a}{v_{\rm jet}}\nonumber\\
\Rightarrow \frac{m}{a}\int\limits_{y_0}^{y_{\rm jet}}
\cosh y^\prime_{\rm jet}\tanh^2 y^\prime_{\rm jet}\,dy^\prime_{\rm jet}&=&t(y_{\rm jet})
\end{eqnarray}
where $y_0$ is the jet's initial rapidity. This relation is equivalent to
\begin{eqnarray}
\label{reference1}
\frac{m}{a}\int\limits_{y_0}^{y_{\rm jet}}
\frac{\sinh^2 y^\prime_{\rm jet}}{\cosh y^\prime_{\rm jet}}\,\,dy^\prime_{\rm jet}&=&
t(y_{\rm jet})\,.
\end{eqnarray}
The integral expression can be solved analytically \cite{GradshteynRyzhik},
\begin{eqnarray}
\int\frac{\sinh^2 y_{\rm jet}}{\cosh y_{\rm jet}}\,\,dy_{\rm jet}&=&
\sinh y_{\rm jet} - \arccos\frac{1}{\cosh y_{\rm jet}}\,,
\end{eqnarray}
thus, one obtains for $t(y_{\rm jet})$ [see Eq.\ (\ref{BetheBloch1})]
\begin{eqnarray}
t(y_{\rm jet}) & = & \frac{m}{a}\, \left[ \sinh y_{\rm jet} 
- \sinh y_0 \right. \nonumber \\
&  & \left. - \arccos \frac{1}{\cosh y_{\rm jet}}
+ \arccos \frac{1}{\cosh y_0} \right]\,,
\end{eqnarray}
which can then be applied to calculate the time dependence of the velocity $v_{\rm jet}(t)$.\\
The location $x_{\rm jet}(t)$, however, is given by
\begin{eqnarray}
x_{\rm jet}(t)&=&x_{\rm jet}(0)+\int\limits_0^t v_{\rm jet}(t)\,dt
\end{eqnarray}
which is, because of $v_{\rm jet}(t)=\tanh y_{\rm jet}$, equivalent to
\begin{eqnarray}
x_{\rm jet}(t)&=&x_{\rm jet}(0)+\int\limits_{y_0}^{y_{\rm jet}} 
\tanh y^\prime_{\rm jet}\,\frac{dt}{dy}dy\,.
\end{eqnarray}
Using Eq.\ (\ref{reference1}) as well as the substitution 
$dy_{\rm jet}\cosh y_{\rm jet}=d\sinh y_{\rm jet}$, 
one obtains
\begin{eqnarray}
x_{\rm jet}(t)&=&x_{\rm jet}(0)+
\frac{m}{a}\int\limits_{\sinh y_0}^{\sinh y_{\rm jet}} 
\frac{\sinh^3 y^\prime_{\rm jet}}{\cosh^3 y^\prime_{\rm jet}} d\sinh y^\prime_{\rm jet}\,.
\end{eqnarray}
With the substitution $x=\sinh y_{\rm jet}$, the integral is 
\begin{eqnarray}
\int\frac{x^3}{(1+x^2)^{3/2}} dx &=&\frac{1+\cosh^2 y_{\rm jet}}{\cosh y_{\rm jet}}\,,
\end{eqnarray}
which can be rewritten using $\cosh y_{\rm jet}=\gamma_{\rm jet}$ and 
$\sinh y_{\rm jet}=\gamma_{\rm jet} v_{\rm jet}$, leading to [cf.\ Eq. (\ref{BetheBloch2})]
\begin{eqnarray}
x_{\rm jet}(t) &=& x_{\rm jet}(0) +
\frac{m}{a}\, \left[ (2-v_{\rm jet}^2)\gamma_{\rm jet} -
(2-v_{0}^2)\gamma_{0} \right] \,.
\end{eqnarray}

%
%
\chapter[Joule Heating]
{Joule Heating}
\label{JouleHeating}

This appendix discusses the connection between a source term and the process of Joule
heating. Starting with the Boltzmann equation, omitting the collision term,
\begin{eqnarray}
\frac{\partial}{\partial t}f +\vec{v}\cdot\vec{\nabla} f
+ \vec{F}\cdot \frac{\partial}{\partial\vec{p}}f 
+ \vec{F}\cdot\vec{v}\frac{\partial}{\partial p_0}f&=&0\,,
\end{eqnarray}
where $f$ is a distribution function, $\vec{v}=\vec{p}/E$ the velocity and $\vec{F}$ an external 
force, one obtains the manifestly covariant expression of the Boltzmann--Vlasov equation by 
multiplying the above equation with $p_0=E=\sqrt{p^2+m^2}$
\begin{eqnarray}
p^\mu\partial_\mu f -ep^\mu F_{\mu\nu}\frac{\partial}{\partial p_\nu}f&=&0\,.
\label{BoltzmannVlasov}
\end{eqnarray}
Here, $F^{\mu\nu}$ is the antisymmetric field tensor. The energy-momentum tensor is given
by
\begin{eqnarray}
T^{\mu\nu}&=&\int \frac{d^3 p}{(2\pi)^3 E}p^\mu p^\nu f(x,p)
=\int d\Gamma p^\mu p^\nu f(x,p)\,,
\end{eqnarray}
implying 
\begin{eqnarray}
\partial_\mu T^{\mu\nu}&=&\int d\Gamma p^\mu p^\nu\partial_\mu f\,.
\end{eqnarray}
Inserting the above Boltzmann equation [Eq.\ (\ref{BoltzmannVlasov})] results in
\begin{eqnarray}
\partial_\mu T^{\mu\nu}&=&e\int d\Gamma p^\mu p^\nu F_\mu^{\;\,\alpha}
\frac{\partial f}{\partial p^\alpha}\,.
\end{eqnarray}
After integration by parts (leading to a vanishing surface integral) and applying that 
$F_\mu^\mu=0$, the expression can be rewritten as
\begin{eqnarray}
\partial_\mu T^{\mu\nu}&=-F_\mu^{\;\,\nu}\, e \int d\Gamma p^\mu f\,.
\end{eqnarray}
Defining the current $J^\mu=-e\int d\Gamma p^\mu f$ \cite{Gatoff:1987uf}, one ends up with
the general expression \cite{Gatoff:1987uf}
\begin{eqnarray}
\partial_\mu T^{\mu\nu}&=&J^\mu F_\mu^{\;\,\nu}\,.
\label{genexp}
\end{eqnarray}
According to Ohm's law, this current is given by
\begin{eqnarray}
\vec{J}&=&\sigma \vec{E}\,,
\end{eqnarray}
which reads in covariant generalization
\begin{eqnarray}
J^\nu&=&F_{\alpha\beta}\sigma^{\alpha\beta\nu}\,.
\end{eqnarray}
The heat produced during this process is called {\it Joule heating}. Thus, Eq.\ (\ref{genexp}) 
becomes
\begin{eqnarray}
\partial_\mu T^{\mu\nu}&=&F_\mu^{\;\,\nu}\sigma^{\mu\alpha\beta}F_{\alpha\beta}\,.
\end{eqnarray}
Therefore, the source term $\mathcal{J}^\nu$ is given by [see Eq.\ (\ref{eqjouleheating})]
\begin{eqnarray}
\partial_\mu T^{\mu\nu}=\mathcal{J}^\nu= F^{\nu\alpha\,a}J_{\alpha}^a&=&
(F^{\nu\alpha\,a}\sigma_{\alpha\beta\gamma}*F^{\beta\gamma\,a})\,.
\end{eqnarray}

%
%
\chapter[Isochronous and Isothermal Freeze-out]
{Isochronous and Isothermal Freeze-out}
\label{IsochronIsothermFreezeout}
\begin{figure}[b]
\centering
\begin{minipage}[c]{4.2cm}
\hspace*{-1.5cm}
  \includegraphics[scale = 0.75]{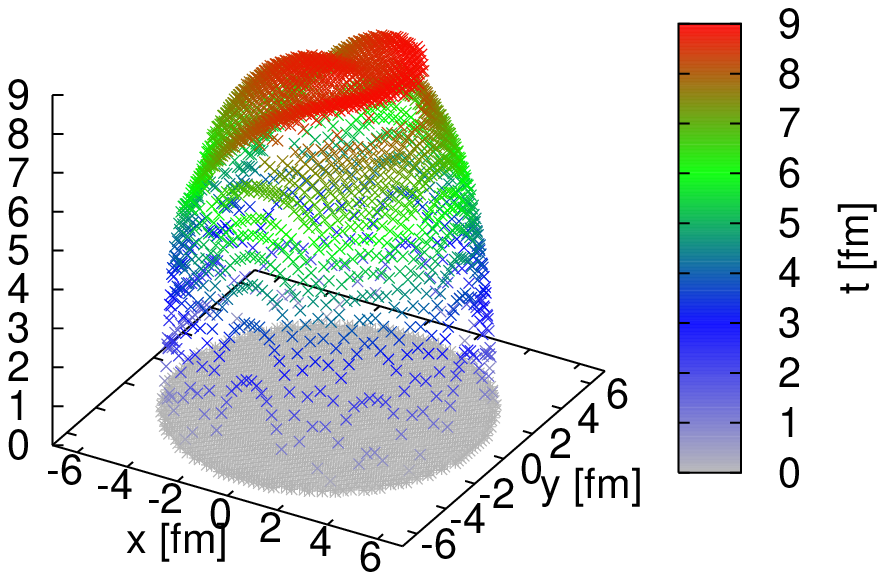}
\end{minipage}
\hspace*{1.0cm}  
\begin{minipage}[c]{4.2cm} 
  \includegraphics[scale = 0.8]{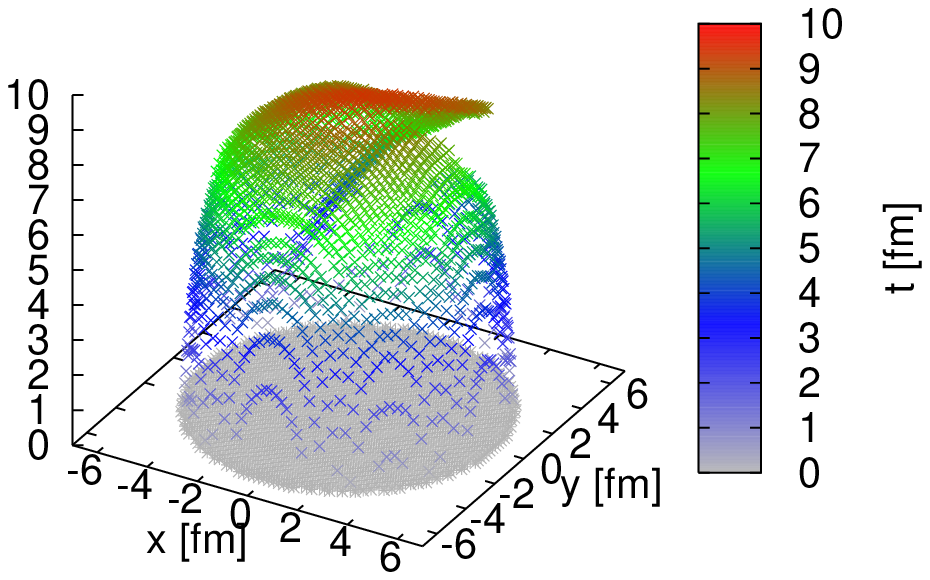}
\end{minipage}
  \caption[The freeze-out hypersurface for a jet moving through the middle of the medium 
  and for a non-central jet.]
  {The freeze-out hypersurface ($T_c=130$~MeV) for a jet moving through the middle 
  of the medium (left panel, Jet 180 according to Fig.\ \ref{JetPaths}) and a non-central jet 
  (right panel, Jet 150). In the right panel a beak-like deformation of the hypersurface 
  due to jet heating is clearly visible.}
  \label{IsoHypersurface}
\end{figure}
\begin{figure}[t]
\centering
\begin{minipage}[c]{4.2cm}
\hspace*{-4.5cm}
  \includegraphics[scale = 0.55]{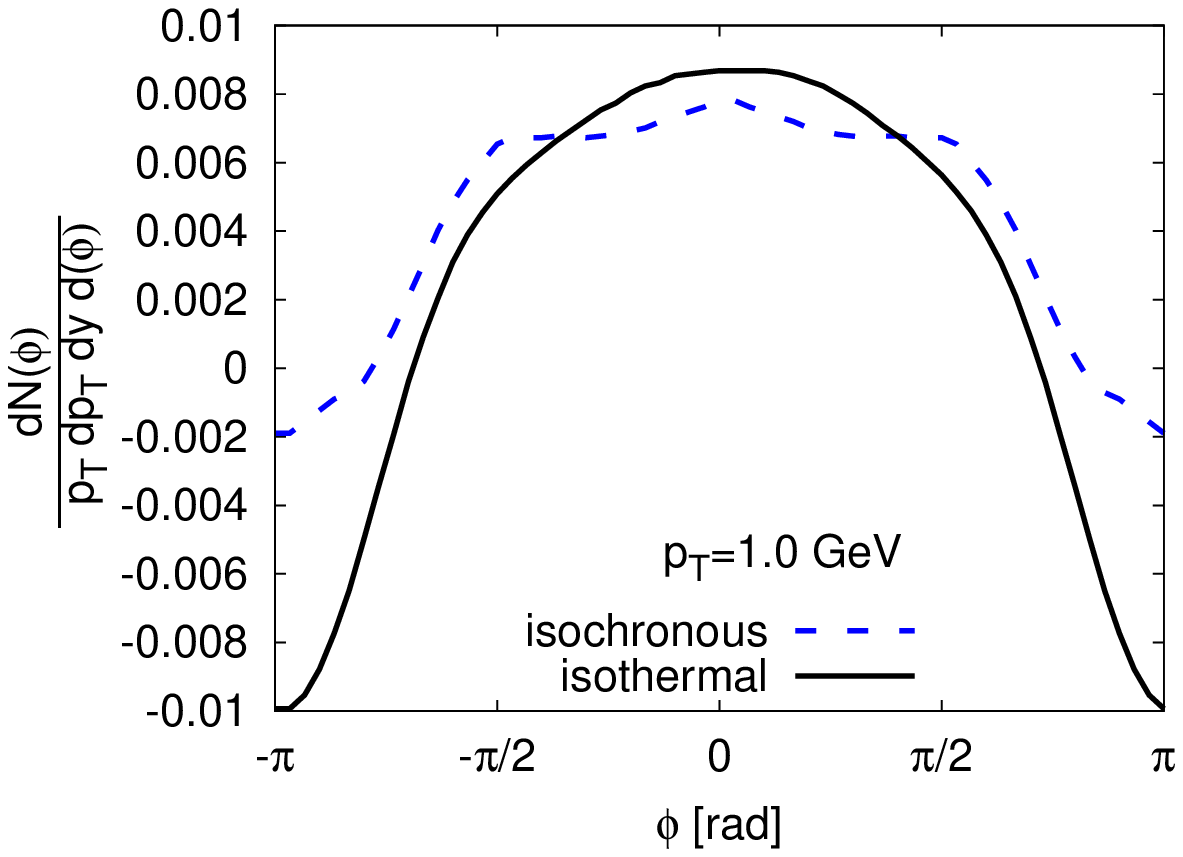}
\end{minipage}
\hspace*{-1.5cm}  
\begin{minipage}[c]{4.2cm} 
  \includegraphics[scale = 0.55]{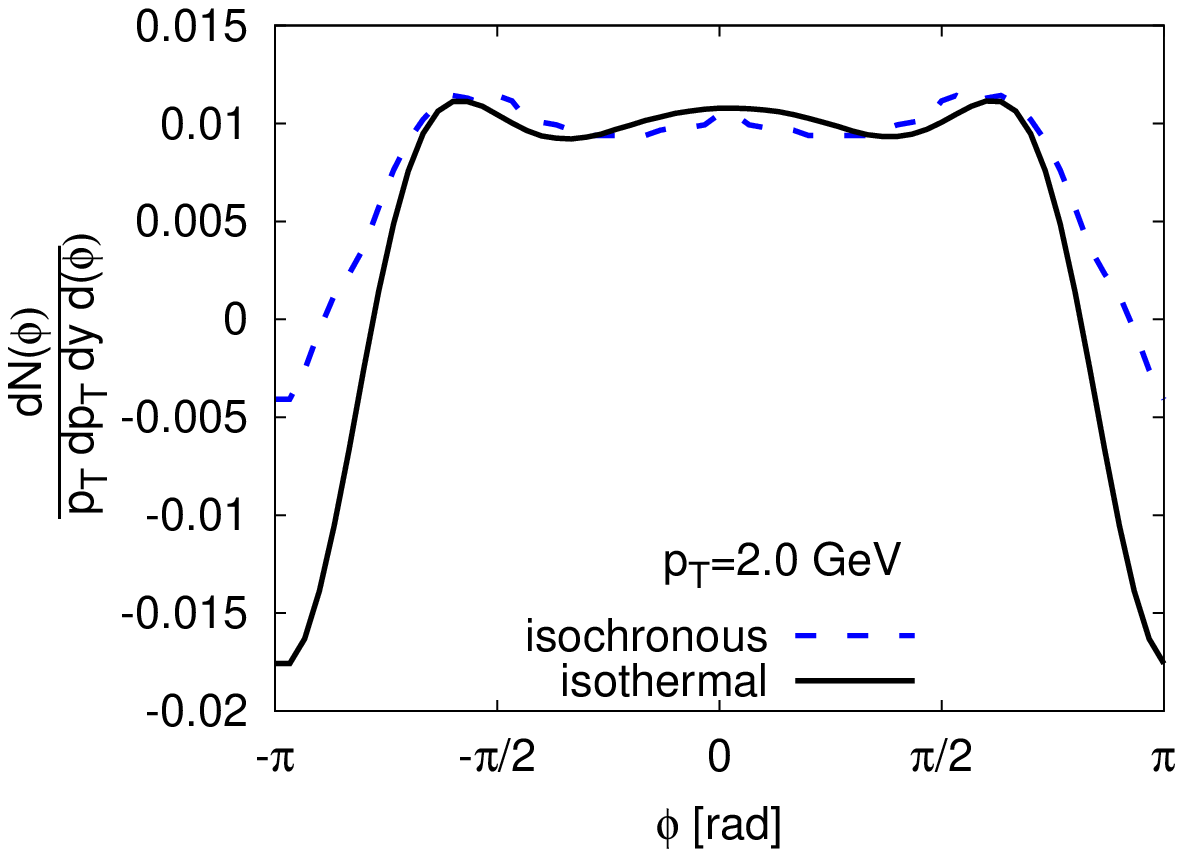}
\end{minipage}
  \caption
  [Background subtracted and normalized azimuthal particle correlations obtained from 
  an isochronous and an isothermal Cooper--Frye freeze-out for $p_T=1$~GeV and $p_T=2$~GeV.]
  {Background subtracted and normalized azimuthal particle correlations obtained from 
  an isochronous (dashed blue line) and an isothermal (solid black line) Cooper--Frye freeze-out
  for $p_T=1$~GeV (left panel) and $p_T=2$~GeV (right panel).}
  \label{CFIsochronIsotherm1}
\end{figure}
\begin{figure}[t]
\centering
\begin{minipage}[c]{4.2cm}
\hspace*{-4.5cm}
  \includegraphics[scale = 0.72]{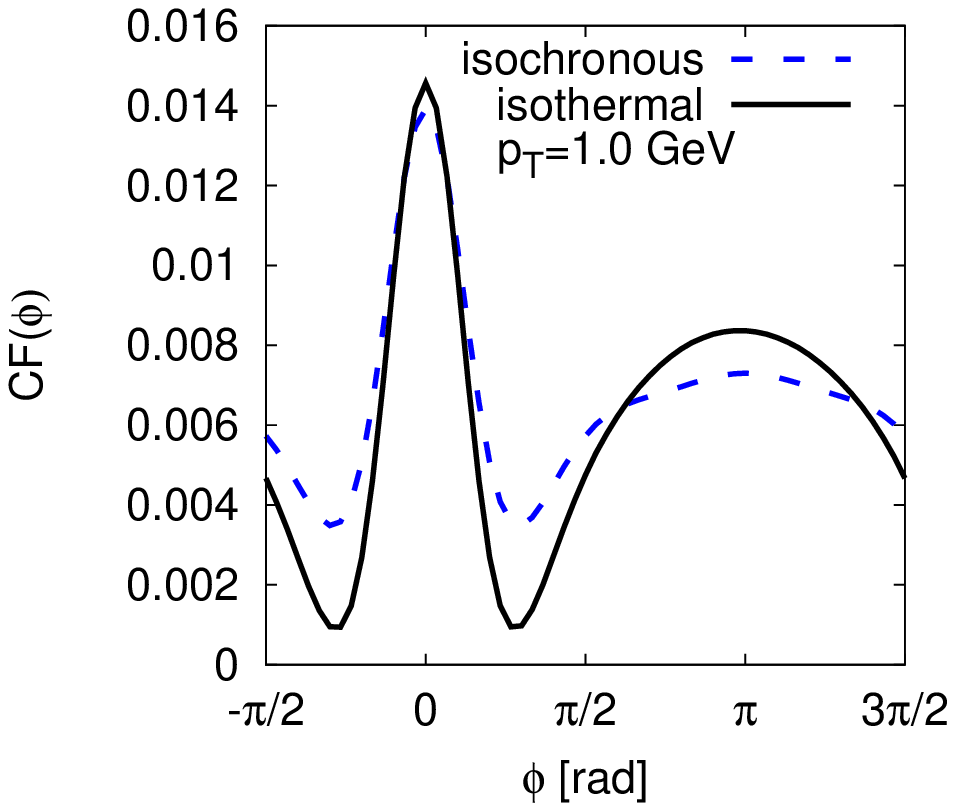}
\end{minipage}
\hspace*{-1.5cm}  
\begin{minipage}[c]{4.2cm} 
  \includegraphics[scale = 0.72]{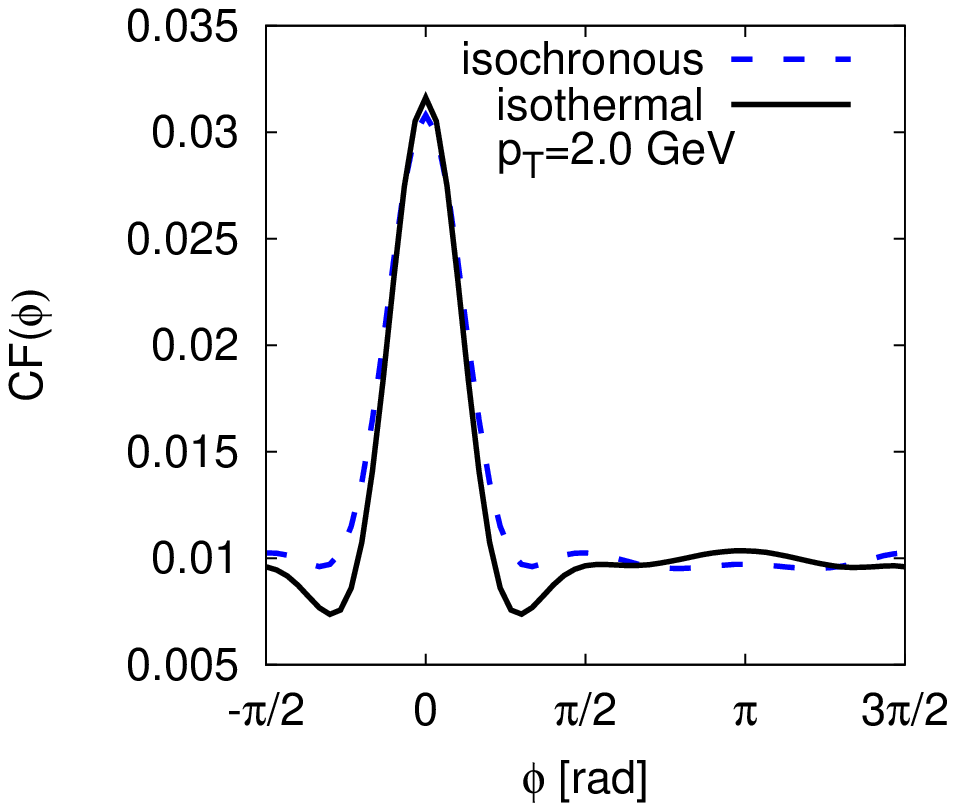}
\end{minipage}
  \caption
  [Background subtracted and normalized azimuthal particle correlations obtained from 
  an isochronous and an isothermal Cooper--Frye freeze-out for $p_T=1$~GeV and $p_T=2$~GeV 
  after convolution with a trigger jet.]
  {Background subtracted and normalized azimuthal particle correlations obtained from 
  an isochronous (dashed blue line) and an isothermal (solid black line) Cooper--Frye freeze-out
  for $p_T=1$~GeV (left panel) and $p_T=2$~GeV (right panel) after convolution with a trigger
  jet according to Eq.\ (\ref{def2pc}).}
  \label{CFIsochronIsotherm2}
\end{figure}

A Mach-cone signal, like any observable from heavy-ion collisions sensitive to 
hydrodynamics, suffers from the problem that the ``fluid'' cannot directly be 
accessed experimentally, but has to be studied from final many-particle correlations 
which are sensitive to all stages of the hydrodynamic evolution, including the late 
(and presumably non-thermalized) stages of freeze-out. \\
A rough approximation assumes that the mean-free path goes from zero to infinity
at a certain locus in space-time $\Sigma^\mu = (t,\vec{x})$. 
This locus can be defined in terms of a local criterion (e.g.\ a freeze-out 
temperature $T_c$), or using a simple global geometry (like an isochronous freeze-out, see
section \ref{Freezeout}). \\
Then, applying the Cooper--Frye formula [Eq.\ (\ref{CooperFormula})],
\begin{eqnarray}
E\frac{dN}{d^3\vec{p}} = \int_{\Sigma} d\Sigma_\mu\, p^\mu\, f(u\cdot p/T)\,,
\end{eqnarray}
which is an ansatz based on energy-momentum and entropy conservation that is repeated here 
for convenience, the fluid fields of momentum $p^\mu$, velocity $u^\mu$ and
temperature $T$ are transformed into particles according to a distribution function $f$.\\
However, the freeze-out hypersurface $\Sigma$ entering the above equation is severely deformed
by a jet propagating through an expanding medium as shown in Fig.\ 
\ref{IsoHypersurface}. A very characteristic, beak-like elongation occurs for non-central jets
(see right panel of Fig.\ \ref{IsoHypersurface}) putting constraints on an isothermal
freeze-out which requires the derivative of time w.r.t.\ spatial coordinates
(cf.\ section \ref{Freezeout}).\\
Nevertheless, the freeze-out results applying an isochronous and an isothermal
freeze-out are rather similar for a central jet (i.e., for jets moving through the middle
of the medium) as Fig.\ \ref{CFIsochronIsotherm1} reveals. Here, the 
particle distribution is obtained according to the above Cooper--Frye formula. Fig.\
\ref{CFIsochronIsotherm2}, however, displays the distribution after convoluting with a trigger 
jet according to Eq.\ (\ref{def2pc}).
\\While there is an excellent agreement for $p_T=2$~GeV, a discrepancy occurs for $p_T=1$~GeV
which could be an effect of larger thermal smearing. \\
Surprisingly enough, these results suggest that applying an iso\-chronous 
freeze-out, which is a very strong model assumption, does not drastically alter the results 
compared to an isothermal conversion (at least for the very central jets).

\clearpage{\pagestyle{empty}\cleardoublepage}
%
%
\chapter[Distortion of Conical Structures due to Background Flow]
{Distortion of Conical Structures due to Background Flow}
\label{AppendixDistortion}
\begin{figure}[t]
\centering
\begin{minipage}[c]{4.2cm}
\hspace*{-6.0cm}
  \includegraphics[scale = 0.8]{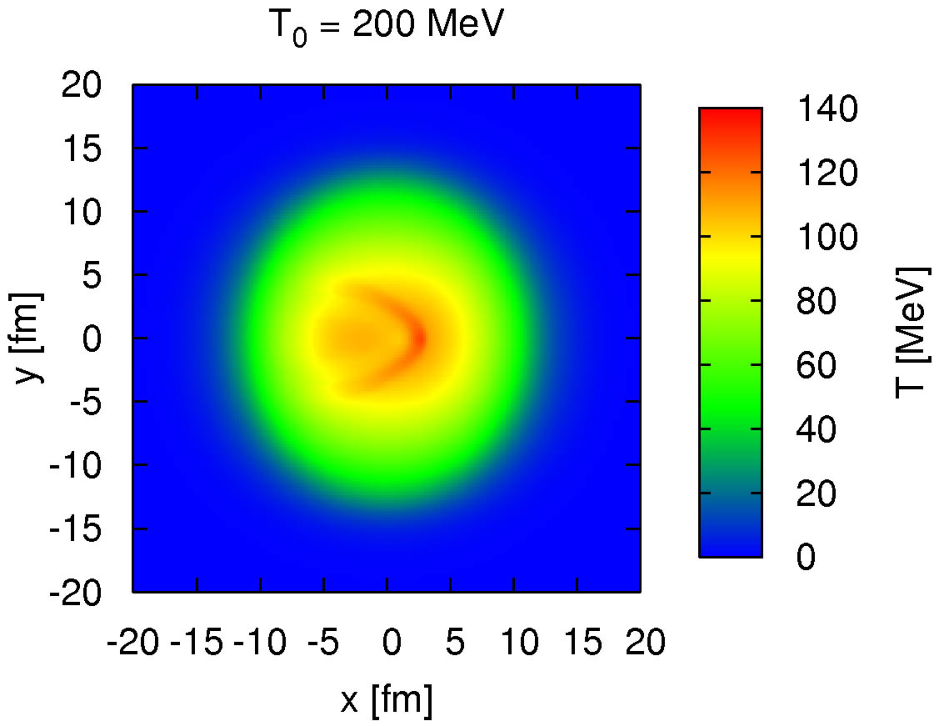}
\end{minipage}
\hspace*{-2.5cm}  
\begin{minipage}[c]{4.2cm} 
  \includegraphics[scale = 0.8]{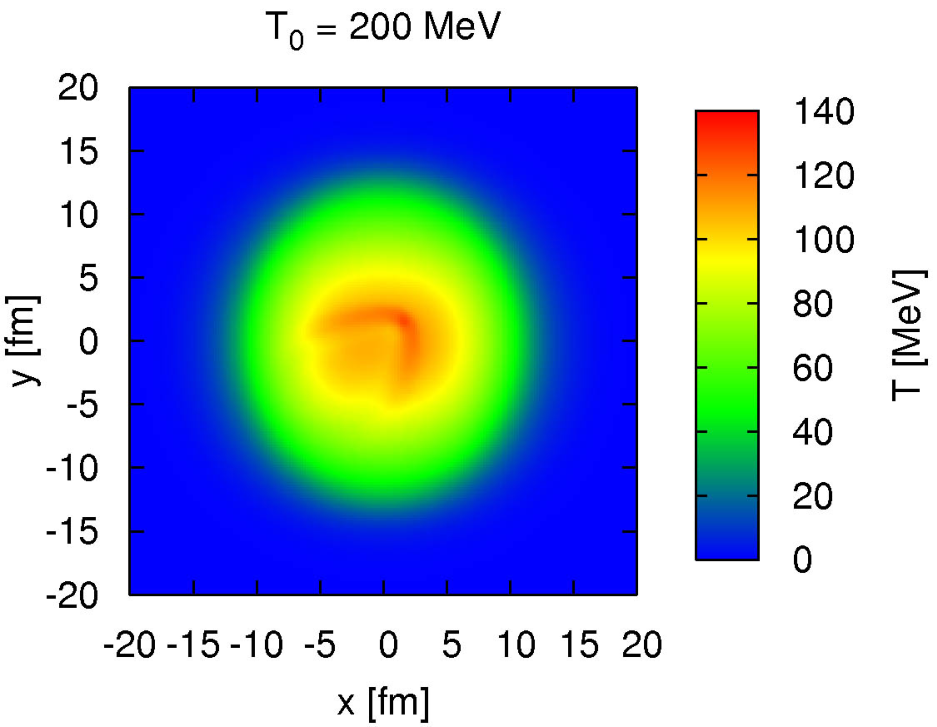}
\end{minipage}\\
\hspace*{-2.5cm}
\includegraphics[scale = 0.8]{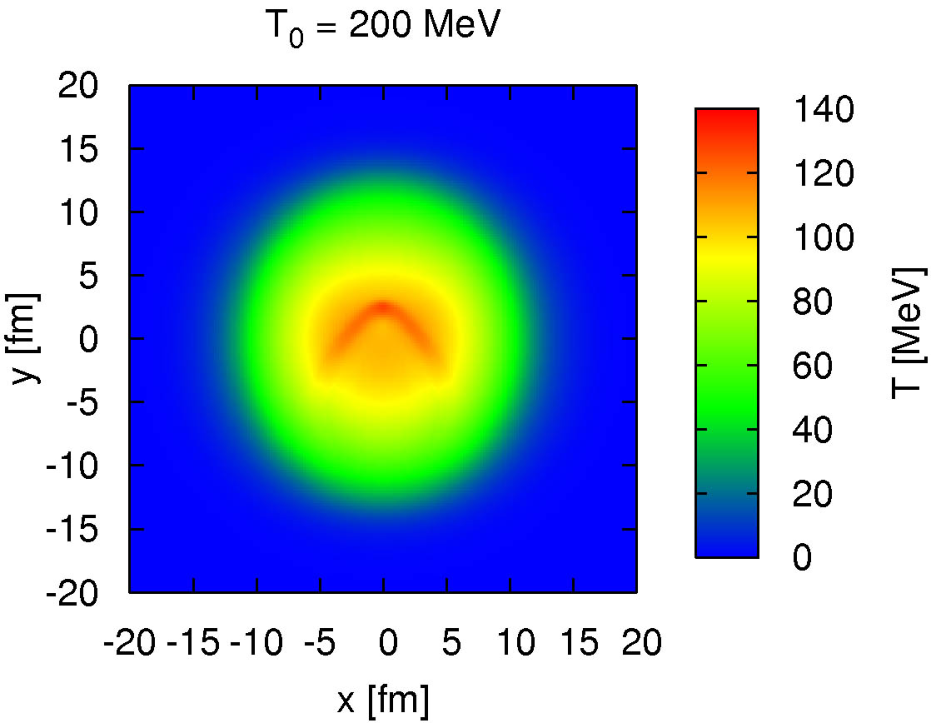}
  \caption[The temperature pattern of three different jets depositing energy and momentum along
  trajectories for varying angles w.r.t.\ the $x$-axis at the moment of 
  freeze-out.]
  {The temperature pattern of three different jets depositing energy and momentum along
  trajectories for varying angles w.r.t.\ the $x$-axis at the moment of 
  freeze-out. The distortion of the conical shape is clearly visible.}
  \label{DegTempEnd}
\end{figure}
Satarov et al.\ \cite{Satarov:2005mv} predicted that a Mach cone is
deformed in an expanding medium due to the interaction of the jet and background
flow patterns. Since a straightforward analytical solution can only be derived 
if the background flow is parallel (or antiparallel) to the direction of jet propagation
(which is not the case for a radially expanding medium), we may just qualitatively
investigate such an effect here.\\
Fig.\ \ref{DegTempEnd} displays the temperature patterns at the moment of freeze-out (i.e.,
when the temperature of all cells has dropped below $T_c=130$~MeV, see chapter 
\ref{ExpandingMedium}) for jets that deposit energy and momentum according to Eq.\ 
(\ref{SourceExpandingMedium}) with $dE/dt\vert_0 = v\, dM/dt\vert_0=1$~GeV/fm into an 
expanding $b=6$~fm medium along trajectories having different angles  w.r.t.\ the $x$-axis. \\
As can be seen, the opening angle of the conical structure produced differs for jets
propagating along the short (upper left panel of Fig.\ \ref{DegTempEnd}) 
or the long axis (lower panel of Fig.\ \ref{DegTempEnd}), with a larger opening angle for
jets encountering a larger background flow gradient. \\
Moreover, for a jet propagating at an angle of $\phi=45$~degrees w.r.t.\ the $x$-axis 
(upper right panel of Fig.\ \ref{DegTempEnd}), the conical structure is distorted
in a way that the opening angle in direction of the larger background flow gradient 
becomes larger than the opening angle in direction
of the smaller background flow gradient. 
This effect, however, remains small for an impact parameter of $b=6$~fm.\\
This is in accord with the prediction of Ref.\ \cite{Satarov:2005mv} 
(see also Fig.\ \ref{MachConeDeformation}). 
There the opening angles for the outer wings of the Mach-cone angles gets larger
since the flow points radially outwards. \\
Though a detailed study about the distortion of Mach cones remains to be done,
the conical structure clearly depends on the centrality of the considered medium.

\end{appendix}
%
%
%
%
\addcontentsline{toc}{chapter}{References}
\bibliographystyle{style/diss_lit}
\bibliography{literature/bib_abbreviations,literature/bib_chargeflucs,literature/bib_charm,literature/bib_clusterhadronization,literature/bib_deuteron,literature/bib_general,literature/bib_neutrinos,literature/bib_pentaquarks,literature/bib_qcd_exotica,literature/bib_qcd_general,literature/bib_qgp_general,literature/bib_qmd_model,literature/bib_strangeness,literature/bib_su6,literature/bib_unsorted,literature/bib_rho}

\begin{thebibliography}{100}
\newcommand{\enquote}[1]{``#1''}
\providecommand{\url}[1]{\texttt{#1}}
\providecommand{\urlprefix}{URL }
\providecommand{\eprint}[2][]{\url{#2}}
\begin{footnotesize}
\addtolength{\itemindent}{+0em}
\addtolength{\itemsep}{-0.4em}


\subsubsection*{Chapter 1: Introduction}

\bibitem{Glashow:1961tr}
  S.~L.~Glashow, ``Partial Symmetries Of Weak Interactions,'' Nucl.\ Phys.\  {\bf 22}, 579 (1961).
  
\bibitem{Weinberg:1967tq}
  S.~Weinberg, ``A Model Of Leptons,'' Phys.\ Rev.\ Lett.\  {\bf 19}, 1264 (1967). 
  
\bibitem{Gross:1973id}
  D.~J.~Gross and F.~Wilczek, ``Ultraviolet Behavior of Non-Abelian Gauge Theories,''
  Phys.\ Rev.\ Lett.\  {\bf 30}, 1343 (1973).  
  
\bibitem{Politzer:1973fx}
  H.~D.~Politzer, ``Reliable Perturbative Results for Strong Interactions?,''
  Phys.\ Rev.\ Lett.\  {\bf 30}, 1346 (1973).  

\bibitem{Collins:1974ky}
  J.~C.~Collins and M.~J.~Perry, ``Superdense Matter: Neutrons Or Asymptotically Free Quarks?,''
  Phys.\ Rev.\ Lett.\  {\bf 34}, 1353 (1975).

\bibitem{Freedman:1976ub}
  B.~A.~Freedman and L.~D.~McLerran, ``Fermions And Gauge Vector Mesons At Finite Temperature And Density. 3. The
  Ground State Energy Of A Relativistic Quark Gas,'' Phys.\ Rev.\  D {\bf 16}, 1169 (1977).  
  
\bibitem{Shuryak:1977ut}
  E.~V.~Shuryak, ``Theory Of Hadronic Plasma,'' Sov.\ Phys.\ JETP {\bf 47}, 212 (1978)
  [Zh.\ Eksp.\ Teor.\ Fiz.\  {\bf 74}, 408 (1978)].  
  
\bibitem{Hofmann:1975by}
  {\it Report of the Workshop on BeV/nucleon Collisions of Heavy Ions: How and Why},
  Bear Mountain, New York, 29 Nov - 1 Dec 1974.  

\bibitem{CERN} 
  U.~W.~Heinz and M.~Jacob, ``Evidence for a new state of matter: 
  An assessment of the results from  the CERN lead beam programme,'' arXiv:nucl-th/0002042.

\bibitem{Arsene:2004fa}
  I.~Arsene {\it et al.}  [BRAHMS Collaboration], ``Quark Gluon Plasma an Color Glass Condensate at RHIC? The perspective from
  the BRAHMS experiment,'' Nucl.\ Phys.\  A {\bf 757}, 1 (2005) [arXiv:nucl-ex/0410020].
  
\bibitem{Adcox:2004mh}
  K.~Adcox {\it et al.}  [PHENIX Collaboration],
  ``Formation of dense partonic matter in relativistic nucleus nucleus collisions at RHIC: Experimental evaluation by the PHENIX  collaboration,''
  Nucl.\ Phys.\  A {\bf 757}, 184 (2005) [arXiv:nucl-ex/0410003].  

\bibitem{Back:2004je}
  B.~B.~Back {\it et al.}, ``The PHOBOS perspective on discoveries at RHIC,'' Nucl.\ Phys.\  A {\bf 757}, 28 (2005)
  [arXiv:nucl-ex/0410022].
  
\bibitem{Adams:2005dq}
  J.~Adams {\it et al.}  [STAR Collaboration], ``Experimental and theoretical challenges in the search for the quark  gluon
  plasma: The STAR collaboration's critical assessment of the  evidence from RHIC collisions,''
  Nucl.\ Phys.\  A {\bf 757}, 102 (2005) [arXiv:nucl-ex/0501009].  

\bibitem{Shuryak:2003xe} 
  E.~Shuryak, ``Why does the quark gluon plasma at RHIC behave 
  as a nearly ideal fluid?,'' Prog.\ Part.\ Nucl.\ Phys.\  {\bf 53}, 273 (2004) [arXiv:hep-ph/0312227].

\bibitem{Kolb:2003dz}
  P.~F.~Kolb and U.~W.~Heinz, ``Hydrodynamic description of ultrarelativistic heavy-ion collisions,''
  arXiv:nucl-th/0305084.  

\bibitem{Romatschke:2007mq}
  P.~Romatschke and U.~Romatschke, ``Viscosity Information from Relativistic Nuclear Collisions: How Perfect is
  the Fluid Observed at RHIC?,'' Phys.\ Rev.\ Lett.\  {\bf 99}, 172301 (2007) [arXiv:0706.1522 [nucl-th]].

\bibitem{Huovinen:2003fa}
  P.~Huovinen, ``Hydrodynamical description of collective flow,'' arXiv:nucl-th/0305064.   

\bibitem{Muronga:2003tb}  
  A.~Muronga, ``Shear Viscosity Coefficient from Microscopic Models,''
  Phys.\ Rev.\  C {\bf 69}, 044901 (2004) [arXiv:nucl-th/0309056].

\bibitem{Shuryak:2004kh}
  E.~Shuryak, ``A strongly coupled quark-gluon plasma,'' J.\ Phys.\ G {\bf 30}, S1221 (2004).

\bibitem{Bali:1992ab}
  G.~S.~Bali and K.~Schilling, ``Static quark - anti-quark potential: Scaling behavior and finite size
  effects in SU(3) lattice gauge theory,'' Phys.\ Rev.\  D {\bf 46}, 2636 (1992).
  
\bibitem{Bethke:2006ac}
  S.~Bethke, ``Experimental tests of asymptotic freedom,'' Prog.\ Part.\ Nucl.\ Phys.\  {\bf 58}, 351 (2007)
  [arXiv:hep-ex/0606035].  
  
\bibitem{Gyulassy:2004zy}
  M.~Gyulassy and L.~McLerran, ``New forms of QCD matter discovered at RHIC,''
  Nucl.\ Phys.\  A {\bf 750}, 30 (2005) [arXiv:nucl-th/0405013].  
  
  
\bibitem{Stephanov:2004wx}
  M.~A.~Stephanov, ``QCD phase diagram and the critical point,'' Prog.\ Theor.\ Phys.\ Suppl.\  {\bf 153}, 139 (2004)
  [Int.\ J.\ Mod.\ Phys.\  A {\bf 20}, 4387 (2005)] [arXiv:hep-ph/0402115].
  
\bibitem{Fodor:2001pe}
  Z.~Fodor and S.~D.~Katz,
  ``Lattice determination of the critical point of QCD at finite T and mu,'' JHEP {\bf 0203}, 014 (2002)
  [arXiv:hep-lat/0106002].

\bibitem{Fodor:2004mf}
  Z.~Fodor and S.~D.~Katz, ``Finite T/mu lattice QCD and the critical point,''
  Prog.\ Theor.\ Phys.\ Suppl.\  {\bf 153}, 86 (2004) [arXiv:hep-lat/0401023].  
  
\bibitem{Karsch:2003jg}
  F.~Karsch and E.~Laermann, ``Thermodynamics and in-medium hadron properties from lattice QCD,''
  arXiv:hep-lat/0305025.    
  
\bibitem{Karsch:2003va}
  F.~Karsch, C.~R.~Allton, S.~Ejiri, S.~J.~Hands, O.~Kaczmarek, E.~Laermann and C.~Schmidt,
  ``Where is the chiral critical point in 3-flavor QCD?,''
  Nucl.\ Phys.\ Proc.\ Suppl.\  {\bf 129}, 614 (2004)
  [arXiv:hep-lat/0309116].  
  
\bibitem{Fodor:2004nz}
  Z.~Fodor and S.~D.~Katz,
  ``Critical point of QCD at finite T and mu, lattice results for physical quark masses,''
  JHEP {\bf 0404}, 050 (2004) [arXiv:hep-lat/0402006].  
  
\bibitem{Rischke:2003mt}
  D.~H.~Rischke, ``The quark-gluon plasma in equilibrium,'' Prog.\ Part.\ Nucl.\ Phys.\  {\bf 52}, 197 (2004) [arXiv:nucl-th/0305030].  

\bibitem{Bailin:1983bm}
  D.~Bailin and A.~Love, ``Superfluidity And Superconductivity In Relativistic Fermion Systems,''
  Phys.\ Rept.\  {\bf 107}, 325 (1984).
  
\bibitem{Ruester:2005jc}
  S.~B.~Rüster, V.~Werth, M.~Buballa, I.~A.~Shovkovy and D.~H.~Rischke,
  ``The phase diagram of neutral quark matter: Self-consistent treatment of quark masses,''
  Phys.\ Rev.\  D {\bf 72}, 034004 (2005) [arXiv:hep-ph/0503184].  

\bibitem{Rischke:1992uv}
  D.~H.~Rischke, M.~I.~Gorenstein, A.~Schäfer, H.~Stöcker and W.~Greiner,
  ``Nonperturbative effects in the SU(3) gluon plasma,'' Phys.\ Lett.\  B {\bf 278}, 19 (1992).
  
\bibitem{Glashow:1967rx}
  S.~L.~Glashow and S.~Weinberg, ``Breaking Chiral Symmetry,'' Phys.\ Rev.\ Lett.\  {\bf 20}, 224 (1968).  
  
\bibitem{Schwarz:2003du}
  D.~J.~Schwarz, ``The first second of the universe,'' Annalen Phys.\  {\bf 12}, 220 (2003)
  [arXiv:astro-ph/0303574].  
  
\bibitem{Bass_HIC_Evolution}
  S.~A.~Bass, ``Microscopic Reaction Dynamics at SPS and RHIC,'' Talk at the Quark Matter Conference 2001, Stony Book, USA (2001).    
  
\bibitem{Soff:1999yg}
  S.~Soff, S.~A.~Bass, M.~Bleicher, H.~Stöcker and W.~Greiner,
  ``Directed and elliptic flow,'' arXiv:nucl-th/9903061.  
  
\bibitem{Rapp:1999ej}
  R.~Rapp and J.~Wambach,
  ``Chiral symmetry restoration and dileptons in relativistic heavy-ion collisions,''
  Adv.\ Nucl.\ Phys.\  {\bf 25}, 1 (2000) [arXiv:hep-ph/9909229].  
  
\bibitem{Rafelski:1982pu}
  J.~Rafelski and B.~Müller,
  ``Strangeness Production In The Quark - Gluon Plasma,'' Phys.\ Rev.\ Lett.\  {\bf 48}, 1066 (1982)
  [Erratum-ibid.\  {\bf 56}, 2334 (1986)].  
  
\bibitem{Andersen:1999ym}
  E.~Andersen {\it et al.}  [WA97 Collaboration],
  ``Strangeness enhancement at mid-rapidity in Pb Pb collisions at 158-A-GeV/c,''
  Phys.\ Lett.\  B {\bf 449}, 401 (1999).  
  
\bibitem{Matsui:1986dk}
  T.~Matsui and H.~Satz, ``J/psi Suppression by Quark-Gluon Plasma Formation,''
  Phys.\ Lett.\  B {\bf 178}, 416 (1986).  
  
\bibitem{Andronic:2006ky}
  A.~Andronic, P.~Braun-Munzinger, K.~Redlich and J.~Stachel,
  ``Statistical hadronization of heavy quarks in ultra-relativistic nucleus-nucleus collisions,''
  Nucl.\ Phys.\  A {\bf 789}, 334 (2007) [arXiv:nucl-th/0611023].  
  
\bibitem{Jeon:2003gk}
  S.~Jeon and V.~Koch, ``Event-by-event fluctuations,'' arXiv:hep-ph/0304012.  
  
\bibitem{Bass:2000az}
  S.~A.~Bass, P.~Danielewicz and S.~Pratt,
  ``Clocking hadronization in relativistic heavy ion collisions with  balance functions,''
  Phys.\ Rev.\ Lett.\  {\bf 85}, 2689 (2000) [arXiv:nucl-th/0005044].  
  
  
\subsubsection*{Chapter 2: The Experimental Search for the QGP}

\bibitem{Poskanzer:1998yz}
  A.~M.~Poskanzer and S.~A.~Voloshin, ``Methods for analyzing anisotropic flow in relativistic nuclear
  collisions,'' Phys.\ Rev.\  C {\bf 58}, 1671 (1998) [arXiv:nucl-ex/9805001].

\bibitem{Heinz:2004pj}
  U.~W.~Heinz, ``Thermalization at RHIC,'' AIP Conf.\ Proc.\  {\bf 739}, 163 (2005) [arXiv:nucl-th/0407067].
  
\bibitem{Xu:2008dv}
  Z.~Xu, C.~Greiner and H.~Stöcker, ``QCD plasma thermalization, collective flow and extraction of shear
  viscosity,'' J.\ Phys.\ G {\bf 35}, 104016 (2008) [arXiv:0807.2986 [hep-ph]].

\bibitem{Alver:2007qw}
  B.~Alver {\it et al.}  [PHOBOS Collaboration], ``Elliptic flow fluctuations in Au+Au collisions at $\sqrt{s_{_{\it NN}}} =$
  200 GeV,'' arXiv:nucl-ex/0702036.

\bibitem{Arnold:2004ti}
  P.~Arnold, J.~Lenaghan, G.~D.~Moore and L.~G.~Yaffe, ``Apparent thermalization due to plasma instabilities in quark gluon
  plasma,'' Phys.\ Rev.\ Lett.\  {\bf 94}, 072302 (2005) [arXiv:nucl-th/0409068].
  
\bibitem{Gribov:1972ri}
  V.~N.~Gribov and L.~N.~Lipatov, ``Deep Inelastic E P Scattering In Perturbation Theory,'' Sov.\ J.\ Nucl.\ Phys.\  {\bf 15}, 438 (1972)
  [Yad.\ Fiz.\  {\bf 15}, 781 (1972)]. 
  
\bibitem{Altarelli:1977zs}
  G.~Altarelli and G.~Parisi, ``Asymptotic Freedom In Parton Language,'' Nucl.\ Phys.\  B {\bf 126}, 298 (1977).
 
\bibitem{Dokshitzer:1977sg}
  Y.~L.~Dokshitzer, ``Calculation Of The Structure Functions For Deep Inelastic Scattering And E+
  E- Annihilation By Perturbation Theory In Quantum Chromodynamics. (In Russian),''
  Sov.\ Phys.\ JETP {\bf 46}, 641 (1977) [Zh.\ Eksp.\ Teor.\ Fiz.\  {\bf 73}, 1216 (1977)]. 
  
\bibitem{Accardi:2004gp}
  A.~Accardi {\it et al.}, ``Hard probes in heavy ion collisions at the LHC: Jet physics,''
  arXiv:hep-ph/0310274.  
  
\bibitem{Cronin:1974zm}
  J.~W.~Cronin, H.~J.~Frisch, M.~J.~Shochet, J.~P.~Boymond, R.~Mermod, P.~A.~Piroue and R.~L.~Sumner,
  ``Production Of Hadrons With Large Transverse Momentum At 200-Gev, 300-Gev, And 400-Gev,''
  Phys.\ Rev.\  D {\bf 11}, 3105 (1975).
  
\bibitem{Antreasyan:1978cw}
  D.~Antreasyan, J.~W.~Cronin, H.~J.~Frisch, M.~J.~Shochet, L.~Kluberg, P.~A.~Piroue and R.~L.~Sumner,
  ``Production Of Hadrons At Large Transverse Momentum In 200-Gev, 300-Gev And 400-Gev P P And P N Collisions,''
  Phys.\ Rev.\  D {\bf 19}, 764 (1979).    

\bibitem{ReviewDavid}
  D.~d'Enterria and B.~Betz, ``High-$p_T$ hadron suppression and jet quenching'', to be published as 
  Springer Lecture Notes in Physics (LNP) for the   QGP Winter School 2008, Springer, Berlin Heidelberg New York 
  (2009); D.~d'Enterria, ``Jet quenching,'' arXiv:0902.2011 [nucl-ex].
  
\bibitem{Adler:2005ig}
  S.~S.~Adler {\it et al.}  [PHENIX Collaboration], ``Centrality dependence of direct photon production in s(NN)**(1/2) =
  200-GeV Au + Au collisions,'' Phys.\ Rev.\ Lett.\  {\bf 94}, 232301 (2005) [arXiv:nucl-ex/0503003].

\bibitem{Adler:2003pb}
  S.~S.~Adler {\it et al.}  [PHENIX Collaboration], ``Mid-rapidity neutral pion production in proton proton collisions at
  $\sqrt{s}$ = 200-GeV,'' Phys.\ Rev.\ Lett.\  {\bf 91}, 241803 (2003) [arXiv:hep-ex/0304038].

\bibitem{Adler:2006hu}
  S.~S.~Adler {\it et al.}  [PHENIX Collaboration], ``Common suppression pattern of eta and pi0 mesons at high transverse
  momentum in Au + Au collisions at s(NN)**(1/2) = 200-GeV,''   Phys.\ Rev.\ Lett.\  {\bf 96}, 202301 (2006)
  [arXiv:nucl-ex/0601037].
  
\bibitem{Adams:2003kv}
  J.~Adams {\it et al.}  [STAR Collaboration], ``Transverse momentum and collision energy dependence of high p(T) hadron
  suppression in Au + Au collisions at ultrarelativistic energies,'' Phys.\ Rev.\ Lett.\  {\bf 91}, 172302 (2003)
  [arXiv:nucl-ex/0305015].  

\bibitem{Adler:2003au}
  S.~S.~Adler {\it et al.}  [PHENIX Collaboration], ``High $p_{T}$ charged hadron suppression in Au + Au collisions at
  $\sqrt{s}_{NN} = 200$ GeV,'' Phys.\ Rev.\  C {\bf 69}, 034910 (2004) [arXiv:nucl-ex/0308006].  
  
\bibitem{VitevGyulassy}
  I.~Vitev and M.~Gyulassy, ``High $p_{T}$ tomography of $d$ + Au and Au+Au at SPS, RHIC, and LHC,''
  Phys.\ Rev.\ Lett.\  {\bf 89}, 252301 (2002) [arXiv:hep-ph/0209161].
  
\bibitem{VitevJet}  
  I.~Vitev, `Jet tomography,'' J.\ Phys.\ G {\bf 30}, S791 (2004) [arXiv:hep-ph/0403089].
  
\bibitem{Muller:2002fa}
  B.~Müller, ``Phenomenology of jet quenching in heavy ion collisions,''
  Phys.\ Rev.\  C {\bf 67}, 061901 (2003) [arXiv:nucl-th/0208038]. 
  
\bibitem{Bjorken:1982qr}
  J.~D.~Bjorken,``Highly Relativistic Nucleus-Nucleus Collisions: The Central Rapidity
  Region,'' Phys.\ Rev.\  D {\bf 27}, 140 (1983). 
  
\bibitem{Gyulassy:1990ye}
  M.~Gyulassy and M.~Plumer, ``Jet Quenching In Dense Matter,'' Phys.\ Lett.\  B {\bf 243}, 432 (1990).  
   
\bibitem{Wang:1991xy}
  X.~N.~Wang and M.~Gyulassy, ``Gluon shadowing and jet quenching in A + A collisions at s**(1/2) =
  200-GeV,'' Phys.\ Rev.\ Lett.\  {\bf 68}, 1480 (1992).  

\bibitem{JetQuenchingSTAR}  
  J.~Adams {\it et al.}  [STAR Collaboration],
  ``Experimental and theoretical challenges in the search for the quark gluon plasma: The STAR collaboration's critical 
  assessment of the  evidence from RHIC collisions,'' Nucl.\ Phys.\  A {\bf 757}, 102 (2005) [arXiv:nucl-ex/0501009].

\bibitem{JetQuenchingPHENIX} 
  K.~Adcox {\it et al.}  [PHENIX Collaboration], ``Suppression of hadrons with large transverse momentum in central Au+Au
  collisions at $\sqrt{s_{NN}}$ = 130-GeV,'' Phys.\ Rev.\ Lett.\  {\bf 88}, 022301 (2002) [arXiv:nucl-ex/0109003].

\bibitem{2pcSTAR} 
   F.~Wang  [STAR Collaboration], ``Soft physics from STAR,'' Nucl.\ Phys.\  A {\bf 774}, 129 (2006)
  [arXiv:nucl-ex/0510068].

\bibitem{2pcPHENIX} 
  A.~Adare {\it et al.}  [PHENIX Collaboration], ``Dihadron azimuthal correlations in Au+Au
  collisions at $\sqrt{s_{NN}}=200$~GeV,'' Phys.\ Rev.\  C {\bf 78}, 014901 (2008) [arXiv:0801.4545 [nucl-ex]].

\bibitem{PunchThroughSTAR} 
  J.~Adams {\it et al.}  [STAR Collaboration], ``Direct observation of dijets in central Au + Au 
  collisions at  s(NN)**(1/2) = 200-GeV,'' Phys.\ Rev.\ Lett.\  {\bf 97}, 162301 (2006)
  [arXiv:nucl-ex/0604018].
  
\bibitem{Ulery:2005cc}
  J.~G.~Ulery  [STAR Collaboration], ``Two- and three-particle jet correlations from STAR,''
  Nucl.\ Phys.\  A {\bf 774}, 581 (2006) [arXiv:nucl-ex/0510055].

\bibitem{Adler:2005ee}
  S.~S.~Adler {\it et al.}  [PHENIX Collaboration], ``Modifications to di-jet hadron pair correlations in Au + Au collisions 
  at s(NN)**(1/2) = 200-GeV,'' Phys.\ Rev.\ Lett.\  {\bf 97}, 052301 (2006) [arXiv:nucl-ex/0507004].

\bibitem{Jia:2008kf}
  J.~Y.~Jia, ``How to Make Sense of the Jet Correlations Results at RHIC?,'' arXiv:0810.0051 [nucl-ex].

\bibitem{CERES_MachCones}
  M.~Ploskon  [CERES Collaboration], ``Two particle azimuthal correlations at high transverse momentum in Pb-Au at
  158 AGeV/c,'' Nucl.\ Phys.\  A {\bf 783}, 527 (2007) [arXiv:nucl-ex/0701023];
  
\bibitem{Kniege_PhD}  
  S.~Kniege, PhD Thesis, University Frankfurt am Main, Germany (2009).
  
\bibitem{vanLeeuwen:2008pn}
  M.~van Leeuwen  [STAR collaboration],``Recent high-pt results from STAR,''
  arXiv:0808.4096 [nucl-ex].  
  
\bibitem{PHENIX_QM09_1}
  C.~Vale, ``Highlights from PHENIX II - Exploring the QCD medium,'' Talk at the Quark Matter Conference
  2009, Knoxville, USA (2009)

\bibitem{PHENIX_QM09_2}  
  W.~G.~Holzmann, ``	Using Two and Three-Particle Correlations in PHENIX to Probe the Response of 
  Strongly Interacting Partonic Matter to High-pT Partons,'' Talk at the Quark Matter Conference
  2009, Knoxville, USA (2009).      

\bibitem{Qiu:2004id}
  J.~W.~Qiu and I.~Vitev, ``Coherent multiple scattering and dihadron correlations in heavy ion
  collisions,'' arXiv:hep-ph/0410218.

\bibitem{Vitev:2005yg}
  I.~Vitev, ``Large Angle Hadron Correlations from Medium-Induced Gluon Radiation,''
  Phys.\ Lett.\  B {\bf 630}, 78 (2005) [arXiv:hep-ph/0501255].
  
\bibitem{Stoecker:2004qu}
  H.~Stöcker, ``Collective Flow signals the Quark Gluon Plasma,''
  Nucl.\ Phys.\  A {\bf 750}, 121 (2005) [arXiv:nucl-th/0406018].

\bibitem{CasalderreySolana:2004qm}
  J.~Casalderrey-Solana, E.~V.~Shuryak and D.~Teaney, ``Conical flow induced by quenched QCD jets,''
  J.\ Phys.\ Conf.\ Ser.\  {\bf 27}, 22 (2005) [Nucl.\ Phys.\  A {\bf 774}, 577 (2006)]
 [arXiv:hep-ph/0411315].  
  
\bibitem{Dremin:2005an}
  I.~M.~Dremin, ``Ring-like events: Cherenkov gluons or Mach waves?,''
  Nucl.\ Phys.\  A {\bf 767}, 233 (2006) [arXiv:hep-ph/0507167].  
  
\bibitem{Majumder:2005sw}
  A.~Majumder and X.~N.~Wang, ``LPM interference and Cherenkov-like gluon bremsstrahlung in dense
  matter,'' Phys.\ Rev.\  C {\bf 73}, 051901 (2006) [arXiv:nucl-th/0507062].
  
\bibitem{Koch:2005sx}
  V.~Koch, A.~Majumder and X.~N.~Wang, ``Cherenkov Radiation from Jets in Heavy-ion Collisions,''
  Phys.\ Rev.\ Lett.\  {\bf 96}, 172302 (2006) [arXiv:nucl-th/0507063].
  
\bibitem{Adler:2002tq}
  C.~Adler {\it et al.}  [STAR Collaboration], ``Disappearance of back-to-back high $p_{T}$ hadron correlations in central
  Au+Au collisions at $\sqrt{s_{NN}}$ = 200-GeV,'' Phys.\ Rev.\ Lett.\  {\bf 90}, 082302 (2003) [arXiv:nucl-ex/0210033].  
  
\bibitem{Adams:2005ph}
  J.~Adams {\it et al.}  [STAR Collaboration], ``Distributions of charged hadrons associated with high transverse  momentum
  particles in p p and Au + Au collisions at s(NN)**(1/2) =  200-GeV,'' Phys.\ Rev.\ Lett.\  {\bf 95}, 152301 (2005)
  [arXiv:nucl-ex/0501016].
  
\bibitem{Ajitanand:2005jj}
  N.~N.~Ajitanand {\it et al.}, ``Decomposition of harmonic and jet contributions to particle-pair correlations at 
  ultra-relativistic energies,'' Phys.\ Rev.\  C {\bf 72}, 011902 (2005) [arXiv:nucl-ex/0501025].
  
\bibitem{Ulery:2007ct}
  J.~G.~Ulery, ``Three-Particle Azimuthal Correlations,'' PoS(LHC07), 036 (2007) [arXiv:0709.1633 [nucl-ex]].  
  
\bibitem{UleryPRL}
  B.~I.~Abelev {\it et al.}  [STAR Collaboration], ``Indications of Conical Emission of Charged Hadrons at RHIC,''
  Phys.\ Rev.\ Lett.\  {\bf 102}, 052302 (2009) [arXiv:0805.0622 [nucl-ex]].
  
\bibitem{Pruneau:2006gj}
  C.~A.~Pruneau, ``Methods for jet studies with three-particle correlations,'' Phys.\ Rev.\  C {\bf 74}, 064910 (2006)
  [arXiv:nucl-ex/0608002].  
  
\bibitem{Ajitanand:2006is}
  N.~N.~Ajitanand  [PHENIX Collaboration], ``Extraction of jet topology using three particle correlations,''
  Nucl.\ Phys.\  A {\bf 783}, 519 (2007) [arXiv:nucl-ex/0609038].  
  
\bibitem{Ulery:2007zb}
  J.~G.~Ulery  [STAR Collaboration], ``Are There Mach Cones in Heavy Ion Collisions? Three-Particle   Correlations
  from STAR,'' Int.\ J.\ Mod.\ Phys.\  E {\bf 16}, 2005 (2007) [arXiv:0704.0224 [nucl-ex]].    
  
\bibitem{Wenger:2008ts}
  B.~Alver {\it et al.}  [PHOBOS Collaboration], ``High $p_T$ Triggered Delta-eta,Delta-phi Correlations over a Broad Range in
  Delta-eta,'' J.\ Phys.\ G {\bf 35}, 104080 (2008) [arXiv:0804.3038 [nucl-ex]].  
  
\bibitem{Adams:2005aw}
  J.~Adams {\it et al.}  [STAR Collaboration], ``Transverse-momentum p(t) correlations on (eta,Phi) from mean-p(t)
  fluctuations in Au - Au collisions at s(NN)**(1/2) = 200-GeV,'' J.\ Phys.\ G {\bf 32}, L37 (2006)
  [arXiv:nucl-ex/0509030].  
  
\bibitem{Putschke:2007mi}
  J.~Putschke, ``Intra-jet correlations of high-$p_t$ hadrons from STAR,'' J.\ Phys.\ G {\bf 34}, S679 (2007)
  [arXiv:nucl-ex/0701074].  
  
\bibitem{Schenke:2008hw}
  B.~Schenke, ``Collective Phenomena in the Non-Equilibrium Quark-Gluon Plasma,'' PhD Thesis, University Frankfurt, Germany (2008), 
  arXiv:0810.4306 [hep-ph].  
  
\bibitem{Dumitru:2008wn}
  A.~Dumitru, F.~Gelis, L.~McLerran and R.~Venugopalan, ``Glasma flux tubes and the near side ridge phenomenon at RHIC,''
  Nucl.\ Phys.\  A {\bf 810}, 91 (2008) [arXiv:0804.3858 [hep-ph]].  
  
\bibitem{Pruneau:2007ua}
  C.~A.~Pruneau, S.~Gavin and S.~A.~Voloshin, ``Transverse Radial Flow Effects on Two- and Three-Particle Angular
  Correlations,'' Nucl.\ Phys.\  A {\bf 802}, 107 (2008) [arXiv:0711.1991 [nucl-ex]].  
  
  
\subsubsection*{Chapter 3: Ideal Hydrodynamics}  
  
\bibitem{Landau:1953gs}
  L.~D.~Landau, ``On the multiparticle production in high-energy collisions,'' Izv.\ Akad.\ Nauk Ser.\ Fiz.\  {\bf 17}, 51 (1953).
  
\bibitem{Stoecker:1986ci}
  H.~Stöcker and W.~Greiner, ``High-Energy Heavy Ion Collisions: Probing The Equation Of State Of Highly
  Excited Hadronic Matter,'' Phys.\ Rept.\  {\bf 137} (1986) 277.

\bibitem{Clare:1986qj}
  R.~B.~Clare and D.~Strottman, ``Relativistic Hydrodynamics and heavy ion reactions,''
  Phys.\ Rept.\  {\bf 141}, 177 (1986).
  
\bibitem{CsernaiBook}
  L.~P.~Csernai, ``Introduction To Relativistic Heavy Ion Collisions,'' John Wiley \& Sons Ltd, Chichester (1994).
  
\bibitem{Bleicher:1999xi}
  M.~Bleicher {\it et al.}, ``Relativistic hadron hadron collisions in the ultra-relativistic quantum
  molecular dynamics model,'' J.\ Phys.\ G {\bf 25}, 1859 (1999) [arXiv:hep-ph/9909407].
  
\bibitem{Xu:2004mz}
  Z.~Xu and C.~Greiner, ``Thermalization of gluons in ultrarelativistic heavy ion collisions by
  including three-body interactions in a parton cascade,'' Phys.\ Rev.\  C {\bf 71}, 064901 (2005)
  [arXiv:hep-ph/0406278].
  
\bibitem{Gutbrod:1988hh}
  H.~H.~Gutbrod, B.~W.~Kolb, H.~R.~Schmidt, A.~M.~Poskanzer, H.~G.~Ritter and K.~H.~Kampert,
  ``A New Component Of The Collective Flow In Relativistic Heavy Ion Collisions,''
  Phys.\ Lett.\  B {\bf 216}, 267 (1989).

\bibitem{Landau} 
 L.~D.~Landau and E.~M.~Lifshitz, ``Fluid Mechanics, Volume 6 (Course of Theoretical Physics) '', Pergamon Press, New York, (1987).
  
\bibitem{deGroot} 
 S.~R.~deGroot, W.~A.~van Leeuwen, and Ch.~G.~van Weert, ``Relativistic Kinetic Theory '', North-Holland, Amsterdam, (1980).  
  
\bibitem{Rischke:1995ir}
  D.~H.~Rischke, S.~Bernard and J.~A.~Maruhn, ``Relativistic hydrodynamics for heavy ion collisions. 1. General aspects and
  expansion into vacuum,'' Nucl.\ Phys.\  A {\bf 595}, 346 (1995) [arXiv:nucl-th/9504018].    
  
\bibitem{BorisBook1}
  J.~Boris and D.~Book, ``Flux corrected transport I: SHASTA a fluid algorithm that works'',
  J.\ Comp.\ Phys.\ {\bf 11}, 38 (1973).  
  
\bibitem{BorisBook2}
  D.~L.~Book, J.~P.~Boris, and K.~Hain, ``Flux-Corrected Transport II - Generalizations of the method'',
  J.\ Comp.\ Phys.\ {\bf 18}, 248 (1975).  

\bibitem{Rischke:1995mt}
  D.~H.~Rischke, Y.~Pürsün and J.~A.~Maruhn, ``Relativistic hydrodynamics for heavy ion collisions. 2. Compression of
  nuclear matter and the phase transition to the quark - gluon plasma,'' Nucl.\ Phys.\  A {\bf 595}, 383 (1995)
  [Erratum-ibid.\  A {\bf 596}, 717 (1996)] [arXiv:nucl-th/9504021].      
  
\bibitem{Rischke:1995pe}
  D.~H.~Rischke, Y.~Pürsün, J.~A.~Maruhn, H.~Stöcker and W.~Greiner,
  ``The phase transition to the quark-gluon plasma and its effects on hydrodynamic flow,'' Heavy Ion Phys.\  {\bf 1}, 309 (1995).  

\bibitem{Schneider:1993gd}
  V.~Schneider, U.~Katscher, D.~H.~Rischke, B.~Waldhauser, J.~A.~Maruhn and C.~D.~Munz,
  ``New algorithms for ultrarelativistic numerical hydrodynamics,'' J.\ Comput.\ Phys.\  {\bf 105}, 92 (1993).
  
\bibitem{Rischke:1998fq}
  D.~H.~Rischke, ``Fluid dynamics for relativistic nuclear collisions,'' arXiv:nucl-th/9809044. 
  
\bibitem{CFLTheorem}
  R.~Courant, K.~O.~Friedrichs, and H.~Lewy,
  ``Über die partiellen Differenzengleichungen der mathematischen Physik,'' Math.\ Ann.\  {\bf 100}, 92 (1928).   
  
\bibitem{LandauModel}
 L.~D.~Landau and S.~Z.~Belenkii, ``\textcyr{Gidrodinamicheskaya teoriya mnozhestvennogo obrazovaniya chastits}'', 
 Uspekhi Fizicheskikh Nauk {\bf 56}, 309 (1955).
  
\bibitem{Cooper:1974qi}
  F.~Cooper, G.~Frye and E.~Schonberg, ``Landau's Hydrodynamic Model Of Particle Production And Electron Positron
  Annihilation Into Hadrons,'' Phys.\ Rev.\  D {\bf 11}, 192 (1975).  
  
\bibitem{LectureMishustin} 
  I.~Mishustin, Lectures on ``Dynamical models of relativistic heavy-ion collisions'', University Frankfurt (2005). 
    
\bibitem{Kolb:2001qz}
  P.~F.~Kolb, U.~W.~Heinz, P.~Huovinen, K.~J.~Eskola and K.~Tuominen, ``Centrality dependence of multiplicity, transverse energy, and elliptic
  flow from hydrodynamics,'' Nucl.\ Phys.\  A {\bf 696}, 197 (2001) [arXiv:hep-ph/0103234].   
  
\bibitem{Luzum:2008cw}
  M.~Luzum and P.~Romatschke, ``Conformal Relativistic Viscous Hydrodynamics:  Applications to RHIC results
  at $\sqrt{s_{NN}} = 200$~GeV,'' Phys.\ Rev.\  C {\bf 78}, 034915 (2008) [arXiv:0804.4015 [nucl-th]].    
  
\bibitem{Arnold:2003rq}
  P.~Arnold, J.~Lenaghan and G.~D.~Moore, ``QCD plasma instabilities and bottom-up thermalization,''
  JHEP {\bf 0308}, 002 (2003) [arXiv:hep-ph/0307325].  
  
\bibitem{Iancu:2002xk}
  E.~Iancu, A.~Leonidov and L.~McLerran, ``The colour glass condensate: An introduction,''
  arXiv:hep-ph/0202270.  
  
\bibitem{Chodos:1974je}
  A.~Chodos, R.~L.~Jaffe, K.~Johnson, C.~B.~Thorn and V.~F.~Weisskopf, ``A New Extended Model Of Hadrons,''
  Phys.\ Rev.\  D {\bf 9}, 3471 (1974).  
  
\bibitem{Serot:1984ey}
  B.~D.~Serot and J.~D.~Walecka, ``The Relativistic Nuclear Many Body Problem,''
  Adv.\ Nucl.\ Phys.\  {\bf 16}, 1 (1986).  
  
\bibitem{Steinheimer:2007iy}
  J.~Steinheimer, M.~Bleicher, H.~Petersen, S.~Schramm, H.~Stöcker and D.~Zschiesche,
  ``(3+1)-Dimensional Hydrodynamic Expansion with a Critical Point from Realistic Initial Conditions,''
  Phys.\ Rev.\  C {\bf 77}, 034901 (2008) [arXiv:0710.0332 [nucl-th]].  
  
\bibitem{NoronhaHostler:2008ju}
  J.~Noronha-Hostler, J.~Noronha and C.~Greiner, ``Transport Coefficients of Hadronic Matter near $T_c$,''
  arXiv:0811.1571 [nucl-th].  
  
\bibitem{Cooper:1974mv}
  F.~Cooper and G.~Frye, ``Comment On The Single Particle Distribution In The Hydrodynamic And
  Statistical Thermodynamic Models Of Multiparticle Production,''
  Phys.\ Rev.\  D {\bf 10}, 186 (1974).
  
\bibitem{Chemout}
  The figure is taken from http://www-rnc.lbl.gov/~ssalur/www/Research3.html.  
  
\bibitem{KolbThesis}
  P.~F.~Kolb, ``Early Thermalization and Hydrodynamic Expansion in Nuclear Collisions at RHIC,''
  PhD Thesis, University Regensburg, Germany (2002).
  
\bibitem{Nonaka:2005aj}
  C.~Nonaka and S.~A.~Bass, ``3-D hydro + cascade model at RHIC,'' Nucl.\ Phys.\  A {\bf 774}, 873 (2006)
  [arXiv:nucl-th/0510038].
    
\bibitem{Petersen:2008dd}
  H.~Petersen, J.~Steinheimer, G.~Burau, M.~Bleicher and H.~Stöcker,
  ``A Fully Integrated Transport Approach to Heavy Ion Reactions with an 
  Intermediate Hydrodynamic Stage,'' Phys.\ Rev.\  C {\bf 78}, 044901 (2008)
  [arXiv:0806.1695 [nucl-th]].
  
\bibitem{Fries:2003vb}
  R.~J.~Fries, B.~Müller, C.~Nonaka and S.~A.~Bass,
  ``Hadronization in heavy ion collisions: Recombination and fragmentation  of
  partons,'' Phys.\ Rev.\ Lett.\  {\bf 90}, 202303 (2003) [arXiv:nucl-th/0301087].
  
\bibitem{Greco:2003xt}
  V.~Greco, C.~M.~Ko and P.~Levai,
  ``Parton coalescence and antiproton/pion anomaly at RHIC,'' Phys.\ Rev.\ Lett.\  {\bf 90}, 
  202302 (2003) [arXiv:nucl-th/0301093].
  
\bibitem{Betz:2008wy}
  B.~Betz, M.~Gyulassy, J.~Noronha and G.~Torrieri,
  ``Anomalous Conical Di-jet Correlations in pQCD vs AdS/CFT,'' 
  Phys.\ Lett.\ B {\bf 675}, 340 (2009) [arXiv:0807.4526 [hep-ph]].
  
\bibitem{Betz:2008ka}
  B.~Betz, J.~Noronha, G.~Torrieri, M.~Gyulassy, I.~Mishustin and D.~H.~Rischke,
  ``Universality of the Diffusion Wake from Stopped and Punch-Through Jets in
  Heavy-Ion Collisions,'' Phys.\ Rev.\  C {\bf 79}, 034902 (2009)
  [arXiv:0812.4401 [nucl-th]].
  
  
\subsubsection*{Chapter 4: Viscous Hydrodynamics}    

\bibitem{Muronga:2001zk}
  A.~Muronga, ``Second order dissipative fluid dynamics for ultra-relativistic nuclear
  collisions,'' Phys.\ Rev.\ Lett.\  {\bf 88}, 062302 (2002) [Erratum-ibid.\  {\bf 89}, 159901 (2002)]
  [arXiv:nucl-th/0104064].

\bibitem{Muronga:2004sf}
  A.~Muronga and D.~H.~Rischke, ``Evolution of hot, dissipative quark matter in relativistic nuclear
  collisions,'' arXiv:nucl-th/0407114.
  
\bibitem{Heinz:2005bw}
  U.~W.~Heinz, H.~Song and A.~K.~Chaudhuri, ``Dissipative hydrodynamics for viscous relativistic fluids,''
  Phys.\ Rev.\  C {\bf 73}, 034904 (2006) [arXiv:nucl-th/0510014].

\bibitem{Baier:2006um}
  R.~Baier, P.~Romatschke and U.~A.~Wiedemann, ``Dissipative hydrodynamics and heavy ion collisions,''
  Phys.\ Rev.\  C {\bf 73}, 064903 (2006) [arXiv:hep-ph/0602249].

\bibitem{Chaudhuri:2006jd}
  A.~K.~Chaudhuri, ``Dissipative hydrodynamics in 2+1 dimension,'' Phys.\ Rev.\  C {\bf 74}, 044904 (2006)
  [arXiv:nucl-th/0604014].

\bibitem{Baier:2006gy}
  R.~Baier and P.~Romatschke, ``Causal viscous hydrodynamics for central heavy-ion collisions,''
  Eur.\ Phys.\ J.\  C {\bf 51}, 677 (2007) [arXiv:nucl-th/0610108].
  
\bibitem{Muronga:2006zw}
  A.~Muronga, ``Relativistic Dynamics of Non-ideal Fluids: Viscous and heat-conducting
  fluids I. General Aspects and 3+1 Formulation for Nuclear Collisions,''
  Phys.\ Rev.\  C {\bf 76}, 014909 (2007) [arXiv:nucl-th/0611090].
  
\bibitem{Muronga:2006zx}
  A.~Muronga, ``Relativistic Dynamics of Non-ideal Fluids: Viscous and heat-conducting
  fluids II. Transport properties and microscopic description of relativistic
  nuclear matter,'' Phys.\ Rev.\  C {\bf 76}, 014910 (2007) [arXiv:nucl-th/0611091].
  
\bibitem{Dusling:2007gi}
  K.~Dusling and D.~Teaney, ``Simulating elliptic flow with viscous hydrodynamics,''
  Phys.\ Rev.\  C {\bf 77}, 034905 (2008) [arXiv:0710.5932 [nucl-th]].

\bibitem{Molnar:2008xj}
  D.~Molnar and P.~Huovinen, ``Dissipative effects from transport and viscous hydrodynamics,''
  J.\ Phys.\ G {\bf 35}, 104125 (2008) [arXiv:0806.1367 [nucl-th]]. 
  
\bibitem{El:2008yy}
  A.~El, A.~Muronga, Z.~Xu and C.~Greiner, ``Shear viscosity and out of equilibrium dissipative hydrodynamics,''
  arXiv:0812.2762 [hep-ph].
  
\bibitem{Hiscock:1985zz}
  W.~A.~Hiscock and L.~Lindblom, ``Generic instabilities in first-order dissipative relativistic fluid
  theories,'' Phys.\ Rev.\  D {\bf 31}, 725 (1985).  
  
\bibitem{IS}
  W.~Israel and J.~M.~Stewart, ``Transient relativistic thermodynamics and kinetic theory,''
  Annals Phys.\  {\bf 118}, 341 (1979).  
  
\bibitem{Hiscock:1983zz}
  W.~A.~Hiscock and L.~Lindblom, ``Stability and causality in dissipative relativistic fluids,''
  Annals Phys.\  {\bf 151}, 466 (1983).    
  
\bibitem{Grad}  
 H.~Grad, ``Statistical mechanics, thermodynamics, and fluid dynamics of systems with an 
 arbitrary number of integrals,'' Commun.\ Pure App.\ Math.\ {\bf 2}, 381 (1949).
 
\bibitem{Mueller}
  I.~Müller, ``Zum Paradoxon der Wärmeleitungstheorie,'' Z.\ Phys.\ {\bf 198}, 329 (1967).
 
\bibitem{Muronga:2003ta}
  A.~Muronga, ``Causal Theories of Dissipative Relativistic Fluid Dynamics for Nuclear
  Collisions,'' Phys.\ Rev.\  C {\bf 69}, 034903 (2004) [arXiv:nucl-th/0309055].   
 
\bibitem{Betz:2008me}
  B.~Betz, D.~Henkel and D.~H.~Rischke, ``From kinetic theory to dissipative fluid dynamics,''
  Prog.\ Part.\ Nucl.\ Phys.\ {\bf 62}, 556 (2009) [arXiv:0812.1440 [nucl-th]]. 

\bibitem{ViscousBetz}  
  B.~Betz, D.~Henkel, H.~Niemi, and D.~H.~Rischke, in preparation.  
 
\bibitem{Song:2008si}
  H.~Song and U.~W.~Heinz, ``Multiplicity scaling in ideal and viscous hydrodynamics,''
  Phys.\ Rev.\  C {\bf 78}, 024902 (2008) [arXiv:0805.1756 [nucl-th]]. 
  
\bibitem{Baier:2007ix}
  R.~Baier, P.~Romatschke, D.~T.~Son, A.~O.~Starinets and M.~A.~Stephanov,
  ``Relativistic viscous hydrodynamics, conformal invariance, and holography,''
  JHEP {\bf 0804}, 100 (2008) [arXiv:0712.2451 [hep-th]].  
  
\bibitem{York:2008rr}
  M.~A.~York and G.~D.~Moore, ``Second order hydrodynamic coefficients from kinetic theory,''
  arXiv:0811.0729 [hep-ph].  
  
\bibitem{Prakash:1993bt}
  M.~Prakash, M.~Prakash, R.~Venugopalan and G.~Welke, ``Nonequilibrium properties of hadronic mixtures,''
  Phys.\ Rept.\  {\bf 227}, 321 (1993).  
  
\bibitem{Danielewicz:1984ww}
  P.~Danielewicz and M.~Gyulassy, ``Dissipative Phenomena In Quark Gluon Plasmas,''
  Phys.\ Rev.\  D {\bf 31}, 53 (1985).
  
\bibitem{Heinz:2001xi}
  U.~W.~Heinz and P.~F.~Kolb, ``Early thermalization at RHIC,'' Nucl.\ Phys.\  A {\bf 702}, 269 (2002)
  [arXiv:hep-ph/0111075].    
  
\bibitem{Song:2007fn}
  H.~Song and U.~W.~Heinz, ``Suppression of elliptic flow in a minimally viscous quark-gluon plasma,''
  Phys.\ Lett.\  B {\bf 658}, 279 (2008) [arXiv:0709.0742 [nucl-th]].    
  
  
\subsubsection*{Chapter 5: Shock Wave Phenomena}  

\bibitem{Taub:1948zz}
  A.~H.~Taub, ``Relativistic Rankine-Hugoniot Equations,''
  Phys.\ Rev.\  {\bf 74}, 328 (1948).  
  
\bibitem{Scheid:1974zz}
  W.~Scheid, H.~Müller and W.~Greiner, ``Nuclear Shock Waves in Heavy-Ion Collisions,''
  Phys.\ Rev.\ Lett.\  {\bf 32}, 741 (1974).  

\bibitem{Baumgardt:1975qv}
  H.~G.~Baumgardt {\it et al.}, ``Shock Waves And Mach Cones In Fast Nucleus-Nucleus Collisions,''
  Z.\ Phys.\  A {\bf 273}, 359 (1975). 
  
\bibitem{Hofmann:1976dy}
  J.~Hofmann, H.~Stöcker, U.~W.~Heinz, W.~Scheid and W.~Greiner,
  ``Possibility Of Detecting Density Isomers In High Density Nuclear Mach Shock Waves,''
  Phys.\ Rev.\ Lett.\  {\bf 36} (1976) 88.
 
\bibitem{Gutbrod:1989wd}
  H.~H.~Gutbrod, A.~M.~Poskanzer and H.~G.~Ritter, ``Plastic Ball Experiments,''
  Rept.\ Prog.\ Phys.\  {\bf 52}, 1267 (1989).  
  
\bibitem{Philip} 
  P.~Rau, J.~Steinheimer, B.~Betz, H.~Petersen, M.~Bleicher and H.~Stocker, to be published.  

\bibitem{Satarov:2005mv}
  L.~M.~Satarov, H.~Stöcker and I.~N.~Mishustin, ``Mach shocks induced by partonic jets in expanding quark-gluon plasma,''
  Phys.\ Lett.\  B {\bf 627}, 64 (2005) [arXiv:hep-ph/0505245].

\bibitem{CasalderreySolana:2006sq}
  J.~Casalderrey-Solana, E.~V.~Shuryak and D.~Teaney, ``Hydrodynamic flow from fast particles,''
  arXiv:hep-ph/0602183.
  
\bibitem{Renk:2005si}
  T.~Renk and J.~Ruppert, ``Mach cones in an evolving medium,''
  Phys.\ Rev.\  C {\bf 73}, 011901 (2006) [arXiv:hep-ph/0509036].  
  
\bibitem{Betz:2006ds}
  B.~Betz, M.~Bleicher, U.~Harbach, T.~Humanic, B.~Koch and H.~Stöcker,
  ``Mini black holes at the LHC: Discovery through di-jet suppression,
  mono-jet emission and a supersonic boom in the quark-gluon plasma in  ALICE,
  ATLAS and CMS,''
  arXiv:hep-ph/0606193.  
  
\bibitem{Bouras:2009nn}
  I.~Bouras {\it et al.}, ``Relativistic shock waves in viscous gluon matter,'' arXiv:0902.1927 [hep-ph].  
 
\bibitem{Rischke:1990jy}
  D.~H.~Rischke, H.~Stöcker and W.~Greiner, ``Flow In Conical Shock Waves: A Signal For The Deconfinement Transition?,''
  Phys.\ Rev.\  D {\bf 42}, 2283 (1990).
 
\bibitem{Courant} 
 R.~Courant and K.~O.~Friedrichs, ``Supersonic Flow and Shock Waves '', Interscience, New York, (1948).
    
\bibitem{TaylorMaccoll1} 
 G.~I.~Taylor and J.~W.~Maccoll, ``The Air Pressure on a Cone Moving at High Speeds. I '', Proc.\ R.\ Soc.\ {\bf A139}, 278 (1933).  
 
\bibitem{TaylorMaccoll2} 
 G.~I.~Taylor and J.~W.~Maccoll, ``The Air Pressure on a Cone Moving at High Speeds. II '', Proc.\ R.\ Soc.\ {\bf A139}, 298 (1933).   
   
  
\subsubsection*{Chapter 6: Jet Energy Loss}    

\bibitem{LPM} 
 A.~B.~Migdal, ``Bremsstrahlung and Pair Production in Condensed Media at High Energies'', Phys.\ Rev.\ {\bf 103}, 1811 (1956).   
  
\bibitem{Yao:2006px}
  W.~M.~Yao {\it et al.}  [Particle Data Group], ``Review of particle physics,''  J.\ Phys.\ G {\bf 33}, 1 (2006).

\bibitem{Wicks:2007am}
  S.~Wicks, W.~Horowitz, M.~Djordjevic and M.~Gyulassy, ``Heavy quark jet quenching with collisional plus radiative energy loss and
  path length fluctuations,'' Nucl.\ Phys.\  A {\bf 783}, 493 (2007) [arXiv:nucl-th/0701063].
 
\bibitem{Qin:2007rn}
  G.~Y.~Qin, J.~Ruppert, C.~Gale, S.~Jeon, G.~D.~Moore and M.~G.~Mustafa, ``Radiative and Collisional Jet Energy Loss in the Quark-Gluon Plasma at
  RHIC,'' Phys.\ Rev.\ Lett.\  {\bf 100}, 072301 (2008) [arXiv:0710.0605 [hep-ph]].
  
\bibitem{Eskola:2004cr}
  K.~J.~Eskola, H.~Honkanen, C.~A.~Salgado and U.~A.~Wiedemann, ``The fragility of high-p(T) hadron spectra as a hard probe,''
  Nucl.\ Phys.\  A {\bf 747}, 511 (2005) [arXiv:hep-ph/0406319].  
  
\bibitem{Loizides:2006cs}
  C.~Loizides, ``High transverse momentum suppression and surface effects in nucleus nucleus
  collisions within the parton quenching model,'' Eur.\ Phys.\ J.\  C {\bf 49}, 339 (2007) [arXiv:hep-ph/0608133].  
  
\bibitem{Adare:2008cg}
  A.~Adare {\it et al.}  [PHENIX Collaboration], ``Quantitative Constraints on the Opacity of Hot Partonic Matter from
  Semi-Inclusive Single High Transverse Momentum Pion Suppression in Au+Au collisions at $\sqrt{s_{NN}}$ = 200 GeV,''
  Phys.\ Rev.\  C {\bf 77}, 064907 (2008) [arXiv:0801.1665 [nucl-ex]].  
  
\bibitem{TECHQM}
  TECHQM,  Theory-Experiment Collaboration for Hot QCD Matter, https://wiki.bnl.gov/TECHQM/index.php/Main\_Page.      
  
\bibitem{Gyulassy:1999zd}
  M.~Gyulassy, P.~Levai and I.~Vitev, ``Jet quenching in thin quark-gluon plasmas. I: Formalism,''
  Nucl.\ Phys.\  B {\bf 571}, 197 (2000) [arXiv:hep-ph/9907461].  

\bibitem{Djordjevic:2003zk}
  M.~Djordjevic and M.~Gyulassy, ``Heavy quark radiative energy loss in QCD matter,'' Nucl.\ Phys.\  A {\bf 733}, 265 (2004)
  [arXiv:nucl-th/0310076].  

\bibitem{Baier:1996kr}
  R.~Baier, Y.~L.~Dokshitzer, A.~H.~Mueller, S.~Peigne and D.~Schiff, ``Radiative energy loss of high energy quarks and gluons in a finite-volume
  quark-gluon plasma,'' Nucl.\ Phys.\  B {\bf 483}, 291 (1997) [arXiv:hep-ph/9607355].  
  
\bibitem{Baier:1996sk}
  R.~Baier, Y.~L.~Dokshitzer, A.~H.~Mueller, S.~Peigne and D.~Schiff, ``Radiative energy loss and p(T)-broadening of high energy partons in
  nuclei,'' Nucl.\ Phys.\  B {\bf 484}, 265 (1997) [arXiv:hep-ph/9608322].  

\bibitem{Zakharov:1996fv}
  B.~G.~Zakharov, ``Fully quantum treatment of the Landau-Pomeranchuk-Migdal effect in QED  and QCD,'' JETP Lett.\  {\bf 63}, 952 (1996)
  [arXiv:hep-ph/9607440].  
  
\bibitem{Wiedemann:2000za}
  U.~A.~Wiedemann, ``Gluon radiation off hard quarks in a nuclear environment: Opacity expansion,'' Nucl.\ Phys.\  B {\bf 588}, 303 (2000)
  [arXiv:hep-ph/0005129].  
  
\bibitem{Qiu:1990xxa}
  J.~W.~Qiu and G.~Sterman, ``Power corrections in hadronic scattering. 1. Leading 1/Q**2 corrections to the Drell-Yan cross-section,''
  Nucl.\ Phys.\  B {\bf 353}, 105 (1991).  
  
\bibitem{Qiu:1990xy}
  J.~W.~Qiu and G.~Sterman, ``Power corrections to hadronic scattering. 2. Factorization,'' Nucl.\ Phys.\  B {\bf 353}, 137 (1991). 
  
\bibitem{Arnold:2001ba}
  P.~Arnold, G.~D.~Moore and L.~G.~Yaffe, ``Photon emission from ultrarelativistic plasmas,'' JHEP {\bf 0111}, 057 (2001)
  [arXiv:hep-ph/0109064].     
  
\bibitem{Majumder:2007iu}
  A.~Majumder, ``A comparative study of jet-quenching schemes,'' J.\ Phys.\ G {\bf 34}, S377 (2007) [arXiv:nucl-th/0702066].  
  
\bibitem{Bass:2008ch}
  S.~A.~Bass, C.~Gale, A.~Majumder, C.~Nonaka, G.~Y.~Qin, T.~Renk and J.~Ruppert, ``Systematic Comparison of Jet Energy-Loss Schemes in a 3D hydrodynamic
  medium,'' J.\ Phys.\ G {\bf 35}, 104064 (2008) [arXiv:0805.3271 [nucl-th]].  
  
\bibitem{Fochler:2008ts}
  O.~Fochler, Z.~Xu and C.~Greiner,
  ``Towards a unified understanding of jet-quenching and elliptic flow within perturbative QCD parton transport,''
  arXiv:0806.1169 [hep-ph].  
  
\bibitem{Neufeld:2008fi}
  R.~B.~Neufeld, B.~Müller and J.~Ruppert, ``Sonic Mach Cones Induced by Fast Partons in a Perturbative Quark-Gluon Plasma,''
  Phys.\ Rev.\  C {\bf 78}, 041901 (2008) [arXiv:0802.2254 [hep-ph]].

\bibitem{CasalderreySolana:2007km}
  J.~Casalderrey-Solana, ``Mach cones in quark gluon plasma,'' J.\ Phys.\ G {\bf 34}, S345 (2007) [arXiv:hep-ph/0701257].
  
\bibitem{Chaudhuri:2005vc}
  A.~K.~Chaudhuri and U.~Heinz, ``Effect of jet quenching on the hydrodynamical evolution of QGP,'' Phys.\ Rev.\ Lett.\  {\bf 97}, 062301 (2006)
  [arXiv:nucl-th/0503028].  
  
\bibitem{Renk:2006mv}
  T.~Renk and J.~Ruppert, ``The rapidity structure of Mach cones and other large angle correlations in
  heavy-ion collisions,'' Phys.\ Lett.\  B {\bf 646}, 19 (2007) [arXiv:hep-ph/0605330].

\bibitem{Neufeld:2008hs}
  R.~B.~Neufeld, ``Fast Partons as a Source of Energy and Momentum in a Perturbative Quark-Gluon Plasma,''
  Phys.\ Rev.\  D {\bf 78}, 085015 (2008) [arXiv:0805.0385 [hep-ph]].
  
\bibitem{Asakawa:2006jn}
  M.~Asakawa, S.~A.~Bass and B.~Müller, ``Anomalous transport processes in anisotropically expanding quark-gluon
  plasmas,'' Prog.\ Theor.\ Phys.\  {\bf 116}, 725 (2007) [arXiv:hep-ph/0608270].  
  
\bibitem{Neufeld:2008dx}
  R.~B.~Neufeld, ``Propagating Mach Cones in a Viscous Quark-Gluon Plasma,'' arXiv:0807.2996 [nucl-th].  
  
\bibitem{Qin:2009uh}
  G.~Y.~Qin, A.~Majumder, H.~Song and U.~Heinz, ``Energy and momentum deposited into a QCD medium by a jet shower,''
  arXiv:0903.2255 [nucl-th].  
   
  
\subsubsection*{Chapter 7: From String to Field Theory: The AdS/CFT Correspondence}    

\bibitem{Strings} 
 M.~B.~Green, J.~H.~Schwarz, and E.~Witten, ``Superstring Theory '', Cambridge University Press, Cambridge, (1987).
 
\bibitem{Aharony:1999ti}
  O.~Aharony, S.~S.~Gubser, J.~M.~Maldacena, H.~Ooguri and Y.~Oz, ``Large N field theories, string theory and gravity,''
  Phys.\ Rept.\  {\bf 323}, 183 (2000) [arXiv:hep-th/9905111]. 
 
\bibitem{'tHooft:1973jz}
  G.~'t Hooft, ``A Planar Diagram Theory For Strong Interactions,'' Nucl.\ Phys.\  B {\bf 72}, 461 (1974). 
  
\bibitem{Maldacena:1997re}
  J.~M.~Maldacena, ``The large N limit of superconformal field theories and supergravity,''
  Adv.\ Theor.\ Math.\ Phys.\  {\bf 2}, 231 (1998) [Int.\ J.\ Theor.\ Phys.\  {\bf 38}, 1113 (1999)] [arXiv:hep-th/9711200].  

\bibitem{Witten:1998qj}
  E.~Witten, ``Anti-de Sitter space and holography,'' Adv.\ Theor.\ Math.\ Phys.\  {\bf 2}, 253 (1998)
  [arXiv:hep-th/9802150].
  
\bibitem{Witten:1998zw}
  E.~Witten, ``Anti-de Sitter space, thermal phase transition, and confinement in  gauge theories,''
  Adv.\ Theor.\ Math.\ Phys.\  {\bf 2}, 505 (1998) [arXiv:hep-th/9803131]. 
  
\bibitem{Klebanov:2000me}
  I.~R.~Klebanov, ``TASI lectures: Introduction to the AdS/CFT correspondence,''
  arXiv:hep-th/0009139.  
  
\bibitem{Chesler:2008uy}
  P.~M.~Chesler, K.~Jensen, A.~Karch and L.~G.~Yaffe,
  ``Light quark energy loss in strongly-coupled N = 4 supersymmetric Yang-Mills
  plasma,'' arXiv:0810.1985 [hep-th].   

\bibitem{Herzog:2006gh}
  C.~P.~Herzog, A.~Karch, P.~Kovtun, C.~Kozcaz and L.~G.~Yaffe, ``Energy loss of a heavy quark moving through N = 4 supersymmetric
  Yang-Mills plasma,'' JHEP {\bf 0607}, 013 (2006) [arXiv:hep-th/0605158].
  
\bibitem{CasalderreySolana:2007qw}
  J.~Casalderrey-Solana and D.~Teaney, ``Transverse momentum broadening of a fast quark in a N = 4 Yang Mills
  plasma,'' JHEP {\bf 0704}, 039 (2007) [arXiv:hep-th/0701123].
  
\bibitem{Gubser:2006bz}
  S.~S.~Gubser, ``Drag force in AdS/CFT,'' Phys.\ Rev.\  D {\bf 74}, 126005 (2006) [arXiv:hep-th/0605182].

\bibitem{Friess:2006fk}
  J.~J.~Friess, S.~S.~Gubser, G.~Michalogiorgakis and S.~S.~Pufu, ``The stress tensor of a quark moving through N = 4 thermal plasma,''
  Phys.\ Rev.\  D {\bf 75}, 106003 (2007) [arXiv:hep-th/0607022].
  
\bibitem{Yarom:2007ni}
  A.~Yarom, ``On the energy deposited by a quark moving in an N=4 SYM plasma,'' Phys.\ Rev.\  D {\bf 75}, 105023 (2007)
  [arXiv:hep-th/0703095].
  
\bibitem{Gubser:2007nd}
  S.~S.~Gubser and S.~S.~Pufu, ``Master field treatment of metric perturbations sourced by the trailing string,''
  Nucl.\ Phys.\  B {\bf 790}, 42 (2008) [arXiv:hep-th/0703090].

\bibitem{Gubser:1996de}
  S.~S.~Gubser, I.~R.~Klebanov and A.~W.~Peet, ``Entropy and Temperature of Black 3-Branes,''
  Phys.\ Rev.\  D {\bf 54}, 3915 (1996) [arXiv:hep-th/9602135].
  
\bibitem{Torrieri:2009mv}
  G.~Torrieri, B.~Betz, J.~Noronha and M.~Gyulassy, ``Mach cones in heavy ion collisions,''
  Acta Phys.\ Polon.\  B {\bf 39}, 3281 (2008) [arXiv:0901.0230 [nucl-th]].
  
\bibitem{Policastro:2001yc}
  G.~Policastro, D.~T.~Son and A.~O.~Starinets, ``The shear viscosity of strongly coupled N = 4 supersymmetric Yang-Mills
  plasma,'' Phys.\ Rev.\ Lett.\  {\bf 87}, 081601 (2001) [arXiv:hep-th/0104066].

\bibitem{Buchel:2003tz}
  A.~Buchel and J.~T.~Liu, ``Universality of the shear viscosity in supergravity,'' Phys.\ Rev.\ Lett.\  {\bf 93}, 090602 (2004)
  [arXiv:hep-th/0311175].

\bibitem{Kovtun:2004de}
  P.~Kovtun, D.~T.~Son and A.~O.~Starinets, ``Viscosity in strongly interacting quantum field theories from black hole
  physics,'' Phys.\ Rev.\ Lett.\  {\bf 94}, 111601 (2005) [arXiv:hep-th/0405231].
  
\bibitem{Kats:2007mq}
  Y.~Kats and P.~Petrov, ``Effect of curvature squared corrections in AdS on the viscosity of the dual
  gauge theory,'' JHEP {\bf 0901}, 044 (2009) [arXiv:0712.0743 [hep-th]].

\bibitem{Brigante:2007nu}
  M.~Brigante, H.~Liu, R.~C.~Myers, S.~Shenker and S.~Yaida,
  ``Viscosity Bound Violation in Higher Derivative Gravity,''
  Phys.\ Rev.\  D {\bf 77}, 126006 (2008) [arXiv:0712.0805 [hep-th]].

\bibitem{Brigante:2008gz}
  M.~Brigante, H.~Liu, R.~C.~Myers, S.~Shenker and S.~Yaida,
  ``The Viscosity Bound and Causality Violation,'' Phys.\ Rev.\ Lett.\  {\bf 100}, 191601 (2008)
  [arXiv:0802.3318 [hep-th]].
 
\bibitem{Noronha:2007xe}
  J.~Noronha, G.~Torrieri and M.~Gyulassy, ``Near Zone Navier-Stokes Analysis of Heavy Quark Jet Quenching in an
  $\mathcal{N}$ =4 SYM Plasma,'' Phys.\ Rev.\  C {\bf 78}, 024903 (2008) [arXiv:0712.1053 [hep-ph]].
 
\bibitem{Gyulassy:2008fa}
  M.~Gyulassy, J.~Noronha and G.~Torrieri, ``Conical Di-jet Correlations from a Chromo-Viscous Neck in AdS/CFT,''
  arXiv:0807.2235 [hep-ph]. 
  
\bibitem{Noronha:2008un}
  J.~Noronha, M.~Gyulassy and G.~Torrieri, ``Di-Jet Conical Correlations Associated with Heavey Quark Jets in anti-de
  Sitter Space/Conformal Field Theory Correspondence,'' Phys.\ Rev.\ Lett.\  {\bf 102}, 102301 (2009)  [arXiv:0807.1038 [hep-ph]].    
  
\bibitem{Chesler:2007sv}
  P.~M.~Chesler and L.~G.~Yaffe, ``The stress-energy tensor of a quark moving through a strongly-coupled N=4
  supersymmetric Yang-Mills plasma: comparing hydrodynamics and AdS/CFT,'' Phys.\ Rev.\  D {\bf 78}, 045013 (2008)
  [arXiv:0712.0050 [hep-th]].  
  
\bibitem{Gubser:2007ga}
  S.~S.~Gubser, S.~S.~Pufu and A.~Yarom, ``Sonic booms and diffusion wakes generated by a heavy quark in thermal
  AdS/CFT,'' Phys.\ Rev.\ Lett.\  {\bf 100}, 012301 (2008) [arXiv:0706.4307 [hep-th]].  
    
\bibitem{Dominguez:2008vd}
  F.~Dominguez, C.~Marquet, A.~H.~Mueller, B.~Wu and B.~W.~Xiao, ``Comparing energy loss and $p_{\perp}$-broadening in perturbative QCD with
  strong coupling $\mathcal{N}=4$ SYM theory,'' Nucl.\ Phys.\  A {\bf 811}, 197 (2008) [arXiv:0803.3234 [nucl-th]].  
  
  
   
  
\subsubsection*{Chapter 8: The Diffusion Wake}    

\bibitem{Neufeld:2008eg}
  R.~B.~Neufeld, ``Comparing different freeze-out scenarios in azimuthal hadron correlations
  induced by fast partons,'' arXiv:0810.3185 [hep-ph].

\bibitem{Chaudhuri:2006qk}
  A.~K.~Chaudhuri, ``Conical flow due to partonic jets in central Au+Au collisions,''
  Phys.\ Rev.\  C {\bf 75}, 057902 (2007) [arXiv:nucl-th/0610121].
  
\bibitem{Chaudhuri:2007vc}
  A.~K.~Chaudhuri, ``Di-jet hadron pair correlation in a hydrodynamical model with a quenching jet,''
  Phys.\ Rev.\  C {\bf 77}, 027901 (2008) [arXiv:0706.3958 [nucl-th]].
  
\bibitem{Bethe}
  H.~Bethe, ``On the theory of passage of fast particles through matter,'' Annalen der Physik {\bf 397}, 325 (1930).
  
\bibitem{Bragg}
  W.~H.~Bragg and R.~Kleemann, ``On the $\alpha$ particles of radium, and their loss of range in passing through various atoms and 
  molecules'', Philos.\ Mag.\ {\bf 10}, 318 (1905).  
  
\bibitem{Chen} 
  G.~T.~Y.~ Chen, J.~R.~Castro and J.~M.~Quivey, ``Heavy charged particle radiotherapy'', 
  Ann.\ Rev.\ Biohys.\ Bioeng.\ {\bf 10}, 499 (1981).  
  
\bibitem{Sihver} 
  L.~Sihver, D.~Schardt and T.~Kanai, ``Depth dose distributions of high-energy carbon, oxygen and neon beams 
  in water'', Jpn.\ J.\ Med.\ Phys.\ {\bf 18}, 1 (1998).  
  
\bibitem{Kraft} 
  U.~Amaldi and G.~Kraft, ``Radiotherapy with beams of carbon ions'', Rep.\ Prog.\ Phys.\ {\bf 68}, 1861 (2005).  
  
\bibitem{WilsonPRL} 
  R.~R.~Wilson, ``Range, Straggling, and Multiple Scattering of Fast Protons'', Phys.\ Rev.\ {\bf 71}, 385 (1947). 
  
\bibitem{pshenichnov1}
  I.~Pshenichnov, I.~Mishustin and W.~Greiner, ``Neutrons from fragmentation of light nuclei in
  tissue-like media: a study with the GEANT4 toolkit'', Phys.\ Med.\ and Biol.\ {\bf 50}, 5493 (2005).
  
\bibitem{pshenichnov2}  
  I.~Pshenichnov, I.~Mishustin and W. Greiner, ``Comparative study of depth-dose distributions from beams of light and heavy nuclei in
  tissue-like media'', Nucl.\ Inst.\ Meth.\ {\bf B 226}, 1094 (2008).
  
\bibitem{Betz:2008js}
  B.~Betz, M.~Gyulassy, D.~H.~Rischke, H.~Stöcker and G.~Torrieri, ``Jet Propagation and Mach Cones in (3+1)d Ideal Hydrodynamics,''
  J.\ Phys.\ G {\bf 35}, 104106 (2008) [arXiv:0804.4408 [hep-ph]].
  
\bibitem{Fries:2003kq}
  R.~J.~Fries, B.~Müller, C.~Nonaka and S.~A.~Bass, ``Hadron production in heavy ion collisions: Fragmentation and  recombination
  from a dense parton phase,'' Phys.\ Rev.\  C {\bf 68}, 044902 (2003) [arXiv:nucl-th/0306027].  
  
\bibitem{Fries:2004hd}
  R.~J.~Fries, S.~A.~Bass and B.~Müller, ``Correlated emission of hadrons from recombination of correlated  partons,''
  Phys.\ Rev.\ Lett.\  {\bf 94}, 122301 (2005) [arXiv:nucl-th/0407102].  
  
\bibitem{Greco:2003mm}
  V.~Greco, C.~M.~Ko and P.~Levai, ``Parton coalescence at RHIC,'' Phys.\ Rev.\  C {\bf 68}, 034904 (2003) [arXiv:nucl-th/0305024]. 
  
\bibitem{Betz:2007kg}
  B.~Betz, M.~Gyulassy and G.~Torrieri, ``Polarization probes of vorticity in heavy ion collisions,'' 
  Phys.\ Rev.\  C {\bf 76}, 044901 (2007) [arXiv:0708.0035 [nucl-th]].  
  
\bibitem{Gyulassy:1993hr}
  M.~Gyulassy and X.~N.~Wang, ``Multiple collisions and induced gluon Bremsstrahlung in QCD,'' Nucl.\ Phys.\  B {\bf 420}, 583 (1994)
  [arXiv:nucl-th/9306003].  
  
\bibitem{Wang:1994fx}
  X.~N.~Wang, M.~Gyulassy and M.~Plumer, ``The LPM effect in QCD and radiative energy loss in a quark gluon plasma,''
  Phys.\ Rev.\  D {\bf 51}, 3436 (1995) [arXiv:hep-ph/9408344].  
  
\bibitem{Wang:2001ifa}
  X.~N.~Wang and X.~F.~Guo, ``Multiple parton scattering in nuclei: Parton energy loss,'' Nucl.\ Phys.\  A {\bf 696}, 788 (2001).
  [arXiv:hep-ph/0102230].
  
\bibitem{Arnold:2001ms}
  P.~Arnold, G.~D.~Moore and L.~G.~Yaffe, ``Photon emission from quark gluon plasma: Complete leading order  results,''
  JHEP {\bf 0112}, 009 (2001) [arXiv:hep-ph/0111107].

\bibitem{Arnold:2002ja}
  P.~Arnold, G.~D.~Moore and L.~G.~Yaffe, ``Photon and Gluon Emission in Relativistic Plasmas,'' JHEP {\bf 0206}, 030 (2002)
  [arXiv:hep-ph/0204343].
  
\bibitem{Liu:2006ug}
  H.~Liu, K.~Rajagopal and U.~A.~Wiedemann, ``Calculating the jet quenching parameter from AdS/CFT,''
  Phys.\ Rev.\ Lett.\  {\bf 97}, 182301 (2006) [arXiv:hep-ph/0605178].  
  
\bibitem{Majumder:2007zh}
  A.~Majumder, B.~Müller and X.~N.~Wang, ``Small Shear Viscosity of a Quark-Gluon Plasma Implies Strong Jet
  Quenching,'' Phys.\ Rev.\ Lett.\  {\bf 99}, 192301 (2007) [arXiv:hep-ph/0703082].  


\subsubsection*{Chapter 9: Polarization Probes of Vorticity}    


\bibitem{Bunce:1976yb}
  G.~Bunce {\it et al.}, ``Lambda0 Hyperon Polarization In Inclusive Production By 300-Gev Protons On Beryllium,''
  Phys.\ Rev.\ Lett.\  {\bf 36}, 1113 (1976).
  
\bibitem{Hoyer}
  P.~Hoyer, ``Particle Polarization As A Signal Of Plasma Formation,''
  Phys.\ Lett.\ B {\bf 187}, 162 (1987).  
  
\bibitem{Panagiotou:1986zq}
  A.~D.~Panagiotou, ``Lambda0 Nonpolarization: Possible Signature Of Quark Matter,'' Phys.\ Rev.\ C {\bf 33}, 1999 (1986).  
  
\bibitem{Andersson:1979wj}
  B.~Andersson, G.~Gustafson and G.~Ingelman, ``A Semiclassical Model For The Polarization Of Inclusively Produced Lambda 0
  Particles At High-Energies,'' Phys.\ Lett.\  B {\bf 85}, 417 (1979).  
  
\bibitem{Bensinger:1983vc}
  J.~Bensinger {\it et al.}, ``Inclusive Lambda Production And Polarization In 16-Gev/C Pi- P Interactions,''
  Phys.\ Rev.\ Lett.\  {\bf 50}, 313 (1983).

\bibitem{Liang:2004ph}
  Z.~T.~Liang and X.~N.~Wang, ``Globally polarized quark gluon plasma in non-central A + A collisions,''
  Phys.\ Rev.\ Lett.\  {\bf 94}, 102301 (2005) [arXiv:nucl-th/0410079].
  
\bibitem{rafpol}
  M.~Jacob and J.~Rafelski, ``Longitudinal Anti-Lambda Polarization, Anti-Xi Abundance And Quark Gluon
  Plasma Formation,'' Phys.\ Lett.\  B {\bf 190}, 173 (1987).  
  
\bibitem{liangvector}
  Z.~T.~Liang and X.~N.~Wang, ``Spin alignment of vector mesons in non-central A + A collisions,''
  Phys.\ Lett.\  B {\bf 629}, 20 (2005) [arXiv:nucl-th/0411101]. 

\bibitem{Becattini:2007zn}
  F.~Becattini and L.~Ferroni, ``The microcanonical ensemble of the ideal relativistic quantum gas with
  angular momentum conservation,'' Eur.\ Phys.\ J.\  C {\bf 52}, 597 (2007) [arXiv:0707.0793 [nucl-th]].

\bibitem{Liangqm2006}
  Z.~T.~Liang, ``Global polarization of QGP in non-central heavy ion collisions at high energies,''
  J.\ Phys.\ G {\bf 34}, S323 (2007) [arXiv:0705.2852 [nucl-th]].  
  
\bibitem{Bravar:1999rq}
  A.~Bravar  [Spin Muon Collaboration], ``Hadron azimuthal distributions and transverse spin asymmetries in DIS  of
  leptons off transversely polarized targets from SMC,'' Nucl.\ Phys.\ Proc.\ Suppl.\  {\bf 79}, 520 (1999).  
  
\bibitem{Liang:1997rt}
  Z.~F.~Liang and C.~Boros, ``Hyperon polarization and single spin left-right asymmetry in inclusive
  production processes at high energies,'' Phys.\ Rev.\ Lett.\  {\bf 79}, 3608 (1997) [arXiv:hep-ph/9708488].  
  
\bibitem{Abelev:2007zk}
  B.~I.~Abelev {\it et al.}  [STAR Collaboration], ``Global polarization measurement in Au+Au collisions,''
  Phys.\ Rev.\  C {\bf 76}, 024915 (2007) [arXiv:0705.1691 [nucl-ex]].
  
\bibitem{jacob}
  M.~Jacob and G.~C.~Wick, ``On the general theory of collisions for particles with spin,''
  Annals Phys.\  {\bf 7}, 404 (1959).
  
\bibitem{Milov:2004sv}
  A.~Milov  [PHENIX Collaboration], ``Centrality and s(NN)**(1/2) dependence of the dE(T)/d eta and dN(ch)/d  eta
  in heavy ion collisions at mid-rapidity,'' J.\ Phys.\ Conf.\ Ser.\  {\bf 5}, 17 (2005) [arXiv:nucl-ex/0409023].  
  
\bibitem{bgk}
  S.~J.~Brodsky, J.~F.~Gunion and J.~H.~Kuhn, ``Hadron Production In Nuclear Collisions: A New Parton Model Approach,''
  Phys.\ Rev.\ Lett.\  {\bf 39}, 1120 (1977).

\bibitem{Adil:2005qn}
  A.~Adil and M.~Gyulassy, ``3D jet tomography of twisted strongly coupled quark gluon plasmas,''
  Phys.\ Rev.\  C {\bf 72}, 034907 (2005) [arXiv:nucl-th/0505004].
  
\bibitem{Busza:2004mc}
  W.~Busza, ``Structure and fine structure in multiparticle production data at high energies,''
  Acta Phys.\ Polon.\  B {\bf 35}, 2873 (2004) [arXiv:nucl-ex/0410035].  

\bibitem{Torrieri:2007qy}
  G.~Torrieri, ``Scaling of v(2) in heavy ion collisions,'' Phys.\ Rev.\  C {\bf 76}, 024903 (2007)
  [arXiv:nucl-th/0702013].

\bibitem{Taub}
  A.~H.~Taub, ``On circulation in relativistic hydrodynamics,'' Arch.\ Rational Mech.\ Anal.\ {\bf 3}, 312 (1959).
  
\bibitem{Florkowski:1992yd}
  W.~Florkowski, B.~L.~Friman, G.~Baym and P.~V.~Ruuskanen,
  ``Convective stability of hot matter in ultrarelativistic heavy ion collisions,''
  Nucl.\ Phys.\  A {\bf 540}, 659 (1992).  
  
\bibitem{Abreu:2007kv}
  N.~Armesto {\it et al.}, ``Heavy Ion Collisions at the LHC - Last Call for Predictions,''
  J.\ Phys.\ G {\bf 35}, 054001 (2008) [arXiv:0711.0974 [hep-ph]].  
  
\bibitem{Barros:2005cy}
  C.~d.~C.~Barros and Y.~Hama, ``Antihyperon polarization in high-energy inclusive reactions,''
  Int.\ J.\ Mod.\ Phys.\  E {\bf 17}, 371 (2008) [arXiv:hep-ph/0507013].  
  
\bibitem{Abelev:2007qg}
  B.~I.~Abelev {\it et al.}  [the STAR Collaboration],
  ``Mass, quark-number, and s(NN)**(1/2) dependence of the second and fourth flow harmonics in ultra-relativistic nucleus nucleus collisions,''
  Phys.\ Rev.\  C {\bf 75}, 054906 (2007) [arXiv:nucl-ex/0701010].  
  
\bibitem{Baran:2003nm}
  A.~Baran, W.~Broniowski and W.~Florkowski, ``Description of the particle ratios and transverse-momentum spectra for
  various centralities at RHIC in a single-freeze-out model,'' Acta Phys.\ Polon.\  B {\bf 35}, 779 (2004)
  [arXiv:nucl-th/0305075].  
  
\bibitem{Torrieri:2000xi}
  G.~Torrieri and J.~Rafelski, ``Search for QGP and thermal freeze-out of strange hadrons,''
  New J.\ Phys.\  {\bf 3}, 12 (2001) [arXiv:hep-ph/0012102].  


\subsubsection*{Chapter 10: Di-Jet Correlations in pQCD vs.\ AdS/CFT}

\bibitem{Gubser:2007ni}
  S.~S.~Gubser and A.~Yarom, ``Universality of the diffusion wake in the gauge-string duality,'' Phys.\ Rev.\  D {\bf 77}, 066007 (2008)
  [arXiv:0709.1089 [hep-th]].

\bibitem{Gubser:2008vz}
  S.~S.~Gubser and A.~Yarom, ``Linearized hydrodynamics from probe-sources in the gauge-string duality,''
  Nucl.\ Phys.\  B {\bf 813}, 188 (2009) [arXiv:0803.0081 [hep-th]].
  
\bibitem{Heinz:1985qe}
  U.~W.~Heinz, ``Quark - Gluon Transport Theory. Part 2. Color Response And Color Correlations In A Quark - Gluon Plasma,''
  Annals Phys.\  {\bf 168}, 148 (1986).  
  
\bibitem{Mueller:2008zt}
  A.~H.~Mueller, ``Separating hard and soft scales in hard processes in a QCD plasma,'' Phys.\ Lett.\  B {\bf 668}, 11 (2008)
  [arXiv:0805.3140 [hep-ph]].  
      
\bibitem{Selikhov:1993ns}
  A.~Selikhov and M.~Gyulassy, ``Color diffusion and conductivity in a quark - gluon plasma,''
  Phys.\ Lett.\  B {\bf 316}, 373 (1993) [arXiv:nucl-th/9307007].      
  
\bibitem{Selikhov:1994xn}
  A.~V.~Selikhov and M.~Gyulassy, ``QCD Fokker-Planck equations with color diffusion,''
  Phys.\ Rev.\  C {\bf 49}, 1726 (1994).  
  
\bibitem{Eskola:1992bd}
  K.~J.~Eskola and M.~Gyulassy, ``Color conductivity and the evolution of the mini - jet plasma at RHIC,''
  Phys.\ Rev.\  C {\bf 47}, 2329 (1993).  
  
\bibitem{Gubser:2006qh}
  S.~S.~Gubser, ``Comparing the drag force on heavy quarks in N = 4 super-Yang-Mills theory and QCD,''
  Phys.\ Rev.\  D {\bf 76}, 126003 (2007)
  [arXiv:hep-th/0611272].  
  
\bibitem{Fleischer:1984dj}
  W.~Fleischer and G.~Soff, ``Bound State Solutions Of The Klein-Gordon Equation For Strong Potentials,''
  Z.\ Naturforsch.\  {\bf 39A}, 703 (1984).

   
\subsubsection*{Chapter 11: Conical Correlations in an Expanding Medium}      

\bibitem{Tomasik:2008dy}
  B.~Tomasik, ``The contribution of hard processes to elliptic flow,''
  arXiv:0812.4071 [nucl-th].
  
\bibitem{Schnedermann:1993ws}
  E.~Schnedermann, J.~Sollfrank and U.~W.~Heinz, ``Thermal phenomenology of hadrons from 200-A/GeV S+S collisions,''
  Phys.\ Rev.\  C {\bf 48}, 2462 (1993) [arXiv:nucl-th/9307020].
  
\bibitem{STAR_FullJet}  
  J.~Putschke, ``Full Jet-Reconstruction in STAR,'' Talk at the RHIC \& AGS Annual Users' Meeting 2009, Brookhaven
  National Laboratory, USA (2009).  
  
\bibitem{PHENIX_FullJet}    
  Y.~S.~Lai, ``Direct jet reconstruction in p+p and Cu+Cu with PHENIX,'' Talk at the RHIC \& AGS Annual Users' 
  Meeting 2009, Brookhaven National Laboratory, USA (2009).  
  
\subsubsection*{Appendices}

\bibitem{DiplomFochler}
  O.~Fochler, ``Energy loss of high-$p_T$ partons in transport simulations of heavy-ion collisions,''
  Diploma Thesis, University Frankfurt, Germany (2006).  
  
\bibitem{Miller:2007ri}
  M.~L.~Miller, K.~Reygers, S.~J.~Sanders and P.~Steinberg, ``Glauber modeling in high energy nuclear collisions,''
  Ann.\ Rev.\ Nucl.\ Part.\ Sci.\  {\bf 57}, 205 (2007) [arXiv:nucl-ex/0701025].  

\bibitem{LeBellac}
  M.~Le Bellac, ``Thermal Field Theory,'' Cambridge Monographs on Mathematical Physics, Cambridge (1996).  
  
\bibitem{DissDirk}
  D.~H.~Rischke, ``Ultrarelativistische Schwerionenphysik - Untersuchungen zur Zustandsgleichung
  heißer und dichter Kernmaterie,'' PhD Thesis, University Frankfurt, Germany (1992).
  
\bibitem{Tsumura:2007ji}
  T.~Tsumura, T.~Kunihiro and K.~Ohnishi, ``Derivation Of Covariant Dissipative Fluid Dynamics In The
  Renormalization-Group Method,'' Phys.\ Lett.\  B {\bf 646}, 134 (2007).    
  
  
\bibitem{GradshteynRyzhik}
  I.~S.~Gradshteyn and I.~W.~Ryzhik, ``Table of Integrals Series and Products,'' Adademic Press Inc.\, New York San Francisco London 
  (1965).  
  
\bibitem{Gatoff:1987uf}
  G.~Gatoff, A.~K.~Kerman and T.~Matsui, ``The Flux Tube Model For Ultrarelativistic Heavy Ion Collisions:
  Electrohydrodynamics Of A Quark Gluon Plasma,'' Phys.\ Rev.\  D {\bf 36}, 114 (1987).
  
  
  
\end{footnotesize}
\end{thebibliography}
\clearpage{\pagestyle{empty}\cleardoublepage}
%
%
%
%
%
%
%
%
%
%
\markboth{Danksagung}{}
\addcontentsline{toc}{chapter}{Acknowledgements}
\thispagestyle{empty}
%
%
%

\chapter*{Acknowledgements}
\vspace*{1\baselineskip}

I would like to thank all those who helped and supported me 
while I did my PhD.\\
First of all, I thank my supervisor Prof.\ Dr.\ Dirk Rischke for his advice, help,
and continuous interest in my work, but also for giving me freedom which
is essential in research.\\
Likewise, I thank Prof.\ Dr.\ Miklos Gyulassy for the great collaboration, 
the enlightning discussions, his many new ideas, and his hospitality during my
stays at Columbia University.\\
Moreover, I thank Prof.\ Dr.\ Horst Stöcker, Prof.\ Dr.\ Carsten Greiner, and
Prof.\ Dr.\ Peter Braun-Munzinger for their interest and strong support of my work
as well as Prof.\ Dr.\ Igor Mishustin and Prof.\ Dr.\ Laszlo Csernai for helpful discussions.\\
Furthermore, I thank Dr.\ Jorge Noronha and Dr.\ Giorgio Torrieri for the stimulating,
inspiring, and extremely fruitful collaboration. I especially thank Dr.\ Jorge Noronha
for being always reachable and for his advice in many different situations.\\
I also thank Daniel Henkel and Andrej El for many useful discussions about
viscous hydrodynamics.\\
A special thank goes to my office mate Dr.\ Joachim Reinhardt for his companionship
during the last 6 years since I started my diploma thesis and for his help in many different 
issues. There has always been a great working environment and his dry humor often helped
to lighten up the atmosphere.\\
I thank Astrid Steidl, Denise Meixler, and Gabriela Meyer for the distractions during 
lunchtime as well as Daniela Radulescu and Veronika Palade for their assistance. \\
I'm especially thankful for the help of Astrid Steidl with many different 
plots and sketches.\\
In addition, I thank the Helmholtz Research School H-QM for financial support, but
especially for bringing together a group of experimental and theoretical physicists,
stimulating discussions, and offering advanced training. In particular, I thank 
Dr.\ Henner Büsching for his continuous enthusiasm, the excellent organization of
the many different events, and for creating a team spirit.\\
I especially thank Prof.\ Dr.\ Dirk Rischke, Prof.\ Dr.\ Peter Braun-Munzinger,
and Prof.\ Dr.\ Miklos Gyulassy for their contribution in my H-QM PhD-Committee.\\
I also thank Prof.\ Dr.\ Azwinndini Muronga, Michael Hauer, Prof.\ Dr.\ Charles Gale, and
Dr.\ Björn Schenke for their hospitality during my stays at Cape Town and Montreal.\\
Besides that I thank the computer administration for the help with all the
technical problems, Dr.\ Stefan Scherer for providing me with the layout for this
thesis, and Oliver Fochler, Mauricio Martinez, Irina Sagert, and Sascha Vogel for
some helpful discussions and explanations.\\
Last but not least I thank my parents for their continuous encouragement 
and for their support that made my studies possible.\\
This work was supported by the Bundesministerium für Bildung und Forschung BMBF.

\vspace*{3\baselineskip}

\noindent
Frankfurt am Main, July 2009 \hfill Barbara Betz

\clearpage{\pagestyle{empty}\cleardoublepage}
%
%
%
%
%
%
%
\end{document}